\documentclass[final,5p,times,twocolumn]{elsarticle}

\usepackage{graphicx}
\usepackage{color}
\usepackage{epstopdf}
\usepackage{amsmath}
\usepackage{gensymb}
\usepackage{morefloats}

\usepackage{tikz}
\usepackage{verbatim}

\usepackage{float}
\usepackage{rotfloat}
\usepackage{textcomp}
\usepackage{amstext}
\usepackage{graphicx}

\usepackage{siunitx}

\makeatletter
\DeclareFontEncoding{LGR}{}{}

\ProvideTextCommand{\~}{LGR}[1]{\char126#1}
\providecommand{\tabularnewline}{\\}
\makeatother
\usepackage{graphicx}
\usepackage{amssymb}
\usepackage{xspace}

\newcommand{\heT}{\ensuremath{^{3}\text{He}}\xspace}

\newcommand{\gF}{\text{Geant4}\xspace}

\usepackage{multirow}
\usepackage{subfigure} %% isn't this deprecated? r.de sangro
\usepackage[percent]{overpic}
\usepackage{caption} 

\journal{Nucl. Instrum. Methods Phys. Res., Sect. A}

\newcommand\Tp{\rule{0pt}{2.6ex}}	% Top strut 
\newcommand\Bt{\rule[-1.2ex]{0pt}{0pt}}	% Bottom strut
\usepackage{array}
\makeatletter
\newcommand{\thickhline}{%
  \noalign {\ifnum 0=`}\fi \hrule height 1pt
  \futurelet \reserved@a \@xhline
}
\newcolumntype{"}{@{\hskip\tabcolsep\vrule width 1pt\hskip\tabcolsep}}
\makeatother

\usepackage{parskip}
\usepackage{booktabs}
\usepackage[backref=page,         % Add page links to..
            pagebackref=false,    % ... not from reference
            hyperindex=true,      % ... in index if present
            colorlinks=true,      %
            citecolor=blue,       %
            filecolor=black,      %
            linkcolor=blue,       %
            urlcolor=blue,        %
            breaklinks=true,      %
            bookmarks=true,       % create bookmarks in pdf for better navigation
            bookmarksopen=false,  %  ... but display only if we want to.
            pdftitle   = {First Measurements of Beam Backgrounds at the SuperKEKB Electron-Positron Collider},
            pdfauthor  = {The BEAST2 group},
            pdfsubject = {The subject}
]{hyperref}

\usetikzlibrary{calc,trees,positioning,arrows,chains,shapes.geometric,%
    decorations.pathreplacing,decorations.pathmorphing,shapes,%
    matrix,shapes.symbols}

\tikzset{
>=stealth',
  punktchain/.style={
    rectangle, 
    rounded corners, 
    draw=black, very thick,
    text width=10em, 
    minimum height=3em, 
    text centered, 
    on chain},
  line/.style={draw, thick, <-},
  element/.style={
    tape,
    top color=white,
    bottom color=blue!50!black!60!,
    minimum width=8em,
    draw=blue!40!black!90, very thick,
    text width=10em, 
    minimum height=3.5em, 
    text centered, 
    on chain},
  every join/.style={->, thick,shorten >=1pt},
  decoration={brace},
  tuborg/.style={decorate},
  tubnode/.style={midway, right=2pt},
}

\def\GEANT {Geant4} % Easy change between Geant4, GEANT4, and small caps GEANT4.
\begin{document}
\begin{frontmatter}


\title{First Measurements of Beam Backgrounds at SuperKEKB}
\author[add3]{P.~M.~Lewis}
\author[add2]{I.~Jaegle} 
\author[add17]{H.~Nakayama}
\author[add11]{A.~Aloisio}
\author[add13]{F.~Ameli}
\author[add4]{M.~Barrett}
\author[add1]{A.~Beaulieu}
\author[add7]{L.~Bosisio}
\author[add14]{P.~Branchini}
\author[add3]{T.~E.~Browder}
\author[add14]{A.~Budano} 
\author[add8]{G.~Cautero}
\author[add12]{C.~Cecchi}
\author[add6]{Y.-T.~Chen}
\author[add6]{K.-N.~Chu}
\author[add4]{D.~Cinabro}
\author[add7]{P.~Cristaudo}
\author[add1]{S.~de~Jong}
\author[add10]{R.~de~Sangro}
\author[add10]{G.~Finocchiaro}
\author[add18]{J.~Flanagan}
\author[add18]{Y.~Funakoshi}
\author[add5]{M.~Gabriel}
\author[add11]{R.~Giordano}
\author[add8]{D.~Giuressi}
\author[add3]{M.~T.~Hedges}
\author[add1]{N.~Honkanen}
\author[add18]{H.~Ikeda}
\author[add18]{T.~Ishibashi}
\author[add18]{H.~Kaji}
\author[add18]{K.~Kanazawa}
\author[add5]{C.~Kiesling}
\author[add6]{S.~Koirala}
\author[add23]{P.~Kri\v{z}an}
\author[add7]{C.~La~Licata}
\author[add7]{L.~Lanceri}
\author[add6]{J.-J.~Liau}
\author[add6]{F.-H.~Lin}
\author[add6]{J.-C.~Lin}
\author[add3]{Z.~Liptak}
\author[add1]{S.~Longo}
\author[add12]{E.~Manoni}
\author[add24]{C.~Marinas}
\author[add21]{K.~Miyabayashi}
\author[add22]{E.~Mulyani}
\author[add18]{A.~Morita}
\author[add17]{M.~Nakao}
\author[add4]{M.~Nayak}
\author[add18]{Y.~Ohnishi}
\author[add14]{A.~Passeri}
\author[add1]{P.~Poffenberger}
\author[add20]{M.~Ritzert}
\author[add1]{J.~M.~Roney}
\author[add12]{A.~Rossi}
\author[add5]{T.~R\"oder}
\author[add19]{R.~M.~Seddon}
\author[add3]{I.~S.~Seong}
\author[add6]{J.-G.~Shiu}
\author[add5]{F.~Simon}
\author[add9]{Y.~Soloviev}
\author[add18]{Y.~Suetsugu}
\author[add5]{M.~Szalay}
\author[add18]{S.~Terui}
\author[add11]{G.~Tortone}
\author[add3]{S.~E.~Vahsen\corref{cor1}}
\cortext[cor1]{Corresponding author. Tel.: +1 808 956 2985.}
\ead{sevahsen@hawaii.edu}
%\author[add3]{S.~E.~Vahsen}
\author[add5]{N.~van~der~Kolk}
\author[add7]{L.~Vitale}
\author[add6]{M.-Z.~Wang}
\author[add5]{H.~Windel}
\author[add21]{S.~Yokoyama}

\address[add24]{University of Bonn, Institute of Physics, Nu{\ss}allee 12, 53115 Bonn, Germany}
\address[add9]{Deutsches Elektronen-Synchrotron, Notkestra{\ss}e 85, 22607 Hamburg, Germany}
\address[add8] {Elettra - Sincrotrone Trieste S.C.p.A., AREA Science Park, 34149 Basovizza, Trieste, Italy}
\address[add2]{University of Florida, Department of Physics, P.O. Box 118440, Gainesville, FL 32611, USA}
\address[add22]{The Graduate University for Advanced Studies (SOKENDAI), 1-1 Oho Tsukuba Ibaraki 305-0801, Japan}
\address[add3]{University of Hawaii, Department of Physics and Astronomy, 2505 Correa Road, Honolulu, HI 96822, USA}
\address[add20]{Heidelberg University, Institute of Computer Engineering, B6, 26, 68159, Mannheim, Germany}
\address[add17]{High Energy Accelerator Research Organization (KEK), Institute of Particle and Nuclear Studies, Oho 1-1, Tsukuba, Ibaraki, 305-0801, Japan}
\address[add18]{High Energy Accelerator Research Organization (KEK), Accelerator Laboratory, Oho 1-1, Tsukuba, Ibaraki, 305-0801, Japan}
\address[add12]{INFN - Sez. di Perugia, Via A. Pascoli, 06123, Perugia, Italy}
\address[add13]{INFN - Sez. ROMA, P.le Aldo Moro, 2 00185, Roma, Italy}
\address[add14]{INFN - Sez. ROMA 3, V. della Vasca Navale, 84, 00146 Roma, Italy}
\address[add23]{J. Stefan Institute, Faculty of Mathematics and Physics, University of Ljubljana, 1000 Ljubljana, Slovenia}
\address[add10]{Laboratori Nazionali di Frascati dell'INFN, Via E. Fermi 40, I-00044, Frascati, Italy}
\address[add5]{Max-Planck-Institut f\"ur Physik, F\"ohringer Ring 6, 80805 M\"unchen, Germany}
\address[add19]{McGill University, Department of Physics, 3600 rue University, Montr\'{e}al, QC  H3A 2T8, Canada}
\address[add11]{Univ. of Naples Federico II \& INFN Sezione di Napoli, Strada Comunale Cintia, 80126 Napoli, Italy}
\address[add21]{Nara Women's University, Nara 630-8506, Japan}
\address[add6]{National Taiwan University, Department of Physics, No.1 Sec.4 Roosevelt Road Taipei 10617, Taiwan}
\address[add7]{University of Trieste, Department of Physics, and INFN, Via Valerio 2, 34127 Trieste, Italy}
\address[add1]{University of Victoria, Department of Physics and Astronomy, 3800 Finnerty Rd., Victoria BC, V8P 5C2, Canada}
\address[add4]{Wayne State University, Department of Physics and Astronomy, 666 W. Hancock, Detroit, MI 48202, USA}

% BEAST logo
%\center{\includegraphics[width=2cm]{img/BeastLogo}}

%A concise and factual abstract is required. The abstract should state briefly the purpose of the research, the principal results and major conclusions. An abstract is often presented separately from the article, so it must be able to stand alone. For this reason, References should be avoided, but if essential, then cite the author(s) and year(s). Also, non-standard or uncommon abbreviations should be avoided, but if essential they must be defined at their first mention in the abstract itself.

\begin{abstract}
The high design luminosity of the SuperKEKB electron-positron collider is expected to result in challenging levels of beam-induced backgrounds in the interaction region. Properly simulating and mitigating these backgrounds is critical to the success of the Belle~II experiment. We report on measurements performed with a suite of dedicated beam background detectors, collectively known as BEAST~II, during the so-called Phase 1 commissioning run of SuperKEKB in 2016, which involved operation of both the high energy ring (HER) of 7~GeV electrons as well as the low energy ring (LER) of 4~GeV positrons. We describe the BEAST~II detector systems, the simulation of beam backgrounds, and the measurements performed. The measurements include standard ones of dose rates versus accelerator conditions, and more novel investigations, such as bunch-by-bunch measurements of injection backgrounds and measurements sensitive to the energy spectrum and angular distribution of fast neutrons. We observe beam-gas, Touschek, beam-dust, and injection backgrounds. As there is no final focus of the beams in Phase 1, we do not observe significant synchrotron radiation, as expected. Measured LER beam-gas backgrounds and Touschek backgrounds in both rings are slightly elevated, on average three times larger than the levels predicted by simulation. HER beam-gas backgrounds are on on average two orders of magnitude larger than predicted. Systematic uncertainties and channel-to-channel variations are large, so that these excesses constitute only 1-2 sigma level effects. Neutron background rates are higher than predicted and should be studied further. We will measure the remaining beam background processes, due to colliding beams, in the imminent commissioning Phase 2. These backgrounds are expected to be the most critical for Belle~II, to the point of necessitating replacement of detector components during the Phase~3 (full-luminosity) operation of SuperKEB.\end{abstract}

\begin{keyword}
\end{keyword}
\end{frontmatter}

%\clearpage
\tableofcontents
%\clearpage

%%\linenumbers

% lead authors: Sven Vahsen, Tom Browder, Hiroyuki Nakayama
 \section{Introduction}
 % file: 		introduction.tex
% authors: 	Vahsen, Browder, Nakayama
%
% contents: introduction to paper
%  	part 1 (Vahsen, Browder): goals of BEAST (Vahsen, Browder)
%  	part 2 (Nakayama): definition of phase 1, how it fits into over Belle II 
%                 and SuperKEKB projects, difference between phases 1,2,3

% file: 	intro_part1.tex
% authors: 	Vahsen, Nakayama, Browder

The SuperKEKB asymmetric-energy electron-positron collider \cite{PTEP:Ohnish} is currently under construction in Tsukuba, Japan. SuperKEKB will produce collisions for the Belle II experiment \cite{Abe:2010gxa}, at an ambitious design luminosity of $8 \times 10^{35}~{\rm cm^{-2}s^{-1}}$. This will be achieved by utilizing the ``nano-beam scheme" proposed by P. Raimondi, where the beams are squeezed to a vertical size of 50~nm at the interaction point \cite{Bona:2007qt}, and by doubling the beam currents with respect to KEKB \cite{KEK:1995sta}. The SuperKEKB full design luminosity currents are 3.2~A for the 7.0~GeV electrons in the high energy (electron) ring (HER) and 2.6~A for the 4.0~GeV positrons in the low energy (positron) ring (LER).

The squeezed beams, higher beam currents, and increased luminosity each increase the rate of background particles generated by the accelerator, which we will collectively refer to as ``beam backgrounds". Beam backgrounds include undesirable particles generated both by single-beam processes such as synchrotron radiation, Touschek scattering, beam-gas scattering, beam-dust scattering, and injection backgrounds as well as by luminosity-dependent processes, such as Bhabha scattering and multi-lepton two-photon events.

Successful operation of SuperKEKB and Belle~II depends critically on limiting and mitigating such beam backgrounds \cite{Nakayama2015}. For example, beam-gas and Touschek scattering cause beam particles to depart from their nominal orbits, causing them to collide with the accelerator beampipe downstream of the scattering location. Beam particles lost in this way reduce the beam current and limit the beam lifetime of the accelerator. Electromagnetic showers resulting from collisions of the lost particles with the beampipe wall also produce secondary particles. Showers produced in or near the interaction region can increase the Belle~II occupancy, and create a challenging event environment for the reconstruction software. Secondaries also increase the neutron fluence and ionizing-radiation dose deposited in detector materials and electronics, shortening their lifetime, increasing their dead time, and increasing the rate of single-event upsets (SEUs). With the present predictions of beam backgrounds from simulations for full design luminosity already showing that several Belle II detector systems are limited in their lifetimes and performance expectations, it is imperative to understand as early as possible the accuracy of these predictions using experiment.

The commissioning of SuperKEKB is performed in three phases. In Phase 1, the machine runs without the final focusing system and without the Belle II detector. Because the beams are unfocused at the interaction point (IP), no collisions occur. From the SuperKEKB perspective, the main goals of this stage are to perform sufficient vacuum scrubbing before Belle II is rolled in, and to carry out basic machine studies such as low-emittance optimization of the optics and feedback system tuning. In Phase 1, we use sixteen beam collimators recycled from KEKB \cite{Suetsugu:2003gb} in the HER and two newly developed collimators \cite{Ishibashi:2017wiw} in the LER. Instead of Belle~II, the BEAST~II beam background detectors are installed in the interaction region.

In Phase 2, the SuperKEKB final focusing system and positron damping ring will be in place, and the Belle II detector, except for the vertex detector, will be rolled in. Instead of the vertex detector, we will deploy another dedicated beam background detector system, which will be described in a future article. With focused beams, SuperKEKB will operate in collision mode for the first time with a Phase 2 target luminosity of $2 \times 10^{34}~{\rm cm^{-2}s^{-1}}$. Before starting Phase 2, we will install three (two horizontal, one vertical) collimators in each ring, near the interaction region. These additional collimators are crucial to protect the Belle II detector at the interaction point.

Phase 3 denotes the final stage, when full luminosity data taking with the complete Belle II detector will occur. Before Phase 3 we need to install additional collimators in each ring to mitigate the increased Touschek background that will result from focusing the beams in the nano-beam scheme.

Table \ref{Table:MachineParamAllPhases} shows the nominal machine parameters during each of these three commissioning stages, as well as the parameters of KEKB, for comparison.

\begin{table*}
\caption{Machine parameters achieved during KEKB and SuperKEKB Phase 1 operation, and nominal machine parameters for SuperKEKB Phases 2 and 3.}\label{Table:MachineParamAllPhases}
\begin{center}
\begin{tabular}{lcccccccc}\toprule
                                                        & \multicolumn{2}{c}{KEKB} &  \multicolumn{2}{c}{Phase 1} & \multicolumn{2}{c}{Phase 2}&\multicolumn{2}{c}{Phase 3} \\
  Ring                                      		& LER    	& HER  	& LER    	& HER  	& LER	& HER	& LER	& HER  \\ 
  \midrule 
  Beam current $I$~[A]      		& 1.64	& 1.19	 & 1.01   	& 0.87   	 & 1.0	& 0.8		& 3.6		& 2.6 \\
  Number of bunches $N_b$        	& 1584	& 1584	 & 1576	& 1576   	 & 1576 	& 1576 	& 2500	& 2500 \\
 % Bunch current $I_b$~[mA]           & 1.0    		& 1.0  	& 0.32	& 0.40	& 1.04	& 1.44 	\\
  Vertical beam size $\sigma_y$~[$\mu \rm m$]  
  							& 0.94	& 0.94	& 110   	& 59 		& 0.25	& 0.39	& 0.048 & 0.062 \\
  %Emittance ratio $\varepsilon_y/\varepsilon_x$ 										& 		& 		& 0.1 	&  0.1	&		&					&\\  
  %Pressure $P$~[nTorr]                 	& 		& 		& 10    	&  10 	& 			& 			& 1.0		& 1.0	\\
  Number of collimators			 & 16		& 16		& 2	  	&   16 	& 	5	&  19 	& 	11 	& 22  \\
  Luminosity~[$10^{34}~{\rm cm}^{-2}{\rm s}^{-1}$]		
  							& \multicolumn{2}{c}{2.1} & \multicolumn{2}{c}{0} & \multicolumn{2}{c}{$2$} & \multicolumn{2}{c}{$80$} \\
  \bottomrule
\end{tabular}
\end{center}
\end{table*}

We report here on measurements performed with the BEAST~II beam background detectors during the Phase 1 operation of SuperKEKB. The main goals of these measurements are to verify that the beam background levels are safe for installation of Belle~II, to provide feedback to the accelerator team on how accelerator parameters affect background conditions at the interaction point, and to validate the beam background simulation. The latter is critical to assess the accuracy of the background models used in determining detector lifetime estimates. Because these lifetime estimates depend on background conditions determined by the beam parameters at full beam currents and full luminosity it is insufficient to compare the total measured background level with simulation. Instead, our goal is to separate and measure the individual components of the total beam background, and to validate that the scaling of each component process with beam parameters is accurately simulated. In this way we determine to which extent the extrapolation of beam background particle rates to full luminosity can be trusted. With focused beams, a significant component of the backgrounds arise from the electron-positron collisions, such as those produced by radiative Bhabha scattering. We use the Phase 1 data to cleanly study backgrounds from Touschek, beam-gas and other sources not associated with collisions.

The remainder of this document is structured as follows: in Section \ref{sec:backgroundsources} we introduce the different beam background sources at SuperKEKB. In Section \ref{sec:beast}
we describe the BEAST~II detectors, their performance, and calibration. The simulation of beam backgrounds is discussed in Section \ref{sec:simulation}. Sections \ref{sec:beamgas_touschek} through \ref{sec:dosimetry} describe a comprehensive program of background measurements, and in most cases a comparison of the results with simulation. The key findings from our measurements are summarized and discussed in Section \ref{sec:findings}, and the resulting implications for Belle II and predicted background rates for Phase 3 are presented in Section \ref{implications}. Some concluding remarks are presented in Section \ref{conclusions}.

% FIXME: create proper references for section numbers quoted here

 % lead author: Hiro Nakayama, w/ language revision by Sven Vahsen
 \section{Beam background sources at SuperKEKB}\label{sec:backgroundsources}
 % original by Hiroyuki Nakayama
% revised for language by Sven Vahsen 
% revised for Tom's corrections by Sven

We begin by giving an overview of the five main beam background sources at SuperKEKB.
We include luminosity-dependent backgrounds such as radiative Bhabha scattering 
and production of two-photon events, which will be important backgrounds during commissioning Phases~2 
and 3, but are absent in Phase~1 because there are no collisions.

\subsection{Touschek scattering}
The first background source is the Touschek effect, which is an intra-bunch scattering process where  
Coulomb scattering of two particles in the same beam bunch changes the particles' energies to deviate from the nominal energy of the bunch. 
One particle ends up with an energy higher than nominal, the other with lower energy than nominal~\cite{Piwinski:1998qs}. The Touschek effect is enhanced at SuperKEKB due to the nano-beam scheme and associated small beam sizes.

The Touschek scattering probability is calculated using Bruck's formula, as described in~\cite{PTEP:Ohnish}. 
The total scattering rate, integrated around the ring, is proportional to the number of filled bunches 
and the second power of the bunch current, and is inversely proportional to the beam size and the third 
power of the beam energy. Simple scaling based on beam size and beam energies predicts that the Touschek 
background at SuperKEKB will be a factor of $\sim$20 higher than at KEKB.

Touschek-scattered particles are subsequently lost at the beampipe inner wall 
after they propagate further around the ring. If the loss 
position is close to the interaction point, the resulting shower might reach the detector.  
To mitigate Touschek background, we utilize horizontal and vertical movable collimators and metal shields.
The collimators, located at various positions around the ring, stop particles that deviate from their nominal 
trajectories and prevent them from reaching Belle II. 
While we had horizontal collimation (in the plane of the rings) only from the 
inner side of the rings at KEKB, Touschek background can be reduced effectively by collimating the beam 
horizontally from both the inner and outer sides. The horizontal collimators located just before the interaction region 
play an important role in minimizing the beam loss rate inside the detector. The nearest LER collimator is 
only 18~m upstream of the interaction point. In Phase~2 and 3, there will also be heavy-metal shields in the vertex detector (VXD) volume
and on the superconducting final focus cryostat, to prevent shower particles from entering the Belle II acceptance region.

%Fake hits generated by the background shower particles deteriorate the detector's physics resolution.
%Radiation dose by gammas or neutrons in the background shower damage the Silicon devices used in the detector.

%The vertical collimator in LER, which is originally installed to reduce the beam-gas Coulomb background explained in the next subsection, 
%also stops the vertically oscillating Touschek scattered particles. 
%Particles scattered in Fuji-area, which is opposite side of IP in the ring, where LER beam orbit is vertically bending to pass under the HER ring.

%The particle loss with various momentum deviations due to the Touschek effect can be evaluated by 
%particle-tracking simulations along each location in the whole ring. 

\subsection{Beam-gas scattering}
The second beam background source is beam-gas scattering, i.e.\ scattering of beam particles by residual gas molecules in
the beampipe. This can occur via two processes: Coulomb scattering, which changes the direction of the beam particle, and bremsstrahlung
scattering, which reduces the energy of the beam particle. The rate of beam-gas scattering is proportional to the beam current and
to the vacuum pressure in the beampipe. At SuperKEKB, the beam currents will be approximately two times higher than
at KEKB, while the vacuum level of 1~nTorr, except for the interaction region, will be similar to that at KEKB. 

The rate of beam-gas bremsstrahlung losses in the detector is effectively suppressed by horizontal collimators 
and is negligible compared to the Touschek loss rate in the detector.
However, the beam-gas Coulomb scattering rate is expected~\cite{BeamGas} to be a factor of $\sim100$ higher than at KEKB,
because the beampipe radius has been reduced from 1.5~cm inside Belle to 1.0~cm inside Belle II, and the maximum vertical beta function is larger.
Beam-gas scattered particles are lost by hitting the beampipe inner wall
while they propagate around the ring, just like Touschek-scattered particles.

The countermeasures used for Touschek background, movable collimators and 
heavy-metal shields, are also expected to effectively reduce the beam-gas background.
In particular, vertical collimators are essential for reducing Coulomb scattering backgrounds. 
However, potential Transverse Mode Coupling (TMC) instabilities caused by vertical collimators 
should be carefully examined, since the vertical beta function is larger than the horizontal beta function. 
Therefore, the collimator width must satisfy two conditions at the same time: first, it must be narrow enough to avoid beam losses in the detector and second, it must be wide enough to avoid TMC instabilities. The only way to satisfy these two conditions is to use vertical collimators with $\sim 2$~mm width in locations at which the vertical 
beta function is relatively small. This is different from horizontal collimators, which are installed where 
the horizontal beta function is large. Further discussion of this can be found in~\cite{BeamGas}.

%The Beam-gas Coulomb scattering probability for a given scattering angle is calculated as shown in~\cite{PTEP}. 

\subsection{Synchrotron radiation}
The third background source is synchrotron radiation (SR) emitted from the beam.
Since SR power is proportional to the beam energy squared and magnetic field strength squared, 
the HER beam is the main source of this type of background.
The energy spectrum of SR photons ranges from a few keV to tens of keV. 

We are particularly careful about this background source because during early running of the KEKB, 
the inner layer of the Belle Silicon Vertex Detector was severely damaged by X-rays with E $\sim$ 2 keV from the HER. 
To mitigate a possible damaging impact of SR, the shape of beampipe in the interaction region 
is designed to avoid direct SR hits at the detector: ridge structures on the inner surface of incoming pipes 
prevent scattered photons from reaching the interaction point. 
To absorb SR photons that would nevertheless reach the Belle II inner detectors (PXD/SVD), the inner surface of the beryllium beampipe is coated with a gold layer

%In SuperKEKB we use two separate quadrupole magnets and 
%both orbits for incoming and outgoing beams are centered in the Q-magnets, 
%which minimize SR emission from off-centered orbits in the Q-magnets.  

In contrast to KEKB, both incoming electron and positron beams are nearly on the magnetic axes of quadrupoles. This reduces the need for shared final focus magnets and reduces SR but requires a much larger crossing angle. Note that PEP-II used magnetic separation to separate incoming and outgoing beams and had large SR backgrounds. 

%Synchrotron radiation is simulated by the physics model implemented in Geant4.
%We estimate the impact of synchrotron radiation on our inner detectors is tolerable.

\subsection{Radiative Bhabha process}
The fourth background source is radiative Bhabha scattering. Photons produced by the radiative Bhabha process propagate nearly along the beam axis and interact with the iron of the accelerator magnets. 
In these interactions, there is copious production of low energy gamma rays 
as well as neutrons (via the giant photo-nuclear resonance mechanism). 
The rate of their production is proportional to the luminosity, which will be 40 times higher at SuperKEKB than it was at KEKB. 
Such neutrons are the main background source for the outermost parts of the Belle II detector, 
the $K^{0}_{L}$ and muon detector (KLM), situated in the return yoke of the experiment’s solenoid magnet. 
Additional neutron shielding in the accelerator tunnel is required to stop these neutrons. 
Low energy gamma rays are, on the other hand, a significant source of background in the central drift chamber (CDC) 
and in the barrel particle identification device (TOP, Time-Of-Propagation counter).

The energies of both the electron and positron decrease after radiative Bhabha scattering. 
KEKB employed shared final focus quadrupole (QCS) magnets for the incoming and outgoing beams, and as a result the scattered particles were over-bent by the QCS magnets. 
The particles then hit the wall of magnets and again electromagnetic showers were generated. In SuperKEKB we use two separate quadrupole magnets and the outgoing beams are centered in their respective quadrupole magnets. We therefore expect the radiative Bhabha background due to over-bent electrons and positrons to be partially mitigated, and only a small fraction of beam particles with large energy loss produce background.
However, since the design luminosity of SuperKEKB is 40 times higher than that of KEKB,
the rate of those particles is not negligible, and will be the most important background source in some of the detectors.

%% too detailed topic? %%
%The transverse kick from the solenoid field due to a finite crossing angle is the crucial and inevitable cause of these beam losses.  
%The intrinsic angular beam divergence at the IP, angular diffusion in the radiative Bhabha process, 
%and leakage fields from the other ring's quadrupole magnets also play a role, but are less crucial than the kick from the Belle II solenoid.

%Scattering is simulated using the generator called ``BBBREM''~\cite{BBBREM} and ``BHWide''~\cite{BHWIDE}.

\subsection{Two-photon process}
The fifth beam background results from very low momentum electron-positron pairs
produced via the two-photon process $ee \to eeee$.
Such pairs can spiral around the solenoid field lines and leave multiple hits in the inner Belle II detectors.

In addition to the emitted pairs, primary particles that lose a large amount of energy or scatter at large angles
can be lost inside the detector, in the same way as for radiative Bhabha backgrounds. 
%Losses within $|s|<65$ cm from the IP are also dangerous.

%Scattering is simulated using the generator called ``BDK(Diag36)''~\cite{BDK}.
  
\subsection{Injection background}
When a charge is injected to a circulating beam bunch, the injected bunch is perturbed and 
a higher background rate is observed in the detector for few milliseconds after the injection.
We apply a trigger veto after each injection to avoid PXD readout bandwidth saturation.
The veto window shape should be determined by measuring time structure of injection background.
The measured injection background rate is also provided to the accelerator group for injection tuning. 
The positron damping ring, which will be very important for good injection efficiency to the low-emittance main ring,
will be available for Phase~2.

 %lead author: Peter Lewis
 \section{BEAST II}\label{sec:beast}
 %     file:	beast_system.tex
%     author: 	Peter Lewis
%
%     contents: overview of BEAST system
%		- support structure and overall physical layout of system at IP
%	     	- list of detectors, DAQ, connection to SuperKEKB via EPICs etc
%
%     Estimated completion date: End o1f September

% Other figures available in BEASTImages directory:
%    BEASTImages/BEAST-Phase1-LineArt-16-9_IsoMetric.pdf_tex [B+W version of two-column-width figure]
%    BEASTImages/BEAST-Phase1-LineArt-4-3_TriMetric.pdf_tex  [same as above but for one-column-width figure]
%    BEASTImages/render.png

\begin{figure*}[bp]
        \centering
        \subfigure{
          \includegraphics[width=1.4\columnwidth]{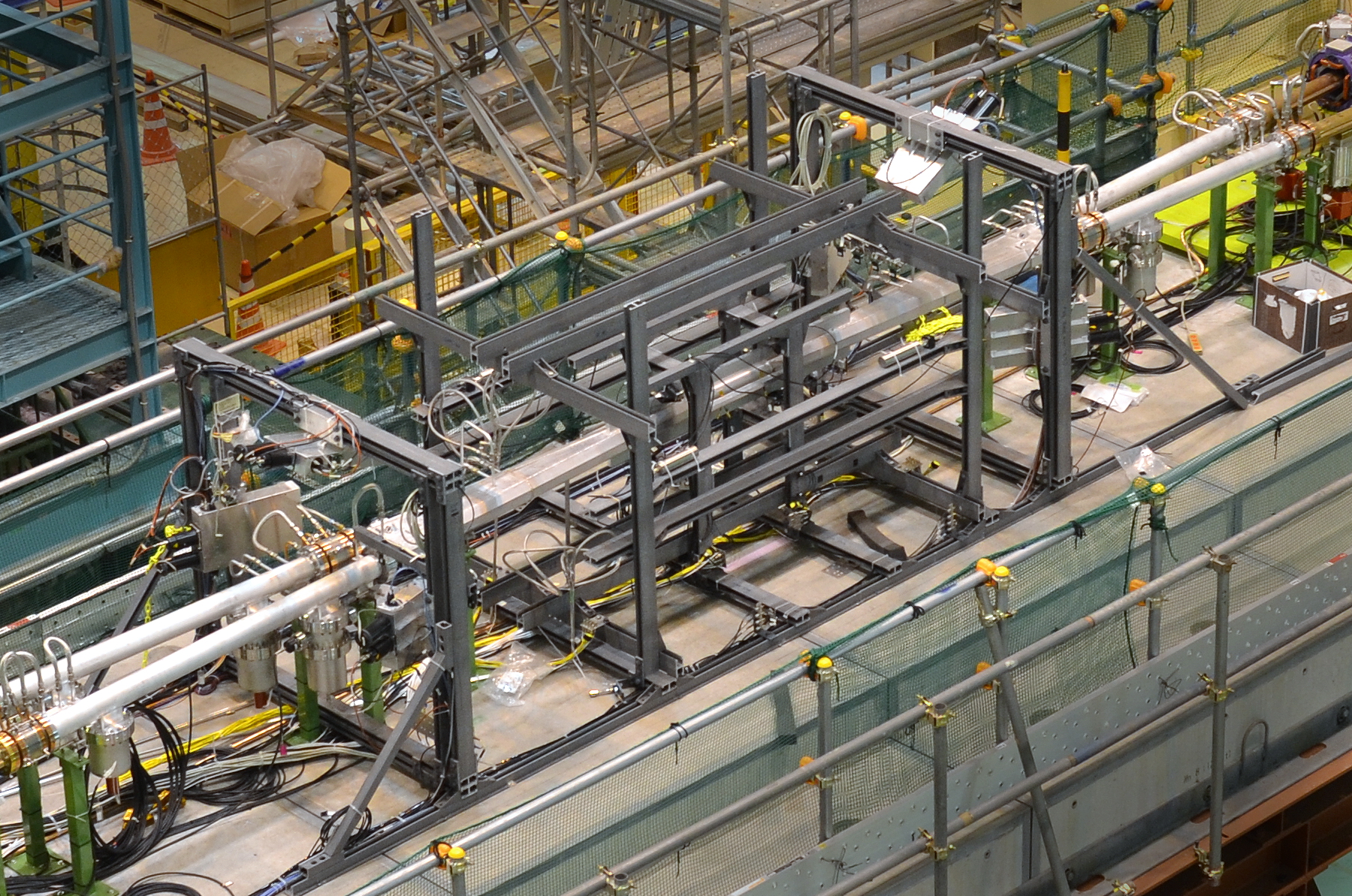}
          \label{fig:BEAST_picture}
        } \\
        \subfigure{
          \includegraphics[width=1.4\columnwidth]{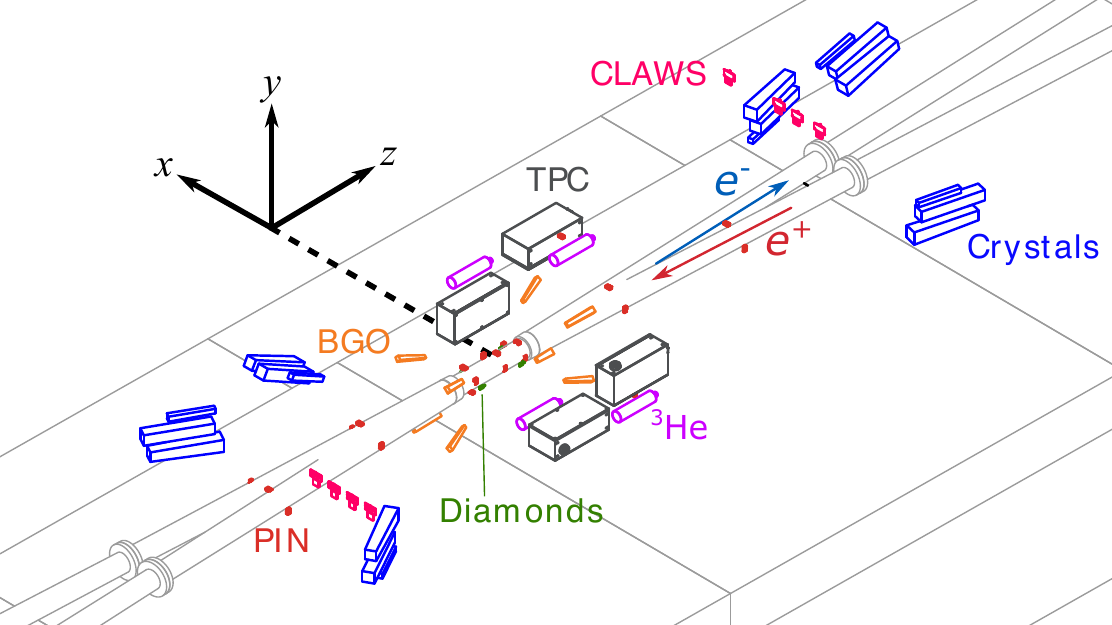}
          \label{fig:BEAST_lines}
        }
        \caption{A photograph (top) of BEAST II, with a CAD rendering (bottom) from the same perspective, prior to installing a concrete shell which surrounds the region approximately delineated by the gray frame. In the rendering, only the platform, beampipe and BEAST II detector volumes are shown, for clarity. The master coordinate system is shown removed from its origin, which is the nominal interaction point.}   
        \label{fig:BEAST}
\end{figure*}

BEAST II consists of eight detector systems, shown in Fig.~\ref{fig:BEAST} and summarized in Table~\ref{tab:beast_detectors}. These systems provide measurements of various properties of the SuperKEKB Phase 1 backgrounds in the interaction region. PIN diodes provide dose rates at various locations in the interaction region, while diamond sensors provide fast dose rates near the IP and will be used later in the beam abort system. The Crystals and BGO systems provide measurements of electromagnetic backgrounds. The CsI and LYSO crystals operate in a fast readout mode to measure the time structure of injection backgrounds. The CLAWS system features plastic scintillators with silicon photomultiplier (SiPM) readout which capture the time structure of injection backgrounds in detail and can be used for accelerator tuning. Neutron backgrounds, which are experimentally challenging, are measured by two systems: \heT tubes for counting thermal neutrons, and time projection chambers (TPCs) for counting and tracking ``fast'' or higher energy neutrons. The QCSS plastic scintillator system is a prototype for beam background monitors that are small enough to mounted between the final focusing magnets and the Belle II detector in SuperKEKB commissioning Phases 2 and 3. The QCSS system was not part of the BEAST II DAQ and did not run during all of Phase 1, and hence is not featured as prominently as the other systems in this article. 

In the remainder of this section, we provide a technical description of BEAST II, starting with the overall system design and proceeding through each detector system.

\begin{table*}[tp] 
\caption{Phase 1 detector system names, detector types, number of channels, and unique measurement or capability provided of each system.}
        \centering
        \begin{tabular}{ p{2cm} p{4cm} p{1cm} p{8cm}}
        \toprule
                System name                           		& Detector Type               		& $\#$    & Unique measurement or capability                       \\ 
        \midrule
                PIN                       				& PIN diodes      			& $64$    & Instantaneous dose rate at many positions \\                  
                Diamond            				& Diamond Sensors            	& $4$     & Near-IP fast dose rate, beam abort prototype  \\
                Crystal   						& CsI(Tl), CsI, LYSO crystals   	& $6+6+6$ & Electromagnetic energy spectrum, injection backgrounds \\
                BGO			              			& BGO crystals         			& $8$     & Electromagnetic dose rate         \\
	    CLAWS						& Plastic scintillators                	& $8$     & Injection backgrounds                     \\

     		\heT                      				&\heT tubes                    		& $4$     & Thermal neutron rate                      \\
                TPC  							& Time Projection Chambers    & $4$     & Fast neutron flux and directionality      \\
                                     QCSS						& Plastic scintillators			& $6$     & Charged particle rates, prototype for Phases 2,3 \\
                \bottomrule
        \end{tabular}
        \label{tab:beast_detectors}
\end{table*}

\subsection{System design}
The BEAST II system is designed to satisfy two primary objectives: first, to provide real-time feedback to shifters and operators, and second, to facilitate analysis of beam backgrounds and their relationship to accelerator conditions, both in data and in simulation. To satisfy the first objective, each detector system shares observables via the Experimental Physics and Industrial Control System (EPICS). The accelerator conditions are also shared via EPICS, allowing BEAST II and SuperKEKB shifters to monitor both in real time. To satisfy the second objective, we generate a unified dataset of BEAST II observables and accelerator conditions for analysis after running. This dataset is based on 1\,s summaries of BEAST II observables to match the update rate of the accelerator conditions monitors. All analysis and results presented in this paper use these 1\,s summaries unless otherwise noted.

\subsubsection{The interaction region}
In Phase 1, the interaction region extends $\pm4$~m from the interaction point along the beam axis, encompassing essentially everything in Fig.~\ref{fig:BEAST} plus a concrete shell not pictured. The central beampipe, or IP chamber, consists of a $40$~cm-long aluminum tube with $4$~cm inner radius and $4$~mm thickness. The IP chamber is is mounted on either end to a pair of aluminum beampipes, one each for the LER and HER beams, with associated cooling, vacuum and support structure. The concrete shell extends along the beam line beyond the limits of the interaction region and, together with the concrete platform that BEAST II sits on, constitutes a hermetic shield with rectangular cross-section and a thickness of $75$~cm.  

\subsubsection{Physical layout}
\label{sec:beastLayout}
Around the interaction region, BEAST II sensors are mounted either directly onto the aluminum beampipe or on a temporary structure, seen in Figure~\ref{fig:BEAST_picture}. This structure is made from off-the-shelf Aikinstrut fiberglass erector components rather than more-common steel or aluminum components to avoid complications with stray magnetic fields and grounding. 

All sensors on the structure are mounted with custom brackets for stability and alignment, to meet a global specification of $1$\,cm position tolerance. Sensors on the beampipe, including some PIN and all diamond sensors, are held in place with nylon cable ties. These sensors are electrically isolated from the beampipe with with a layer of kapton tape. 

A cable length of $37$\,m separates the sensors at the interaction region from the readout electronics, located in a radiation-safe counting room below the beam line. With the exception of a few front-end amplifiers and digitizers attached to sensors in the interaction region, all detector amplification, digitization and processing occurs in these racks. 

\subsubsection{Coordinate system}
\label{sec:beastCoordinates}
The BEAST II coordinate system, identical to the Belle II system with respect to the nominal interaction point, is represented in Figure~\ref{fig:BEAST}. The $x$-axis is horizontal (in the plane of the rings) and points towards the outside of the accelerator tunnel, which is approximately north-east. The $y$-axis is vertical, and points upwards. The $z$-axis is the Belle II solenoid axis, which is the bisector of the two beams; it points approximately towards the direction of electron beam and passes through the nominal interaction point.

The azimuthal angle $\phi$ about the $z$-axis is defined so that $\phi = 0$ corresponds to $(x, y, z) = (1, 0, 0)$ and $\phi = +90$\,degrees corresponds to $(x, y, z) = (0, 1, 0)$. The polar angle $\theta$ is measured with respect to the $z$-axis.

 % subsection{System design}
 %\clearpage

 %lead author: Minakshi Nayak
 \subsection{PIN detector system}\label{sec:pins}
 %     file:		pins.tex
%     author: 	Minakshi Nayak, Dave Cinabro
%
%     contents:  Functional description of PIN diode system, to include the following:
%		physical description: size of sensitive volume, number of detectors etc
%		process by which radiation is detected. How signal is amplified, and read out
%		quantitative performance spec: radiation sensitivity, threshold, timing capability, noise rates, etc
%		description of in-situ calibration procedure
We monitor ionization radiation dose by using 64 PIN diodes in 32 locations. A PIN diode has a thick, intrinsic (I) semiconductor region 
between a p-type (P) semiconductor and an n-type (N) semiconductor forming a sandwich 
of~P,~I,~and~N layers.
Ionizing radiation leaves a trail of
free electrons and holes, effectively causing an increase
in the dark current from such diodes.  This current is passively
amplified and its integral is proportional to the ionizing
radiation dose.  The diodes are not biased, simplifying the associated electronics.
Such a system was used at CLEO and CESR as a beam background monitor
and beam tuning aid to minimize beam induced radiation.  At
CLEO half of the diodes were behind a thin layer of high-Z shielding consisting of a layer of gold paint, and half were unshielded.  X-rays from
synchrotron radiation are considerably reduced on the shielded
diodes while particle radiation from beam-gas scattering and radiative
Bhabha events are not.  Thus the difference between a shielded and unshielded
diode pair gives a direct measure of the synchrotron radiation
component of the dose.  At CLEO and CESR this made it easy to map the
location and extent of synchrotron radiation and backscattering fans
caused by the beams passing through the final focusing elements and
X-rays scattering off of shielding elements \cite{Minakshi:CBN-007}.

\subsubsection{PIN system physical description}

\begin{figure}
\centering
 \includegraphics[width=\columnwidth]{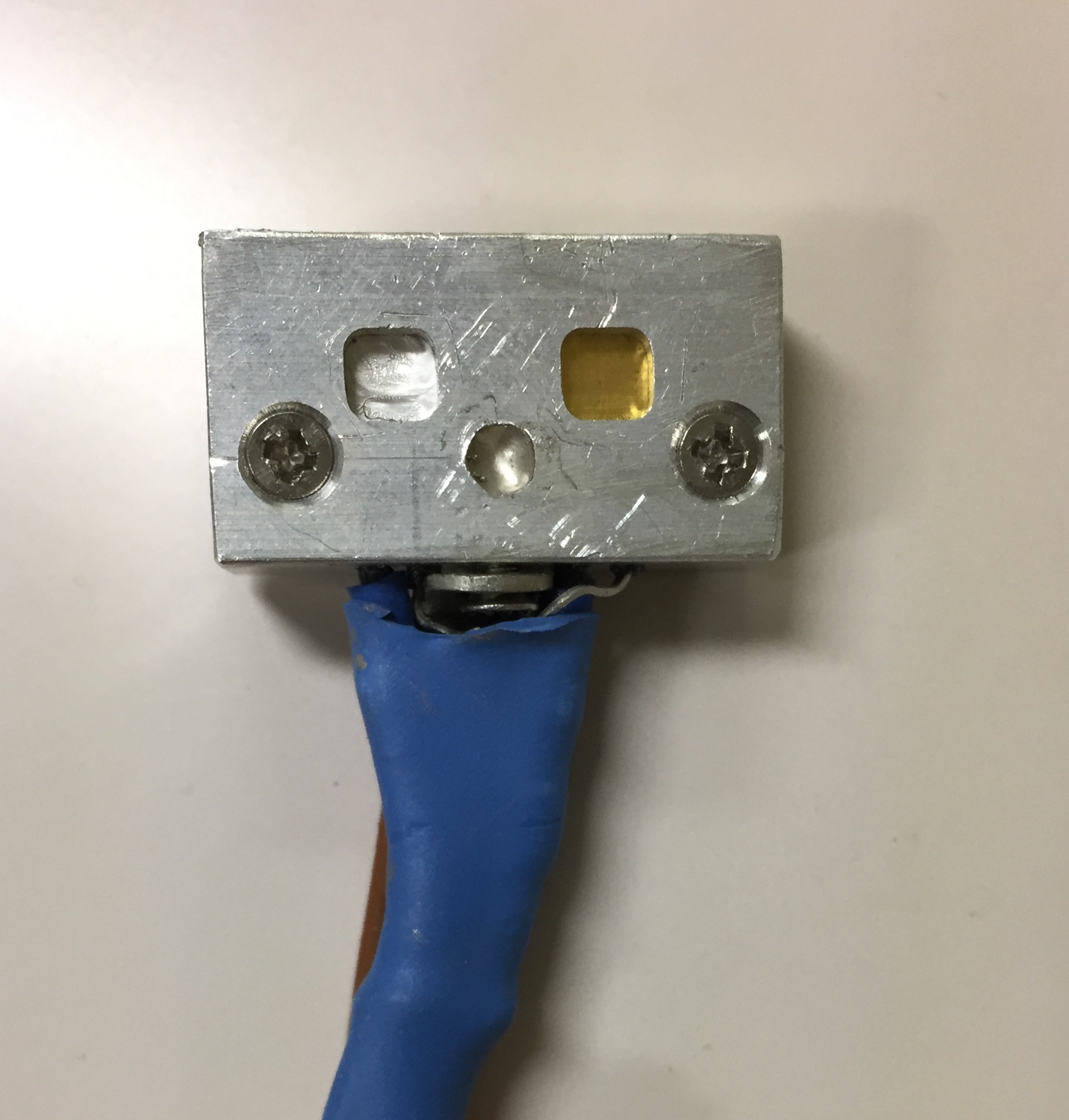}
 \caption{A PIN diode module containing two diodes and a thermocouple. The active area of the diodes is exposed, showing the aluminum foil covering one and the gold foil covering the other to reduce the 
 X-rays from synchrotron radiation.}
\label{fig:PINdiode}
\end{figure}

A PIN diode radiation monitor module consists of a pair of Siemens-SFH206K type  \cite{Minakshi:sfh206k} photodiodes 
with active volume $(2.65 \times 2.65 \times 0.2) \rm ~mm^{3}$ and a 
thermocouple sensor packaged in an aluminum 
block of size $(2 \times 3 \times 1) \rm ~cm^3$, with holes drilled 
through that leave the sensitive surface of the diodes exposed.
A picture of the aluminum block containing a gold and aluminum 
foil covered diodes is shown in Figure~\ref{fig:PINdiode}.
The covering foils are 0.1~mm thick, and the gold foil reduces
10~keV X-rays, typical for the synchrotron radiation we expect,
by more than a factor of 100.  
In each aluminum block is a thermocouple-based
temperature monitor.  This is used to correct for
the temperature dependence of the thermal dark current.
Four blocks located on top, underneath, inside, and outside the accelerator ring, 
form a basic measuring unit of 8 individually read channels at a single location along the beam line.  

The diodes are connected to the the PIN amplifier system by a short length of thin coaxial cables, 
followed by 37~m long coaxial cables. 
Each thermocouple sensor is connected to a digital multimeter by a 37~m long 
thermocouple cable.
During BEAST II Phase 1 commissioning, we use 32 PIN diode modules to monitor radiation and temperature. Of the 32 PIN modules, 28 are installed at different 
$z$ and $\phi$ locations along the beampipe. The remaining four PIN 
modules corresponding to Channel number 56 to 63 are attached to the TPC plates at locations above, below, and on either
side of the IP. Sensors and their locations are 
summarized in Table~\ref{table:pinlocation_beampipe}.
\begin{table}
\caption{Channel numbers and locations of 32 PIN diode modules mounted on the beampipe and near the TPCs. Each module contains two diodes, one covered with gold and one covered with aluminum. The diode module positions are also shown graphically, and the BEAST II coordinate system is defined, in Figure~\ref{fig:BEAST} (bottom).}
\label{table:pinlocation_beampipe}
\centering
 \begin{tabular}{cccc}
\toprule
\multicolumn{2}{c}{Channel number} &   &   \\  
Gold & Aluminum & $z$~[cm]  & $\phi [\degree]$   \\  

 \midrule

 35   & 39   & -128.0 & 0\\
 24   & 28   & -128.0  & 90 \\ 
 25   & 29   & -128.0 & 180 \\ 
 34   & 38   & -128.0 & 270\\ 

 32   & 36   & -70.0 & 0\\
 26   & 30   & -70.0  & 90 \\ 
 27   & 31   & -70.0  & 180 \\ 
 33   & 37   & -70.0 & 270\\ 
 
 19   & 22   & -11.0 & 0 \\ 
 2   	& 3   &  -11.0 & 90 \\  
 4   	& 5   & -11.0  & 180 \\ 
 18   & 23   &  -11.0 & 270 \\ 

 16 	& 17 & 3.0  & 0  \\ 
 8 	& 9 & 3.0  & 90  \\  
 0 	& 1 & 3.0  & 180 \\ 
 20 	&  21   & 3.0  & 270  \\ 
 
 14   & 15   &  15.0 & 0 \\ 
 7   	& 10   & 15.0 & 90 \\ 
 6   	& 11   & 15.0 & 180 \\ 
 12   & 13   & 15.0  & 270 \\ 

 40   & 47   & 69.0 & 0\\
 48   & 52   & 69.0 & 90\\
 49   & 53   & 69.0 & 180\\
 41   & 46   & 69.0 & 270\\ 
 
 42   & 45   & 134.0 & 0\\
 50   & 54   & 134.0 & 90\\
 51   & 55   & 134.0 & 180\\
 43   & 44   & 134.0 & 270\\ 
 
 61   & 63   & +33.5 &  0  \\
 60   & 62   & +32.4 &  90  \\ 
 59   & 58   & +33.5 &  180  \\
 56   & 57   & +35.8 &  270  \\
 \bottomrule

\end{tabular}

\end{table}

\subsubsection{PIN system principle of operation}
A commercial charge-to-voltage amplifier from Cremat, Inc. is used
to amplify the PIN diode dark current.  The basic amplifier of
the system and the associated readout circuit are taken from a Cremat specification sheet \cite{Minakshi:cr110}.

The current observed from a PIN diode consists of a radiation signal in the form of ionization current, plus thermal dark current. The thermal dark current pedestal increases with temperature.
To amplify the small signal current, which is in the nA range, 
the CR-110 preamplifier module is used, which operates in the direct 
coupled (DC) mode with a gain of 200 mV/nA. 

The pedestal-subtracted voltage output is proportional to the dose rate 
with the calculated calibration factor of $(0.75 \pm 0.15)$~rad/s/V.. 

This system, besides measuring the ionizing radiation
dose rate and total dose, gives a low energy resolution
view of any sharp X-ray features incident on the beampipe at the 
longitudinal position of a diode unit.  These radiation features are broadened
as they scatter out of the beampipe, thus making a higher resolution
view not useful.  We expect, based on our background simulation discussed
below, that we will not see any X-ray features.

\subsubsection{PIN system performance}
We tested the system in the lab with low-activity ionizing radiation
sources and verified that it returns voltages proportional to the source activity. For a $\beta$ source of known activity the response agrees well with
the calculated energy deposition in the sensor.  The calculation assumes 
that the $\beta$
source produces minimum ionizing particles in the silicon of the PIN diodes, liberating
a number of electrons given by the known ionization energy of silicon and the geometry of
the silicon, 0.20 mm thick.  Given the activity of the source, distance between
the diode and the source, and the amplification of the preamp this gives a current 
out of the Cremat system.  The observed signal agrees with the expectation within 20\%.
The system gives a clear signal above noise in the presence of the
source, no signal when the source is blocked by lead foil, and the signal showed
a one over distance squared dependence when the distance between the source and
diode is varied.  Scatter among the response of different channels of the system is less than 10\%.

We also tested the \emph{in situ} performance of the system at
KEK using Sr-90 $\beta$ sources of known activities.  There is a clear signal above noise in 
presence of sources and the system output tracks the source activity until the output saturates at 3~V, as shown in Table~\ref{table:2}. 
As expected, the signal shows $1/{\rm distance^2}$ dependence when sources are moved away from sensor.  
\begin{table}
\caption{PIN diode system amplifier output using Sr-90 sources of various activities. The source is positioning 1~cm away 
from the PIN diode, which is covered with 0.1~mm thick Aluminum foil.}
\centering
 \begin{tabular}{lc} 
 \toprule
 Activity (MBq) & Signal observed (mV) \\  
 \midrule
0.35  & 10  \\ 
17  & 500 \\
212  & 3000 (saturation voltage) \\ 
 \bottomrule
\end{tabular}
\label{table:2}
\end{table}
This test was similar to the lab tests described above.
The output showed no appreciable dependence on cable length.

The gains of the 64 amplifiers are tested using a supply voltage of 0.2-2.0~V passing through a
$100 \rm ~M\ohm$ resistor giving input currents of 2-20~nA. The output voltage 
is proportional to input current as expected.
For a known input current, the ouput voltage is measured for all 64 channels 
and the gain variation is within $15\%$ of nominal. 

\subsubsection{PIN system calibration}
\begin{figure}
 \begin{center}
  \includegraphics[width=\columnwidth]{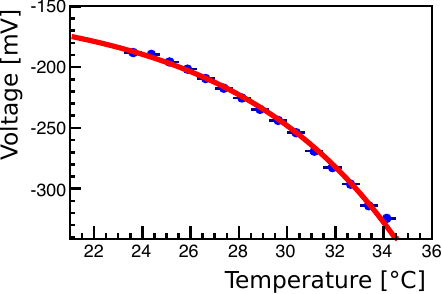}
  \end{center}
    \caption{(color online) Pedestal voltage versus temperature for a PIN diode with no radiation and before Phase 1 running. In this plot, blue points with error bars
represent data while the fit is shown with the solid red curve.}
      \label{fig:pin_calibration}
\end{figure}

The pedestal given by the diode dark current is very sensitive to temperature. To measure it, we used a heat gun to slowly heat the diodes from room temperature while measuring the diode temperature. We fit the resulting dependence of the output on diode temperature $T$ with:
\begin{equation}
V = p_0 - e^{(p_1 + p_2 T)}, 
\end{equation}
where $p_0$ gives the pedestal independent of temperature and $p_1$ and $p_2$ give the temperature dependence.
Figure~\ref{fig:pin_calibration} shows a fit of diode to this 
functional form.

\begin{figure}
 \centering
  \includegraphics[width=\columnwidth]{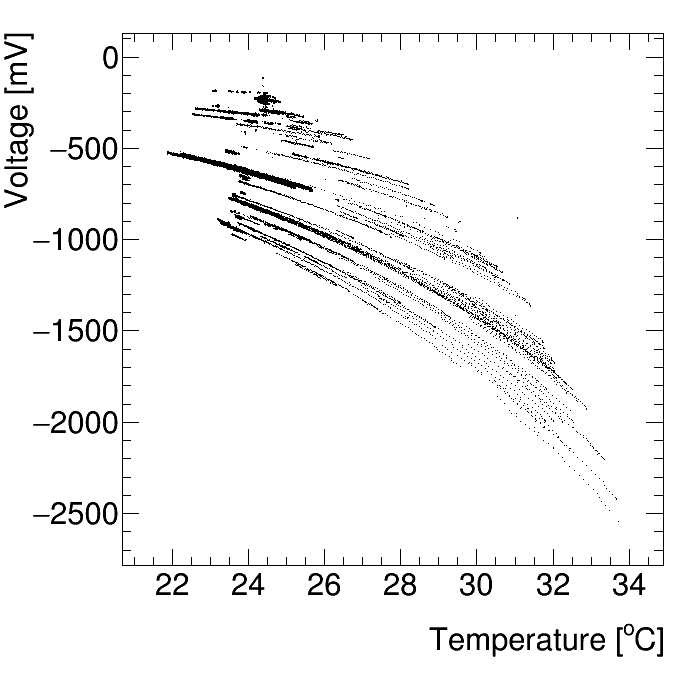}
    \caption{Pedestal voltage versus temperature for a single PIN diode 
during times of no beam for all of the Phase 1 running. Here each pedestal voltage versus temperature band corresponds to one time period.}
      \label{fig:pin_pedestal}
\end{figure}

\begin{figure}
\centering
 \includegraphics[width=\columnwidth]{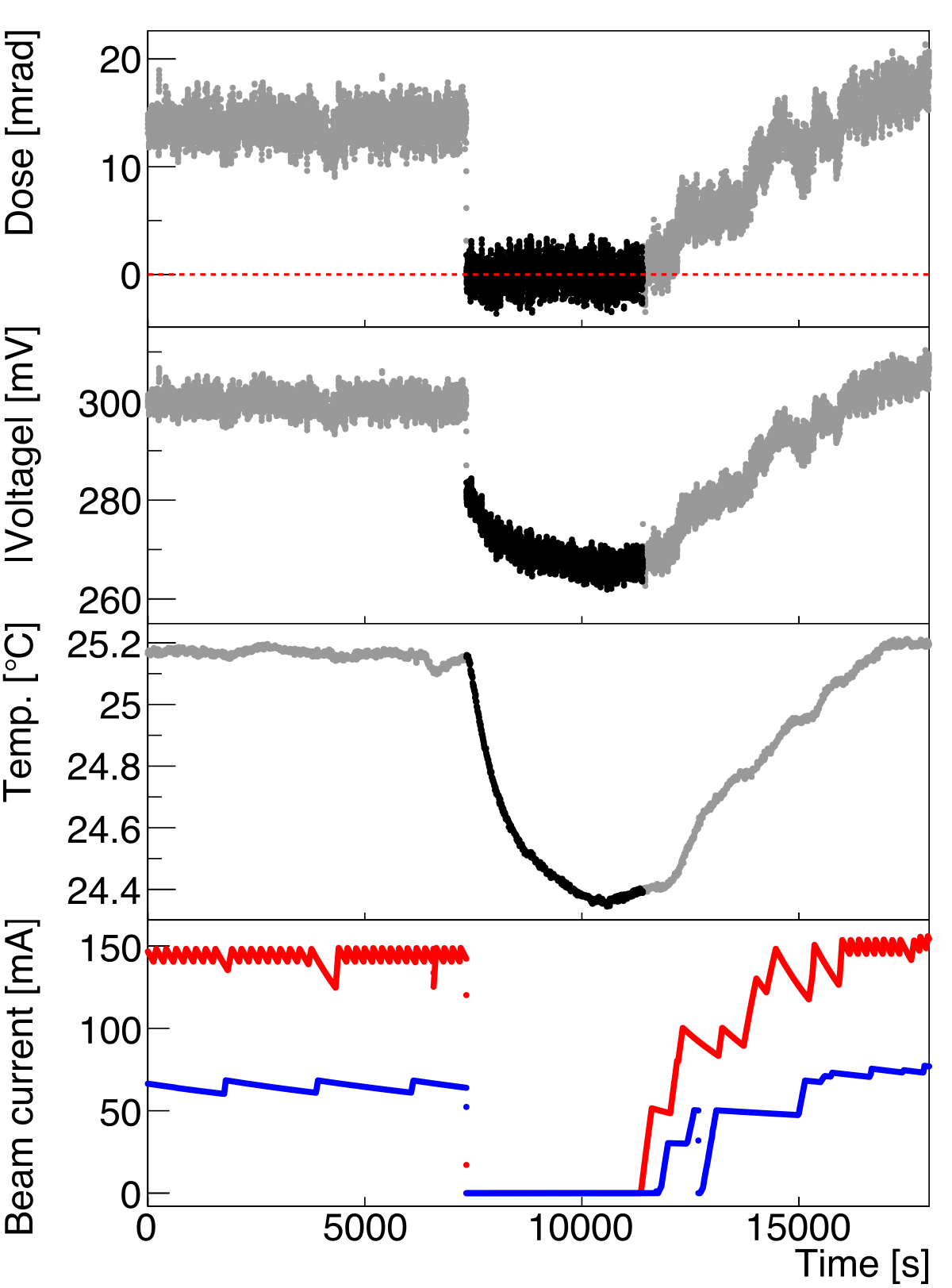}
 \caption{(color online) An illustration of the PIN diode calibration during 5 hours of Phase 1 running. Starting from the bottom, we first see the LER current (red) and HER current (blue) during stable running and also after a dual beam abort event. Next we see the PIN module temperature, which is elevated when the beams are on (gray) and slowly returns to ambient temperature after the beam abort (black). The absolute value of the raw amplifier voltage output, next, shows response to both temperature and dose from the beam. The final calibrated PIN dose, top, shows zero dose throughout the abort period, indicating that the calibration succesfully removed dynamic dark current effects.}
\label{fig:pin_calibration-1}
\end{figure}

With increasing integrated radiation dose on the diodes their 
dark current pedestal, observed when no beam was present, rose. 
This is shown Figure~\ref{fig:pin_pedestal} which is the voltage
output of one diode during times when there is no beam versus temperature
for the entirety of the Phase~1 running.  While the form of the temperature
dependence changes little, the voltage at a fixed temperature became
lower with increasing dose.
Thus we were forced to do a daily calibration during times when there 
was no beam. We check this procedure by calibrating the pedestal with only the first half day of data and fixing the 
pedestal for the rest of the day.  We then measure the dose deposited when there is no beam in the second half day, find it consistent with zero, 
and use the standard deviation of the scatter around zero as our uncertainty on the dose due to uncertainty in the pedestal. To illustrate how the daily calibration procedure solves the 
problem of pedestal increase, we compare the voltage and 
dose output during 5 hours of Phase 1 running. 
Figure~\ref{fig:pin_calibration-1} shows the beam current, temperature, voltage, and calibrated dose
for a single diode which clearly shows a high output voltage due to dark current, while the dose is as expected.

While the PIN system operated successfully during Phase 1 operation, it
was hampered by a complex calibration procedure.  The system would work 
much better if the sensors had been actively temperature controlled.
The combination of dark current pedestal dependence on temperature and base 
pedestal drift with increased observed dose made it challenging to keep the 
pedestal up to date during operations.

 %\clearpage
  
 % lead author: Livio Lanceri
 \subsection{Diamond detector system}
 %     file:		diamonds.tex
%     author:  	Livio Lanceri
%     updated    2017-07-30, addressing comments by Robert Seddon (2017-06-06)
%                      2017-07-31, as above
%                      2017-10-07, "final" editing before October B2GM
%
%     contents:  Functional description of diamond sensor system, to include the following:
%		physical description: size of sensitive volume, number of detectors etc
%		process by which radiation is detected. How signal is amplified, and read out
%		quantitative performance spec: radiation sensitivity, threshold, timing capability, noise rates, etc
%		description of in-situ calibration procedure

The pixel detector (PXD) and silicon vertex detector (SVD), together forming the inner vertex detector (VXD) of Belle II, will be exposed to the largest radiation doses and radiation damage during the life of the experiment. For this reason, instantaneous and integrated radiation doses will be monitored by a system made of single-crystal diamond detectors. Their readout electronics will provide both the continuous monitoring of radiation doses and also beam abort signals, whenever radiation due to beam losses will increase to excessive levels.

\subsubsection{Diamond system physical description}

In Phase 1, we mounted four prototype sensors on the beampipe, as already shown in Fig.~\ref{fig:BEAST}. In this section we briefly describe the diamond detectors, their location, their readout electronics and the calibration procedures.
%More details can be found elsewhere~\cite{ref:diamonds_paper}.
%\begin{figure}[ht!]
%	\centering
%		\includegraphics[width=\columnwidth]{diamondsImages/fig_diamonds_setup.png}
%	\caption{Diamond detectors in Phase 1. The positions of the four detectors on the beampipe are highlighted by circles, with labels; their coordinates are also shown at the top and bottom. Arrows at the left and right indicate the approximate directions of the incoming and outgoing beams.}
%	\label{fig:diamonds_setup}
%\end{figure}

\begin{figure}[ht!]
	\centering
		\includegraphics[width=\columnwidth]{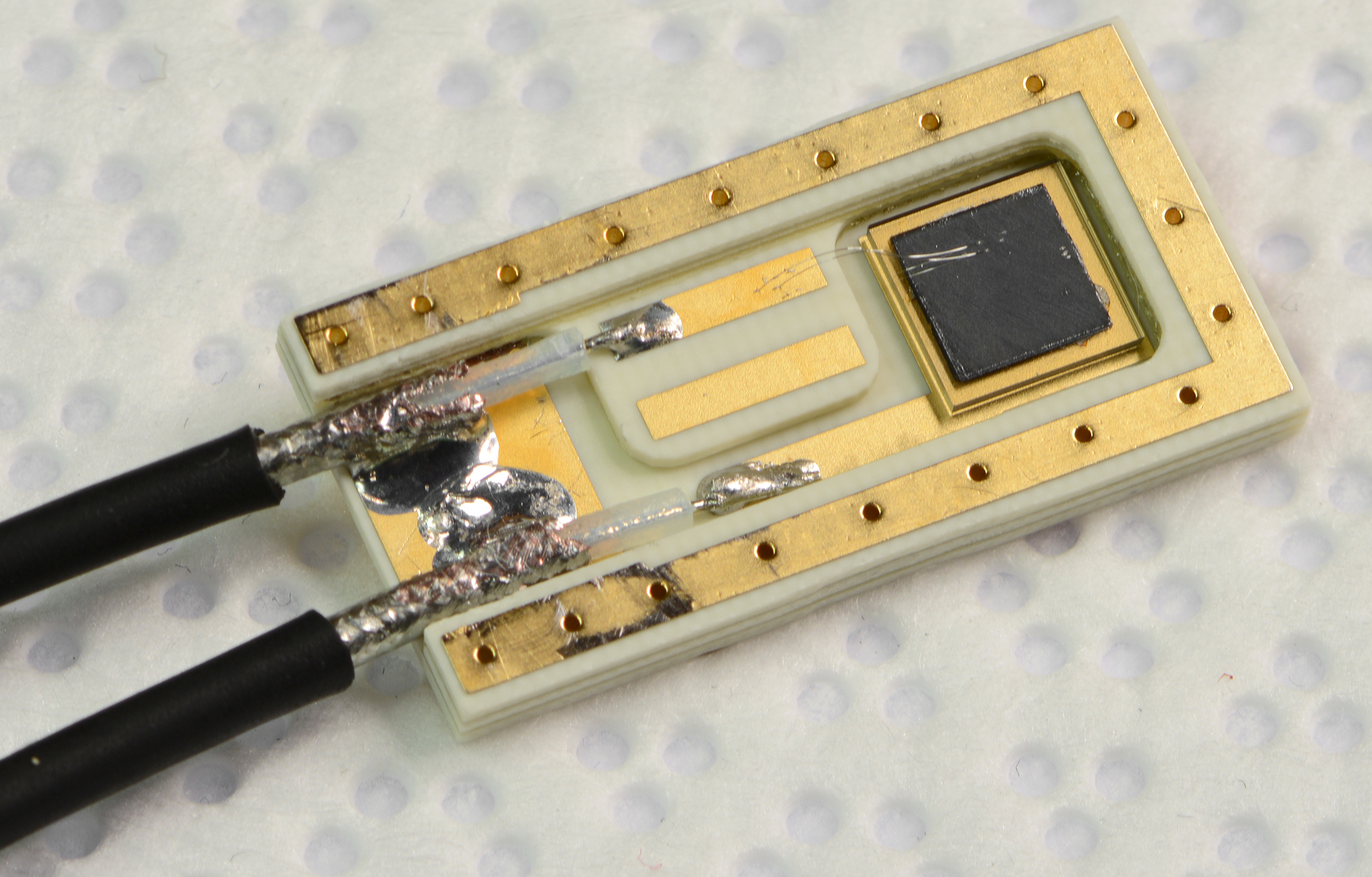}
		\includegraphics[width=\columnwidth]{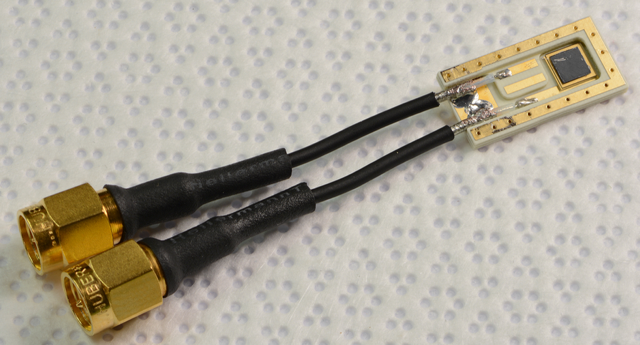}
	\caption{(color online) One of the $(4.5 \times 4.5 \times 0.5)$~mm$^{3}$ diamond sensors, mounted in its package with short, thin coaxial cables and SMA connectors. In the enlarged image, the golden wire bonds are visible on the upper electrode, deposited on the transparent diamond crystal.}
	\label{fig:diamond_package}
\end{figure}

Diamond crystals are artificially grown~\cite{ref:diamonds_E6} as single- or poly-crystals by the Chemical Vapour Deposition (CVD) technique, and classified as sCVD and pCVD respectively. Electrodes are deposited by metallisation procedures on two opposite faces of the diamond. Each sensor is mounted on a small printed circuit board (Fig.~\ref{fig:diamond_package}), providing electrical connections and insulation; a small aluminum cover completes the electrical screening. One electrode is connected to a pad on the printed circuit by electrically conductive glue; the other electrode is wire-bonded. Two thin coaxial cables are soldered to the printed circuit. We refer to the complete package as the ``diamond detector''.

We mounted four $(4.5 \times 4.5 \times 0.5)$~mm$^{3}$ diamond detectors on opposite sides of the beampipe, approximately in the horizontal plane, at $y = 0$, close to the nominal interaction point. The detector labels, sensor labels, sensor types (sCVD or pCVD), metallisation providers (Micron~\cite{ref:diamonds_Micron} or CIVIDEC~\cite{ref:diamonds_Cividec}), and locations ($z$-coordinate) are summarized in Table~\ref{tab:diamonds_loc}.

The detector labels refer to the forward (FW) or backward (BW) position with respect to the nominal position of the Interaction Point (IP), and to the position in the horizontal plane as identified by the $\phi$ angle ($0$ or $180$ degrees) in the Belle II coordinate system.

We connected the detectors to the readout electronics by $2.5$~m long, thin coaxial cables, followed by $30$~m long high-quality coaxial cables with double-layer external conductor, the same configuration planned for the following phases of Belle II.

\begin{table}[ht]
	\centering
	\caption{Labels, types, providers and $z$-coordinates of the diamond detectors. IP is at $(0,0,0)$, $z$ runs parallel to the beampipe, as shown in Fig.~\ref{fig:BEAST}.}

	\begin{tabular}{ lllll }
		\toprule
		Detector Label & Type & Provider & $z$~[cm] 	\\  \midrule
		FW-180 &  pCVD  & Micron     & $+9.5$ \\
		FW-0     &  sCVD & CIVIDEC  & $+9.5$ \\
		BW-180 &   sCVD  & Micron     & $-13.2$ \\
		BW-0     &   sCVD  & Micron     & $-13.2$ \\		\bottomrule
	\end{tabular}
	\label{tab:diamonds_loc}
\end{table}

\subsubsection{Diamond system principle of operation}

The sCVD and pCVD diamond sensors are most commonly available from the manufacturer~\cite{ref:diamonds_E6} in the standard size mentioned above. Sensors provided by Micron~\cite{ref:diamonds_Micron} have aluminum electrodes, while CIVIDEC sensors~\cite{ref:diamonds_Cividec} have Ti-Pt-Au electrodes.

The sensors are polarized by a voltage difference applied to the electrodes, generating electric fields typically in a range up to about $1$~V/\si{\micro}m, and behave approximately as solid state ionization chambers, detecting the crossing of charged particles, that generate one electron-hole pair per 13 eV of deposited energy on average. Electrons and holes drift towards the opposite electrodes, inducing currents that can be measured by external circuits. Depending on the properties of the metal-diamond interface, the electrodes may have blocking or ohmic behaviour. In the second case, charge injection from the electrodes is the origin of  voltage-dependent photoconductive gain~\cite{ref:diamonds_photoconductive_gain}.

With a fast amplifier located close to the sensor, individual current pulses can be detected, corresponding to the drift and collection of electron-hole pairs generated by single particles. In our application, we measure instead the global effect of many particles, measuring a variable current that is proportional to the instantaneous dose rate (deposited energy/sensor mass per second). The two electrodes of each sensor are connected to the electronics by coaxial cables: one High Voltage (HV) for the polarising voltage, the other (signal) connected to the readout instrument measuring the current.

\paragraph{Electronics}

The readout system was specifically designed and built as a prototype for Phase 1. Its functionality includes all the features required by the radiation monitoring and beam abort for Phase 3 of the Belle II experiment: amplification, digitization, signal processing, readout, and individual HV supply for the four diamond sensors.

\paragraph{Amplification}
Trans-impedance amplifiers convert the current signals to voltage, with a cut-off frequency matched to the required $10$~\si{\micro}s time resolution, corresponding to the revolution period of the electron and positron beams. Two current ranges ($5$~nA and $10$~\si{\micro}A) can be selected by nMOS transistors.

\paragraph{Digitization and signal processing}
\label{sec:diamond:signal_processing}
The digitization is based on the 16-bit LTC2208 ADC from Linear Technology Corporation~\cite{ref:diamonds_ADC}, with maximum sampling frequency of $130$~MHz. The digital signal processing of four input channels is performed by a Stratix III FPGA. Noise reduction is achieved by four levels of moving averages of the input data, in moving time windows of $10$~\si{\micro}s, $1$~ms, $50$~ms and $1$~s. Comparisons with programmable ``abort'' and ``warning'' thresholds generate signals that will be used in the Beam Abort system of SuperKEKB in Phases 2 and 3. The continuously updated current averages are stored in four revolving buffers with programmable depth.

\paragraph{Data readout}

One Ethernet port is used for continuous monitoring readout, at $10$~Hz, of individual currents, time-averaged over $0.1$~s. A second Ethernet port is devoted to the initialisation of the system and to download the contents of the four buffer memories after a beam abort, for ``post mortem'' analysis of beam losses versus time, with $10$~\si{\micro}s time resolution.

\paragraph{High voltage}

Four separate HV supplies are programmable in the $0$~V to $1000$~V range.

\paragraph{Offsets and noise}

The entire system is designed for the measurement of very small currents with stable offsets (pedestals) and low noise. With the $5$~nA conversion range selected during Phase 1, the intrinsic noise of the complete readout chain, including sensors and long cables, was typically at the level of a few pA in the $0.1$~s time averages, during all operation phases of SuperKEKB.

\subsubsection{Diamond system performance}

\begin{figure}[ht!]
	\centering
		\includegraphics[width=\columnwidth]{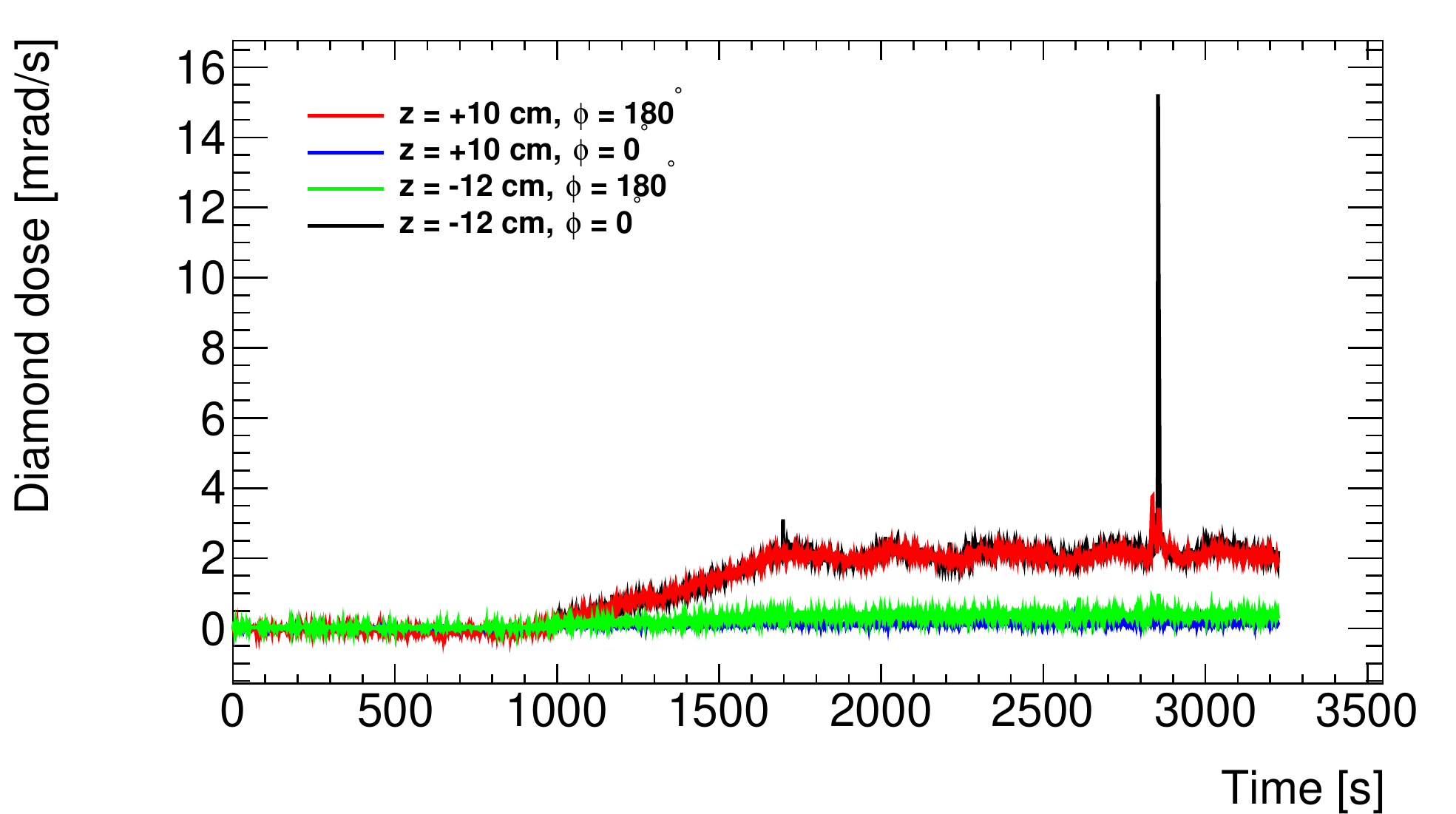}
	\caption{Example of dose rates in diamond detectors, due to beam losses during the initial vacuum scrubbing by the positron LER beam at about $18$~mA. This data sample corresponds to about one hour; the initial beam injection and the top-off at intervals of 4 minutes can be clearly identified. Occasional large beam loss spikes (so called ``beam-dust events'') are seen in coincidence by more than one diamond detector.}
	\label{fig:diamonds_InitialScrubbing}
\end{figure}

The operation of the diamond system was efficient and smooth over the Phase 1 commissioning of SuperKEKB. For diamond currents averaged over 0.1~s the noise was in the range of a few pA. Pedestals were also stable at the pA level, with the exception of one partially damaged channel (FW-180) that developed a larger but controllable pedestal drift. These small uncertainties allowed very sensitive measurements of beam losses: diamond currents were typically of the order of $1$~nA during the last phase of SuperKEKB vacuum scrubbing, with beam currents between $0.7$ and $1$~A. Correlations were observed between the currents measured over each diamond sensor and the currents of the two circulating beams.

As an example, Fig.~\ref{fig:diamonds_InitialScrubbing} shows diamond dose rates, measured in about one hour in February 2016, during initial vacuum scrubbing by the positron LER beam at only approximately $18$~mA.

\subsubsection{Diamond system calibration}

%	Testing of the diamond detectors at the INFN-Trieste laboratories included $I$-$V$ measurements in the dark, measurements of Charge Collection Efficiency (CCE) with minimum-ionizing particles from a radioactive $\beta$ source, Transient Current Technique (TCT) characterisation of diamond crystal quality and transport parameters with an $\alpha$ source, and finally calibrations of instantaneous dose measurements with the $\beta$ source.

	Testing of the diamond detectors included $I$-$V$ measurements in the dark, measurements of Charge Collection Efficiency (CCE) with minimum-ionizing particles from a radioactive $\beta$ source, Transient Current Technique (TCT) characterisation of diamond crystal quality and transport parameters with an $\alpha$ source, and finally calibrations of instantaneous dose measurements with the $\beta$ source.

\paragraph{$I$-$V$ measurements}

After glueing the metallized diamond crystal onto its printed-circuit support and after wire-bonding the upper electrode, the complete sensor is tested in the dark with no irradiation, to measure the dark current $I$ as a function of the voltage $V$ applied to the electrodes, up to $1000$~V. At the typical operating voltage of $100~$V dark currents are well below the pA range.

\paragraph{Charge collection efficiency}

\begin{figure}[ht!]
	\centering
		\includegraphics[scale=0.45]{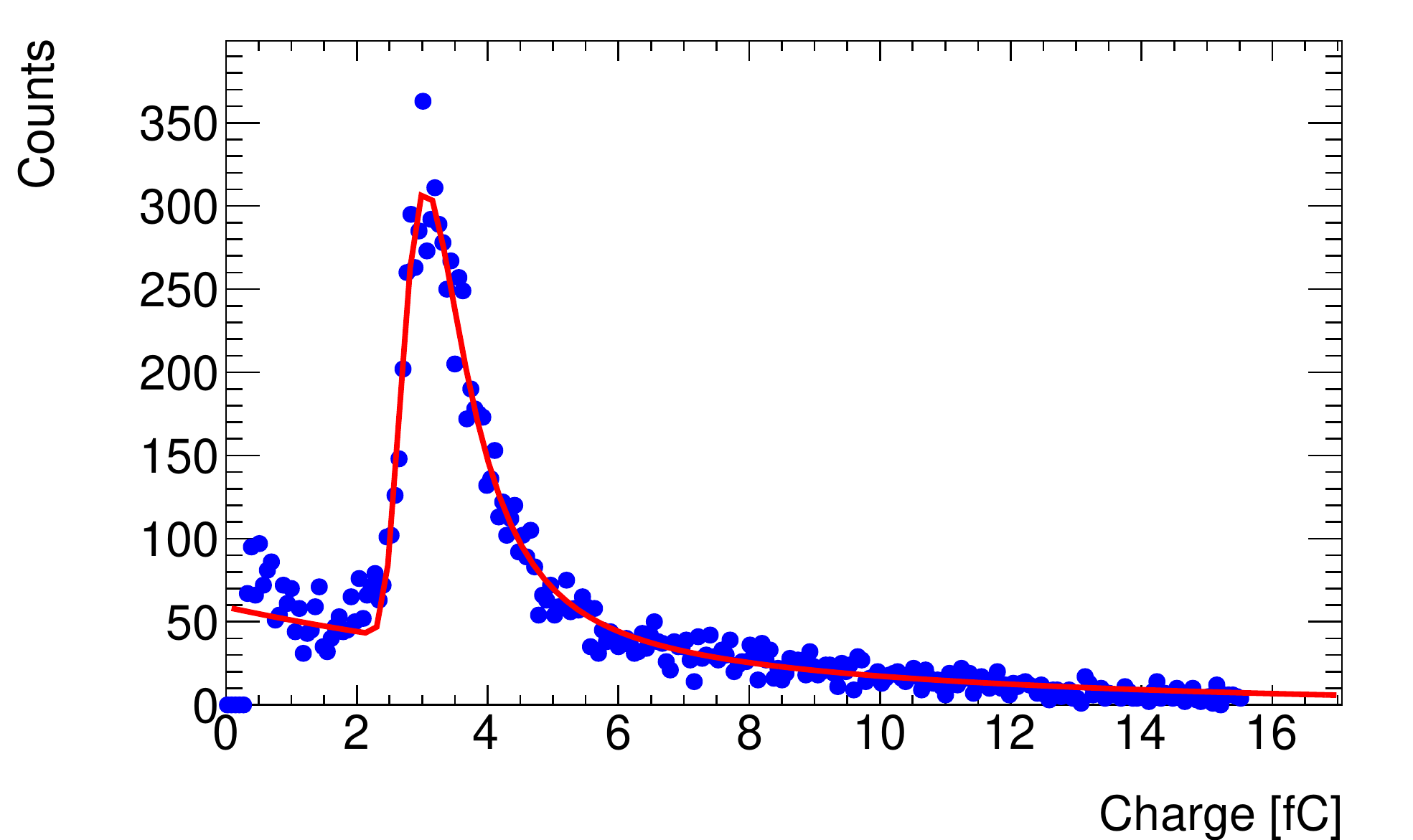}
	\caption{(color online) Example of distribution of signals in a diamond sensor from minimum-ionizing particles, in the experimental set-up described in the text. After calibration of the readout chain, the signal charge is expressed in femtocoulombs. Blue dots represent data, showing a clean Landau peak above a smooth exponential background. The red curve represents the fit, from which the most-probable value is obtained and used to evaluate the Charge Collection Efficiency.}
	\label{fig:landau}
\end{figure}

\begin{figure}[ht!]
	\centering
		\includegraphics[scale=0.45]{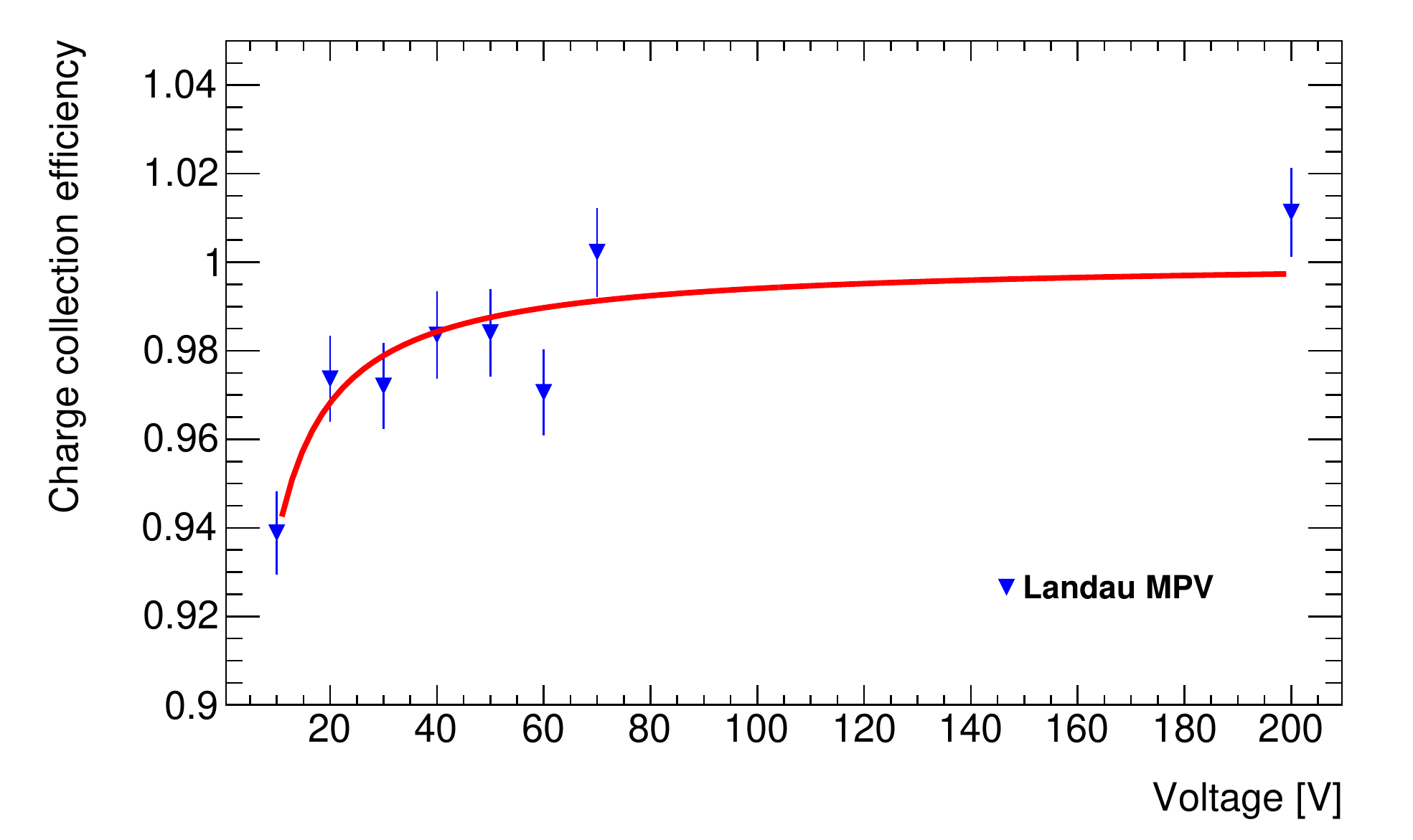}
	\caption{(color online) The measured Charge Collection Efficiency (CCE) of a single diamond detector as a function of the applied voltage. The red line represents the fitting function described in the text.}
	\label{fig:CCE_vs_V_expandedScale}
\end{figure}

A $^{90}$Sr $\beta$-source is coupled to a magnet and collimators to select a beam of minimum-ionizing electrons in the MeV energy range. A low-noise spectroscopy amplifier and digitizer chain is used to measure the signals from individual particles crossing the diamond sensor. After charge calibration of the amplifier chain, the most probable value of the observed Landau distribution (Fig.~\ref{fig:landau}) gives a clean measurement of the collected charge. Charge Collection Efficiency (CCE) is obtained by comparing this value with the expected most probable ionization by the minimum-ionizing electrons. As a function of the applied HV, CCE approaches unity at about $50$~V (Fig.~\ref{fig:CCE_vs_V_expandedScale}); at higher HV values, the position of the Landau peak does not change significantly, indicating a stable, full efficiency for the collection of electrons and holes generated by the passage of minimum-ionizing particles. The dependence of CCE on HV is well described by the following fitting function:
\begin{equation}
CCE = \frac{Q_{collected}}{Q_{generated}} = \frac{v\tau}{d} (1 - e^{-\frac{d}{v\tau}}),
\end{equation}
where $v$ is drift velocity of charge carriers, $\tau$ their lifetime, and $d$ the detector thickness.

%\begin{figure}[ht!]
%	\centering
%		\includegraphics[scale=0.45]{diamondsImages/cce.pdf}
%	\caption{Example of Charge Collection Efficiency (CCE) as a function of the applied voltage, measured for one of the diamond sensors.}
%	\label{fig:CCE_vs_V}
%\end{figure}

\paragraph{Transient current technique}

\begin{figure}[ht!]
	\centering
		\includegraphics[width=\columnwidth]{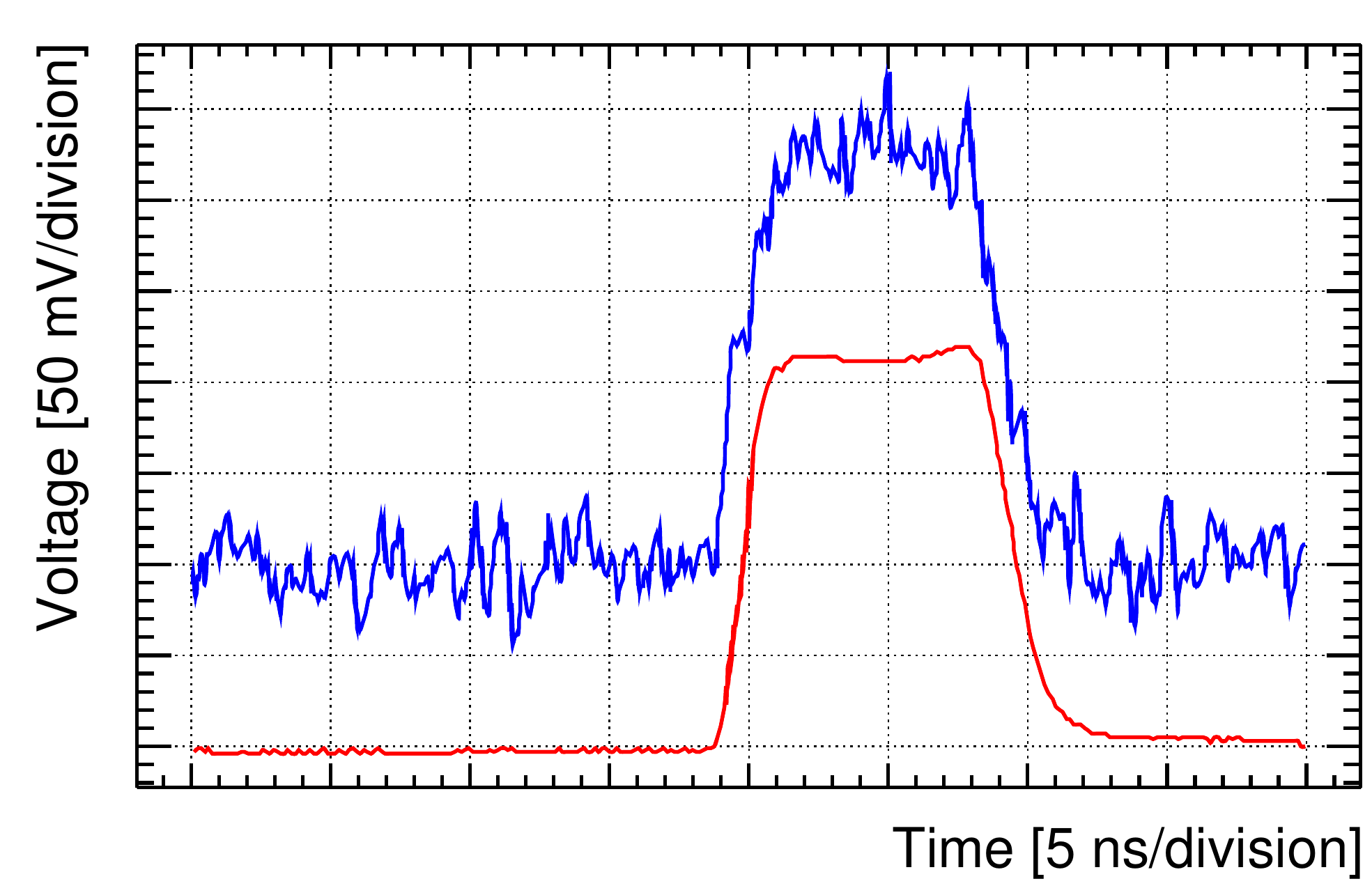}
	\caption{Snapshot of signal from a diamond channel generated by $\alpha$ particles, processed by a fast amplifier and a large bandwidth oscilloscope. The upper line is the signal due to a single event, while the lower, smoother line is the result of an average of 1000 events.}
	\label{fig:TCT_signal}
\end{figure}

The Transient Current Technique (TCT) test consists of exposing one side of the detector to $\alpha$ particles from an $^{241}$Am radioactive source. The  $\alpha$ particles release all their energy within a few microns of the surface, inside the diamond sensor, creating a concentrated excess of electron-hole pairs. Depending on the sign of the HV bias, carriers of one type are immediately collected by the nearby electrode, while carriers of the other type drift towards the opposite electrode, traverse the full $500$~\si{\micro}m sensor thickness, and induce a current pulse that can be processed by a fast amplifier and a large bandwidth oscilloscope. The pulse width is related to the drift time and gives a measurement of the carrier mobility (Fig.~\ref{fig:TCT_signal}); the pulse shape is rectangular for perfect crystals and no carrier losses by trapping. Deviations from the ideal behaviour can be identified and related to crystal imperfections and impurities.

\paragraph{Current-dose calibration}
\label{sec:Diamond:Current-Dose-Calibration}
The response of diamond sensors as dosimeters is not uniform: differences stem from the CVD crystal growing process, and from the properties of the diamond-electrode interface, both rather complicated at the microscopic level. The charge collection efficiency can vary for several reasons. Different amounts of imperfections corresponding to traps can capture charge carriers. Moreover, ohmic contacts between electrodes and diamond may inject charge into the diamond bulk, inducing a photoconductive gain that may exceed unity~\cite{ref:diamonds_photoconductive_gain}.

For these reasons, an individual calibration of diamond sensors is needed to relate the measured currents to the deposited radiation doses. We performed this calibration exposing each sensor to electrons from a pointlike radioactive $^{90}$Sr $\beta$-source of known $3.2$\,MBq activity. We explored both a range of HV bias values, from $100$~V to $300$~V in steps of $50$~V, and a range of distances from $d = 2$~mm to $d = 50$~mm between sensor and source, changing the accepted solid angle and the electron flux in a controlled and reproducible way.

We used the FLUKA~\cite{ref:diamonds_FLUKA} simulation program to describe the entire geometry and the materials of the experimental set-up, and to compute the $\Delta E / \Delta t$, the average energy deposited per second in the diamond crystal by particles coming from the pointlike source located at a fixed distance $d$.

The current in the simulated sensor, at the distance $d$ from the source, can be predicted as:
\begin{equation}
I_{FLUKA} = \frac{\Delta E[\mathrm{eV}]}{\Delta t} \frac{1}{E_{eh}[\mathrm{eV}]} q_{e} \times CCE_{FLUKA} \times G_{ph, FLUKA}
\end{equation}
where $E_{eh} = 13$~eV is the average energy required to generate an electron-hole pair, $q_{e}$ is the electron charge, $\Delta E$ is also expressed in eV and we
assume in the simulation full charge collection efficiency $CCE$ and unity photoconductive gain $G_{ph}$:
\begin{equation}
CCE_{FLUKA} = 1,
\end{equation}
\begin{equation}
G_{FLUKA} = 1.
\end{equation}
In a distance range from about $d = 5$~mm to $d = 50$~mm the predicted current, expressed in nA, can be parameterized as:
\begin{equation}
I_{FLUKA} [\mathrm{nA}] = A + \frac{B}{(d - d_{0})^{C}}
\end{equation}
with the coefficients $A = 0.35$~pA, $B = 24.9$~nA~mm$^{-C}$, $C = 2.07 \pm 0.06$, $d_{0} = -37.4 \pm 0.26$~mm.

The combination of $CCE$ and $G_{ph}$ being difficult to simulate accurately, we prefer to empirically determine measured values of their product, by comparing measured $I_{meas}$ and simulated $I_{FLUKA}$ currents in a ratio $G$, where the real $\Delta E$, deposited energy, and $E_{eh}$, energy per electron-hole pair, may be safely assumed to cancel with their simulated counterparts:
\begin{equation}
G = \frac{I_{meas}}{I_{FLUKA}} = \frac{CCE_{meas} G_{ph, meas}}{CCE_{FLUKA} G_{ph, FLUKA}} = CCE_{meas} G_{ph, meas};
\end{equation}
we use $G$ in the following as an empirical overall gain factor.

\begin{figure}[ht!]
	\centering
		\includegraphics[scale=0.45]{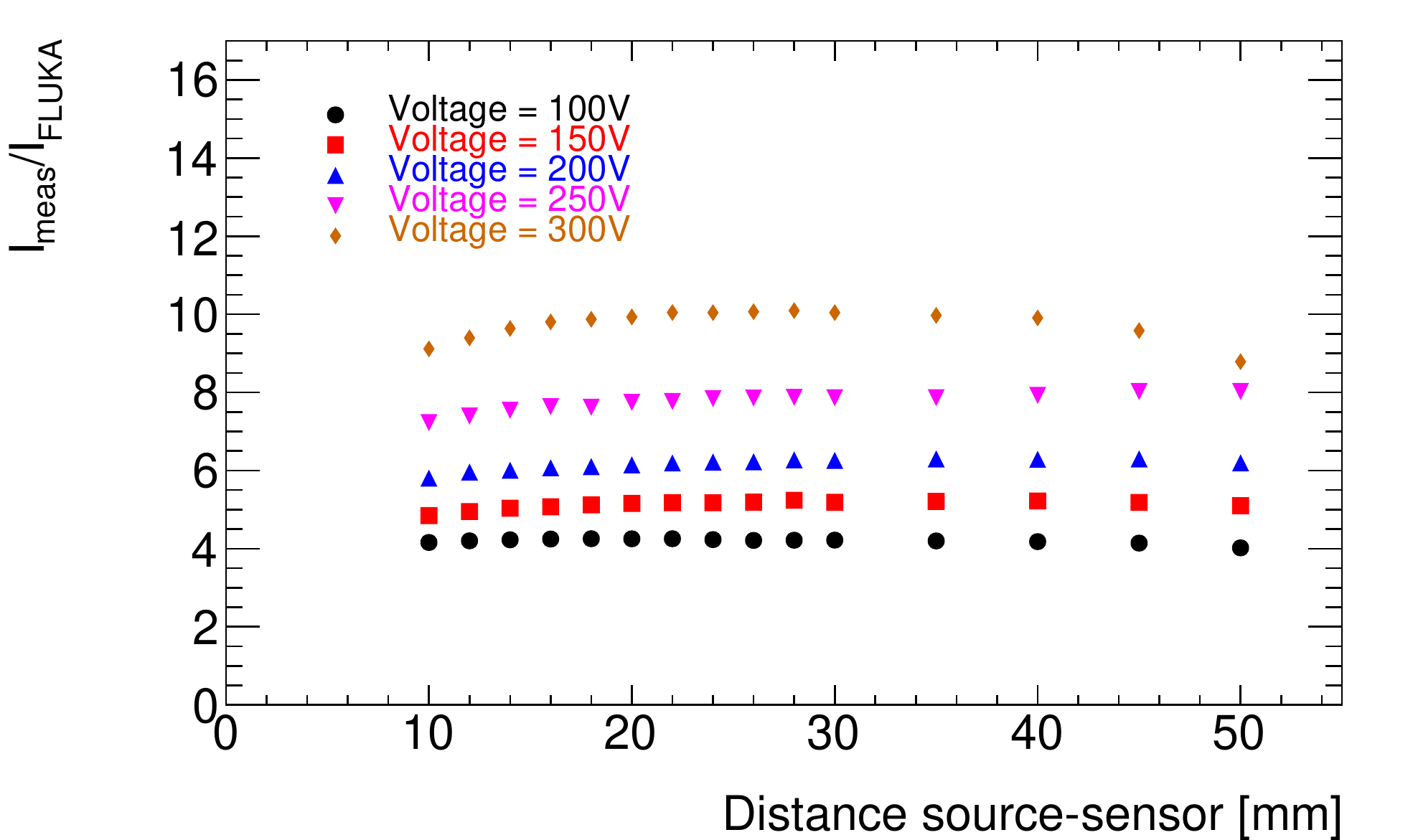}
	\caption{Example of the ratio of measured to predicted current $G = I_{meas}/I_{FLUKA}$ for one of the three single-crystal diamond sensors, as a function of the distance $d$, for different HV bias values. The ratio increases with HV bias but it is stable for a given voltage applied.}
	\label{fig:diamonds_calibration_1}
\end{figure}

\begin{figure}[ht!]
	\centering
		\includegraphics[scale=0.45]{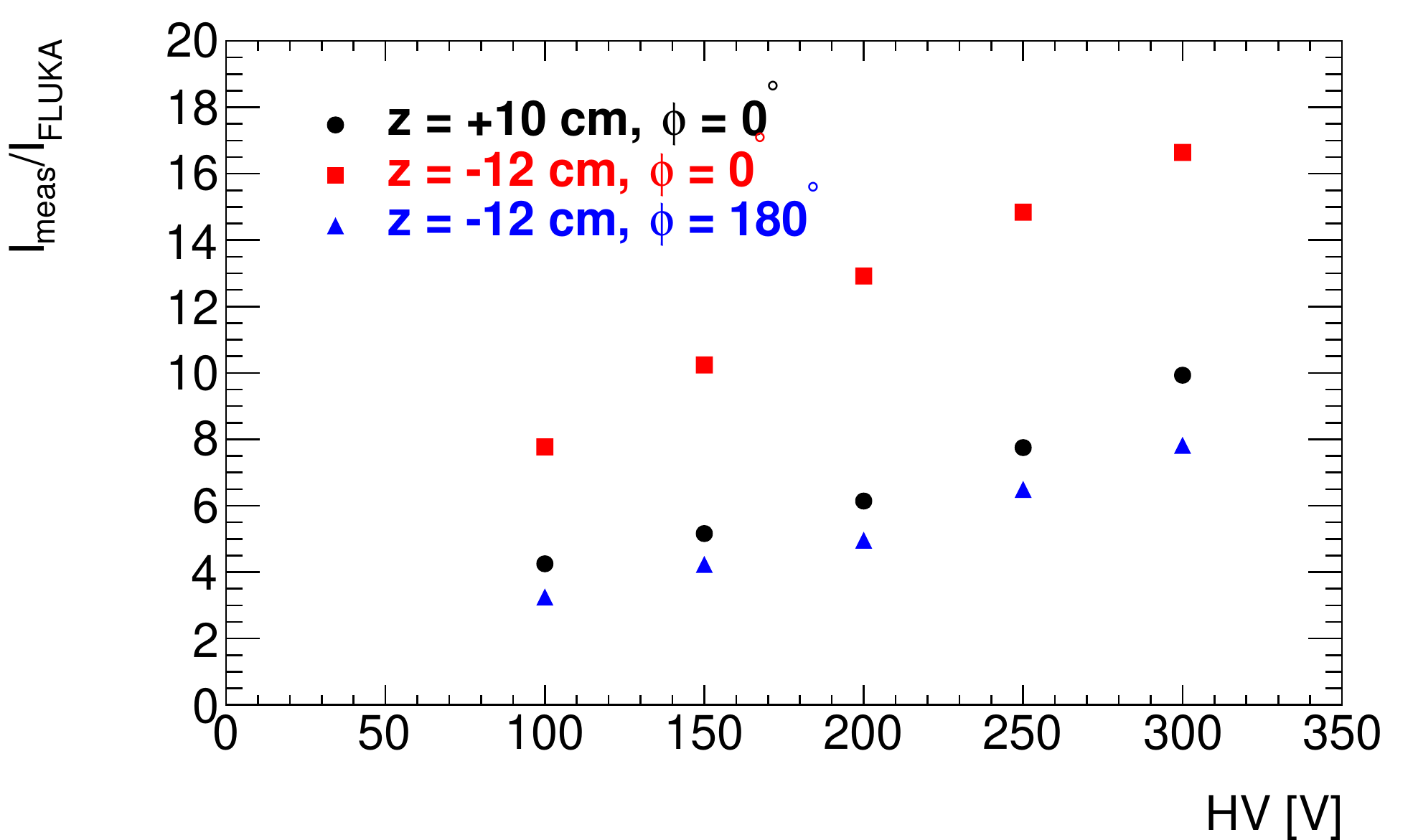}
	\caption{Gain factors $G$ for the three single-crystal sensors, as a function of the HV bias.}
	\label{fig:diamonds_calibration_2}
\end{figure}

Fig.~\ref{fig:diamonds_calibration_1} shows the ratio of measured to predicted current $G = I_{meas}/I_{FLUKA}$ for one of the diamonds as a function of the distance $d$, for different HV bias values. Excluding the small distance region where the $2.5$~nA measurement range saturates and the fit function describes less accurately the simulated current values, the values of $G$ at fixed HV are fairly constant as a function of $d$, implying stable sensor current response, linear with respect to the expected particle flux and dose rate.
Choosing $d = 20$~mm as a good reference position for dose rate, the corresponding values for the gain factor $G$ are shown in Fig.~\ref{fig:diamonds_calibration_2} as a function of the HV bias, for the three single-crystal sensors FW-0, BW-0 and BW-180. The operation at $100$~V HV bias guaranteed a stable and reproducible gain for them. The polycrystalline FW-180 sensor had a different, non-linear behaviour, with dose-rate dependent gain factor.

Taking into account the sensor volume $V = 4.5 \times 4.5 \times 0.5$~mm$^{3}$ and density $\rho = 3.52$~g~cm$^{-3}$, with a mass $M = \rho V = 3.56 \times 10^{-5}$~kg, the absorbed radiation-dose rate $\Delta D / \Delta t $ (deposited energy/mass in Gy) is:
\begin{equation}
\frac{\Delta D}{\Delta t} = \frac{1}{M}\frac{\Delta E[\mathrm{J}]}{\Delta t}
\end{equation}
A conversion factor $F_{cd}$ from simulated current $I_{FLUKA}$ to dose rate $\Delta D / \Delta t $ can be defined, and computed as:
\begin{eqnarray}
F_{cd} & = & \frac{\Delta D}{\Delta t} \frac{1}{I_{FLUKA}} = \\
           & = & \frac{1}{M} \frac{\Delta E[\mathrm{J}]}{\Delta t} \frac{\Delta t}{\Delta E[\mathrm{J}]} \frac{E_{eh}[\mathrm{J}]}{q_{e}[C]} = \\
           & = & \frac{E_{eh}[\mathrm{J}]}{M[\mathrm{kg}] q_{e}[\mathrm{C}]} = \\
           & = & 365 \frac{[\si{\micro}\mathrm{Gy/s}]}{[\mathrm{nA}]} = 36.5 \frac{[\mathrm{mrad/s}]}{[\mathrm{nA}]}
\end{eqnarray}

Therefore the measured dose rates $\Delta D / \Delta t$ can be obtained from the measured currents $I_{meas}$ using the conversion factor $F_{cd}$ and the overall gain factor $G$ measured for the given sensor:
\begin{equation}
\frac{\Delta D}{\Delta t} = F_{cd} I_{FLUKA} = F_{cd} \frac{I_{meas}}{G}
\end{equation}
The gain factors $G$ for the four diamond sensors at their operating HV bias values are shown in Table~\ref{tab:diamonds_gain}. Statistical errors in their determination are negligible with respect to systematic uncertainties, summarized in Table~\ref{tab:diamonds_gain_systematics}. The polycrystalline FW-180 sensor, operated at $300$~V, had a non-linear response in calibration tests. For this last case we determined its conversion factors as weighted averages over the typical observed current ranges.

\begin{table}[ht]
	\centering
	\caption{Measured diamond sensor gain factors $G$. The corresponding overall calibration factors $F_{cd}/G$ are expressed in (mrad/s)/nA at the specified HV operating values. The quoted uncertainties are dominated by the systematic uncertainties described in Table~\ref{tab:diamonds_gain_systematics}. Detector and sensor labels are those introduced in Table~\ref{tab:diamonds_loc}.}
	\begin{tabular}{llllc}
	\toprule
		Detector & HV [V] & $G$  &  $F_{cd}/G$ 	\\
		              &               &          & [(mrad/s)/nA]   \\                     \midrule
		FW-180 &   300     & $4.0 \pm 0.7$       & $9.1 \pm  1.6$ \\
		FW-0     &   100     & $4.3 \pm 0.7$  & $8.5 \pm  1.4$  \\
		BW-180 &   100      & $3.2 \pm 0.5$ & $11.4 \pm  1.8$  \\
		BW-0     &  100      & $7.8 \pm 1.3$ &  $4.7 \pm  0.8$ \\		\bottomrule
	\end{tabular}

	\label{tab:diamonds_gain}
\end{table}

\begin{table}[ht]
	\centering
	\caption{Contributions to the systematic uncertainties on the diamond sensor gain factor $G$.}
	\begin{tabular}{ lc }
		\toprule
		Origin of systematic uncertainty& $\delta G / G$	[\%]  \\  \midrule
		Source-diamond sensor distance  & $10$ \\
		Diamond sensor active volume     & $10$ \\
		Diamond sensor priming or pumping  & $5$ \\
		Source activity & $7$ \\
		FLUKA simulation statistics & $1$ \\
		HV reproducibility & $1$ \\ \hline
		Combination in quadrature & $17$ \\  \bottomrule
	\end{tabular}
	\label{tab:diamonds_gain_systematics}
\end{table}

 %\clearpage
 
 %lead author: Alex Beaulieu
 \subsection{Crystals detector system}\label{sec:crystals}
 %     file:		crystals.tex
%     author:  	Alexandre Beaulieu
%     timeline: September 10th
%
%     contents:  Functional description of Crystal system, to include the following:
%		physical description: size of sensitive volume, number of detectors etc
%		process by which radiation is detected. How signal is amplified, and read out
%		quantitative performance spec: radiation sensitivity, threshold, timing capability, noise rates, etc
%		description of in-situ calibration procedure
BEAST II contains an inorganic scintillator electromagnetic calorimeter, which we call the ``Crystals''. The main motivation for adding these components to BEAST II is their ability to measure the rate and spectrum of electromagnetic background radiation at the position corresponding to the innermost part of the Belle II calorimeter. With positions and detection technology similar to the Belle II electromagnetic calorimeter (ECL), their measurements therefore correspond to what would have been observed by the ECL had it been present in Phase 1. The Crystals have a precise enough time resolution to measure bunch-by-bunch beam-induced backgrounds, therefore they will also provide a measurement of the importance of the injection background relative to the Touschek and beam-gas components. 

\subsubsection{Crystals system physical description}
The Crystals system is made of six identical units, each containing three crystals read out with photo-multiplier tubes. The three crystal types are thallium-doped caesium iodide, CsI(Tl), pure caesium iodide, CsI(pure), and cerium-doped lutetium yttrium orthosilicate, LYSO. The CsI(Tl) crystals are 300~mm-long trapezoidal prisms where the small face measures approximately $45 \times 45 \text{~mm}^2$, and the large face measures approximately $60 \times 60 \text{~mm}^2$. The CsI(pure) and LYSO crystals are rectangular prisms measuring respectively $50 \times 50 \times 300 \text{~mm}^3$, and $25 \times 25 \times 200 \text{~mm}^3$.  The crystals are over 2.5 radiation lengths across and 15 radiation lengths long. Each crystal is wrapped with a diffuse reflector made of DuPont\texttrademark\, Tyvek\textsuperscript{\textregistered}~\cite{ref:dupont_tyvek}, and a light-tight wrapping of either aluminum foil or black adhesive tape.

\begin{figure}[ht!]
\centering
\includegraphics[width=\columnwidth]{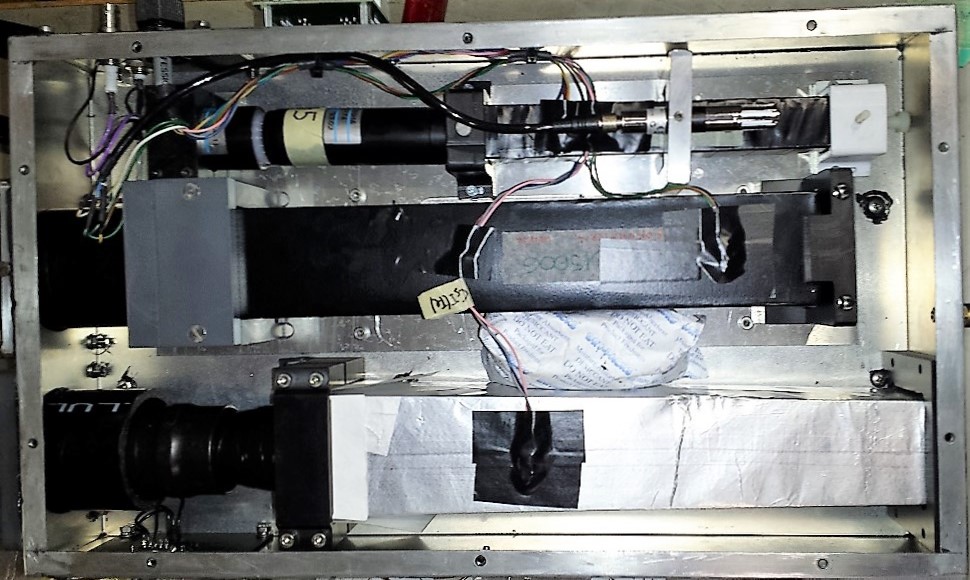}
\caption[Photograph of a Crystal calorimeter detector unit]{Photograph of a Crystal calorimeter detector unit. The LYSO crystal, smaller than the other two, is visible at the top. The undoped CsI crystal is in the center of the image, and the CsI(Tl) crystal is at the bottom. The box is instrumented with three temperature sensors and one relative humidity sensor.}
\label{fig:crystalsDarkBoxInside}
\end{figure}

A picture showing one unit of the Crystal system is shown in Figure~\ref{fig:crystalsDarkBoxInside}.
The dark boxes also contain environment monitoring sensors: one for relative humidity and three temperature probes.
%, which are described in more detail in section~\ref{sec:usop}.
The boxes are sealed to ensure they are light and air tight, and a packet of silica gel is added to control the relative humidity inside.

The locations of the six boxes correspond to the positions of the end-cap crystals in the Belle II electromagnetic calorimeter, with three units in the backward positions at $z=-1240\pm25\text{~mm}$, and three units in the forward positions at $z=2155\pm25\text{~mm}$. The boxes are at $\phi$ angles of $0^\circ$, $90^\circ$ and $180^\circ$, for both the backward and forward sides and at $\theta$ angles of $29\pm1^\circ$ and $14\pm1^\circ$, equivalent to the tilt angle of the Belle II end-cap crystals in the backward and forward sides, respectively. The coordinate system of the BEAST II detector is described in section \ref{sec:beastLayout}.

\subsubsection{Crystals system principle of operation}\label{sec:cryReadout}
The crystals are inorganic scintillators that act as electromagnetic calorimeters in which electrons and photons interacting with the crystal generate a shower and produce visible light proportional to the deposited energy. The scintillation light is collected and converted to an electronic signal by photomultiplier tubes: Hamamatsu model R580\cite{Hamamatsu} for the CsI(Tl) channels, Photonis XP2262 \cite{Photonis} for the CsI(pure) channels, and Hamamatsu R1355HA \cite{Hamamatsu} for the LYSO channels. The signals of each tube are connected to CAEN model V1730 digitizers \cite{CAEN}. The LYSO and CsI pure channels have an in-line  attenuator --- 20~dB for LYSO, 12~dB for CsI(pure) --- to match the signal amplitude to the 2~V input range of the digitizer. The V1730 is a VME module with 500~MS/s sample rate and 14-bit resolution. We use one 16-channel and one 8-channel otherwise identical boards to readout all our signals. Both digitizer boards are equipped with 5.12~MS memory, which enables recording up to 10~ms of data continuously during injection noise measurement studies.

The digitizers also exploit firmware-level digital pulse processing (DPP) algorithms to calculate and record, for each signal pulse, the integrated current (the charge) over two different gate lengths, the baseline level measured before each pulse, and the trigger time. This dramatically reduces data throughput compared to recording the full signal waveforms. The acquisition starts independently on each channel, self-triggering on the signals. The triggers are gated by a nominal 10 ms time window synchronized with the injection signal delivered by SuperKEKB and a nominal 1 ms time-window at 2~Hz ---  uncorrelated to the injection signal --- to pre-scale the acquisition by a factor of 500 between injections. The two gates are added together to create a logical OR.

\begin{figure}
	\centering
        \subfigure[]{
          \includegraphics[width=\columnwidth]{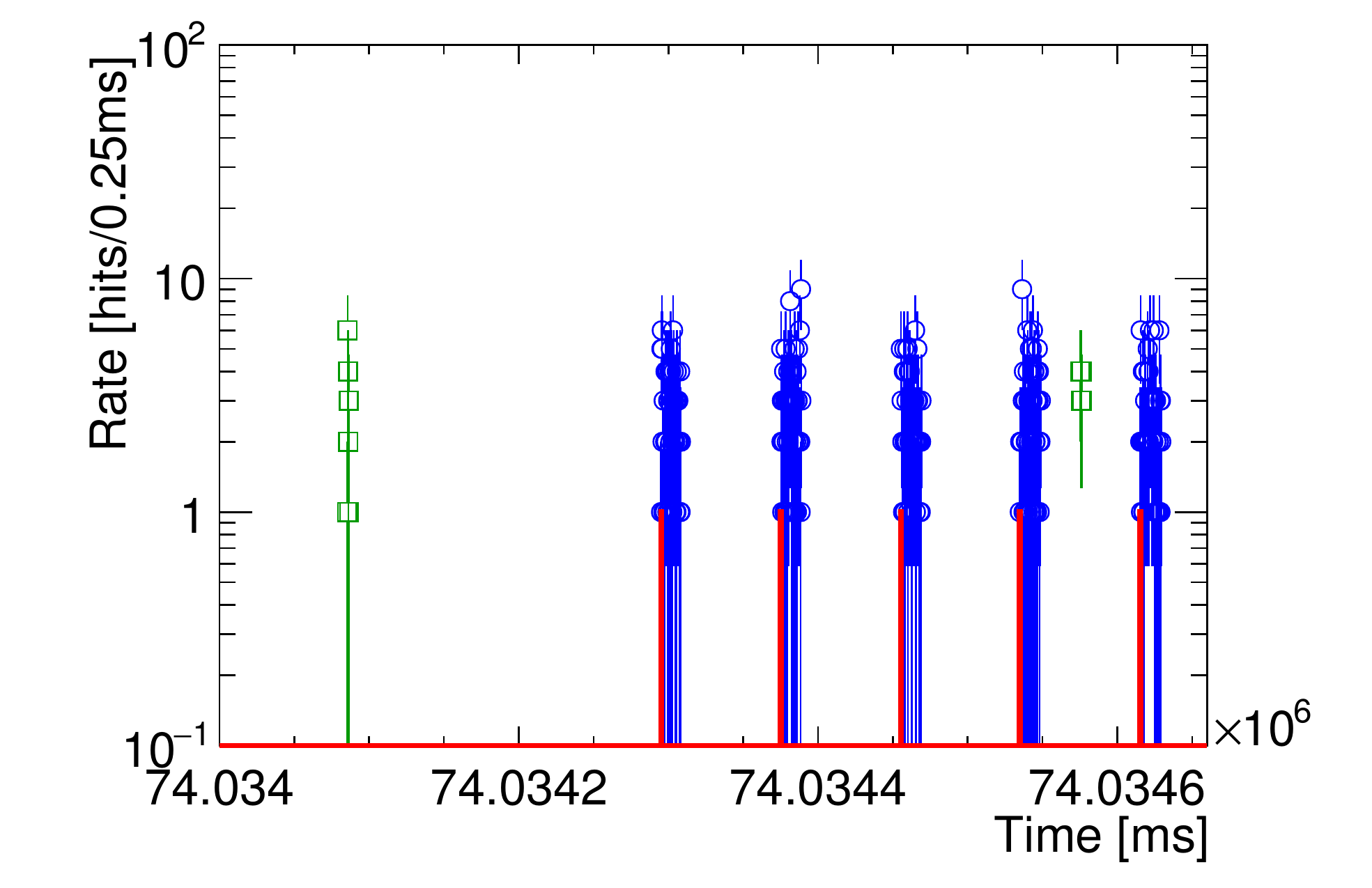}
          \label{fig:timeStruc-1}
        }
        \subfigure[]{
                \includegraphics[width=\columnwidth]{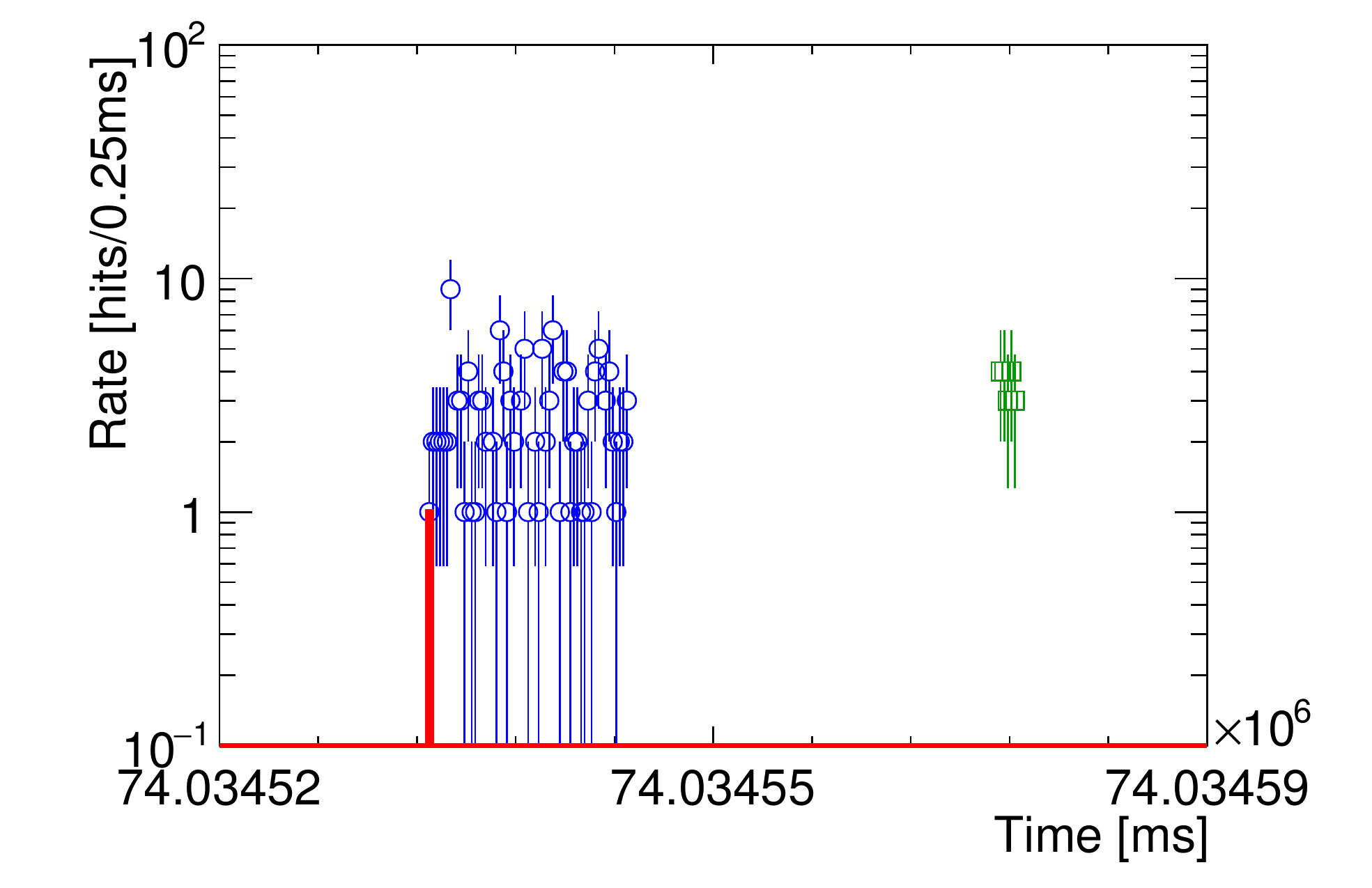}
                \label{fig:timeStruc-2}
        }
	\caption{
                (a) (color online) Plot of the hit rate in CsI crystals as a funtion of the acquisition time showing the time structure due to the data acquisition settings described in the text.
                Green squares represent the rate collected during the nominal 1~ms gates generated asynchronously every 500~ms, whereas the blue circles represent the rate collected during 
                the nominal 10~ms gates in coincidence with the arrival of the injection signal from SuperKEKB that are, in turn, indicated by the red histogram (12.5~Hz injection rate in this case).
                (b) (color online) Zoomed in version of the plot in (a), where the different width of the acquisition gates can be seen more clearly.
        }
\end{figure}

The time structure of the data acquisition with its 2~Hz, 1~ms gates off-injection and 10~ms gates on-injection is evident in Figs.~\ref{fig:timeStruc-1} and~\ref{fig:timeStruc-2}, in which the hit rate is plotted as a function of the hit time.

Moreover, the fast signals from LYSO and CsI (pure) are also directed to a scaler unit (CAEN model V830 \cite{CAEN}) in order to obtain instantaneous counts independently of the signal digitization. This feature is useful to provide a normalization for the energy spectra, and also to keep track of larger expected rates during injections and pressure bump experiments.

The digitizers and scalers are driven by a Motorola VME controller \cite{MotorolaVME}, in turn connected to a standard PC which serves as an interface to the rest of the BEAST II data acquisition platform described in section~\ref{sec:beast}.

\subsubsection{Crystals system performance}
The performance specification depends on the crystal material for each channel type. The key nominal metrics are reported in Table~\ref{tab:CsI-Specs} as a function of the scintillator material.

\begin{table}[ht!]
\caption{Nominal performance specifications for each of the Crystal materials used. The energy resolution quoted is at 1.3~MeV.}\label{tab:CsI-Specs} 
% Data from CrystalsCalibration-2016-01-28.xlsx
\begin{tabular}{lccc}
\toprule
 & CsI(Tl) & CsI(pure) & LYSO\\
\midrule
Energy threshold [MeV]                  &    0.3 &    0.8 &   0.1\\ 
Energy range [MeV]                      &  330   & 3700   & 610  \\
Energy resolution [\%]       &   20   &   15   &   7  \\
% \quad equivalent noise (MeV) & \\
Signal decay time\footnotemark[1] [ns] & 1000   &   16   &  41  \\ 
\bottomrule
\end{tabular}
\parbox{\columnwidth}{\footnotemark[1]{\footnotesize According to manufacturer data \cite{stgobain-lyso, stgobain-csi, stgobain-csitl}}}
\end{table}

\subsubsection{Crystals system calibration}\label{sec:CrystalCal}
\paragraph{Initial calibration}
We conducted two calibration campaigns: the first immediately after the installation of the experimental apparatus, and the second during data-taking after changing the PMT supply voltages. We used two radioactive sources, $^\text{137}$Cs with 273~kBq activity (one photo-peak at 0.662~MeV), and $^\text{60}$Co with 431~kBq activity (two photo-peaks at 1.173~MeV and 1.333~MeV) to obtain a four-point calibration curve where
\begin{itemize}
	\item $E_1 = 0$ from signal at random triggers;
	\item $E_2 = 0.662$~MeV from the $^\text{137}$Cs photo-peak;
	\item $E_3 = 1.253$~MeV from the average of the two  $^\text{60}$Co photo-peaks;
	\item $E_4 \approx 30$~MeV (depending on the crystal size and orientation) from the energy of a cosmic ray minimum-ionizing particle (MIP) passing through the crystal\footnote{The acquisition is self-triggered on the individual signals. We use the most probable value of the deposited energy distributions as the calibration point, and assume for calculations that this energy value corresponds to a cosmic-ray particle coming from the zenith.}.
\end{itemize}	

We placed the sources outside the dark boxes, separated from the crystals by 1.6~mm of aluminum alloy and approximately 2~cm of air. %, as seen in Figure~\ref{fig:CalibSource}.

\begin{figure}[ht!]
\centering
\includegraphics[width=\columnwidth]{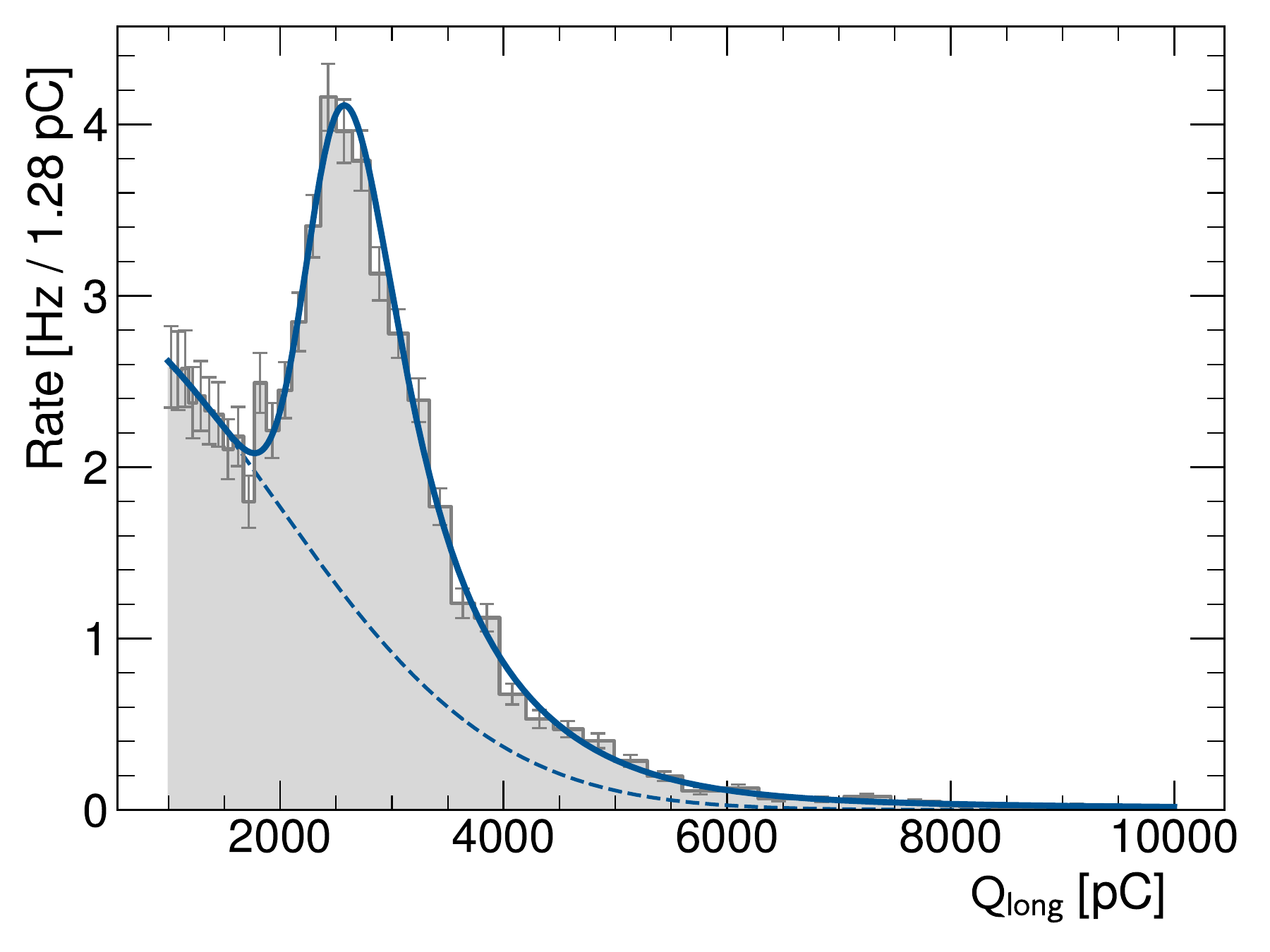}
\caption[Example of the charge distribution of cosmic ray muon signals in CsI(Tl) used for calibrations]{Example of the charge distribution of cosmic ray muon signals in CsI(Tl) used for energy calibration. Uncertainties on the bin contents are statistical only. The optimal parameters of Equation~\ref{eqn:Crystals_MIP_Peak} for this distribution are $a_{bkg}=\left(2.98 \pm 0.12\right)\text{ Hz/pC}$, $b_{bkg} = \left(1.31\times 10^{-6} \pm 2.9\times 10^{-8}\right)\text{ pC}^{-2}$, $MPV = \left(2.62\times 10^{3} \pm 77\right)\text{ pC}$, $\eta = \left(208 \pm 9\right)\text{ pC}$, $\sigma = \left(226 \pm 30\right)\text{ pC}$, $A = \left(2.9\pm0.1\right)\text{ Hz/pC}$ with a fit $\chi^2/\text{ndf} = 21.8/34$. The dashed line represents the contribution of the Gaussian-tail background only. }
\label{fig:CsITl_MIP_Calib_Fit}
\end{figure}

For the calibration data using test sources, we fitted a Gaussian distribution to each spectrum to obtain the average charge and charge as a function of the deposited energy. For data utilizing minimum ionizing cosmic ray muons, we modeled the charge distribution $f_\text{cosmic}(q)$ as the sum of a signal part and a background part. The signal is represented as a Landau distribution, peaking at $MPV$ with a scale parameter $\eta$, convoluted with a Gaussian centered at 0 with standard deviation $\sigma$. The background contribution is represented by a Gaussian tail with width parameter $b_{bkg}$:
\begin{multline}\label{eqn:Crystals_MIP_Peak}
f_\text{cosmic}(q) = a_{bkg} \cdot \text{exp}\left( -b_{bkg}q^2\right) + \\ A\cdot\text{Landau}\left(q, MPV, \eta\right) \ast \text{Gaussian}\left(q, \mu=0, \sigma\right).
\end{multline}
In this formulation, parameters $a_{bkg}$ and $A$ represent the relative contributions of the background and signal components, respectively. Figure~\ref{fig:CsITl_MIP_Calib_Fit} shows an example of such a fit.

The equation used to calculate the deposited energy corresponding to a integrated charge $Q_\text{long}$ read out from the digitizers includes the energy calibration:
\begin{equation}\label{eqn:CsiCalibration}
E~\left[\text{MeV}\right] = \frac{a Q_\text{long} - c}{b}
\end{equation}
where the parameter $a$ comes from the setting of the DSP algorithm in the digitizer, and parameters $b$ and $c$ are obtained from a linear fit to the energy points described above. The measured parameters from the initial calibration campaign are reported in Table~\ref{tab:CsI-JanCalibration}.
\begin{table}[ht!]
\caption{Crystals system calibration parameters versus crystal channel number (\#) resulting from the initial calibration campaign. Parameter $a$ is a hard-coded value in the DSP settings. The values and uncertainties for $b$ and $c$ are obtained by the linear least squares algorithm.}
\centering
\label{tab:CsI-JanCalibration}
\begin{tabular}{cccr@{$\pm$}lr@{$\pm$}l}
\toprule
\#& Material &  $a$~[pC/LSB]    & \multicolumn{2}{c}{$b$~[pC/MeV]}    & \multicolumn{2}{c}{$c$~[pC]}\\ 
 \midrule
 0 & CsI(Tl)   & 1.28 & 167.4  & 0.7 & -14.1 & 11\\
 1 & CsI(pure) & 0.32 &   5.45 & 0.3 &  0.74 & 0.3\\
 2 & LYSO      & 0.32 &  15.8  & 0.6 &   1.2 & 0.8\\
 3 & CsI(Tl)   & 1.28 & 102.6  & 0.3 &  -9.1 & 5\\
 4 & CsI(pure) & 0.32 &   5.78 & 0.3 &  1.13 & 0.3\\
 5 & LYSO      & 0.32 &  19.2  & 0.8 &   1.5 & 1\\
 6 & CsI(Tl)   & 1.28 & 132.9  & 0.4 &  -0.6 & 6\\
 7 & CsI(pure) & 0.32 &   5.81 & 0.3 &  0.78 & 0.3\\
 8 & LYSO      & 0.32 &  18.4  & 0.7 &  -0.8 & 1\\
 9 & CsI(Tl)   & 1.28 & 129.6  & 0.4 & -11.1 & 7\\
10 & CsI(pure) & 0.32 &   5.76 & 0.3 &  1.11 & 0.3\\
11 & LYSO      & 0.32 &  17.3  & 0.7 &   0.2 & 1\\
12 & CsI(Tl)   & 1.28 & 114.1  & 0.1 &   2.8 & 2\\
13 & CsI(pure) & 0.32 &   5.39 & 0.3 &  1.36 & 0.3\\
14 & LYSO      & 0.32 &  15.1  & 0.6 &     4 & 1\\
15 & CsI(Tl)   & 1.28 & 128.6  & 0.5 &  -7.6 & 8\\
16 & CsI(pure) & 0.32 &   5.39 & 0.3 &  1.29 & 0.3\\
17 & LYSO      & 0.32 &  17.3  & 0.7 &  0.64 & 0.9\\
\bottomrule
\end{tabular}
\end{table}

\paragraph{Dose dependence of the gain}
Unfortunately, due to radiation damage, the calibrations of Table~\ref{tab:CsI-JanCalibration} changed during the Phase 1 operating period. The gain of all channels was degraded, sometimes significantly, by a combination of damage to the crystals themselves --- most likely for the CsI(Tl) channels --- and to the PMTs.

In order to recover from this, we use data points recorded when neither beam was circulating to measure the position of the MIP peak. On a daily basis, we populate histograms such as the one presented in Figure~\ref{fig:CsITl_MIP_Calib_Fit} with these these ``beam-off'' events, and attempt parameter estimation. We then use successful results, defined based on a list of criteria such as convergence of the algorithm, estimated parameters away from boundaries and adequate signal-to-noise ratio, to measure the shift of the peak's position as a function of total integrated current, referred to here as ``beam dose''. We take the ratio between the measured peak position and the value recorded during the initial calibration period, reported in Table~\ref{tab:CsI-JanCalibration}, and call the resulting quantity the ``relative gain'' of the channel.

The relative gain is parametrized as a linear function of $\log_{10}$ of the total beam dose, and therefore equation~\ref{eqn:CsiCalibration} is modified to include this term:
\begin{equation}\label{eqn:CrystalsGainLogDepedence}
E\left(Q_\text{long}, I_\text{int}\right)~\left[\text{MeV}\right] = \frac{a Q_\text{long} - c}{b} \frac{1}{p_0 + p_1\log_{10} I_\text{int}}
\end{equation}
where $I_\text{int}$ is the sum of the integrated beam currents in both beams, and $p_0$, $p_1$ are parameters adjusted to data, specific to each channel but constant throughout the experiment.

\begin{figure}[ht!]

\centering
\includegraphics[width=\columnwidth]{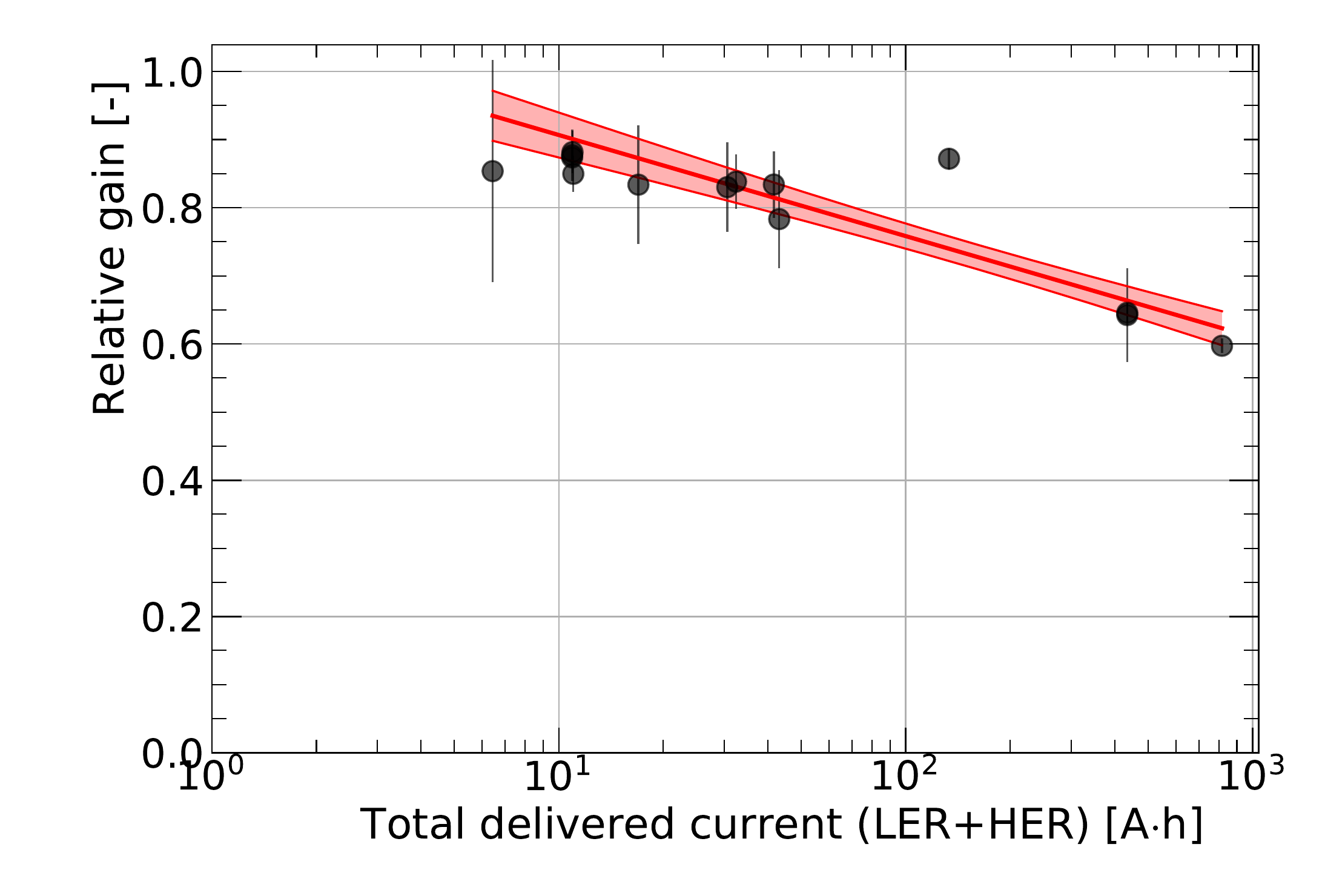}
\caption[Example of the dose dependence of a Crystal channel's relative gain]{(color online) Example of the dose dependence of a Crystal channel's relative gain. Channel 11: LYSO at position B1 is presented. Black markers show measurements, with error bars indicating the uncertainty on the position of the MIP peak output by the {\tt HESSE} algorithm. The red line shows the best fit, with the shaded area indicating a 1-$\sigma$ confidence interval around the best-fit curve. The optimal parameters for the logarithmic dependence of Equation~\ref{eqn:CrystalsGainLogDepedence} are $p_0=1.06\pm0.05$, $p_1=-0.148\pm0.024$. }
\label{fig:CsI-ExCalVsDose}
\end{figure}

An example result is presented in Figure~\ref{fig:CsI-ExCalVsDose}. %Full results are reported in appendix~\ref{app:Crystal_Cal}. 

 %\clearpage

 % lead author: Lin Jie-Cheng (Jason)
 \subsection{BGO detector system}
 \label{sec:BGOs}

The bismuth germanium oxide (BGO) detector system is designed for monitoring real-time beam backgrounds in the
form of electrons and gammas. It is also capable
of monitoring the luminosity of the collider by counting Bhabha event
rates if the beams are focused.

\subsubsection{BGO system physical description}

\begin{figure*}[tp]
\centering\includegraphics[width=14cm]{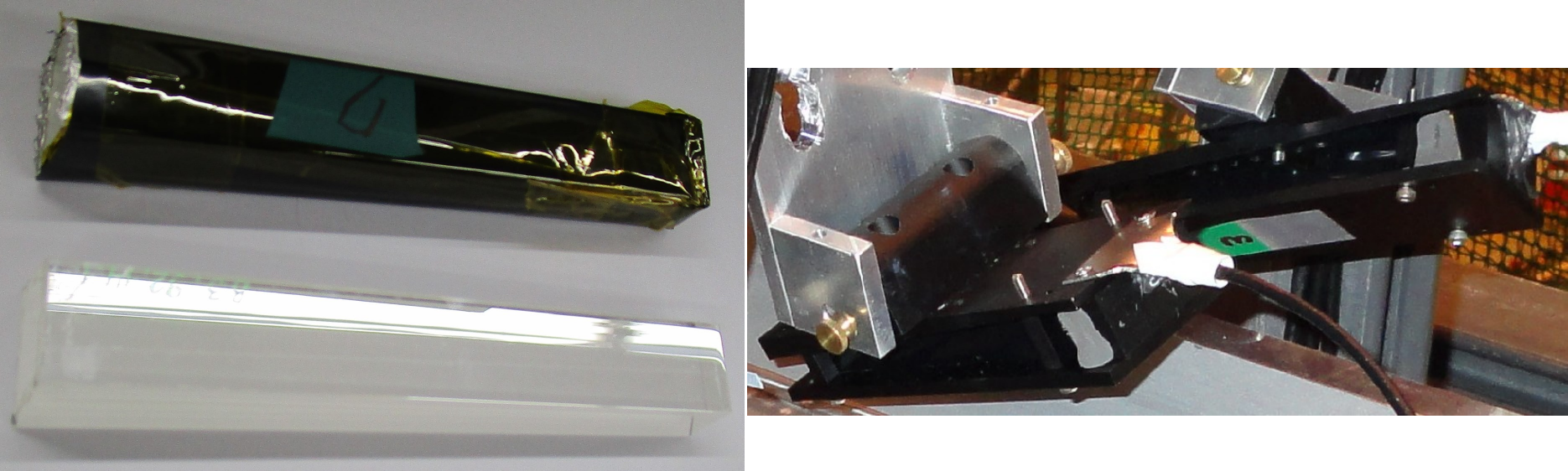}

\caption[Left: a BGO crystal with (top) and without (bottom) the wrapping material.
Right: the BGO crystals with the supporting structures.]{Left: a BGO crystal with (top) and without (bottom) the wrapping
material. Right: the BGO crystals with the supporting structures.}
\label{fig:Left: a BGO crystal with (top) and without (bottom) the wrapping material. Right: the BGO crystals with the supporting structures} 
\end{figure*}

From the extreme forward calorimeter (EFC) \cite{EFC} of the Belle
detector, the BGO detector system reuses the
scintillating BGO crystals as its sensors (see Figure \ref{fig:Left: a BGO crystal with (top) and without (bottom) the wrapping material. Right: the BGO crystals with the supporting structures}).
Each BGO crystal occupies approximately $2\times2\times13$~cm$^{3}$,
and has a mass of about 0.3135~kg. Light-tight treatments are applied
to the crystals to achieve maximum light-collection efficiency and
also prevent leakage of light from the environment. We installed eight
BGO crystals around the IP of the Belle II experiment:
four in the forward region and four in the backward region.

\noindent 

%\noindent %\begin{figure*}[tp]
%\centering\includegraphics[width=14cm]{BGOImages/BGO_system}

%\caption[Eight BGO sensors installed around the IP.]{Eight BGO sensors installed around the IP. (a) Photograph of the
%sensors. (b) CAD rendering of the sensors.}
%\label{fig:Eight BGO sensors installed around the IP}
%\end{figure*}

\subsubsection{BGO system principle of operation}

\begin{figure}[tbph]
\centering\includegraphics[width=9cm]{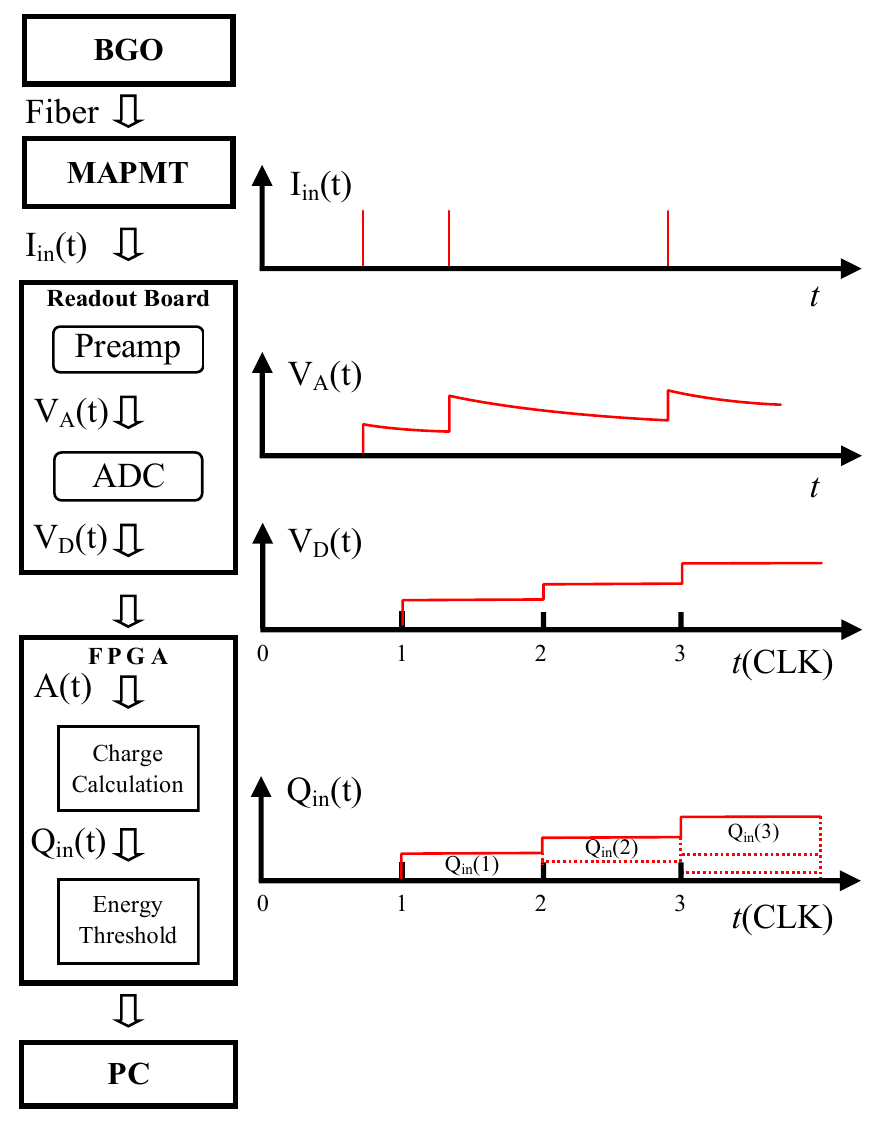}\caption[Signal-flow graph of the BGO detector system.]{Signal-flow graph of the BGO detector system.

\textbf{$I_{in}(t)$}: the charge signal from the MAPMT to the preamplifier
(in Amperes).

$V_{A}(t)$: the analog voltage signal from the preamplifier to the
ADC (in Volts).

$V_{D}(t)$: the digitized voltage signal from the ADC to the FPGA
(in ADUs; analog-to-digital units).

$A(t)$: the digital signal in the FPGA (in ADUs).

$Q_{in}(t)$: the accumulated input charge in the increment of the
25~ns clock period (in ADUs).}
\label{fig:Signal-flow graph of the BGO detector system}
\end{figure}

A signal-flow graph of the BGO detector system is shown in Figure \ref{fig:Signal-flow graph of the BGO detector system}.
For gamma rays traveling
inside a BGO crystal, scintillation photons with wavelengths in the
range from 375~nm to 650~nm are emitted. The scintillation light
is then guided to a Hamamatsu H7546 MAPMT by a single optical fiber. The
MAPMT, which has a spectral interval of 300\textendash 650~nm, converts
the scintillation light to the charge signal. Then, the charge signal
is fed to an FPGA-based readout system. The readout system
consists of an eight-channel readout board and an FPGA board. The
preamplifier in the readout board converts the charge signal to the
voltage signal with a 187~ns shaping time. The voltage signal is
then transmitted to the FPGA after digitization by a 10-bit pipeline
analog-to-digital converter (ADC). With the 25~ns clock period of
the FPGA and the 187~ns shaping time of the preamplifier, we use
the following algorithm to obtain the input charge:

\begin{eqnarray}
Q_{in}(t) & = & A(t)-A(t-1)\times e^{\left(-\frac{25}{187}\right)},\\
 & \approx & A(t)-A(t-1)\times\frac{7}{8}.
\end{eqnarray}
We also apply a 5~ADU energy threshold to \textbf{$Q_{in}(t)$} to
exclude random electronic noise in the DAQ system. The FPGA communicates with a PC via RS-232 protocol, and the software of the BGO system is integrated into the BEAST II DAQ system.
The full 66Hz accumulated background energy information is saved to disk for offline analysis, while a 1~Hz rolling average of the same information is provided to online monitors over EPICS.

\noindent %\begin{figure}[tbph]
%\centering\includegraphics[width=9cm]{BGOImages/Schematic_diagram2}\caption[Schematic of the BGO detector system.]{Schematic of the BGO detector system.}
%\label{fig:Schematic of the BGO detector system}
%\end{figure}

\noindent %\begin{figure*}[tp]%\centering\includegraphics[width=19cm]{BGOImages/Circuit_diagram}\caption[Circuit diagram of the eight-channel readout board of the BGO detector%system.]{Circuit diagram of the eight-channel readout board of the BGO detector%system. Each channel consists of a preamplifier circuit and a 10-bit%ADC chip. For simplicity, only one channel is shown in this figure.}%\label{fig:Circuit diagram of the eight-channel readout board of the BGO detector system}%\end{figure*}

\subsubsection{BGO system performance}
Radiation damage causes losses in light yield of BGO crystals. This was studied in Belle and as part of the BEAST II BGO effort \cite{EFC,BGO}. The results
show that the light outputs drop by 30\textendash 45\% after the crystals
receive doses of 2\textendash 4 krad and remain stable afterwards.
On the basis of these results, we assume that the radiation damage
to the BGO crystals in Phase 1 was negligible. Since the dose rates
of the BGO crystals in Phase 1 were very low compared with those in the dose damage tests, the rates of the recombination and thermal release
(relaxation process) of the trapped charge carriers could suppress
the radiation damage of the BGO crystals. This assumption agrees with
our observations.

We define the radiation sensitivity for one channel of the detector
as the calibrated value for the dose of the BGO crystal normalized
to one ADU count. Radiation sensitivities depend on the impurities
in the BGO crystals, the wrapping methods, and the quality of the
fibers for transmission. In our case, the major reason for the variation
of the radiation sensitivities is the quality of the fibers installed
at KEK. The radiation sensitivities of the BGO detector channels are
summarized in Table~\ref{table:Radiation sensitivity of each BGO detector system channel, with the MAPMT operating at 700 V.}.
All channels functioned well except for one dead channel (Channel
\#0) where the optical fiber was fractured on the side of the BGO during installation.
Electronic noise was the major source of noise in the BGO detector
system. However, with the 5-ADU integrated energy threshold, the noise
from the electronics was less than 1~ADU per clock period.

\noindent 
\begin{table}[tbph]
\centering{}\caption[Radiation sensitivity of each BGO detector system channel, with the
MAPMT operating at 700~V.]{Radiation sensitivity of each BGO detector system channel, with the
MAPMT operating at 700~V.}
\label{table:Radiation sensitivity of each BGO detector system channel, with the MAPMT operating at 700 V.}
\begin{tabular}{cccc}
\toprule 
Channel \# & $z$~{[}cm{]}  & $\phi$~{[}\textdegree {]}  & Sensitivity {[}Gy/ADU{]} \tabularnewline
\midrule 
0 & 31.2 & 323.0 & No response\tabularnewline
1 & 34.0 & 119.8 & $(4.58\pm0.59)\times10^{-11}$\tabularnewline
2 & 32.8 & 242.0 & $(1.18\pm0.17)\times10^{-11}$\tabularnewline
3 & 32.8 & 73.3 & $(2.28\pm0.30)\times10^{-10}$\tabularnewline
4 & -20.7 & 35.6 & $(1.19\pm0.17)\times10^{-10}$\tabularnewline
5 & -23.9 & 130.3 & $(6.11\pm1.51)\times10^{-11}$\tabularnewline
6 & -22.6 & 248.5 & $(3.82\pm0.95)\times10^{-10}$\tabularnewline
7 & -22.5 & 296.9 & $(1.14\pm0.21)\times10^{-10}$\tabularnewline
\bottomrule
\end{tabular}
\end{table}

\subsubsection{BGO system calibration}

The BGO calibration procedures and findings before the BEAST II installation
are described in detail in Ref. \cite{BGO}. We determine the gain factor for each
channel of the MAPMT by shining pulsed LED light onto
each pixel with the operation voltage of the MAPMT set at 700\,V.
We then fit the accumulated charges of each pixel to a Poisson-like
functional form in order to ascertain the gain factor in units of
ADU/p.e. (photoelectron). Table~\ref{table:Gain factor for each channel of the MAPMT operating at 700 V}
summarizes the gain factor for each channel of the MAPMT.

\begin{table}[tbph]
\centering{}\caption[Gain factor for each channel of the MAPMT operating at 700 V.]{Gain factor for each BGO system MAPMT channel, when operating at
700~V \cite{BGO}.}
\label{table:Gain factor for each channel of the MAPMT operating at 700 V}
\begin{tabular}{cc}
\toprule 
Channel \# & Gain {[}ADU/p.e.{]}\tabularnewline
\midrule 
0 & $10.90\pm0.40$\tabularnewline
1 & $13.18\pm0.43$\tabularnewline
2 & $10.82\pm0.42$\tabularnewline
3 & $12.65\pm0.54$\tabularnewline
4 & $10.58\pm0.46$\tabularnewline
5 & $11.93\pm0.64$\tabularnewline
6 & $10.51\pm0.42$\tabularnewline
7 & $11.62\pm0.61$\tabularnewline
\bottomrule
\end{tabular}
\end{table}

We obtain the p.e. yield (p.e./GeV) of the BGO detector system from
the results of cosmic ray tests. A simple setup with one BGO crystal
sandwiched between two triggering plastic scintillators was constructed
in a temperature-controlled room. We fit the obtained charge distribution
with a Landau-Gaussian function, and matched its peak value to that
from a simulation. The p.e. yield is determined to be $27.1\pm1.5$
p.e./GeV with ideally zero meter fibers for transmission.

To ensure the success of the data taking in the Phase 1 commissioning,
we tested the BGO detector system in an irradiation facility in LongTan,
Taiwan. The $^{60}$Co source used for irradiation had an activity
of around $1.11\times10^{14}$ Bq. We monitored the doses received
by our BGO crystals in real time and compared the total accumulated
doses with the commercial dosimeters placed on top of each BGO crystal.
The results agree within uncertainty. We also observed a trend of
the light yield dropping by up to 40\% which was due to the radiation
damage to the BGO crystals.

\begin{figure}[tbph]
\centering\includegraphics[width=9cm]{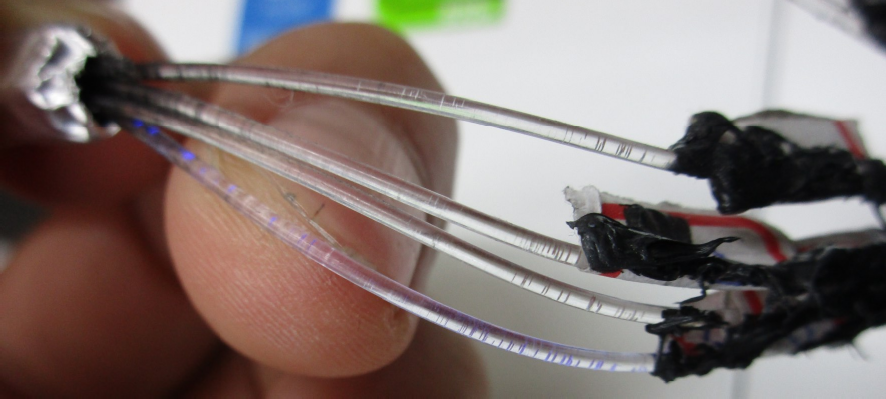}\caption[Scars on the fibers close to the MAPMT of the BGO detector system.]{Scars on the fibers close to the MAPMT of the BGO detector system.}
\label{fig:Scars on the fibers close to the MAPMT of the BGO detector system} 
\end{figure}

During the Phase 1 commissioning, the signals of the BGO detector
system were considerably weaker than we had expected. We examined
the detector, and discovered some scars on the fibers close to the
MAPMT that might have been caused during the installation. These scars,
as shown in Figure \ref{fig:Scars on the fibers close to the MAPMT of the BGO detector system},
were verified as a leading cause of the huge uncertainty of the signals.
Hence, we performed an \emph{in situ} recalibration after the Phase 1 commissioning by measuring the attenuation
of the signal in each 37~m long fiber with a 450~nm blue light LED
pulsed by a portable generator. Each pulse would give a burst of photons
into the fiber, and some of them would reach the MAPMT. Then, we cut
each fiber into 3 segments: 
\begin{description}
\item [{$A$:}] the fiber's end on the side of the MAPMT, 
\item [{$B$:}] the fiber's 37~m long major trunk, 
\item [{$C$:}] the fiber's end on the side of the BGO crystal. 
\end{description}
Under the same amount of pulse energy, we measured the pulse heights
with $A$ segments, $B$ segments, and ideally zero meter fibers for
transmission, respectively. We did not measure the attenuation of
the signals in $C$ segments because the remaining fibers on the side
of the BGO crystals were too short. We assume that the damage to the
fibers on the side of the BGO crystals was negligible. This assumption
agrees with our observations. The attenuation of the signal in each
fiber is obtained by choosing the results obtained from the ideally
zero meter fibers as the reference, as given in Table~\ref{table:Attenuation of the signals in the fibers of the BGO detector system}.
The results show unevenness of attenuation across different channels,
implying the damage to the fibers on the side of the MAPMT was severe,
especially for Channel \#3 and Channel \#6. The significant fiber attenuation is corrected by scaling up the measured doses.

%\begin{figure}[tbph]
%\centering\includegraphics[width=9cm]{BGOImages/Equipment_LED_test}
%\caption[Equipment for measuring the attenuation of the signals in the fibers
%of the BGO detector system.]{Equipment for measuring the attenuation of the signals in the fibers
%of the BGO detector system. (a) The portable function generator and
%the 450 nm blue light LED. The portable function generator is equipped
%with an FPGA board which is capable of pulsing the LED at 0.1 MHz.
%(b) A wooden plug which functions as a collimator and fixture.}
%\label{fig:Equipment for measuring the attenuation of the signals in the fibers of the BGO detector system}
%\end{figure}

\begin{table}[tbph]
\centering{}\caption{Signal attenuation in the optical fibers of each BGO detector system
channel. $A_{fiber}$, $A_{fiber}^{body}$, and $A_{fiber}^{MAPMT}$
are defined as the percentage of signal remaining after the attenuation
in the entire fiber, the fiber's 37\,m major trunk, and the fiber's
end on the side of the MAPMT, respectively. They are related by the
formula: $A_{fiber}=A_{fiber}^{body}\times A_{fiber}^{MAPMT}$.}
\label{table:Attenuation of the signals in the fibers of the BGO detector system}
\begin{tabular}{crrr}
\toprule 
Channel \# & $A_{fiber}$~{[}\%{]}  & $A_{fiber}^{body}$~{[}\%{]}  & $A_{fiber}^{MAPMT}$~{[}\%{]}\tabularnewline
\midrule 
0 & $22.79\text{\ensuremath{\pm}}2.05$  & $33.01\pm2.35$  & $69.04\pm3.79$\tabularnewline
1 & $3.12\text{\ensuremath{\pm}}0.24$  & $25.25\pm1.37$  & $12.34\pm0.66$\tabularnewline
2 & $14.78\text{\ensuremath{\pm}}1.61$  & $30.09\pm1.72$  & $49.12\pm4.55$\tabularnewline
3 & $0.65\text{\ensuremath{\pm}}0.08$  & $26.19\pm2.95$  & $2.49\pm0.14$\tabularnewline
4 & $1.49\text{\ensuremath{\pm}}0.16$  & $30.07\pm2.69$  & $4.97\pm0.31$\tabularnewline
5 & $2.58\text{\ensuremath{\pm}}0.31$  & $33.63\pm3.26$  & $7.67\pm0.53$\tabularnewline
6 & $0.47\text{\ensuremath{\pm}}0.11$  & $27.73\pm6.43$  & $1.69\pm0.09$\tabularnewline
7 & $1.42\text{\ensuremath{\pm}}0.23$  & $32.64\pm4.84$  & $4.36\pm0.31$\tabularnewline
\bottomrule
\end{tabular}
\end{table}

 %lead author: Frank Simon
 \subsection{CLAWS detector system}
 %     file:		claws.tex
%     author:  	Frank Simon
%
%     contents:  Functional description of CLAWS system, to include the following:
%		physical description: size of sensitive volume, number of detectors etc
%		process by which radiation is detected. How signal is amplified, and read out
%		quantitative performance spec: radiation sensitivity, threshold, timing capability, noise rates, etc
%		description of in-situ calibration procedure

\label{sec:CLAWS}

The goal of the CLAWS system is to measure background levels, in particular those connected to the injection, with a time resolution of better than the bunch crossing frequency (250 MHz).
The detectors measure the total rate and exact arrival time of minimum-ionizing particles (MIPs).

\subsubsection{CLAWS system physical description}
The CLAWS system consists of eight independent plastic scintillator tiles read out with silicon photomultipliers.
These detectors are primarily sensitive to charged particles, but in principle also show responses to  high-energy photons and to MeV neutrons.
The detector design and the full readout chain is based on the CALICE-T3B experiment \cite{Simon:2013zya, Adloff:2014rya} used to measure the time structure of hadronic showers in a tungsten-scintillator calorimeter. 

\begin{figure}[htb]
\centering
\includegraphics[width =\columnwidth]{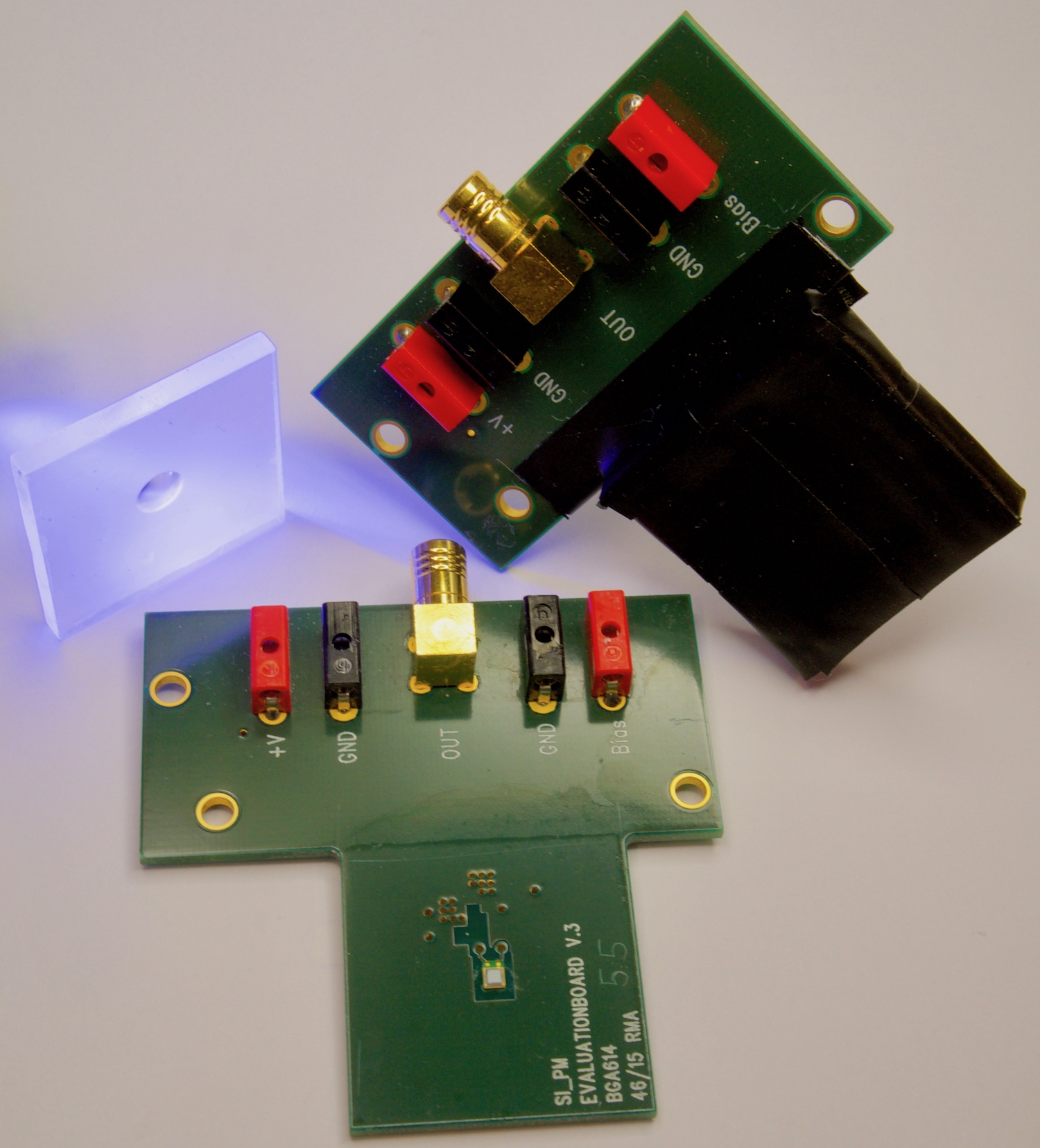}

\caption{Photograph of one CLAWS detector (top right) as well as of the main components, a 30 $\times$ 30 $\times$ 3 mm$^2$
scintillator tile (top left) and the PCB with photon sensor, preamplifier and connectors for power and signal (bottom middle). The plastic scintillator tile, wrapped in 
reflective foil\textcolor{blue}{,} is mounted with the dimple located on top of the photon sensor on the PCB. The full detector is enclosed in light-tight wrapping. 
\label{fig:CLAWS:Detector}} 
\end{figure}

Figure \ref{fig:CLAWS:Detector} illustrates the main components of a CLAWS detector. Each of the detectors consists of a 30 $\times$ 30 $\times$ 3 mm$^3$ scintillator tile directly coupled to a Hamamatsu multi-pixel photon counter (MPPC) S13360-1325PE silicon photomultiplier (SiPM).
This photon sensor is coupled to the center of the tile, which has a specifically designed dome at the coupling position to achieve a uniform response over the full active area of the detector.
This design has been developed for surface-mounted photon sensors \cite{Liu:2015cpe} for the CALICE Analog Hadronic Calorimeter, and is a further development of the scintillator tiles used in the CALICE-T3B experiment, which employs SiPMs coupled to the side face of scintillators \cite{Simon:2010hf}. 

As described in \cite{Simon:2010hf}, the photon sensor is connected to a preamplifier which amplifies the signal and matches the impedance to 50 $\Omega$ for transmission over longer distances and for further amplification and digitization.
The preamplifier circuit is located on the PCB that is holding the photon sensor and the scintillator tile, with the amplifier input located only a few millimeters from the SiPM on the back side of the board to ensure minimum noise pickup. 
From the preamplifier board, the signal is carried on a coaxial cable over a distance of three meters to an additional amplifier (Mini Circuits ZFL-500), which then drives the signal over the cable length of 37 m to the DAQ room. 

\begin{figure}[htb]
\centering
\includegraphics[width =\columnwidth]{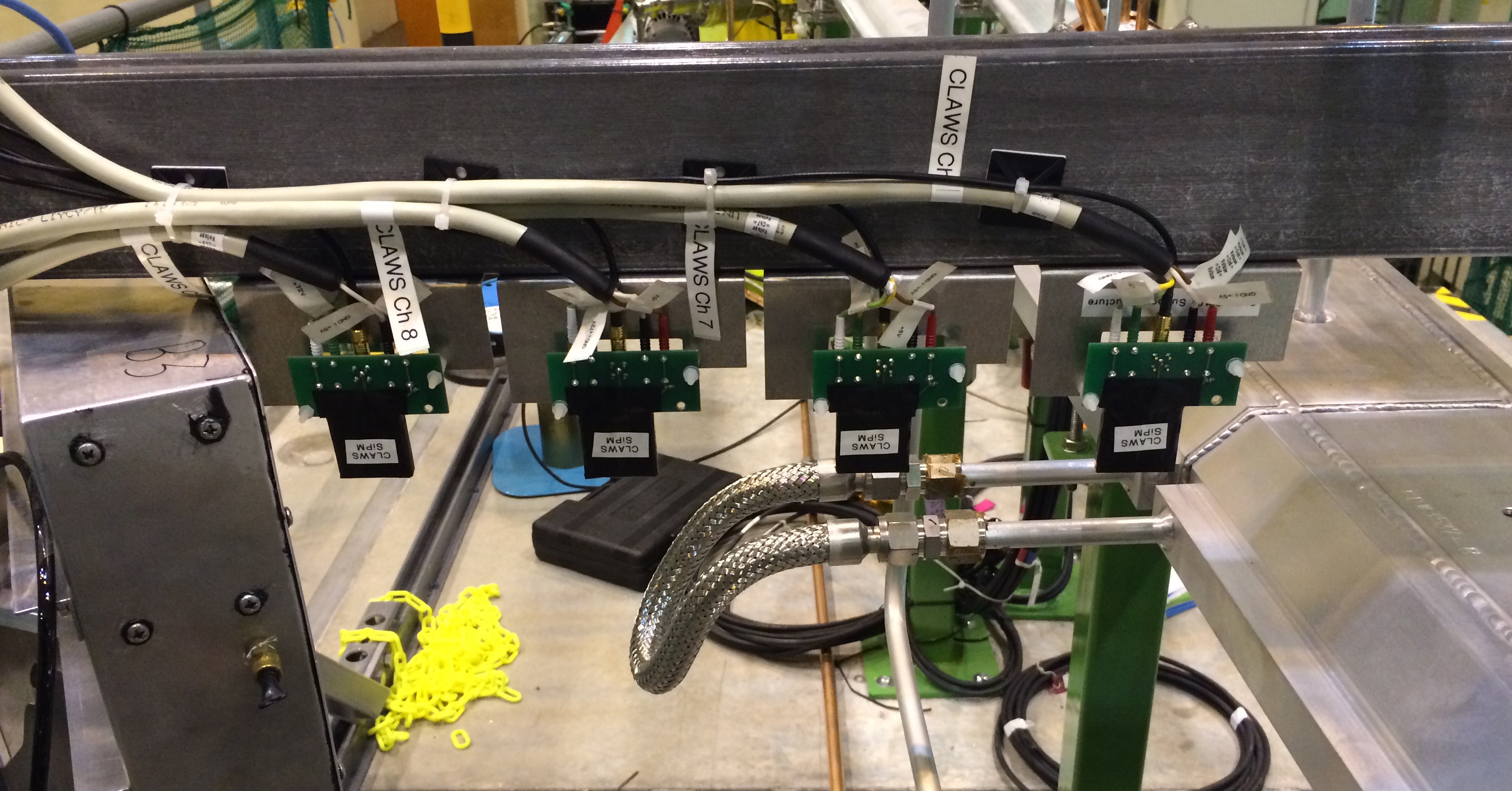}
\caption{Photograph of the backward CLAWS station installed in the BEAST II system taken from the position of the IP. Four CLAWS detectors are mounted on the BEAST frame in the plane  of the accelerator with increasing radial distance from the beampipe, which is shown on the right. On the left, one of the Crystals boxes as discussed in Section \ref{sec:crystals} can be seen.
\label{fig:CLAWS:Installation}} 
\end{figure}
The eight CLAWS detectors are arranged in two stations with four detectors each, one on each end of the BEAST II setup. The four detectors in each station are arranged in a line roughly perpendicular to the beam with a spacing of around 10 cm between each detector.
To cover the full range of expected rates, the detectors of one station are installed in the region with the highest background rates predicted by simulations, on the forward side outside of the accelerator ring in the accelerator plane.
The detectors of the other station are installed in the region with the lowest background prediction, on the backward inside of the accelerator ring.
Figure \ref{fig:CLAWS:Installation} shows a photograph of the four backward CLAWS detectors as installed in the BEAST II system.

\subsubsection{CLAWS system principle of operation}
\label{sec:CLAWS:Operation}

Under typical operating conditions, the signal yield is one photon-equivalent (p.e.) per 30 keV deposited energy, corresponding to approximately 15 p.e.\ for the most probable value of a through-going MIP at perpendicular incidence.

The signals are digitized with two 4-channel PC-based oscilloscopes (Picotech PicoScope 6404D), which sample each channel at 1.25 GHz and 8 bit resolution, and can record continuous waveforms for up to 50 ms.
This allows uninterrupted monitoring of particle rates over up to 5000 consecutive turns in SuperKEKB.
The oscilloscopes are controlled via USB-3 by a PC running a custom-made LabVIEW program, which records and stores the waveforms from the detectors.
A full offline analysis is performed for each recorded waveform, with the goal of determining the time-dependent rate of the injection background.
In addition, amplitude and signal decay time information from a fast online analysis within the DAQ software are directly made available via EPICS for the global BEAST II system.

The CLAWS system is triggered by the SuperKEKB injection trigger. This trigger signal is distributed to both readout oscilloscopes.
The data is thus time-stamped relative to the injection trigger, which has a fixed (but \emph{a priori} not precisely known) time offset to the time of arrival of the injection bunches at the IP. 
This offset can be determined from CLAWS data and be used to define the time region of interest for the injection bunches in the data.
In addition to this external trigger, an automatic self-trigger is available, which starts the data acquisition once a pre-defined waiting time after the previous trigger has elapsed.

\subsubsection{CLAWS system performance}

Since a key feature of CLAWS is the capability to resolve particle signals on the nanosecond level, the time resolution of the detectors has been studied in the laboratory.
With a simple waveform analysis taking into account the peak height of the signal, a resolution of approximately 500 ps for MIPs is observed.
This is consistent with the results obtained by the CALICE-T3B experiment, which used a very similar setup \cite{Simon:2013zya}. 

At the MIP level, the detectors are essentially noise-free. 
The single p.e.\ noise level of the detectors is in the 70~kHz range,
which very quickly falls off towards higher amplitudes due to the low cross-talk levels of the latest generation of Hamamatsu SiPMs.
At amplitudes above 3 p.e.\ (20\% of a MIP) the noise rate is below 1~Hz. For the sensors closest to the beampipe, which receive a non-negligible amount of radiation, the single p.e.\ noise rate increased by approximately two orders of magnitude during the data taking period.
Since the pixel-to-pixel cross talk level is not negatively affected by the radiation damage to the device, the noise level at relevant signal amplitudes remains negligible. 

\subsubsection{CLAWS system calibration}
\label{sec:CLAWS:Calibration}
\begin{figure}[htb]
\centering
\includegraphics[width =0.9\columnwidth]{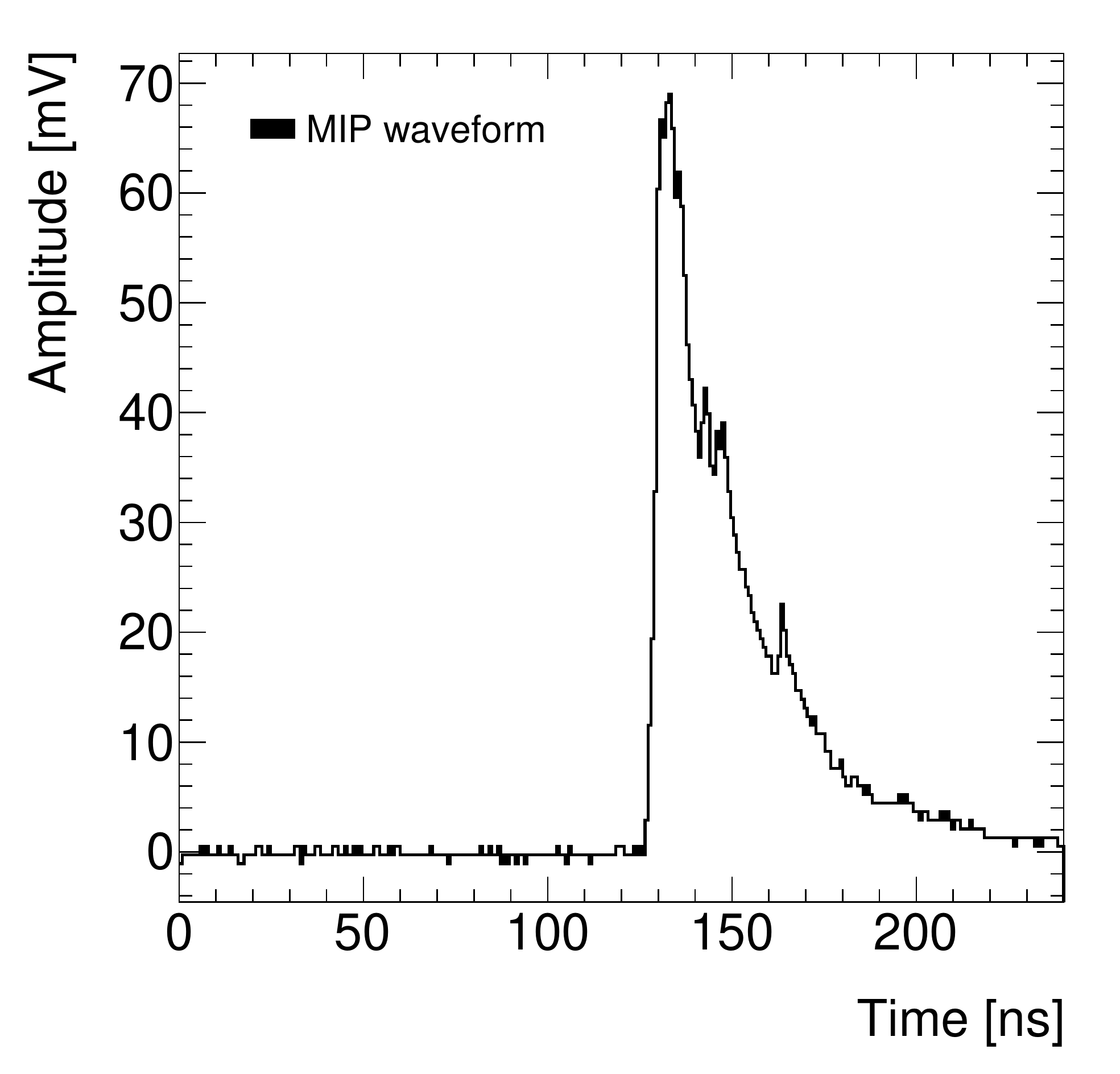}
\caption{
Typical waveform (pedestal subtracted) of a through-going cosmic muon recorded with one of the CLAWS detectors during calibration measurements in the laboratory.
To maximize calibration speed and limit data volume, only short waveforms with 250 ns duration are recorded for calibration.
The region before the signal rise is used for pedestal determination. \label{fig:CLAWS:Waveform}} 
\end{figure}
An example of a typical waveform in one of the CLAWS detectors from a cosmic muon is shown in Figure \ref{fig:CLAWS:Waveform}.
The signal pulse is a superposition of the pulses of single firing pixels, each corresponding to one p.e.\
On the falling slope, additional single p.e.\ pulses are visible, either due to delayed photons from the scintillator or due to afterpulsing of the photon sensor.
The shape of the pulses is characterized by a very fast rise and a slower fall-off, primarily determined by the electrical properties of the SiPMs.
The waveforms recorded in the CLAWS detectors during run time are a sequence of such signal pulses generated by MIPs.
To determine the arrival time of the p.e.\ and, therefore, the particles, an analysis based on the iterative subtraction of single p.e.\ pulses is applied, inspired by the waveform decomposition used in the CALICE-T3B experiment \cite{Simon:2010hf}.
Two types of calibration measurements, one taken in the laboratory and one continuously performed at run time, are needed as an input to the analysis.

Calibration measurements in the laboratory determine the number of p.e.\ which correspond to one MIP under nominal operating conditions in an individual CLAWS detector.
Therefore, events with cosmic muons are recorded after the completion of Phase 1 data taking.
An external trigger is provided by two spare CLAWS detectors mounted with minimal separation above and below the one under study, accepting a wide range of incident angles.
\begin{figure}[htb]
\centering
\includegraphics[width =0.9\columnwidth]{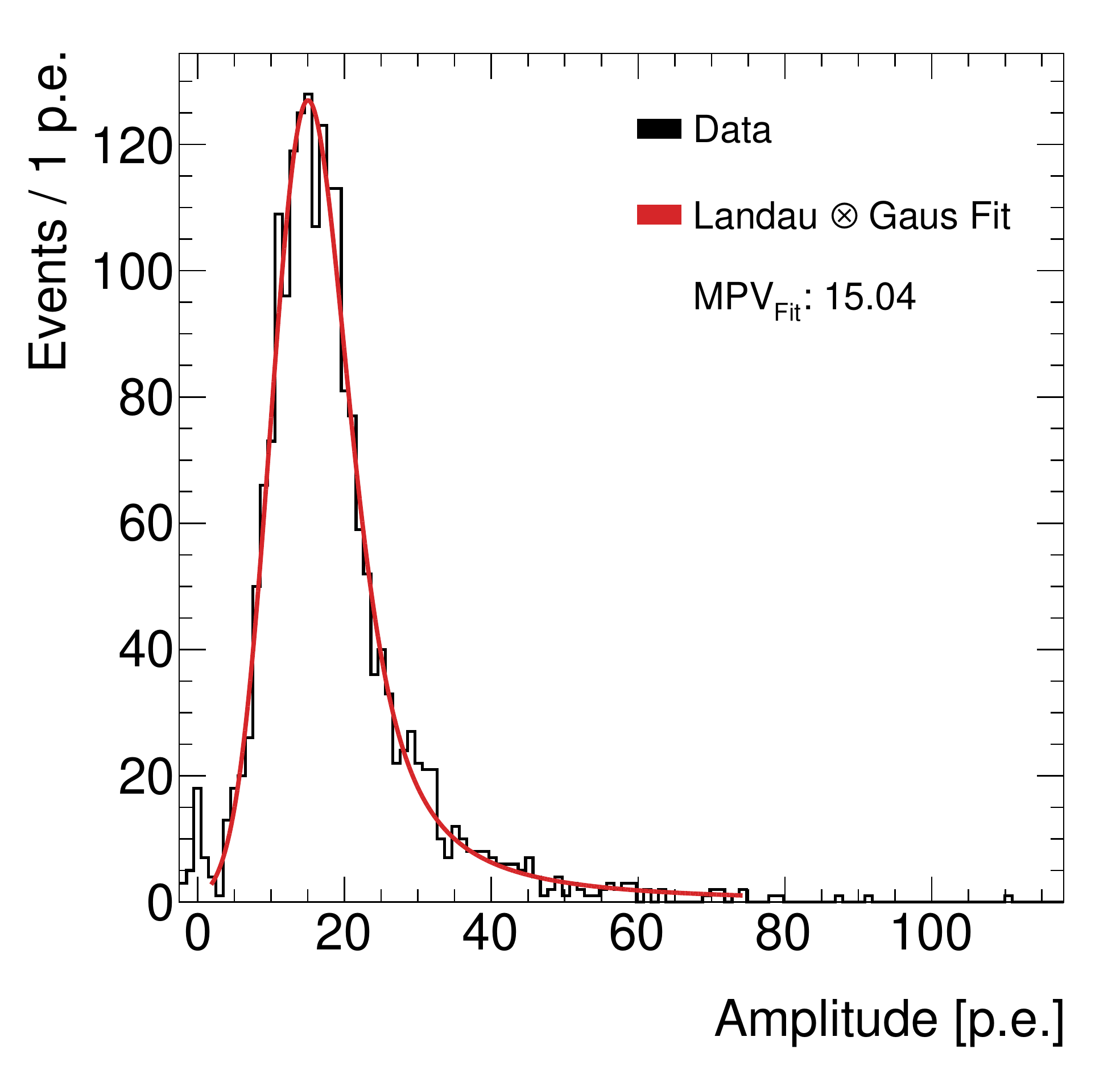}
\caption{Distribution of the signal amplitude for
cosmic muons in one CLAWS detector. 
The most probable value (MPV$_{\text{Fit}}$), used for the detector-to-detector calibration of the CLAWS system, is determined by a fit with a Landau convolved with a Gaussian. The fit result is 15.0 $\pm$ 0.2 p.e.\label{fig:CLAWS:MIP}} 
\end{figure}
Since for the calibration exact time information was irrelevant, it used a simplified analysis based on the comparison of the signal integrals for MIPs and single p.e.
Figure \ref{fig:CLAWS:MIP} shows the distribution of recorded p.e.\ from 2000 muon events.
The most probable value (MPV) of this distribution is then used to convert the number of p.e.\ to the number of recorded MIPs.
It is  determined by fitting a Landau convolved with a Gaussian to the distribution.
The arithmetic mean of the MPVs over all detectors in CLAWS is 14.6 p.e., with a standard deviation of 1.2 p.e., demonstrating a high degree of uniformity among the utilized elements.

In addition to the ones in the laboratory, calibration measurements taken during run time provide a continuous gain calibration of the SiPMs which allows correction for non-standard operating conditions like temperature variations.
Single p.e.\ pulses from dark noise of the photon sensors are recorded in dedicated calibration events taken in between standard physics events. 
One thousand calibration events for each detector are recorded at the end of each run\footnote{The time to record a full run varied but usually was around 30 minutes.}.

%In the first step of the analysis of the waveforms, a pedestal subtraction is applied, which is calculated individually for each event\footnote{Similarly done for calibration and physics events.} from the part of the waveform not corresponding to a signal pulse.
A pedestal subtraction, which is calculated individually for each event\footnote{Similarly done for calibration and physics events.} from the part of the waveform not corresponding to a signal pulse, is applied in the first step of the analysis of the waveforms.
Subsequently, an average waveform of the pulse of a single p.e.\ is determined individually for each detector by averaging over the calibration events. 
The number and the precise arrival time of p.e.\ in the physics waveforms are then reconstructed by iteratively subtracting this one p.e.\ waveform.
The signal seen in units of MIPs is given by converting the number of p.e.\ using the calibration constant obtained in the laboratory.

 %lead author: Sam de Jong
 \subsection{\heT detector system}
 %     file:		he3.tex
%     author:  	Sam de Jong
%
%     contents:  Functional description of He-3 system, to include the following:
%		physical description: size of sensitive volume, number of detectors etc
%		process by which radiation is detected. How signal is amplified, and read out
%		quantitative performance spec: radiation sensitivity, threshold, timing capability, noise rates, etc
%		description of in-situ calibration procedure
% 

 The purpose of the \heT tube system is to measure the rate of thermal neutrons (neutrons with kinetic energy of $\sim$0.025~eV).

\subsubsection{\heT system physical description}

\begin{figure}
	\centering
		\includegraphics[width=\columnwidth]{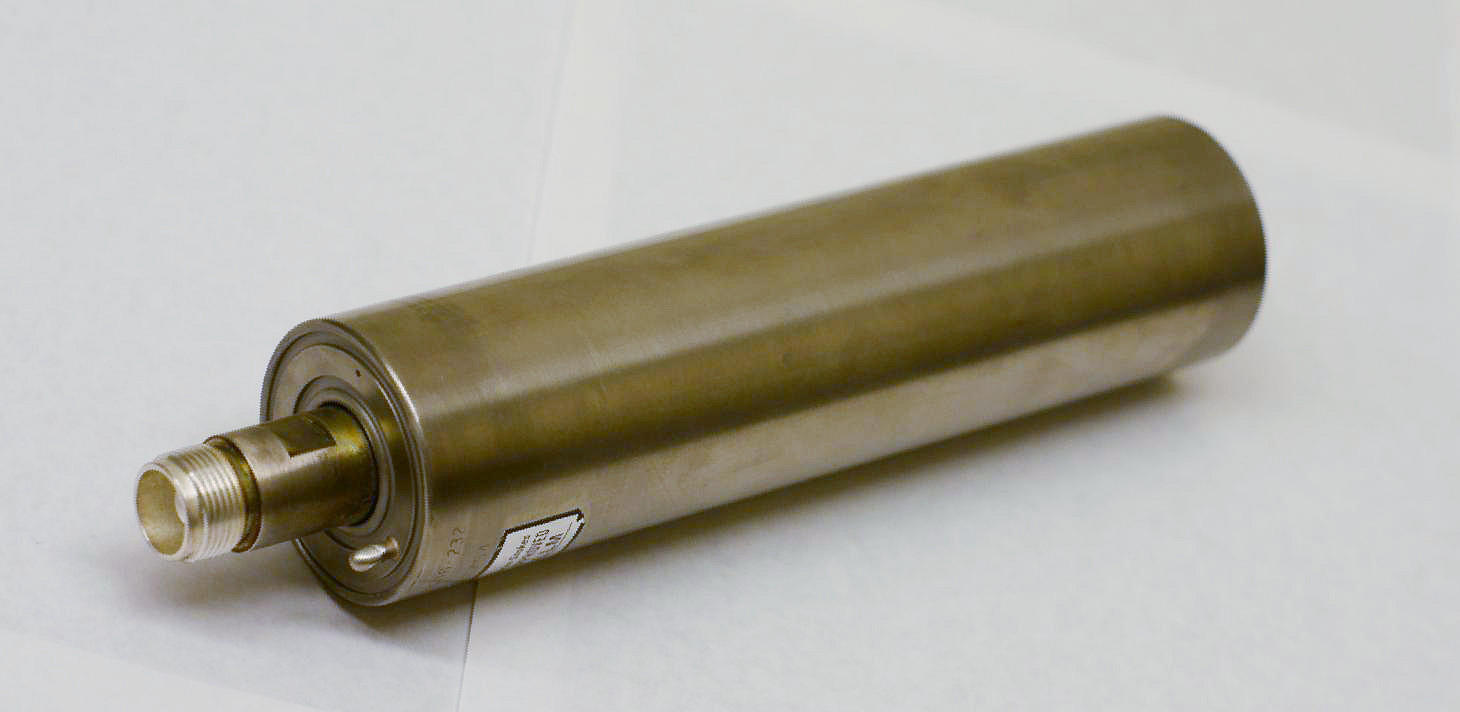}
	\caption{Photograph of a \heT tube.}	
	\label{fig:he3Photo}
\end{figure}

	The $^3$He proportional tube system consists of four detectors manufactured by General Electric Reuter-Stokes. Each detector is a stainless steel cylinder 5.08~cm in diameter and 20.38~cm long, filled with \heT and a small amount of CO$_{2}$ at 4~atm of pressure. In the center of the tube, there is a sense wire, which is set to a high voltage of 1.58~kV. The active length is 15.24~cm, for a sensitive volume of 0.309~L for each counter. A photo of the one of the detectors is shown in Fig.~\ref{fig:he3Photo}. 

\paragraph{Locations}

\begin{figure}[htb]
	\centering
	\includegraphics[width=\columnwidth]{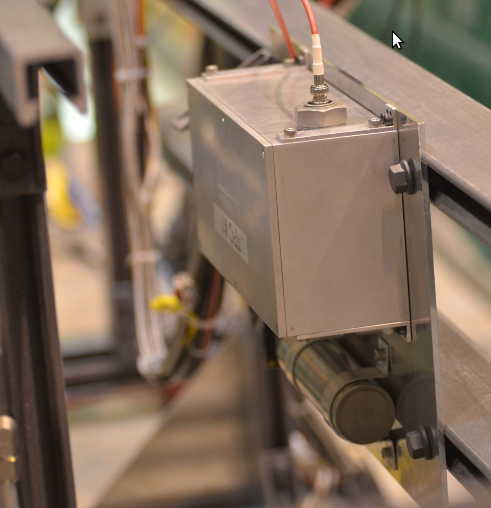}
	\caption{TPC mounting plates with TPC above and \heT tube below, for tube located at 0$^{\circ}$. }	
	\label{fig:he3TPCPlate}
\end{figure}

	The \heT tubes are attached to the same plates as the TPCs (see Fig~\ref{fig:he3TPCPlate}), at locations above, below, and on either side of the IP. The locations of the tubes can be found in Table~\ref{tab:he3Loc}.

\begin{table}[ht]
	\caption{Locations of the \heT tube thermal neutron detectors.}
	\centering	
	\begin{tabular}{ crrrr }
\toprule
	Channel	&	$x$ [m]	&	$y$ [m]	&	$z$ [m]	&	$\phi$ [$^{\circ}$] (approximate)	\\ 
\midrule
	0	&	0.439	&	0.073	&	0.469	&	0	\\
	1	&	-0.130	&	0.469	&	0.517	&	90	\\
	2	&	-0.477	&	-0.083	&	0.485	&	180	\\
	3	&	0.052	&	-0.451	&	0.470	&	270	\\ 
\bottomrule
	\end{tabular}
	\label{tab:he3Loc}
\end{table}

\subsubsection{\heT system principle of operation}

\begin{figure}
	\centering
		\includegraphics[width=\columnwidth]{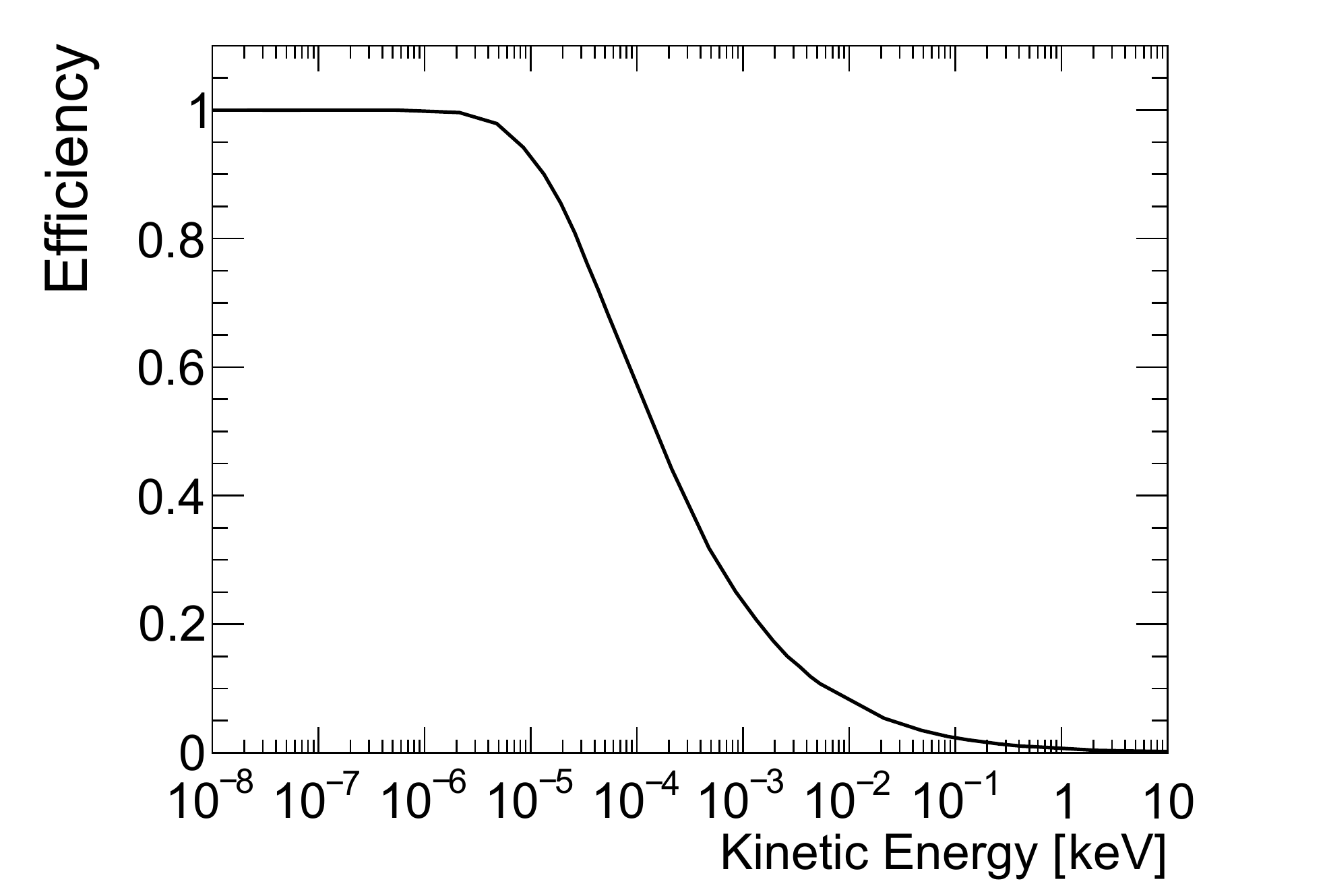}
	\caption{Efficiency of neutron detection vs neutron kinetic energy for \heT tubes, from a \gF simulation.}	
	\label{fig:he3Eff}
\end{figure}

	The purpose of this system is to detect thermal neutrons, which are neutrons with kinetic energy below about 0.025~eV (which corresponds to a momentum of 6.8~keV). This is achieved using the following process \cite{Oed200462}:
\begin{equation}
		{^3\mathrm{He}+\mathrm{n}\rightarrow~  ^3\mathrm{H}+\mathrm{p}+764~\mathrm{keV}}
\end{equation}
The cross-section for this process falls rapidly as a function of neutron kinetic energy, as demonstrated in Fig~\ref{fig:he3Eff} , which makes it useful for detecting thermal neutrons. The proton and tritium are emitted in opposite directions and ionize the gas, which produces a signal on the sense wire. The momentum of the proton and tritium is much higher than the momentum of the incident neutron, so measurement of the neutron momentum is impossible. The \heT tubes are therefore used to count the number of thermal neutrons, not their spectrum.

\paragraph{Amplification}
	The amplification electronics consist of two devices: an amplifier module that attaches directly to the \heT tube itself, and a signal receiver box contained within a NIM module. The amplifier uses two op-amps to amplify the signal from the \heT tube. A differential line driver is then used to drive the signal down a twisted pair cable (a 39~m CAT-6 cable) to the signal receiver box. The receiver box extracts the difference signal from the CAT-6 cable, and sends it to the digitizer. The receiver box also sends low voltage down the CAT-6 cable to power the amplifier circuitry.

%\begin{figure}
%	\centering
%	\begin{tabular}{cc}
%		\subfigure[Amplifier Front]{\includegraphics[scale=0.12]{He3Images/preamp_front}\label{fig:he3ampfront}} & 
%		\multirow{-3}[18]{*}{\subfigure[Reciever box]{\includegraphics[height=6.2cm]{He3Images/RecieverBox}\label{fig:he3Rec}}} \\
%		\subfigure[Amplifier Rear]{\includegraphics[scale=0.12]{He3Images/preamp_rear}\label{fig:he3ampread}}\\
%	\end{tabular}
%	\caption{\heT tube amplifier module and receiver box}
%	\label{fig:he3ampRec}
%\end{figure}

\paragraph{Digitizer}

	The signal from the receiver box is sent to a CAEN V1726 8-channel VME64 digitizer, which triggers on signals larger than 250~mV, which is well below the signal size, and also well above the electronic noise. This digitizer firmware measures the pulse height of the input signal, as well as the time of the trigger. These data are sent through the VME backplane to a CAEN V1718 VME64-USB bridge, which relays the data to a PC with the data acquisition software installed on it.

\paragraph{High Voltage}

	A Bertan model 323 HV power supply is used to supply 1.58~kV of high voltage to the \heT tubes. A single 39~m cable runs from the power supply in the DAQ room to the IP, where it is connected to a splitter, which provides each \heT tube with high voltage.

\subsubsection{\heT system performance}

	The background from non-neutron events is essentially zero, since the energy deposited by the proton and tritium nuclei is very large compared to the energy deposited by any particle that traverses the detector. In order to deposit a comparable amount of energy, a particle would be travelling too slowly to penetrate the outside of the tube. Testing of the system away from a neutron source has confirmed this.

	The twisted pair approach used between the amplifier and receiver minimized the common mode induced noise over the long cables.

	In the absence of a source of thermal neutrons, the \heT tubes produce no triggers.

\subsubsection{\heT system calibration}
\label{sec:heTCalibration}
	Testing of the detectors was done at the University of Victoria. The university has an $^{241}$AmBe neutron source, with an activity of 168~GBq (measured at 185~GBq in 1966). Neutrons are produced by the following process \cite{barschall1983neutron}:
\begin{subequations}
\begin{align}
		{^{241}_{95}\mathrm{Am}\rightarrow ~^{237}_{93}\mathrm{Np} + ~^4_2\mathrm{He} + \gamma}\\
		{^9_4\mathrm{Be}+~^4_2\mathrm{He}\rightarrow ~^{12}_6\mathrm{C}+~^1_0\mathrm{n}+\gamma}
\end{align}
\end{subequations}
The source produces 1.0$\times10^{7}$~neutrons/s \cite{neutronFluxPaper} isotropically, and is surrounded by a cube of graphite 1.85~m per side, which thermalizes the neutrons.

\begin{figure}
	\centering
		\includegraphics[width=\columnwidth]{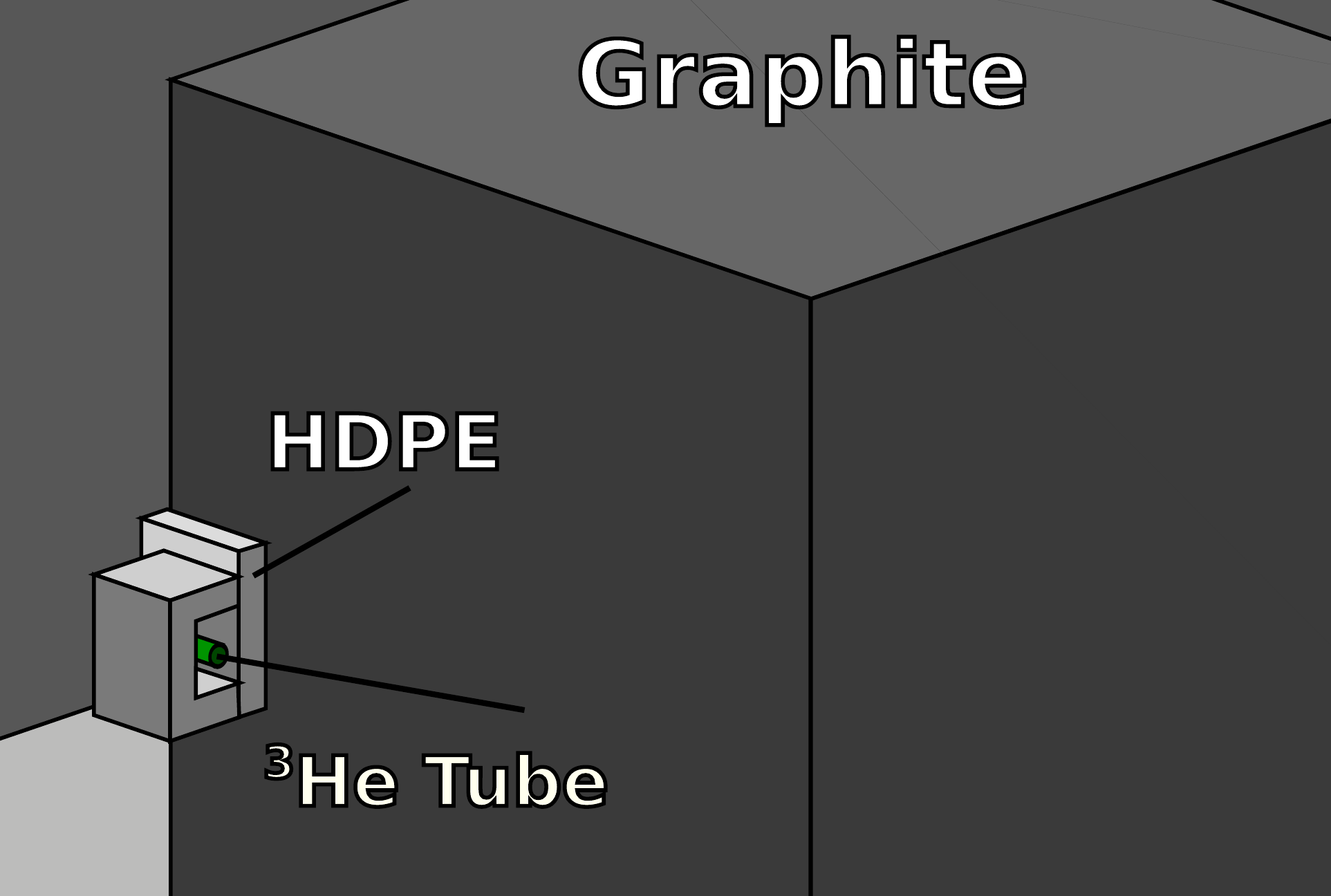}
	\caption{Experimental setup for \heT tube calibration. The AmBe source is in the center of the graphite cube.}	
	\label{fig:he3Par}
\end{figure}

To calibrate the \heT tubes, we placed each tube one at a time into a cradle made of high density polyethylene (HDPE). The polyethlene reduced the thermal neutron flux in the source room to a rate similar to that observed in BEAST II. The rate in each \heT tube was recorded, then the cradle was moved to a position farther away from the source and the process was repeated. The arrangement of the \heT tube relative to the graphite cube can be see in Fig~\ref{fig:he3Par}.

\begin{figure}
	\centering
		\includegraphics[width=\columnwidth]{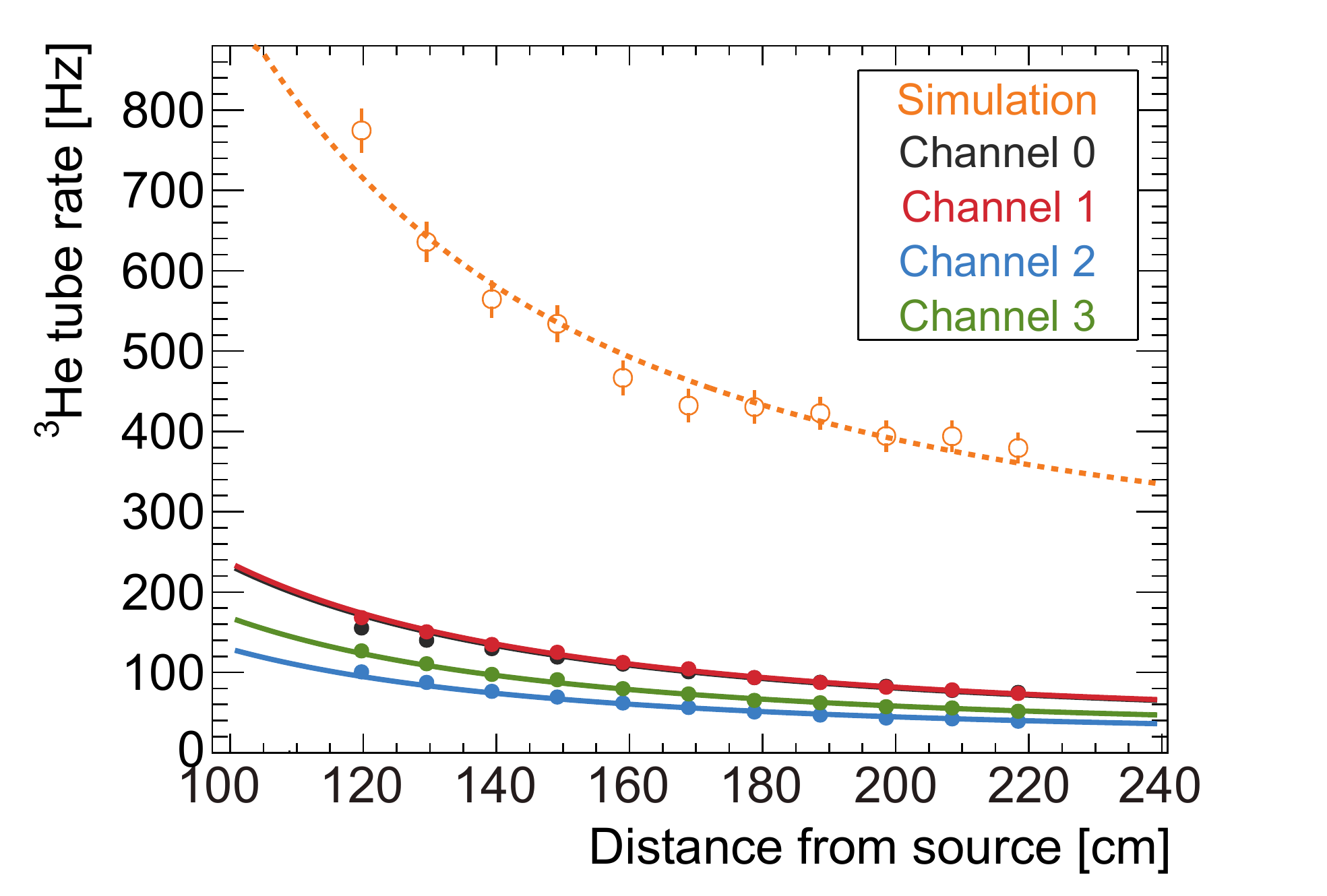}
	\caption{\heT tube rate vs distance from thermal neutron source with fits to Eq.~\ref{eqn:invSq}, showing relative and and absolute efficiency of \heT tubes }	
	\label{fig:he3calibrate}
\end{figure}

The rate in each \heT tube as a function of the distance from the source can be found in Fig~\ref{fig:he3calibrate}.

We performed a simple \gF simulation of the expected rate at the locations the tubes were placed. The expected rates are also shown in Fig~\ref{fig:he3calibrate}.

We fit both the measured rate and the simulated rate simultaneously to an inverse square relationship:
\begin{equation}
	{R = A_{n}\times\left(\frac{B}{(r-r_{0})^2}+C\right)}
	\label{eqn:invSq}
\end{equation}	
The parameters $B$, $r_{0}$ and $C$ are the same for the four different tubes, while $A_{n}$ varies from tube to tube.

The room that holds the AmBe source has various equipment that can reflect neutrons that would otherwise leave the room back into the \heT tubes, and is very difficult to simulate correctly. This will manifest as an additional background in the room, and should only affect the `$C$' part of the fit, so a modified version of Eq.~\ref{eqn:invSq} is used:
\begin{equation}
	{R = A_{\mathrm{sim}}\times\left(\frac{B}{(r-r_{0})^2} + C\right)+C_{\mathrm{sim}}}
\end{equation}
where the parameters $B$, $r_{0}$, and $C$ are the same as the fits to data. Here $A_{\mathrm{sim}}$ was fixed to be 1. The values of $A_{n}$ will then be the efficiency of each tube relative to the simulation. The values of these efficiencies can be found in Table~\ref{tab:he3Calib}. A detailed description of this fitting procedure can be found in \cite{samsThesis}. Note that these efficiencies cover the entire readout system: amplifiers, digitizer, and \heT tubes.

\begin{table}[htb]
	\caption{\heT tube experimental efficiencies, relative to simulation, and their uncertainties.}
	\centering
	\begin{tabular}{ cccc}
\toprule
Channel	&	$A_{n}$	&	$\sigma_{+}^{\mathrm{Tot}}$	&	$\sigma_{-}^{\mathrm{Tot}}$	\\	
\midrule
0	&	0.278	&	0.023	&	0.021	\\		
1	&	0.282	&	0.021	&	0.029	\\		
2	&	0.154	&	0.017	&	0.013	\\		
3	&	0.201	&	0.016	&	0.015	\\			
\bottomrule
	\end{tabular}
	\label{tab:he3Calib}
\end{table}

 %lead author: Michael Hedges
 \subsection{TPC detector system}
 %     file:		tpcs.tex
%     author:  	Michael Hedges
%
%     contents:  Functional description of TPC system, to include the following:
%		physical description: size of sensitive volume, number of detectors etc
%		process by which radiation is detected. How signal is amplified, and read out
%		quantitative performance spec: radiation sensitivity, threshold, timing capability, noise rates, etc
%		description of in-situ calibration procedure
The goal of the BEAST II TPC system is to provide direction and energy measurements of fast neutron recoils produced by the various beam backgrounds present during the commissioning phases.  Additionally, we aim to use the directional measurements to discriminate prompt neutrons originating from the Phase 1 beam-pipe from neutrons originating elsewhere.

\subsubsection{TPC system physical description} 
\label{subsec:tpcs_intro}
The TPCs are a system of four independent Time Projection Chambers (TPCs) oriented around the IP at $\phi= 0^{\circ}, 90^{\circ}, 180^{\circ},$ and $270^{\circ}$, corresponding to TPCs 1 through 4. The TPCs provide detailed 3D measurements of charge density distributions via micro pattern gas detectors. The BEAST TPCs are second generation detectors that will be described in a forthcoming publication \cite{beasttpc}. The first generation prototype is described in Ref.~\cite{Vahsen:2014fba}.

In Phase 1, we experienced difficulties with the TPC data acquisition, gas delivery, and high voltage systems, making operation and readout of more than two of the available detectors at KEK infeasible. As a consequence, only results from TPCs 3 and 4 are shown here. We use a full detector simulation suite for detailed comparisons between experimental data and Monte Carlo. Further details of the TPC simulation will be provided in a separate, forthcoming publication \cite{JaegleTPCSimulator}.

In Phase 1, TPC 3 was mounted in the horizontal plane at $\phi = 180^{\circ}$, while TPC 4 was mounted in the vertical plane at $\phi = 270^{\circ}$. Both types of detectors are of the same design and worked equally well.

\subsubsection{TPC system principle of operation}
The TPCs detect fast neutrons with a target-gas mixture of helium and carbon dioxide (70\% He, 30\% CO$_{2}$) contained within a $2.0\times1.68\times10.0$~cm$^3$ active volume. Elastic scattering of neutrons off of target gas nuclei leads to nuclear recoils, which produce clouds of ionization.  The clouds are reconstructed in $3$D as follows: the charge traverses a drift field of $530$~V/cm produced by a field cage. The drifted charge is then amplified by two sequential Gas Electron Multipliers (GEMs) \cite{Sauli:1997qp} and detected by an ATLAS FE-I4B pixel ASIC (or ``chip'') which digitizes the collected charge signal.

Detailed documentation on the design and performance of the chip can be found elsewhere \cite{ATLAS:fei4b, Aad:2008zz}.
The details can be summarized as follows: the chip collects the amplified charge exiting the GEMs with high spatial and temporal resolution in units of Time Over Threshold (TOT) calculated using a $40$~MHz clock. The readout is spatially segmented into $26880$ individual $250\times50~\mathrm{\si\micro m}^2$ rectangular pixels organized into 80 columns and 336 rows. The electron drift direction lies parallel to the global BEAST II $z$ axis.  Quantization in the drift direction for our setup is $250~\mathrm{\si\micro m}$ defined by $25~\mathrm{ns}$ time-bins and the drift velocity of the charge in the drift field, as calculated by Magboltz \cite{magboltz}.  The pixel chip is read out using the USBPix2 and SEABAS2 DAQ systems, controlled by the pyBAR software package\cite{multiio,usbpix,seabas,pybar}.

\subsubsection{TPC system performance}
\label{tpc_performance}

Various detector performance studies, including angular resolution of detected recoils and gain resolution, will be presented in a forthcoming publication \cite{beasttpc}.  In summary, the angular resolution for a 2~$\mathrm{cm}$ alpha track segment in the azimuthal angle $\phi$ and the polar angle $\theta$ is $\sim1^{\circ}$.

\begin{figure}[hbt]
\begin{center}
\includegraphics[width=\columnwidth]{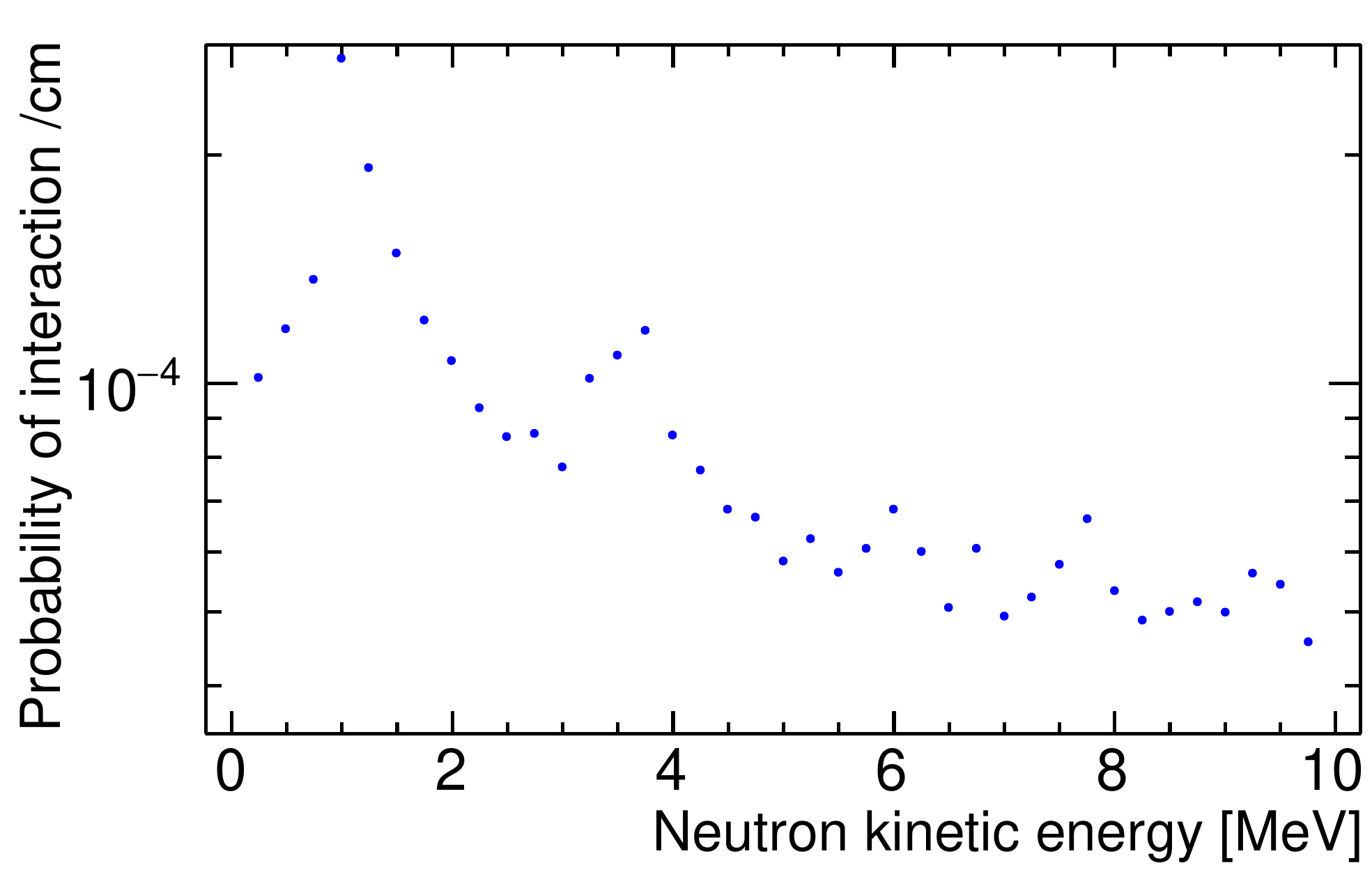}
\caption[single column]{Simulated fast neutron interaction probability per $\mathrm{cm}$ as a function of neutron energy in $1~\mathrm{atm}$ of pressure of the $70\%$ helium and $30\%$ carbon dioxide TPC target gas at room temperature. Below $2~\mathrm{MeV}$, the scattering is almost exclusively elastic.}
\label{fig_app_eff_neutron}
\end{center}
\end{figure}

The efficiency of detecting neutrons is determined by the interaction probability of a fast neutron with the target gas, which depends on the scattering cross-section versus neutron energy and the gas pressure in the detector, typically near $1~
\mathrm{atm}$.  The interaction probability versus the neutron energy is shown in Figure \ref{fig_app_eff_neutron}.  The interaction probability for fast neutrons entering the active volume of one of the TPCs is $\sim10^{-4}$.

\begin{figure}[hbt]
\begin{center}
\includegraphics[width=\columnwidth]{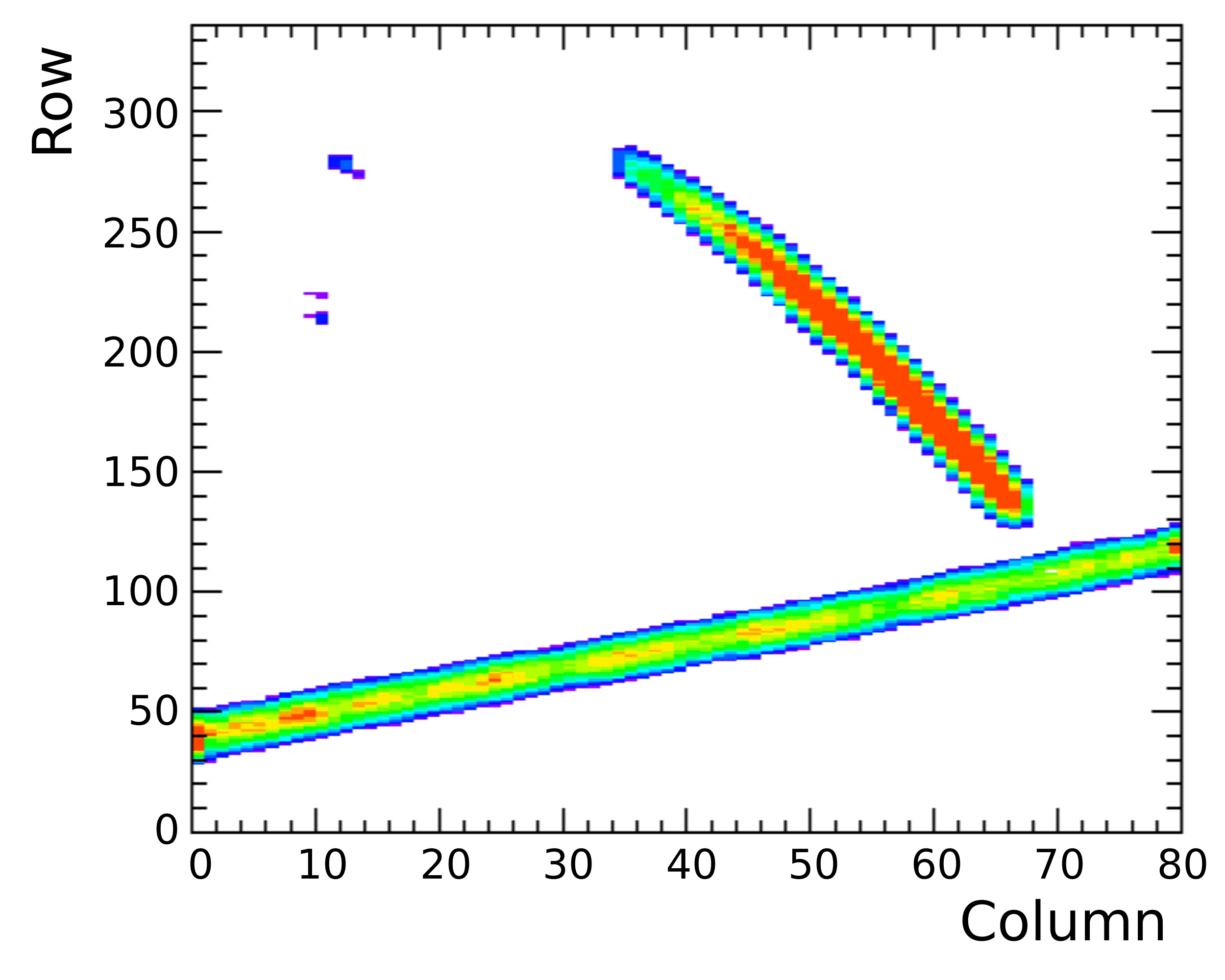}
\caption[single column]{(color online) Three separate events detected by a TPC, superimposed in the same event display.  The display is an occupancy plot of all of the pixels that triggered in the events, organized by row and column number. The color indicates the amount of charge collected in each pixel. The small isolated clusters are from X-rays, the long continuous track spanning the entire width of the pixel chip is from an MeV energy-scale alpha particle emitted from a $^{210}$Po calibration source. The track completely contained within the chip area is our signal: the resulting nuclear recoil from a fast neutron elastically scattering off of a nucleus in the target gas.}
\label{fig_evt_display}
\end{center}
\end{figure}

The high 3D spatial resolution of the ionization charge cloud produced by nuclear recoils allows for very robust particle identification based on a measurement of $dE/dx$.  This provides powerful discrimination of signal from background, and is immediately apparent even when inspecting the projection of tracks on a visual level.  The three most common types of events are shown in Figure \ref{fig_evt_display}.

While X-rays are easy to reject at the analysis level, triggering on many X-rays can lead to significant detector dead-time.  We avoid this by implementing a trigger-level veto for X-ray events. The trigger length of an event corresponds to the total length of time from when the integrated charge in any pixel is first larger than a configured threshold until the measured charge on all pixels is under threshold.  The trigger veto rejects events where the trigger length is less than a set length. We expect that an X-ray event will have significantly shorter trigger length than a nuclear recoil, because a nuclear recoil will have a far larger charge density per pixel than an X-ray event.  The trigger veto was tuned to veto X-ray events while accepting events from nuclear recoils.

Additionally, we have verified from the distribution of the time-difference between subsequent events that neither dead-time nor downtime affected the rate significantly.

\subsubsection{TPC system calibration}
\label{subsubsec:tpc_calibration}

\begin{figure}[hbt]
	\begin{center}
	\includegraphics[width=\columnwidth]{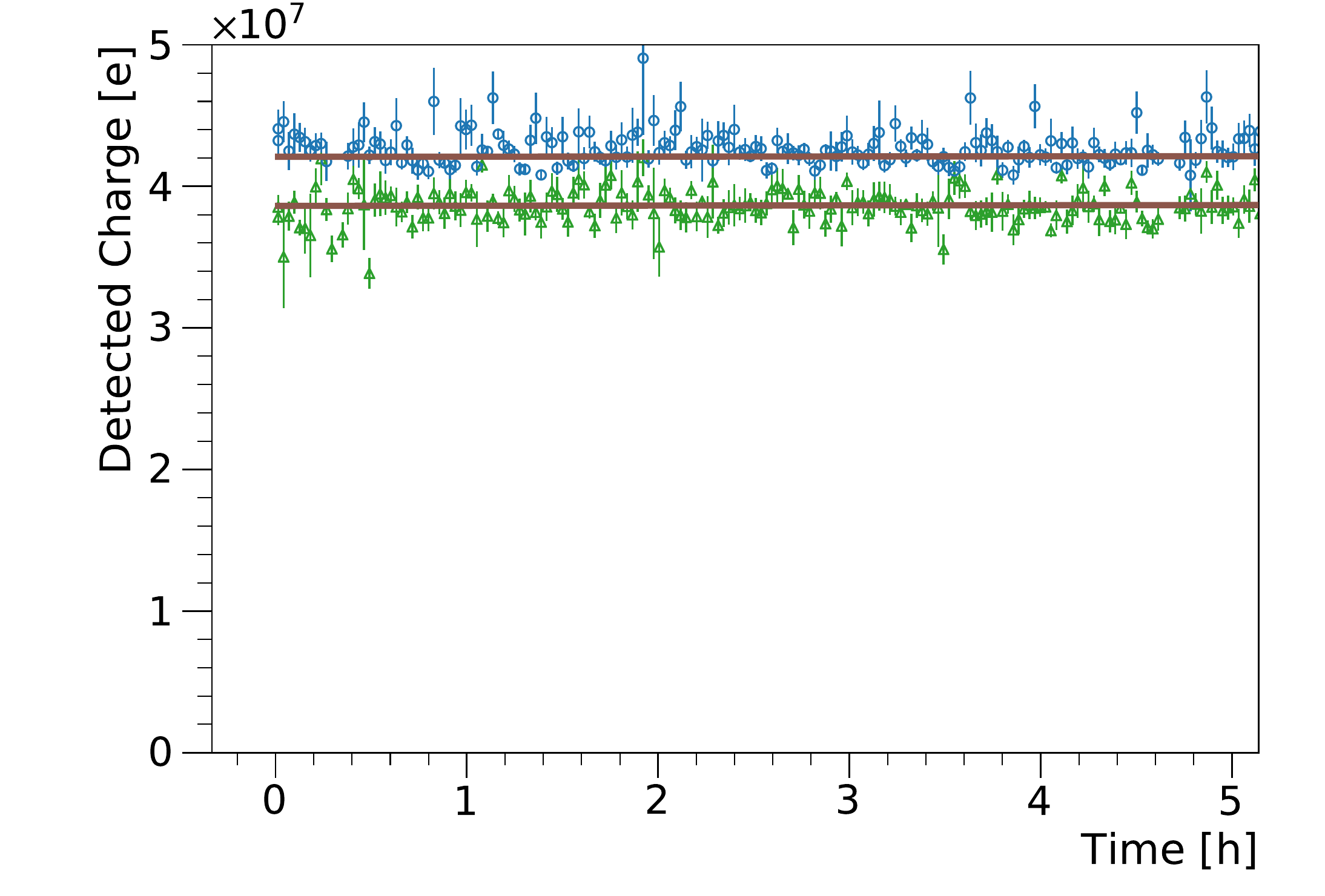}
	\includegraphics[width=\columnwidth]{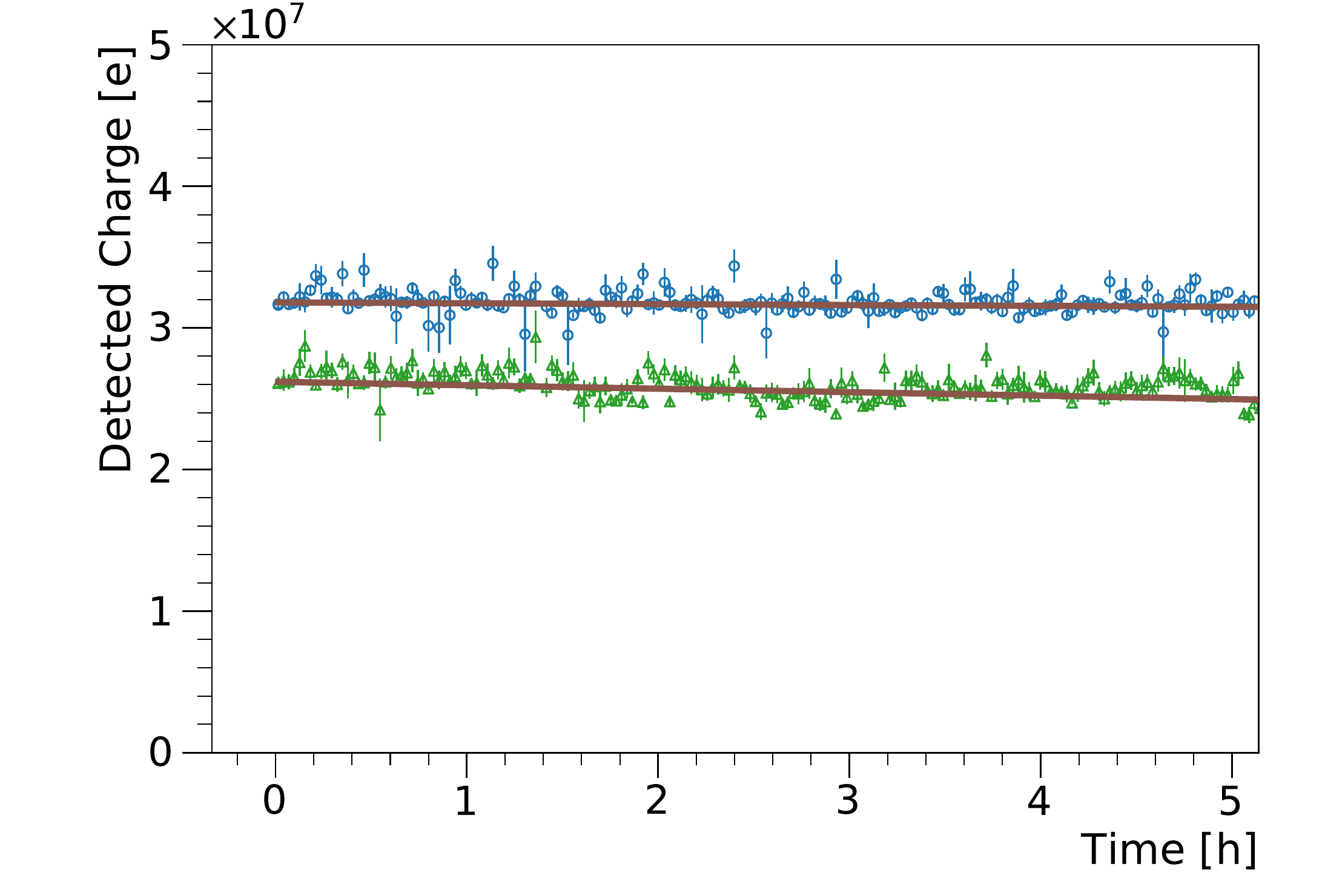}
	\caption[single column]{(color online) Detected charge of alpha particle calibration events in TPCs 3 (upper plot) and 4 (lower plot), versus time.  The blue circles in each plot correspond to the bottom $^{210}$Po source, corresponding to the source at smaller drift distance, and the green triangles correspond to the top calibration source at larger drift distance. Each line represents the fit of the change of energy over time of events from each internal calibration source.}
	\label{tpcfig_gainstability1}
	\end{center}
\end{figure}

\begin{figure*}[hbt]
	\begin{center}
	\includegraphics[angle=90,width=\textwidth]{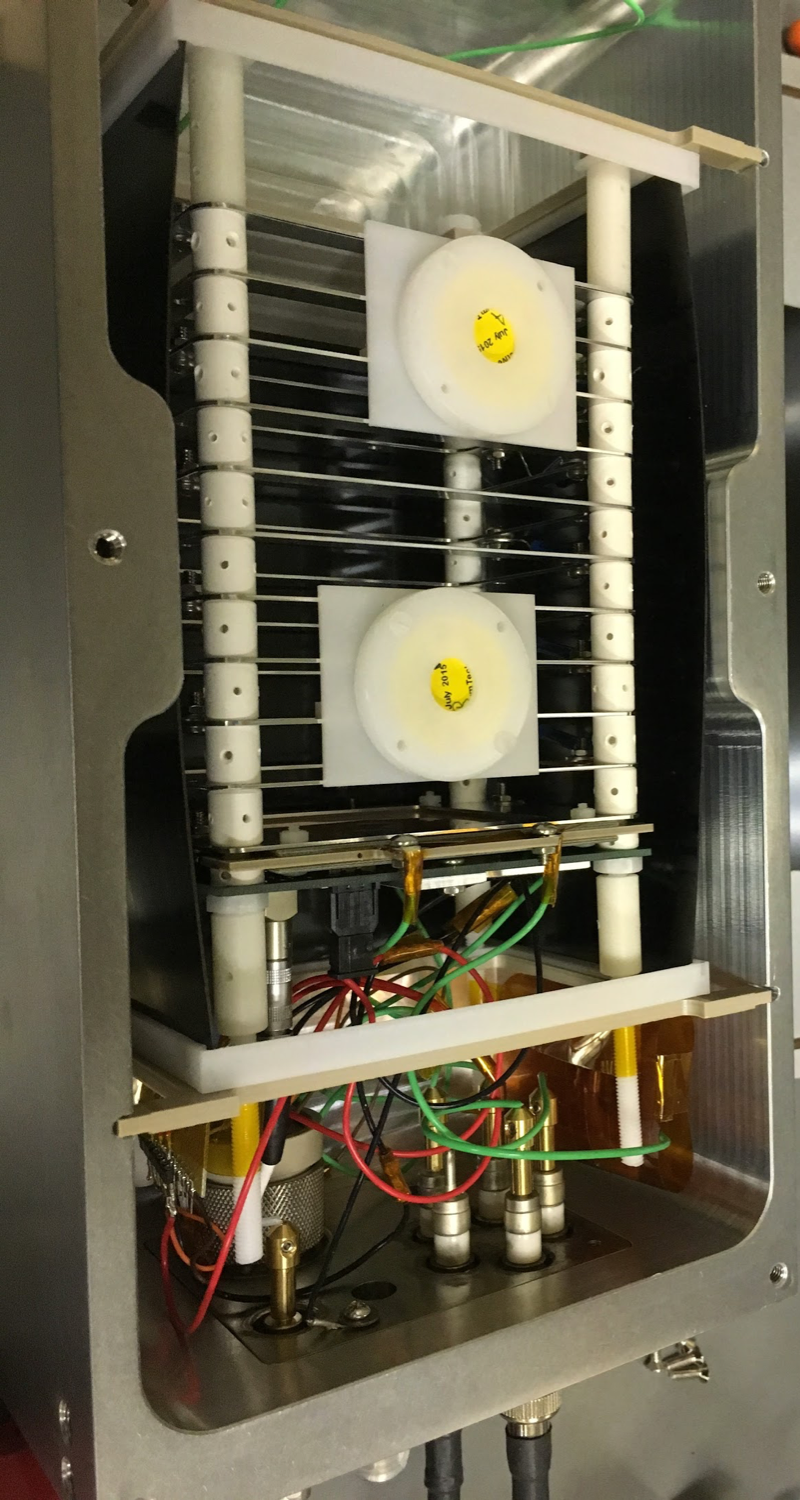}
	\caption[single column]{(color online) A photo of the inside of a TPC showing the internal calibration sources. The white containers with the yellow centers in the left half of the photo are the $^{210}$Po source holders. The source holder at the left of the photo is the ``top'' source; the source at largest drift distance, and the source holder towards the right of the photo, closest to the green wires, is the ``bottom'' source; the source at smallest drift distance.}
	\label{fig:tpc_internals}
	\end{center}
\end{figure*}

The TPC response must be calibrated so that the detected primary ionization in neutron recoil events can be obtained from the TOT values of individual detected pixel hits. This procedure consists of two steps.

First, we achieve a uniform pixel chip response by calibrating each individual pixel with a test-pulser built into the pixel chip itself: test pulses of varying charges are injected repeatedly into each pixel, while the digitally-programmable controls of the charge-threshold and charge-integration time of each pixel are automatically adjusted in the offline DAQ software until a uniform charge-threshold and integration time distribution across the whole pixel chip is achieved. The result is a mean threshold of approximately $2000$ electrons for both TPCs. This calibration also yields a measurement of the mean of the pixel noise distributions, which is of the order of a few hundred electrons; well below the pixel threshold. After this first calibration step, we can calculate the detected avalanche charge in a cluster of pixel hits, based on the TOT values of the individual pixel hits in the cluster. 

Second, we convert the charge detected by the pixel chip into primary ionization. To do so, we need to determine the double-GEM effective gain, which is primarily determined by the GEM high voltage, but also affected by charge reabsorption as the initial ionization drifts through the TPC. The reabsorption in turn depends quite strongly on gas purity, which can vary during detector operation. We therefore measure the effective gain in situ, versus time, using two $10 ~\mathrm{nCi}$ $^{210}$Po alpha-emitting calibration sources per TPC. The two sources in each TPC are positioned so that the emitted alpha particles enter the sensitive volume at different (local) $x$ and $z$ coordinates.  A photo of the source locations relative to the GEMs is shown in Figure \ref{fig:tpc_internals}.

We can easily identify events from the $^{210}$Po sources from the fact that the emitted alpha particles traverse the entire width of the pixel chip.  The ionization track of the alpha particles, as shown in Figure \ref{fig_evt_display}, traverse the entire width of the pixel chip with a very large amount of total ionization collected in the event. After selecting alpha events that match these criteria, we use the $x$-position of the alpha tracks to deduce which of the two sources the alpha was emitted from, which in turn tells us the $z$-position of the track. In summary, this allows us to monitor the gain versus drift distance and time. Figure \ref{tpcfig_gainstability1} shows the detected charge from the calibration sources, versus time, during the dedicated TPC Touschek run reported in Section \ref{fast_neutron_results}.

From these plots, we note several observations.  First, we note that for both TPCs, the ``bottom" source, located at small drift distance, results in events with higher charge, because a shorter drift length results in less charge reabsorption, and lower charge below threshold (due to reduced transverse diffusion). Secondly, the average observed charge for each calibration source in TPC 4 is smaller than the observed charge for the analogous source in TPC 3, and the difference between the charge from the two sources is greater in the former. This suggests that the gas purity in TPC 4 was worse during this period. Thirdly, the data for each source are fit to a line, where the slope of the line indicates the change in detected charge from the source over time.  We find that the maximum change in detected charge over the $6 ~\mathrm{hr}$ period for any individual calibration source is approximately $5 \mathrm{\%}$.  We consider this to be a negligible effect and thus we treat the gain as constant over time throughout our analyses.

\begin{figure}[hbt]
	\begin{center}
	\includegraphics[width=\columnwidth]{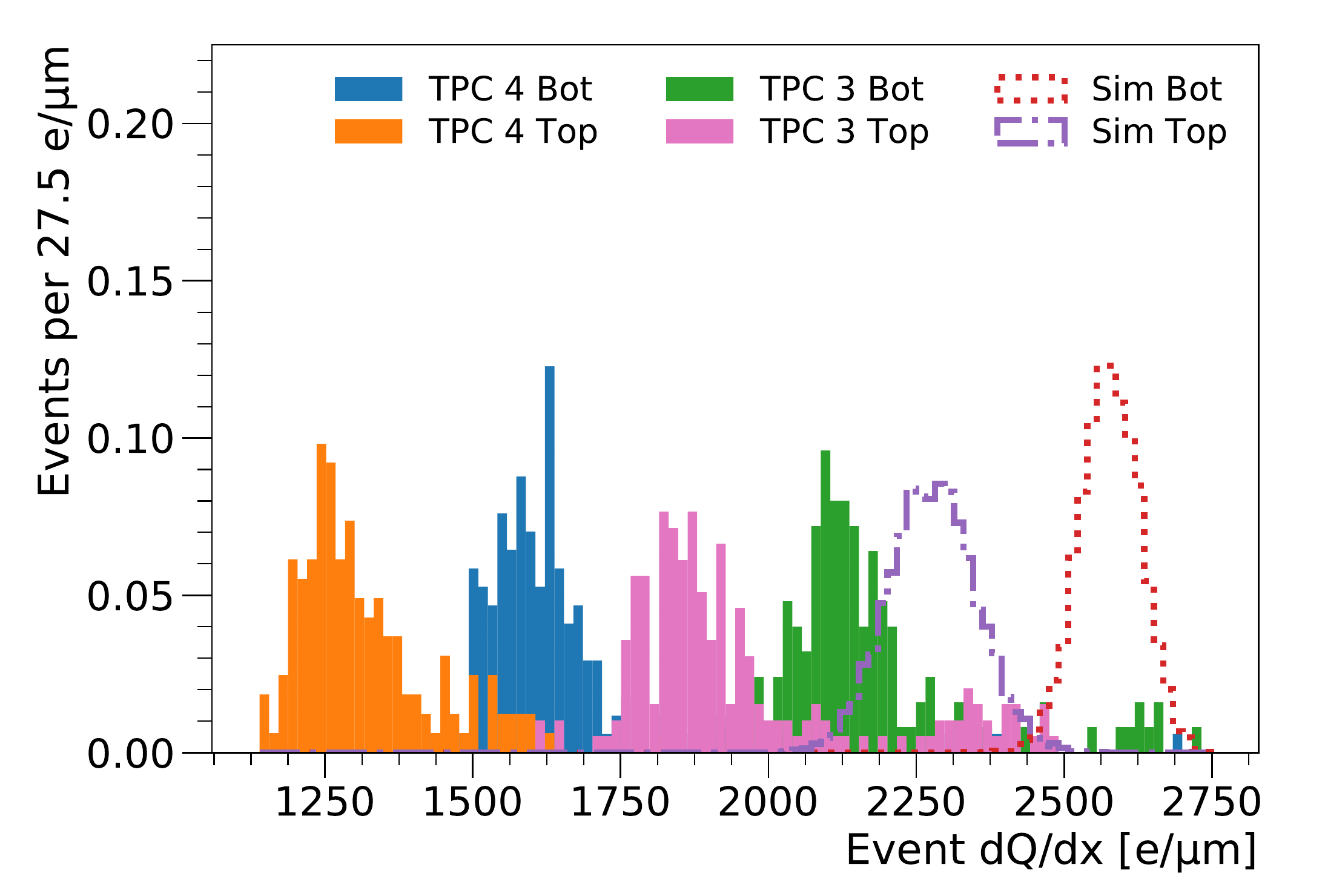}
	\caption{Histograms of the ratio of detected charge to track length [$dQ/dx$] for events from internal $^{210}$Po calibration alpha sources in experimental and Monte Carlo data. The vertical axis shows the total number of events from both sources, normalized to 1, for TPC 3, TPC 4, and Monte Carlo separately. The mean value of each peak is then used as an input in calculating the correction factors shown in Table \ref{tab:tpc_alpha_energy_calibration}.}
	\label{fig:tpc_alpha_energy_calibration}
	\end{center}
\end{figure}

We calculate the effective gain of each TPC by comparing the measured ratio of detected charge to track length ($dQ/dx$) of each alpha source in a TPC to a dedicated Geant4 Monte Carlo simulation \cite{JaegleTPCSimulator}.  The simulation assumes a known, fixed double-GEM gain value of 1500, and includes charge diffusion due to drift while excluding gas reabsorption effects. We select ideal alpha tracks in calibration data and Monte Carlo as having local $\theta$ and $\phi$ values within $1^{\circ}$ of $90^{\circ}$ and $0^{\circ}$, respectively, in order to select tracks of similar topology, total energy deposition, and track length.  The $dQ/dx$ distributions for each source in both TPCs and in Monte Carlo are shown in Figure \ref{fig:tpc_alpha_energy_calibration}. After obtaining a suitable sample in both experimental and Monte Carlo data, we compare the average $dQ/dx$ for the top and bottom $^{210}$Po sources of each TPC to the average $dQ/dx$ of the top and bottom $^{210}$Po sources in the simulation. This average value describes the expected measurement of $dQ/dx$ for an alpha track traversing through the midpoint between the top and bottom $^{210}$Po source positions in the drift axis, which by design is approximately the drift distance for a track in the middle of the TPC detector volume, i.e.\ the average amount of charge diffusion due to drift for a charge cloud of ionization in the TPC. We then calculate the ratio of this average $dQ/dx$ for each TPC in experimental data to the corresponding value for the Monte Carlo simulation assuming ideal gain conditions.  This ratio corresponds to a multiplicative correction factor for the overall energy scale in each TPC to be used in analysis.  The results are shown in Table \ref{tab:tpc_alpha_energy_calibration}.  Validation of this method is shown in Section \ref{neutrons_analysis}, where a comparison between the detected charge versus track length from simulated nuclear recoils from fast neutrons and nuclear recoil candidates selected in data are shown in Figure \ref{fig_tpc_dQdx}. We find excellent agreement between the Monte Carlo simulation and experimental data using this calibration method.

\begin{table}[ht]
	\centering
	\caption{Table of values of $dQ/dx$ in TPC 3, TPC 4, and Monte Carlo simulation and resulting conversion factors. A mean value of $dQ/dx$ obtained from averaging the $dQ/dx$ of each of the two $^{210}$Po calibration sources in Monte Carlo, TPC 3, and TPC 4, shown in Figure \ref{fig:tpc_alpha_energy_calibration}, is calculated separately and shown in the second column of the table.  The third column shows the ratio of the obtained mean in each TPC to the mean calculated from the Monte Carlo simulation. This ratio is then used as a multiplicative correction to the detected recoil energies presented in Sections \ref{neutrons_analysis} and \ref{fast_neutron_results}. \newline}
	\begin{tabular} { lcr }
	\toprule & Average $dQ/dx$ [e/\si\micro m]& Correction Factor \\ \midrule
	Simulation & 2426 & 1.0 \\
	TPC 3 & 2056 & 1.18 \\
	TPC 4 & 1480 & 1.64 \\ \bottomrule
	\end{tabular}
	\label{tab:tpc_alpha_energy_calibration}
\end{table}
  
 %lead author: Sae Yokoyama
 \subsection{QCSS detector system}
 % file:         scintillators.tex 
% lead author:  Sae Yokoyama, Kenkichi Miyabayashi, Hiro Nakayama

\label{sec:scintillators}

Plastic scintillator-based detectors can have a small footprint, yet detect charged particles generated by beam background showers with high efficiency. 
During SuperKEKB Phase 2 and Phase 3, we plan to install scintillators around the QCS cryostat, where space is very limited,
to monitor background distributions. The distributions  can differ depending on the beam background production process.
The resulting information is important for adjusting collimators to minimize 
beam losses inside Belle II. Therefore, we decided to install this type of scintillator in Phase 1,
to confirm its performance and gain operational experience. To distinguish this system from other scintillators used in BEAST II, we refer to it as the QCS scintillator (QCSS) system.

\subsubsection{Physical description}
\begin{figure}
	\begin{center}
		\includegraphics[width=\columnwidth]{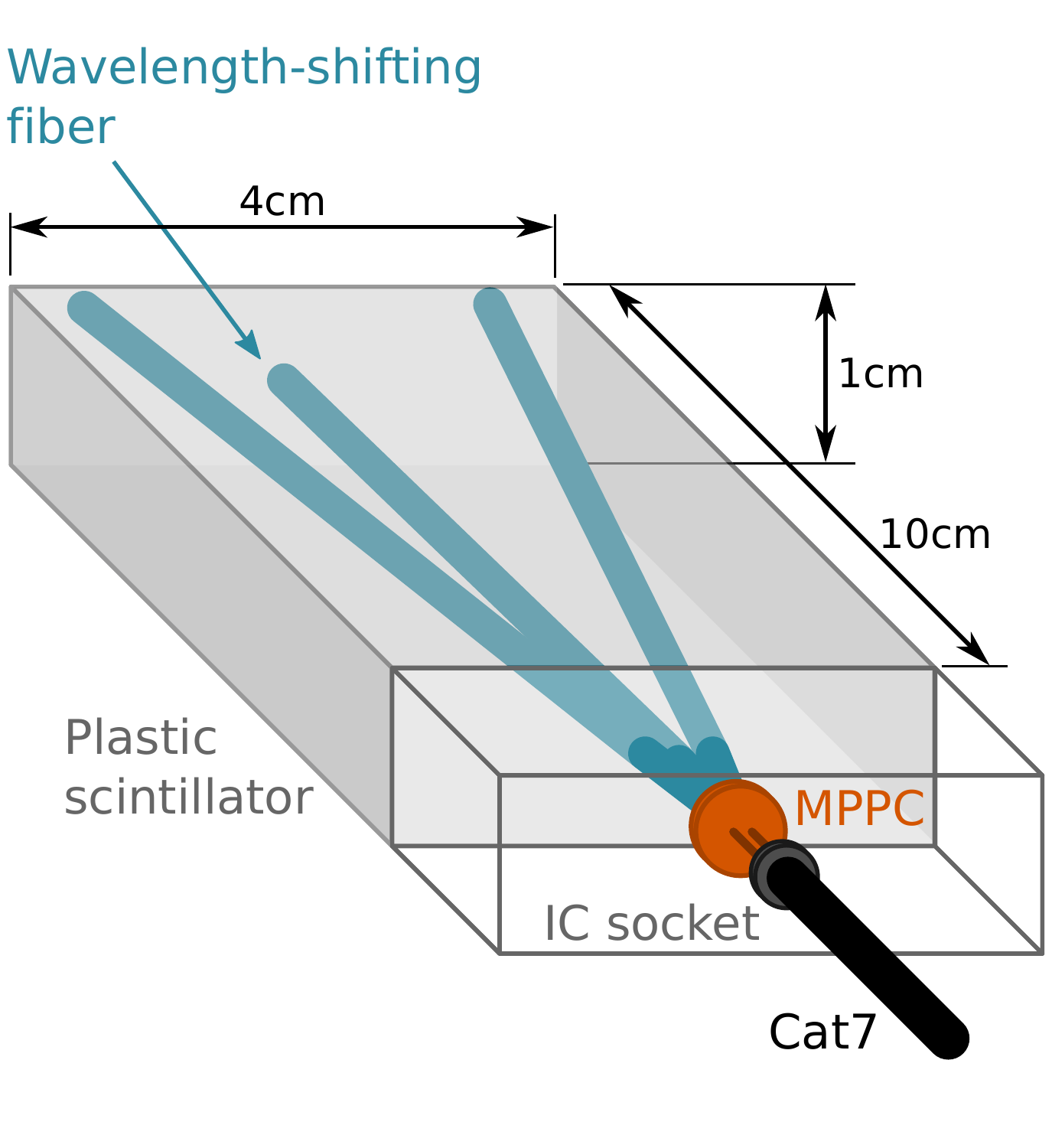}
		\caption{Diagram of a QCSS detector with embedded wavelength shifting fibers, MPPC and integrated circuit socket.}
		\label{fig:structure}
		%	\end{center}
		%\end{figure}
		
		%\begin{figure}[h]
		%	\begin{center}
%		\includegraphics[width=\columnwidth]{SCIImages/fixing}
%		\caption{cable structure}
%		\label{fig:fix}
	\end{center}
\end{figure}

In Phase~1, we installed two plastic scintillators with silicon photomultipliers near the IP. Each detector consists of a $100 \times 40 \times 10 ~\mathrm{mm}^3$ plastic scintillator bar with three embedded wavelength-shifting fibers.
The fiber ends are attached to a Hamamatsu photonics S12572-050C $3 \times 3 ~\mathrm{mm}^2$ MPPC.
A schematic view of the detector is shown in Figure~\ref{fig:structure}.

\subsubsection{Principle of operation}
%The readout system diagram is shown in Figure~\ref{fig:diagram}.
The QCSS detectors are mainly sensitive to the charged particles in beam background showers. The MPPC raw signal is delivered to the BEAST II DAQ room via 20~m-long Cat7 cables. The raw signal is received by the readout NIM module called EASIROC~\cite{EASIROC}, which can amplify up to 64 MPPC raw signals and also provides MPPC bias voltages.

The amplified signal is divided into two signals.
One signal goes to a system with a discriminator and a scaler (Contec CNT24-2(USB)GY),
to measure the hit rates versus machine conditions during Phase~1 machine studies.
The scaler records the hit rate every 10 seconds. Its saturation limit is about 500~kHz.
The discriminator threshold was set to be about half of a MIP signal, see
Table~\ref{tab:discri}.

\begin{figure}
	\begin{center}
		\includegraphics[width=\columnwidth]{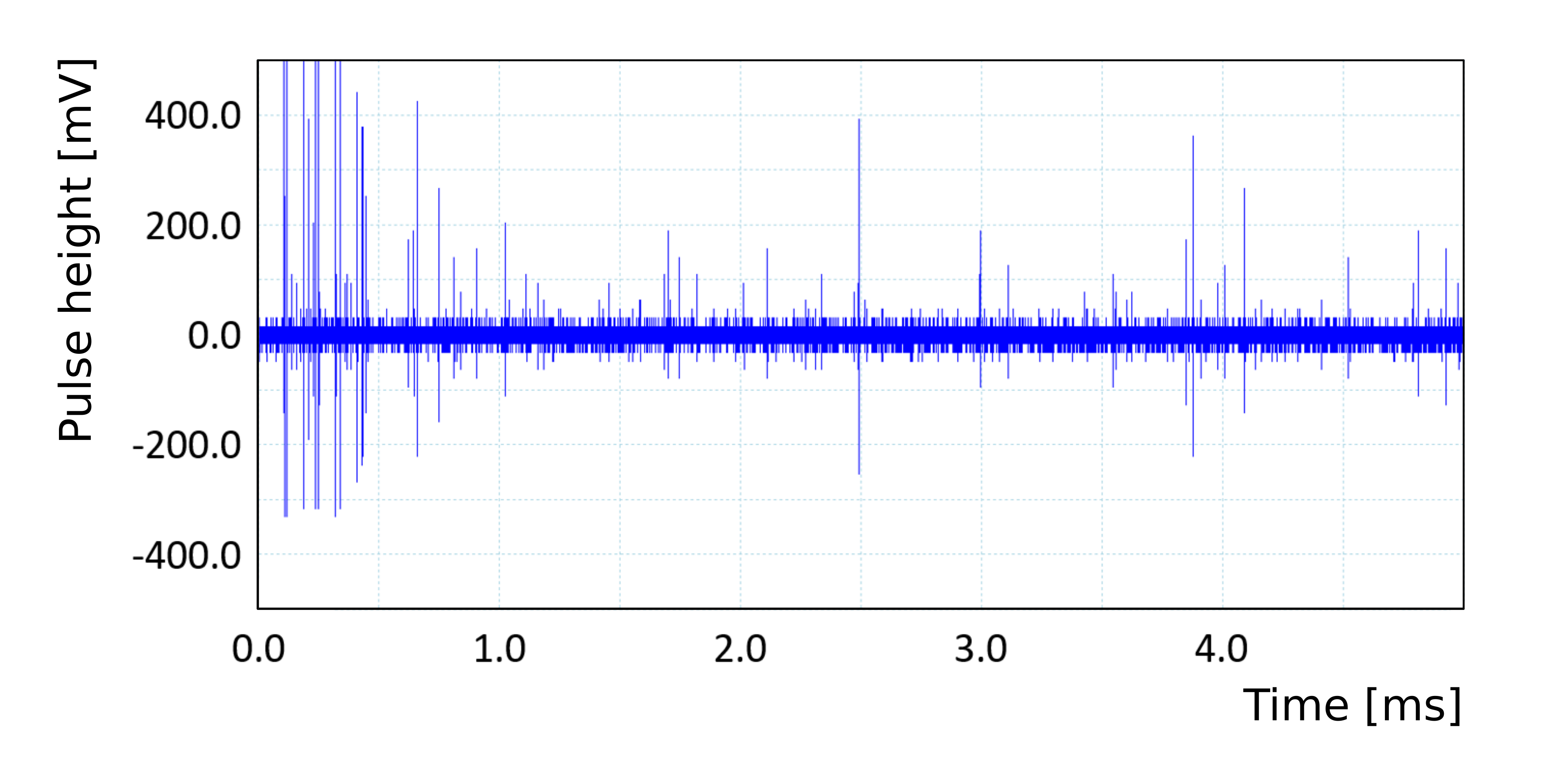}
		\caption{Typical scintillator waveform recorded by the PicoScope oscilloscope.
                         Sampling intervals are 51.2~ns.
                         The injected bunch arrives at t $\sim$0.1ms, and noisy spikes are seen after the injection.}
		\label{fig:waveform}
	\end{center}
\end{figure}

The other signal goes to a PC-based oscilloscope (PicoScope 6402C),
in order to record fast waveforms to observe injection background hits.
The oscilloscope is triggered by bunch injection timing signals.
We store 5~ms-long waveforms with a sampling frequency of $\sim$50~ns.
See Figure~\ref{fig:waveform} for typical waveforms recorded after injection.

The scaler and the oscilloscope are controlled by a PC, which can be accessed remotely.

%\begin{figure}
%	\begin{center}
%		\includegraphics[width=\columnwidth]{SCIImages/diagram}
%		\caption{Block diagram of readout system.}
%		\label{fig:diagram}
%	\end{center}
%\end{figure}

\begin{table}
	\begin{center}
		\caption{Discriminator threshold settings in the forward (FWD) and backward (BWD) QCSS detectors for each BEAST II beam study run.
                The threshold was adjusted to be around the 0.5~MIP level, but sometimes slightly changed. 
                The BWD detector was installed during the break between run 5003 and 5004.
                Unfortunately the threshold for the BWD detector after run 6004 was not recorded.}
                \begin{tabular}{lll} \toprule
			 Run range & FWD threshold [mV] & BWD threshold [mV]\\ \midrule
			2001 - 4008  & 550 &  no sensor \\ 
			5001 - 5003  & 640-720\footnotemark[1] & no sensor \\ 
			5004 - 5100  & 760 & 640-720\footnotemark[1] \\
			6001 - 13011 & 760 &  not recorded\footnotemark[2] \\ 
			\bottomrule
		\end{tabular}
		\label{tab:discri}
	\end{center}
\end{table}

\footnotetext[1]{It was adjusted using an analogue knob, so only rough values were recorded.}
\footnotetext[2]{We only know it was adjusted so that BWD and FWD rates are similar at run 6001.}

%\subsubsection{Performance}
%\subsubsection{Calibration}

 \clearpage

 % lead authors: Hiro Nakayama, Yuri Soloviev, Igal Jaegle
 \section{Simulation of beam backgrounds and detectors}\label{sec:simulation}
 %     file:		simulation.tex
%     authors:  	Hiro Nakayama, Igal Jaegle, Yuri Soloviev
%
%     contents:  
%		- Background generators (Hiro)
%			- SAD(Hiro)
%			- SR (Yuri)
%		- Detector Simulation (Igal)
%		        - Geant4, BEAST geometry, digitization (Igal)
%               - Simulation results and re-weighting procedure (Igal)
%                       

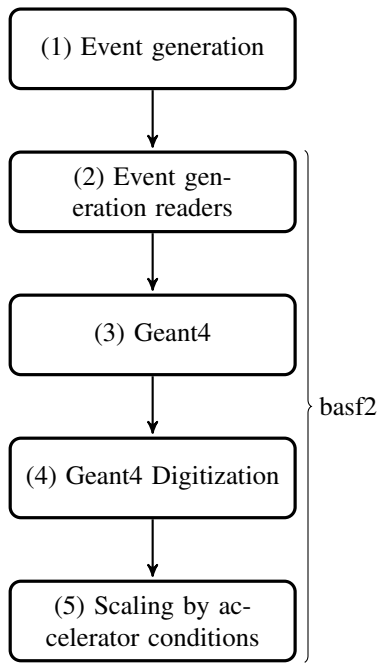
\begin{figure}[ht!]
\begin{center}
   \begin{tikzpicture}
    [node distance=0.8cm, start chain=going below, auto]
    \node[punktchain, join] (evt)   {(1) Event generation};
    \node[punktchain, join] (red)   {(2) Event generation readers};
    \node[punktchain, join] (gea)   {(3) {\GEANT}};
    \node[punktchain, join] (dig)   {(4) {\GEANT} Digitization};
    \node[punktchain, join] (wei)   {(5) Scaling by accelerator conditions};
    \draw[tuborg, decoration={brace}] let \p1=(red.north), \p2=(wei.south) in
    ($(2, \y1)$) -- ($(2, \y2)$) node[tubnode] {basf2};
    \end{tikzpicture}
\end{center}
\caption{The main steps of the simulation software pipeline.}
\label{fig:SimulationFlowChart}
\end{figure}

The BEAST II simulation produces simulated data files that can be compared directly with experimental data files. The simulation software pipeline involves several main steps, illustrated in Figure~\ref{fig:SimulationFlowChart} and described further below: (1) generation of primary particles from beam-induced backgrounds (Section~\ref{simulation_generation}); (2) handoff of the primary particles to {\GEANT} (Section~\ref{simulation_readers}); (3) modeling of the setup and the interaction and transport of the primary and secondary particles in {\GEANT} (Section~\ref{simulation_geant4}); (4) simulation of the detector response and digitization (Section~\ref{simulation_digitization}); and (5) scaling of the detector response with accelerator conditions to produce the simulated BEAST II data files (Section~\ref{simulation_scaling}). Steps (2) through (5) are done within the Belle II analysis software framework basf2 \cite{basf2}.

We discuss the validation of the detector response in Section~\ref{simulation_validation}, the simulation results in Section~\ref{simulation_results} and the systematic uncertainties on these results in Section~\ref{simulation_systematics}. 

% Hiroyuki Nakayama and Yuri Soloviev
\subsection{Event generation\label{simulation_generation}}
%     file:		simulation.tex
%     authors:  	Hiro Nakayama, Igal Jaegle, Yuri Soloviev
%
%     contents:  
%		- Background generators (Hiro)
%			- SAD(Hiro)
%			- SR (Yuri)

% Igal Jaegle
During Phase 1 there were no beam-beam collisions. Therefore, we only generate single-beam induced backgrounds: Touschek, beam-gas (bremsstrahlung and Coulomb), and synchrotron radiation. We do not simulate injection backgrounds. 

% Hiroyuki Nakayama
\subsubsection{Touschek and beam-gas scattering}
%     file:		simulation_generators_sad.tex
%     authors:  	Hiro Nakayama
%
% suggested outline:
%    - describe what SAD is and how it works
%    - describe how different background types are generated
%

%\subsubsection{Touschek and beam-gas event generators}
\label{simulation_generators_sad}

To estimate the Phase~1 beam loss rate due to Touschek and beam-gas (Coulomb and bremsstrahlung) scattering, we use the ``Strategic Accelerator Design'' (SAD) software framework \cite{SADHP} to track scattered beam particles. SAD is a versatile accelerator tracking code developed by the KEK accelerator group; it was used for the optics design and operation of KEKB, and it is now used for SuperKEKB. The default Phase~1 beam loss simulation assumes the machine parameters summarized in Table~\ref{Table:MachineParamForSim}. Other beam conditions are simulated by rescaling this default simulation, see Section~\ref{simulation_scaling}. We simulate synchrotron radiation in a different way, as described in Section~\ref{simulation_generators_sr}.

\paragraph{Definition of scattering position, loss position, slice, and SAD envelope}
 We here introduce a number of specific terms that are used in the following description of the beam background simulation. We denote the location where a beam particle undergoes Touschek, Coulomb, or bremsstrahlung scattering as the ``scattering position''. We generate sets of scattered beam particles for each section of the two SuperKEKB rings and propagate them using matrix-element calculations in SAD. When a simulated particle exits the physical aperture, determined by the vacuum pipes and the movable collimators, it is considered lost and SAD records the so-called ``loss position'' and the 4-momentum of the particle. The detailed procedure for determining whether a particle has in fact exited the physical aperture involves SAD comparing the transverse position of beam particles against a SAD specific envelope (hereafter referred to as the ``SAD envelope''), whenever the particle has travelled one fixed step size, which we refer to as the ``slice'' size.  The choice of slice size is a compromise between the computational power requirements and the accuracy of the SAD loss distribution. The SAD envelope is an approximation and hence differs slightly from the actual physical aperture implemented in the {\GEANT} model. If the loss position is in the vicinity of the interaction point, the loss position and 4-momentum are handed to the {\GEANT} simulation framework. {\GEANT} then runs a shower development simulation using the lost particle as a primary particle (see Section~\ref{simulation_geant4}).
 
SAD output files contain events weighted by scattering probability, which is calculated based on the beam optics parameters and vacuum pressure at the scattering position, and the particle energy and/or direction change resulting from the scattering process. The loss rate of each scattering process is calculated using the formulae described in \cite{PTEP:Ohnish}.

\begin{table}
\caption{Machine parameters used for the Phase 1 simulation.}
\label{Table:MachineParamForSim}
\begin{center}
\begin{tabular}{lcc}
  \toprule
  Machine parameters             & HER    & LER  \\ \midrule
  Beam current $I$~[A]            & 1.0    & 1.0  \\
  Number of bunches $N_b$        & 1000   & 1000 \\
  Bunch current $I_b$~[mA]        & 1.0    & 1.0  \\
  Vertical beam size $\sigma_y$~[\si{\micro}m]           & 59     & 110\\
  Emittance ratio $\varepsilon_y/\varepsilon_x$ & 0.1 & 0.1\\  
  Pressure $P$~[nTorr]            &  10    &  10\\
  \bottomrule
\end{tabular}
\end{center}
\end{table}

\paragraph{Simulation of collimators}
The Phase 1 collimators are implemented in the SAD tracking code. Any simulated particle that hits a collimator and is far from the interaction point is simply removed from the simulation. Tip-scattering off collimators is currently not included, but should be added in the future. 

The LER collimators called D06H3 and D06H4 are horizontal collimators that scrape the beam tail from both the inner and outer side of the ring. The eight HER collimators named DH09H1 through DH09H4 and D12H1 through D12H4 are horizontal collimators that scrape the beam tail from only the inner ring side. The eight HER collimators named D09V1 through D09V4 and D12V1 through D12V4 are vertical collimators that scrape the beam tail from only the upper or lower side. All the collimators in the simulation are set to be in ``fully-open'' positions, which are 25(20) $\mathrm{mm}$ from the beam orbit for the LER (HER).  

\begin{table}
\caption{SuperKEKB Phase 1 total ring loss rates and IR loss rates predicted by SAD.  Note that the two columns use different units.}
\label{Table:SADresults}
\begin{center}
\begin{tabular}{lcc}
\toprule
                      & Ring loss  & IR loss  \\ 
                      & rate [GHz] & rate [MHz] \\ \midrule
LER Coulomb           &      0.79       &      3.71 \\
LER Bremsstrahlung    &      2.92       &      5.07 \\
LER Touschek          &      3.48       &      4.59 \\
HER Coulomb           &      2.12       &      1.35 \\
HER Bremsstrahlung    &      3.55       &      1.94 \\
HER Touschek          &      3.75       &      0.05  \\
\bottomrule
\end{tabular}
\end{center}
\end{table}

\paragraph{Results of SAD simulation}
Here, we summarize select results of the SAD tracking simulation. Please note that all figures in this section use the SAD accelerator coordinates $x$, $y$, and $s$, which differ from both {\GEANT} and the BEAST II coordinates: the SAD $x$-axis is horizontal and points toward the outer ring, the SAD $y$-axis is vertical and points downward, and the SAD $s$-axis points along the beam orbit of each ring in the same (opposite) direction as the positron (electron) movement. At the interaction point $s$ is defined to be zero for both rings. The interaction region (IR) is defined as the region within $\pm4$~m of the IP in $s$.

\begin{figure*}[p]
        \centering
                \includegraphics[width=\columnwidth]{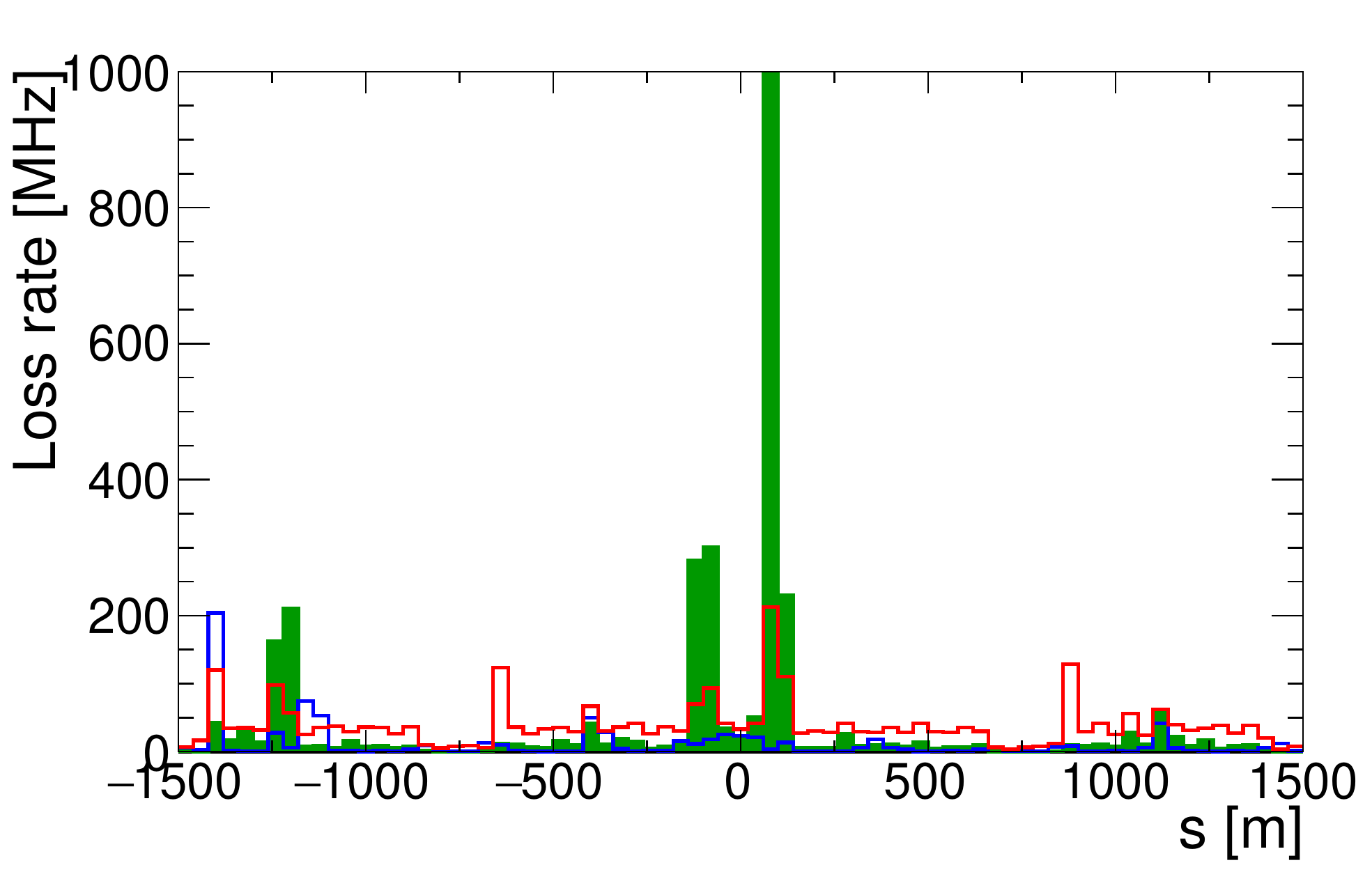}
                \includegraphics[width=\columnwidth]{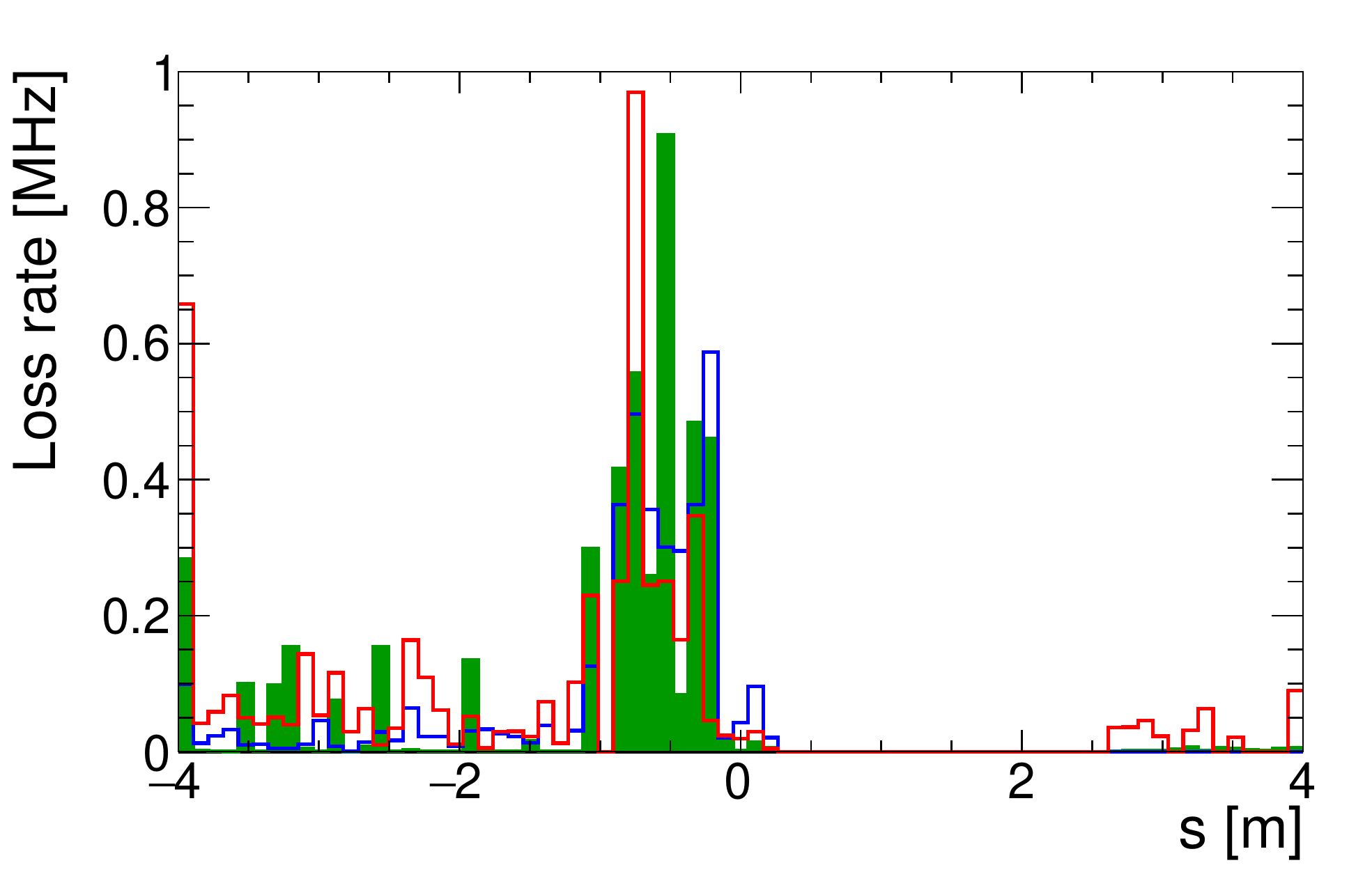}
        \caption{(color online) LER loss rate distributions from SAD, with contributions from Touschek (solid green), Coulomb (blue) and bremsstrahlung (red). For the LER, positive $s$ is downstream. Upper: full ring. Lower: Interaction region only.}
        \label{fig:ler_loss}
%\end{figure}

%\begin{figure}[ptb!]
%        \centering
                  \includegraphics[width=\columnwidth]{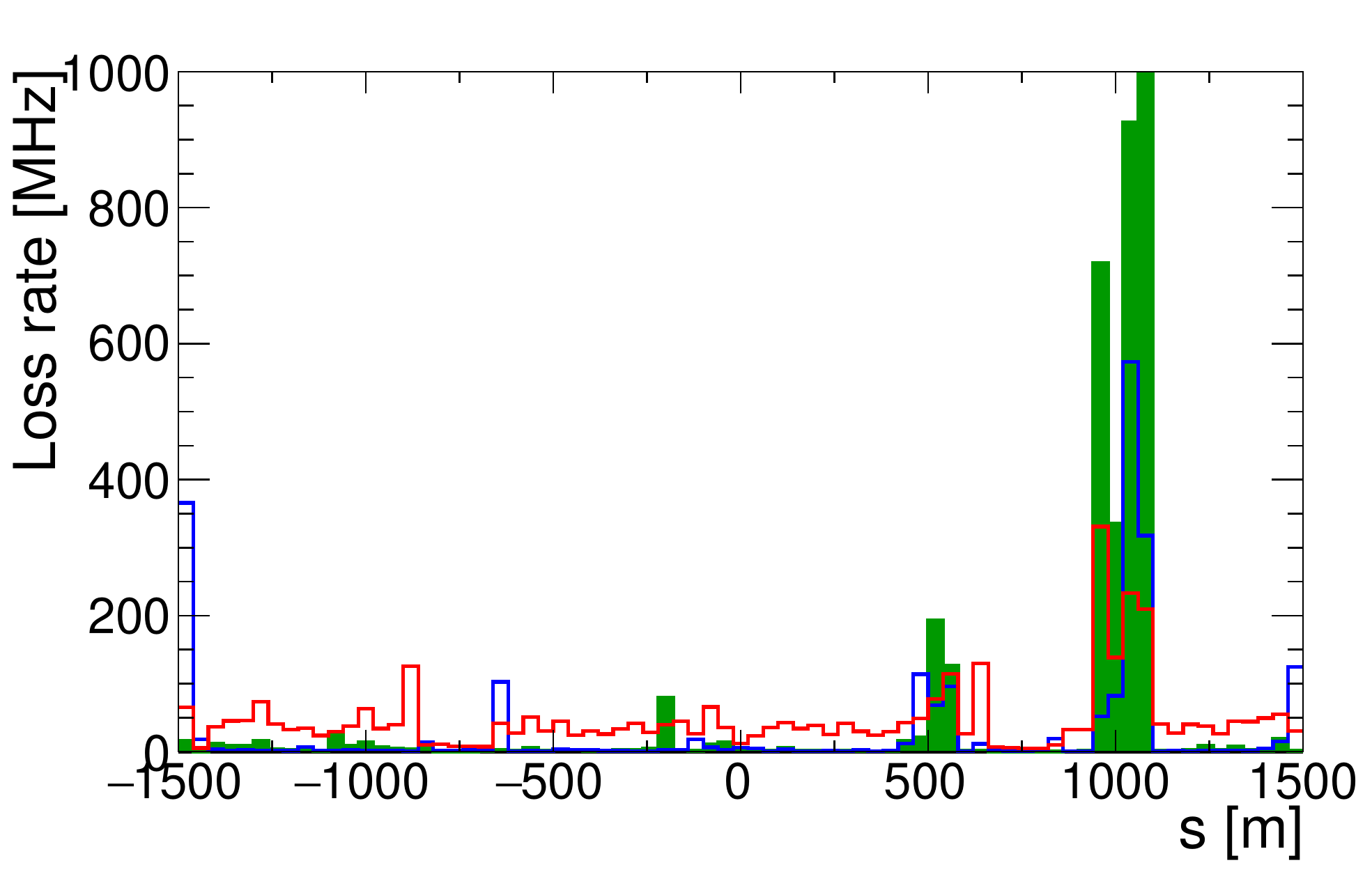}
                   \includegraphics[width=\columnwidth]{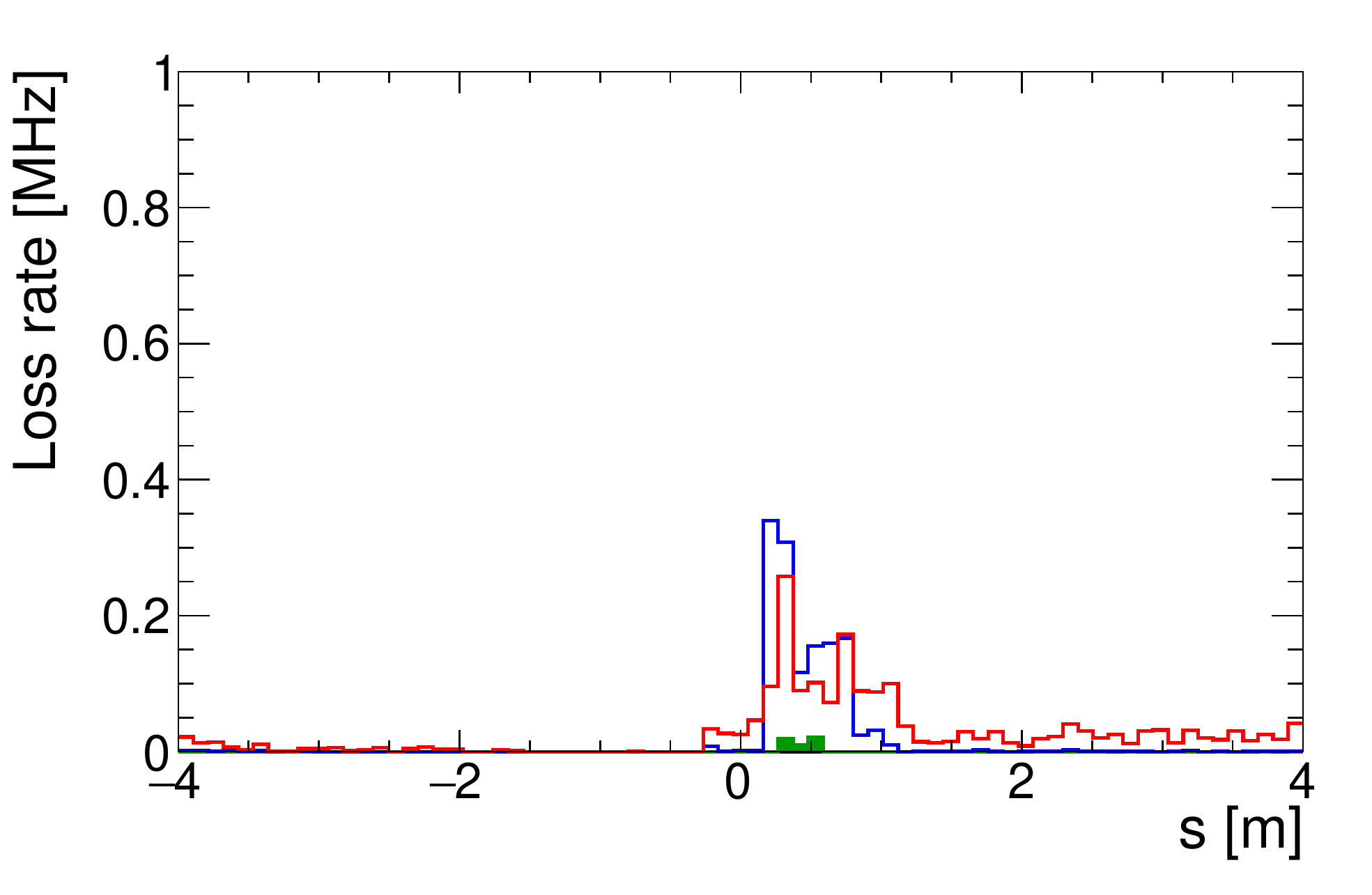}
        \caption{(color online) HER loss rate distributions from SAD, with contributions from Touschek (solid green), Coulomb (blue) and bremsstrahlung (red). For the HER, positive $s$ is upstream. Upper: full ring. Lower: Interaction region only.}
        \label{fig:her_loss}
\end{figure*}

Table~\ref{Table:SADresults} shows the predicted loss rates for the default SAD simulation settings in Table~\ref{Table:MachineParamForSim}. Figures~\ref{fig:ler_loss} and \ref{fig:her_loss} show the predicted loss position distribution in the LER and HER, respectively. Generally speaking, off-orbit particles tend to hit the beam pipe wherever the aperture decreases. As a result, the IR loss positions are mainly upstream of the IP in each ring; downstream losses are very small. In the LER Touschek loss distribution there are also spikes around $s \sim -1200~\mathrm{m}$, which correspond to losses at the D06H3 and D06H4 collimators. The spikes around $ |s| \sim 70~\mathrm{m}$ and $120~\mathrm{m}$ correspond to positions where the horizontal beam dispersion is large. In the HER loss distributions there are spikes around $s \sim 1000~\mathrm{m}$, corresponding to losses at the D12H1 through D12H4 collimators. 

\begin{figure}[tb]
        \centering
        \subfigure[LER.]{
                \includegraphics[width=\columnwidth]{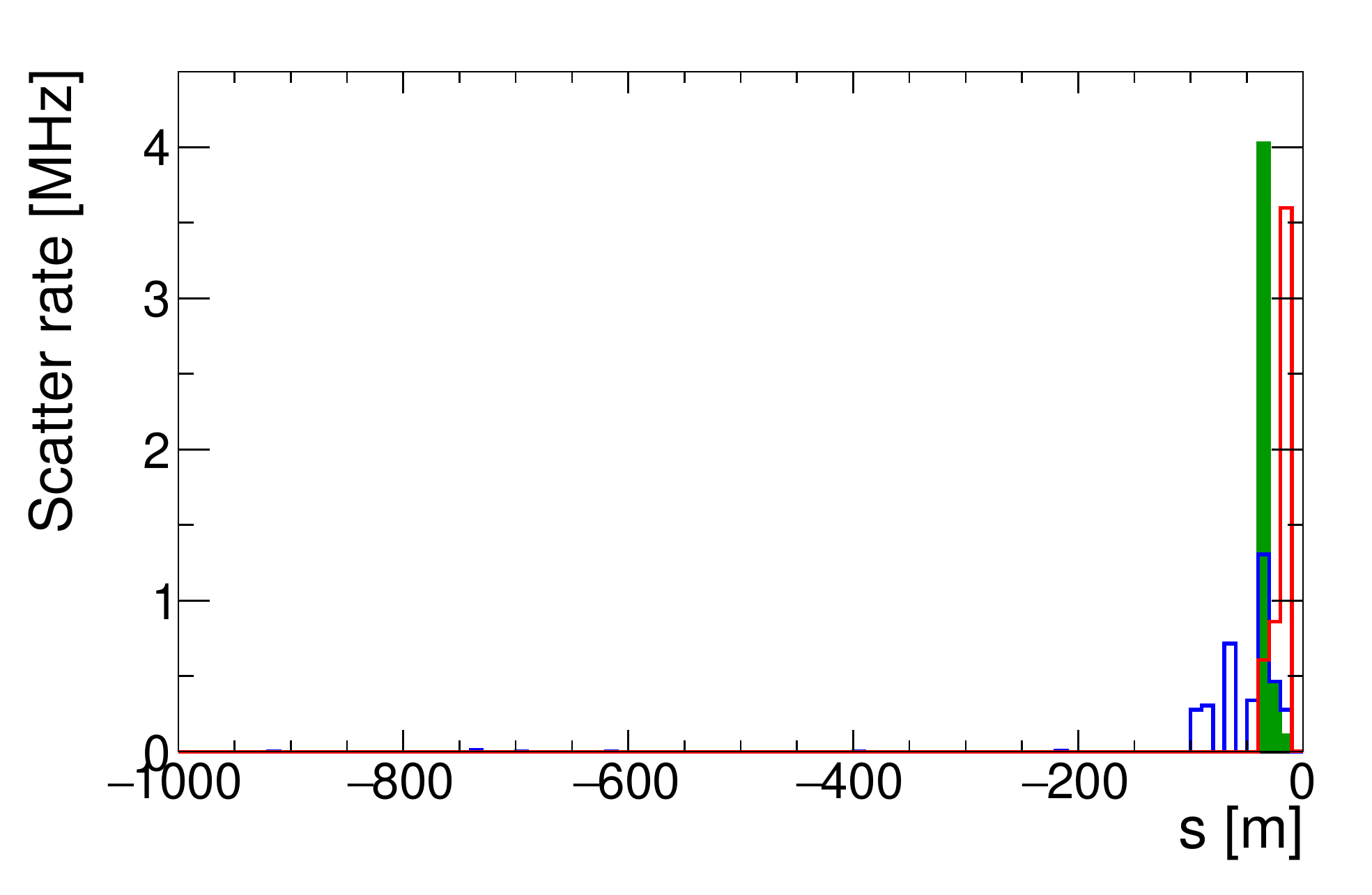}
                \label{fig:ringscatter_LER}
        } 
        \subfigure[HER.]{
                \includegraphics[width=\columnwidth]{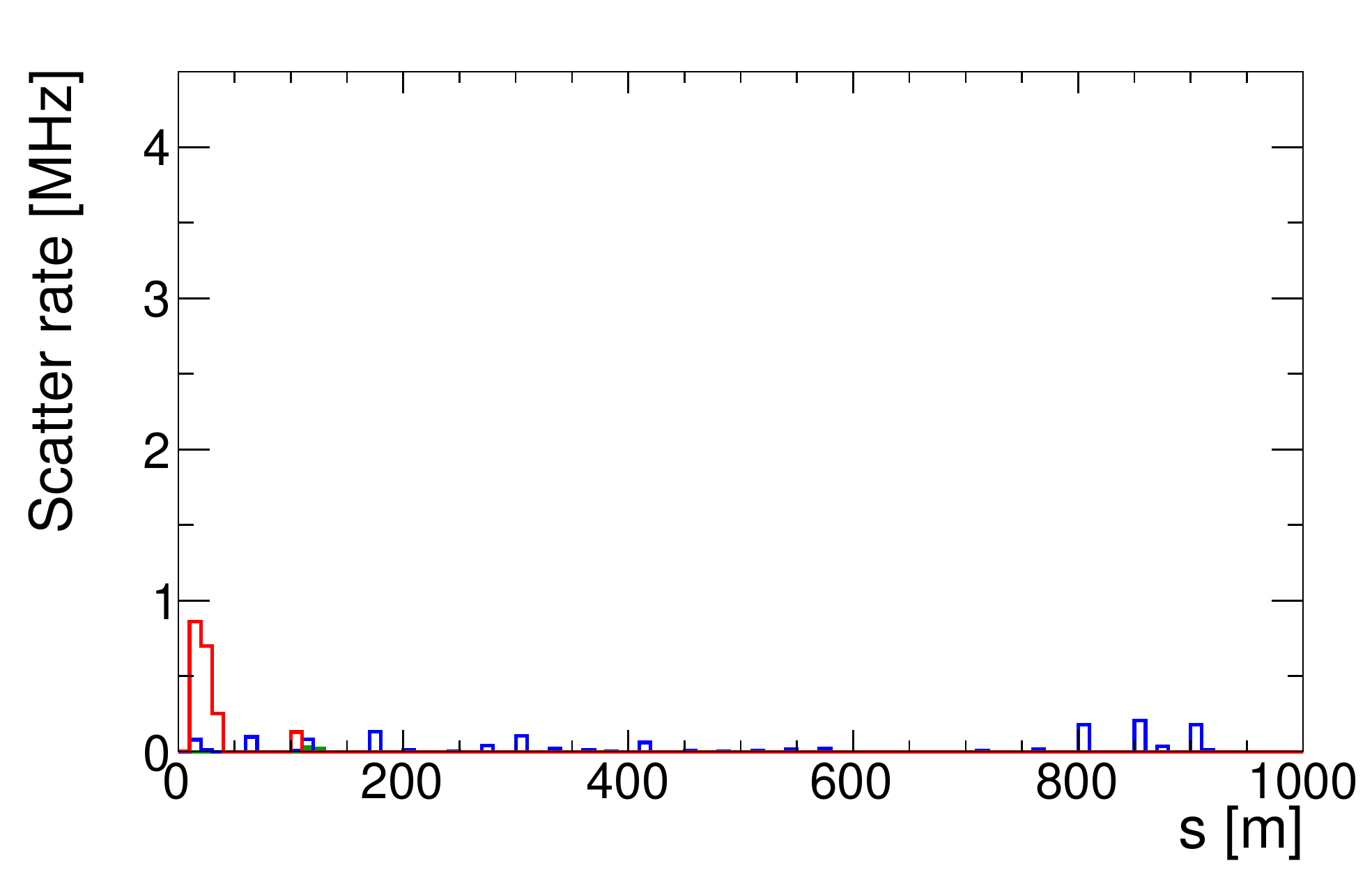}
                \label{figure:ringscatter_HER}
        }
        \caption{(color online) Scattering distributions from SAD for beam particles lost within $\pm4$~m of the IP, with contributions from Touschek (solid green), Coulomb (blue) and bremsstrahlung (red). The region within 1km upstream of the interaction point is shown. Note that for the LER, upstream is negative $s$, while or the HER upstream is positive $s$.}
        \label{fig:scattering_dist}
\end{figure}

\begin{figure*}
\begin{center}
\includegraphics[width=\textwidth]{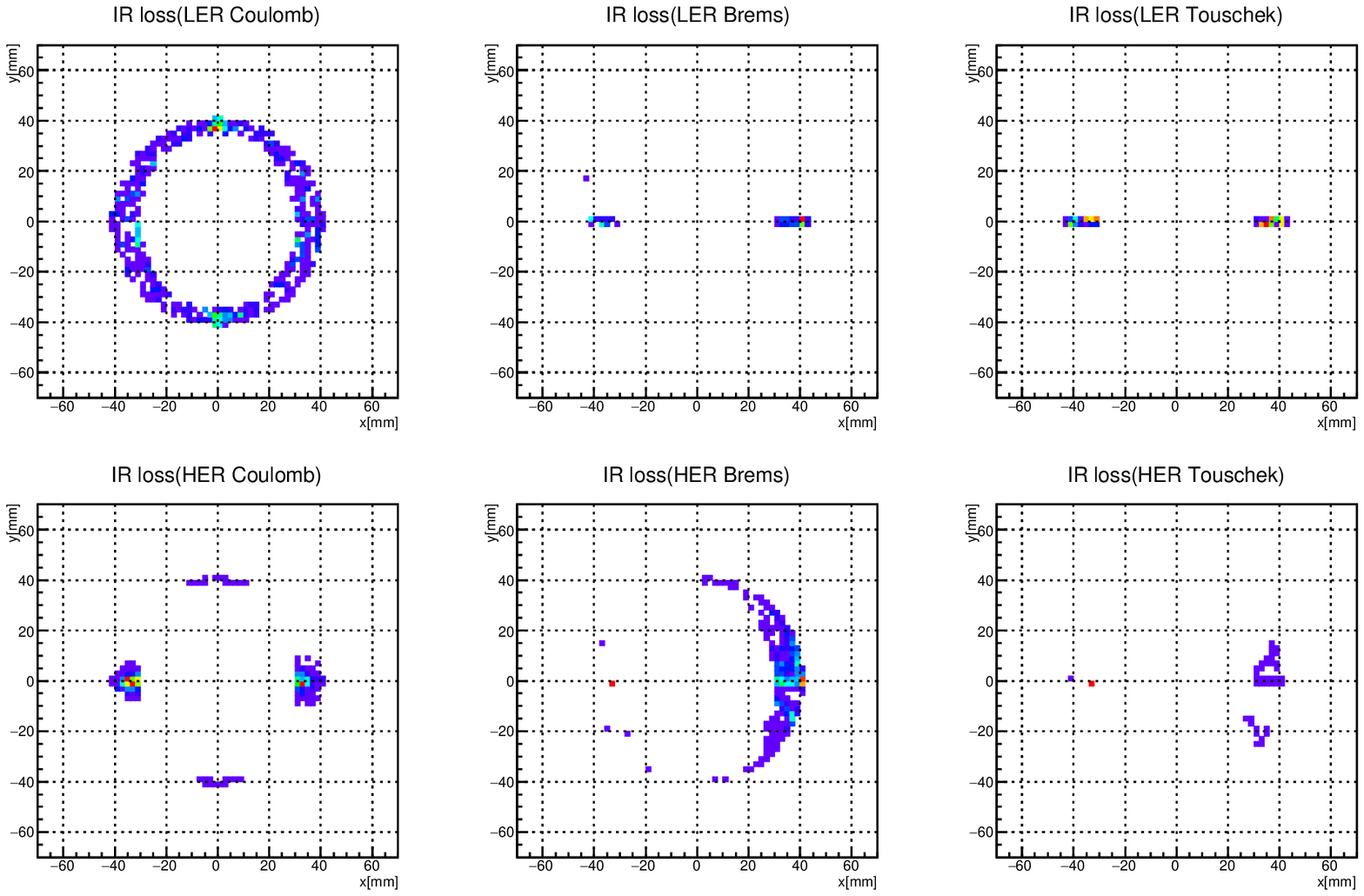}
\caption{The $x$-$y$ distribution of the particles lost within $\pm4$~m of the IP.}
\label{Figure:xy_ir}
\end{center}
\end{figure*}

Figure~\ref{Figure:xy_ir} shows the $x$-$y$ distributions of particle loss positions in the IR. Touschek and bremsstrahlung losses are mainly horizontal because these scattering processes change particle energies and SuperKEKB has horizontal dispersion.
LER Coulomb losses are mainly vertical while HER Coulomb losses are mainly horizontal; this is because the beta functions at the IP are $(\beta_x, \beta_y)=(25, 25)~\mathrm{m}$
for LER and $(40, 5.5)~\mathrm{m}$ for HER. Figure~\ref{fig:scattering_dist} shows the scattering position distributions for LER and HER particles that are lost in the IR.

\paragraph{Dependence on radiation damping and collimator settings}
In all simulation results presented in this article, the simulation of radiation damping in SAD is disabled by default. This could bias estimated loss rates if particles were lost after many turns. When we include radiation damping in the simulation, however, the total loss rate for the whole ring decreases only by $\sim\mathcal{O}$(10\%), while the IR loss rate does not change, because IR losses are determined by single turn effects. When we vary the apertures of the collimators by $\pm 0.1$~mm, the total loss rate and IR loss rate change by less than 0.1\%.

% dp/p0 plots and other plots will be included in Appendix?

% Yuri Soloviev
\subsubsection{Synchrotron radiation}
%     file:		simulation_generators_sr.tex
%     authors:  	Yuri Soloviev
\label{simulation_generators_sr}
%place for definitions and newcommands
\def\belletwo {Belle~II} % Consistent with usage in other sections of the paper.
\def\SR {SR}

The generation of primary charged particles and simulation of synchroton radiation ({\SR}) photon emission 
during propagation through the magnetic field requires a large amount of CPU time to get a full picture of the synchrotron radiation background. It is necessary to generate at least a few ms of integration time to observe hits of {\SR} photons in detectors, which is not realistic in terms of CPU requirements when the primary simulation is performed by {\GEANT} inside basf2.

Instead, we adopt the approach of generating only {\SR} photons, using vertex and momentum information from the primary beam particle simulation \cite{pv}. This results in a significant increase in the speed of the simulation. The primary simulation has to be redone only when we modify aspects of the geometry, such as the lattice or beampipe design. We transform the output of the primary simulation into the HEPEvt format, which is then used as input to the subsequent, {\GEANT} based steps of the simulation software pipeline.
%\subsubsection{Primary simulation}
%\subsubsection{Synchroton radiation event generators}

\paragraph{Ideal alignment}
The approach we use for generating beam particles in the primary simulation is described in several places, e.g. \cite{as}, \cite{on}. 
The simulation utilizes 2D phase space (transverse emittance only) i.e. [$(x, x'), (y, y')$] (where $(x, y)$ is the position and $(x', y')$ is the
direction of the beam particle), assuming an uncoupled beam matrix and a Gaussian beam shape.  In the co-moving coordinate system the 2D beam phase space ellipse 
can be parametrized by the Courant-Snyder invariant: ${\gamma}x^{2}+2{\alpha}xx'+{\beta}x'^{2}=\epsilon$, where $\epsilon$ is the transverse single particle emittance and $\alpha, \beta, \gamma$ are the Twiss parameters.
The particle positions and directions are most conveniently generated using normalized coordinates [$(\chi,\chi'), (\psi,\psi')]$ given by:

\begin{center}
$ x=\sqrt{\beta}\chi,\hspace{0.1cm}  x'=-\frac{\alpha}{\sqrt{\beta}}\chi - \frac{1}{\sqrt{\beta}}\chi'$, \\
$ y=\sqrt{\beta}\psi,\hspace{0.1cm}  y'=-\frac{\alpha}{\sqrt{\beta}}\psi - \frac{1}{\sqrt{\beta}}\psi'$, 
\end{center}
where the phase space can be simply described by a circle with radius $\sqrt{\epsilon}$ and then transformed to the physical 
coordinates in the {\belletwo} coordinate system using the inverse 2D beam matrix.
We determine the initial values for transverse position and direction of the generated $(e^{\pm})$ beam particles in the {\belletwo} coordinate
system by tracking them back from the IP upstream to the point of generation (assumed to be the entrance of the QC2 magnet) with a step length of 0.5~mm using {\GEANT} inside the basf2 framework.

In the next step the beam particles are then transported through the magnetic field of the beam elements, and we simulate the emission of synchrotron radiation using {\GEANT}. Only events with photons that hit the \emph{Target region} are selected. For Phase~1 the \emph{Target region} used is $\vert z \vert \le 200$~cm, which corresponds approximately to the extent of the BEAST~II support structure.

To simulate SR primaries, we modified a number of elements of the standard basf2 simulation used for Belle II, which will not be detailed here. To improve the usage of CPU time, we also optimized the simulation step size, physics list, and compilation options, and minimized printout. This improved the SR simulation performance by two orders of magnitude, with a final performance of 15~ms per event. 

\paragraph{Misalignment}
The central beampipe is connected to the vacuum system of the last beam elements of the machine through a system of bellows, which are located approximately $\pm500$~mm from the IP and are designed to absorb up to $\pm 0.5$~mm of misalignment. 
%installation of the central part of the {\belletwo} detector ($\pm 482mm$ from IP) doesn't exclude misalignment of beampipe 
%relative to the beam orbit which can reach $\pm 0.5mm$ at the bellows positions ($\pm 482mm$ from IP).
We consider two extreme cases of misalignment: rotation and longitudinal displacement. In the case of a rotation, a shift of $\pm 0.5$~mm (transverse to the beam line) at the bellows positions would result in an angle of rotation of $\sim 0.06^{\circ}$, which would propagate to a shift of $\pm 0.1$~mm at the ``sensitive'' part of the central beampipe (within 10~cm of the IP).
A longitudinal ($z$) displacement of $\pm 0.5$~mm in the bellows results in the same shift for the entire beampipe. 
A displacement is found to be the most dangerous type of misalignment. 
The effect of a misalignment in the vertical plane can be neglected due to the very narrow vertical beam size; our simulation confirms this assumption.
\paragraph{Beam tails}
The tails of the beam can produce a sizable contribution of {\SR} photons leading to backgrounds around the IP.
We can estimate the size of the beam tails with a detailed beam-beam simulation \cite{ko} for SuperKEKB and seperately from the TDR for KEKB \cite{KEK:1995sta}. 
The fraction of particles in beam tails beyond $10\sigma_{x}$ is $5\times10^{-7}$ according to beam-beam simulations, while the fraction beyond $30\sigma_{y}$ is up to $10^{-5}$ according to the KEKB TDR.
We use the more conservative estimate from the KEKB TDR in the following to evaluate the contribution of beam tails to the {\SR} background.
A uniform distribution of the beam tails with a half width of $20\times\sigma_{core}$ is assumed in the primary simulation.

\paragraph{Obtaining detector occupancies due to SR}

We use the HEPEvt file for each type of geometry as an input file to subsequent steps in the simulation pipeline. To simulate {\SR} background in detectors for the misaligned geometry we modify the HEPEvt file for the final simulation such that the $x$, $y$ coordinates of {\SR} vertex position are displaced by $\mp 0.5$~mm, i.e. where the sign is opposite to the assumed displacement of the beampipe. As a cross-check, we repeat the primary simulation for the modified beampipe geometry (displaced by $\pm 0.5$~mm) and we use the resulting HEPEvt file with {\SR} parameters for the final simulation instead. The results we obtain with both methods are consistent within statistical errors. We include atomic de-excitation processes Fluorescence (FLUO) and Particle Induced X-Ray Emission (PIXE) in the {\GEANT} physics list for all final simulations, though preliminary estimation did not show significant contributions from these processes.

\begin{table}[ht]
	\caption{Magnets included in the simulation of synchroton radiation.}
	\centering	
	\begin{tabular}{lcll}
\toprule
	Magnet name			&	Ring 		& 	Magnet type	& extent in $s$ [m] \\	
\midrule
	ZHQLC2LE			& HER		& bending 		& 5.248 to 5.593\\
 	QLC2LE				& HER		& quadrupole 		& 5.839 to 6.398\\
 	BLC1LE				& HER 		& bending	 		& 7.505 to 11.105\\
	BLCWRP	  			& LER 		& bending 		& 5.538 to 7.768\\
 	BLC1RP	  			& LER		& bending 		& 8.268 to 10.499\\
\bottomrule
	\end{tabular}
	\label{tab:magnets}
\end{table}

\paragraph{Phase 1 magnets and beam pipe}
We take several special features of the Phase~1 setup into account in the simulation for the case of ideal alignment and gaussian bunch shape. We include the magnets listed in Table \ref{tab:magnets}. The effect of the corrector magnets is neglected. To account for the effect of fringe fields the effective length of the magnets is approximated by their physical length plus half the distance at which the field drops to zero on both sides. While for the bending magnets a constant field value is assumed, for the quadrupole magnets a position dependent field value  $B = K \times X$, with field gradient $K$-field and deviation from the central orbit $X$ inside the effective length is used. 

The thickness of the aluminum beampipe in Phase~1 is 4~mm. For such an amount of material the fraction of transmitted intensity for {\SR} photons with energy less than $14$~keV is lower than $\sim 10^{-6}$. Conservatively, photons with energy above 10~keV (corresponding to a fraction of transmitted intensity of $\sim 5\times10^{-13}$) are kept in the simulation.

Beam particles are generated at $z = \pm \sim 11.6$~m from the IP with the parameters determined from the optics files. We assume beam currents of 1.0~A and 2500 bunches for both the LER and the HER. 

\paragraph{SR simulation results}

\begin{figure}[h]
\centering
\includegraphics[width=\columnwidth]{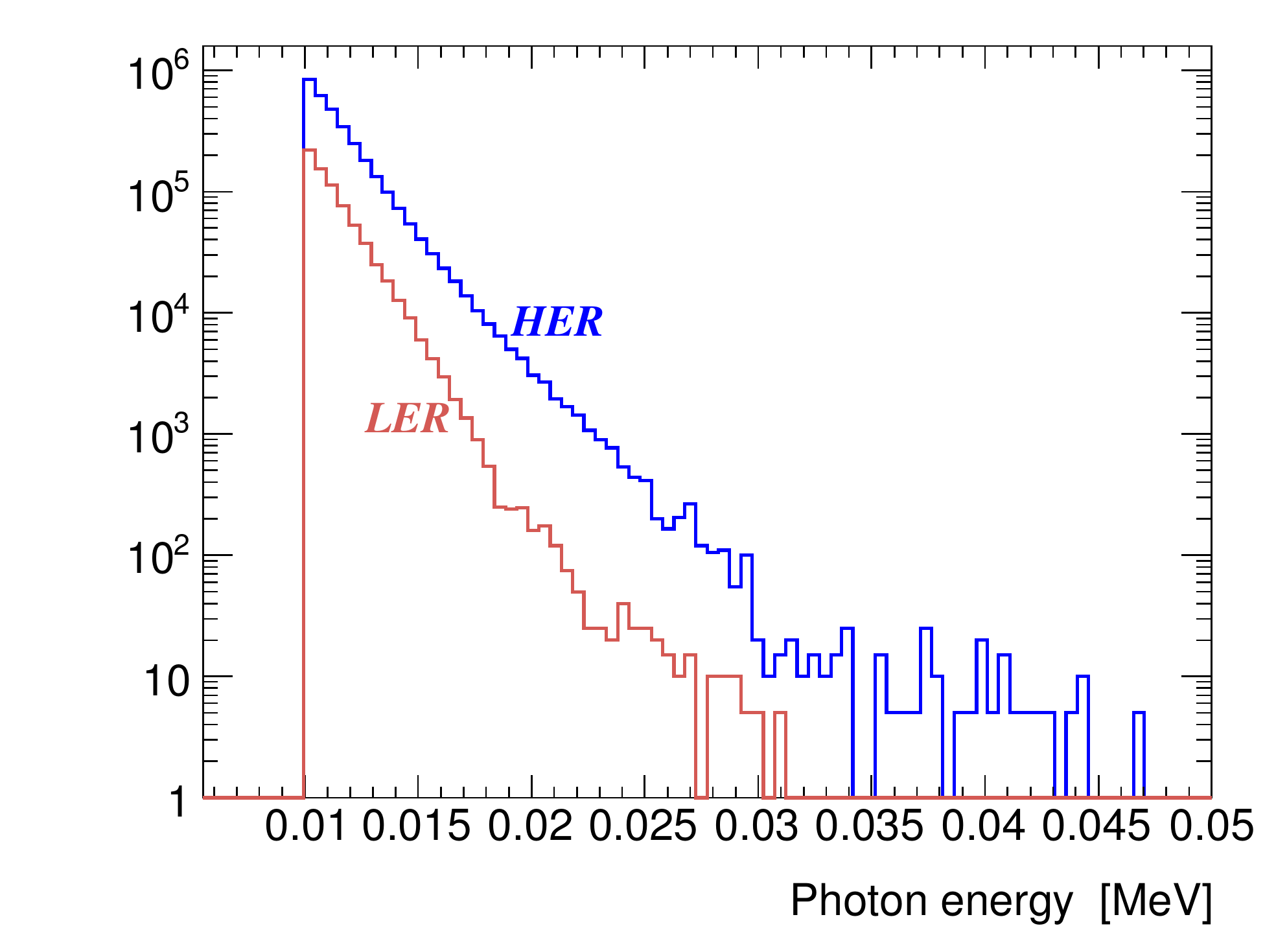}
\caption{Energy spectrum of generated {\SR} photons.}
\label{fig:SREnerHER}
\end{figure}

In the primary beam particle simulation, $2.5\times10^{10}$ initial particles (corresponding to one bunch at 1A of beam current) are generated. Most of the {\SR} photons that hit the target region originate from the BLC1LE magnet (HER) and from the BLCWRP magnet (LER).
Figure~\ref{fig:SREnerHER} shows the energy distribution of the generated photons.

\begin{figure}[h]
\centering
\includegraphics[width=\columnwidth]{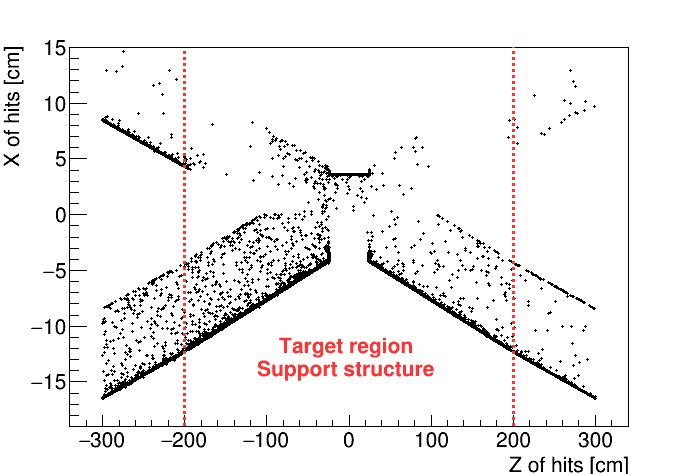}
\caption{Projection of hit map of SR photons on the horizontal plane}
\label{fig:XZSRHits}
\end{figure}

We generated a sample corresponding to 2~ms integration time in the final detector simulation for both HER and LER with the following BEAST II detector systems included: PIN, Diamond, Crystals, BGO, CLAWS, $^{3}$He, and TPC. We observed no simulated hits in any of these detectors. Most of the hits of {\SR} photons in the support structure area are on the inner side of the beampipe (towards the center of the ring in the negative $x$-direction) (see Figure~\ref{fig:XZSRHits}). We perform a cross-check simulation that shows that all {\SR} photons in the nominal simulation are absorbed in the beampipe walls. We derive an upper limit of 500~Hz for the hit rates from {\SR} background in all BEAST~II detectors for Phase~1.

% Igal Jaegle
\subsection{Event generation readers\label{simulation_readers}}
% File: simulation_readers.tex
% Lead author: Igal Jaegle

The Phase 1 beam loss positions, momenta, and rates in the IR are calculated with SAD and the synchroton radiation software as described in Section~\ref{simulation_generation}. In addition, the SAD loss positions are given at the exit of the slice and do not correspond to the exact loss position, which occurs upstream. Therefore we transform into the {\GEANT} coordinate system and correct the SAD coordinates in order to be equal to or smaller than the aperture size set in SAD by using the (exact) momentum provided by SAD. This can be achieved iteratively by moving upstream by a fixed step size of $ds$ = 10~\si{\micro}m, which we find to result in the smallest systematic uncertainty on the SAD loss rate (for more details read~\ref{simulation_systematics}).
The number of (primary) particles launched by {\GEANT} at each loss position corresponds to the calculated loss rate attached to the particle lost scaled by a time equivalent to the running accelerator beam time, which is typically 1~s for each background type and ring. The SAD loss rate has non-negligible statistical uncertainty that depends on the number of particles launched at the injection point and the number of SAD simulations done and introduces a systematic error on the simulated {\GEANT} sensor rates and doses.

% Igal Jaegle
\subsection{Geant4}
% File: simulation_geant4.tex
% Lead author: Igal Jaegle

\label{simulation_geant4}

We use the {\GEANT} toolkit \cite{g4} to simulate the passage of particles through the Phase~1 setup. We use {\GEANT} version 10.02 with the standard database for interaction cross sections. 

\subsubsection{Geometry}

\begin{figure*}[htp]
  \centering
  \includegraphics[width=1.5\columnwidth,page=4]{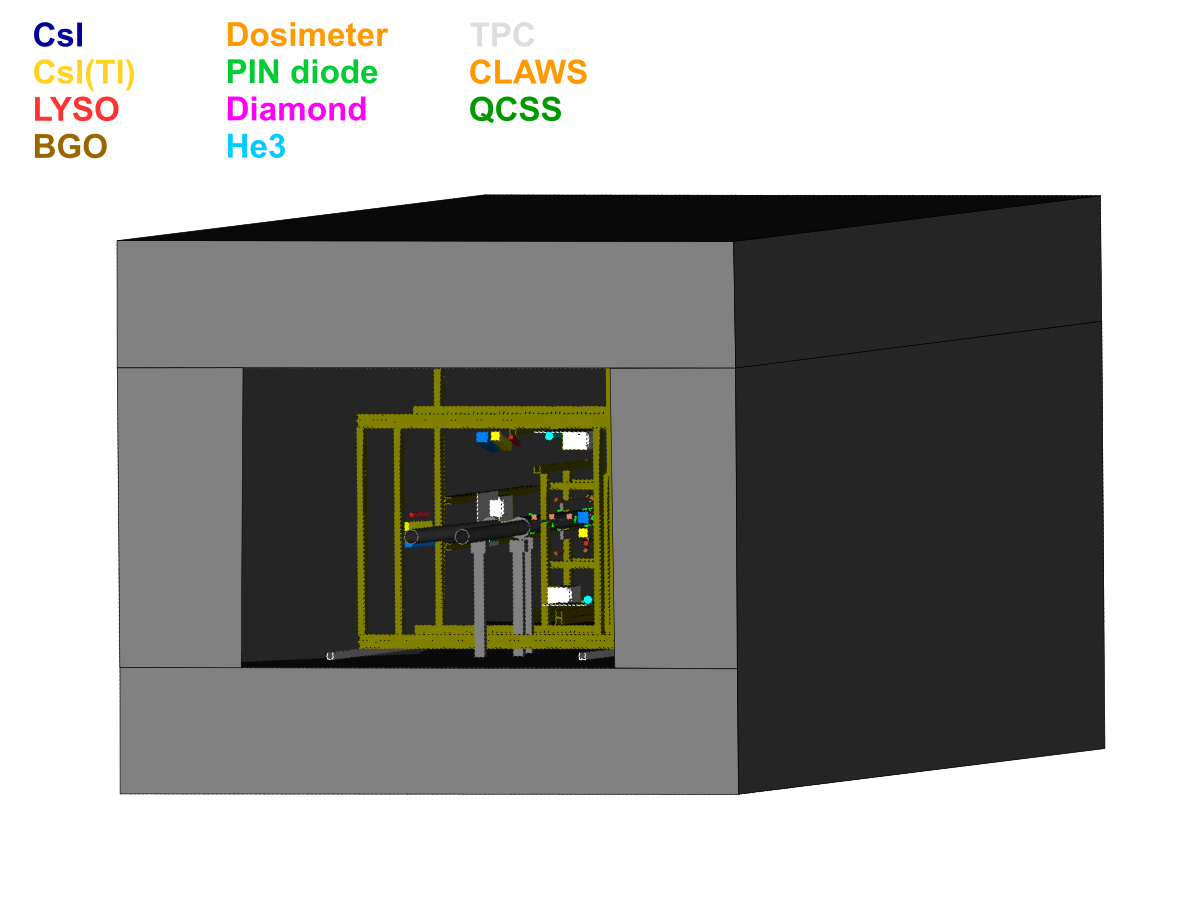}
  \caption{(color online) Color-coded visualization of the {\GEANT} model. The largest structure is a cement shield, which surrounds the BEAST II setup and SuperKEKB IP chamber.}
  \label{fig:Geant4Cave}
\end{figure*}

\begin{figure*}[hbp]
  \centering
  \includegraphics[width=1.5\columnwidth,page=3]{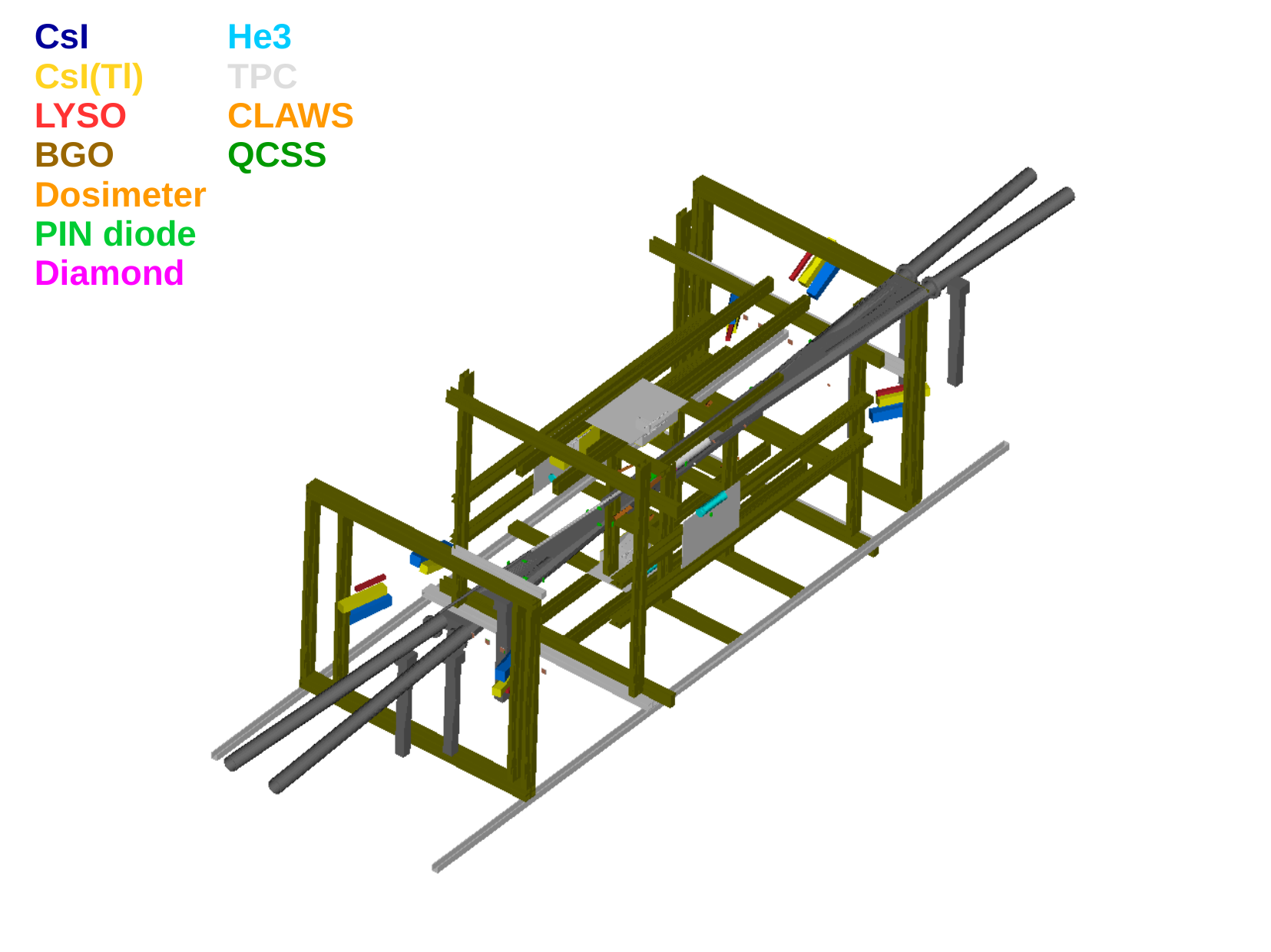}
  \caption{(color online) Color-coded visualization of the {\GEANT} model of the BEAST II detectors and support structure. Compare to Fig.~\ref{fig:BEAST}.}
  \label{fig:Geant4woCave}
\end{figure*}

\begin{figure*}[htp]
  \centering
  \includegraphics[width=1.5\columnwidth,page=2]{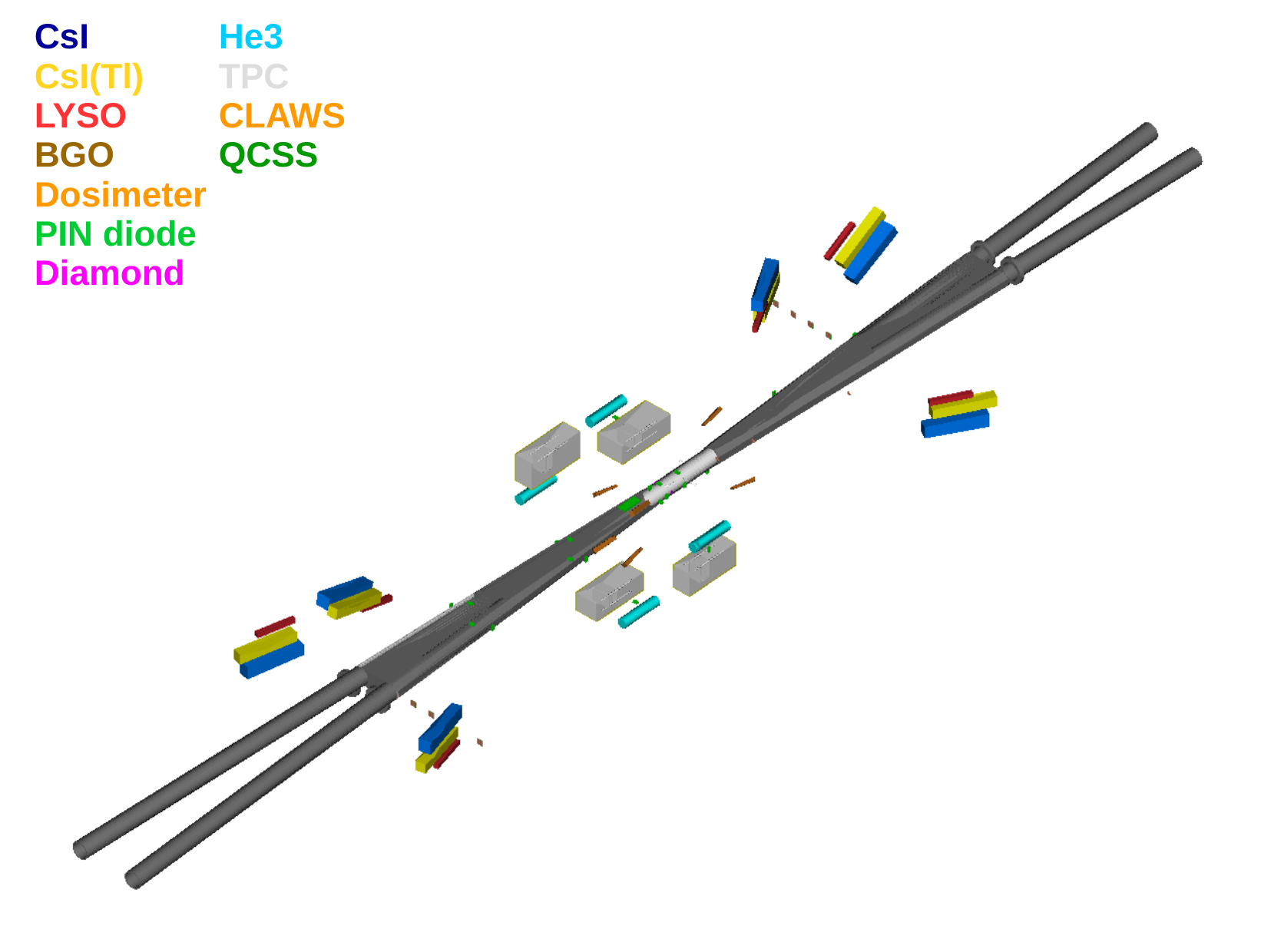}
  \caption{(color online) Color-coded visualization of the {\GEANT} model of the BEAST II sensors and the SuperKEKB IP chamber. Compare to Fig.~\ref{fig:BEAST}.}
  \label{fig:GeantwSensorsOnly}
\end{figure*}

\begin{figure*}[hbp]
  \centering
  \includegraphics[width=1.5\columnwidth]{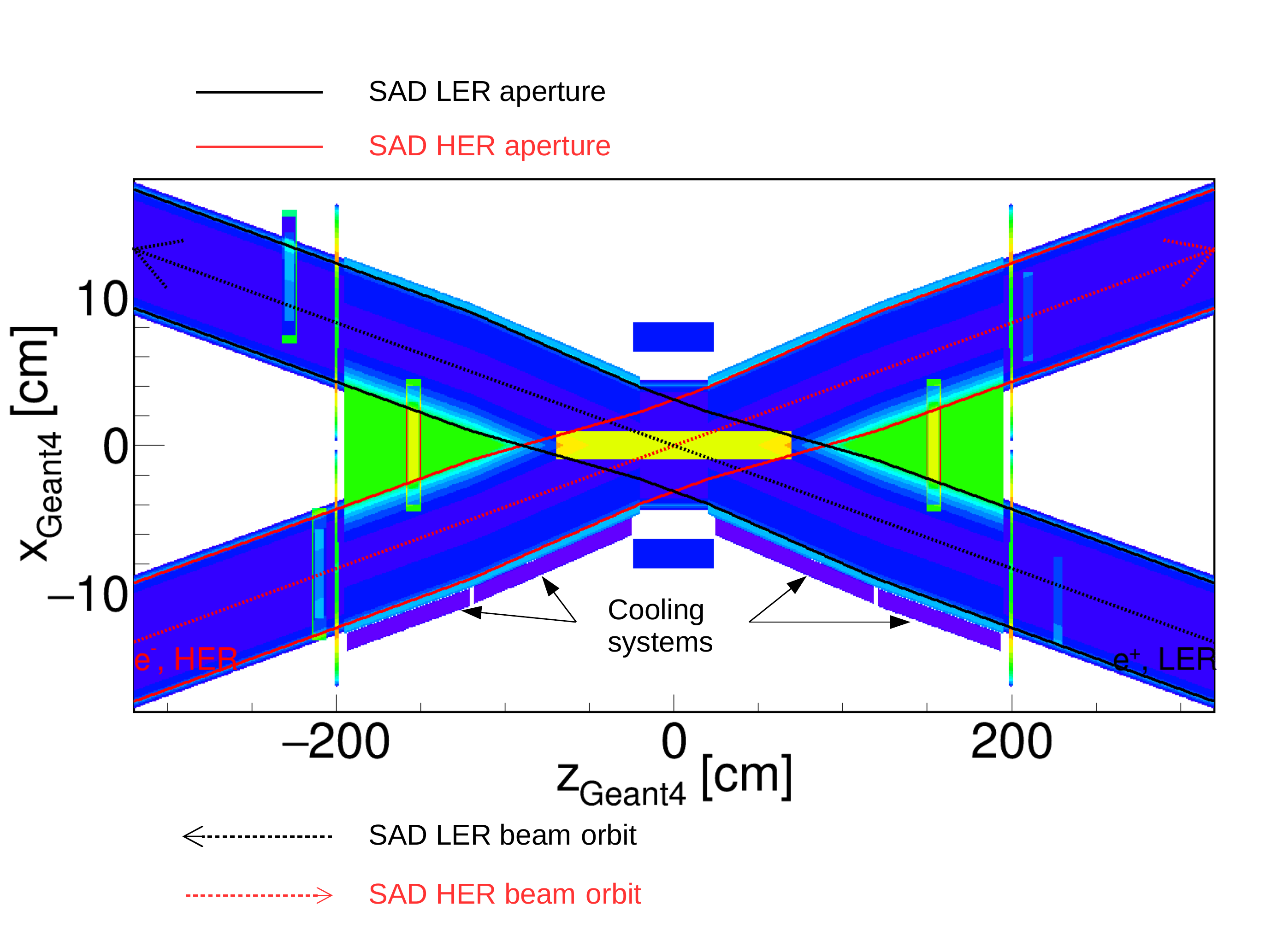}
  \caption{(color online) {\GEANT} IP chamber material scan with the SAD aperture and beam orbits overlaid. The arrows indicate the positron beam and electron beam directions. The cooling systems are placed where the loss rates are expected to be the highest.}
  \label{fig:GeantwOnlyV7Chber}
\end{figure*}

The geometry and materials of the BEAST II sensors, mounting brackets, beampipes, and cement shields surrounding the interaction region are modeled in {\GEANT} based on an as-built survey with a precision of 1~cm. Precise measurements of the positions of the fiberglass support structure are not available and hence approximate values are used in the simulation. In addition, not all the small components, such as attachment screws and mounting brackets, could be included. 

Figure~\ref{fig:Geant4Cave} shows the complete {\GEANT} model. The most prominent feature is the 180 metric ton cement shield surrounding the BEAST II setup. The IP chamber is also visible. The cement shield has walls of 37.5~cm thickness and contains rebar, which was approximated in the model by adding iron as an extra element to the cement. Figure~\ref{fig:Geant4woCave} shows the model with the shield removed. Figure~\ref{fig:GeantwSensorsOnly} shows a closer view of just the BEAST II sensors and the IP chamber. 

Figure~\ref{fig:GeantwOnlyV7Chber} shows a material scan of the {\GEANT} IP chamber model with the SAD aperture and beam orbits overlaid. 

\begin{figure*}[htp]
\centering
\includegraphics[width=1.5\columnwidth]{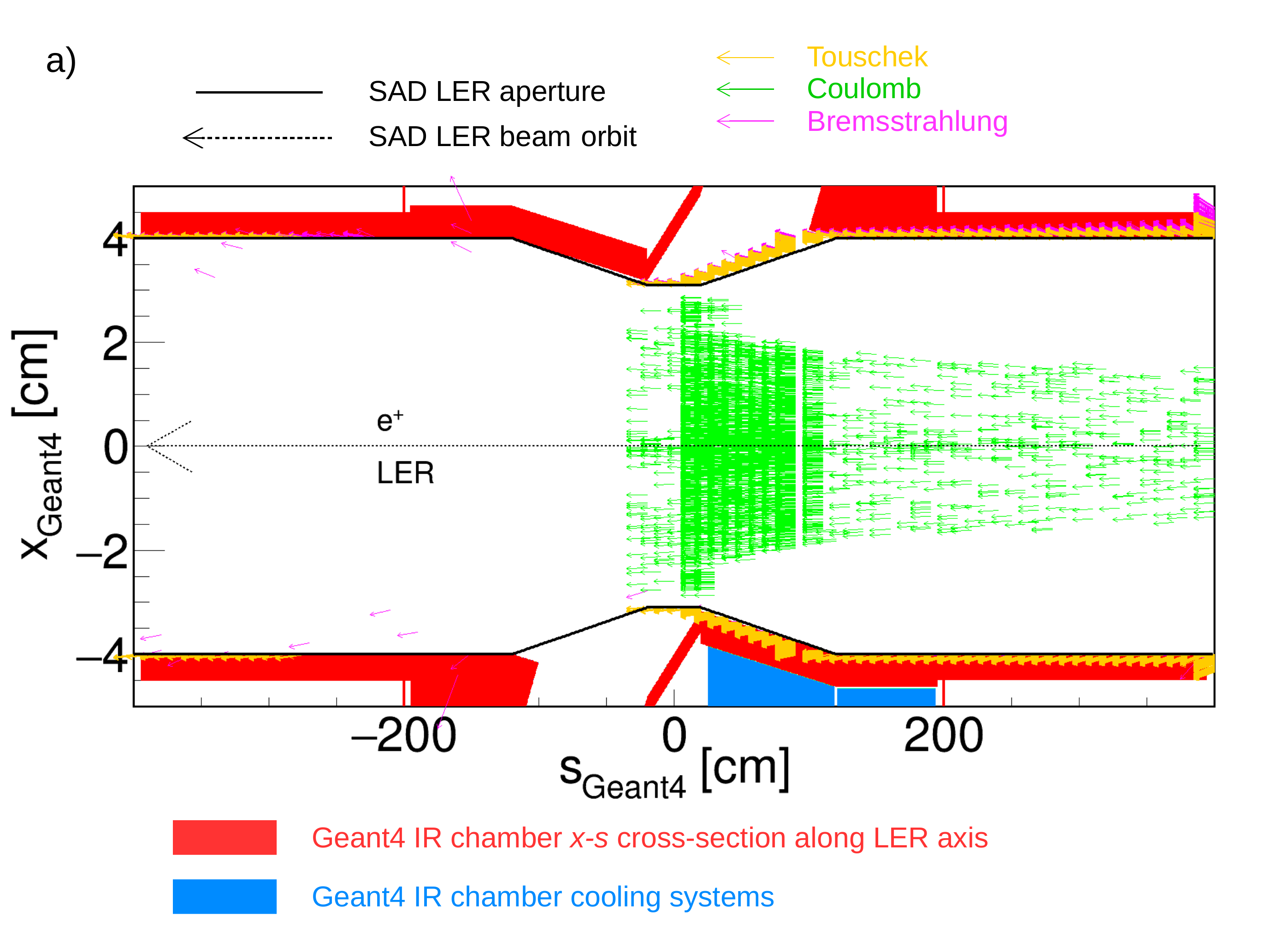}
\includegraphics[width=1.5\columnwidth]{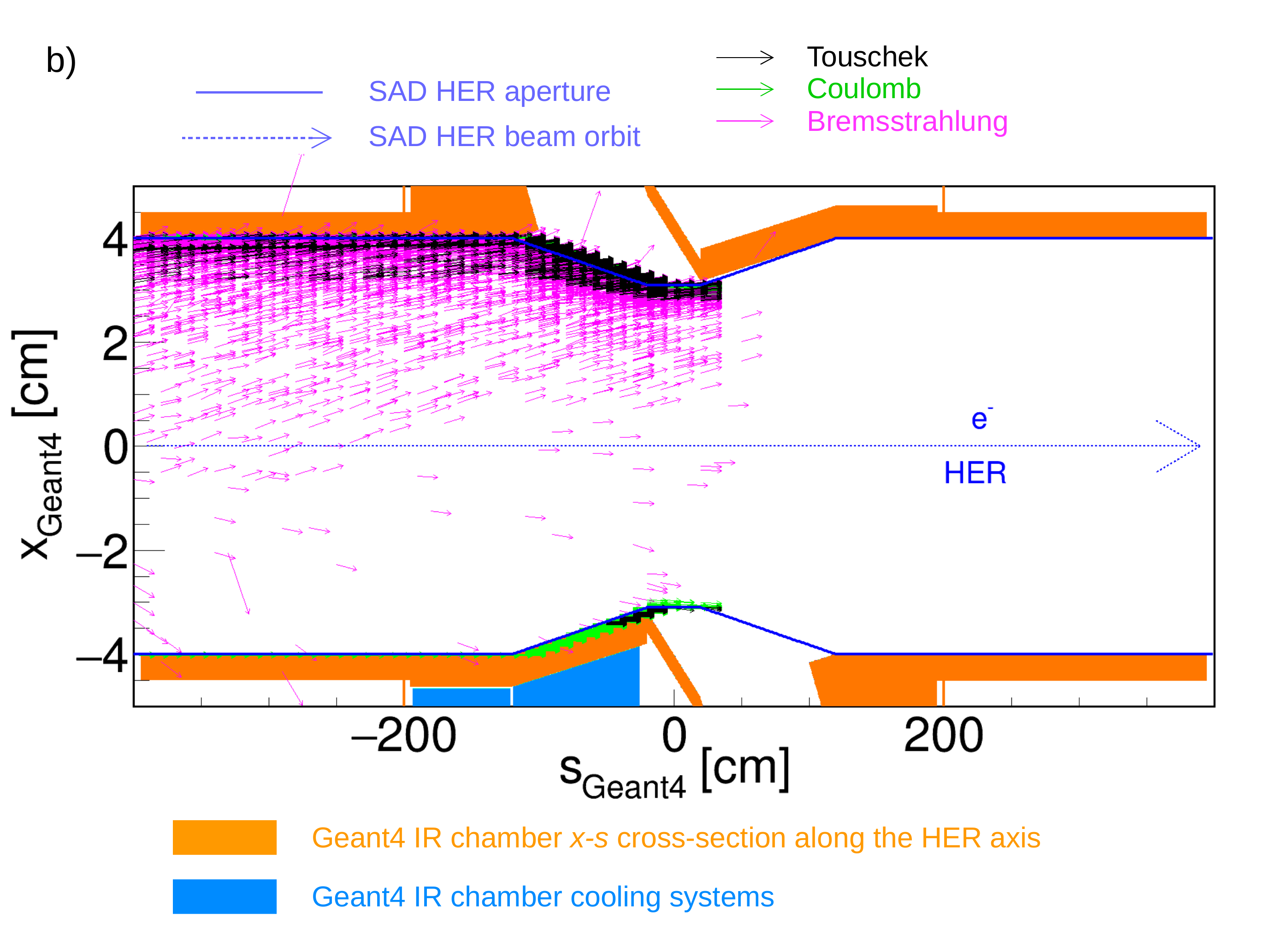}
\caption{(color online) {\GEANT} IP chamber horizontal cross section versus (a) the LER axis and (b) the HER axis, $s_{Geant4}$, with the SAD aperture and beam orbit overlaid. The arrows indicate the positron beam direction and particles lost direction due to Touschek, bremsstrahlung, and Coulomb.}
\label{fig:v7ChberXS} 
\end{figure*}

\begin{figure*}[htp]
\centering
\includegraphics[width=1.5\columnwidth]{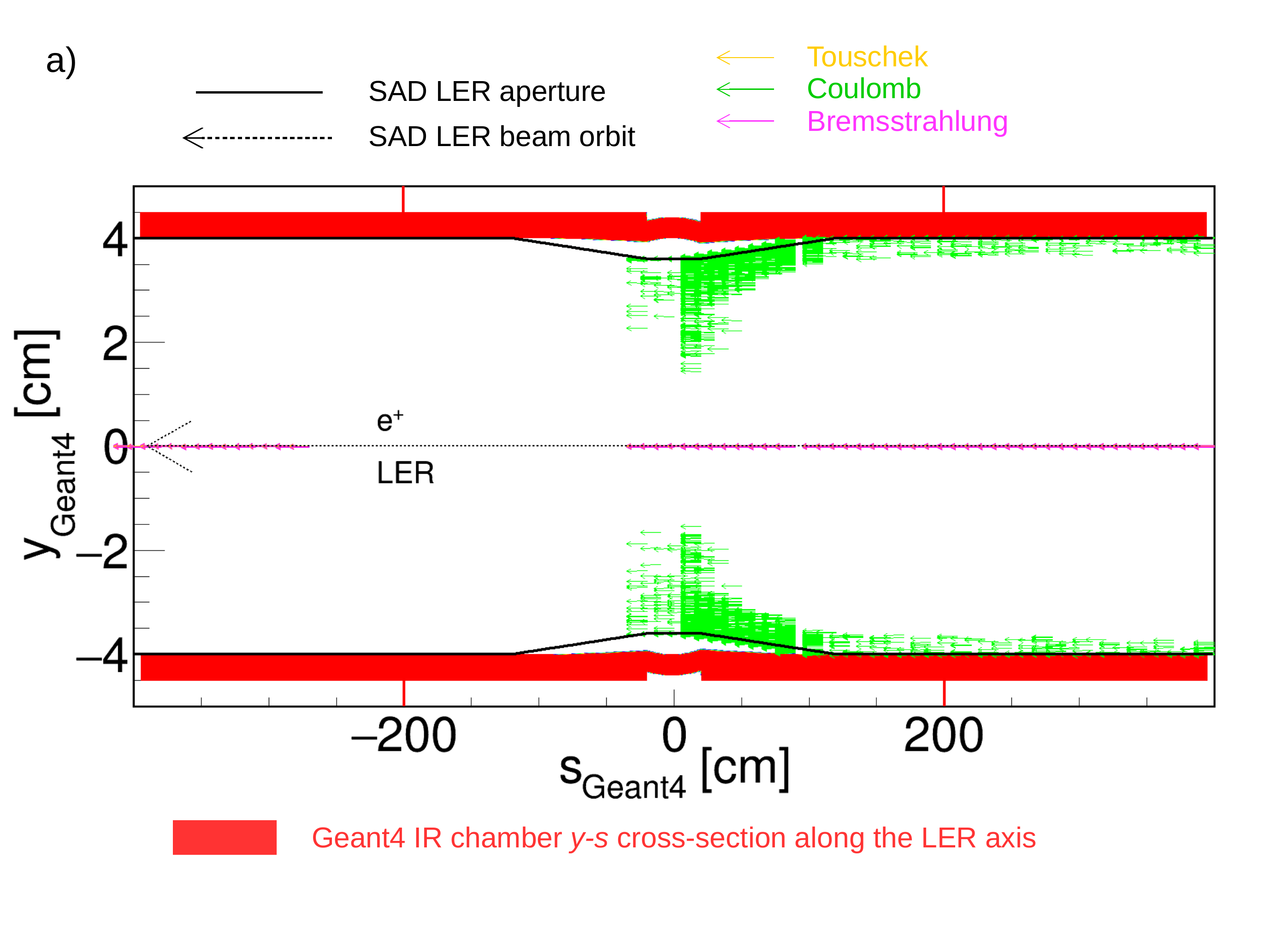}
\includegraphics[width=1.5\columnwidth]{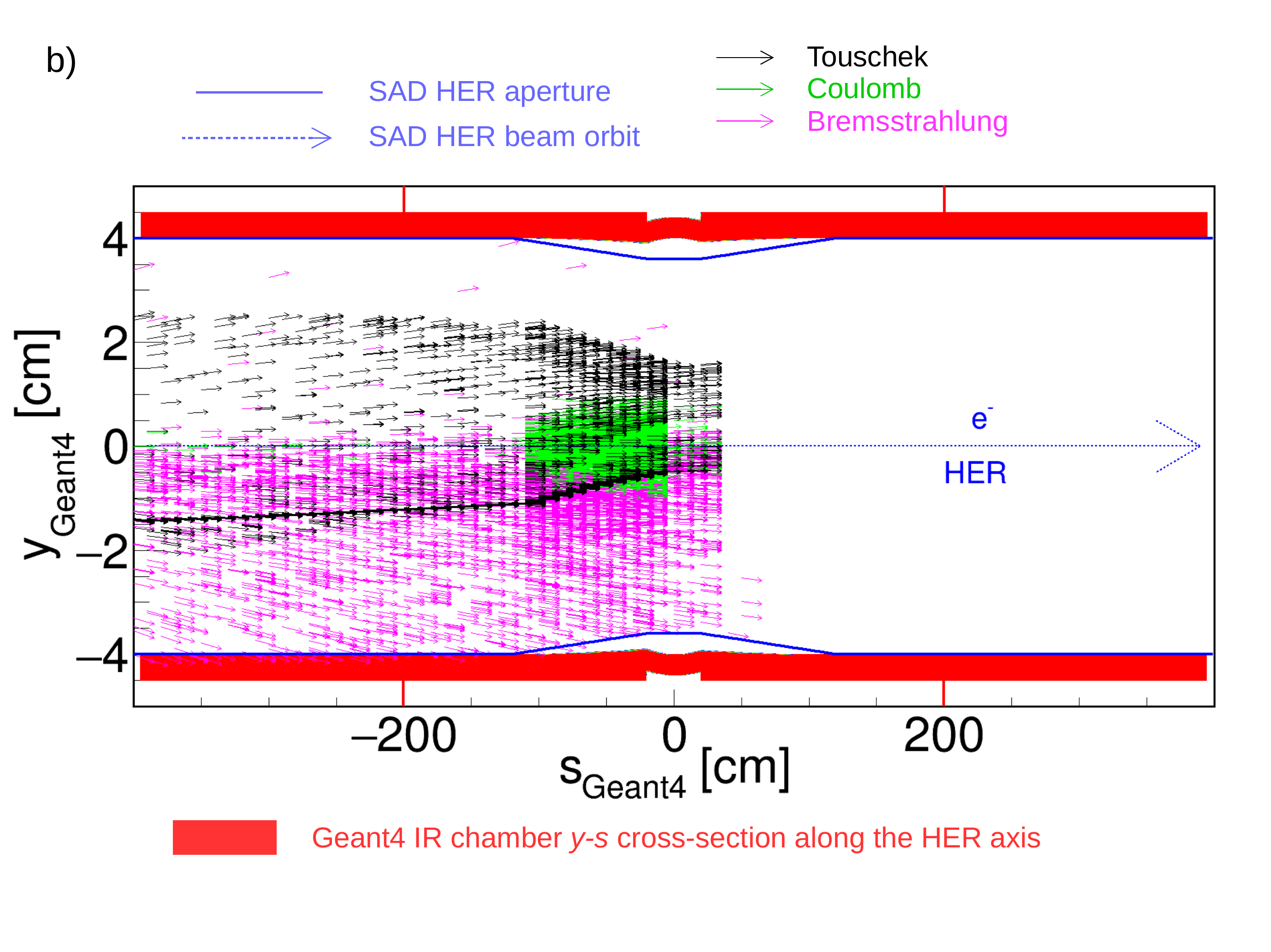}
\caption{(color online) {\GEANT} IP chamber vertical cross section versus (a) the LER axis and (b) the HER axis, $s_{Geant4}$, with the SAD aperture and beam orbit overlaid. The arrows indicate the electron beam direction and particles lost direction due to Touschek, bremsstrahlung, and Coulomb.}
\label{fig:v7ChberYS} 
\end{figure*}

The {\GEANT} Aluminum 5083 alloy IP chamber consists of: (1) a 40~cm length central beampipe, with inner radius of 4~cm and outer radius of 4.4~cm, centered around a virtual Interaction Point (which is offset by 7.5~cm in the +$x$-direction compared to Phase 2 and 3 IP), (2) between 20~cm and 2~m in $|z|$, an octagon-shape containing two pipes of 4~cm inner radius, which cross each other with an angle of 4.8$^o$ and 5.7$^o$ between 1.2~m and 4~m and 20~cm and 1.2~m, respectively. The octagon shape has a constant height of 90~cm, as can be seen in Figures~\ref{fig:v7ChberYS} and~\ref{fig:v7ChberYS}, and an increasing width as function of the crossing angles. In the $x$- and $y$-directions, i.e. perpendicular to each pipe's axis, the wall width is 5~mm and constant (as can be seen in Figures~\ref{fig:v7ChberYS} and ~\ref{fig:v7ChberXS}).  (3) Between 2~m and 4~m in $|z|$, the pipes have a 4~cm inner and a 4.5~cm outer radius. The LER has in addition a layer of 200~nm TiN coated on its inner walls. There is also a water cooling system, visible in Figures~\ref{fig:GeantwOnlyV7Chber},~\ref{fig:v7ChberXS}, and~\ref{fig:v7ChberXS}, which was simplified in {\GEANT} as four external Aluminum bars with water contained in a tube of 6~mm radius. The IP chamber is supported by six pillars and four reinforcement bars, which show up as brighter-colored rectangles in the material scan in Figure~\ref{fig:GeantwOnlyV7Chber}. 

\subsubsection{Physics lists}

We simulate particle propagation and interaction with matter by testing three different {\GEANT} physics lists: (1) \textit{QGSP\_BERT\_HP} hadronic model with the data-driven 
high precision neutron package (HP), for neutrons below 20 MeV down to thermal energies; (2) \textit{FTFP\_BERT\_HP} hadronic model with a different string model compared to
\textit{QGSP\_BERT\_HP} hadronic model but the same HP neutron package, and (3) \textit{Shielding} model with uses \textit{FTFP\_BERT} hadronic model for high energetic particles and for lower
energetic particles the best possible GEANT4 options can offer. All three physics lists are used as-is except when we simulate the synchroton radiation (see Section~\ref{simulation_generators_sr}) and give the same results within $1\%$. Consequently, we use the default Belle II physics list \textit{FTFP\_BERT\_HP}.

\subsubsection{Primary particles}

\begin{figure}[ht!]
\centering
\includegraphics[width=1.0\columnwidth]{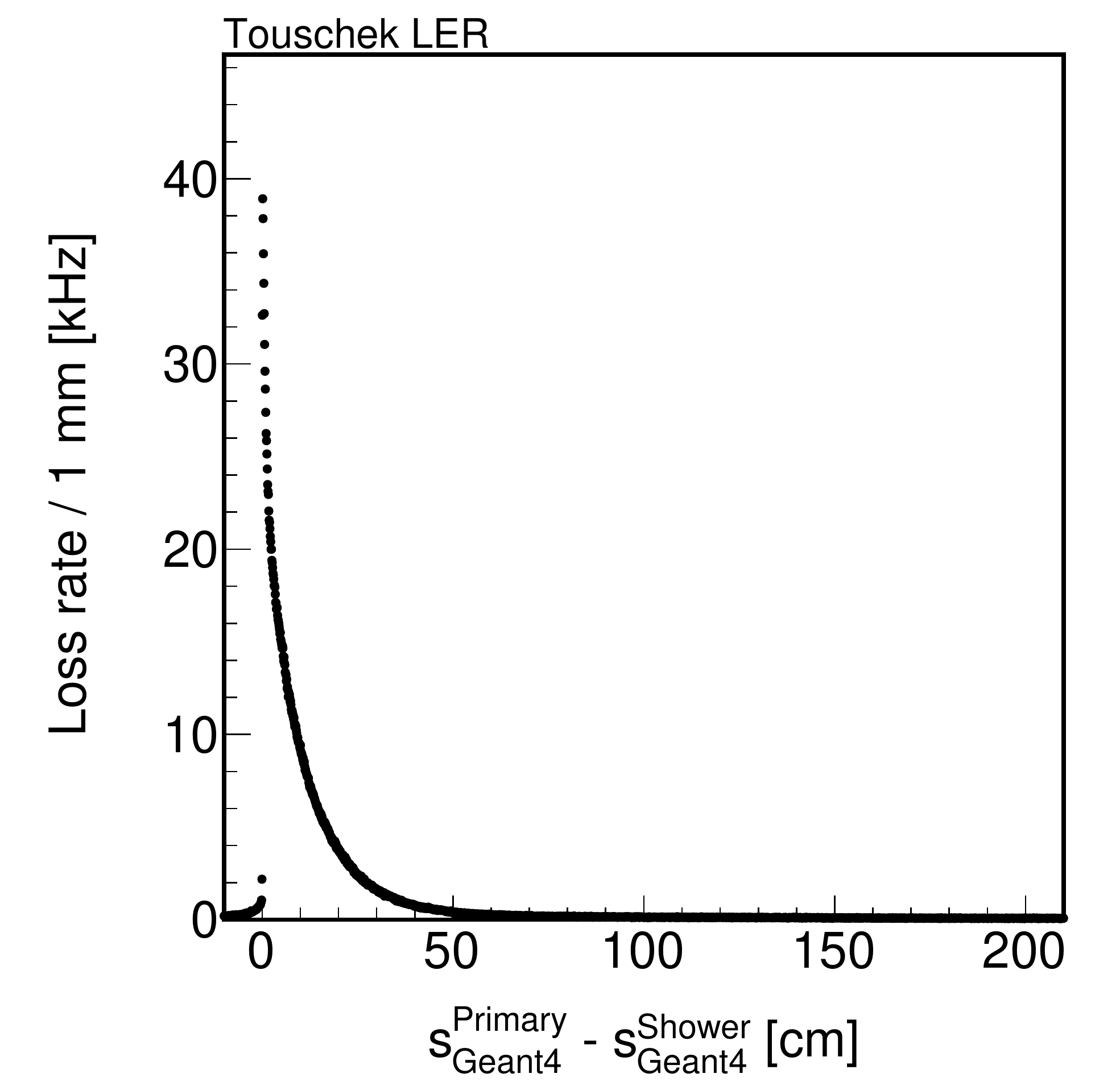}
\caption{Position difference along the LER axis between the Touschek primary particle and the first shower production location.}
\label{fig:LER_T_PosDif} 
\end{figure}

All {\GEANT} primary particles are SAD particles that hit the beampipes between $\pm$~4m around the Phase 1 IP. However, the SAD loss position does not necessarily correspond to the vertex production of the shower in the beampipes and the IP chamber as shown in Figure~\ref{fig:LER_T_PosDif}. This difference can be as large as 50~cm, particularly between $\pm$1.2~m. 

The primary particle energy distribution depends on scattering type: for Touschek it is spread around the electron/positron beam energy, for Coulomb it is essentially a delta function at the electron/positron beam energy, and for bremsstrahlung it peaks at the electron/positron beam energy but with a significant number of particles with much lower energies.

\subsubsection{Particle Shower}
\begin{figure}[ht!]
\centering
\includegraphics[width=1.0\columnwidth]{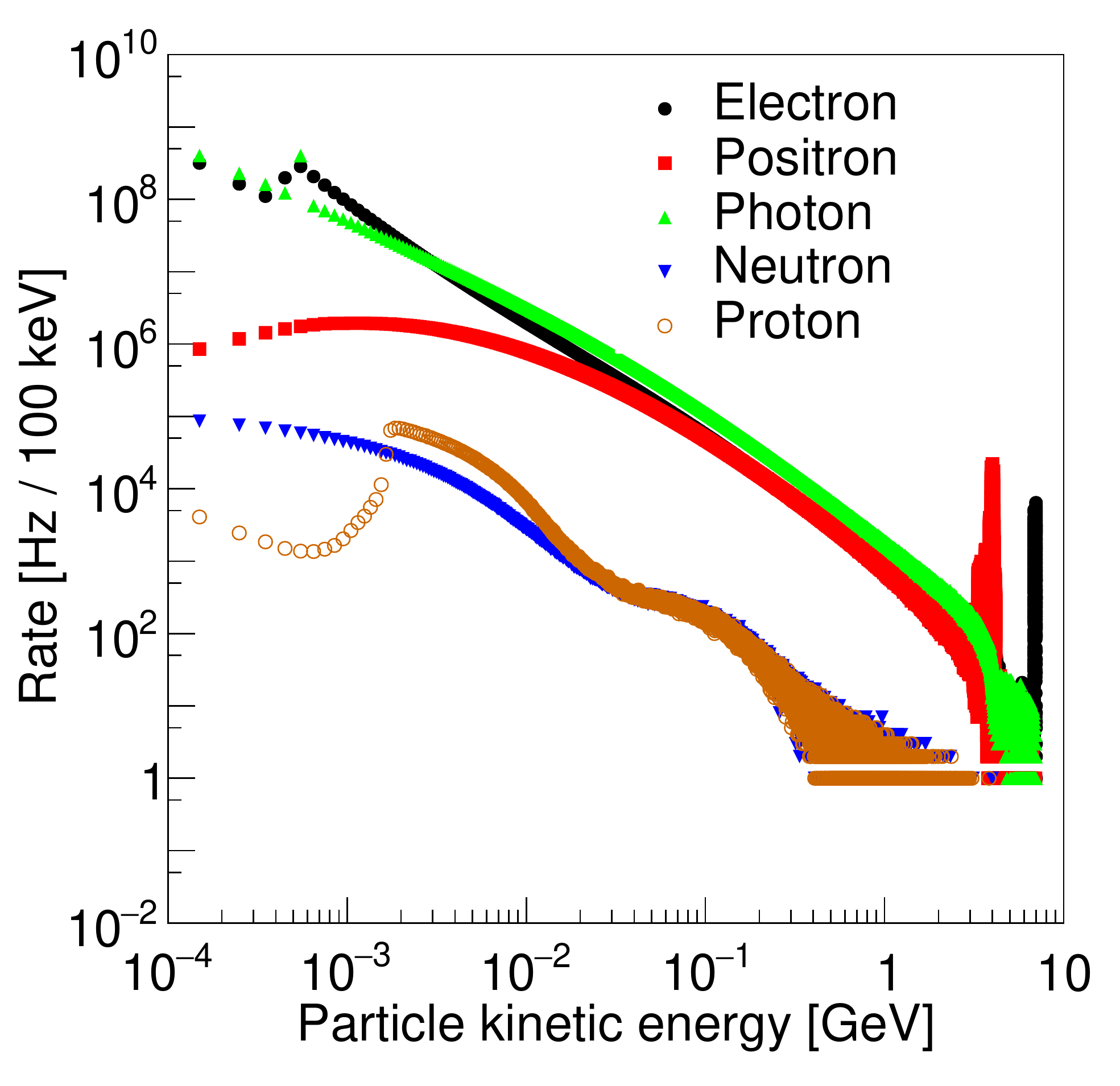}
\caption{Kinetic energy distrubtion of the particles created in IP chamber showers between $\pm$ 4~m around the IP.}
\label{fig:Shower1} 
\end{figure}

Of the particles in showers generated by primary particles in the IR, $99.96\%$ are electromagnetic in nature.  The remaining 0.04$\%$ are mostly protons and neutrons with 25$\%$ more protons than neutrons. This is illustrated in Figure~\ref{fig:Shower1}.

Initially, the primary particles interacting in the IP chamber walls produce similar numbers of protons and neutrons; these are produced directly and indirectly by bremsstrahlung photons via Giant Dipole Resonance (GDR) processes.  The GDR protons and neutrons have the same energy and angular distributions.  Most of the protons (${\sim}90\%$ of the total number) are produced by electromagnetic particles interacting in the concrete shields. The GDR protons also lose all their energy in the concrete shields while the GDR neutrons re-scatter until they thermalize. 

Roughly 40$\%$ of the hit rates in the Crystals, CLAWS, and QCSS are due to IP chamber showers. For the doses, the contribution from showers
created in the IP chamber vary between 15$\%$ to 100$\%$ depending the sensors' proximity to the cement shield walls, location in $z$, and background type. The contribution
from backscattering radiation is significant in all sensors which are at least 20~cm away from the IP chamber. The energy of backscattered particles is in
general lower than the energy of particles coming directly from IP chamber showers.

% Igal Jaegle
\subsection{Geant4 digitization}
% File: simulation_digitization.tex
% Lead author: Igal Jaegle

\label{simulation_digitization}

We simulate a detailed detector response, including detection, amplification, and digitization, for the TPC, CLAWS, and QCSS detectors. For all other sensors (BGO, PIN, Diamonds, Crystals, and $^3$He tubes) this is not necessary and we calculate the hit rate or dose directly from the {\GEANT} deposited energy using the calibrations presented in Section~\ref{sec:beast}. Here we describe in more detail how this is done for each detector. 

\subsubsection{Electromagnetic calorimeters}
For the Crystals system we simulate two observables: a hit rate from a scaler, and a dose rate from a digitizer. Each channel and device has an energy threshold that changes daily with crystal degradation and device settings. We determine these thresholds experimentally and apply them to the simulated observables, so that the hit rate corresponds to the number of {\GEANT} hits in a second above the energy of the daily scaler threshold, and the dose rate corresponds to the sum of the deposited energy of all {\GEANT} hits above the energy of the daily digitizer threshold, converted to mrad/s. 

For the BGO system we apply no energy threshold, and the dose is simply the sum of the energy of all {\GEANT} hits in a second, converted into rad/s. 
 
\subsubsection{Ionization sensors}

For the ionizing radiation sensors, the PIN diodes and Diamond sensors, the dose is directly calculated from the {\GEANT} energy deposited in each second, converted into rad/s. 

\subsubsection{Gaseous detectors}

For the $^{3}$He tubes, we apply an inefficiency correction to the simulated count rate.  We determine this inefficiency correction by comparing the simulated and measured count rates from a known neutron source (described in section~\ref{sec:heTCalibration}). 

For the TPC system, we give a short summary here; a detailed description can be found in ~\cite{JaegleTPCSimulator}. The energy deposited in the {\GEANT} simulation is converted into a number of charged particles, $N_{e^-}$, if the energy deposited is above the work function $W$, subject to the following:
\begin{equation}
N_{e^-}^{mean} =  E_{dep}^{G4} / W
\end{equation}
where $N_{e^-}^{mean}$ is the mean number of charged particles.  We deduce the resolution, $\sigma$, from the Fano factor, $F$, and the mean number of charged particles according to:
\begin{equation}
\sigma = \sqrt{F \times N_{e^-}^{mean} }.
\end{equation}
Finally, we randomly select the number of charged particles, $N_{e^-}$, according to a Gaussian probability distribution function with mean $N_{e^-}^{mean}$ and width $\sigma$. 

Table~\ref{tab:gas} gives the work function and Fano factor used for 1~atm He:CO$_2$ with a $70:30$ mixture.

\begin{table}
  \caption{Work function and Fano factor for the TPC system.}
  \centering
  \label{tab:gas}
  \begin{tabular}{lll}
    \toprule
    Sensor&Work function [eV]&Fano factor\\
    \midrule
    $He:CO_2 (70:30)$ & 35.075 & 0.19\\
    \bottomrule
  \end{tabular}
\end{table}

In addition to these steps, we simulate the drift of the charge from ionization to the amplification stage (GEMs) and readout plane with a fast Monte Carlo simulation which uses electron diffusion and drift velocity values calculated by the software package MAGBOLTZ ~\cite{magboltz}.  
We then simulate the digitization of the avalanche charge after GEM amplification that takes place in the pixel chip, which for each hit produces a pixel column and a pixel row number, the charge above threshold, and the relative hit time with respect to the first hit in the event. For each TPC, we use fixed gas parameters determined by MAGBOLTZ from the experimental field cage and GEM HV settings. Since the pixel thresholds and GEM gains were not precisely known at the time the simulation was produced, we simulate a threshold of 2600 electrons and a combined double-GEM gain of 1500.

\subsubsection{Plastic scintillators}

For the plastic scintillators, QCSS and CLAWS, we use the light saturation, or Birk's law, to calculate the (attenuated) energy deposited in {\GEANT} simulations.  The attenuated energy deposited is converted into a number of Minimum Ionization Particles (MIPs) using conversions derived from simulation. The number of MIPs is then converted into photoelectrons (p.e.) per second, using the conversion factors in Table~\ref{tab:sci}. 

%Split column title to better fit document column width.
\begin{table}
  \caption{Conversion factors from {\GEANT} deposited energy to photoelectrons (p.e.) for CLAWS and QCSS.}
  \centering
  \label{tab:sci}
  \begin{tabular}{ccccc}
    \toprule
    Sensor & keV/MIP & MIP/p.e.\ & Threshold  & Time window \\ 
           &         &        &            & [ns]  \\
    \midrule
    CLAWS & 467.11 & 12 to 16 & 1 [MIP] & 8\\
    QCSS &  1628.20 & 15 & 0.5 [p.e.]& -\\
    \bottomrule
  \end{tabular}
\end{table}

% Igal Jaegle
\subsection{Scaling by accelerator conditions}
% File: simulation_scaling.tex
% Lead author: Igal Jaegle

\label{simulation_scaling}

To compare with experimental results, we scale the simulation to the accelerator conditions according to theoretical formulae described here. 

\subsubsection{Model}

We expect the simulated observable for each background type ($B$ for bremsstrahlung, $C$ for Coulomb, and $T$ for Touschek) to follow these functional forms:
\begin{eqnarray}
\label{eqn:SimObs1}
\mathcal{O}_{B} & = &I \sum_{i} B_{i} P_{i} f_B(Z_i),  \\
\label{eqn:SimObs2}
\mathcal{O}_{C} & = &I \sum_{i} C_{i} P_{i} f_C(Z_i), \\
\label{eqn:SimObs3}
\mathcal{O}_{T} & = &T \frac{N_bI_b^2}{\sigma_y},
\label{eqn:SimObs}
\end{eqnarray}
where $I$, $P_i$, and $Z_i$ are the beam current, the average pressure of the ring section section $i$, and the atomic number in the ring section $i$, respectively, and the sum is over the 12 beam sections. The bremsstrahlung and Coulomb coefficients $B_i$ and $C_i$ map losses in section $i$ to the detected observables $\mathcal{O}_{B}$ and $\mathcal{O}_{C}$.
 
The atomic number scalings used by SAD take the form:
\begin{eqnarray}
\label{eqn:f_B}
f_B (Z_i) & = & Z_i^2\left[ln\left(\frac{a_B}{Z_i^{1/3}}\right) + b_B\right],\\
\label{eqn:f_C}
f_C (Z_i) & = & Z_i^2\left[\frac{1}{a_C + \left(b_CZ_i^{1/3}\right)^2}\right]^2,
\end{eqnarray}
where the parameters $a_B$, $b_B$, $a_C$, and $b_C$ are not constant but depend on the initial particle energy for bremsstrahlung and the scattering particle energy and angle for Coulomb\cite{chao:2013}. Therefore, these parameters are different for the LER compared to the HER and for the entire ring compared to the IR. 

In order to accurately scale from SAD to actual beam conditions we need to determine the appropriate atomic number scaling parameters $a_B$, $b_B$, $a_C$, and $b_C$ as well as the section-by-section beam-gas coefficients $B_i$ and $C_i$ and the ring Touschek coefficient $T$.

\subsubsection{Atomic number scaling parameters}
\begin{figure}[ht!]
\centering
\includegraphics[width=1.0\columnwidth]{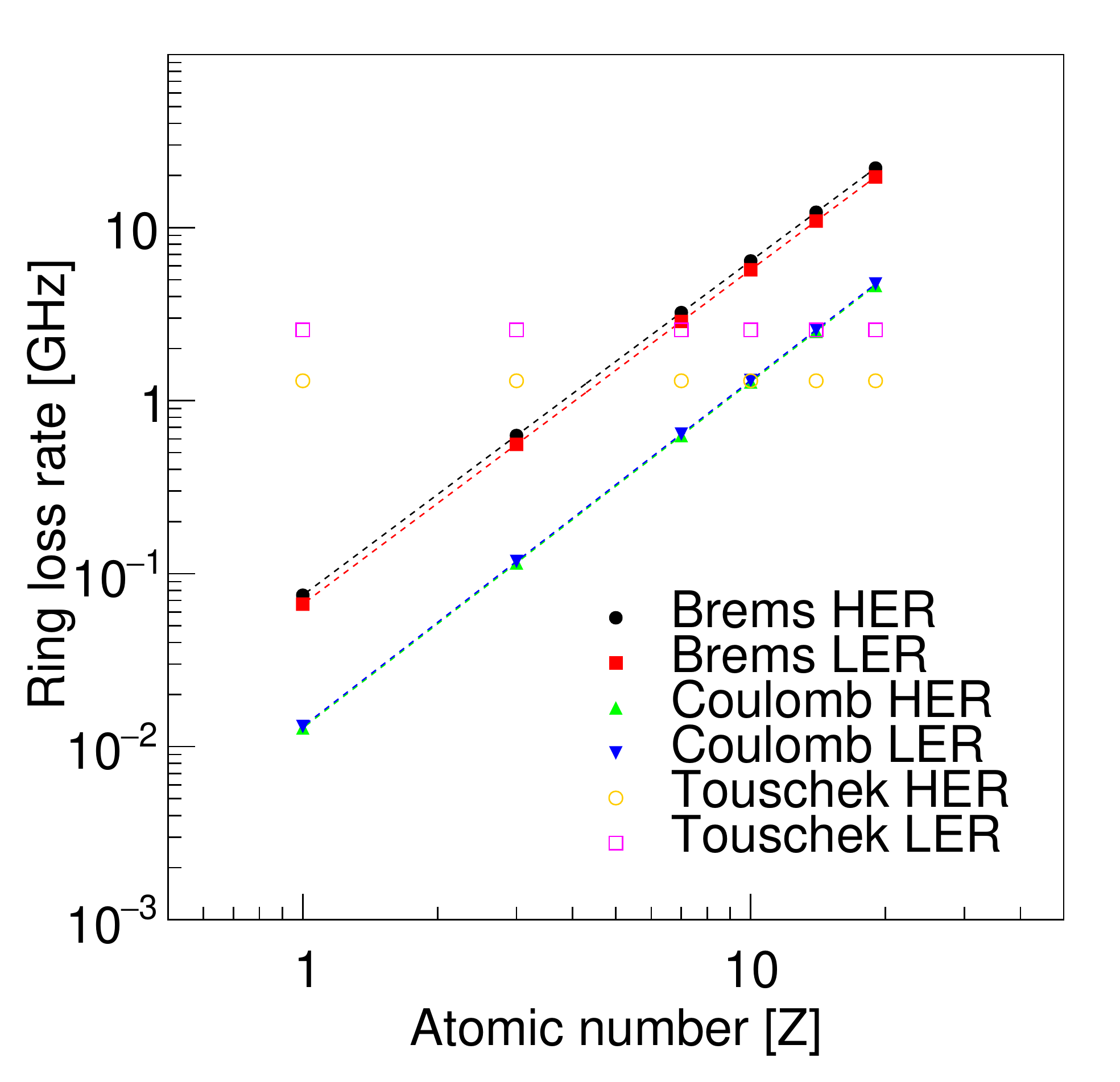}
\caption{Ring loss rates versus atomic number for all background types in each ring.}
\label{fig:RingLossRate}
\end{figure}

\begin{figure}[ht!]
\centering
\includegraphics[width=1.0\columnwidth]{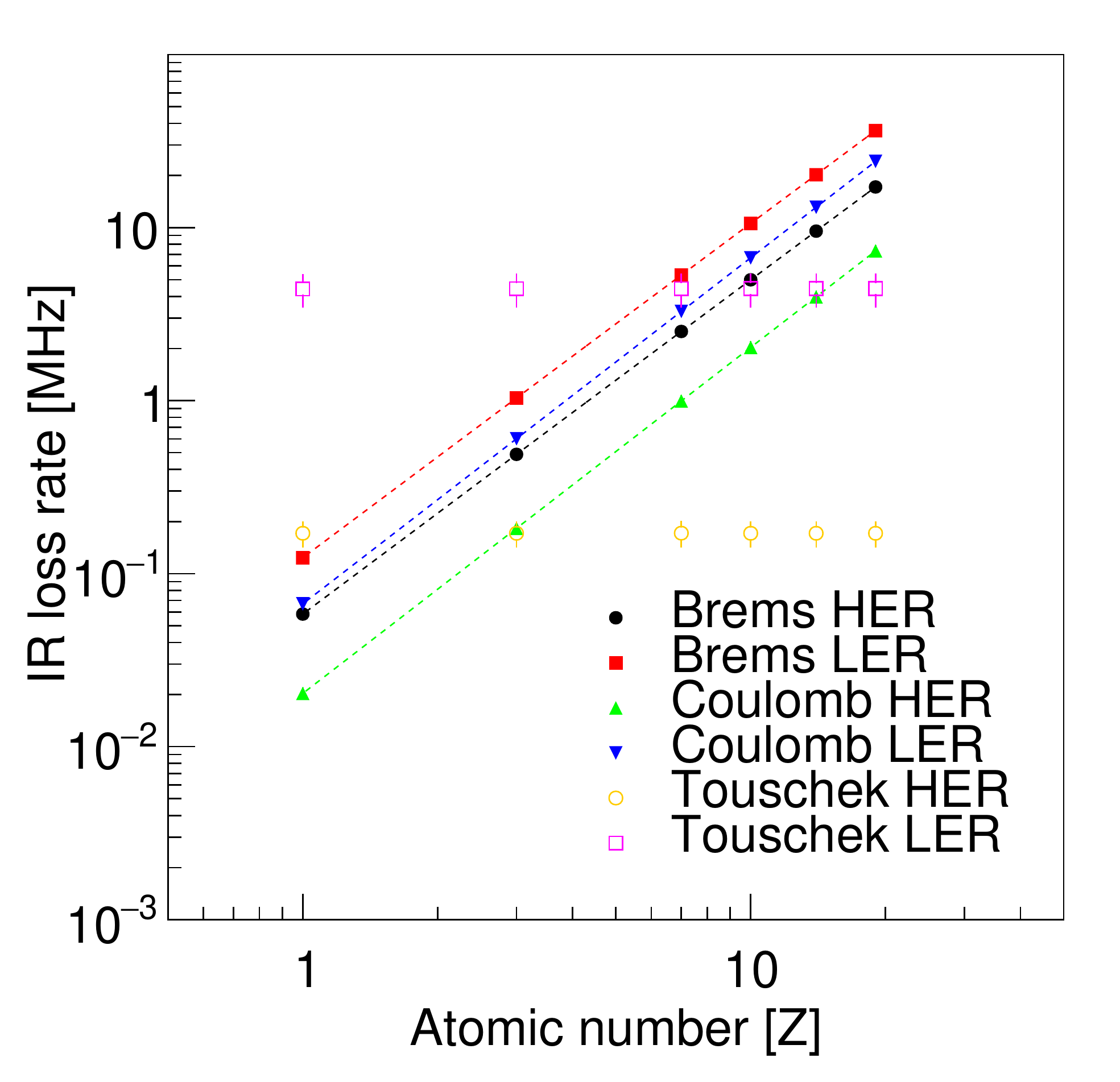}
\caption{IR loss rates versus atomic number for all background types in each ring. The dashed lines are fits.}
\label{fig:RingLossRateIR}
\end{figure}

To determine the atomic number scaling parameters we performed SAD simulations at six different atomic numbers $Z$. We fit the resulting loss rates versus $Z$ distributions with Equations~\ref{eqn:f_B} and \ref{eqn:f_C} with the scaling parameters free. The results of these fits are shown in Figures~\ref{fig:RingLossRate} and \ref{fig:RingLossRateIR} and summarized in Table~\ref{tab:SADpara}. 

These parameters are in principle sensor-dependent, but we find that the ratios $f_B(Z)/f_B(Z')$ and $f_C(Z)/f_C(Z')$ are the same for the entire LER and HER rings and also for the IR and presumably also for each sensor. Therefore we scale the simulation by $f_B(Z_{e})/f_B(Z=7)$ and $f_C(Z_{e})/f_C(Z = 7)$, where $Z_{e}$ is the experimental effective atomic number. For reference, in Tables~\ref{tab:SADparaZb} we show $f_B$ and $f_C$ values at $Z=2.7$, representing typical conditions during Phase 1.  

\begin{table}
  \caption{Atomic number scaling parameters $a_B$, $b_B$, $a_C$, and $b_C$ for the entire rings and for the IR, determined from SAD.}
  \centering
  \label{tab:SADpara}
  \begin{tabular}{ccccc}
    \toprule
     Ring & $a_B$ & $b_B$ & $a_C$ & $b_C$\\ 
    \midrule
     HER (entire) & 2.23 & 4.51 & 3.41 & 0.00826\\ 
     LER (entire) & 2.28 & 4.49 & 2.47 & -0.00766\\
     HER (IR) & 11.8 & 2.79 & 2.51 & -0.000404\\       
     LER (IR) & 1.72 & 4.59 & 3.38 & -0.000433\\
    \bottomrule
  \end{tabular}
\end{table}

\begin{table}
  \caption{Bremsstrahlung and Coulomb atomic number scaling function $f_B$ and $f_C$ values for $Z$ = 2.7 for the entire rings and for the IR.}
  \centering
  \label{tab:SADparaZb}
  \begin{tabular}{ccccc}
    \toprule
      & \multicolumn{2}{c}{$f_B(Z=2.7)$} & \multicolumn{2}{c}{$f_C(Z=2.7)$} \\ 
      & Ring & IR & Ring & IR \\
    \midrule
     HER & 36.4  & 37.0 & 0.628 & 1.15 \\
     LER & 36.3  & 36.0 & 1.19  & 0.637 \\
    \bottomrule
  \end{tabular}
\end{table}

\subsubsection{Beam-gas and Touschek scaling coefficients}
\begin{figure}[ht!]
\centering
\includegraphics[width=1.0\columnwidth]{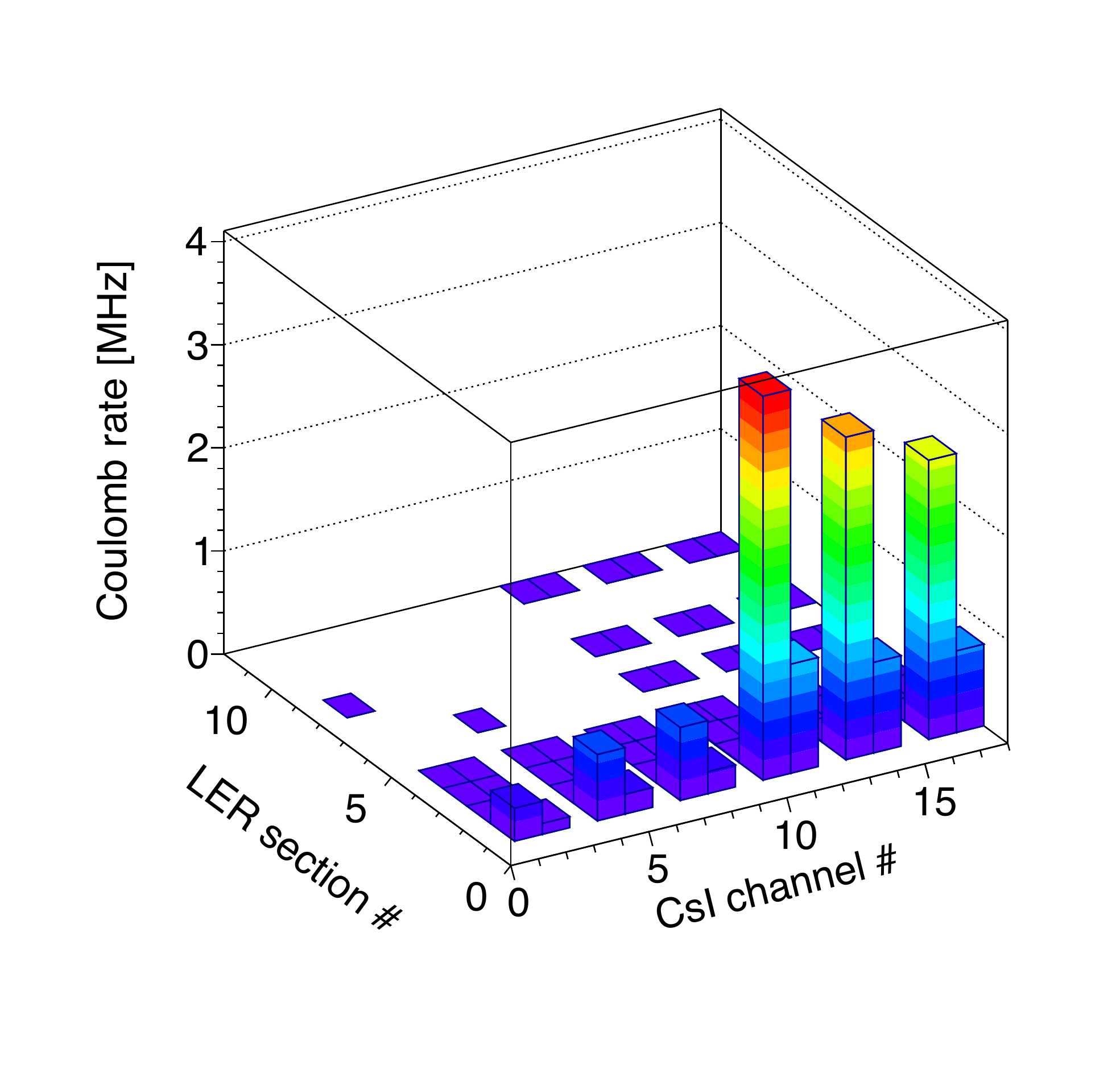}
\caption{Crystals hit rate as function of the Coulomb scattering LER ring loss position. The rings have been divided in 12 sections, each of 250~m. The dashed lines are fits.}
\label{fig:MCHit_Rate_RS2}
\end{figure}

We determine the section-by-section bremsstrahlung and Coulomb scaling coefficients $B_i$ and $C_i$ and the ring Touschek scaling coefficients $T$ on a channel-by-channel basis by measuring the amount of the observable $\mathcal{O}$ that is generated by losses in each section $i$. To illustrate, Figure~\ref{fig:MCHit_Rate_RS2} shows the simulated Crystals hit rate for LER Coulomb scattering separated by scattering section. For a single channel, the relative height of the 12 bins labeled by section number correspond to the coefficients $C_i$.

\subsubsection{The final simulated observable}
After determining the scaling coefficients and atomic number parameters we generate a final scaled observable for each channel:
\begin{eqnarray}
\mathcal{O} & = & \sum_i\left(\left[B_if_B(Z_i) + C_if_C(Z_i)\right] \cdot IP_i + T \frac{N_bI_b^2}{\sigma_y}\right),
\label{eqn:SumOfSimObs}
\end{eqnarray}
where the sum $i$ is over the ring sections. We add contributions from both rings to obtain the final simulated observable, which constitues a prediction of the experimental observables for any channel at any time during Phase 1.

% Igal Jaegle
\subsection{Simulation validation}
% File: simulation_analysis.tex
% Lead author: Igal Jaegle

\label{simulation_validation}

We validate the modeling of the detector response by comparing the simulation to laboratory measurements with calibration sources where available. 

\begin{figure}[ht!]
\centering
\includegraphics[width=\columnwidth]{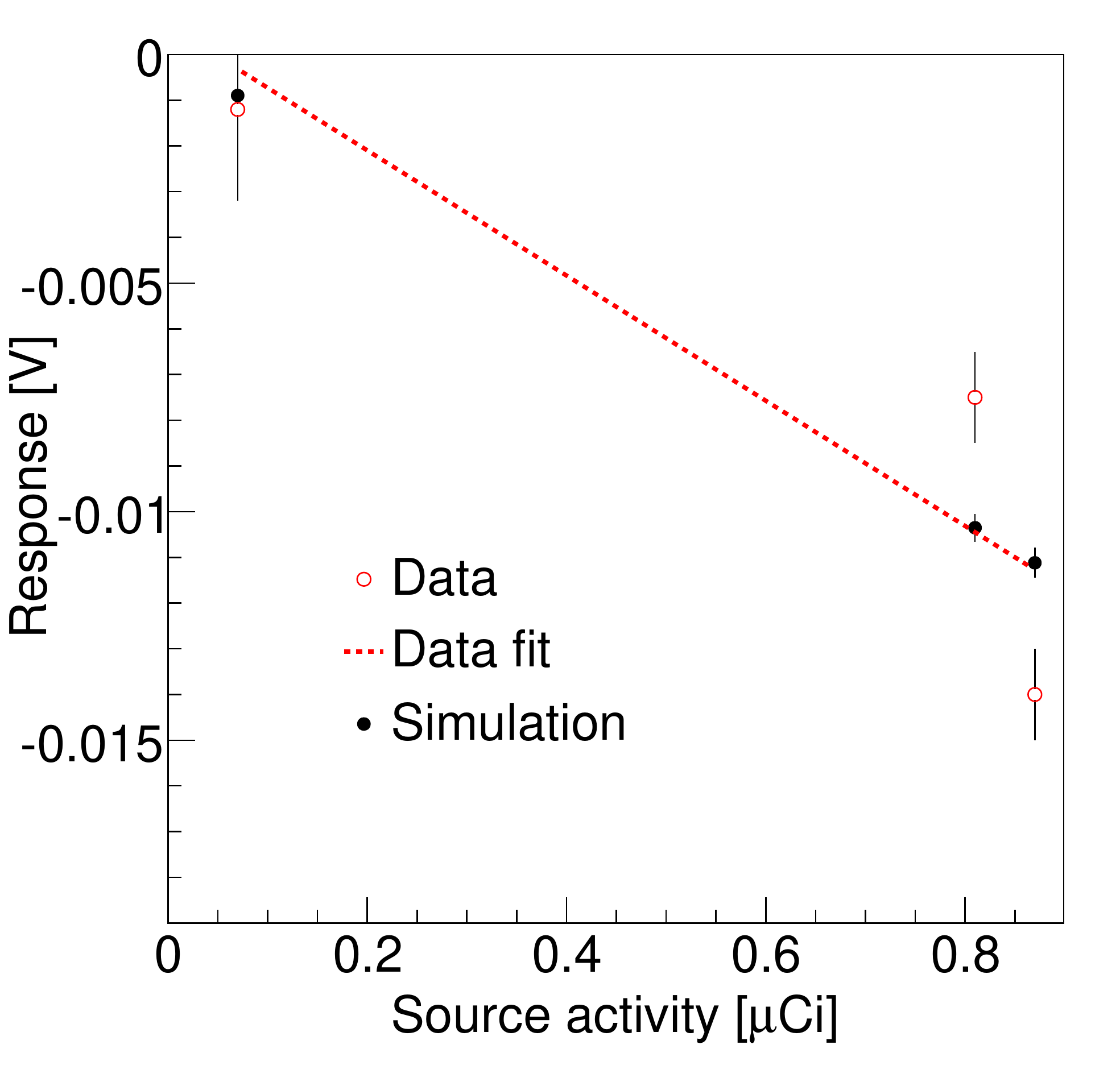}
\caption{PIN calibration: output versus source activity for a PIN diode with both experimental (open circles) and simulated (closed circles) data.}
\label{fig:PINvalidation} %Removed underscore from label name to attempt to resolve non-linking of figure
\end{figure}

For the PIN diodes, we simulate measurements done with three $^{60}$Co sources, each having a different activity. Figure \ref{fig:PINvalidation} shows that the simulation is in good agreement with the data.

\begin{figure}[ht!]
\centering
\includegraphics[width=\columnwidth]{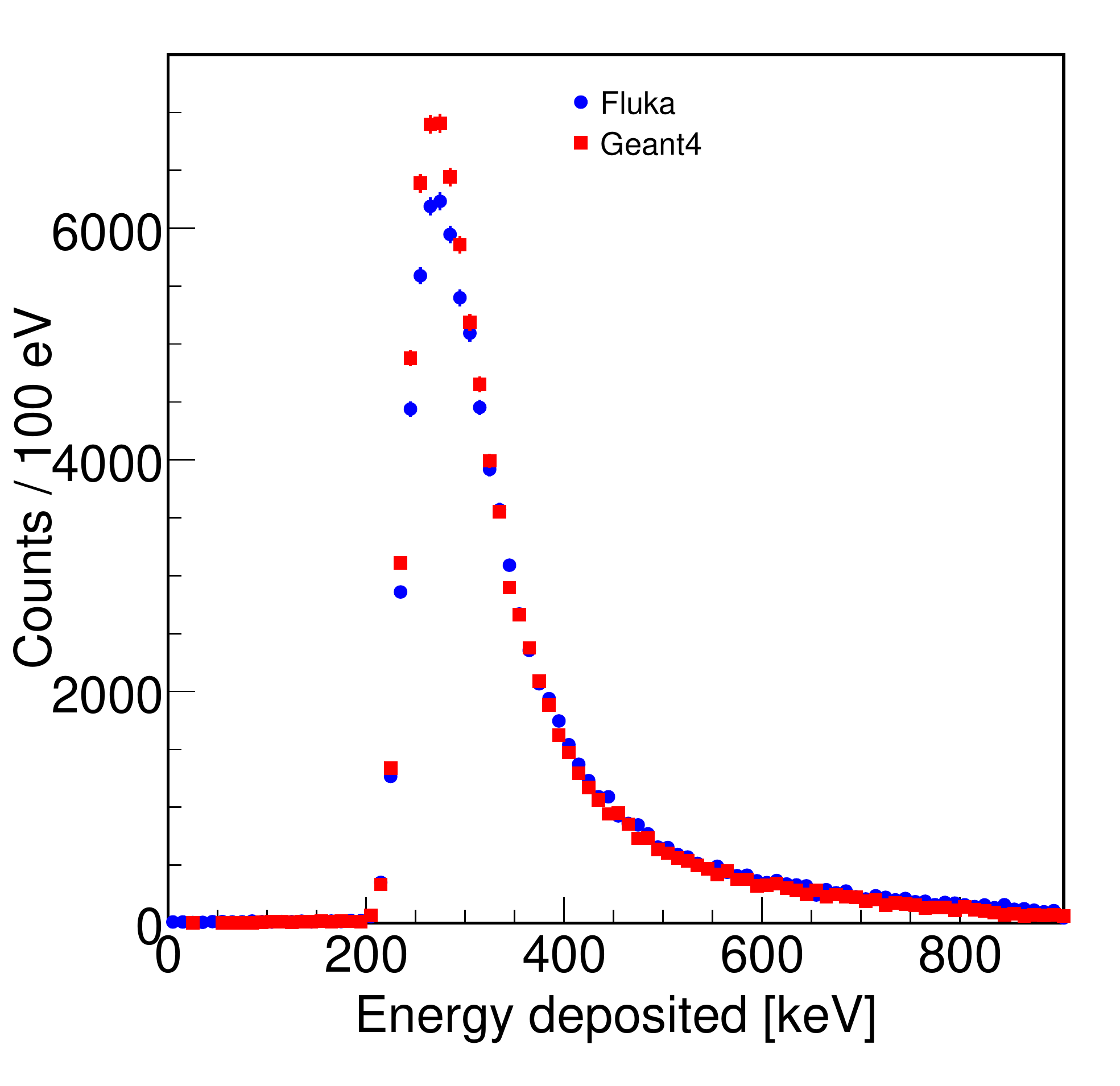}
\caption{(color online) Diamond calibration: energy deposited by a 1~MeV normally incident electron beam, as simulated by Fluka (blue points) and {\GEANT} (red squares).}
\label{fig:DIAvalidation} %removed underscore from label name to attempt to resolve non-linking of figure
\end{figure}

To validate the simulation of the diamond response, we compare the energy deposited by a 1~MeV normally incident electron beam simulated by Fluka and {\GEANT}. Figure~\ref{fig:DIAvalidation} shows that Fluka and {\GEANT} are in good agreement.

To validate the simulation of the TPC response, we compare the measured and simulated $dE/dx$ of the internal calibration alpha sources. We also compare the ionization energy versus track length for neutron recoil condidates in experimental data and simulation. In both these cases we find good agreement, as described in detail in Section~\ref{neutrons_analysis}. Further details about the TPC detector simulation and validation can be found in Ref.~\cite{JaegleTPCSimulator}.

% Igal Jaegle
\subsection{Simulation results}
% File: simulation_analysis.tex
% Lead author: Igal Jaegle

\label{simulation_results}

We show here results of the simulation of key BEAST II detector observable $\mathcal{O}$ (either hit rates or doses) for each of the six beam-induced background components (Touschek, bremsstrahlung and Coulomb for each ring). We present these results at a fixed set of accelerator parameters, which are summarized in Table~\ref{Table:MachineParamForSim}, and versus channel number.

\begin{figure}[ht!]
\centering
\includegraphics[width=1.0\columnwidth]{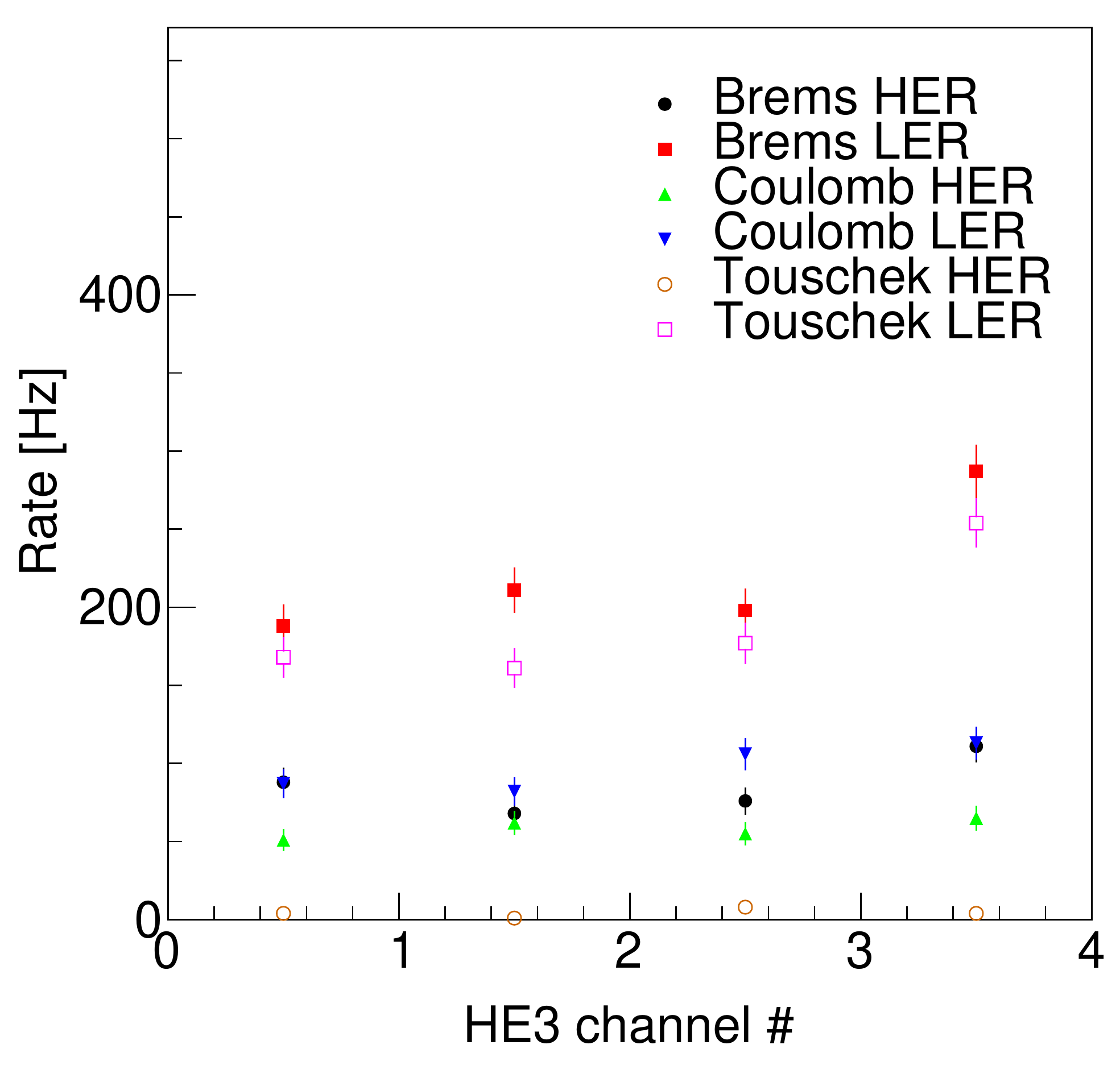}
\caption{$^3$He tube rates for each backgroud type and ring.}
\label{fig:HE3Hit}
\end{figure}

Figure~\ref{fig:HE3Hit} shows the simulated hit rates in the $^3$He tubes, which are dominated by LER bremsstrahlung and Touschek for all channels. 

\begin{figure}[ht!]
\centering
\includegraphics[width=1.0\columnwidth]{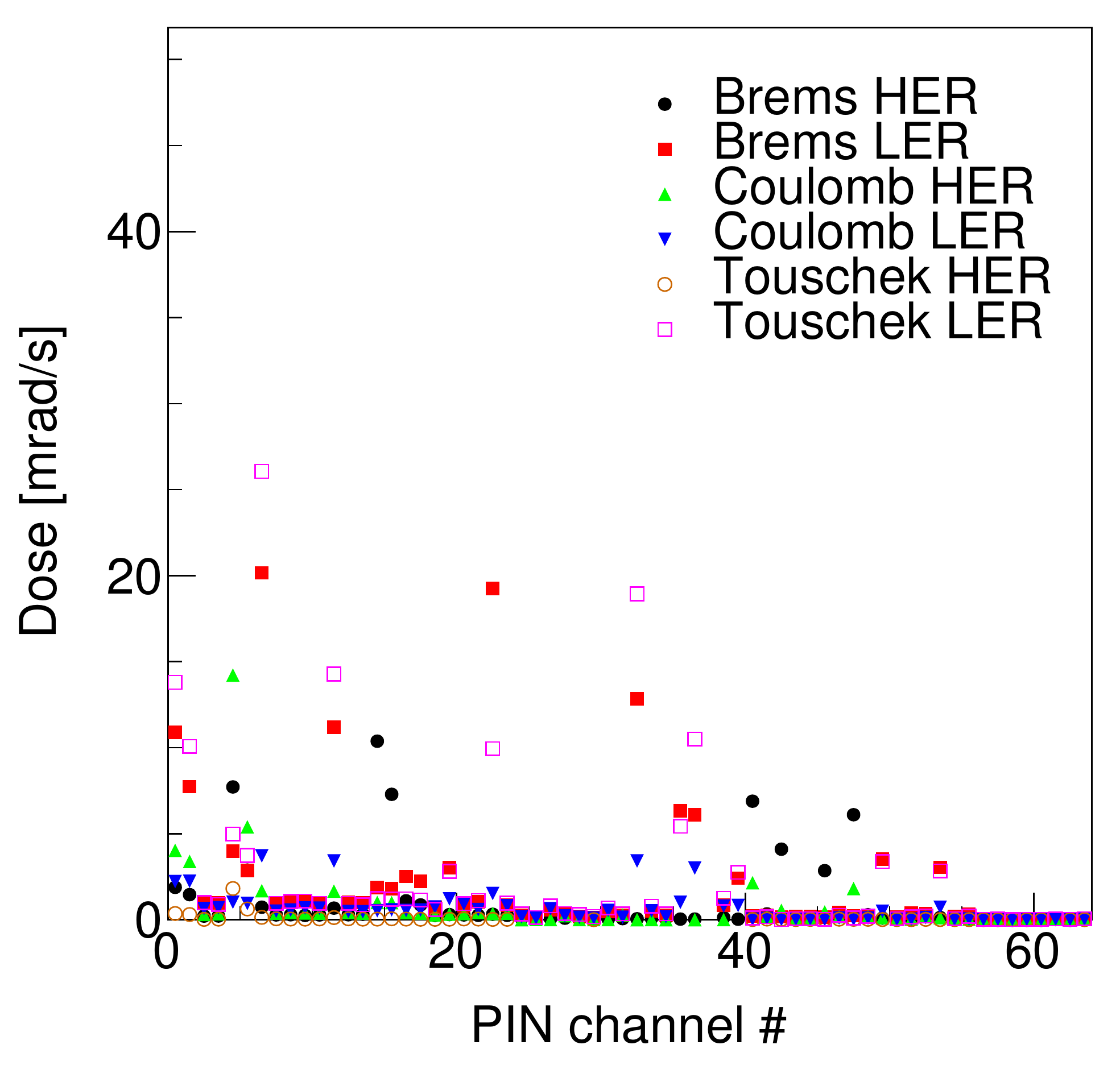}
\caption{PIN diode doses for each backgroud type and ring.}
\label{fig:PINDose}
\end{figure}

Figure~\ref{fig:PINDose} shows the simulated PIN doses, which show clear channel-by-channel sensitivity to both total dose and scattering type.

\begin{figure}[ht!]
\centering
\includegraphics[width=1.0\columnwidth]{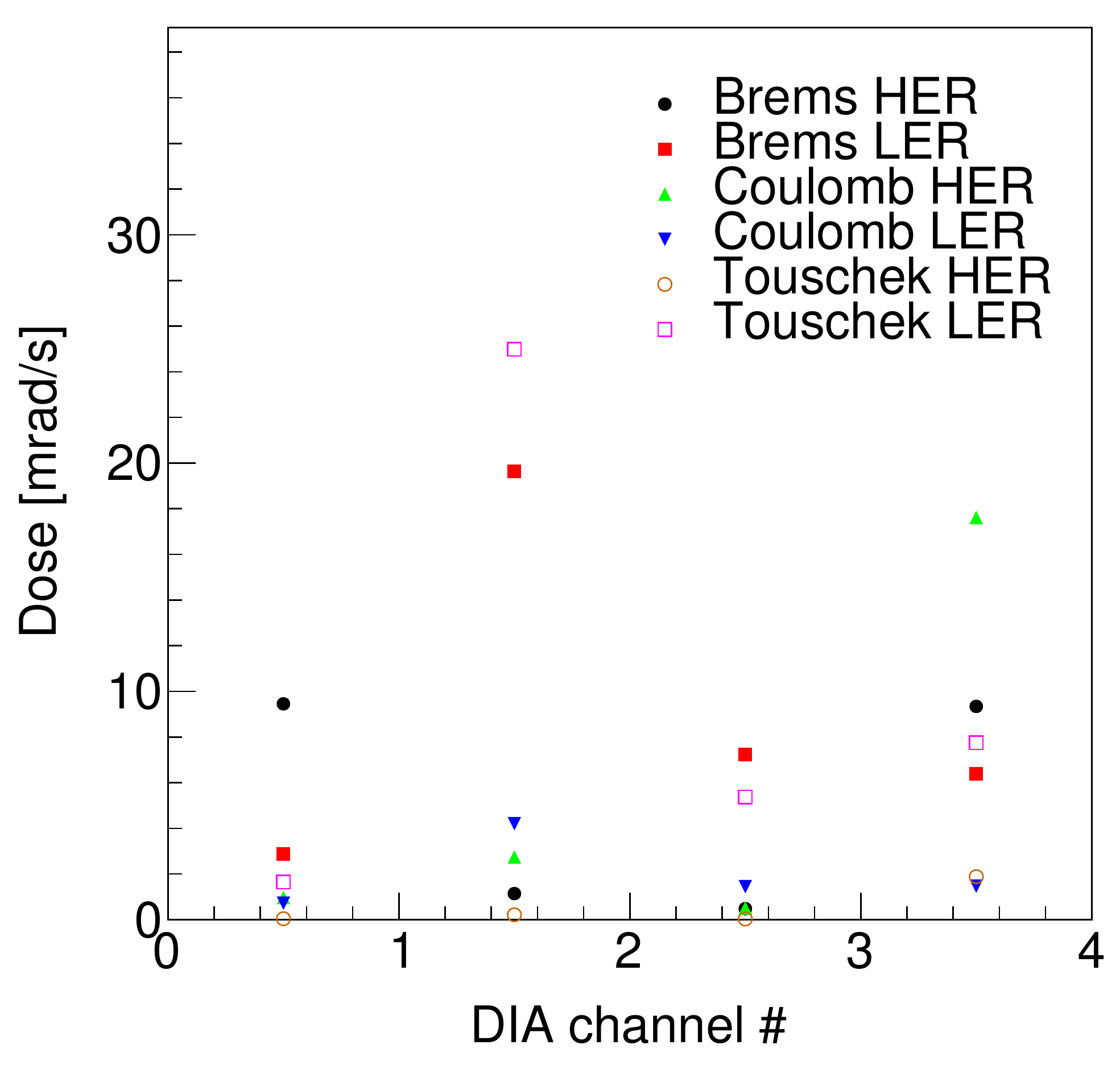}
\caption{Diamond doses for each backgroud type and ring.}
\label{fig:DIADose}
\end{figure}

Figure~\ref{fig:DIADose} shows the simulated Diamond doses, which show some channel-to-channel sensitivity to scattering type.  

% Igal Jaegle
\subsection{Simulation systematics}
% File: simulation_analysis.tex
% Lead author: Igal Jaegle

\label{simulation_systematics}

The main sources of the simulation systematic errors are:

\begin{itemize}
\item The event generator and more precisely its usage
\item The geometry description and more precisely sensor positions
%\item The Z dependency
\end{itemize}

The systematic uncertainty of the event generator is dominated by the rate statistical uncertainty and is estimated by using a tracking code where the slice size can be decreased up to 5~mm. We then compare the {\GEANT} BEAST sensor rates and doses between a simulation done with 5~mm slice size and another of 10~cm slice size where in both cases the loss position is recalcultated with a precision of 10~\si{\micro}m. Table~\ref{tab:sim_sys} summarizes the results.

The systematic uncertainty due to the position uncertainty is also summarized for each sensor in Table~\ref{tab:sim_sys}. This uncertainty is estimated by simulating the {\GEANT} sensor observable after moving by $\pm$ 1~cm in $x$, $y$, and $z$, within physical constraints.

\begin{table}
  \caption{Systematic uncertainties due to the rate statistical and position uncertainties.}
  \centering
  \label{tab:sim_sys}
  \begin{tabular}{ccc}
    \toprule
     & \multicolumn{2}{c}{Syst. uncertainty [$\%$]}\\ 
     & Rate & Position \\
    \midrule
    PIN & 40 & 100 \\
    Diamond & 40 & 100 \\
    Crystal & 15 & 5 \\
    BGO & 15 & 10 \\
    TPC & 40 & 10 \\
    $^3$He & 40 & 10 \\
    CLAWS & 15 & 5 \\
    QCSS &  15 & 5 \\
    \bottomrule
  \end{tabular}
\end{table}

In principle, the $Z$ dependency correction should lead to additional systematic error as we are using the average $Z$ dependency insteasd of a sensor dependency. However, because
this correction is done by using the ratio $f_{B/C}(Z)/f_{B/C}(Z')$, the systematic errors cancel.

 \clearpage

 % Each analysis (such as beam-gas or ) should have two files here, where appropriate:
 % <analysis>_analysis.tex and <analysis>_results.tex
 %
 % AUTHORS PLEASE READ: 	Please put the plots for this section in (Results)
 %

 \section{SuperKEKB conditions monitors}\label{sec:superkekb_conditions}
 In order to simulate and understand beam backgrounds at the interaction region, it is essential to understand the beam and accelerator conditions that contribute to the backgrounds. In this section we describe the monitoring systems that provide the key conditions measurements that will be used in subsequent analyses.

\subsection{Beam size} % Flanagan, Mulyani
X-ray monitors (XRM) have been installed in each SuperKEKB ring, primarily for vertical beam size measurement. Both rings have been commissioned in Phase 1, and several XRM calibration studies have been carried out, detailed here. 

\subsubsection{XRM Apparatus}

In SuperKEKB, we have chosen to use X-ray beam size monitors (XRM) in each ring, the LER and the HER. Each monitor uses the X-ray component of synchrotron radiation from a bending magnet in its ring, and will eventually have the capability for single shot (single bunch, single turn) vertical beam size measurements. Two types of optical elements are used in the XRM: single-slit (pinhole) and multi-slit optical elements (coded aperture or CA)\cite{Dicke}. The CA technique was developed by X-ray astronomers using a mask to modulate incoming light. An open aperture of 50\% gives high flux throughput for bunch-by-bunch measurements.  

\subsubsection{Beamline}
\begin{figure}[tb]
	\centering 
	\includegraphics[width=\columnwidth]{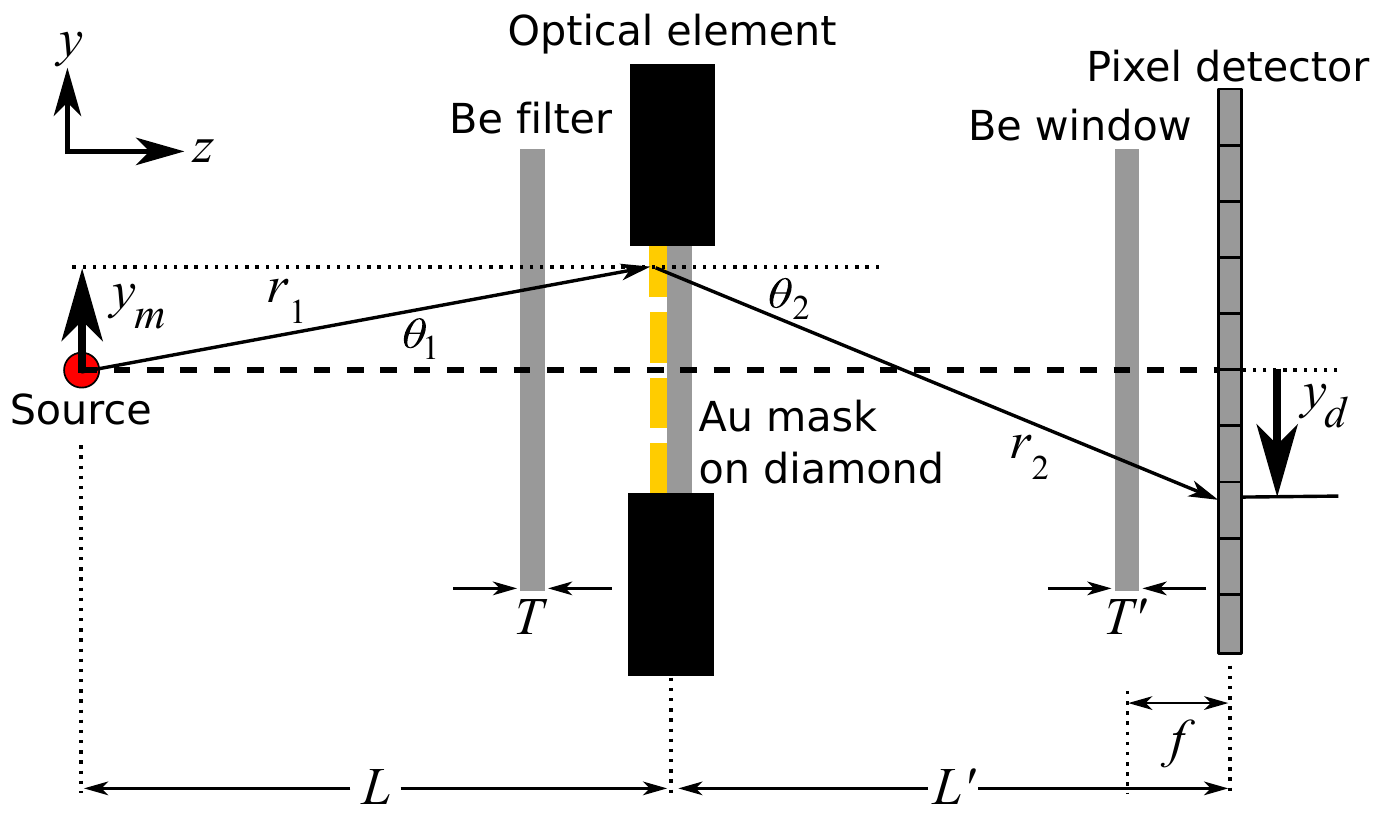} 
	\caption{Schematic of the XRM beam line at each ring during Phase 1 commissioning of SuperKEKB operation (not to scale). The beam line consists of a Beryllium filter placed upstream of the optics to reduce the heat load, three set of optical elements (a pinhole and two sets of coded apertures), Beryllium extraction window, and detector system (a 141~\si{\micro}m-thick YAG:Ce scintillator with CCD camera as an imaging system).}
	\label{fig:XRMscheme} 
\end{figure}

A simplified schematic of the XRM setup is shown in Fig.~\ref{fig:XRMscheme}, with relevant dimensions in Table \ref{tab:beamline}. There are separate installations for electrons (in the HER) and positrons (in the LER). Each of the SuperKEKB rings has four straight sections and four arc-bends. The X-ray sources are the last arc-bends located immediately upstream of the straight sections in Fuji (LER) and Oho (HER). The beamlines are about \SI{40}{\meter} from the source points to the detectors.  A list of the parameters for the beamlines are shown in Table~\ref{tab:beamline}. The optical elements (pinhole and coded apertures) are located in optics boxes 9-10~m from the source points, for geometrical magnification factors of $\sim3$ for both lines. Beryllium filters are placed between source points and optics boxes to reduce the incident power levels for both lines. A \SI{0.2}{mm}-thick Be window is also placed at the end of each beamline to separate vacuum (in beamline) and air (in detector box).

\begin{table}[hbt]
	\centering
	\caption{Dimensions of the XRM beamlines used to measure the vertical beam size in the LER and HER. Also see Fig~\ref{fig:XRMscheme}.}
	\begin{tabular}{lll}
		\toprule
		Parameter							&LER  	& HER\\
		\midrule
		Energy [GeV] 							&4 		&7  	    \\ 
		Source to optics ($L$) [m]					&9.26       &10.3	      \\
		Optics to detector ($L'$) [m]				&31.8      &32.7	        \\
		Air gap ($f$) [cm]						&10		&10				\\
		Thickness of Be filter ($T$) [mm]			&0.5		&16			\\
		Thickness of Be window ($T'$) [mm]			&0.2		&0.2			\\
		\bottomrule
	\end{tabular}
	\label{tab:beamline}
\end{table}

\subsubsection{Optical Element and Detection System}
\begin{figure}[!htb]
	%   \vspace*{-.5\baselineskip}
	\centering
	\includegraphics*[width=\columnwidth]{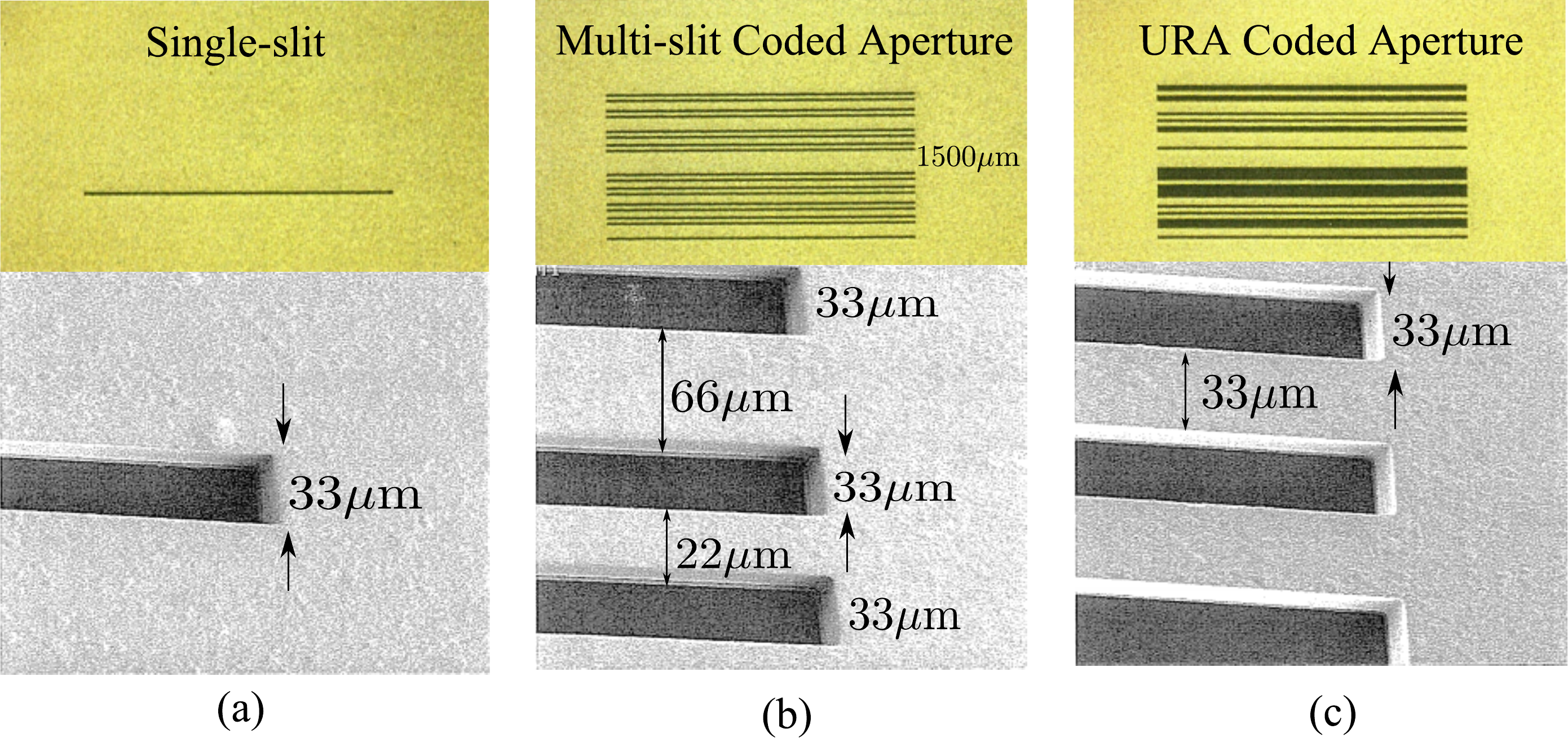}
	\caption{The optical elements of the XRM at 70$\times$ magnification and 1000$\times$ Scanning Electron Microscope (SEM): (a) single-slit, (b) multi-slit coded aperture, and (c) URA coded aperture.}
	\label{fig:optical-elements}
	%   \vspace*{-\baselineskip}
\end{figure}

Optical elements have been designed and installed in each ring: a single slit, a multi-slit coded aperture (17 slits) and a Uniformly Redundant Array (URA) coded aperture (12 slits), as shown in Fig.~\ref{fig:optical-elements} \cite{Mulyani:IBIC2015-TUPB025}.
These optical elements consist of 18-20~\si{\micro}m-thick gold masking material on \SI{600}{\si{\micro}m}-thick CVD diamond substrates.

For Phase 1 of SuperKEKB commissioning, a cerium-doped yttrium-aluminum-garnet (YAG:Ce) scintillator is combined with a CCD camera for the X-ray imaging system \cite{IBIC2016}.
%\begin{figure}[!htb]
%	\centering
%	\includegraphics*[width=\columnwidth]{XRMImages/detector.pdf}
%	\caption{XRM detection system for Phase 1 of SuperKEKB commissioning. Inside the detector box is a Be extraction window and a \SI{141}{\si{\micro}m} thick YAG:Ce scintillator combined with CCD camera.}
%	\label{fig:detector}
%\end{figure}

%------------------------------------------------------------------------------
\subsubsection{Calibration Studies} \label{sec:xrm_corrections}
Some calibration studies during Phase 1 of SuperKEKB commissioning have been carried out, such as geometrical scale checks and emittance knob ratio measurements~\cite{N.Iida}. The geometrical scale factors seem to be well understood for both rings~\cite{IBIC2016}. From the emittance knob ratio method, the value for the LER
is close to the design value ($\sim10$~pm), but is much higher
than design for the HER. To investigate this discrepancy, a
study of smearing factors (point spread functions) was made
using beam lifetime data. \\

If a beam of initial size $\sigma_{y_0}$ is convolved with a Gaussian smearing function of size $\sigma_{s}$ to make a measured beam size $\sigma_{y}^{meas}$, then the measured beam size can be represented by adding the real beam size and the smearing size in quadrature as shown in Eq.~\ref{eq:smear}.
\begin{align}
\label{eq:smear}
\sigma_{y}^{meas}=\sqrt{{(\sigma_{y_0})}^{2}+(\sigma_{s})^{2}} \nonumber \\
\sigma_{y_0}=\sqrt{{(\sigma_{y}^{meas})}^{2}-(\sigma_{s})^{2}}
\end{align}
If we consider just the Toushek effect (the only beam decay mechanism related to the beam size) then the correlation between  lifetime $\tau$ and measured beam size $\sigma_{y}^{meas}$ becomes~\cite{IBIC2016}
\begin{align}
\label{eq:toushek-smear}
\tau=\alpha\sigma_{y_0}=\alpha\sqrt{{(\sigma_{y}^{meas})}^{2}-(\sigma_{s})^{2}}
\end{align}

\begin{figure}[!htb]
	\centering
	\includegraphics*[width=\columnwidth]{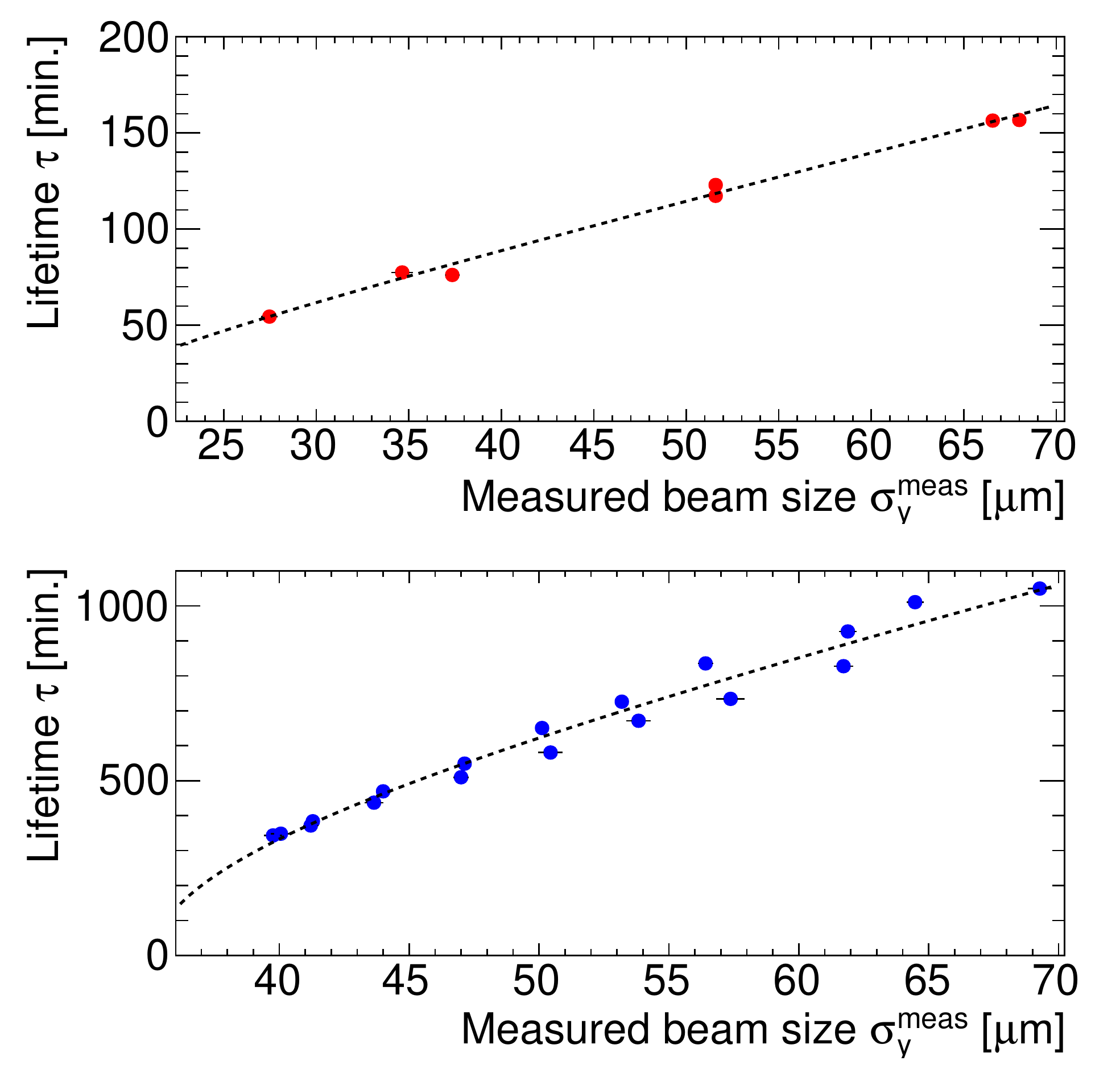}
	\caption{Relation between lifetime and beam size for multi-slit mask of the XRM at LER (top) and HER (bottom) fitted by Eq.~\ref{eq:toushek-smear}.}
	\label{fig:t-vs-sigma}
\end{figure}

Fitting the $\tau$ vs $\sigma_{y}^{meas}$ data via Eq.~\ref{eq:toushek-smear} with $\alpha$ and $\sigma_s$ as free parameters gives results like those shown (for the multi-slit mask) in Fig.~\ref{fig:t-vs-sigma}.  By using the correlation between beam parameters in Table~\ref{tab:beam-param}, we can calculate the true minimum beam size $\sigma_{y_0}$ from the smallest measured beam size $\sigma_{y}^{meas}$, and corresponding vertical emittance $\epsilon_{y_0}$. The average values over measurements made with all three optical elements, for $\sigma_{s}$, $\sigma_{y_0}$ and $\epsilon_{y_0}$ are shown in Table~\ref{tab:smearing}.  We see that the smearing function for the HER is much larger than that for the LER.  Also, even after accounting for this smearing function, the HER emittance is about 4 times larger than the design value.

\begin{table}[hbt]
	\centering
	\caption{Beam parameters during the XRM calibration studies.}
	\begin{tabular}{lll}
		\toprule
		Parameter 								&LER  		& HER	 \\
		\midrule
		$\Delta{\epsilon_y}$ for unit knob change	[pm] 		&70.0946		&43.0096 \\ 
		$\beta_y$ [m] 								&67.1721		&7.63647	     \\
		\bottomrule
	\end{tabular}
	\label{tab:beam-param}
\end{table}

\begin{table}[hbt]
	\centering
	\caption{Average smearing factor, $\sigma_{s}$, minimum vertical beam size, $\sigma_{y_0}$, and minimum vertical emittance, $\epsilon_{y_0}$, measured with all three XRM optical elements.}
	\begin{tabular}{lll}
		\toprule
		Parameter 						&LER 	& HER	 \\
		\midrule
		$\sigma_{s}~[\si{\micro}$m]			&$12.1 \pm 2.1$	& $32.8 \pm 0.4$   \\ 
		$\sigma_{y_0}~[\si{\micro}$m]			&$23.5 \pm 0.3$	& $17.8 \pm 0.8$	     \\
		$\epsilon_{y_0}$~[pm]				&$\sim8$			&$\sim41$	       \\
		
		\bottomrule
	\end{tabular}
	\label{tab:smearing}
\end{table}

\subsubsection{XRM Conclusion}
We have conducted some calibration studies during the Phase 1 of SuperKEKB commissioning. The geometrical magnification factors seem to be well understood for both LER and HER. The overall performance is reasonable for the LER, and results are consistent with expectations based on the optics estimation with $\sim$8~pm of vertical emittance ($\epsilon_y$). For the HER, the vertical emittance $\epsilon_y$ is $\sim$41~pm, 4$\times$ higher than the optics estimation. In addition, some smearing is observed, not all of which is fully accounted for yet. In the future, we plan to study possible sources of smearing either at the X-ray source point or in the beamline.

\subsection{Gas composition} % Alex and Sam
\label{sec:Zeff}
% file:         beamgas_touschek_zeff.tex
% lead author:  Sam de Jong and Alex Beaulieu
% This file is included at level subsection

% First draft written by Alex Beaulieu
The simplest model of beam-gas interactions assumes that the losses, and hence the recorded background rate, are proportional to the product of the beam average pressure and beam current. We get a better description by including a gas composition term derived from residual gas analyzer (RGA) data.

The SuperKEKB accelerator is instrumented with two RGAs: one near the BEAST II detector on the positron storage ring, and the other  --- approximately diametrically opposed  in the storage ring --- near the positron injection site. These RGAs are mass spectrometers that provide the detected amplitudes for gas ion fragments with mass-to-charge ratios $m/z$ between 1 and 50. In order to compare measurements with the simulations generated assuming a pure gas of atomic number $Z=7$, we need to extract information from the detected amplitudes to determine the relative importance of each gas constituent, and calculate an effective $Z$ for this gas, noted $Z_e$.  The analysis method can be broken down in three steps: definition of the gas model, calculation of the proportion of each gas constituent, and calculation of the corresponding $Z_e$.

\subsubsection{Gas model} The raw information from the RGAs need to be processed in order to interpret the detected amplitudes, one for each $m/z$ value, as abundances for given molecules in the residual gas \cite{gross:2011}. The general idea is to find a list of standard gas spectra that could form a basis in which we can decompose the measured distributions. Some initial assumptions about the nature of the residual gas are therefore required. We first added di-hydrogen and air constituents to the list of gases as the default hypothesis: H$_2$, H$_2$O, N$_2$, O$_2$, CO, CO$_2$, Ar. Then, we added light hydrocarbons progressively, starting from CH$_4$, until all features of the measured spectra could be described: CH$_4$, C$_2$H$_6$, C$_2$H$_4$, C$_3$H$_4$, C$_3$H$_6$, C$_3$H$_8$.

Finally, none of these standard gases predict a peak at $m/z=3$ as strong as the one observed by the RGAs in BEAST II. Two hypotheses could explain this feature: deuterium-hydrogen molecules DH and tri-hydrogen atoms $\text{H}_3$. It is worth noting that both these species are relatively exotic on Earth, however the SuperKEKB vacuum chamber, with its very low pressure, hydrogen-rich residual gas and high levels of ionizing and neutron radiation provides conditions favorable to their creation. 

\subsubsection{Calculation of the proportion of each gas}
The proportion of each gas species in the residual gas is then found by calculating the optimal proportions to explain the measured spectra in the least-squares sense. The problem is expressed as solving  
\begin{equation}
\arg\min_\mathbf{x}  \left\lVert{\mathbf{Ax} - \mathbf{y}}\right\rVert _2, ~\mathbf{x} \geq 0
\end{equation}
where $\mathbf{y}$ is a column vector of the observed relative abundances for each $m/z$ peak, $\mathbf{A}$ is a matrix whose columns each correspond to the standard spectrum for a gas model constituents, and $\mathbf{x}$ is a column vector of the relative proportion (in number of molecules) of each gas in the mix. The vector of the optimal proportions of each gas, $\hat{\mathbf{x}}$, is therefore 
\begin{equation}
\hat{\mathbf{x}} = \left(\mathbf{A}^{\rm {T}} \mathbf{A}\right)^{-1}\mathbf{A} ^{\rm{T}}\mathbf{y}.
\end{equation}

\subsubsection{Calculation of an effective $Z$ for this gas mix} \label{sec:calculating_zeff}
The gas proportions $\hat{x_i}$, together with their molecular formulae, are then used to calculate the number $b_j$ of atoms of element $Z_j$ by simply multiplying each $\hat{x}_i$ by the number of atoms of $Z_j$ in the gas molecules. Assuming that the probability of interaction between a beam electron and an atom $Z_j$ is proportional to $Z_j^2$ (see Section~\ref{simulation_scaling}), the effective atomic number $Z_e$, is expressed as a weighted average of $Z_j^2$: 
\begin{eqnarray}
\left< Z^2\right>  &=& \frac{\sum_j{ Z_j^2 b_j} }{\sum_j{b_j}}, \label{eqn:Ze2}\\
Z_e  &=&  \sqrt{\left< Z^2\right>}. \label{eqn:Ze}
\end{eqnarray}
It is ``effective'' in the sense that this $Z_e$ is the atomic number of a pure gas that would produce the same level of beam-gas interactions as the gas mix found in the vacuum chamber. This number can then readily be used to scale the simulation that has been generated with a single value of $Z$.

\subsubsection{Dedicated experiments}
Dedicated ``pressure bump'' experiments were conducted to artificially enhance beam-gas interactions and study their scaling relationships with respect to operating parameters. The general idea of a pressure bump experiment is to heat the non-evaporable getters (NEGs) in a given section of the vacuum chamber in order to reach an increase of pressure of at least 100.

A typical pressure bump experiment lasts approximately 30-40 min. The NEGs are first heated and maintained at an intermediate temperature, to allow the release of the heavier molecules, then the temperature is raised again to a second plateau for approximately 10 minutes until the target vacuum chamber pressure is reached. The study of the resulting change in gas composition is presented in more detail in Section~\ref{sec:RGAresults}.

\subsection{Sample results of residual gas analysis and effective atomic number}
\label{sec:RGAresults}
Using pressure bump experiments, we can probe the agreement between our models and data in two different ways: observing the time series of pressure and background readings during a beam bump experiment, and measuring the slope ratio between the two bumps within one experiment.

\subsubsection{Time series of pressure readings in beam bump experiments}\label{sec:RGAresultsTimeSeries}

\begin{figure}[h]
 \centering
 \includegraphics[width=\columnwidth]{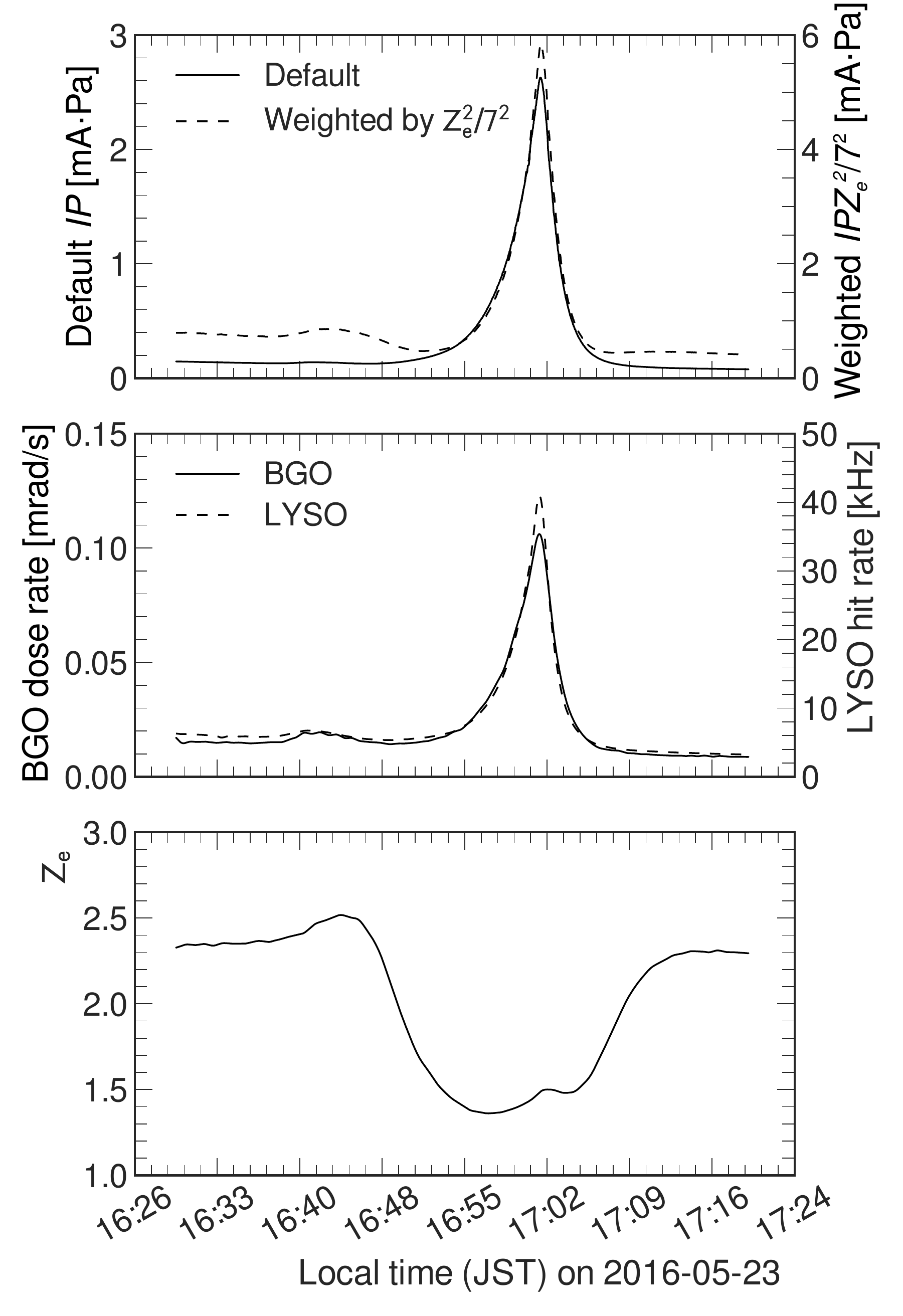}
 \caption{Time series of a pressure bump experiment and typical results. The top panel shows the product of pressure and current $IP$ before and after effective atomic number re-weighting. The applied weight is $Z_e^2 / 7^2$ since all simulation is generated with a fixed $Z=7$. The central panel shows recorded background rates for one typical channel of two different subsystems, showing qualitatively better agreement with the re-weighted pressure than with the raw reading. The effective atomic number $Z_e^2$ for this particular experiment is presented in the bottom panel.}
 \label{fig:PressureTimeSeries}
 \end{figure}

\begin{figure}[h]
	\centering
	\subfigure[Unweighted $IP$]{\includegraphics[width=\columnwidth]{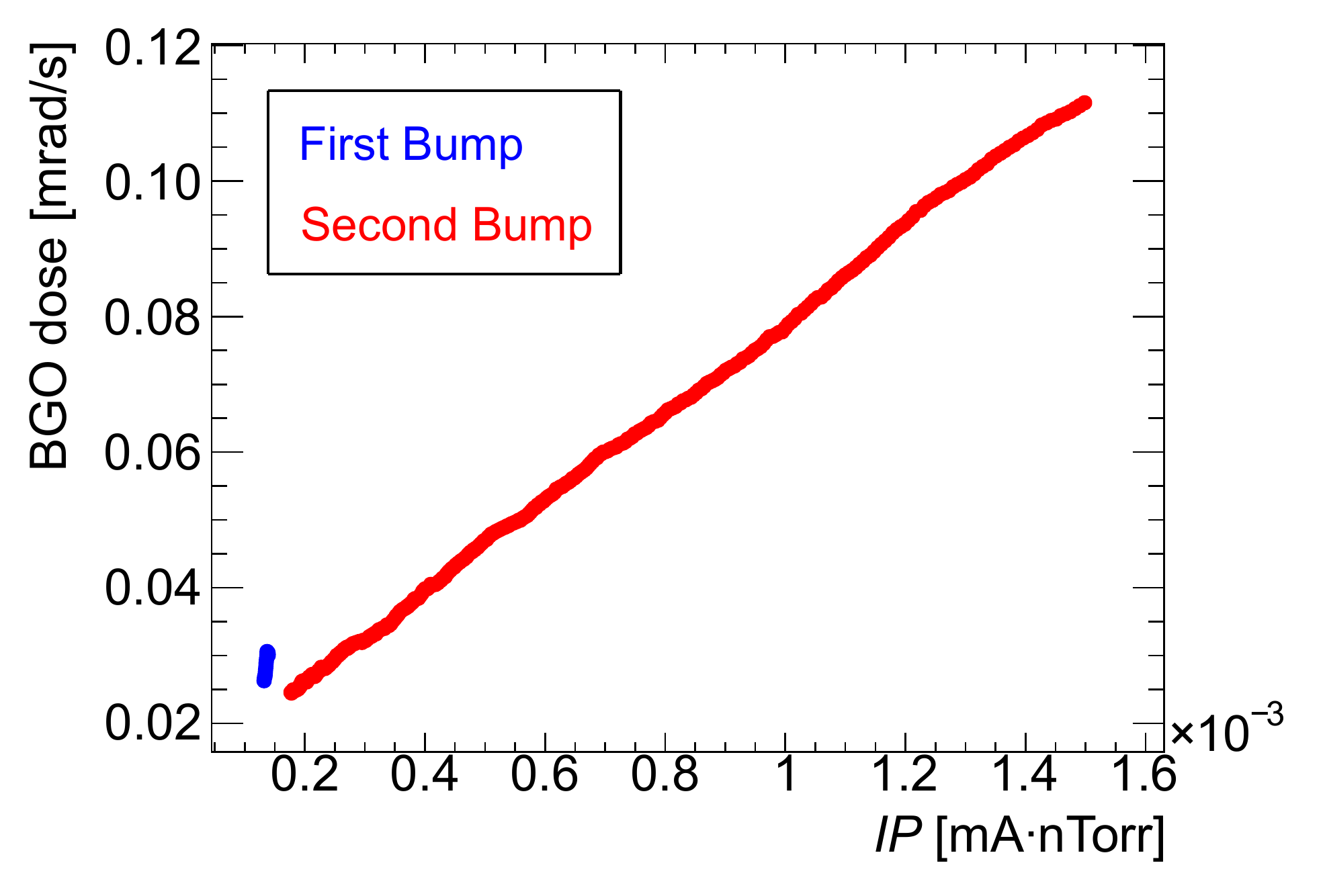}\label{fig:noMassSpec}}  
	\subfigure[Weighted $IP$]{\includegraphics[width=\columnwidth]{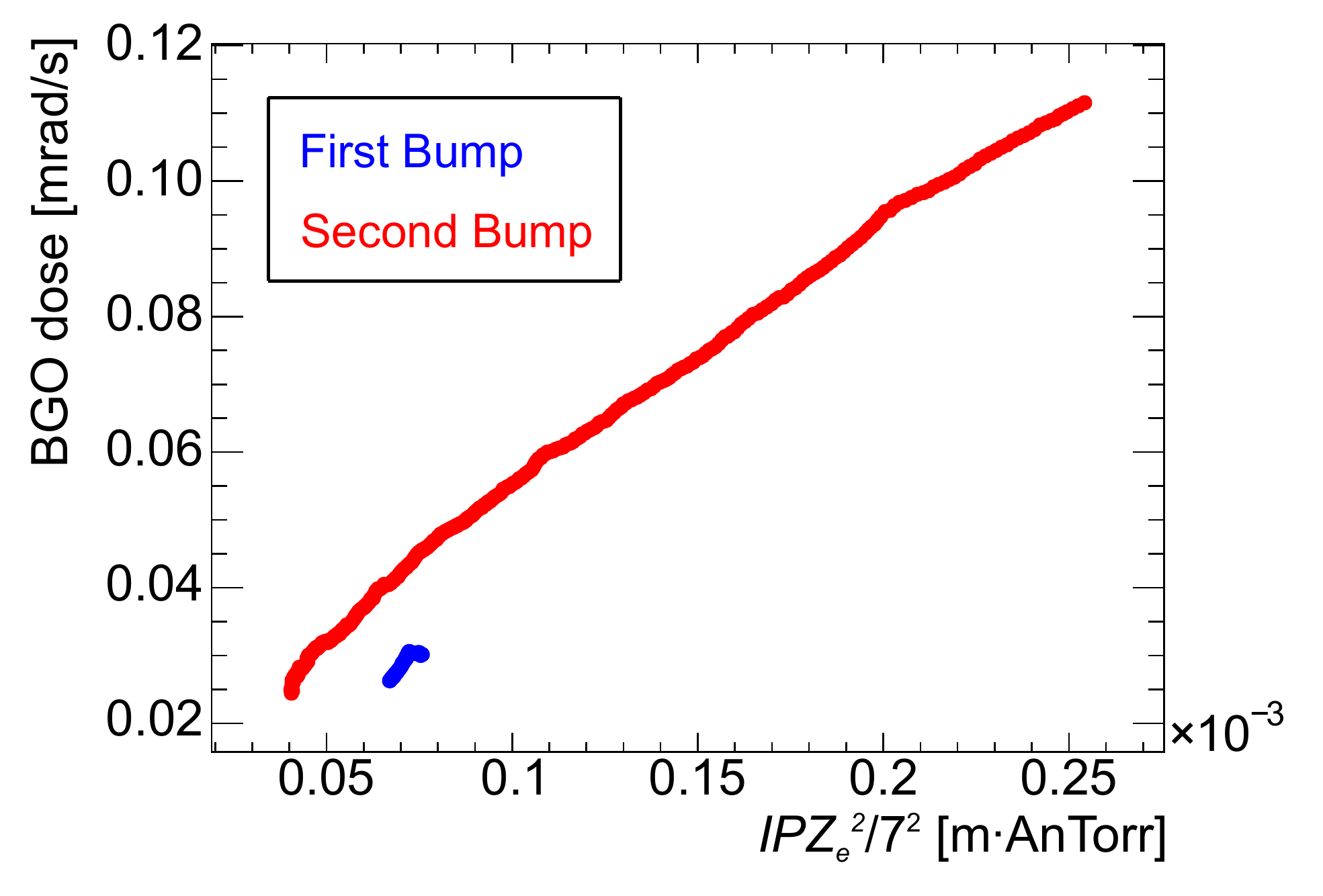}\label{fig:MassSpec}}
	
	\caption{(color online) Demonstration of dose rate in BGO channel 7 as a function of weighted and un-weighted $IP$, showing the improvement produced by including $Z_{e}$. Data from the first bump is blue, and data from the second bump is red.}
	\label{fig:massSpecExample}
\end{figure}

Figure~\ref{fig:PressureTimeSeries} shows example results obtained from the beam-gas constituents analysis.
Qualitatively, we observe that the recorded backgrounds track $IPZ^2$ better than $IP$. There is an increase of the hit rates around 16:42 which is not associated with any notable increase of the average pressure. This behavior is explained by the $Z^2$-dependence of beam-gas interactions. The heavier elements, such as carbon and oxygen, are released first in the experiment. Their large atomic numbers therefore have a notable impact on the effective $Z$ of the gas mixture, despite them not contributing significantly to the net pressure increase. 
Figure~\ref{fig:massSpecExample} shows the dose rate in BGO channel 7 as a function of both weighted and un-weighted $IP$ for the rising portion of both the first and the second pressure increases. In the un-weighted plot (Figure~\ref{fig:noMassSpec}), the two different bumps have a very different response. The weighted plot (Figure~\ref{fig:MassSpec}) however shows a similar response to both bumps.

\subsubsection{Slope ratio}
\begin{figure}[H]
 \centering
 \includegraphics[width=\columnwidth]{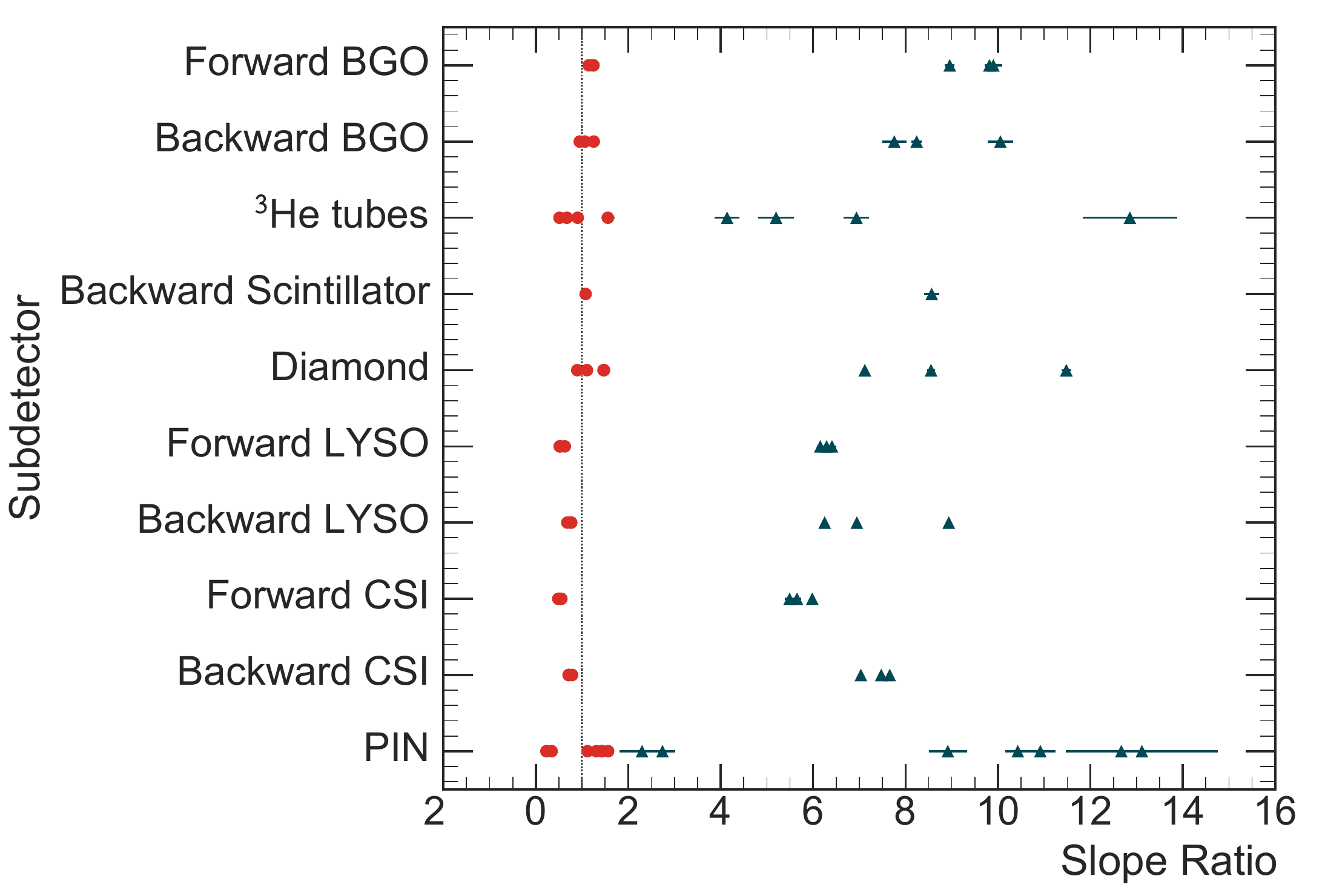}
 \caption{(color online) Ratios of the slope --- increase rate of the observable per unit increase in $IP$ --- in the initial phase to the slope in the final phase in pressure bump experiment, as defined by Eq~\ref{eqn:slopeRatio}. Each point corresponds to a single detector channel. Blue triangles denote the slope ratios calculated without including $Z_{e}$, and red circles represent the slope ratio when the $Z_{e}$ correction is applied. Including $Z_{e}$ brings the slope ratio closer to one, showing a better understanding of the effect of the residual gas constituents.}
 \label{fig:SlopeRatio}
\end{figure}

The improvement due to applying the correct gas composition weighting from the residual gas mixture is also notable when we plot each detector channel measurement as a function of $IP$, fit a straight line to these data, and calculate the ratio of the slope of the first bump to the slope of the second bump.
\begin{equation}
 	\text{Slope ratio} = \frac{m_\text{first}}{m_\text{second}}
 	\label{eqn:slopeRatio}
\end{equation}
If our beam-gas model is correct, the ratio will be equal to one. Figure~\ref{fig:SlopeRatio} shows this slope ratio for each detector channel without weighting $IP$ (in blue) and with the weighting (in red). It is observed that weighting $IP$ by $Z_{e}^2$ makes the slope ratio consistent with unity, therefore indicating that the effect of a different gas mixture in the two bumps is compensated appropriately in this model.

\subsection{Pressure} % Peter
% file:    beamgas_touschek_pressures.tex
% author : Peter
% \subsection{Pressure} was declared just before this

In order to understand backgrounds in the interaction region, particularly those generated by beam-gas interactions, we need to understand the vacuum pressure throughout the trajectories of the beams. In this section we discuss the pressure measurements at SuperKEKB, and introdce corrections to the measured pressures and a method to select the pressure gauges that best predict the measured observables.

\subsubsection{Pressure corrections}
\label{sec:pressureCorrections}
In SuperKEKB, cold cathode gauges (CCGs) are located roughly every 10\,m around each ring at the end of ~1\,m ducts shared with sputter ion pumps. Due to the physical proximity of the gauges to the pumps and their separation from the beam line, the measured pressure is lower than that seen by the beam. SuperKEKB simulations show that the dynamic component of the pressure is roughly a factor of $3$ lower in the vicinity of the CCGs than it is in the beampipe. The ``base'' pressure, due to residual gasses that remain in the beampipe long after the beams have stopped circulating, is not subject to this correction. Therefore we obtain
\begin{equation}
  P_{CCG} = P_{base} + \frac{1}{3} \cdot P_{dynamic},
\end{equation}
where $P_{CCG}$, $P_{base}$ and $P_{dynamic}$ are the measured, base and dynamic pressures, respectively. The quantity of interest is the pressure seen by the beam, $P_{beam}$, given by:
\begin{equation}
  P_{beam} = P_{dynamic} + P_{base},
\end{equation}
and we substitute in order to express $P_{beam}$ in terms of measurable quantities:
\begin{equation}
  P_{beam} = 3 \cdot P_{CCG} - 2 \cdot P_{base}.
\end{equation}
For each CCG, we find the minimum pressure recorded during the last multi-hour period with no current in either beam and call this the base pressure $P_{base}$.

In practice, the dynamic pressure in the LER is much larger than the base pressure, so $P_{beam} \approx 3 \cdot P_{CCG}$, but the base pressure has a large influence in the HER. The CCG minimum reading is $10^{-8}$~Pa, so we assume this value as the base pressure when the CCG reading is out-of-range. 

From this point forward, all pressures are assumed to be the corrected pressure $P_{beam}$.

\label{sec:superkekb_conditions_pressures}

 \clearpage

 % lead author: Peter Lewis
 \section{Beam-gas and Touschek backgrounds in BEAST II}\label{sec:beamgas_touschek} % Peter
 The purpose of this section is to demonstrate a tractable method for disentangling and measuring beam-gas and Touschek backgrounds in the interaction region on a channel-by-channel basis. To do this, we first propose parameterizations of both backgrounds using accelerator and beam conditions and validate using dedicated beam studies. Finally, we use these measurements to test the accuracy of beam-gas and Touschek simulation independently. We present here two largely complementary approaches. 

\subsection{Background parameterization}
In order to disentangle beam-gas and Touschek backgrounds, we here propose a simplified parameterization of these two background types using their expected behavior in relation to the accelerator and beam conditions described in Section~\ref{sec:superkekb_conditions}.

\subsubsection{Beam-gas parameterization}
Beam-gas background is due to two distinct underlying processes: bremsstrahlung and Coulomb scattering. We combine these two processes into a simplified parameterization given by
\begin{equation} \label{eqn:beamgas_param}
  \mathcal{O}_{bg} = S_{bg} \cdot I P Z_{e}^2,
\end{equation}
where $\mathcal{O}_{bg}$ is the quantity of a BEAST II sensor's observable that can be attributed to beam-gas backgrounds, $S_{bg}$ is a constant of proportionality we call the beam-gas \textit{sensitivity}, $I$ is the beam current, $P$ is the vacuum pressure, and $Z_{e}$ is the effective atomic number of the gas, as described in Section~\ref{sec:Zeff}. Note that the true bremsstrahlung and Coulomb dependencies on $Z$ are more complicated than $Z^2$ (see Equations \ref{eqn:f_B} and \ref{eqn:f_C}). However, for typical gas compositions these functions are roughly proportional to $Z^2$ and we subsume the constants of proportionality into the sensitivities $S$.  

This relation is simple and is physically motivated. However, the physical motivation refers to the gas pressure and composition at the scattering location, which is unknown and variable. The gas properties in the ring are highly local, therefore we cannot assume that an average pressure or composition will adequately describe the backgrounds. 

\paragraph{A more-precise parameterization}
Pressure, particularly in the LER, is dominated by dynamic pressure caused by desorption of gasses from the beampipe walls due to collisions by off-orbit beam particles or radiation emitted from beam particles. These collisions are highly position-dependent, sensitive to the details of the beam trajectory, magnetic fields, and geometry around the ring. During beam storage, pressures measured in two adjacent gauges roughly 10\,m apart disagree by a factor that is typically 2-5 but sometimes exceeds 100.  

Composition of the gas, expressed as the effective atomic number $Z_e$, is affected by the relative local contributions of photon-stimulated and electron-stimulated desorption, which vary along the beam line. The $Z_e$  measured at two positions on the LER separated by roughly a kilometer disagree by $5-25\%$ during a typical run (see Fig.~\ref{fig:tousZef}).

In the ideal case we would know the continuous pressure and gas composition throughout the rings, and we could weight these by the scattering position distribution. However, we have only coarse pressure readings and two measurements of the gas composition. Instead, we consider an ``effective pressure'' $P_e$:
\begin{equation} \label{eq:effective_pressure}
  P_e = \frac{\sum\limits_i^{CCG} P_i w_i}{\sum\limits_i^{CCG}w_i},
\end{equation}
where the weights $w_i$ reflect the relative likelihood of scattering in the vicinity of CCG $i$ leading to measurable IR losses. The instantaneous backgrounds from beam-gas interactions can then be written as
\begin{equation}
  \mathcal{O}_{bg} = S_{bg} \cdot I P_e Z_e^2,
\end{equation}
where $Z_{e}$ must be estimated or interpolated to compensate for the lack of measurements at each CCG. The challenge then becomes to select the correct weights; we will see later that it is sufficient to choose the single CCG that gives the best agreement between data and the parameterization. 

\subsubsection{Touschek parameterization}
Although Touschek scattering depends on beam energy and the number of filled bunches in the ring, in practice these do not change during routine operation. Therefore, we expect the detected observable generated by Touschek scattering to depend only on the beam current $I$ and the vertical beam size $\sigma_y$:
\begin{equation}
  \mathcal{O}_{T} = S_{T} \cdot \frac{I^2}{\sigma_y},
\end{equation}
where $S_T$ is the Touschek sensitivity, in analogy with the beam-gas sensitivity.

\subsubsection{The beam-gas and Touschek combined heuristic} \label{sec:combined_heuristic}
\begin{figure}[H]
	\centering
	\includegraphics[width=\columnwidth]{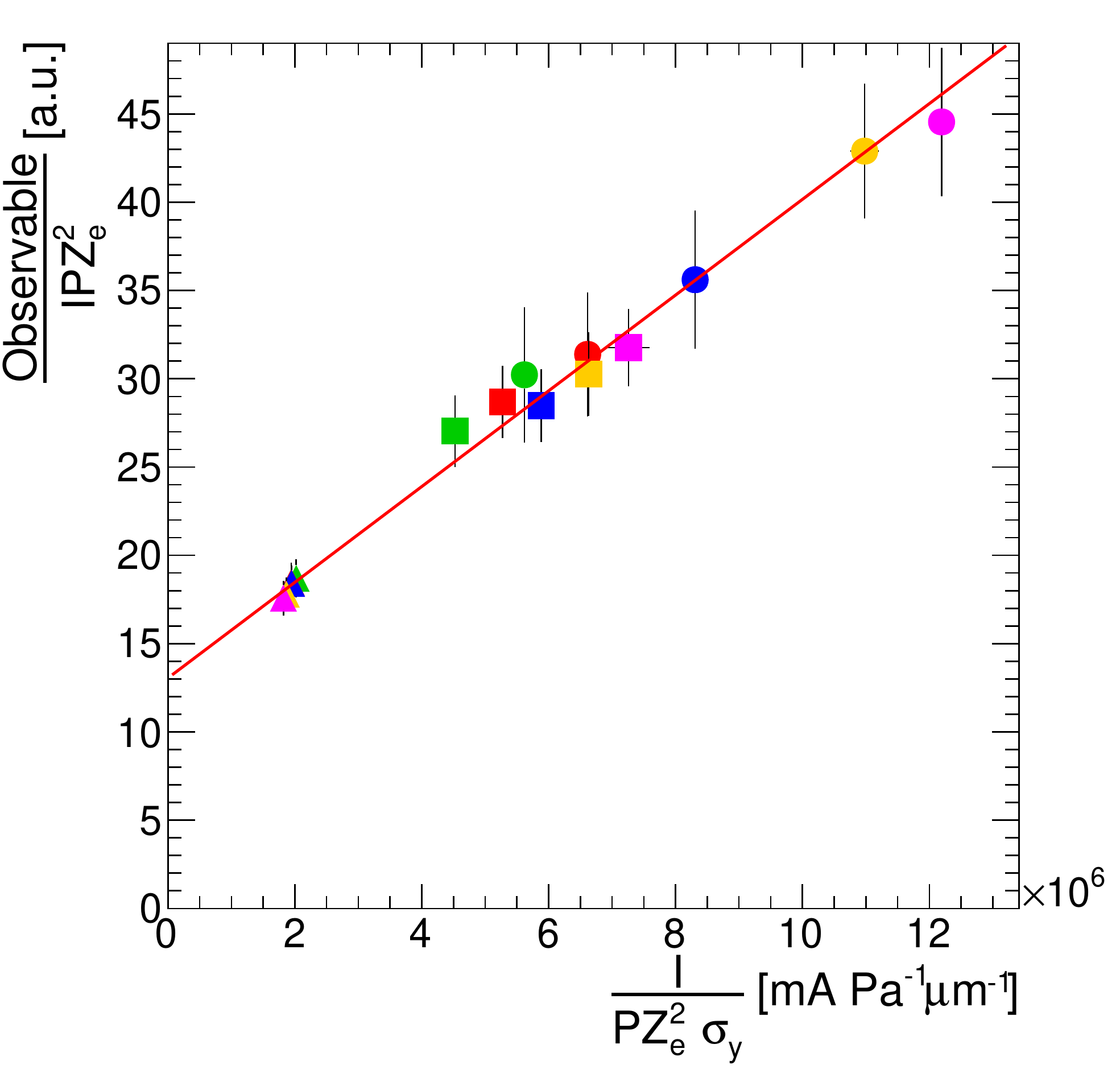}
	\caption{(color online) An example of fitting the heuristic beam-gas and Touschek model of Eq.~\ref{eq:heuristic_divided} for all three LER size sweeps. A single BGO channel provides the observable, and quantities on both axes have been averaged within a subrun. Shapes correspond to currents and colors to beam size settings. The offset of the best-fit line contains the beam-gas contribution, and the slope indicates changes in the Touschek contribution as the beam size is varied.}	
	\label{fig:touschek_demo_fit}
\end{figure}

We are now ready to write a parameterization for combined beam-gas and Touschek backgrounds that should explain the large majority of measured observables during non-injection runs:
\begin{equation} \label{eqn:combined_heuristic}
  \mathcal{O} = S_{bg} \cdot I P_e Z_e^2 + S_{T} \cdot \frac{I^2}{\sigma_y}.
\end{equation}
We refer to this as the combined heuristic. The beam-gas and Touschek sensitivities $S_{bg}$ and $S_{T}$ are unique to each channel and should be constant when accelerator conditions are fixed, with independent LER and HER values. For visualization purposes, it is convenient to rewrite the model as
\begin{equation} \label{eq:heuristic_divided}
	\frac{\mathcal{O}}{I P_e Z_e^2} = S_{bg} + S_{T} \cdot \frac{I}{P_e Z_e^2\sigma_y}
\end{equation}
and plot $\mathcal{O}/(I P_e Z_e^2)$ vs.\ $I/(P_e Z_e^2\sigma_y)$. On such a scatterplot the data should fall on a line with offset indicating the beam-gas sensitivity $S_{bg}$ and slope equal to the Touschek sensitivity $S_T$. Figure ~\ref{fig:touschek_demo_fit} shows a fit of the heuristic model to a single channel of the BGO detector using data described in Section~\ref{sec:beamgas_touschek_beam_studies}. Each point represents the mean values of $\text{Observable}/(IPZ_e^2)$ and $I/(PZ_e^2\sigma_y)$ for one combination of beam size and current. The linearity of this distribution despite large variations in current, pressure, and beam size validates the heuristic model. 

\label{sec:beamgas_touschek_parameterization}

\subsection{Dedicated beam studies}
During Phase 1, BEAST II and SuperKEKB performed a number of dedicated beam study runs to measure the beam-gas and Touschek backgrounds in both rings. These runs consist of five short subruns, each at a different beam size setting, topped off to a fixed current before each subrun begins. We refer to this as a size sweep. For each ring, LER and HER, we performed three sweeps at different top-off currents, summarized in table ~\ref{tab:touschek_runs}. 

The methods used for controlling beam sizes in the two rings differed. In the LER, we varied the strength of a skew quadrupole magnet to introduce more $x-y$ coupling. This allowed us to control the beam size over the whole ring while not disturbing the beam orbit itself. In the HER, we shifted the beam orbit vertically in an isolated location using a pair of bending magnets. This introduced vertical dispersion in the beam.

\begin{table}[ht]
	\caption{Size sweep runs. The beam size is a measurement from the X-ray monitor. Although each subrun corresponds to a different beam size setting, full control of the beam size was not in practice achievable.}
	\centering	
	\begin{tabular}{ ccc }
        \toprule
	Run       & Subrun \# & Beam size (\si{\micro}m)	\\ 
        \midrule
	HER 320~mA & 1	     & 85			\\
	          & 2	     & 68			\\
	          & 3	     & 39			\\
	          & 4	     & 44			\\
	          & 5	     & 45			\\
	HER 480~mA & 1	     & 91			\\
	          & 2	     & 66			\\
	          & 3	     & 47			\\
	          & 4	     & 32			\\
	          & 5	     & 41			\\
	HER 640~mA & 1	     & 121			\\
	          & 2	     & 74			\\
	          & 3	     & 46			\\
	          & 4	     & 40			\\
	          & 5	     & 56			\\
	LER 360~mA & 1	     & 81			\\
	          & 2	     & 65			\\
	          & 3	     & 51			\\
	          & 4	     & 38			\\
	          & 5	     & 32			\\
	LER 540~mA & 1	     & 95			\\
	          & 2	     & 72			\\
	          & 3	     & 67			\\
	          & 4	     & 58			\\
	          & 5	     & 51			\\
	LER 720~mA & 1	     & 148			\\
	          & 2	     & 147			\\
	          & 3	     & 141			\\
	          & 4	     & 145			\\
	          & 5	     & 146			\\
        \bottomrule
	\end{tabular}
	\label{tab:touschek_runs}
\end{table}

\label{sec:beamgas_touschek_beam_studies}

\subsection{Direct analysis} 
% file: touschek_analysis.tex
% lead author: Peter Lewis
%
% Estimated completion date: End of September (hahaha)
% like.... September 2018?

We now show how the parameterization of beam-gas and Touschek backgrounds described in Sec.~\ref{sec:combined_heuristic} and SuperKEKB condition measurements subsequently described can be used to measure beam-gas and Touschek backgrounds in size sweeps. This analysis can be used to probe the accuracy of simulation and measure integrated doses in Phase 1 for beam-gas and Touschek backgrounds simultaneously. 

\subsubsection{Analysis procedure}
Using the dedicated Touschek size sweeps of Table~\ref{tab:touschek_runs} we perform fits similar to that shown in Fig.~\ref{fig:touschek_demo_fit} for each channel in every BEAST II detector. For these fits, we use the measured current $I$; the corrected beam size, $\sigma_y$, as described in Sec.~\ref{sec:xrm_corrections}; the effective atomic number, $Z_e$, as described in Sec.~\ref{sec:calculating_zeff}; and the most-predictive CCG as described shortly. For both the LER and HER runs we use a constant $Z_e$ equal to the average derived from the RGA in the immediate-upstream LER arc section during LER Touschek size sweeps. This approximation has no effect on the predictive power of the fit. 

The fit gives us two unique parameters $S_{bg}$ and $S_T$. These parameters, combined with the identity of the most-predictive CCG, allow us to write the observable expected for any current, pressure, beam size and gas composition in either ring, as given by Eq.~\ref{eqn:combined_heuristic}.

\begin{figure}[H]
	\centering
	\includegraphics[width=\columnwidth]{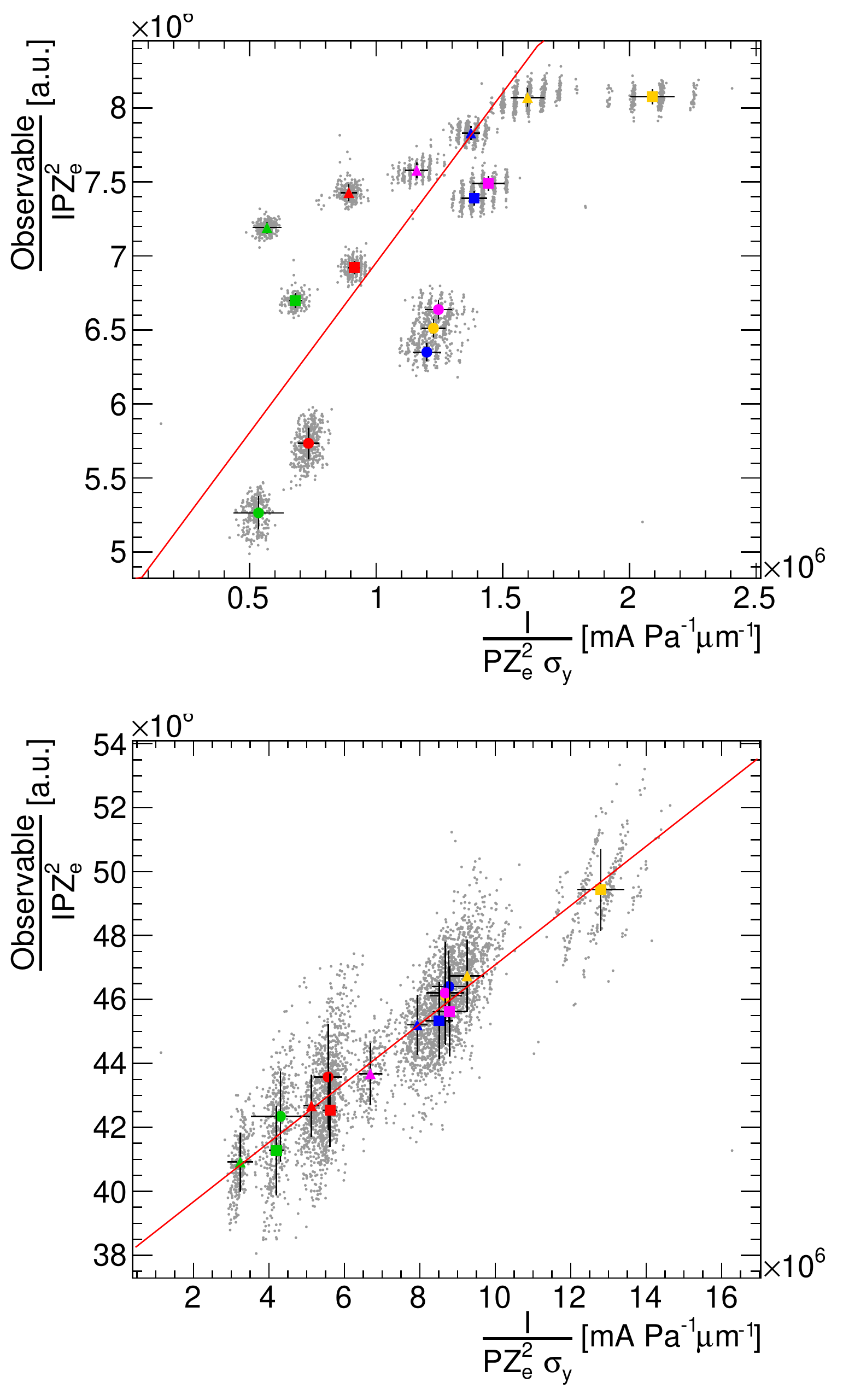}
	\caption{(color online) A demonstration of the effect of CCG choice on the quality of the heuristic fit. The observable is the CsI hit rate from a single channel during a full set of HER Touschek size sweeps, where color corresponds to subruns (beam sizes) and shapes correspond to runs (currents). The top plot is the result when using the nearest upstream CCG. The bottom plot shows the result when using the most ``predictive'' CCG, that is, the one that gives the lowest normalized $\chi^2$.}	
	\label{fig:touschek_ccg_choice_demo}
\end{figure}

\paragraph{Weighting CCGs} \label{sec:choosing_ccg}
In this section we demonstrate a method for choosing the CCG weights of Eq.~\ref{eq:effective_pressure}. Without knowing the scattering distribution, we instead try to find the CCGs with the greatest predictive power by maximizing the linearity of the combined heuristic fit as defined in Sec.~\ref{sec:combined_heuristic}. Maximum linearity is defined as occurring when the normalized $\chi^2$ is smallest. Fig.~\ref{fig:touschek_ccg_choice_demo} illustrates the sensitivity of the heuristic fit to CCG weighting. The optimal CCG weighting will not necessarily indicate the dominant scattering positions.

Simulation predicts that almost all LER beam-gas scattering occurs near the IP, suggesting that the nearest-upstream CCG is always the correct choice. Indeed, this CCG is always highly predictive of observables in LER runs and therefore we use it for all channels. In HER runs, different detector channels are sensitive to backgrounds from different scattering positions, consistent with simulation predictions. We therefore look for CCG weightings that are unique to each channel. In these runs, it is always possible to find a single CCG that produces a parameterization with a normalized $\chi^2$ less than 1. We conclude that there is no gain to be achieved by looking for linear combinations of CCG pressures and therefore we focus on finding the single most-predictive CCG for each channel.

Our CCG selection procedure is as follows. First, using data from a HER size sweep, we fit the heuristic for each channel for each of 75 CCGs upstream of the IP identified by simulation as contributing substantially to beam-gas losses in the IR. We select the fit with the lowest $\chi^2$ to identify the most-predictive CCG. We use the beam-gas and Touschek sensitivities and most-predictive CCG from this fit to parameterize the beam-gas and Touschek observable in terms of beam conditions. 

\paragraph{Selections}
We include all BEAST II detector data from size sweep runs in the fits with the following exceptions. To avoid CsI/LYSO hit rate saturation effects, we require $I_{LER}<500~\text{mA}$ only for those observables. We include only physically plausible beam sizes with $35$~\si{\micro}m$<\sigma_y<400$~\si{\micro}m. We ignore data during injection and for ten seconds afterward. 

\paragraph{Comparing experimental and simulated data} \label{sec:beamgas_touschek_data_mc_analysis}
Using the parameterization with the combined heuristic allows us to make direct comparisions with simulation without relying on complicated reweighting procedures. We perform the simulation with fixed and uniform pressure, current, gas composition and beam size; essentially a single point in the 2D space of Fig.~\ref{fig:touschek_demo_fit}. Once we have measured the beam-gas and Touschek sensitivities, we can predict the observable under the simulated conditions with the combined heuristic. 

Given the beam conditions assumed by SAD ($I^{SAD}$, $P^{SAD}$, $Z^{SAD}$, $\sigma_{y}^{SAD}$) and including the pressure corrections described above, we can use the results of the heuristic fit to predict the value of the observable we would see in experimental data under the same conditions used in the simulation:
\begin{equation}
  \mathcal{O}_{bg}^{\mathrm{exp}} = S_{bg} \cdot I^{SAD} P_e^{SAD} Z_e^2,
  \label{eq:predicted_observable}
\end{equation}
and analogously for the value of the Touschek observable. Here, $P_e^{SAD}$ indicates the value we expect in the most-predictive CCG when the ring average pressure is equal to $P_{SAD}$, obtained from comparing the CCG pressure to ring average pressures across a range of pressures. 

For the value of the simulated observable we use the raw observables obtained after {\GEANT} digitization, but rescale the bremsstrahlung and Coulomb observables by $f_B(Z_{e})/f_B(Z=7)$ and $f_C(Z_{e})/f_C(Z=7)$, respectively, as described in Sec.~\ref{simulation_scaling}. This explains why we use $Z_{e}$ instead of $Z_{SAD}$ in Eq.~\ref{eq:predicted_observable}, and it corrects for the differences between the simplistic $Z^2$ beam-gas scaling used on experimental data and the true scalings of Eqs.~\ref{eqn:f_B} and \ref{eqn:f_C}.

\subsubsection{Results}
\begin{figure}[H]
	\centering
	\includegraphics[width=\columnwidth]{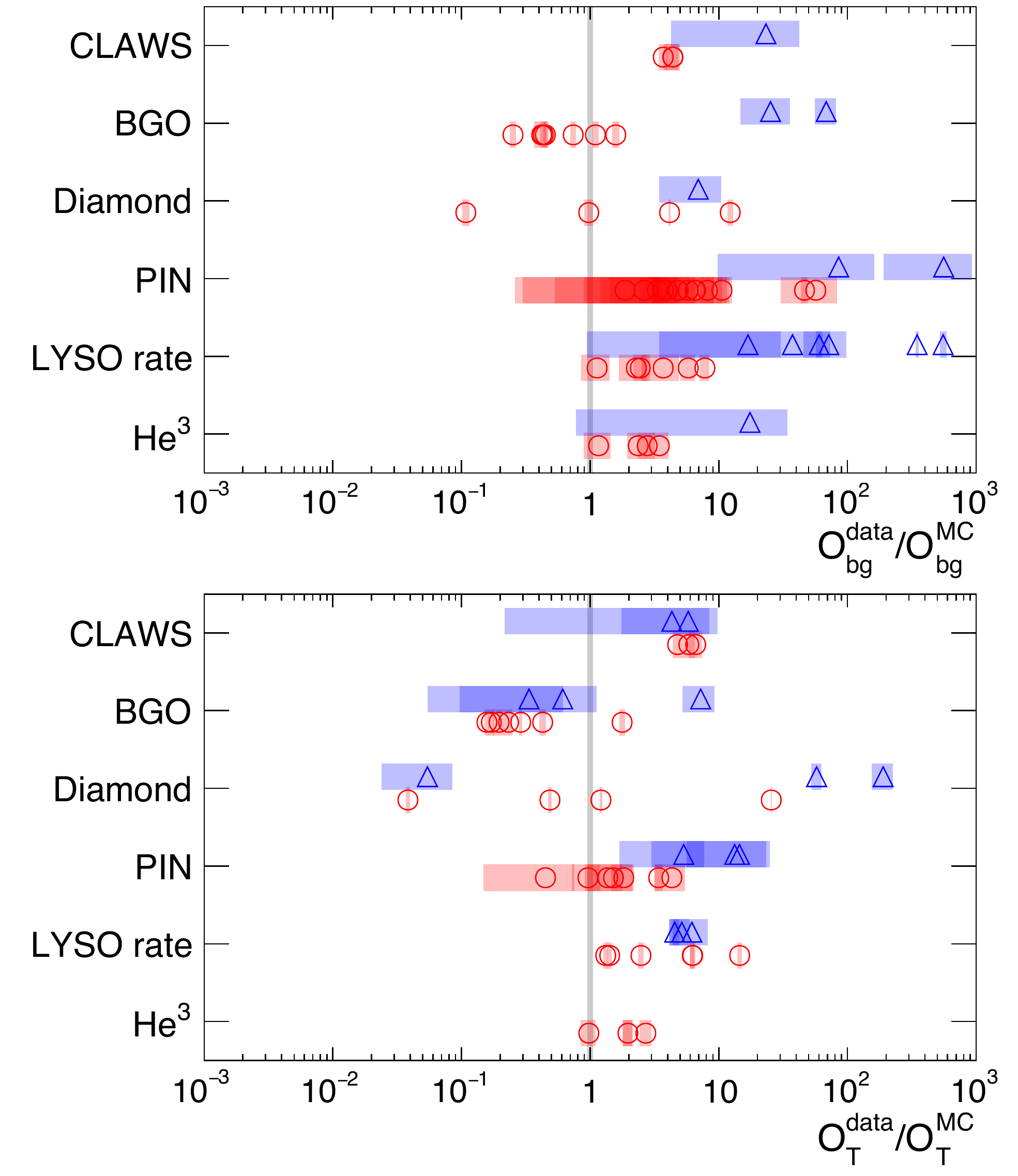}
	\caption{(color online) Experimental/simulated observable ratios derived from Touschek size-sweep runs as described in Sec.~\ref{sec:beamgas_touschek_data_mc_analysis}. Each point corresponds to a set of LER (red circles) or HER (blue triangles) size sweep runs for a single channel. Shaded bars indicate the errors on the ratio derived from the fit errors to the heuristic.}	
	\label{fig:beamgas_touschek_ratios}
\end{figure}

The primary aim of the Touschek size-sweeps and combined analysis is to compare experimentally derived observables with simulation based on the same beam conditions, namely $I^{SAD}=1.0$~A, $Z^{SAD}=7$, $\sigma_y^{SAD}=59$~\si{\micro}m (HER) and $110$~\si{\micro}m (LER) with $P^{SAD}=1.33\times 10^{-6}~$Pa, as described in Sec.~\ref{sec:beamgas_touschek_data_mc_analysis}. We obtain predictions for the values of the beam-gas and Touschek observables for the simulated conditions by inserting these values into the combined heuristic Eq.~\ref{eqn:combined_heuristic} with the sensitivities derived from the size-sweep runs. Uncertainties in the sensitivities originate from the parameter errors in the fits. For each channel, we calculate the experimental/simulated ratio, shown in Fig.~\ref{fig:beamgas_touschek_ratios}.

\paragraph{CCG selection results}
The selected most-predictive CCG for each channel for HER size sweeps may indicate the dominant scattering positions for beam-gas scattering for those channels. The results of this selection are summarized here for the channels most sensitive to HER Touschek backgrounds:
\begin{itemize}
\item All He$^3$ channels converge on a single CCG located 22~m upstream of the IP.
\item All sensitive BGO channels converge on CCGs between 4 and 25~m upstream.
\item All diamond channels converge on CCGs between 201 and 277~m upstream.
\item 9 of the 12 useable CsI/LYSO channels converge on CCGs between 193 and 228~m upstream.
\item Using near-upstream (4-25~m) CCGs for the CsI/LYSO channels and mid-upstream (193-277~m) CCGs for the He$^3$ and BGO channels yields poor fits. 
\end{itemize}

Although we cannot prove that the most predictive CCG occurs at the dominant scattering position, the patterns in best-CCG selection may hint at underlying patterns in the sensitivities of each system to various scattering processes. First, the He$^3$ and BGO detector systems may be most sensitive to bremsstrahlung, which SAD predicts to be the dominant process generating losses in the IR from near-upstream scattering. Second, the diamond and CsI/LYSO detectors may be most sensitive to Coulomb scattering, which can generate IR losses from scattering in the upstream arc section (roughly 250~m upstream) and beyond. These suggestions are at odds with the simulation predictions. For example, a plurality of Coulomb scattering leading to losses in the IR in simulation occurs in the D10 section, 727 to 1006~m upstream. No CCG from this section was selected as the most-predictive gauge by any channel. Similarly, although CCG selection suggests that all channels from a particular detector are sensitive to scattering from the same region upstream, simulation shows large variation in the relative contributions of different sections between channels in a single detector. Our conclusion is that CCG selection is weakly correlated, at best, with the scattering distribution. 

\paragraph{Angular distributions}
One potential explanation for the large variation in the level of agreement between experiment and simulation is that the angular distributions of backgrounds in simulation are wrong. However, we see no significant correlation between the ratio and sensor position. We conclude that errors in angular distributions in simulation cannot account for the large channel-to-channel variations in the experiment/simulation ratio.

\paragraph{Systematic uncertainties}
The error bars shown in Fig.~\ref{fig:beamgas_touschek_ratios} are derived from the uncertainties on the parameters from the fit combined with uncertainty due to the CCG choice procedure. To measure this effect, we fit the heuristic using all CCGs that have a normalized $\chi$-squared less than 1.5. The RMS of the beam-gas and Touschek parameters over all of these CCG choices is taken as the parameter uncertainty. For the HER, this uncertainty dominates over fit parameter uncertainty. 

In Fig.~\ref{fig:beamgas_touschek_ratios}, the detector-to-detector and channel-to-channel variations are large compared to the error bars. This must be due to additional unknown errors in the simulation, detector calibrations, and analysis methodology. A complete quantification of these factors is impractical, but by repeating this analysis on size sweeps performed on different days, we can test its reproducibility. To this end, we performed an analogous series of size sweeps one month after the series used in these results with a subset of the detectors. We found that the mean of each detector did not change significantly, while values for the individual channels did. These results suggest additional errors comparable to the variation among channels within each detector. However, we cannot conclude whether the remaining detector-to-detector differences are legitimate experiment/simulation disagreement or whether they are due to detector-level systematic biases present in both series. 

\paragraph{Combined results}
In order to determine the overall level of agreement between experiment and simulation, we combine results from all detectors and channels. The systematic uncertainties of Fig.~\ref{fig:beamgas_touschek_ratios} are incomplete and cannot be used to weight channels in a global average. Furthermore, the variation of the points is much larger than the single-channel uncertainty. Consequently we discard the uncertainties and calculate the unweighted mean of the common logarithm of the channel ratios. The uncertainty then is the standard error on the mean. Finally, we convert the logarithms back to simple ratios and obtain our combined ratios with asymmetric errors.

We obtain the following combined experiment/simulation ratios: 
\begin{itemize}
\item LER beam-gas: $2.8^{+3.4}_{-2.3}$,
\item LER Touschek: $1.4^{+1.8}_{-1.1}$,
\item HER beam-gas: $108^{+180}_{-64}$,
\item HER Touschek: $4.8^{+8.2}_{-2.8}$.
\end{itemize}

\paragraph{Beam-gas and Touschek discussion}
We have presented a method for disentangling beam-gas and Touschek backgrounds using size sweeps in order to probe simulation. For the LER beam-gas and Touschek components, we see an excess of less than 1-$\sigma$ in experiment compared to simulation, where $\sigma$ is the uncertainty in the combined ratio. For HER beam-gas we see a 1-to-2 order-of-magnitude excess with a significance of 1.7$\sigma$. For HER Touschek, we see a small excess with significance $1.4\sigma$. 

While the combined ratios constitute some evidence for an excess of backgrounds in experiment compared to simulation, this analysis is limited by very large systematic effects. Prominent among these is the highly local nature of gas conditions within the beam-pipe, confounding both experiment (due to limited instrumentation) and simulation (due to assumed uniform gas conditions). 

\label{sec:beamgas_touschek_direct_analysis}

\subsection{Constrained analysis}

Here we present an alternative simulation weighting scheme and analysis that we perform on most detector systems. The experimental data we use here is the same as in Section~\ref{sec:beamgas_touschek_direct_analysis}, as shown in Table~\ref{tab:touschek_runs}. As mentioned in Section~\ref{sec:superkekb_conditions_pressures}, there is a scale factor of approximately three between the measured and actual pressures. The consistency and accuracy of this factor is unknown. Additionally, while there was information on the gas mixture in the LER beampipe (see Section~\ref{sec:Zeff}), there was no information on the HER beampipe's gas mixture, as there are no RGAs there. The goal of this analysis is to additionally account for the difference between measured and actual pressure and, for the HER, account for the gas mixture in the beampipe. We achieve this by forcing the ratio of the Touschek observable to the beam-gas observable in simulation to be the same as in the experimental data. We present this as a complementary analysis to that presented in Section~\ref{sec:beamgas_touschek_direct_analysis}.

\subsubsection{Simulation scaling}
\label{sec:samsim}

\begin{figure}[htb]
	\includegraphics[width=\columnwidth]{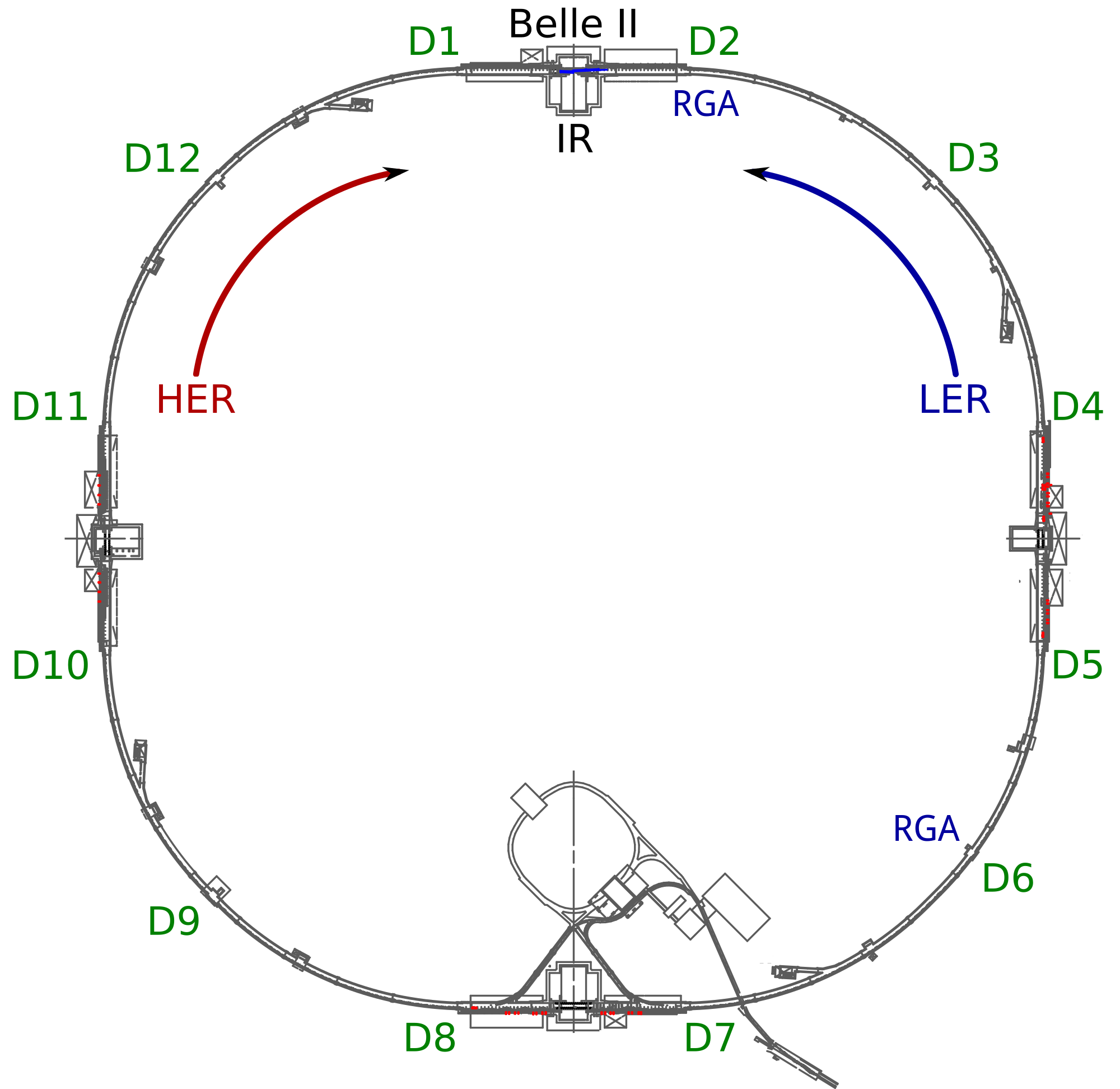}
	\caption{SuperKEKB ring showing `D' sections~\cite{SKBgroup}.}
	\label{fig:SKB}
\end{figure}

We simulate the beam-gas and Touschek components of the background separately for fixed beam parameters, as seen in Section~\ref{simulation_scaling}. We then scale each component to match the measured beam parameters by re-weighting the events produced with SAD and \gF by the scale value for that moment in time. We use the same simulated events for all beam settings.

We scale the simulated (``sim'') beam-gas observable using the ratio of the experimental (``exp'') to SAD beam-gas factors $IPZ^2$:
\begin{equation}
	{\mathcal{O}_{bg}^{\mathrm{scaled\:sim}} = \sum _{i=1}^{12}(\mathcal{O}^{\mathrm{B}}_i+\mathcal{O}^{\mathrm{C}}_i)^{\mathrm{sim}}\cdot\frac{C_{\mathrm{scale}}(I P_i Z_{e,~i}^{2})^{\mathrm{exp}}} {(I P Z^{2})^{\mathrm{SAD}}}},
\label{eq:simScale}
\end{equation}
where $\mathcal{O}^{\mathrm{B}}_{i}$ and $\mathcal{O}^{\mathrm{C}}_{i}$ are the components of the inelastic and elastic beam-gas simulated observables produced in each detector from interaction at the $i$th section of the HER and LER rings (see Fig. \ref{fig:SKB}), $P_{i}$ is the average pressure in each `D' section, and $Z_{e,~i}$ is the effective atomic number of the gas in each section (see Eq.~\ref{eqn:Ze}), if available. If $Z_{e,~i}$ is not available for a section, we use a value of 2.7 for the LER (since this was near the mean value of $Z_{e}$ during the experiment, see Section~\ref{sec:TousExp}). $C_{\mathrm{scale}}$ is a scale factor which absorbs any scale factor difference between experimental and simulated data. This scale factor will be discussed further in Section~\ref{sec:TousExp}.

The scaled Touschek component of the simulation is given by:
\begin{equation}
	\mathcal{O}_{T}^{\mathrm{scaled\:sim}} = \mathcal{O}^{\mathrm{sim}}_{T}\cdot\frac{\left(\frac{I^2 N_{\mathrm{bunch}}}{\sigma_{y}}\right)^{\mathrm{exp}}}{\left(\frac{I^2 N_{\mathrm{bunch}}}{\sigma_{y}}\right)^{\mathrm{sim}}},
\end{equation}
where $\mathcal{O}^{\mathrm{sim}}_{T}$ is the Touschek simulated observable, $\sigma_{y}$ is the vertical beam size, and $N_{\mathrm{bunch}}$ is the number of filled bunches in the ring.

These scaled components are combined to obtain the predicted observable in each detector,
\begin{equation}
	\mathcal{O}^{\mathrm{scaled\:sim}} = \mathcal{O}_{bg}^{\mathrm{scaled\:sim}}+\mathcal{O}_{T}^{\mathrm{scaled\:sim}}.
\end{equation}

\subsubsection{Analysis procedure}
\label{sec:TousExp}

\begin{figure}
	\centering
		\includegraphics[width=\columnwidth]{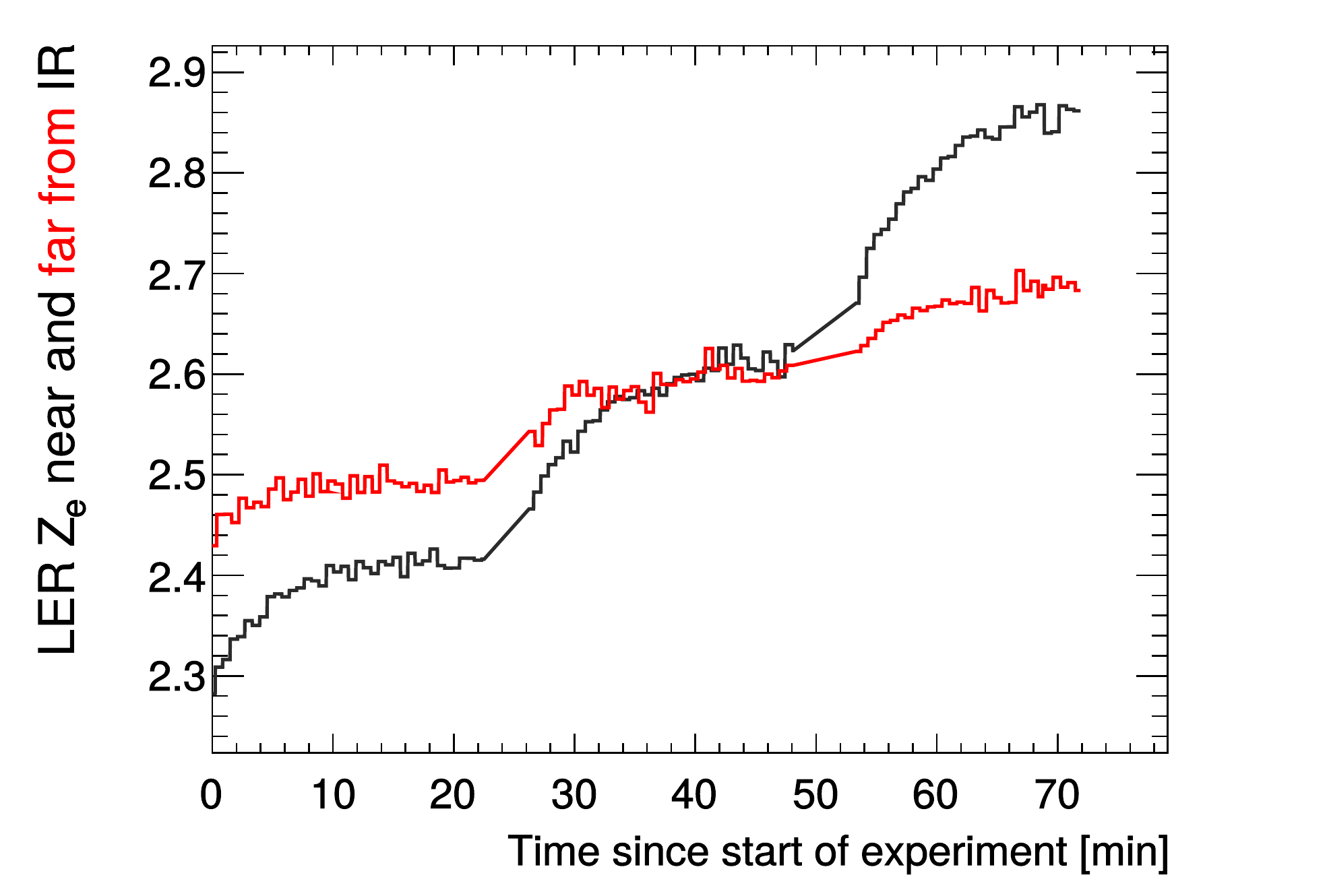}
	\caption[$Z_{e}$ during LER beam size runs]{$Z_{e}$ during LER beam size runs examined in this analysis, showing that $Z_{e}$ evolves over the course of the experiment. Data were recorded on May 17, 2016.}	
	\label{fig:tousZef}
\end{figure}

	In order to separate the beam-gas and Touschek components of the detector observable during the size sweeps, we fit both the measured and simulated observable to Eq.~\ref{eqn:combined_heuristic}, restated here:
\begin{equation}
	{\mathcal{O}^{\mathrm{exp}} = S_{bg}\cdot IP Z_{e}^{2}+S_{T}\cdot \frac{I^{2}}{\sigma_{y}}},
	\label{eqn:TousFit}
\end{equation}
where $P$ is the average pressure in the entire beampipe and $Z_{e}$ is the effective atomic number of the beampipe gas taken from the RGA closest to the IR. Fig.~\ref{fig:tousZef} demonstrates how $Z_{e}$ changed during the size sweeps, showing the importance of including it in the fit. For the LER, the value of $Z_{e}$ measured for D02 is used since it is closest to the IR. This technique is described in detail in Section~\ref{sec:Zeff}.

For each subrun, the average of the detector observable, we calculate $IP Z_{e}^{2}$, and $\frac{I^{2}}{\sigma_{y}}$. These are the input parameters of the fit in Eq.~\ref{eqn:TousFit}.

In order to remove channels with little to no Touschek contribution, we ignore any channel where $S_T/\sigma_{S_T}<1$.

\begin{figure}
	\centering
	\subfigure[Fitting Eq.~\ref{eqn:TousFit} to data.]{
		\begin{overpic}[width=\columnwidth]{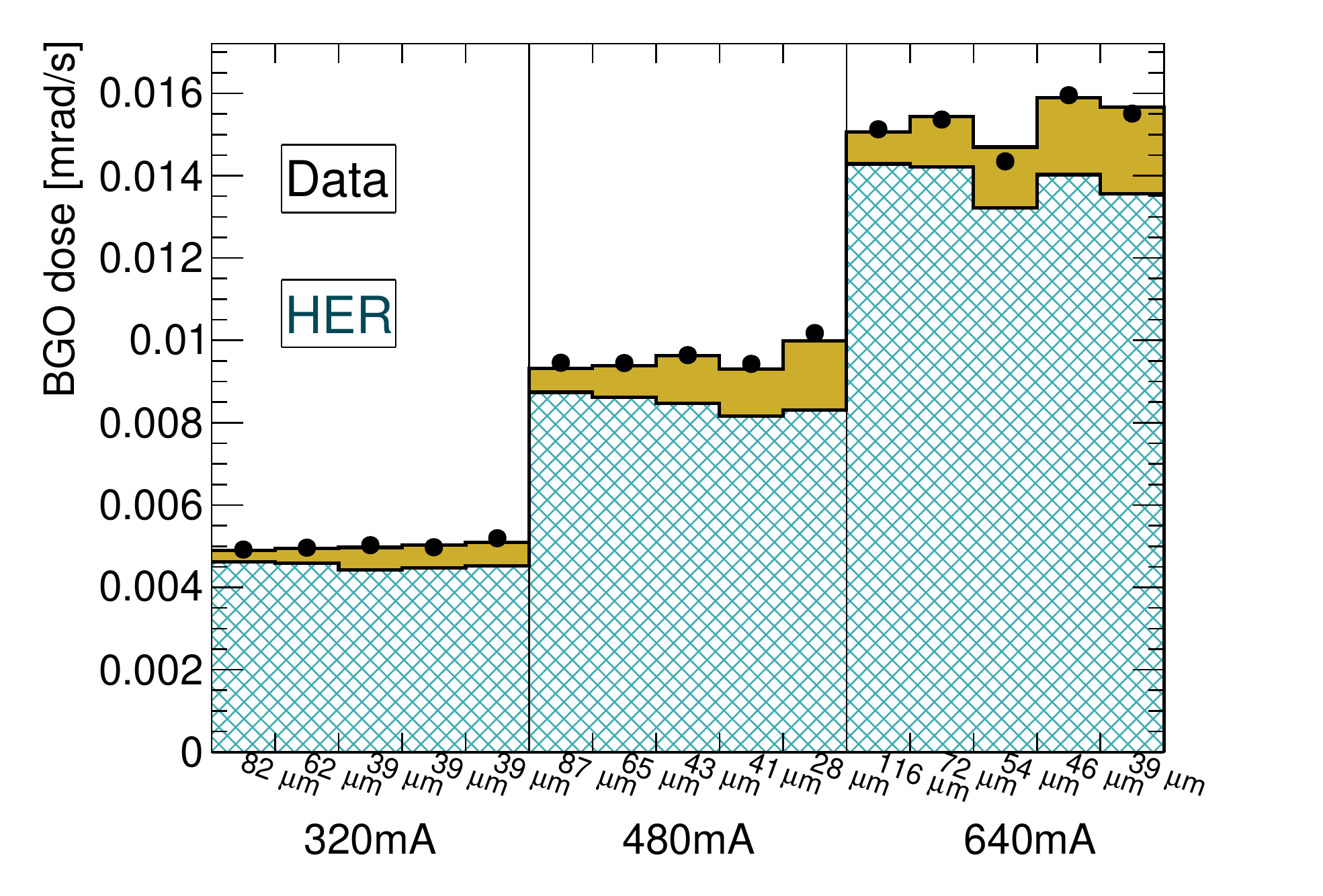}
		\end{overpic}
		\label{fig:HERTousData10}
	}
	\subfigure[Fitting Eq.~\ref{eqn:TousFit} to simulation. Note the major difference between the  Touschek and beam-gas fit contributions in data and simulation.]{
		\begin{overpic}[width=\columnwidth]{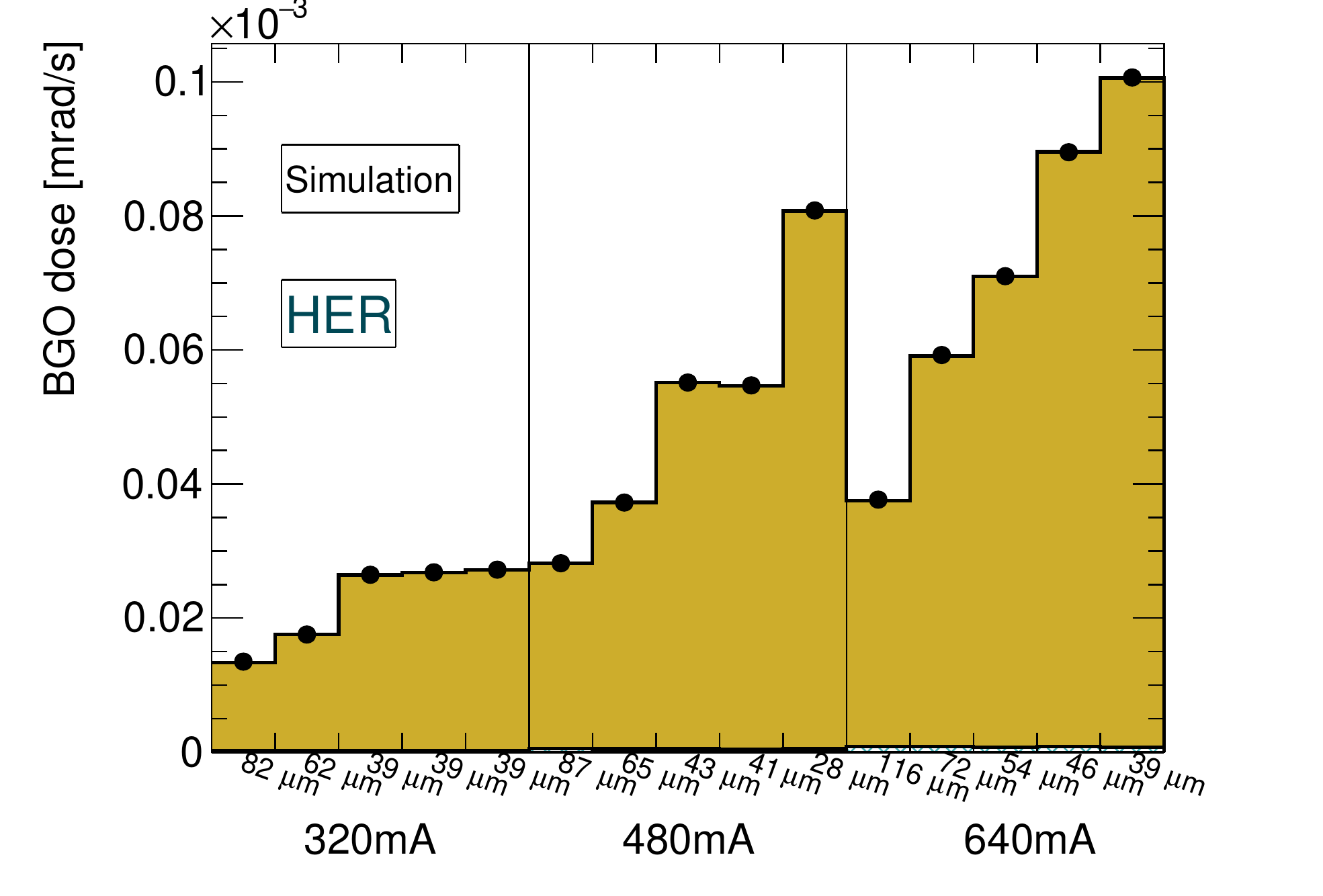}
		\end{overpic}
		\label{fig:HERTousSim10}
	}
	\subfigure[Fitting Eq.~\ref{eqn:TousFit} to simulation after reweighting by $C_{\mathrm{scale}}Z_{e}^{2}$. Note that the Touschek and beam-gas contributions in simulation match the data better than without the reweighting.]{
		\begin{overpic}[width=\columnwidth]{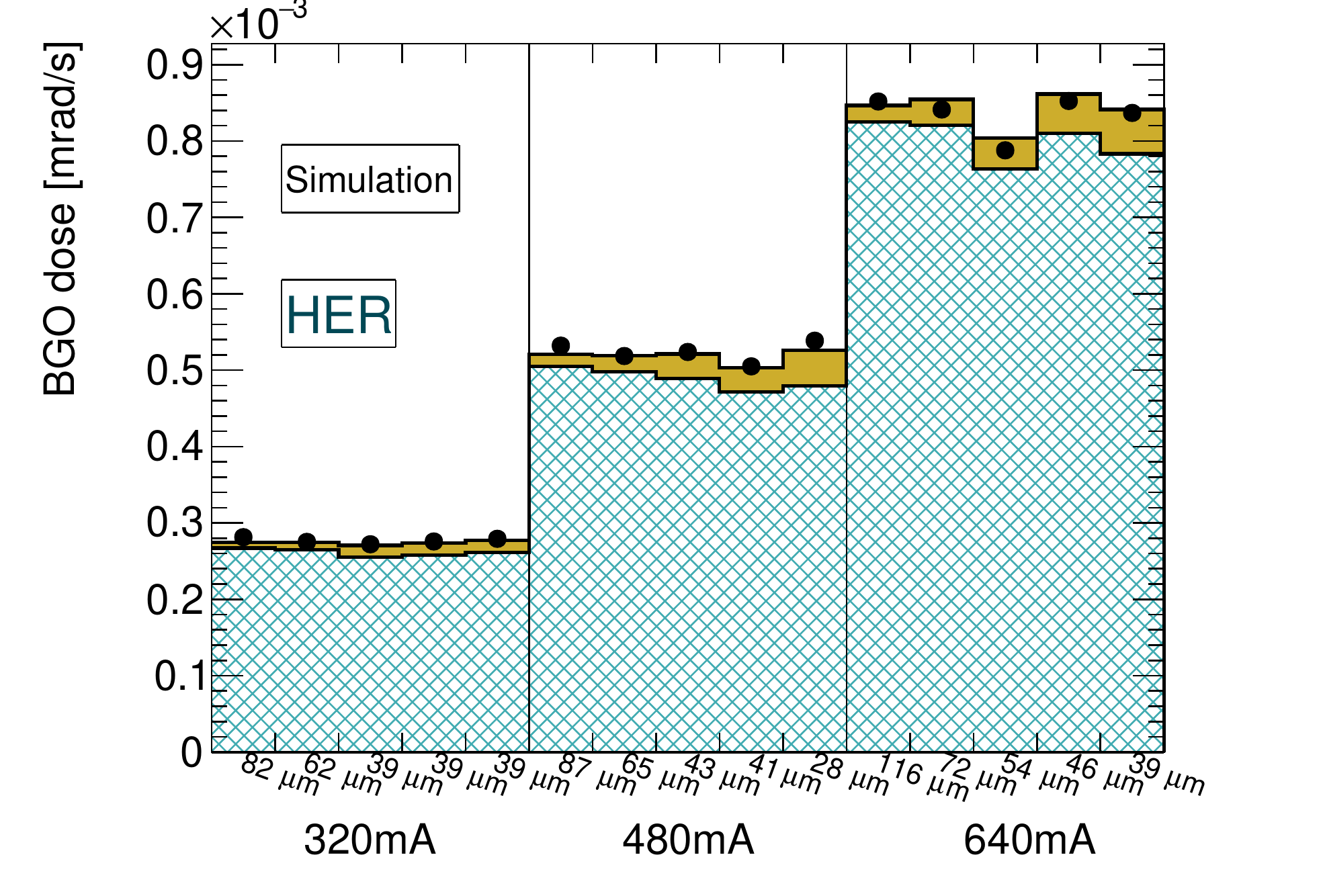}
		\end{overpic}
		\label{fig:HERTousSim20}
	}
	\caption{(color online) Result of fitting Eq.~\ref{eqn:TousFit} to the BGO dose in channel 4 during HER Touschek experiments. Each bin corresponds to one sub run, where the black point is the mean of the measured BGO dose in that subrun, solid yellow is the Touschek component of the fit, as given by Eq.~\ref{eqn:TousGraph}, and blue (crosshatched) is the beam-gas component of the fit, as given in Eq.~\ref{eqn:BGGraph}. A similar, but less significant effect is observed in the LER simulation.}
	\label{fig:HERTous10}
\end{figure}

The results of this fit for experiment and simulation for the HER (in BGO channel 4) are shown in Fig.~\ref{fig:HERTous10}. The experiment is divided into three runs, each with five subruns. The runs had different beam currents, and the subruns each had different vertical beam sizes. In these figures, the black points are the average measured detector observables in each subrun. The Touschek component, in solid yellow, is given by:
\begin{equation}
	\mathcal{O}_{T} = S_{T}\cdot \frac{I^{2}}{\sigma_{y}},
	\label{eqn:TousGraph}
\end{equation}
and the beam-gas component, in blue (crosshatched), is given by:
\begin{equation}
	\mathcal{O}_{bg} = S_{bg}\cdot IP Z_{e}^{2}.
	\label{eqn:BGGraph}
\end{equation}
The error bars shown are the standard deviation of the detector observable for that subrun. The pressure in the beampipe changes over the course of the experiments and is included in the fit through the $S_{bg}\cdot IP Z_{e}^{2}$ component, but for simplicity the pressure changes during the subruns are not shown explicitly in the figures. 

%-----------------------------------------------------------------------------------------------

As evident from Fig.~\ref{fig:HERTousSim10}, the simulation greatly underestimates the beam-gas component in the HER. This is due to two factors: there is no estimate of $Z_{e}$ in the HER, and $C_{\mathrm{scale}}$ is not estimated for either beam. Using a value of 1 for these parameters does not give the correct beam-gas component. To compensate, we ensure that the ratio of $\mathcal{O}_{bg}$ to $\mathcal{O}_{T}$ is the same for experiment and MC, by using a scale factor:
\begin{equation}
	{C_{\mathrm{scale}}\cdot\left(\frac{\mathcal{O}_{bg}}{\mathcal{O}_{T}}\right)^{\mathrm{sim}}=\left(\frac{\mathcal{O}_{bg}}{\mathcal{O}_{T}}\right)^{\mathrm{exp}}}.
\label{eqn:Pscale2}
\end{equation}
Solving for $C_{\mathrm{scale}}$:
\begin{equation}
{C_{\mathrm{scale}}=\frac{ \mathcal{O}_{bg}^{\mathrm{exp}}/ \mathcal{O}_{bg}^{\mathrm{sim}}}{\mathcal{O}_{T}^{\mathrm{exp}}/ \mathcal{O}_{T}^{\mathrm{sim}}}}.
\end{equation}
Substituting in Eqs.~\ref{eqn:TousGraph} and \ref{eqn:BGGraph} yields:
\begin{equation}
{C_{\mathrm{scale}}=\frac{ S_{bg}^{\mathrm{exp}}/ S_{bg}^{\mathrm{sim}}}{S_{T}^{\mathrm{exp}}/ S_{T}^{\mathrm{sim}}}}.
\end{equation}
Which can be rearranged to:
\begin{equation}
	{C_{\mathrm{scale}} = \frac{(S_{bg}/S_{T})^{\mathrm{exp}}}{(S_{bg}/S_{T})^{\mathrm{sim}}}}.
\label{eqn:Pscale1}
\end{equation}
The motivation for this is that all the parameters of the Touschek component of the fit are relatively well known, while the pressure and $Z_{e}$ in the beam-gas component are not estimated. We use the beam-gas to Touschek ratio in the experiment to determine $C_{\mathrm{scale}}Z_{e}^{2}$ to be used in the simulation weighting. Note that this does not constrain the overall total prediction of the background from the simulation or the absolute individual contributions.

\begin{figure}
	\centering
	\subfigure{
		\begin{overpic}[width=\columnwidth]{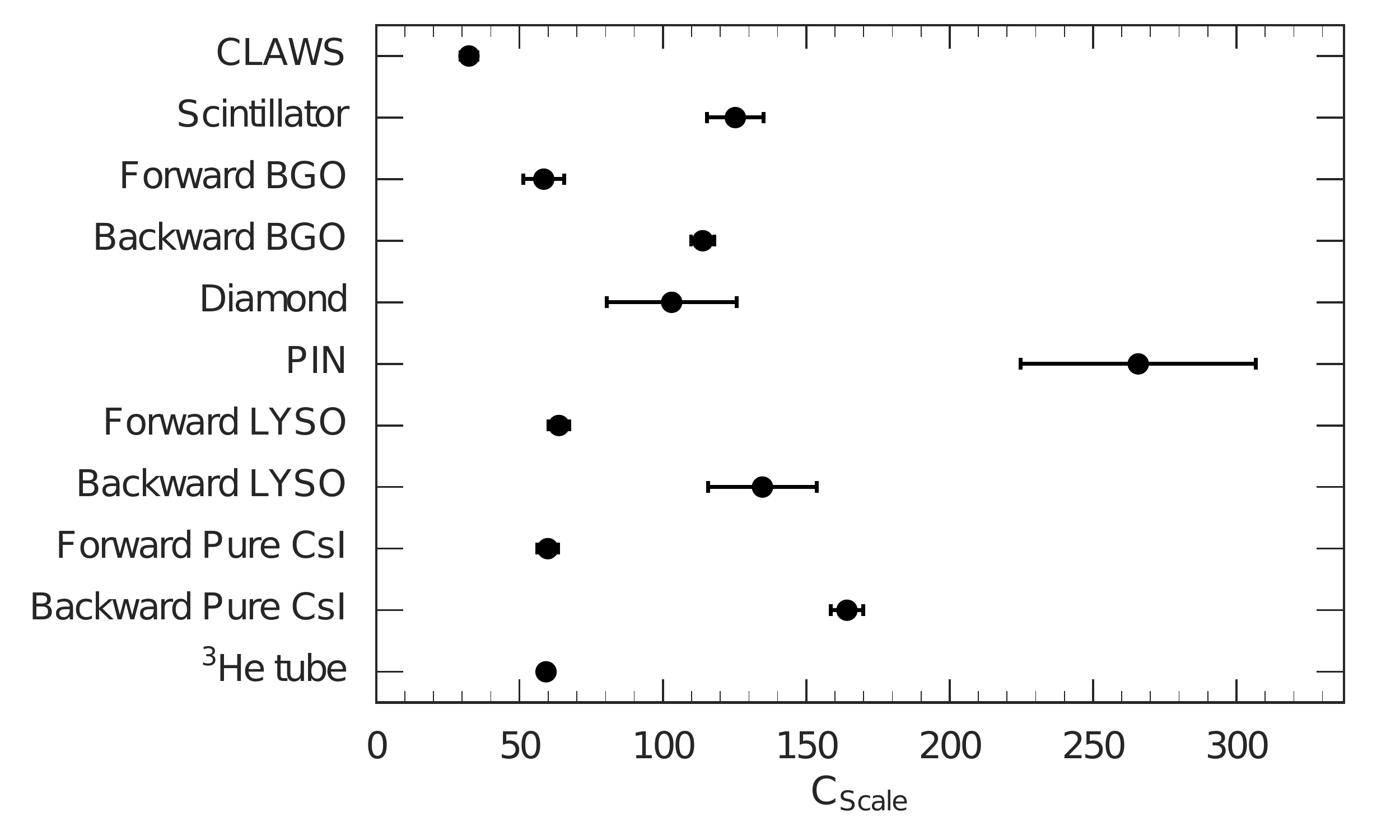}
		\end{overpic}
		\label{fig:LERPScale}
	}
	\subfigure{
		\begin{overpic}[width=\columnwidth]{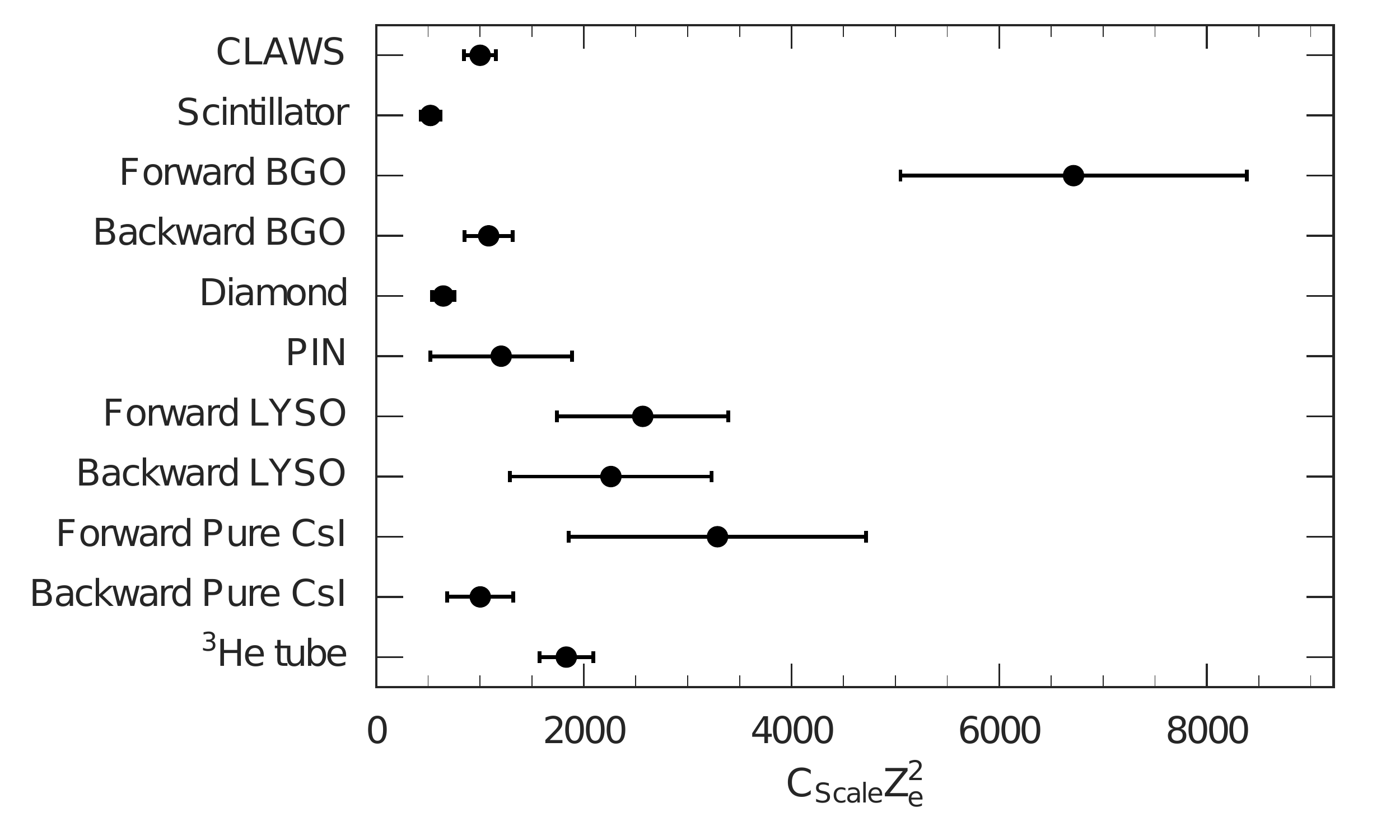}
		\end{overpic}
		\label{fig:HERPScale}
	}
	\caption{Measured values of $C_{\mathrm{scale}}$ for the LER (top) and $C_{\mathrm{scale}}Z_{e}^{2}$ for the HER (bottom). There is some agreement between the values for $C_{\mathrm{scale}}$ and $C_{\mathrm{scale}}Z_{e}^{2}$ for some detectors, and a large disagreement for others.}	
	\label{fig:PScaleZsq}
\end{figure}

Fig.~\ref{fig:PScaleZsq} shows the value of $C_{\mathrm{scale}}Z_{e}^{2}$ for several BEAST II subdetectors. Note that the values are significantly different between the LER and HER. This is because there is an estimate of $Z_{e}$ for the LER, but not the HER. %In the LER, there is good agreement between the $^3$He tubes and the forward CsI, LYSO, and BGO detectors. The backwards CsI, LYSO, and BGO are relatively close to one another, but not in agreement. There is much less agreement between the values of $C_{\mathrm{scale}}Z_{e}^{2}$ for the different detectors.

We repeated the simulation for each detector using its $C_{\mathrm{scale}}Z_{e}^{2}$ value. The results of the fit for the HER (in BGO channel 4) are shown in Fig.~\ref{fig:HERTousSim20}.

\subsubsection{Comparing data to simulation}

In order to verify the overall accuracy of the simulation, we define a ratio of the values from experiment to reweighted simulation: 
\begin{subequations}
\begin{align}
		\frac{\mathcal{O}^{\mathrm{exp}}}{\mathcal{O}^{\mathrm{scaled\:sim}}} = \frac{\sum\limits_{i=0}^{n_{\mathrm{subruns}}} \mathcal{O}^{\mathrm{exp}}_{i}}{\sum\limits_{i=0}^{n_{\mathrm{subruns}}}   \mathcal{O}_{i}^{\mathrm{scaled\:sim}}},
\end{align}
\end{subequations}
where $\mathcal{O}_{i}$ is the average observable in the $i$th subrun.

The more accurate the simulation, the closer these ratios will be to 1.  The ratios show how much of the experiment/simulation discrepancy is not accounted for by a constant scale for beam-gas.

\paragraph{Estimating uncertainty}

\begin{table}[htb]
	\caption{Uncertanties on the beam current and the beam size in each ring~\cite{samsThesis}.}
	\centering
	\begin{tabular}{ lll }
\toprule
	&	LER	&	HER	\\	
\midrule
I [mA]	&	0.03	&	0.03	\\		
$\Delta \sigma_{y}/\sigma_{y}$	&	1.37\%	&	3.71\%	\\			
\bottomrule

	\end{tabular}
	\label{tab:paramUncert}
\end{table}

To estimate the systematic uncertainty from the beam current, beam size and $C_{\mathrm{scale}}Z_{e}^{2}$, we re-weight the simulation with each quantity separately adjusted by +1$\sigma$ and -1$\sigma$, where $\sigma$ is the uncertainty on each quantity. We then redo the full analysis with these modified quantities. For example:
\begin{subequations}
\begin{align}
		{I\rightarrow I+0.03\mathrm{~mA}}, \\
		{I\rightarrow I-0.03\mathrm{~mA}},
\end{align}
\end{subequations}
and similarly for $\sigma_{y}$ and $C_{\mathrm{scale}}$Z$^{2}$ (for more details, see~\cite{samsThesis}). The uncertanity on the current and beam size can be found in Table~\ref{tab:paramUncert}. For the uncertanties on $C_{\mathrm{scale}}Z_{e}^{2}$, see Fig.~\ref{fig:PScaleZsq}. We repeat the analysis described in Section~\ref{sec:TousExp} to get new values of ($\mathcal{O}^{\mathrm{exp}}/\mathcal{O}^{\mathrm{scaled\:sim}}$). We used these to calculate the systematic error for each quantity:
\begin{subequations}
\begin{align}
	{\sigma_{(\mathcal{O}^{\mathrm{exp}}/\mathcal{O}^{\mathrm{scaled\:sim}})_{+}}^{I} = \left. \frac{\mathcal{O}^{\mathrm{exp}}}{\mathcal{O}^{\mathrm{scaled\:sim}}}\right|_{I=I+0.03\mathrm{~mA}} - \left. \frac{\mathcal{O}^{\mathrm{exp}}}{\mathcal{O}^{\mathrm{scaled\:sim}}}\right|_{\mathrm{nominal}}}, \\
	{\sigma_{(\mathcal{O}^{\mathrm{exp}}/\mathcal{O}^{\mathrm{scaled\:sim}})_{-}}^{I} = \left. \frac{\mathcal{O}^{\mathrm{exp}}}{\mathcal{O}^{\mathrm{scaled\:sim}}}\right|_{\mathrm{nominal}} - \left. \frac{\mathcal{O}^{\mathrm{exp}}}{\mathcal{O}^{\mathrm{scaled\:sim}}}\right|_{I=I-0.03\mathrm{~mA}}}.
\end{align}
\end{subequations}

The uncertainty contribution from each quantity, the statistical uncertainty, and the fitting uncertainty are added in quadrature to get the total uncertainty.

%------------------------------------------------------------------------------------------------

\subsubsection{Results}
\label{thermal_neutrons_results}

\begin{figure}[htb]
	\includegraphics[width=\columnwidth]{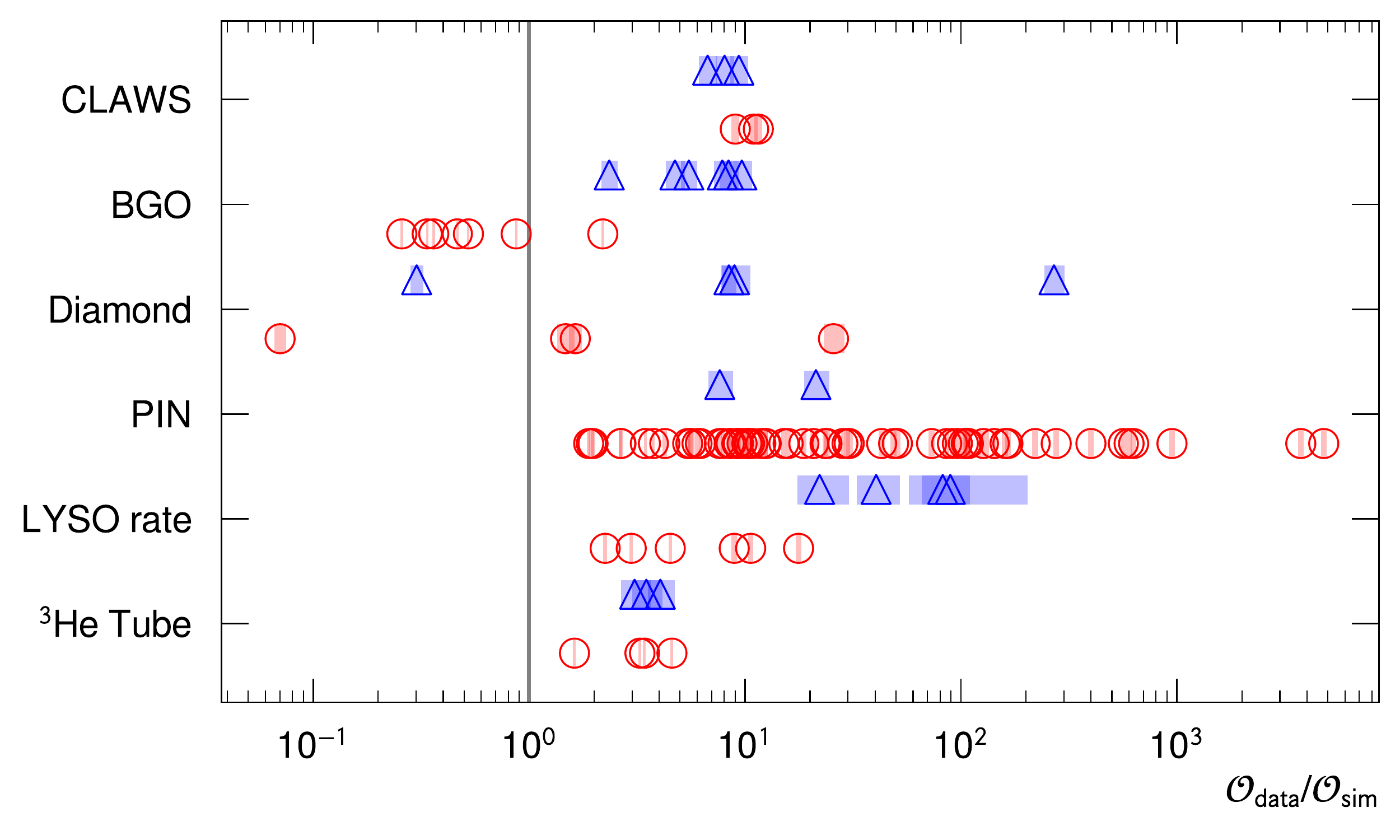}
	\caption{(color online) LER and HER experiment/simulation ratios for some BEAST II detectors. Each point corresponds to a set of LER (red circles) or HER (blue triangles) size sweep runs for a single channel. Shaded bars indicate the errors as described in Sec.~\ref{sec:TousExp}. }	
	\label{fig:RatioWithErrors}
\end{figure}

In Fig.~\ref{fig:PScaleZsq}, the values of $C_{\mathrm{scale}}Z_{e}^{2}$ for different detector systems are shown. In Fig.~\ref{fig:PScaleZsq} (top/LER), there is good agreement between the forward BGO, forward LYSO, forward pure CsI, and the \heT tubes, with CLAWS close, but not in agreement. The other detectors show large spread. The uncertainty on the scale factor for the Diamond and PIN systems is significantly larger than for the other systems. The scale factors presented in Fig.~\ref{fig:LERPScale} (bottom/HER) show much larger uncertanties than for the LER. The detector systems which had similar $C_{\mathrm{scale}}$ values for the LER do not show agreement for the HER.

Fig.~\ref{fig:RatioWithErrors} shows $\mathcal{O}^{\mathrm{exp}}/\mathcal{O}^{\mathrm{scaled\:sim}}$ for both LER and HER. Most detectors are within two orders of magnitude of agreement with unity, which shows a good agreement between experiment and simulation. There is a large spread in $\mathcal{O}^{\mathrm{exp}}/\mathcal{O}^{\mathrm{scaled\:sim}}$ for most of the detector systems, which is shows how the positional distribution of the particle flux differs between experiment and simulation.

\label{sec:beamgas_touschek_constrained_analysis}

\subsection{Effect of Collimators} % Hiro
% file: collimators.tex
% lead author: Hiro Nakayama

\begin{figure}
        \centering
                \includegraphics[width=\columnwidth]{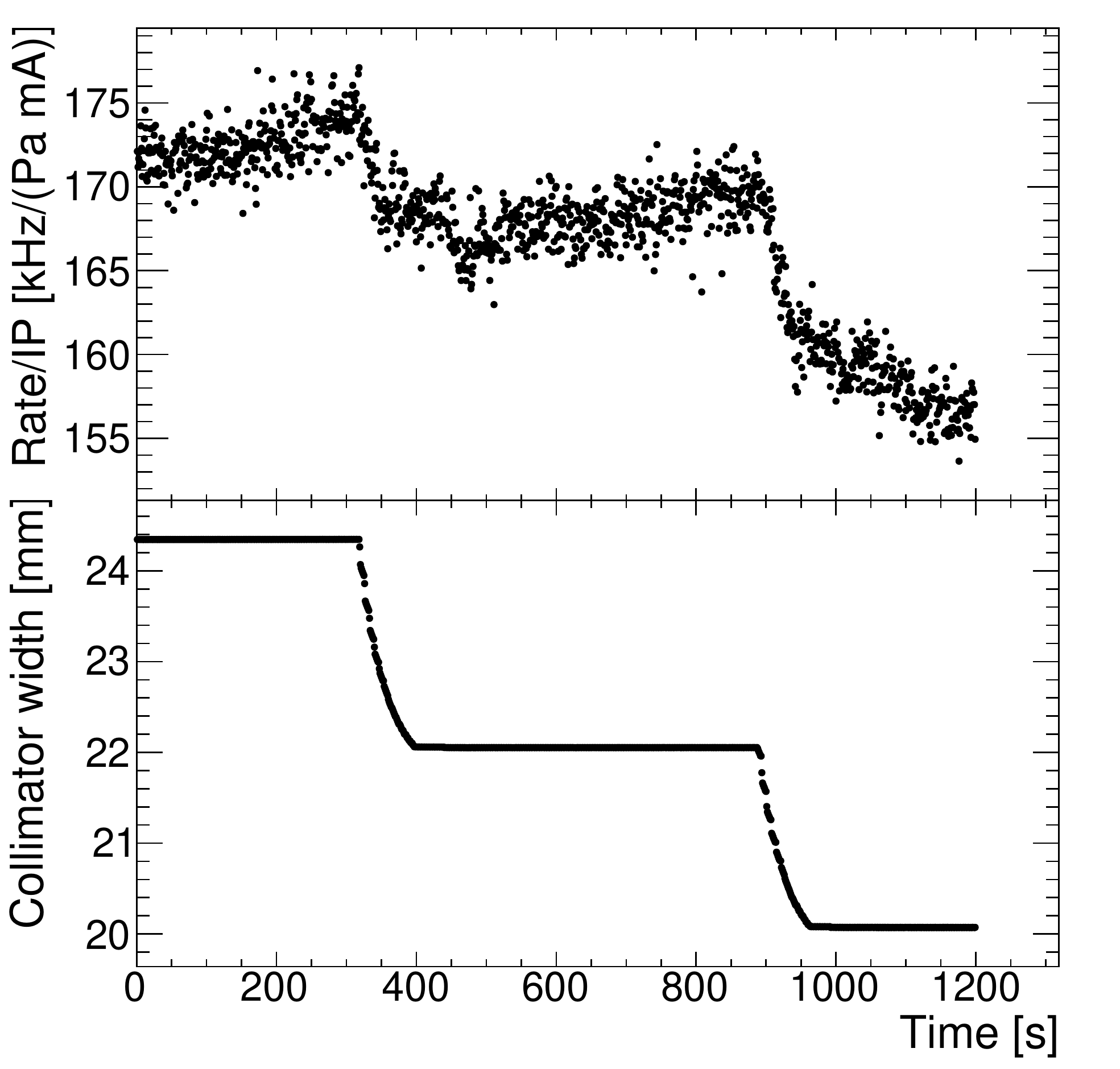}
                \caption{Results of a collimator study using the beam-gas-normalized rate from a pure CsI crystal as the observable and monitoring its response to adjustments of the position of the D06H3OUT collimator. This demonstrates that the D06H3OUT collimator effectively mitigates IR loss.}
        \label{fig:colstudy}
\end{figure}

During the collimator study machine time, we varied collimator widths to
observe how BEAST II background rates and beam lifetime changed.
Fig.~\ref{fig:colstudy} shows the result of run 4005,
where we changed the width of a collimator called `D06H3OUT'. 
As we changed the D06H3OUT width from 22~mm to 17~mm,
we observed the BEAST II CsI hit rate was slightly reduced. 
When the collimator was further narrowed to 16mm, we observed that the beam lifetime started to decrease,
and we had to stop narrowing it to avoid a beam abort.
We also changed the width of the other collimators
but did not observe any mitigation of BEAST II CsI hit rate.  

\begin{figure}
        \centering
                \includegraphics[width=\columnwidth]{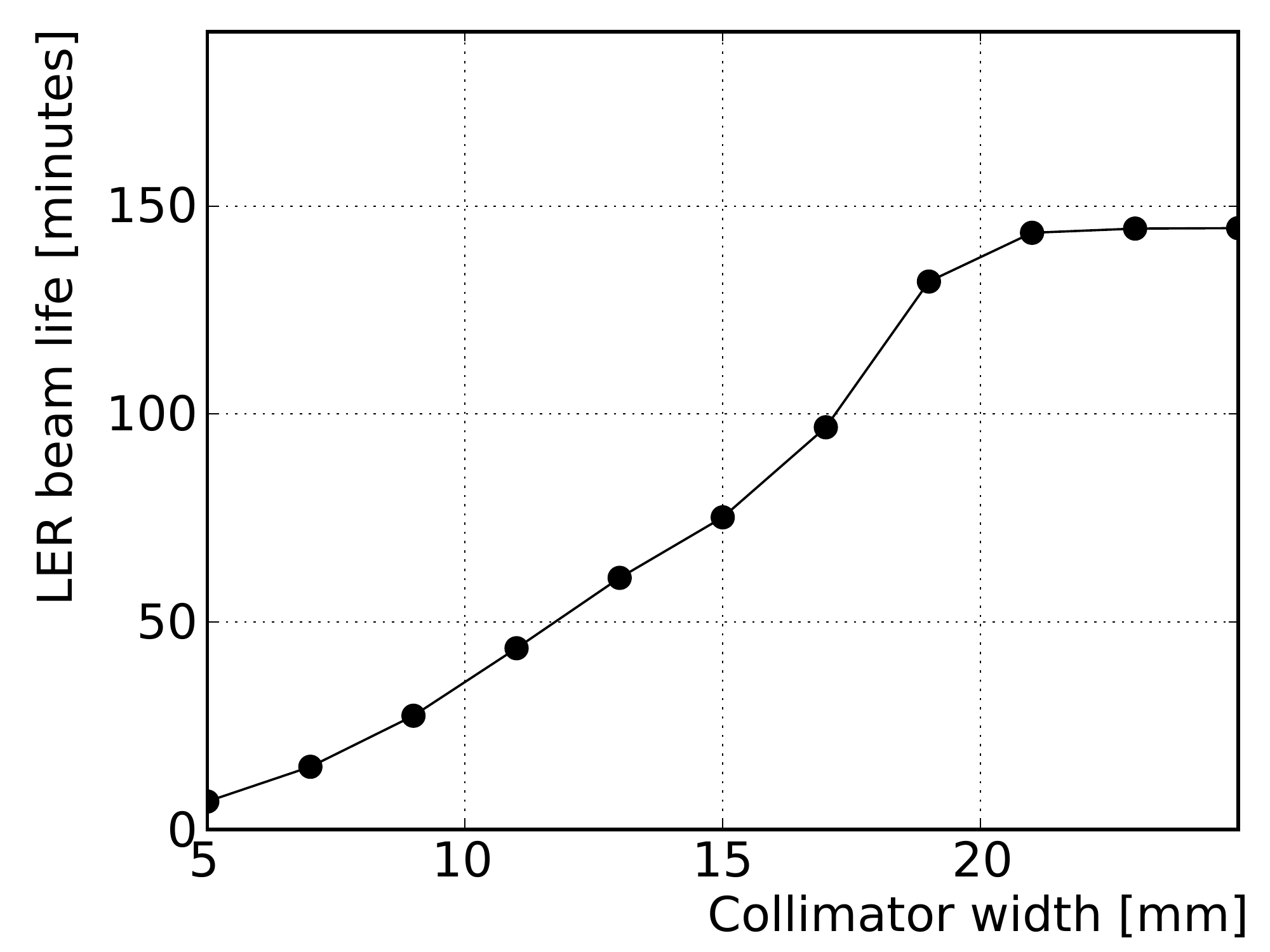}
                \caption{LER beam lifetime vs. D06H3OUT width, simulated by SAD.}
        \label{fig:colsimlife}
\end{figure}

\begin{figure}
        \centering
                \includegraphics[width=\columnwidth]{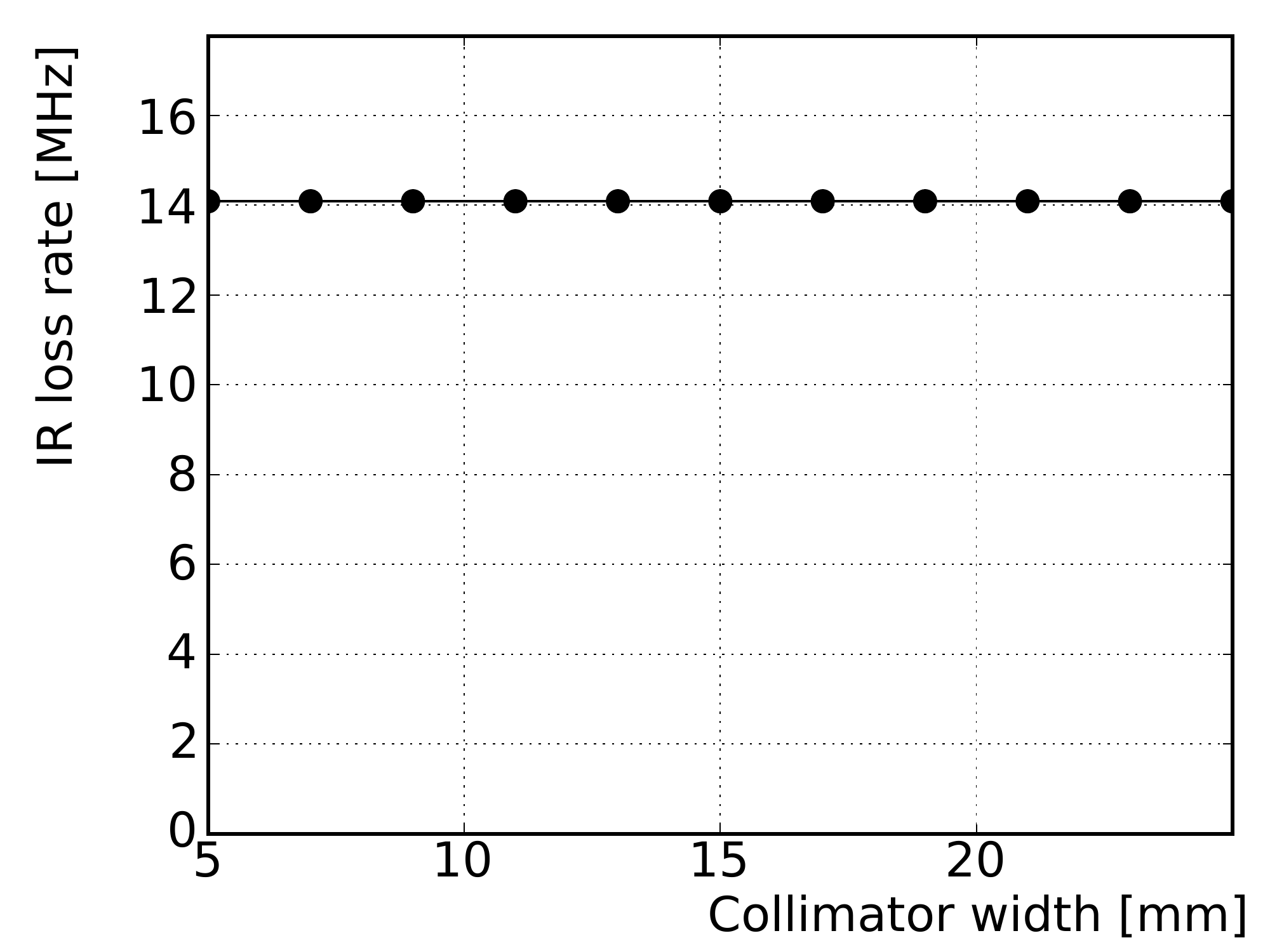}
                \caption{LER IR loss rate vs. D06H3OUT width, simulated by SAD.}
        \label{fig:colsimir}
\end{figure}

We run the SAD simulation with various collimator widths,
to check if the simulation  can reproduce the collimator study measurement.
Fig.~\ref{fig:colsimlife} and Fig.~\ref{fig:colsimir} show the effect of varying the D06H3OUT width in the SAD simulation on the LER beam lifetime and LER IR loss rate, respectively. The simulated lifetime decreased with decreasing collimator width as we observe in the machine study.

 \clearpage

 % lead author: Hiro and Igal
 \section{Beam-gas and Touschek loss rates from SuperKEKB lifetime}
 % File: lifetime_analysis.tex
% Lead author: Igal Jaegle

We test the validity of our SAD simulation by comparing the simulated beam lifetimes to those measured by SuperKEKB\cite{lifetime}. 

\subsection{Experimental loss rates}
\begin{figure}[ht!]
\centering
\includegraphics[width=1.0\columnwidth]{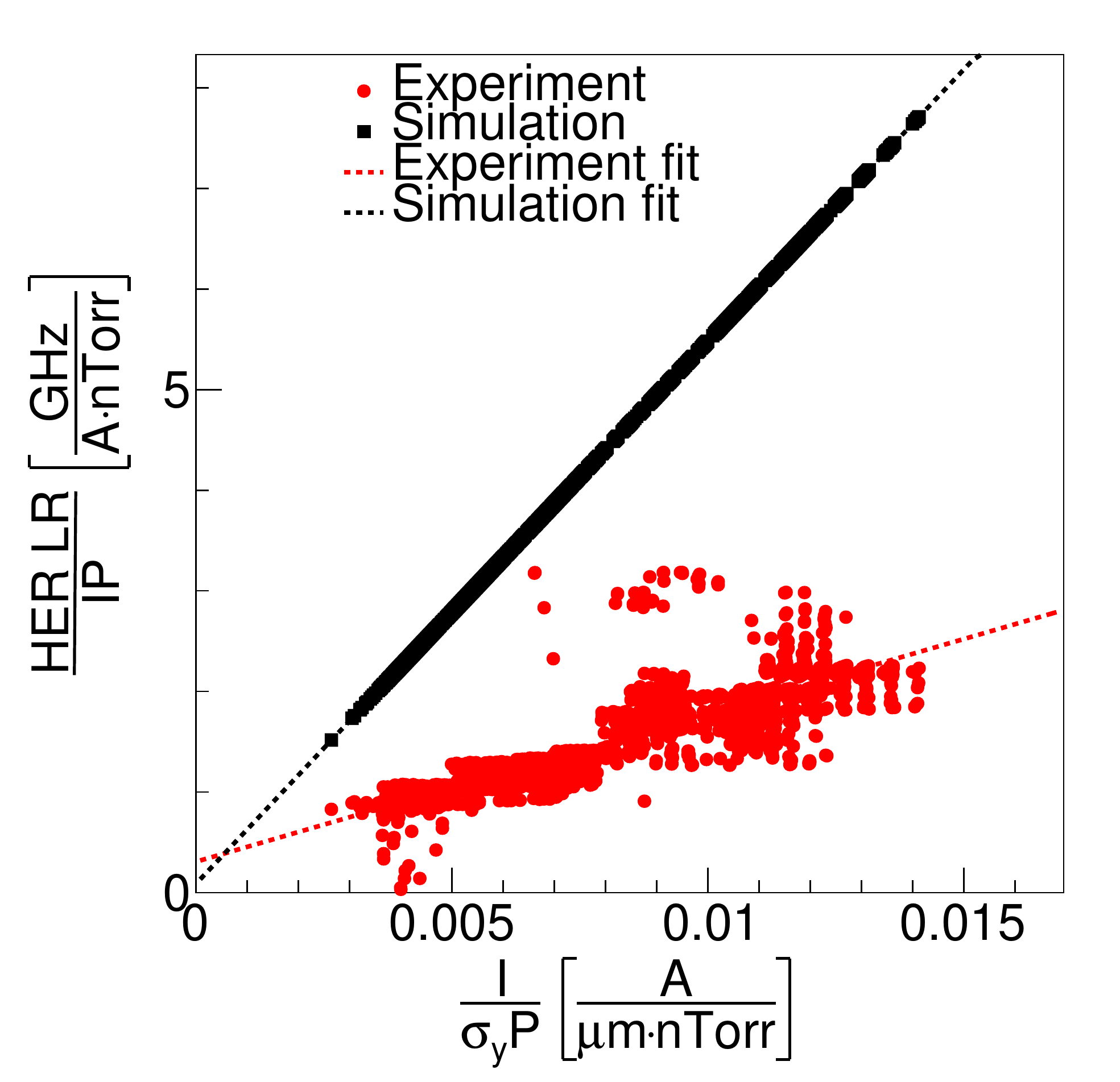}
\caption{(color online) Fit of the heuristic model for the first HER Touschek size sweep series with experimental (red) and simulated (black) data.}
\label{fig:HERLossRateRun2000sSummary}
\end{figure}

\begin{figure}[ht!]
\centering
\includegraphics[width=1.0\columnwidth]{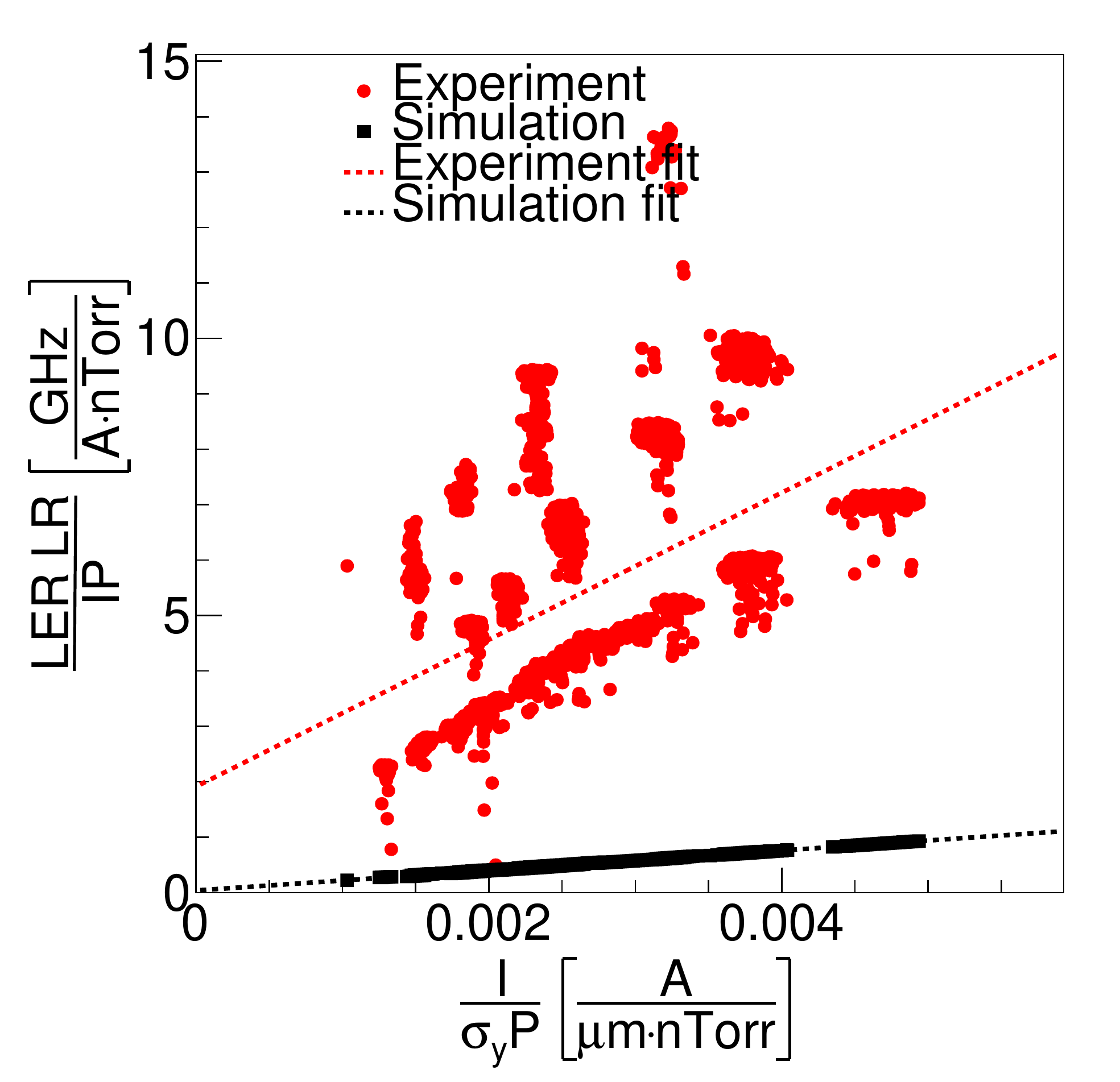}
\caption{(color online) Fit of the heuristic model for the last LER Touschek size sweep series with experimental (red) and simulated (black) data.}
\label{fig:LERLossRateRun3000sSummary}
\end{figure}

The measured lifetime $\tau$ can be converted into the total loss rate ($TLR$):
\begin{equation}
  TLR = \frac{IL}{c}\frac{1}{\tau},
\end{equation} 
where $I$, $L$, and $c$ are the current, the ring length, and the speed of light, respectively. We can write the total loss rate as the sum of loss rates $LR$ due to beam-gas ($bg$) and Touschek ($T$) scattering:
\begin{equation}
TLR = LR_{bg} + LR_{T}.
\end{equation}
Using the fitting method described in Section~\ref{sec:beamgas_touschek_parameterization} with $TLR$ as the observable, we can separately measure the beam-gas and Touschek loss rates. Figures~\ref{fig:HERLossRateRun2000sSummary} and \ref{fig:LERLossRateRun3000sSummary} demonstrate fits to the combined heuristic for two of the sweep series of Table~\ref{tab:touschek_runs}. For the pressure $P$ we use the ring-average pressure and for the atomic number $Z$ we use $Z_{e}$ from the nearest RGA for the LER and 2.7 for the HER (see Section~\ref{sec:Zeff}).

\subsection{Simulated loss rates}
For the simulation, we take the $TLR$ generated by SAD and scale it to the conditions of the size sweep series using the procedure described in Sec.~\ref{simulation_scaling}. 

\subsection{Comparing experimental to simulated loss rates}
The ratios between experimental data and simulation for $TRL$, $LR_{bg}$, and $LR_{T}$ are summarized in Table~\ref{tab:LR}.

\begin{table}
  \caption{Ratio of experimental to simulated loss rates for the three LER and four HER size sweep series.}
  \centering
  \label{tab:LR}
  \begin{tabular}{cccc}
    \toprule
          & \multicolumn{3}{c}{Experiment/simulation ratio} \\
          & $TLR$ & $LR_{bg}$ & $LR_{T}$\\ 
    \midrule
    LER 1 & 7.08 $\pm$ 0.89 & 9.08 $\pm$ 1.14 & 6.09 $\pm$ 0.77 \\ 
    LER 2 & 6.07 $\pm$ 0.70 & 19.1 $\pm$ 1.5 & 2.92 $\pm$ 0.22 \\ 
    LER 3 & 10.9 $\pm$ 3.8 & 49 $\pm$ 17 & 7.4 $\pm$ 2.5 \\
    HER 1 & 0.34 $\pm$ 0.05 & 3.50 $\pm$ 0.43 & 0.27 $\pm$ 0.03 \\ 
    HER 2 & 0.33 $\pm$ 0.01 & 23.92 $\pm$ 0.63 & 0.100 $\pm$ 0.003 \\ 
    HER 3 & 0.59 $\pm$ 0.42 & 11.0 $\pm$ 5.1 & 0.45 $\pm$ 0.20 \\ 
    HER 4 & 0.65 $\pm$ 0.16 & 9.8 $\pm$ 2.5 & 0.56 $\pm$ 0.14 \\ 
    \bottomrule
  \end{tabular}
\end{table}

We observe large variations in the level of agreement between experiment and simulation throughout Phase 1, with a consistent overall excess of beam-gas background in experiment, in agreement with the results of the two analyses in Sections~\ref{sec:beamgas_touschek_direct_analysis} and \ref{sec:beamgas_touschek_constrained_analysis}. We conclude that improvements in our implementation of SAD are likely needed to adequately simulate beam-gas backgrounds. 

The poor linearity of the LER fit in Figure~\ref{fig:HERLossRateRun2000sSummary} may explain some of the large variation in the agreement, and it suggests that using the average ring pressure and single-RGA $Z_e$ may be insufficient for total ring losses. We expect this to be the case if gas conditions are highly localized so that the ring beam-gas loss rate is dominated by isolated pockets of high-$P$ or high-$Z$ gas. This is consistent with the implications of Section~\ref{sec:beamgas_touschek_direct_analysis}, implying that better gas instrumentation is needed to accurately simulate beam-gas losses.

 \clearpage

 % lead author: Alex Beaulieu
 \section{Improvement in SuperKEKB conditions}
 % file:        scrubbing.tex
% lead author: Alexandre Beaulieu
The BEAST II experiment is also valuable to study long-term performance improvements of the accelerator as the commissioning program is ongoing. Two aspects are particularly useful: the vacuum scrubbing process and the time structure of so-called ``beam-dust" events, both contributing to backgrounds seen in BEAST II and expected in Belle II. Vacuum scrubbing is the general term used to describe outgassing of the vacuum chamber impurities promoted by beam circulating in the accelerator. Beam-dust events, on the other hand, correspond to rapid increases in a few local pressure readings and may trigger beam aborts. Thought to be the result of beam particles vaporizing microscopic or macroscopic particles, they were previously  observed and studied at many accelerator laboratories such as KEK \cite{Saeki1991, Tanimoto2009}, CERN \cite{Papotti:2016, Baer:2011} and DESY \cite{Zimmermann:1993}.

In both cases, we are interested in the long-term behavior of these processes, and in how they are related to operating conditions such as instantaneous and integrated currents. More precisely, the time-integrated current, also known as the ``beam dose" and often given in A$\cdot$h, is the common measure of progress of the accelerator commissioning. A low beam dose corresponds to early commissioning, while a high beam dose corresponds to late commissioning. The following sections describe the approaches taken by the BEAST II group to characterize these phenomena as well as their respective results.

\subsection{Vacuum scrubbing}\label{sec:ScrubbingMethodology}
\subsubsection{Phenomenology}
In the particular context of particle accelerators, the typical thermal outgassing  process is augmented by radiation-induced effects called electron-induced desorption (ESD) and photon-induced desorption (PSD). While the former is more typical of electron-positron colliders such as SuperKEKB, and can arise from machine-induced electron multipacting \cite{Flanagan:2005}, the latter is the result of sychrotron photons from the accelerated charged beams irradiating the vacuum chamber material. The energy of these synchrotron photons can be as high as a few MeV \cite{Grobner:1983}.  

With both phenomena, the desorption rate is a function of the integrated radiation dose on the vacuum chamber material and follows the same  behaviour \cite{Oleg:2012}. After an initial electron or photon dose $D_0$, the out-gassing rate $\eta$ follows a power-law dependence with respect to the integrated beam dose $D$:
\begin{equation}
\label{eqn:psd_esd}
\eta = \eta_0 \left( \frac{D}{D_0} \right)^{\beta},
\end{equation}
where $\eta_0$ is the desorption rate at the initial dose $D_0$, and the coefficient $\beta$ is a function of the irradiation energy spectrum, the material characteristics, and the physical process (ESD or PSD) behind the scrubbing.

Because such scrubbing releases gas molecules into the vacuum chamber, it should be directly observable in terms of a dynamic pressure $dP/dI$ component. Moreover, the beam-gas interactions between the impurities and the charged beams produce increased particle losses around the accelerator. Such losses are observed with the BEAST II systems as increased radiation levels. These two techniques are used to assess the rate of vacuum scrubbing, and provide a statement on the adequacy of the vacuum expected for Belle II operation. 

\subsubsection{Dynamic pressure measurement}

\begin{figure}[ht!]
\centering
\includegraphics[width=\columnwidth]{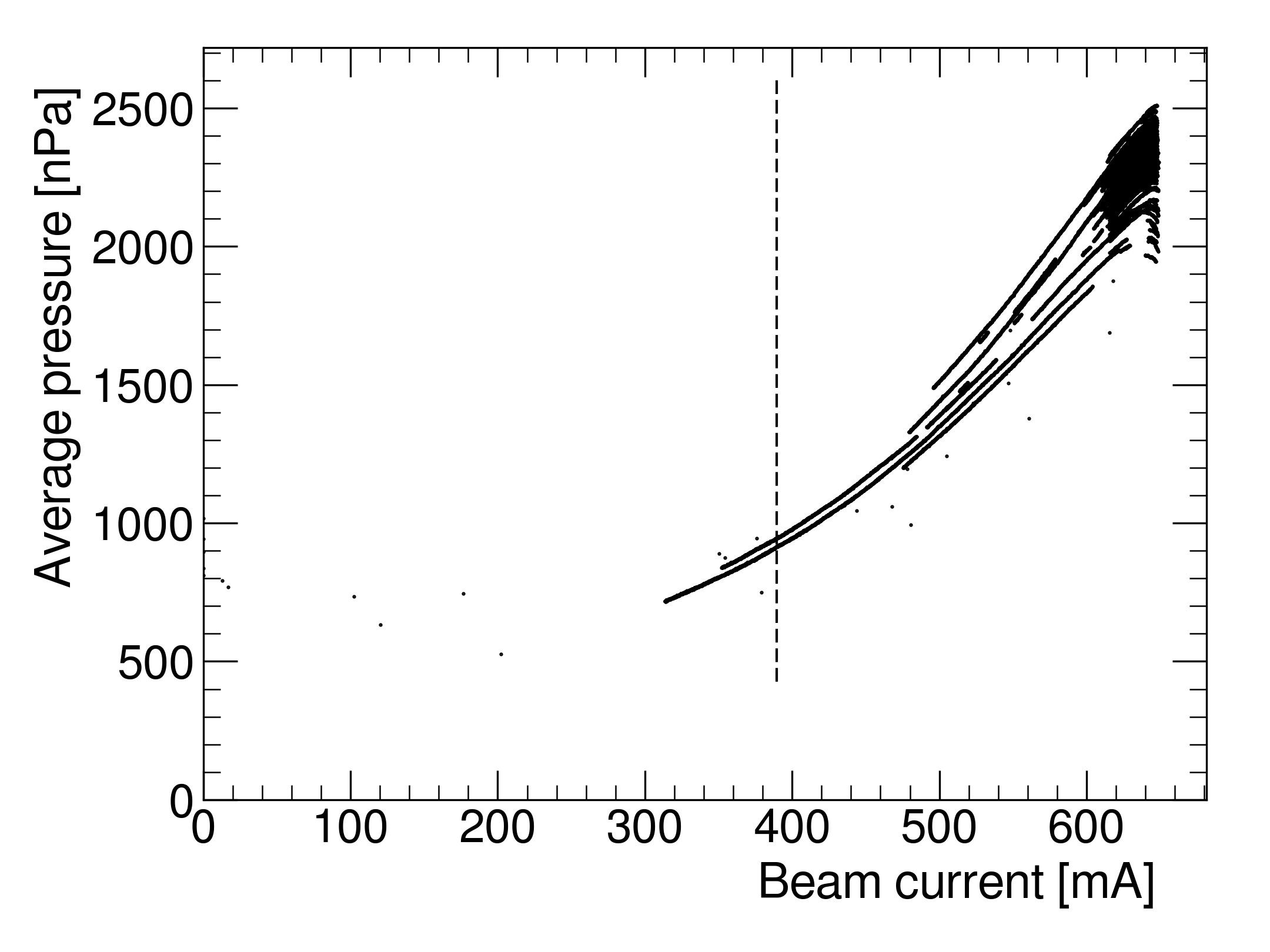}
% PDF version also in the folder if needed (slows down page display considerably)
\caption[Dynamic and static pressure measurements: example of P-I curves on a full day]{Dynamic and static pressure measurements: example of P-I curves on a full day. Data shown are for the LER on 2016-05-08. The dashed line represents 60\% of the maximum current value. Beam injection periods are removed. }
\label{fig:Ex_PvsI}
\end{figure}

The dynamic pressure  is a measurement of the rate of gas released from the material per unit current circulated in the beampipe. Therefore it is expected to track the desorption rate:
\begin{equation}
\frac{dP}{dI} \sim \eta
\end{equation}
This dynamic pressure is the fundamental quantity used by the SuperKEKB group to quantify the rate of vacuum scrubbing \cite{Suetsugu:2017}. However powerful, using a single $dP/dI$ value for a given set of beam parameters remains an approximation. This fact is illustrated in Figure~\ref{fig:Ex_PvsI}.
This example shows that the residual gas pressure is not a function of the current alone. More importantly, it  shows that $dP/dI$ depends on other quantities such as the actual operating current --- the relationship is not linear between 300~mA and 500~mA --- and the time after the last injection. The data points at the far right, where currents are the highest, were taken immediately after an injection period and exhibit a short-term pressure increase for decreasing currents (negative $dP/dI$).

With these warnings in mind, it is nonetheless useful to estimate this dynamical pressure term on a continuous basis. One approach is to neglect the equilibrium pressure inside the chamber and assume linearity between pressures and currents. The equilibrium pressure is also referred to as the ``base pressure'', and is the steady-state pressure reached when no beam is circulating. Examining only currents above 60\% of the maximum values to reduce the effects of base pressure, the dynamic pressure is estimated by
\begin{equation}\label{eqn:dPdI_PovI}
\frac{dP}{dI}\text{(est.)} = \frac{P}{I}.
\end{equation}

A second approach to obtain the dynamic pressure is to consider the effect of the base pressure, which can be evaluated every time the beams are off for a sufficient period of time. A settling time of one hour is considered adequate to measure the base pressure. Using this definition of the base pressure $P_\text{base}$, 
\begin{equation}\label{eqn:dPdI_P-Pbase_ov_I}
\frac{dP}{dI}\text{(daily)} =  \frac{ \left(P - P_\text{base}\right)}{I}.
\end{equation}

\subsubsection{Dynamic pressure from machine-induced background measurement}
The BEAST II detectors also provide insight into the rate of vacuum scrubbing. An additional difficulty here lies in the fact that it is non-trivial to disentangle the background contribution due to beam-gas from those due to intra-bunch effects such as Touschek and electron cloud.  According to the heuristic model of Eq.~\ref{eqn:combined_heuristic}, we can use the ratio $\mathcal{O}/I^2$ as a proxy for the Touschek-subtracted beam background contribution. As long as the beam-gas contribution dominates the overall background radiation,  
\begin{equation} \label{eqn:OovI2simeta}
	\frac{\mathcal{O}}{I^2} \sim \eta,
\end{equation}
can be used in the study of the vacuum process. In Eq.~\ref{eqn:OovI2simeta} the observable $\mathcal{O}$ is a dose rate or hit rate, depending on the sub-detector of interest.

\subsection{Vacuum scrubbing results}

\subsubsection{Measurement based on the dynamic pressure}

\begin{figure}[ht!]
\centering
\includegraphics[width=\columnwidth]{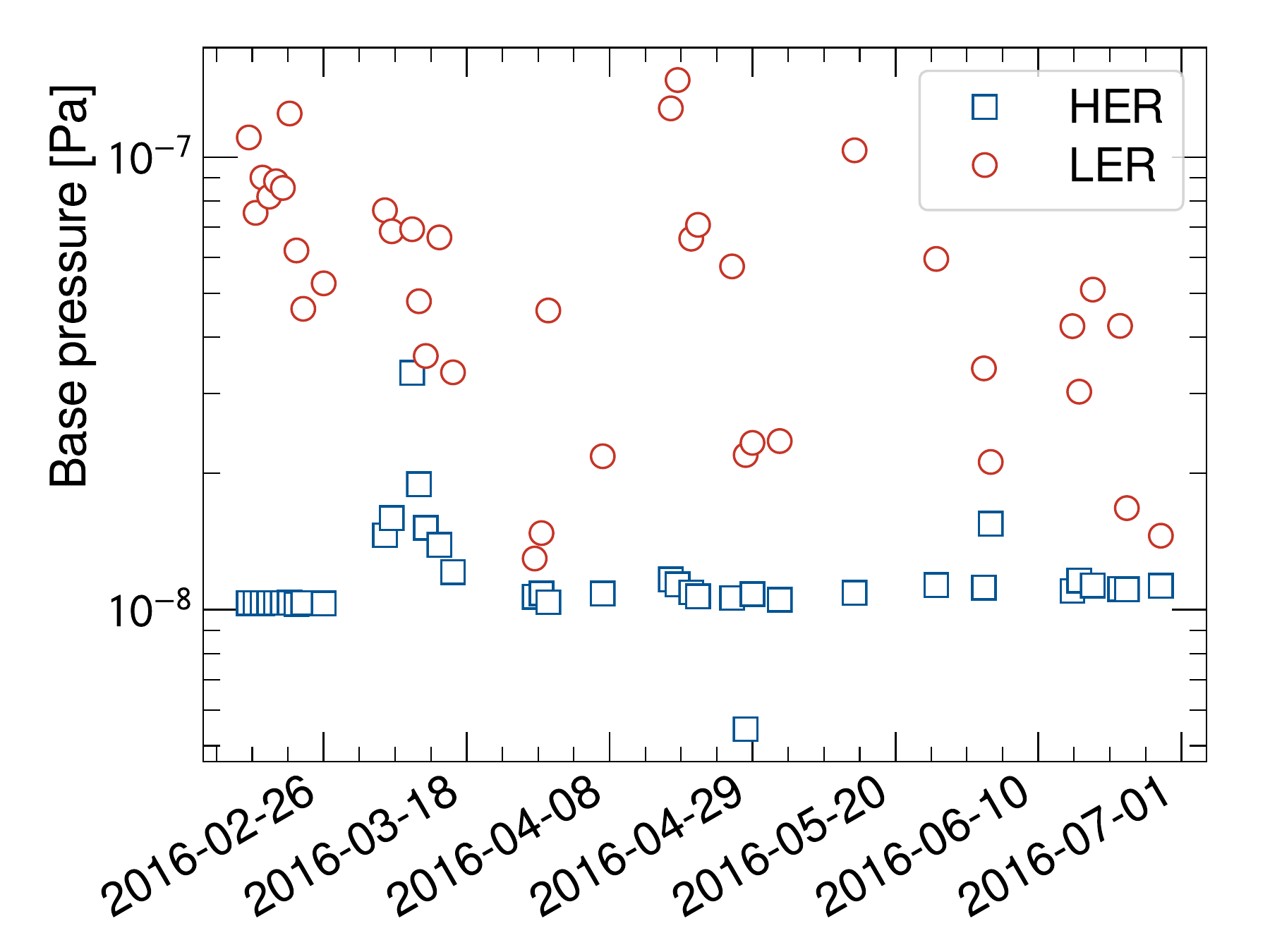}
\caption[Dynamic and static pressure measurements: base pressure measured as a function of date]{(color online) Dynamic and static pressure measurements: base pressure measured as a function of date. Red circles represent LER measurements, while blue squares represent HER measurements.}
\label{fig:Scrubbing_BasePressure}
\end{figure}

Figure~\ref{fig:Scrubbing_BasePressure} shows the evolution of the base pressure during Phase 1 operation.
While the HER quickly reaches the equilibrium value of $1\times 10^{-8}\text{ Pa}$ after it was turned on in March, the LER shows no appreciable asymptotic behaviour, with the minimum recorded pressure varying between $1\times 10^{-8}\text{ Pa}$ and $1\times 10^{-7}\text{ Pa}$. Even if the daily variation is more dramatic than with the HER, the values remain well below the pressures observed during operation, which are on the order of $1\times 10^{-6}\text{ Pa}$ or more during operation at full nominal current. Such variability should produce negligible effects on the dynamic pressure measurement.

\begin{figure}[ht!]
\centering
\includegraphics[width=\columnwidth]{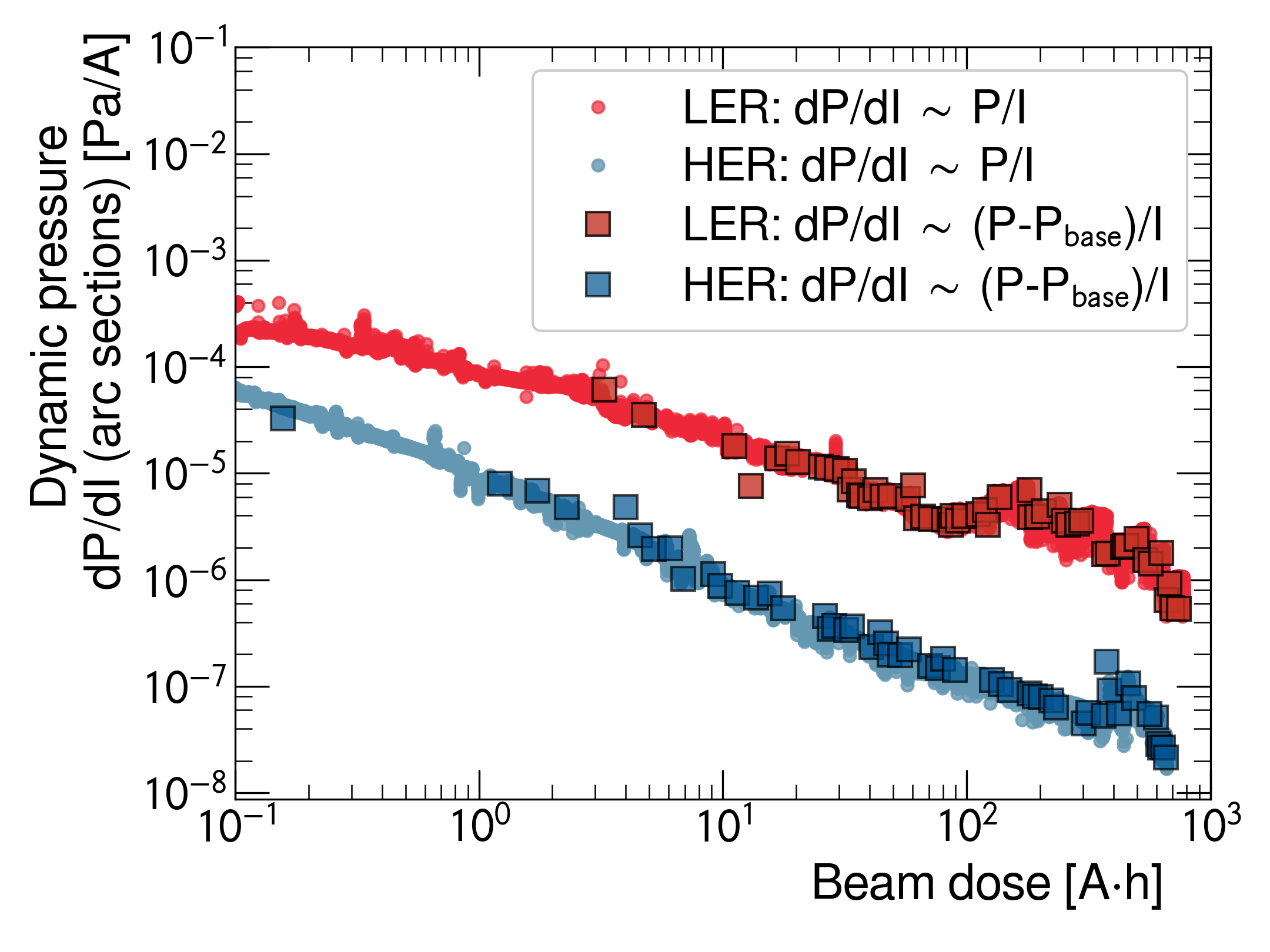}
% PDF version also in the folder if needed (slows down page display considerably)
\caption[Comparison of two methods used to estimate the dynamic pressure]{Dynamic and static pressure measurements: comparison of two methods used to estimate the dynamic pressure. The circles are the SuperKEKB group results, obtained using Eq.~~\ref{eqn:dPdI_PovI} whereas the squares were obtained with BEAST II data using Eq.~~\ref{eqn:dPdI_P-Pbase_ov_I}. Blue points represent HER data and the red ones, LER data}
\label{fig:dPdI_Comparison}
\end{figure}

Figure~\ref{fig:dPdI_Comparison} shows a comparison of the two different estimates for $dP/dI$.
Both the methods of Eq.~\ref{eqn:dPdI_PovI} and Eq.~\ref{eqn:dPdI_P-Pbase_ov_I} are in good agreement, exhibiting a power-law behavior over more than 3 decades. This is the expected behavior when the base pressure is negligible compared to the dynamic component. The slope between 100~$\text{A}\cdot\text{h}$ and 1000~$\text{A}\cdot\text{h}$ is $\eta_\text{LER} = -0.9$ for the LER and $\eta_\text{HER} = -0.6$ for the HER.

Finally, the result in Figure~\ref{fig:dPdI_Comparison} shows that for operating currents reaching Amperes at the end of Phase 1, the HER dynamic pressure contribution is of comparable scale with the base pressure. However, for the LER, the dynamic contribution dominates the base pressure by a factor of at least 10. Should $dP/dI$ keep following the same power-law behaviour, the LER should be operated for more than $1\times10^{4}\text{ A}\cdot\text{h}$  in order for the dynamic pressure to reach the same level as the base pressure, at the design 3.6~A beam current.

\subsubsection{Measurement based on BEAST II detectors}
\begin{figure}[ht!]
\centering
\subfigure[Electron beam (HER) \label{fig:Scrubbing_HER_BEAST}]{
\includegraphics[width=\columnwidth]{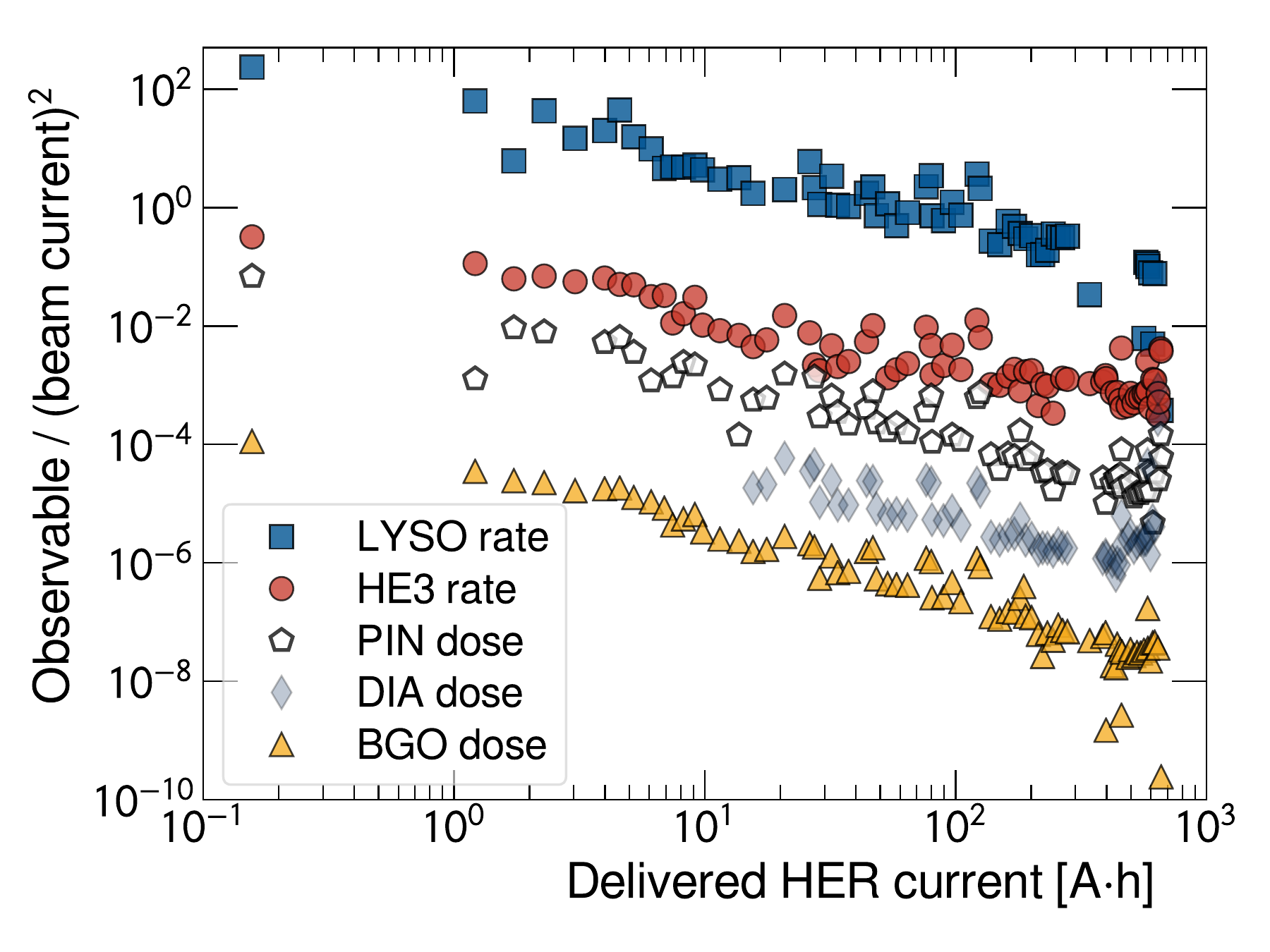}
}

\subfigure[Positron beam (LER) \label{fig:Scrubbing_LER_BEAST}]{
\includegraphics[width=\columnwidth]{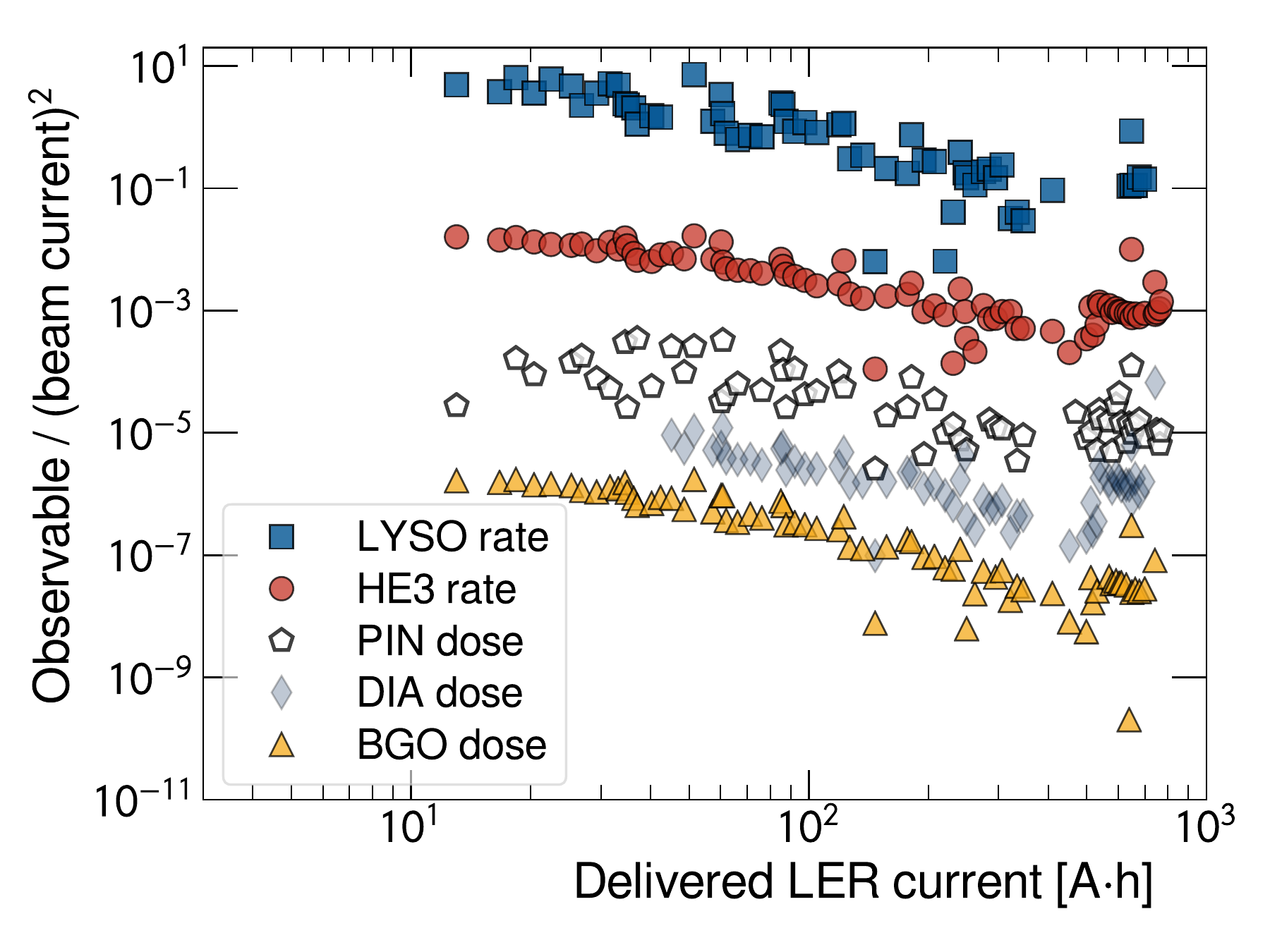}
}
\caption{(color online) Vacuum scrubbing evolution as a function of delivered current. The observable units are arbitrary: they have been adjusted to offset each data series for more clarity.}
\label{fig:Scrubbing_BEAST}
\end{figure}

Figure~\ref{fig:Scrubbing_BEAST} shows the scrubbing process as seen by the BEAST II detectors for the HER and LER scrubbing processes. The same general power-law dependence is observed across all detectors. 

For the HER scrubbing shown in Figure~\ref{fig:Scrubbing_HER_BEAST}, the LYSO, \heT, PIN diodes and BGO are all in good agreement with the power-law model across four decades. However, the numerical values of the slopes are not compatible with the $dP/dI$ value. They are also inconsistent with each other except for the LYSO and PIN diodes, at $-1.09\pm0.08$ and $-0.98\pm0.05$, respectively. For the LER scrubbing shown in Figure~\ref{fig:Scrubbing_LER_BEAST}, LYSO, \heT, PIN diodes and BGO are all in agreement with the power-law model across one and a half decade. Quantitatively, the discrepancy between the slopes is more pronounced than with the HER case.

In both Figures~\ref{fig:Scrubbing_HER_BEAST} and \ref{fig:Scrubbing_LER_BEAST}, there is plateauing or an increase of the rates beyond 400 A$\cdot$h observed with all subdetectors.  A possible explanation is related to conditioning of the non-evaporable getters (NEG) that happened during the associated period. Such conditioning is known to release heavier elements in the vacuum chamber, which produces considerably more background due to the $Z^2$ dependence. 

Otherwise, the most significant improvement to the accelerator during this period is the addition of permanent magnets to the uncoated aluminium bellows meant to reduce electron multipacting at large currents. Measurements of the beam size by the SuperKEKB group showed that this effectively reduced the electron-cloud effect without changing beam orbit and optics~\cite{Suetsugu:2017}.

 % file: 		transients_analysis.tex
% lead author:	Alex Beaulieu
% timeline:		September 17th (this is embarassing)
\subsection{Beam-dust events}\label{sec:Transients}

\subsubsection{General description and motivation}\label{sec:Transients_intro}

During commissioning of the accelerator, one concern was the observation of localized pressure bursts and accompanying background spikes. The prevalent hypothesis for these observations is collisions between the beam electrons and positrons, and small particles such as dust coming off the vacuum chamber material \cite{Suetsugu:2017} \cite{Baer:2011} \cite{Papotti:2016}. We will therefore refer to these as ``beam-dust'' events in this report.

Such beam-dust events are important during commissioning and running of an accelerator, since the corresponding increase in observed background often results in beam aborts and loss of operation time. The SuperKEKB group monitors these events by measuring pressure peaks around the beampipe. The questions we address in the present report are 
\begin{itemize}
\item What is the time structure of these beam-dust events during Phase 1 of SuperKEKB, and does it change as vacuum scrubbing progresses?
\item Do the pressure bursts measured by SuperKEKB correspond to background peaks seen in the BEAST II detectors?
\item Could the radiation resulting from these events be damaging to the detectors?
\end{itemize}

\subsubsection{Analysis methodology}
\label{sec:Transients_method}
Qualitatively, beam-dust events present sharp peaks in BEAST II detector observables, such as the deposited energy rate in the BGO or the hit rate in the CsI crystals. These peaks are much higher than the typical signal, and last less than 2 seconds.

The analysis therefore consists of finding peaks that are at least six standard deviations above the mean signal, calculated using 60-second running time windows. To further exclude electronic noise peaks in the count, we require such peaks to be seen in at least four channels in at least two different detector systems. The coincidence time window is set to 3 seconds since BEAST II data coming from different sub-detectors is not perfectly aligned in time. These requirements were adjusted on a small sample of clearly identifiable peaks, then extended to the complete Phase 1 data set. 

The results are compared to the list coming from the SuperKEKB accelerator group, who define beam-dust events as pressure bursts where any cold cathode pressure gauge value is 15\% larger than the previous 1-second average. From this list, only burst events located in regions D01 and D02, the straight sections on either side of the BEAST II detector, are considered. This amounts to 338 distinct burst events. The other regions were excluded since vacuum bump experiments outside this range showed no corresponding increase of the BEAST II observables, and therefore these events would be impossible to correlate with BEAST II data.

The time resolution of the events provided by SuperKEKB is one minute so we discard the second information from the BEAST II events. We calculate the cross correlation between the two lists to ensure there is no time misalignment between the two datasets. A peak at lag=0 indicates the correct time alignment, and the cross-correlation value outside the peak corresponds to the number of accidental coincidences due to the finite time resolution.

Finally, we aim to provide a statement on the dose resulting from such events. To achieve this, we compare the peak level dose to the running average of the previous 60 seconds, and multiply by the probability of occurrence of such a peak for the given sub-detector that recorded it.

 % file: 		transients_results.tex
% lead author:	Alex Beaulieu
% timeline:		September 17th (this is embarassing)

\subsection{Beam-dust results} \label{sec:transients_results}
\subsubsection{Observation in BEAST II}
\begin{figure}[ht!]
\centering
\includegraphics[width=\columnwidth]{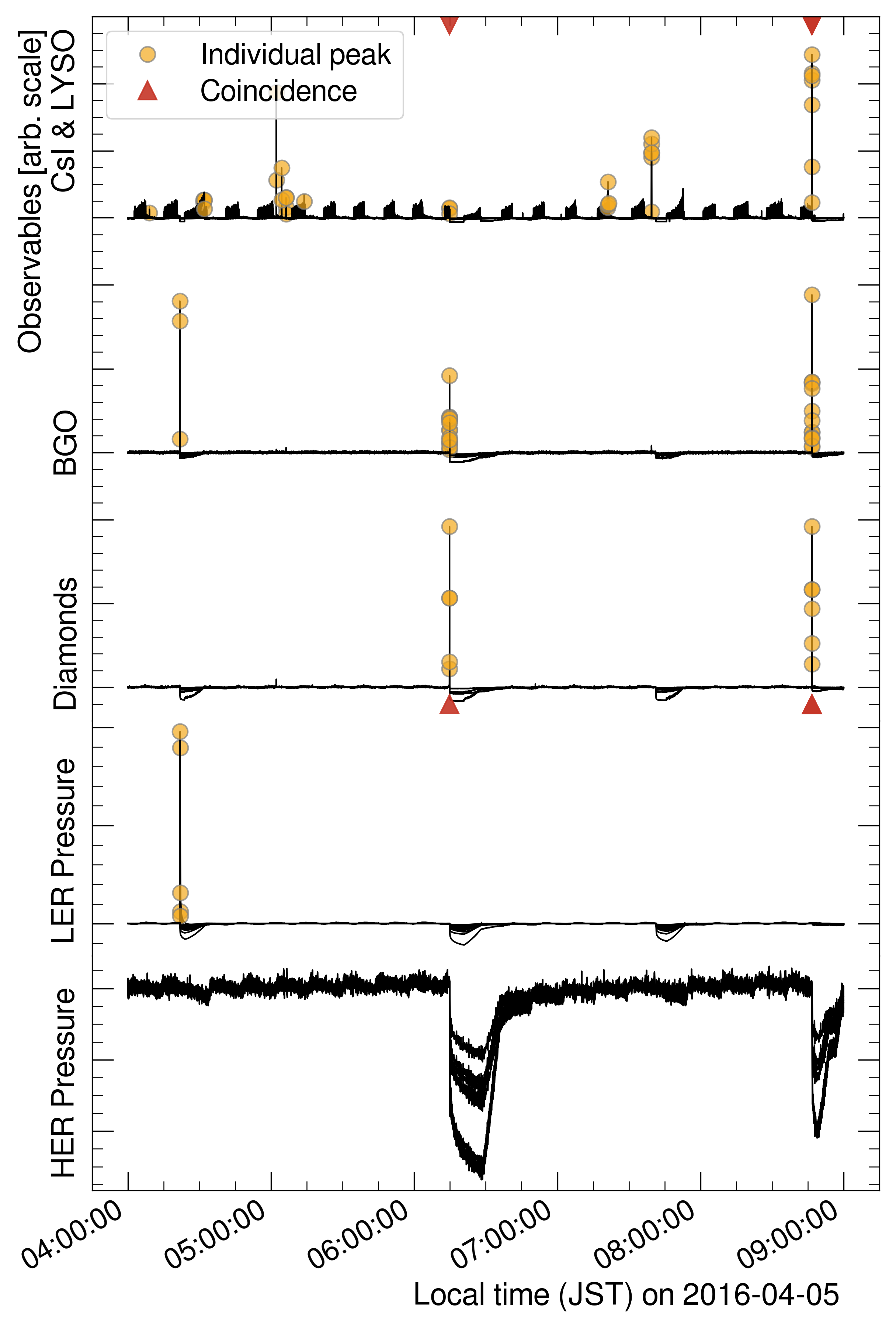}
% PDF version also in the folder if needed (slows down page display considerably)
\caption{Example of beam-dust events and corresponding pressure readings. A coincidence is defined as a peak seen at least four channels, which are in at least two different sub-detector systems. The vacuum chamber pressures are reported for reference only, and do not count in the coincidences.}
\label{fig:UFO_peak_with_pressures}
\end{figure}

Figure~\ref{fig:UFO_peak_with_pressures} shows examples of beam-dust events together with the results of the peak-finding algorithm described in section~\ref{sec:Transients_method}. 
The algorithm functions as expected by selecting large-amplitude signal peaks while rejecting fluctuations that can be attributed to noise on some channels. A total of 598 beam-dust events were identified using this algorithm between 2016-04-05 and 2016-06-28.

\subsubsection{Comparison with the SuperKEKB list}
\begin{figure}[ht]
\centering
\includegraphics[width=\columnwidth]{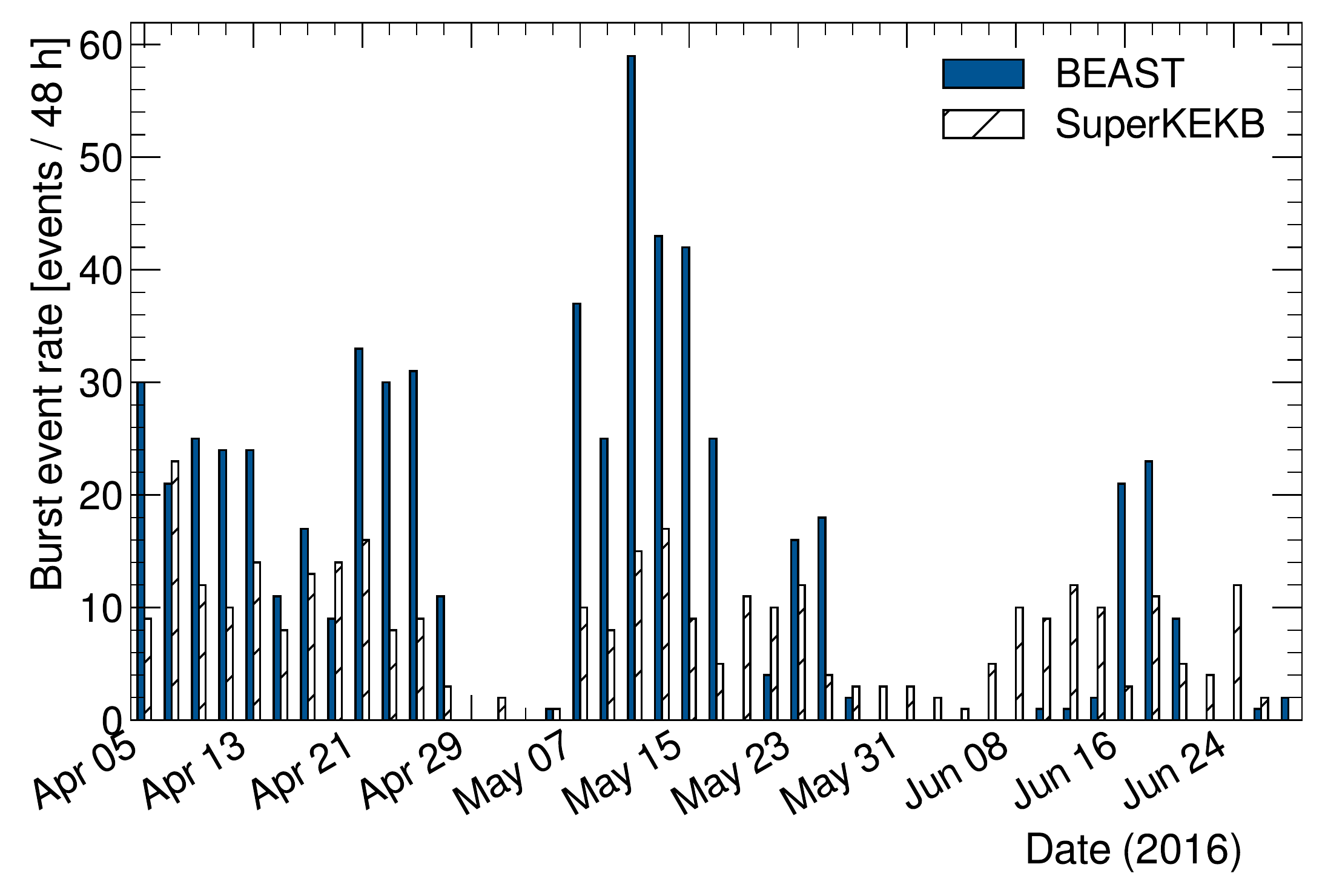}
\caption[Comparison of the time structure of beam-dust events between SuperKEKB and BEAST II data]{(color online) Comparison of the time structure of beam-dust events between SuperKEKB and BEAST II data.}
\label{fig:UFO_BEAST_SKB_rate_vs_time}
\end{figure}

Figure~\ref{fig:UFO_BEAST_SKB_rate_vs_time} shows a comparison of the time structure of vacuum burst events measured by SuperKEKB compared to the beam-dust events measured in BEAST II.
From this figure we see that there is no obvious steady reduction in the 48-hour rate of these events as scrubbing progresses, and more operation time would be needed to provide a statement on this aspect. 

\begin{figure}[ht]
\centering
\includegraphics[width=\columnwidth]{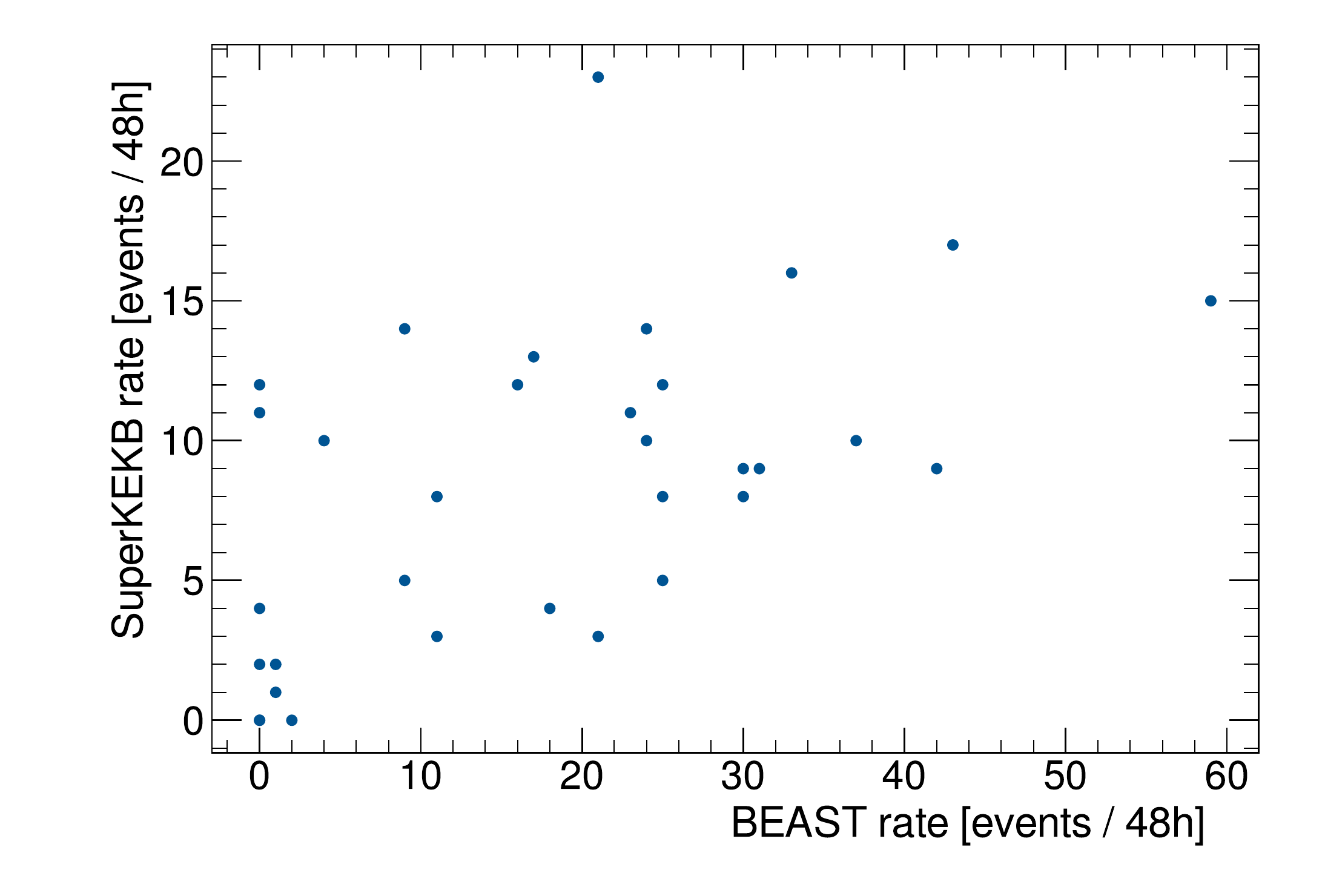}
\caption{48-hour rate of beam-dust events from the SuperKEKB and BEAST II lists. Data from both sets are only weakly correlated, with a coefficient of correlation  $r_\text{beam-dust} = 0.54$.}
\label{fig:UFO_BEAST_vs_SKB_rate}
\end{figure}

The second question to address is whether or not the pressure bursts correlate with BEAST II's observation of the so-called beam-dust events. The answer is best expressed by displaying the data of Figure~\ref{fig:UFO_BEAST_SKB_rate_vs_time} as a cloud of points of the 48-hour rate seen by the SuperKEKB pressure gauges against the rate observed by BEAST II, and then calculating the Pearson product-moment correlation coefficient between the two sets. The result is shown in Figure~\ref{fig:UFO_BEAST_vs_SKB_rate}.
The coefficient of correlation is $r_\text{beam-dust} = 0.54$,
which indicates a weak correlation. The relative weakness of this correlation is not fully understood. 

\begin{figure}[ht!]
\centering
\includegraphics[width=\columnwidth]{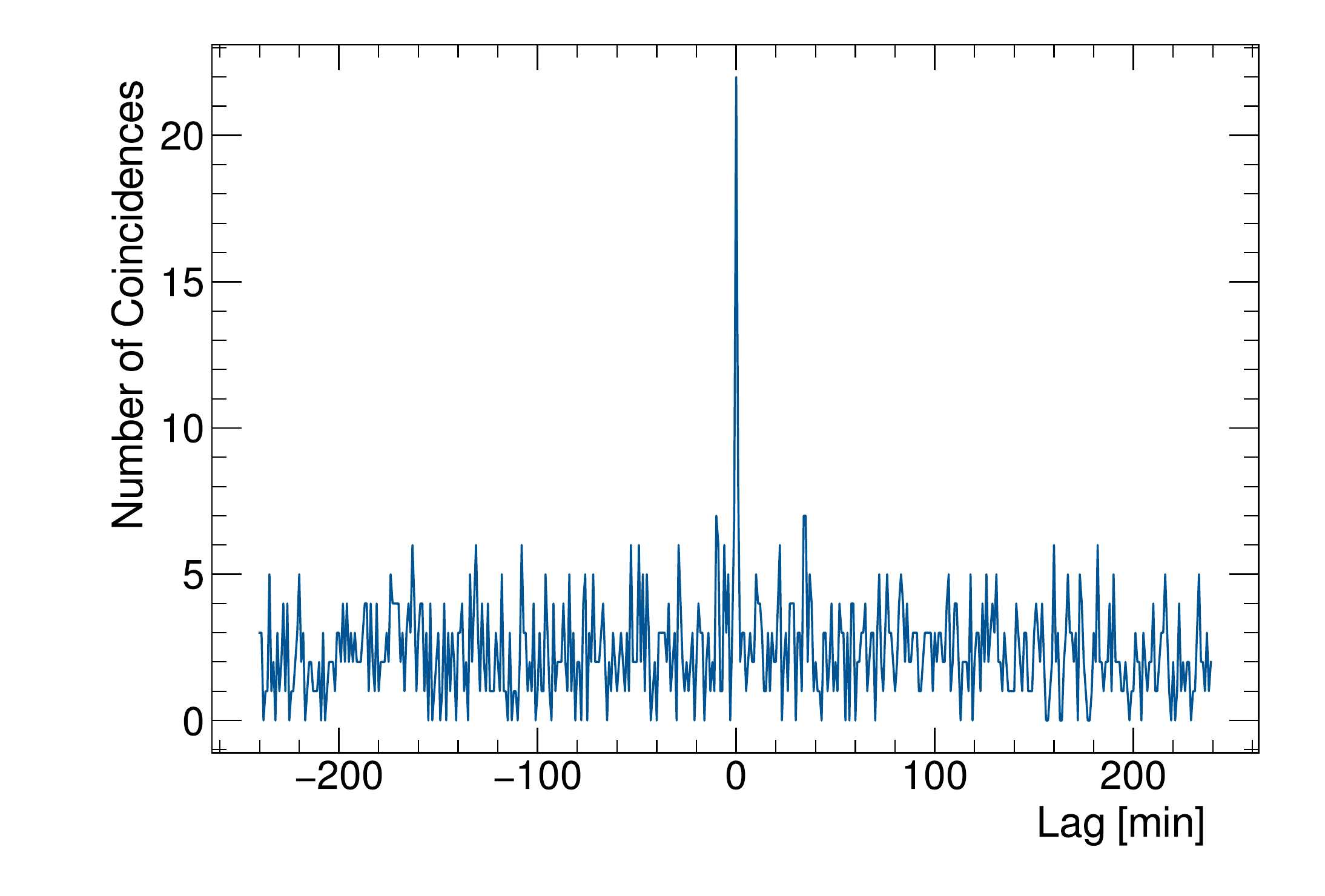}
\caption{Cross-correlation between the beam-dust events from SuperKEKB and BEAST II datasets. The sharp peak at lag = 0~min implies that the data sets are aligned in time, and the relative level of the side-bands indicate the number of accidental correlations expected if the two lists were completely independent.}
\label{fig:UFO_SKB_BEAST_Beamdust_xcorr}
\end{figure}

Looking at the cross-correlation between the two lists, shown in Figure~\ref{fig:UFO_SKB_BEAST_Beamdust_xcorr}, gives us a different angle on the situation.
The maximum correlation is obtained at lag~=~0, and therefore the relative weakness of the correlation of the scatter plot in Figure~\ref{fig:UFO_BEAST_vs_SKB_rate} cannot be attributed to a misalignment of the data coming from different sources. 

Also of particular interest in Figure~\ref{fig:UFO_SKB_BEAST_Beamdust_xcorr} is the relative height of the central peak with respect to the side bands. There are 22 coincidence events in the joint data taking period, whereas the random coincidence level, given by the side bands,  is $2.3\pm1.5$. In other words, while 22 coincidences between BEAST II and SuperKEKB is a low fraction of the total burst events --- 6.5\% of the SuperKEKB list or 3.7\% of the BEAST II list --- it is still significantly larger than what one would expect if they were uncorrelated.

An hypothesis to explain this effect is that the impact of beam-dust events, in terms of the background radiation generated, are very localized. This locality would explain why events contained in the BEAST II list do not always result in beam aborts from beam loss monitors, and that a small fraction of beam aborts resulted from an increase of background that was also observed in BEAST II. Data supporting this hypothesis is displayed in Figure~\ref{fig:UFO_peak_with_pressures}.  Here we show the results of the peak finder algorithm also including the local pressure time series. The LER pressure burst seen around $t=$04:15:00 is only observed in the BGO detectors, thus not recognized as a BEAST II vacuum burst event. The converse is also true: the clear BEAST II events found at $t=$~06:15:00 and $t=$~8:45:00 do not necessarily show a pressure increase in the nearby detectors. This is consistent with the hypothesis of a local generation of background from these beam-dust events.

\subsubsection{Relative dose of beam-dust events}
\begin{figure}[ht!]
\centering
\includegraphics[width=\columnwidth]{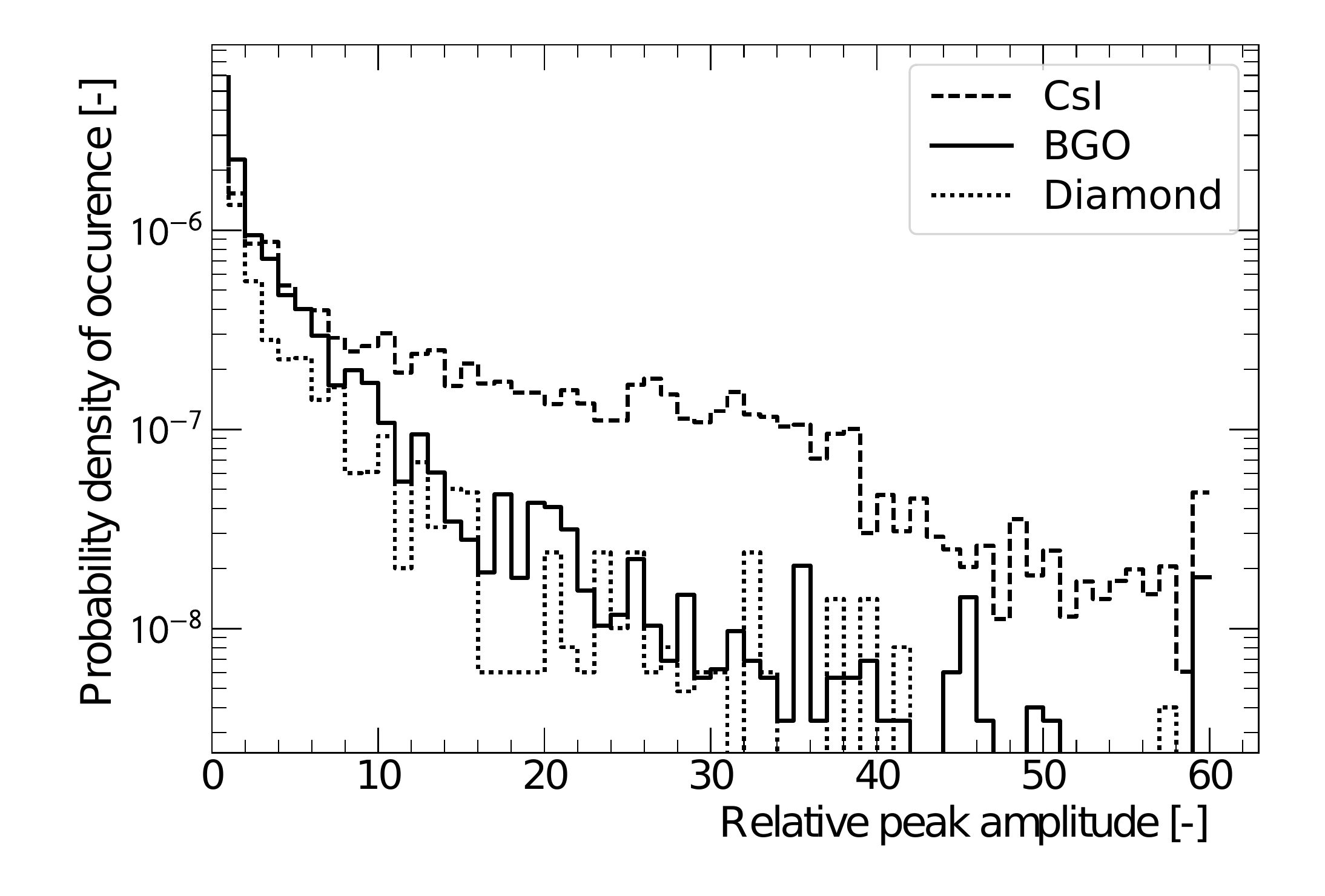}
\caption{Probability distributions of the beam-dust events amplitude relative to the previous 60-second running average for the Crystals (CsI), BGO, and Diamonds (DIA) systems. These distributions show that the Crystals are more sensitive to these beam-dust events in the sense that they are detected more often, and that a larger fraction of the detected events have large amplitude relative to the nominal background levels. The total probability of observing one beam-dust event in a 1-second sample is 0.015\% for CsI, 0.012\% for BGO, and 0.007\% for DIA.}
\label{fig:BeamDust_PeakAmplitudeDistros}
\end{figure}

Figure~\ref{fig:BeamDust_PeakAmplitudeDistros} shows the probability distributions of the peak height amplitude of the beam-dust events for the Crystal, BGO and Diamond systems. For the Crystals, the beam-dust events represent 0.2\% of the total dose. It corresponds to 0.05\% and 0.03\% of the total dose for the BGO and the diamonds sub-detectors, respectively. 

 \clearpage
 
 % lead author: Riccardo De Sangro
 \section{Injection background measurements}
 % file: 		injection.tex
% lead author:  Riccardo De Sangro
%\subsubsection{Introduction}

During normal running of the SuperKEKB accelerator, we expect to have continuous injection at 100~Hz. 
Since during injection the electrons' orbit parameters of the injected bunch are perturbed, a higher rate
of lost particles is present around the machine arcs and brackground is increased as a result. 
The orbits tend to stabilize after a few milliseconds and the background returns 
to the level normally produced by circulating beams coasting with the given beam current.
A direct measurement of the background occurring during the injection phase is necessary because it is very difficult 
to simulate and may be orders of magnitude higher than the one generated during beam coasting.

In this section we describe the measurement of injection backgrounds using the Crystals, CLAWS and QCSS detector systems, 
taken during two dedicated, three-day run periods (the first starting May 16 and the second starting May 23, 2016), 
in which we deliberately moved the injection parameters away from their optimal values.

\subsection{Injection backgrounds in the Crystals}

At the end of the data taking period of the Belle experiment at KEKB, a dedicated run was taken to study the injection noise in the electromagnetic calorimeter (ECL). 
The results are discussed in the Belle II Technical Design Report \cite{Abe:2010gxa}, where it is shown that, during injection, only the injected bunch(es)
produce higher backgrounds in the detector lasting a few milliseconds after the injection. 
We made a similar study using BEAST II data from the Crystals, which we present in the following sections.

\subsubsection{Measured quantities and data set}
Lost beam particles near the interaction point (IP) may hit the beampipe or other machine elements producing secondary particles (mainly photons and electrons) 
of relatively low energy (few MeV) that eventually hit the crystals. These particles transfer all or part of their energy to the crystal material, which reacts by 
emitting a proportional amount of scintillation light that is measured with a photo-multiplier tube (PMT).

As described in Section\,\ref{sec:cryReadout}, while the CsI(Tl) crystal photo-multipliers are read out only with a digitizer, we make two independent measurements 
of the signals from the LYSO and pure CsI crystals; the signals are split, one signal is fed to the digitizer and the other, after discrimination, to a scaler. 

For the study of injection background described here, we slightly modified the DAQ used for normal running that we described in Sec.\,\ref{sec:cryReadout}: 
the digitizer's acquisition time window  was shortened from 10 to 1 ms, recording each hit's charge and arrival time with 2~ns precision, while the scaler time base 
was shortened to record the hit rate every $3$~\si{\micro}s within a 5~ms time window. Both acquisition gates are triggered by the injection signal provided by SuperKEKB. 
To precisely record the time of the injection, we feed the injection signal coming from SuperKEKB to the digitizer.
The 2~Hz asynchronous 1~ms gate, used to record the background while the beam is coasting between injections, was left unchanged.

Most of the results presented here are from the data taken during the second dedicated period of injection background studies. 
During this time we changed, in turn, the values of four parameters 
used to tune the injection, to study their effect on the observed background; they are listed in Table\,\ref{tab:injdata}. %where we give the corresponding EPICS name.
They regulate: the phase with wich new particles are injected in the bunch (Phase Shift); the vertical incidence angle (Vertical Steering 1 and 2); the incident angle in the 
horizontal plane (Septum Angle).

%  Run number mapping
% 14 LER-InjRef
\def\runFourteen{LER-Reference}
%  3 LER-InjPhase
\def\runThree{LER-Phase}
%  6 LER-InjVer2
\def\runSix{LER-Vertical2}
% 17 LER-InjHor
\def\runSeventeen{LER-Septum}
% 10 HER-InjRef
\def\runTen{HER-Reference}
%  9 HER-InjPhase
\def\runNine{HER-Phase}
% 12 HER-InjVer1a
\def\runTwelve{HER-Vertical1a}
% 13 HER-InjVer1b
\def\runThirteen{HER-Vertical1b} 
  \begin{table*}[ht]
    \centering
    \caption{List of the HER and LER injection parameters setting during the injection background study. In bold faces (red online) the parameter that was changed, in each given run, with respect to its 
      nominal value. For the injection efficiency we quote the mean value of the distribution 
      of all the measurements taken during the run, with its RMS as the error. In the last line, we quote the initial and final beam current in the run.}
    \vskip 5mm
      \begin{tabular}{ llllll } 
        \toprule
        LER Inj. Param.              & \runFourteen  & \runThree               & \runSix               & \runSeventeen \\
        \midrule
        Phase Shift [$^\circ$]    &  1.0      & {\bf\color{red}31.0}  & 1.0                   & 1.0       \\			
        Vertical Steering 1 [mrad]  &  $-0.378$   & $-0.378$                & $-0.378$                & $-0.378$    \\			
        Vertical Steering 2 [mrad]  &  0.12     & 0.12                  & {\bf\color{red}0.043} & 0.12      \\			
        Septum Angle [mrad]      &  5.51     & 5.51                  & 5.51                  & {\bf\color{red} 5.39} \\			
        Injection Efficiency [\%]  & $76\pm 9$ & $51\pm 10$           & $17\pm 14$            & $39\pm 17$ \\			
        Current Ramp [mA]        & 0-290     & 330-500               & 260-350               & 0-200    \\
        \bottomrule	

        \multispan{5}{\rule{0pt}{3ex}}  \\ 
        \toprule
        HER Inj. Param. & \runTen & \runNine         & \runTwelve         & \runThirteen \\ 
        \midrule
        Phase Shift [$^\circ$]    &     258     & {\bf\color{red}305}  & 258                    & 258                   \\			
        Vertical Steering 1 [mrad]  &     $-0.385$  &  $-0.385$              & {\bf\color{red}$\mathbf{-0.465}$} & {\bf\color{red}$\mathbf{-0.435}$} \\			
        Vertical Steering 2 [mrad]  &     0.08    &  0.08                &  0.08                  &  0.08                  \\			
        Septum Angle [mrad]      &     2.35    &  2.35                &  2.35                  &  2.35                  \\			
        Injection Efficiency [\%]             &  $93\pm 13$ & $72\pm 11$           & $74\pm 9$              & $75\pm 12$             \\			
        Current Ramp [mA]        &    0-150    & 270-450              & 210-300                & 300-400             \\ 
        \bottomrule	
      \end{tabular}
    \label{tab:injdata}
  \end{table*}

The injection backgrounds in the high energy and low energy rings were studied separately. 
In Table\,\ref{tab:injdata} we also list the injection parameters' set values in each run, and the value of the measured injection efficiency. % ({\color{blue}\it{recall definition here...}})
In run number \runFourteen\ (\runTen), that we use as a reference, the LER (HER) injection parameters were set at their optimal values, chosen so as to yield the highest possibile injection efficiency.   

\subsubsection{Injection background time structure}
\begin{figure*}[htb]
  \centering      
%%      \includegraphics[width=0.95\textwidth]{CrystalImages/8oct2017-scalerInj_run010_HER-Reference_scaPlot_LER_refRun14-log.pdf}
%%   \caption{Scaler rates (number of hits) as a function of time after injection recorded in CsI crystals in forward positions F1 through F3 and backward positions B1 through B3. 
%% In each plot, we show data from HER injection taken with the injection parameters set at their nominal values (run \runTen, red open circle) and similarly data from LER injection 
%% (run \runFourteen, blue open squares). The rates are normalised to the beam current at the time each scaler reading was taken (colors online).}
      \includegraphics[width=0.95\textwidth]{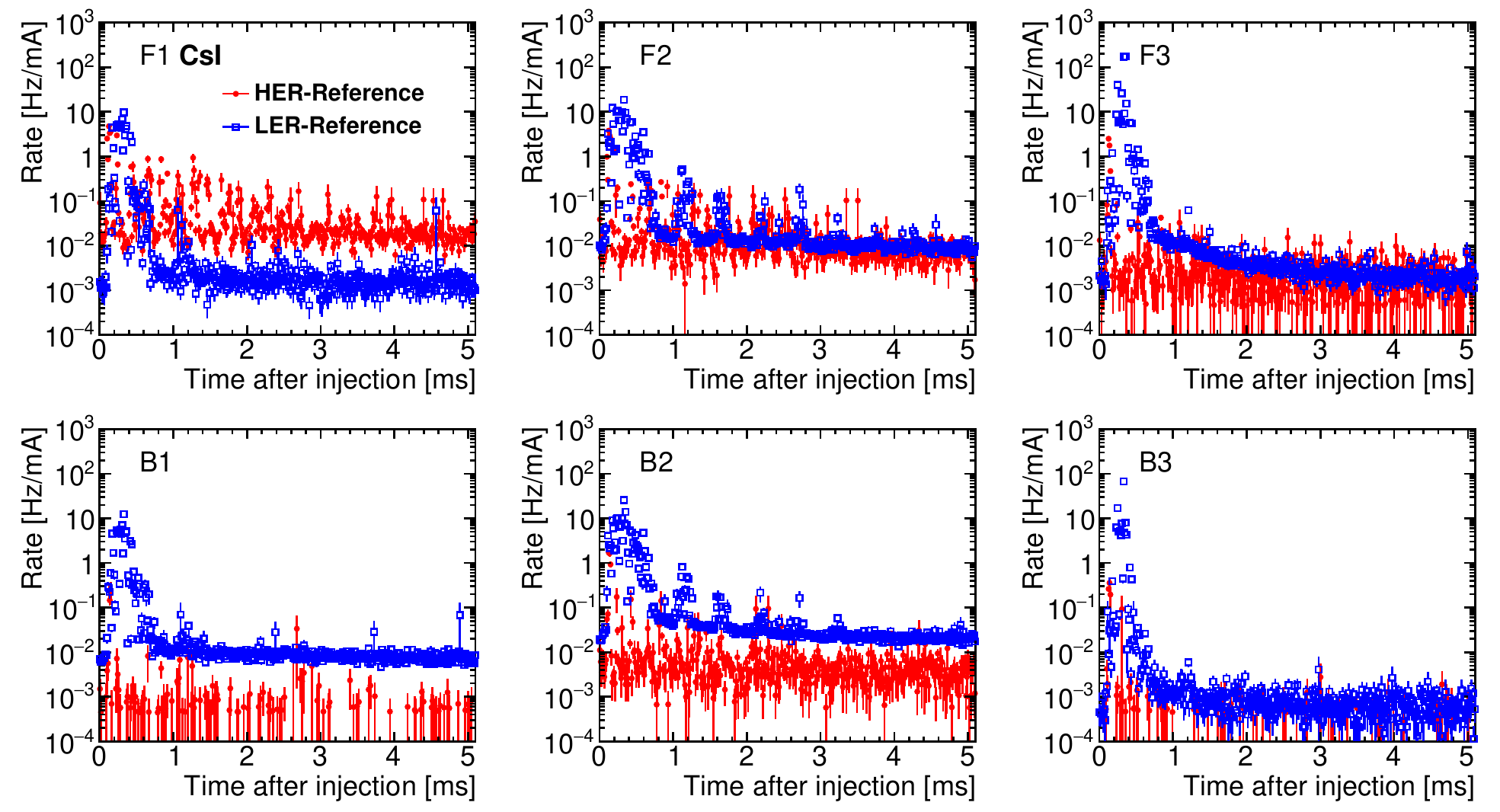}
      \caption{(color online) Scaler rates as a function of time after injection recorded in CsI crystals in forward positions F1 through F3 and backward positions B1 through B3. 
        In each plot, we show data from HER injection taken with the injection parameters set at their nominal values (run \runTen, red open circle) and similarly data from LER injection 
        (run \runFourteen, blue open squares). The rates are normalized to the beam current at the time each scaler reading was taken (colors online). 
%        The frequency is obtained rescaling the histograms to the average rate recorded in the 5~ms gate (total number of hits per mA of beam current in each histogram, divided by 5~ms).}
        To express the rate in Hz, we normalized the histograms to the average rate recorded in the 5~ms gate (total number of hits per mA of beam current in each histogram, divided by 5~ms).}
      \label{fig:scalerInj-HERLER}
\end{figure*}

The knowledge of the time structure of the injection background is of great importance for the design of a vetoing scheme for the data acquisition systems of the Belle II experiment, and in
particular for the electromagnetic calorimeter.
In the Belle experiment, where the continuous injection rate was 1~Hz, the calorimeter DAQ was inhibited for $\approx 3.5~$ms after the injection for all bunches.
At the 100~Hz rate of continuous injection foreseen during SuperKEKB normal operation, this scheme would lead to an unacceptably high deadtime level of 35\%, therefore a more sophisticated veto must be adopted.
One possibility is to limit the veto to the bunches which are most affected by the injection, for example the injected bunch and its nearest neighbors. To 
implement such a veto, one has to study with high precision when the backgrounds occur in the detector with respect to time of the injected bunch.

We performed such study by looking at the hit rate as a function of the time after injection for each crystal. 
To have sufficient statistics in each bin, we sum the data of all the injection gates in a given run. Runs usually lasted several minutes, including ramp-up time from its initial to its final values listed in Tab.\,\ref{tab:injdata}, and contain up to a few thousands injection gates delivered at a rate from a few Hz up to 25~Hz. 
The duration of each run was short enough to ensure that the beam conditions were reasonably constant within the measurement time.
As the background hit rate increases with increasing beam currents, we normalize the measured rates to the value of the beam current at the time of each injection gate;
to correctly compare data from different runs, we normalize each plot to the number of injection gates in the run.

We have analysed the data taken in each of the six positions, three forward and three backward, for each of the three crystal species CsI(Tl), CsI and LYSO, and for each of the 8 runs 
listed in Table\,\ref{tab:injdata}. In the following we shall present a selection of the most interesting results.
 
Firstly, we compare HER and LER injection, using data from each ring's reference run.
In Fig.\,\ref{fig:scalerInj-HERLER} we show a set of plots of the normalized hit rate measured with the scalers in CsI crystals, as a function of the time after injection, 
for each of the three forward positions: F1 ($\phi=0^\circ$, horizontal plane
ring's outside), F2 ($\phi=90^\circ$, vertical plane), F3 ($\phi=180^\circ$, horizontal plane, ring's inside) and the corresponding backward positions 
B1, B2, B3. 

It would be reasonable to expect that backgrounds originating near the interaction point during HER injection produce slightly higher hit rates in the forward crystals, i.e.\ in the 
direction the beam is travelling, than in the backward direction and vice versa during LER injection.
This seems to be confirmed by the data shown in these plots, which also indicate that, for a given beam current the backgrounds 
are generally higher in the LER (run \runFourteen, blue squares) than in the HER (run \runTen, red circles).

\begin{figure}[htb]
  \centering      
  %% \includegraphics[width=0.975\columnwidth]{CrystalImages/8oct2017-scalerInj_run003_LER-Phase_scaPlot_LER_Cell4_refRun14-log.pdf}
  %% \caption{Effect of changing the injection phase angle on the measured injection background in crystals. We compare the normalized scaler rate as a function of 
  %%   time after injection recorded by the CsI crystal in position F2 in run \runThree\ (red circles) with those of run \runFourteen\ (blue squares, colors online).
  %%   The non optimal injection phase angle induces a much higher background in the crystals, as expected.
  %%   The peaks in rates observed in the first about 3~ms have a period consistent with being due to synchrotron oscillations.}
  \includegraphics[width=0.975\columnwidth]{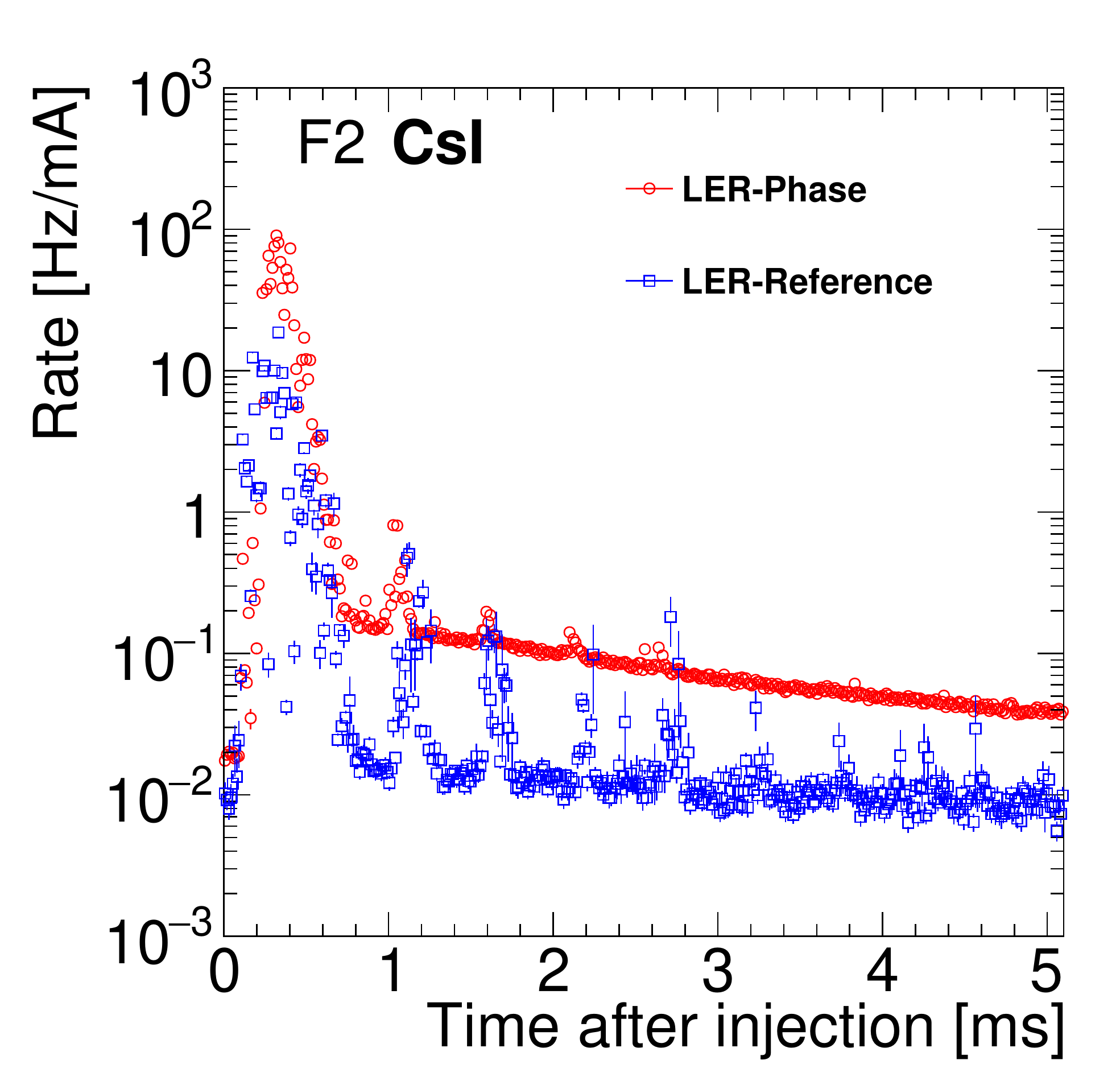}
  \caption{(color online) Effect of changing the injection phase angle on the measured injection background in the Crystals detector system. We compare the normalized scaler rate as a function of 
    time after injection recorded by the CsI crystal in position F2 in run \runThree\ (red circles) with those of run \runFourteen\ (blue squares).
    The non optimal injection phase angle induces a much higher background in the Crystals, as expected.
    The peaks in rates observed in the first about 3~ms have a period consistent with being due to synchrotron oscillations.
%%    The frequency is obtained rescaling the histogram to the average rate recorded in the 5~ms gate (total number of hits per mA of beam current in the histogram, divided by 5~ms).}
    To express the rate in Hz, we normalized the histogram to the average rate recorded in the 5~ms gate (total number of hits per mA of beam current in the histogram, divided by 5~ms).}
  \label{fig:scaInjrun3-F2}
\end{figure}

In Fig.\,\ref{fig:scaInjrun3-F2} we show the effect of changing the LER injection phase shift from its nominal value of $1^\circ$ to the value of $31^\circ$.
The plot shows the normalized scaler rates in the CsI crystal in position F2 as recorded during run \runThree, compared to that of reference run \runFourteen. 
We observe that the background is higher when the injection parameter is changed from its optimal value.
Another interesting difference is that the background hit rate in run \runThree\ decays with a slower time constant with respect to run \runFourteen.

In the case of the HER, 
changing the injection phase shift from its nominal value of 258$^\circ$ (reference run \runTen) to 305$^\circ$ (run \runNine) has a much less pronounced effect than that of the LER.

However, in both HER and LER data we see several peaks in the hit rate appearing at regular time intervals %at $\approx 400~\mu s$ (about 40 turns around the machine arcs), 
as late as 3~ms after injection; these peaks are also seen in the reference run, but they tend to disappear within the first 1~ms. 
These periodic surges in the background hit rate may be related to longitudinal oscillations of the beam particles within the RF bucket (synchrotron oscillations).
These oscillations cause the particles to fall out of phase with the accelerating field and thus change their energy. 
%This oscillation is self dumping, but 
In the highly perturbed injected bunch, the change in energy and momentum may occasionally be so large that the particles are lost from the beam and generate the high backgrounds detected by the Crystals with a characteristic period.

Synchrotron oscillations in SuperKEKB have a nominal period of 50.5 turns in the LER and 40.7 turns for the HER\,\cite{Funakoshi}.
Since the actual values in each run may vary from these nominal values for different machine parameter settings, a tracking simulation of the beam orbits was performed\,\cite{Funakoshi} using the
same machine parameters in use during data runs \runThree, \runFourteen, \runNine\ and \runTen, to obtain a prediction for the synchrotron oscillations period that is directly comparable to the data.
To measure the period of the oscillations, we have rebinned the plots of the rate as a function of the time after injection in order to make the peaks position more defined, 
 taken the average time difference between adjacent peaks in the hit rate and converted it to number of turns. 

The results for all runs are summarized in Table\,\ref{tab:syncRes}, where we list our measurements together with  their predictions from simulation, finding excellent agreement.  

\begin{table}[ht]
  \caption{Measured values of the synchrotron oscillation period using injection background Crystals data. The measurements are compared with those obtained from a tracking-based 
    simulation of the beam orbits, using for the machine parameters the same values in use during each run.} \vskip 5mm
  \centering
  \begin{tabular}{ lcc } 
    \toprule
    Run    &   Experiment [\# turns]  &   Simulation [\# turns] \\  
    \midrule
    \runFourteen  &     52.5$\pm$0.5              &      52.3               \\			
    \runThree   &       53.1$\pm$1.4            &        53.3             \\			
    \runTen       &     40.6$\pm$0.7              &      40.6               \\			
    \runNine     &       41.7$\pm$1.0            &        42.4          \\
    \bottomrule
  \end{tabular}
  \label{tab:syncRes}
\end{table}

\begin{figure}[htb]
  \centering      
  %% \includegraphics[width=0.975\columnwidth]{CrystalImages/8oct2017-scalerInj_run017_LER-Septum_scaPlot_LER_Cell13_refRun14-log.pdf}
  %% \caption{
  %%   Effect of changing the injection septum angle on the measured injection background in crystals. We compare the normalized scaler rate as a function of 
  %%   time after injection recorded by the CsI crystal in position B2 in run \runSeventeen\ (red circles) with those of run \runFourteen\ (blue squares, colors online).
  %%   The non optimal injection septum angle induces a higher background in the crystals mostly in the first ms. 
  %%   The synchrotron oscillation peaks in rates are enhanced with respect to the reference run.
  %% }
  \includegraphics[width=0.975\columnwidth]{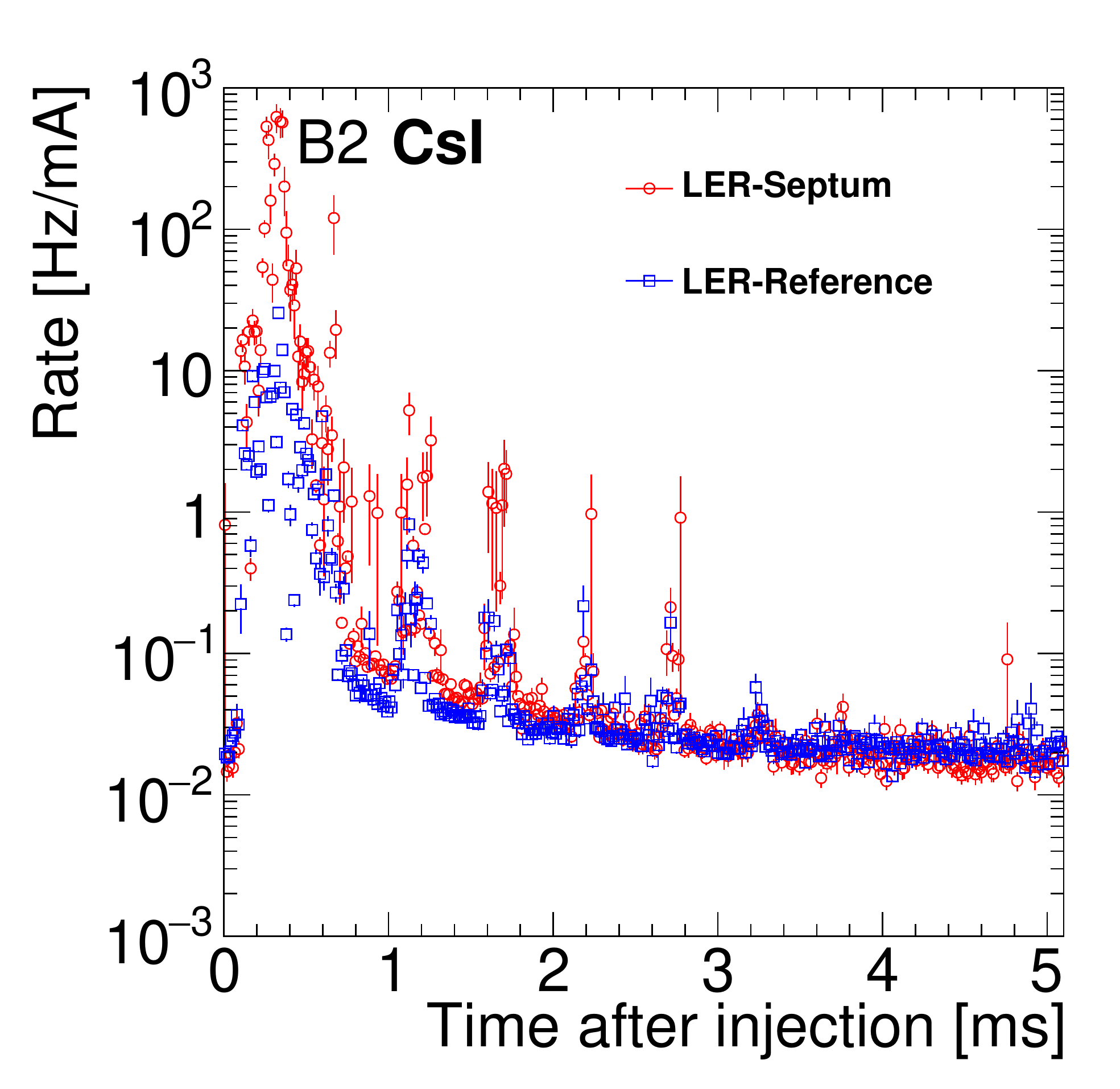}
  \caption{
    (color online) Effect of changing the injection septum angle on the measured injection background in Crystals. We compare the normalized scaler rate as a function of 
    time after injection recorded by the CsI crystal in position B2 in run \runSeventeen\ (red circles) with those of run \runFourteen\ (blue squares).
    The non optimal injection septum angle induces a higher background in the Crystals mostly in the first ms. 
    The synchrotron oscillation peaks in rates are enhanced with respect to the reference run.
    To express the rate in Hz, we normalized the histogram to the average rate recorded in the 5~ms gate (total number of hits per mA of beam current in the histogram, divided by 5~ms).}
  \label{fig:LERSeptum}
\end{figure}

We now present the results obtained by changing the other injection parameters. 
In Fig.\,\ref{fig:LERSeptum} we show the effect of changing the septum angle in the LER. In this case, as well as for the HER (not shown), the effect on injection background is 
also clearly visible, occurring mostly in the first ms after injection, in all crystals, of which we only show CsI for brevity. Also the synchrotron oscillation peaks are enhanced with respect to the reference run.

Changing the vertical steering angle in both rings produced instead a limited overall increase of the injection background in both the LER and HER;
perhaps the amount of change in the parameters' values was not large enough to produce any significant change in the background levels.

In general, the duration and amount of injection background showed great variations throughout the BEAST II running. 
We also observed large variations in the same runs at different positions and for different crystals. 
A common feature however, is that most of the injection related activity damps down by about two (or three) orders of magnitude within about 1-1.5~ms after the injection.

\begin{figure*}
  \subfigure{
    \includegraphics[width=0.975\columnwidth]{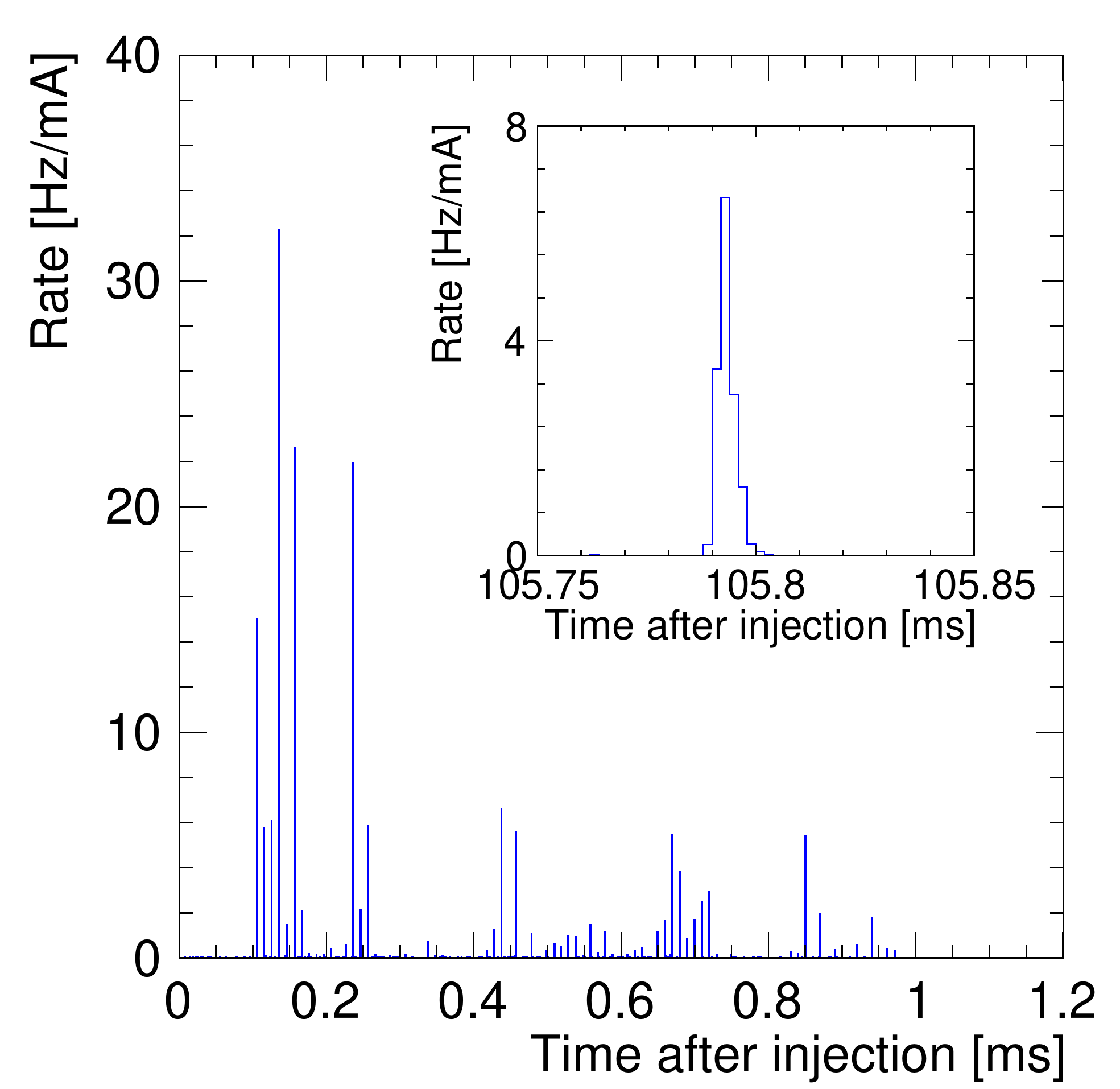}
  }
  \subfigure{
  \includegraphics[width=0.975\columnwidth]{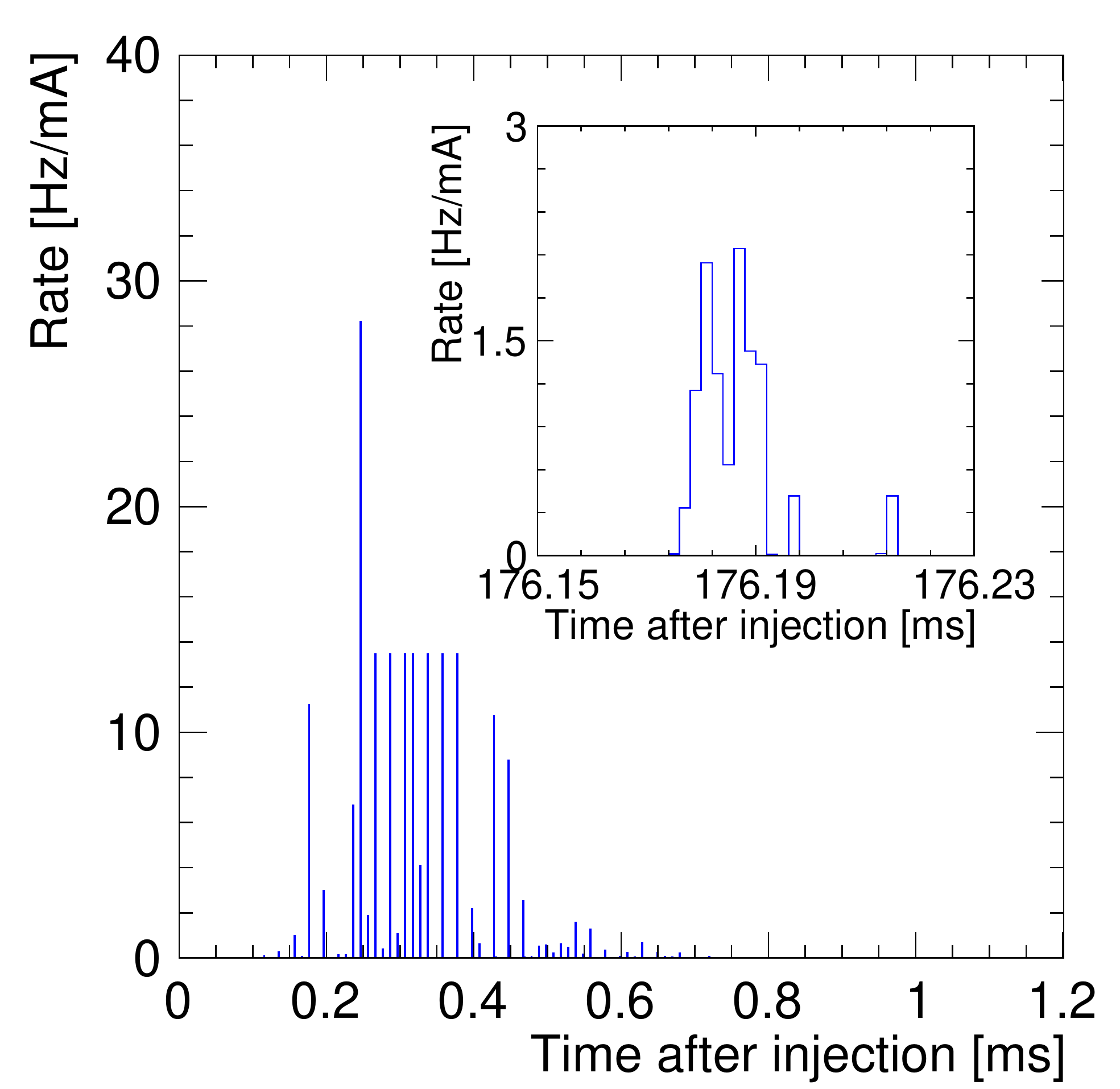}
  }
  \caption{
Injection background rate in the Crystals measured with high time resolution using the digitizer. 
 We show the normalized rate recorded in the first 1~ms after injection with a time bin of 1~\si{\micro}s. 
The high background peaks are typically spaced about 10~\si{\micro}s, which corresponds to 1 turn. In the inset we show a zoomed-in view around the first background peak with a bin size of 2 ns.
The plot on the left shows data taken during HER injection with nominal parameters (run \runTen) with the CsI crystal in position F1; in the zoomed in view a $<2$~ns wide peak is clearly visible,
showing that the background is coming from one bunch. The plot on the right similarly shows data from LER injection (run \runFourteen) with the CsI crystal in position B3; this time the
zoomed in view shows two $<2$~ns wide peaks separated by $6$~ns, which is consistent with there being two consecutive bunches contributing to this particular background spike.
To express the rate in Hz, we normalized the histogram to the average rate recorded in the 1~ms digitizer acquisition gate (total number of hits per mA of beam current in the histogram, divided by 1~ms).
}
  \label{fig:timeHiRes}
\end{figure*}

In the study of injection background discussed in the Belle II TDR mentioned at the beginning of this section, it was found that typically most noise hits occurred mostly right after injection, 
and that they are highly correlated in time with the time of passage of the injected bunch near the interaction point.
Using the high resolution timing data recorded by the digitizer we have looked for a similar behaviour in the BEAST II data. The digitizers record the time of each hit with 2 ns precision, 
so each hit is assigned a well defined time after the injection, and we can resolve the finest details of the injection background's time distribution.
As the radio frequency bucket size of SuperKEKB rings is 1.965~ns, and the bunch were spaced three buckets apart ($\approx$6~ns), 
at this resolution it is possible to resolve the background contribution of the single bunch.  
An example of this is given in Fig.\,\ref{fig:timeHiRes} 
where we show the distribution of the hit rate in the first ms after 
HER or LER injection with a bin size of 1~\si{\micro}s; in the insets, a blow-up view of one of the peaks with a time bin of 2~ns is shown.
The plot on the left shows that the background hits are associated to a single bunch, which is observed in almost all the data;
the plot on the right shows instead a rarer case in which the hits are associated to two consecutive bunches.

The bunch revolution time around the machine arcs is $T_{rev}$=10.0614~\si{\micro}s. This means that 
every given bunch crosses the interaction point every $T_{rev}$~\si{\micro}s, followed a few ns later by the bunch right behind it, followed in turn by the next bunch and so on. The time interval between
subsequent bunches depends on the fill configuration pattern. 
After $T_{rev}$~\si{\micro}s (one complete turn) all the bunches filled in the machine will have passed through the interaction point, and the next turn begins repeating this pattern. 
Each bunch crosses the IP at its own time within the time $T_{rev}$ of one turn, depending on its position in the train of bunches filling the machine. 
So indicating with $T_{inj}$ the time after injection recorded by the digitizers in ns, we compute the time within one turn as $T_{turn}=T_{inj}\bmod T_{rev}$. In this way, if one particular bunch 
crossing the IP (i.e.\ the injected bunch) generates some background that produces hits in the Crystals, the hits will all have the same $T_{turn}$ of that particular bunch.

\begin{figure*}
	\centering
        \includegraphics[width=0.975\textwidth]{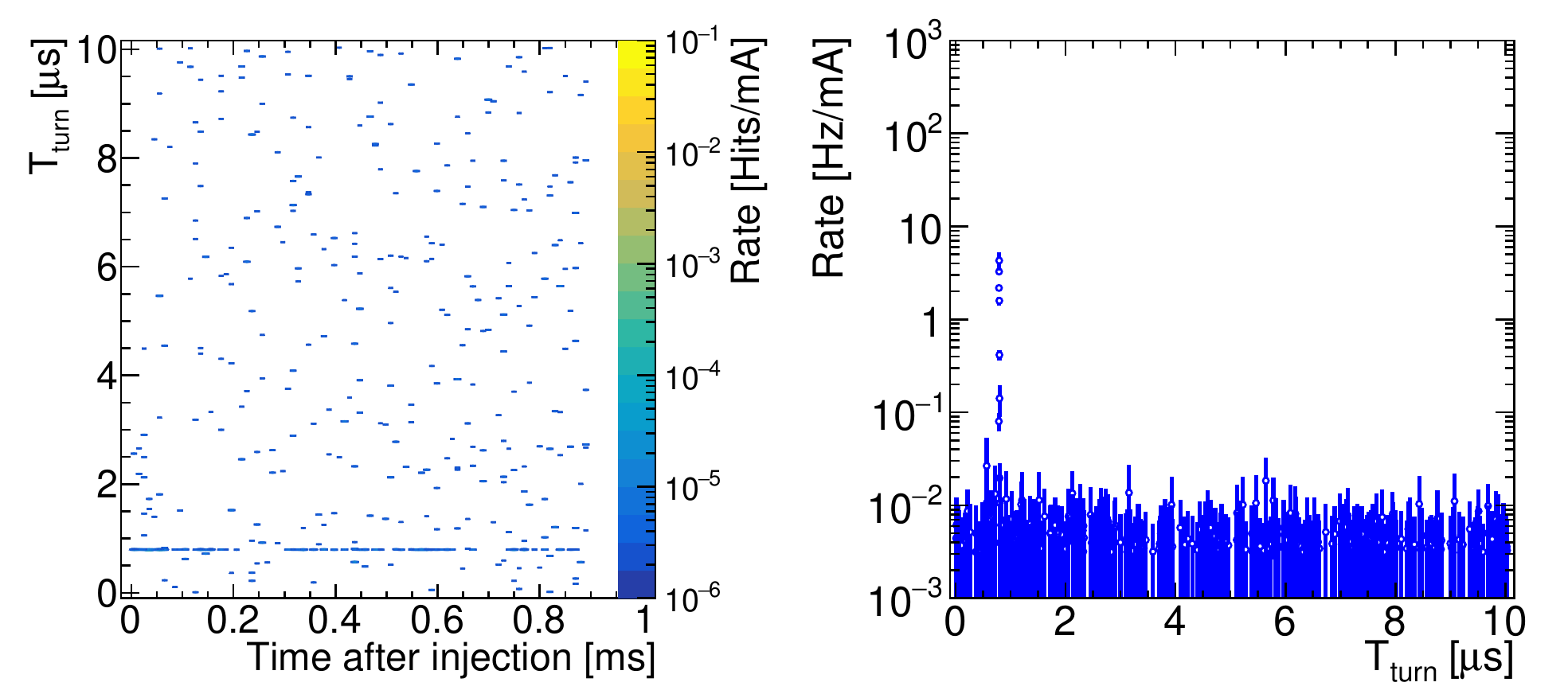}
	\caption{
          Detailed time structure of HER background rates in the Crystals measured with high time resolution using the digitizer.
          We show (left) a plot of the hit time after injection in 1-turn bins (10~\si{\micro}s) versus the hit time within one turn ($T_{turn}$) in 2~ns bins, for the CsI crystal in position F3 taken
          during HER injection with nominal injection parameters (run \runTen), showing an accumulation of hits around the same value of $T_{turn}$. 
          This is clearly evidenced in the projection along the $T_{turn}$ axis (right), which shows a peak of a few ns width. The background hits closely correlated in
          time with the injected bunch are 2-3 order of magnitudes more frequent than those occurring at different times. This timing structure is present in data from all crystals.
          To express the rate in Hz, we normalized the histogram to the average rate recorded in the 1~ms digitizer acquisition gate (total number of hits per mA of beam current in the histogram, divided by 1~ms).
        }
        \label{fig:TinTurnHER}
\end{figure*}

\begin{figure*}
	\centering
        \includegraphics[width=0.975\textwidth]{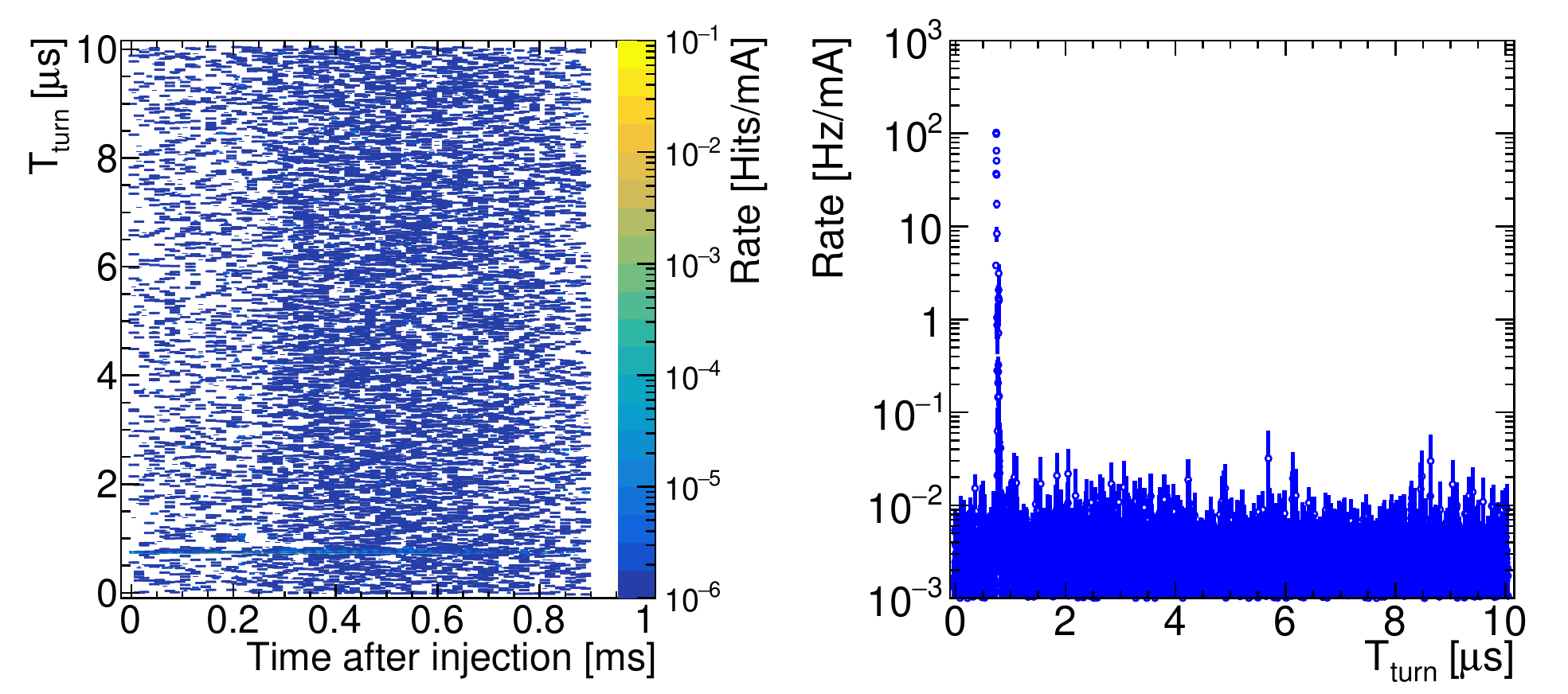}
	\caption{
          Detailed time structure of LER background rates in the Crystals measured with high time resolution using the digitizer.
          We show (left) a plot of the hit time after injection in 1-turn bins (10~\si{\micro}s) versus the hit time within one turn ($T_{turn}$) in 2~ns bins, for the CsI crystal in position F3 taken
          during LER injection with nominal injection parameters (run \runTen), showing an accumulation of hits around the same value of $T_{turn}$. 
          This is clearly evidenced in the projection along the $T_{turn}$ axis (right), which shows a peak of a few ns width. The background hits closely correlated in
          time with the injected bunch are 2-3 orders of magnitude more frequent than those occurring at different times. This timing structure is present in data from all crystals.
          Although the overall injection background level is higher than in the HER, the timing structure is very similar as that shown in Fig.\,\ref{fig:TinTurnHER}.
          To express the rate in Hz, we normalized the histogram to the average rate recorded in the 1~ms digitizer acquisition gate (total number of hits per mA of beam current in the histogram, divided by 1~ms).
        }
        \label{fig:TinTurnLER}
\end{figure*}

We show in Fig.\,\ref{fig:TinTurnHER} (\ref{fig:TinTurnLER}) a plot of the time after injection $T_{inj}$ versus $T_{turn}$ recorded by the crystals in position F3 during the injection of the HER (LER). 
As the data show, is clear that the background hits are correlated in time with one bunch; 
projecting the data on the $T_{turn}$ axis, this feature appears as a few nanosecond wide peak in the distribution. The background hits closely correlated in 
time with the injected bunch are 2-3 order of magnitudes more frequent than those occurring at different times. This timing structure is present in data from all crystals.
%The different widths reflect the different time characteristics of the scintillation light emission of the different crystals.
The difference in overall hit rate between HER and LER injection observed using the digitizer data, confirms the observation made earlier looking at the scalers data that 
the injection of the LER produces higher backgrounds. 

\begin{figure*}
	\centering
       	\includegraphics[width=0.975\textwidth]{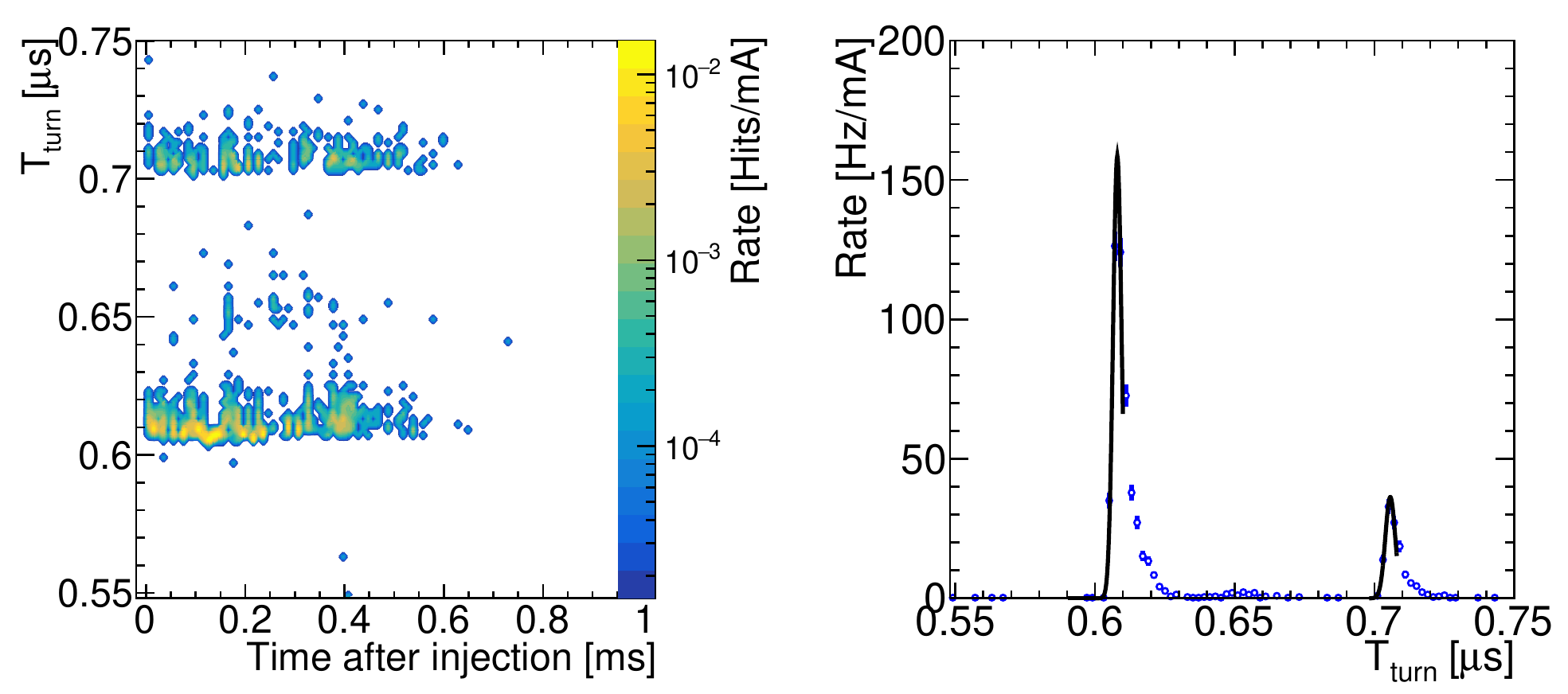}
	\caption{
          Injection background in the Crystals for two-bunch injection in LER. We show a plot of the hit time after injection versus the hit time within one turn $T_{turn}$ (left) recorded by the LYSO crystal in position F2 during 
          a run in which two bunches were injected at the same time in the LER. 
          The apperance of two peaks indicates that the background hits are highly correlated in time with the injected bunches. 
Using the projection along the $T_{turn}$ axis (right), 
          we find that the two peaks are separated by $\approx 97$~ns, in good agreement with the expected time separation of the two injected bunches, which confirms this indication.
          To express the rate in Hz, we normalized the histogram to the average rate recorded in the 1~ms digitizer acquisition gate (total number of hits per mA of beam current in the histogram, divided by 1~ms).
        }
       	\label{fig:twoBunchLYSO}
\end{figure*}

A confirmation that the bunch to which the background is correlated in time is indeed the one being injected is obtained by looking at a run in which two bunches, separated
by 49 RF buckets (96.3 ns), are injected in the LER. 
The data is shown Fig.\,\ref{fig:twoBunchLYSO} where the $T_{turn}$ plots axis is zoomed in to show more details; the rate of hits accumulates clearly around two distinct values
of $T_{turn}$ forming two peaks with a separation of $\approx 97$, in good agreement with the expectation.

To synthetically describe the features of the injection background, based on the observations presented above and following previous injection background studies, 
we define two ratios: the fraction $F_{<1ms}$ defined as the total number of hits occurring within the first millisecond after injection ($T_{inj}<$1~ms) over the total number of hits
within the 5 ms acquisition gate; 
the fraction $F^{R}_{bunch}$ of all the hits occurring within the first millisecond after injection ($T_{inj}<$1~ms) that are also
associated to the injected bunch. We define hits as being associated to the injected bunch those whith $T_{turn}$ inside a window of 
$|T_{turn}-T_{bunch}|<\pm$7~ns for CsI and LYSO crystals, and $|T_{turn}-T_{bunch}|<\pm20$~ns for CsI(Tl), to take into account the different characteristic
emission time of the scintillation light of the different crystals.
We compute the fractions $F_{<1ms}$ using the scalers data that is acquired for a full 5~ms after injection, therefore we can do this only for the CsI and LYSO crystals, 
whereas to compute the fractions $F^R_{bunch}$ we use the higher resolution time data from the digitizers, so we compute them for all crystals.
As the digitizer also provides a measurement of the hits energy, we can also define the fraction of energy associated
to the bunch $F^{E}_{bunch}$ obtained summing the energy of all the hits satisfying the same timing criteria that define $F^{R}_{bunch}$.

In Table~\ref{tab:LER-Rate-Energy_CsI-ALLboxes_Fracs.txt} we list the rate fractions recorded in the CsI crystal in all positions during the injection study of the LER, 
and in Table~\ref{tab:HER-Rate-Energy_CsI-ALLboxes_Fracs.txt} the same for the HER injection. As the data shows, using the nominal injection parameters 
about $\approx$95\% (75\%) of the injection background hits and associated energy in the LER (HER) occurs within the first 1~ms from the injection, 
and of these hits, $\approx$90\% (75\%) are strictly correlated in time with the injected bunch. This would allow one to reduce significantly the dead time associated 
with a veto by vetoing the DAQ for a few ms, but only for a few tens of ns around the injected bunch.

We clearly see from the table that when the injection parameters are away from their nominal values, the background tends to be more spread out as the $F^R_{bunch}$ and $F^E_{bunch}$ 
fractions are reduced. While in most cases the background is still contained in the first ms after injection as the $F_{<1ms}$ remain very high, there are few in
which this fraction also become low, thus making a possible injected bunch veto less effective. 
%% /////////////////////
\begin{sidewaystable*}
  \begin{subtable}{}
    %  \centering
    \caption{LER - Rate and energy injection background fractions in CsI crystals for the different injection study runs. The fraction $F_{1<ms}$ was computed using only scalers data.
      % which was not available for CsI(Tl) crystals. 
      $F^R_{bunch}$ represents the fraction of hits associated to the injected bunch, while $F^E_{bunch}$
      represents what fraction of the total energy recorded by the crystals is associated to the bunch, according to the same criteria used to compute $F^R_{bunch}$ described in the text.}
    \vskip 5mm
    \footnotesize
    \begin{tabular}{ llllllllllllll }
      \toprule
      &   & \multicolumn{3}{c}{\runFourteen}   & \multicolumn{3}{c}{\runThree}   & \multicolumn{3}{c}{\runSix}   & \multicolumn{3}{c}{\runSeventeen}  \\ 
      Crystal & Position  & \multicolumn{1}{c}{$F_{<1ms}$(\%)\Bt } & \multicolumn{1}{c}{$F^R_{bunch}$(\%)} & \multicolumn{1}{c}{$F^E_{bunch}$(\%)}  & \multicolumn{1}{c}{$F_{<1ms}$(\%)\Bt } & \multicolumn{1}{c}{$F^R_{bunch}$(\%)} & \multicolumn{1}{c}{$F^E_{bunch}$(\%)}  & \multicolumn{1}{c}{$F_{<1ms}$(\%)\Bt } & \multicolumn{1}{c}{$F^R_{bunch}$(\%)} & \multicolumn{1}{c}{$F^E_{bunch}$(\%)}  & \multicolumn{1}{c}{$F_{<1ms}$(\%)\Bt } & \multicolumn{1}{c}{$F^R_{bunch}$(\%)} & \multicolumn{1}{c}{$F^E_{bunch}$(\%)} \\ \hline
      \multirow{6}{*}{CsI} & F1 \Tp & 98.8 $\pm$ 0.005& 96.1 $\pm$0.6 & 99.7 $\pm$0.01 & 98.4 $\pm$ 0.004& 83.5 $\pm$0.6 & 98.1 $\pm$0.02 & 90.5 $\pm$ 0.05& 87.0 $\pm$0.7 & 98.9 $\pm$0.02 & 99.7 $\pm$ 0.0006& 32.2 $\pm$ 2 & 34.6 $\pm$0.2 \Bt \\ 
      & F2 & 95.5 $\pm$ 0.006& 94.3 $\pm$0.4& 99.3 $\pm$0.008 & 97.0 $\pm$ 0.002& 69.1 $\pm$0.3& 92.2 $\pm$0.03 & 78.1 $\pm$ 0.04& 75.3 $\pm$0.5& 96.8 $\pm$0.02 & 98.8 $\pm$ 0.001& 34.1 $\pm$ 1& 34.0 $\pm$0.07 \Bt \\ 
      & F3 & 99.7 $\pm$ 0.0009& 88.4 $\pm$0.9& 98.7 $\pm$0.02 & 94.1 $\pm$ 0.004& 17.6 $\pm$0.09&  1.4 $\pm$0.0007 & 97.3 $\pm$ 0.009& 66.0 $\pm$0.7& 95.8 $\pm$0.03 & 99.9 $\pm$ 0.0002& 31.1 $\pm$ 2& 34.0 $\pm$0.1 \Bt \\ 
      & B1 & 96.1 $\pm$ 0.008& 83.8 $\pm$ 1& 98.9 $\pm$0.03 & 60.5 $\pm$ 0.04& 57.4 $\pm$0.4& 91.1 $\pm$0.06 & 69.9 $\pm$ 0.06& 62.1 $\pm$0.7& 95.8 $\pm$0.04 & 98.6 $\pm$ 0.003& 29.8 $\pm$ 2& 33.6 $\pm$0.2 \Bt \\ 
      & B2 & 93.1 $\pm$ 0.007& 90.8 $\pm$0.5& 99.2 $\pm$0.001 & 93.0 $\pm$ 0.002& 32.3 $\pm$0.2&  5.0 $\pm$0.0006 & 71.4 $\pm$ 0.03& 57.3 $\pm$0.4& 93.4 $\pm$0.002 & 99.3 $\pm$ 0.0004& 31.8 $\pm$ 1& 34.2 $\pm$0.008 \Bt \\ 
      & B3 & 99.8 $\pm$ 0.002& 90.2 $\pm$ 1& 79.6 $\pm$0.07 & 91.5 $\pm$ 0.03& 68.5 $\pm$ 1& 25.1 $\pm$0.07 & 98.0 $\pm$ 0.02& 70.2 $\pm$ 1& 38.0 $\pm$0.05 & 99.9 $\pm$ 0.0005& 31.0 $\pm$ 3& 22.6 $\pm$0.1 \Bt \\ 
      \bottomrule
    \end{tabular}
    \label{tab:LER-Rate-Energy_CsI-ALLboxes_Fracs.txt}
  \end{subtable}

  \vskip 2cm

  \begin{subtable}{}
    %  \centering
    \caption{HER - Rate and energy injection background fractions in CsI crystals for the different injection study runs. The fraction $F_{1<ms}$ was computed using only scalers data.
      %, which was not available for CsI(Tl) crystals. 
      $F^R_{bunch}$ represents the fraction of hits associated to the injected bunch, while $F^E_{bunch}$
      represents what fraction of the total energy recorded by the crystals is associated to the bunch, according to the same criteria used to compute $F^R_{bunch}$ described in the text.
      In some cases the recorded data was insufficient to determine some of the fractions.}
    %  \caption{HER - Rate-Energy injection fractions.}
    \vskip 5mm
    \footnotesize
    \begin{tabular}{ llllllllllllll }
      \toprule
      &   & \multicolumn{3}{c}{\runTen}   & \multicolumn{3}{c}{\runNine}   & \multicolumn{3}{c}{\runTwelve}   & \multicolumn{3}{c}{\runThirteen}  \\ 
      Crystal & Position  & \multicolumn{1}{c}{$F_{<1ms}$(\%)\Bt } & \multicolumn{1}{c}{$F^R_{bunch}$(\%)} & \multicolumn{1}{c}{$F^E_{bunch}$(\%)}  & \multicolumn{1}{c}{$F_{<1ms}$(\%)\Bt } & \multicolumn{1}{c}{$F^R_{bunch}$(\%)} & \multicolumn{1}{c}{$F^E_{bunch}$(\%)}  & \multicolumn{1}{c}{$F_{<1ms}$(\%)\Bt } & \multicolumn{1}{c}{$F^R_{bunch}$(\%)} & \multicolumn{1}{c}{$F^E_{bunch}$(\%)}  & \multicolumn{1}{c}{$F_{<1ms}$(\%)\Bt } & \multicolumn{1}{c}{$F^R_{bunch}$(\%)} & \multicolumn{1}{c}{$F^E_{bunch}$(\%)} \\ \hline
      \multirow{6}{*}{CsI} & F1 \Tp & 65.9 $\pm$ 0.046& 92.9 $\pm$1.9 & 96.3 $\pm$0.16 & 43.9 $\pm$ 0.05& 76.8 $\pm$ 7 & 91.8 $\pm$0.72 & 33.4 $\pm$ 0.054 & \hfill & \hfill & 30.6 $\pm$ 0.065& 33.9 $\pm$1.5 & 65.9 $\pm$0.77 \Bt \\ 
      & F2 & 75.2 $\pm$ 0.06& 94.1 $\pm$2.9& 97.5 $\pm$0.22 & 49.7 $\pm$ 0.072& 79.1 $\pm$10& 90.7 $\pm$1.1 & 35.5 $\pm$ 0.08& \hfill& \hfill & 32.5 $\pm$ 0.096& 37.9 $\pm$2.4& 73.6 $\pm$0.75 \Bt \\ 
      & F3 & 85.5 $\pm$ 0.087& 88.8 $\pm$21& 74.1 $\pm$9.4 & 56.6 $\pm$ 0.13& 75.0 $\pm$3.8& 84.5 $\pm$0.6 & 42.3 $\pm$ 0.15& 46.3 $\pm$7.2& 61.6 $\pm$8.8 & 38.6 $\pm$ 0.18& 37.3 $\pm$5.9& 66.1 $\pm$1.1 \Bt \\ 
      & B1 & 83.6 $\pm$ 0.25& 83.8 $\pm$39& 87.9 $\pm$7.5 & 63.8 $\pm$ 0.41& 76.8 $\pm$18& 81.5 $\pm$3.9 & 53.1 $\pm$ 0.47& 57.6 $\pm$33& 66.8 $\pm$ 7 & 59.9 $\pm$ 0.52& 57.2 $\pm$27& 69.5 $\pm$5.4 \Bt \\ 
      & B2 & 72.4 $\pm$ 0.1& 68.7 $\pm$20& 79.3 $\pm$0.24 & 40.9 $\pm$ 0.11& 45.9 $\pm$1.9& 61.8 $\pm$0.17 & 32.3 $\pm$ 0.11& 20.1 $\pm$4.5& 33.1 $\pm$0.93 & 31.1 $\pm$ 0.13& 14.1 $\pm$2.7& 31.5 $\pm$0.077 \Bt \\ 
      & B3 & 96.5 $\pm$ 0.12& 75.4 $\pm$54& 21.1 $\pm$23 & 75.7 $\pm$ 0.32& 69.1 $\pm$7.5& 64.1 $\pm$0.59 & 69.7 $\pm$ 0.39& 43.1 $\pm$36& 43.9 $\pm$2.8 & 73.5 $\pm$ 0.43& 33.7 $\pm$27& 35.5 $\pm$2.1 \Bt \\ 
      \bottomrule
    \end{tabular}
    \label{tab:HER-Rate-Energy_CsI-ALLboxes_Fracs.txt}
  \end{subtable}
\end{sidewaystable*}

%%%%%%%%%%%%
\subsubsection{Injection background energy}
As we have mentioned, we measure the energy of the radiation incident on each crystal by converting the charge of the hits recorded by the digitizers to energy using the calibration described in Sec.\,\ref{sec:CrystalCal}.

Measuring the energy (charge) associated with the injected bunches is very important for the the Belle II electromagnetic calorimeter DAQ vetoing scheme, because one has to make sure that no saturation occurs 
in the crystals front end electronics during the time in which the DAQ is vetoed. 
This measurement can be accomplished by exploiting the digitizer capability to measure both energy and time of crystal hits. 

\begin{figure*}[htb]
  \centering
  \subfigure{
    \includegraphics[width=0.975\columnwidth]{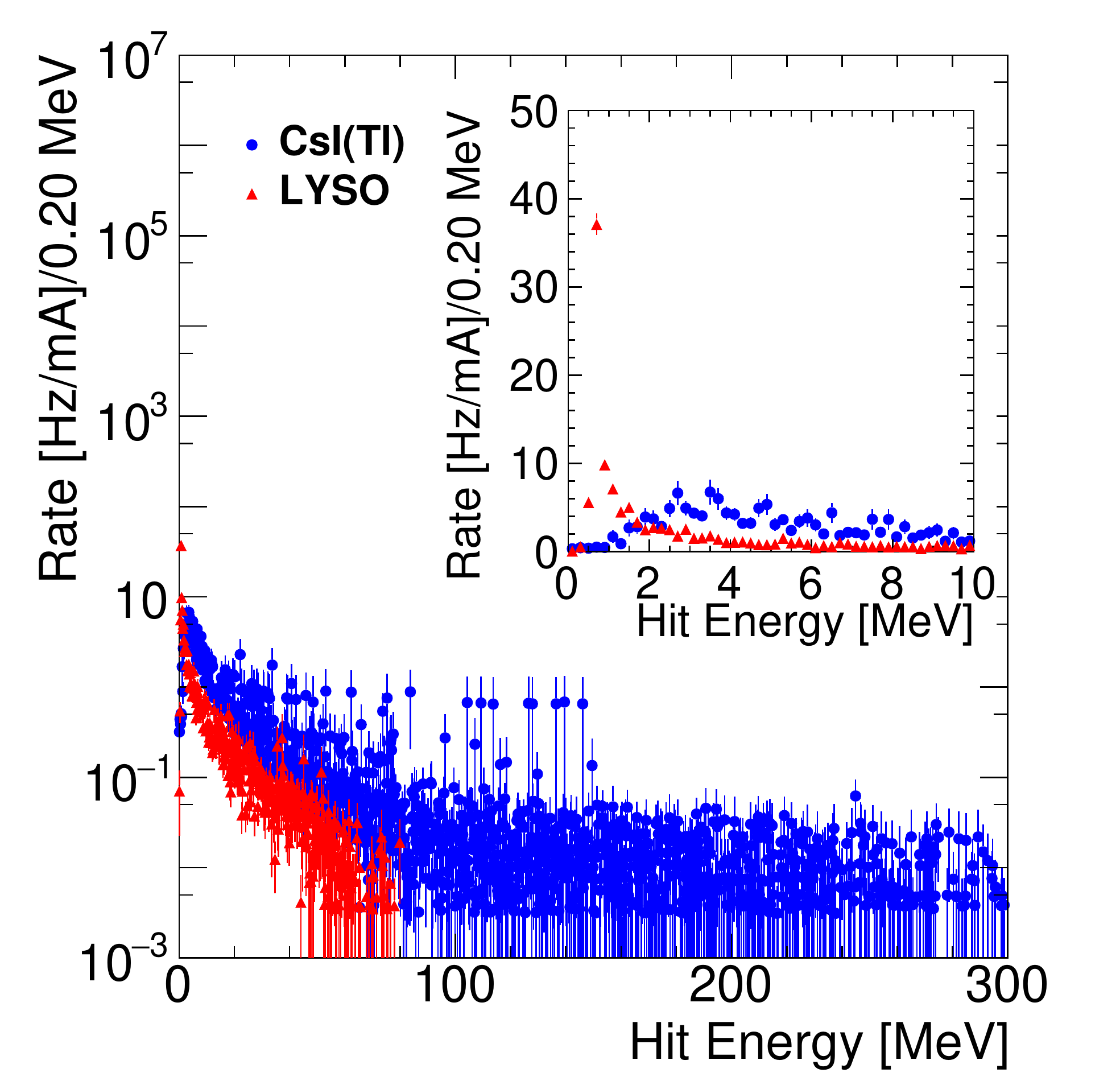}
  }
  \subfigure{
    \includegraphics[width=0.975\columnwidth]{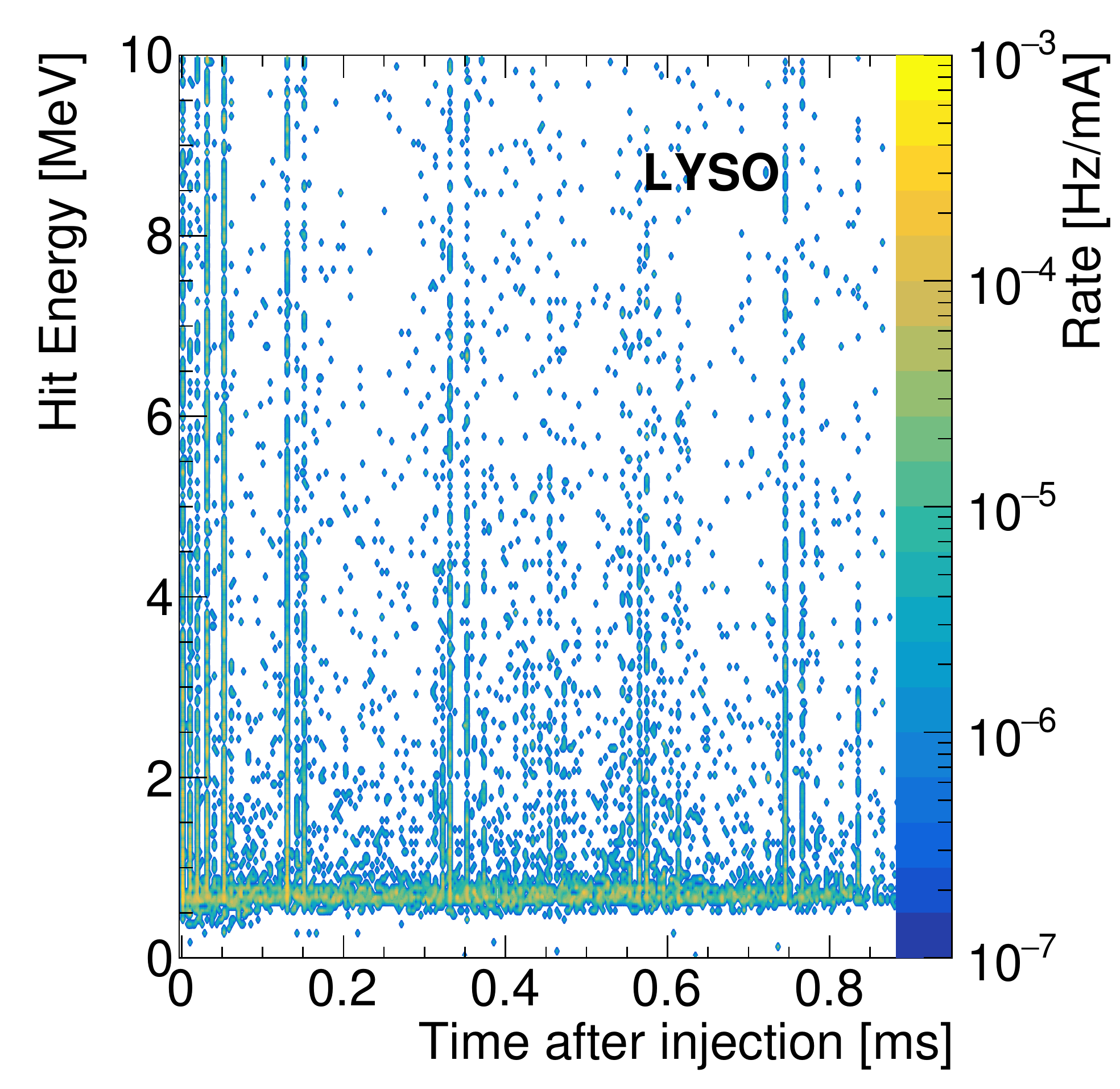}
  }
  \caption{
    (color online) Injection background hit energy spectrum measured with the Crystals for the HER. We show on the left a plot of the normalized rate as a function of the hit energy recorded in the LYSO (red triangles) and CsI(Tl) (blue circles) 
    crystals in position F3 during HER injection with nominal parameters (run \runTen). The inset shows a zoomed in view of the low energy end of the spectrum; the observed difference between LYSO and CsI(Tl) 
    is due to the different rate capabilities of the two crystals, the LYSO being much faster than the CsI(Tl).
    On the right, we show a plot of the hit energy as a function of time after injection for the CsI(Tl) crystal; each $10$~\si{\micro}s time bin corresponds to about 1 machine turn; 
    the color scale represent the rate per mA of beam current.
    To express the rate in Hz, we normalized the histogram to the average rate recorded in the 1~ms digitizer acquisition gate (total number of hits per mA of beam current in the histogram, divided by 1~ms).
  }
  \label{fig:cryEneHER}
\end{figure*}

\begin{figure*}
  \centering
  \subfigure{
    \includegraphics[width=0.975\columnwidth]{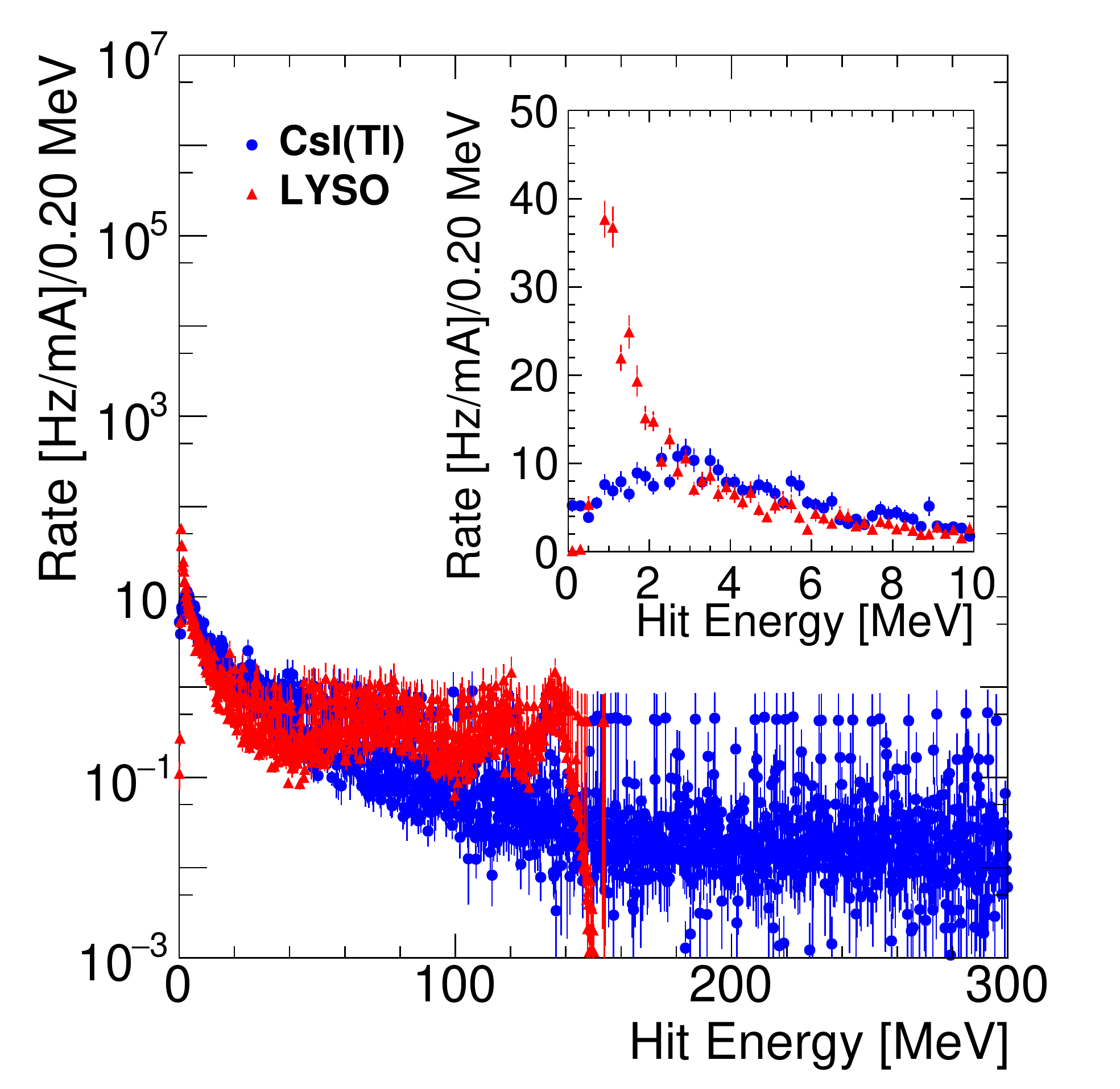}
  }
  \subfigure{
    \includegraphics[width=0.975\columnwidth]{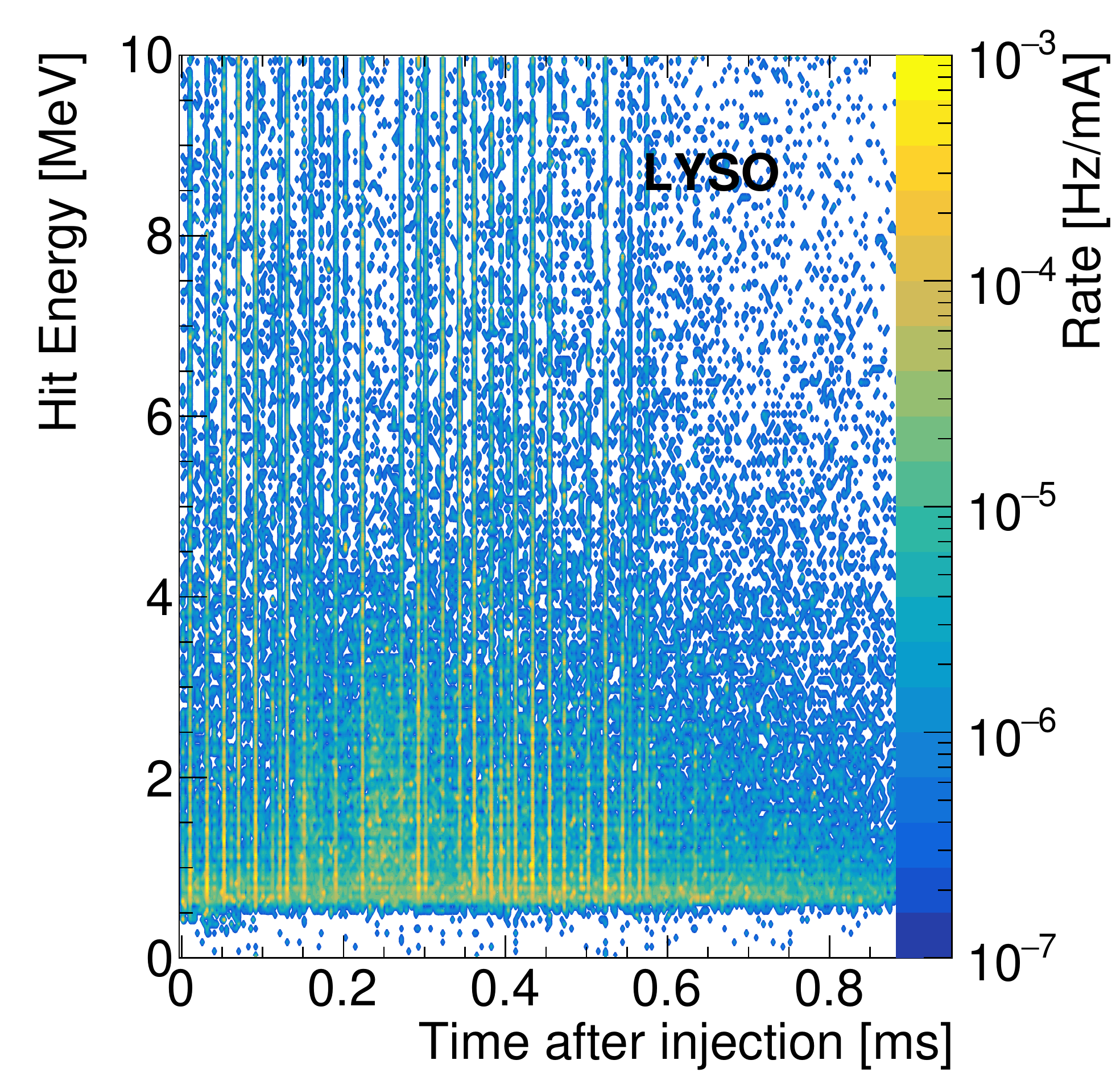}
  }
  \caption{
    (color online) Injection background hit energy spectrum measured with the Crystals for the LER. We show on the left a plot of the normalized rate as a function of the hit energy recorded in the LYSO (red triangles) and CsI(Tl) (blue circles) 
    crystals in position F3 during LER injection with nominal parameters (run \runFourteen). The inset shows a zoomed in view of the low energy end of the spectrum; the observed difference between LYSO and CsI(Tl) 
    is again due to the different rate capabilities of the two crystals, the LYSO being much faster than the CsI(Tl).
    On the right, we show a plot of the hit energy as a function of time after injection for the CsI(Tl) crystal; each $10$~\si{\micro}s time bin corresponds to about 1 machine turn; the color scale represent the rate per mA of beam current.
    To express the rate in Hz, we normalized the histogram to the average rate recorded in the 1~ms digitizer acquisition gate (total number of hits per mA of beam current in the histogram, divided by 1~ms).
  }
  \label{fig:cryEneLER}
\end{figure*}

In Fig.\,\ref{fig:cryEneHER} (\ref{fig:cryEneLER}) we show on the left the energy distribution of the hits recorded during HER (LER) injection in the LYSO and CsI(Tl) crystals in position F3, and on the right the
hit energy as a function of time after injection for the CsI(Tl) crystal, which show that higher energy hits tend to occur at earlier times.
As the energy shown in these plots is all due to hits occurring within $T_{Inj}<1$~ms, the data in Tables\,\ref{tab:LER-Rate-Energy_CsI-ALLboxes_Fracs.txt} and \ref{tab:HER-Rate-Energy_CsI-ALLboxes_Fracs.txt} tells us that 
most of it is also associated to the bunch and thus would be vetoed. 
The energy spectra that we measured show that, at least in the conditions of this study, most of the injection background hits deposit only a few 10~MeV in the CsI(Tl) crystal, which is low enough to 
avoid saturation of the front end readout electronics.

 % file: 		injection_CLAWS.tex
% lead author:  Frank Simon
%
\renewcommand{\topfraction}{.9}

\clearpage
\subsection{Injection backgrounds in CLAWS}

\begin{figure*}[t]
\centering
\subfigure[HER reference injection (\runTen).]
  {
    \includegraphics[width =0.95\columnwidth]{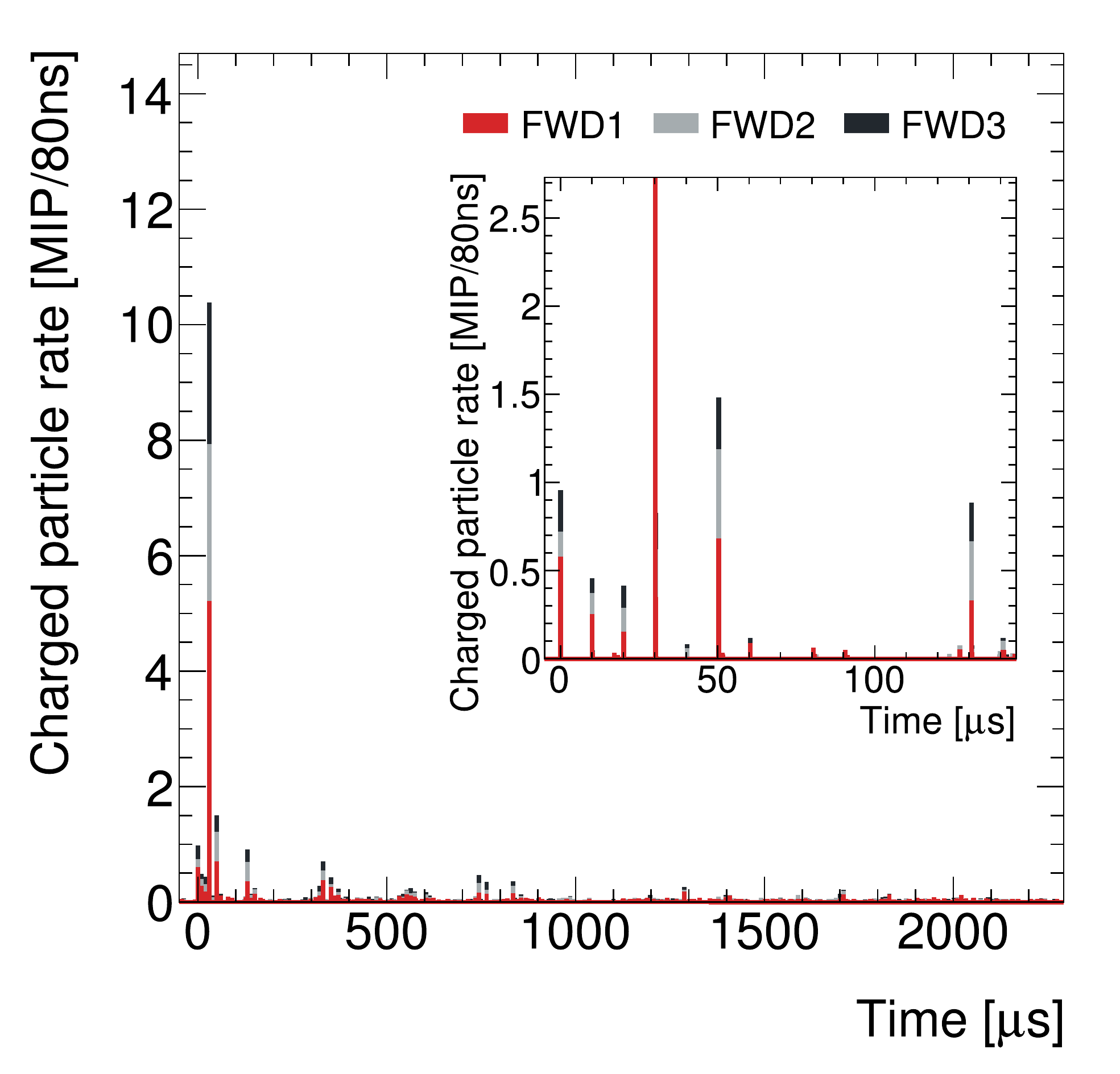}
    \label{fig:CLAWSInjection:RefRunHER}
  } 
\subfigure[LER reference injection (\runFourteen).]
  {
    \includegraphics[width = 0.95\columnwidth]{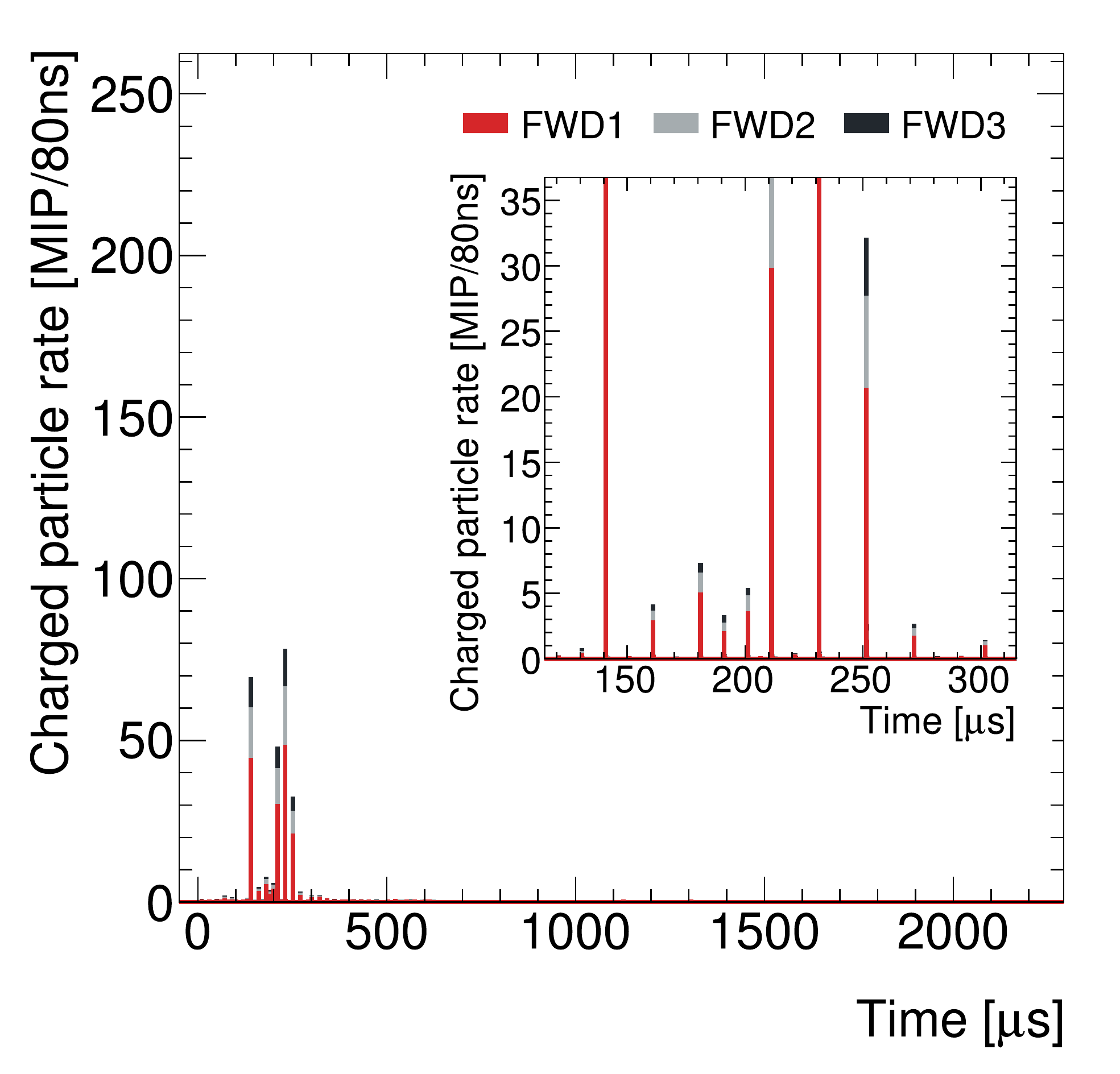}
    \label{fig:CLAWSInjection:RefRunLER}
  }
\subfigure[HER injection with a changed phase shift (\runNine).]
  {
    \includegraphics[width = 0.95\columnwidth]{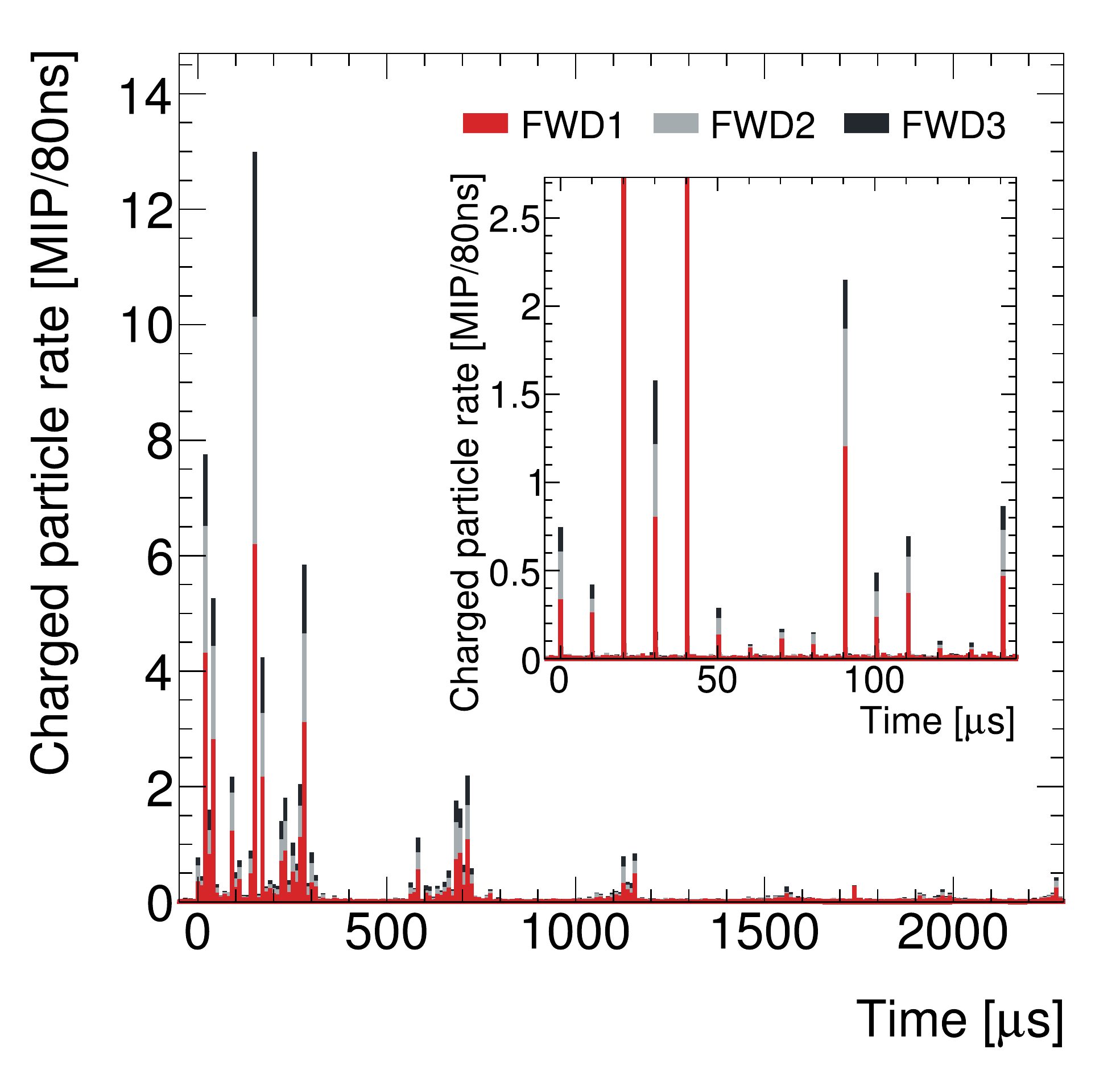}
    \label{fig:CLAWSInjection:PhaseShiftHER}
  }
\subfigure[LER injection with a changed phase shift (\runThree).]
  {
    \includegraphics[width = 0.95\columnwidth]{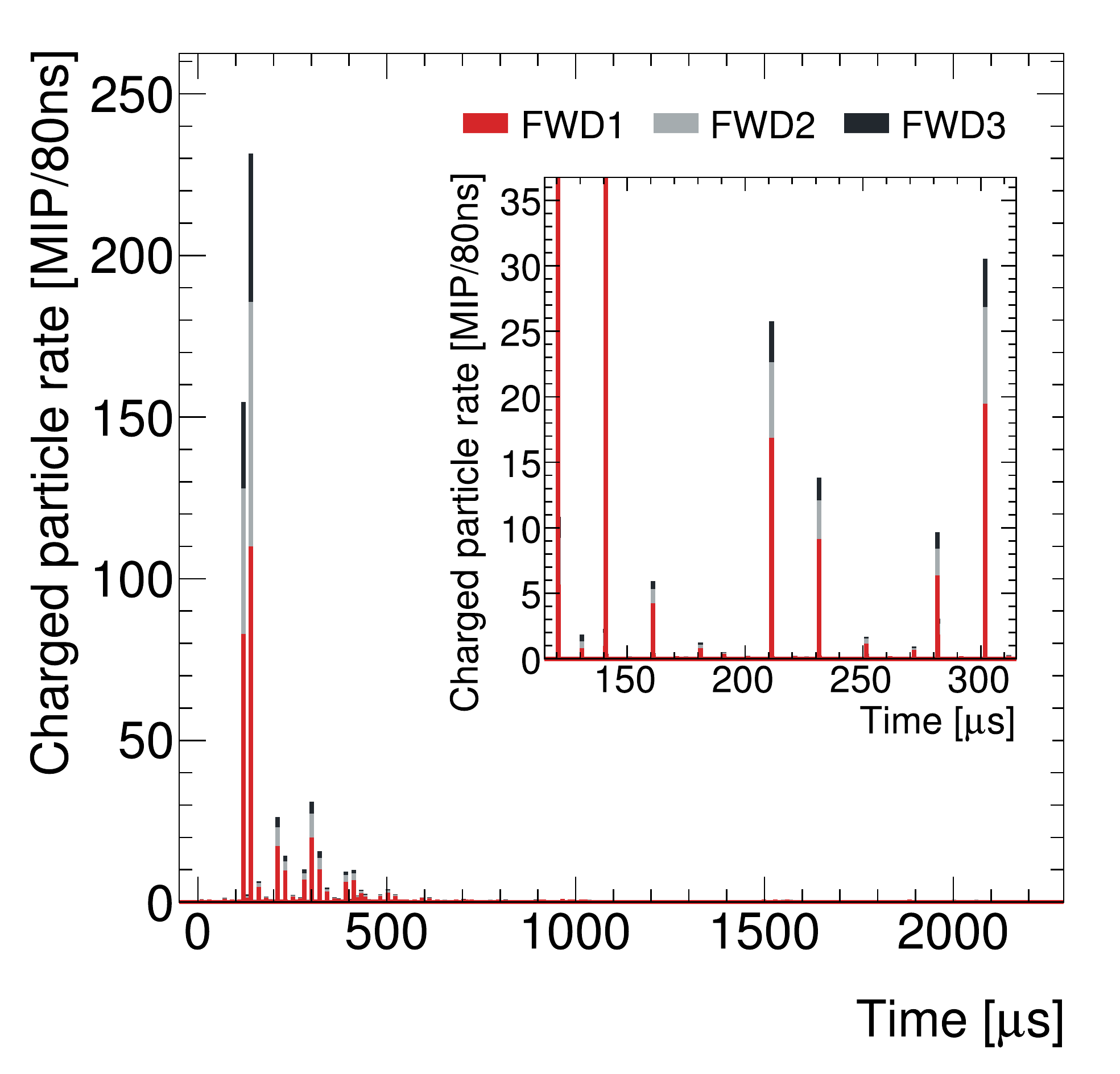}
    \label{fig:CLAWSInjection:PhaseShiftLER}
  }
\caption{Particle rates in the CLAWS detectors in the forward station during different types of injection as described in Table \ref{tab:injdata}. Reconstructed signals from the three innermost FWD sensors are shown as stacked histograms. The more detailed time structure can be seen in the insets, which show a close-up view of the time period shortly after injection.} \label{fig:CLAWSInjection:Injections}
\end{figure*}

The CLAWS system, introduced in Section \ref{sec:CLAWS}, is specifically designed to study the time structure of charged particle injection background.
For data taking during the injection background campaign on May 25, 2016 the system was triggered on the injection signal, and recorded 2.5 ms long continuous waveforms, corresponding to 250 turns of the accelerator.
These studies add further details on the timing properties of the injection background presented in the previous subsection. 

In general, the FWD CLAWS sensors located on the outside of the ring in the direction of the outgoing LER see a somewhat higher signal than the BWD sensors in the inside of the ring on the outgoing HER. 
Since the timing patterns observed are consistent between the two detector stations, only results of the FWD sensors are presented in the following for brevity.
The signals in MIP-equivalents, as described in Section \ref{sec:CLAWS:Calibration}, of the three forward sensors are combined to a stacked histogram with the contribution from the innermost sensor shown in red (FWD1), the second in light grey (FWD2) and the third in black (FWD3).
For illustration purposes, the distributions are rebined by a factor of 100, gathering all recorded MIPs within a 80 ns window in a single bin.
A saturation correction for very large signals is not performed; signals in excess of 50 MIP-equivalents should be regarded as lower limits of the full amplitude.

The trigger signal for an injection is recorded several micro seconds prior to the arrival of the first particles at the interaction point. 
This offset is determined by averaging the detection time of the first photoelectron over the three innermost sensors. From the HER reference run, labeled \runTen\ %\#10
 in Table \ref{tab:injdata}, a trigger delay of 107.448~\si{\micro}s is obtained and subsequently subtracted in all runs presented in the following.

The time structure of the charged particle background for the HER reference injection and the LER reference injection, labeled \runFourteen\ %run \#14 
in Table \ref{tab:injdata}, is shown in Figure \ref{fig:CLAWSInjection:RefRunHER} and Figure \ref{fig:CLAWSInjection:RefRunLER}, respectively.
The substantially higher level of background in the LER is immediately apparent from the one order of magnitude difference in signal amplitudes. In terms of timing, LER and HER show distinctly different behavior.
In the HER case, signals are clearly visible right after injection, while in the LER case the signal is below the background rejection threshold for the MIP - based analysis and thus not visible.
While the background particles from the HER injection are detected primarily in the first 6 turns (60~\si{\micro}s), the first sizeable signals (relative to the maximum signal) from the LER injection occur 14 turns (140~\si{\micro}s) after the first passage of the injection bunch. 
80$\%$ of the particles are detected within 1 ms after injection in the HER and 87$\%$ within 1.5 ms. For the LER the decay is somewhat faster with 97$\%$ and 98$\%$ of the particles being recorded within 1 ms and 1.5 ms, respectively. 

The largest effect of varying the injection parameters, as outlined in Table \ref{tab:injdata}, is observed for changes of the phase shift.
Runs \runNine\ %\#9 (phase shift change in HER) 
and \runThree\ %\#3 (phase shift change in LER) 
are shown in Figures \ref{fig:CLAWSInjection:PhaseShiftHER} and \ref{fig:CLAWSInjection:PhaseShiftLER}, respectively.
The changed phase shift substantially increases the background level in both cases, while also changing the timing properties of the signal. 

In the HER, signals are again visible immediately following injection, with signals typically visible in each turn.
In addition to this 10~\si{\micro}s period, a longer period of approximately 13 turns (130~\si{\micro}s) is also apparent.
Furthermore, a third time pattern can be observed, with a period of around 40 turns (400~\si{\micro}s) and small signals visible up until the end of the recording window of 2.5 ms. 

The LER signal is again characterized by a substantial delay of the first appearance of particles, which are then primarily detected in two turns separated by 20~\si{\micro}s approximately 12 turns after the injection.
As already partially apparent in Figure \ref{fig:CLAWSInjection:RefRunLER}, there is a general 2-turn pattern of the LER background, with 20~\si{\micro}s separating larger signals. With the changed phase shift, a longer period of 90~\si{\micro}s also becomes apparent. 

The change of the septum angle, performed in run \runSeventeen %\#17 
for the LER, results in an increase of the peak amplitudes of the signal and a somewhat faster decay, reducing the number of turns where sizable background signals are observed.
Changes of the vertical steering parameter show little effect on the background structure observed in CLAWS.

\begin{figure}[htb]
\centering
\includegraphics[width =0.9\columnwidth]{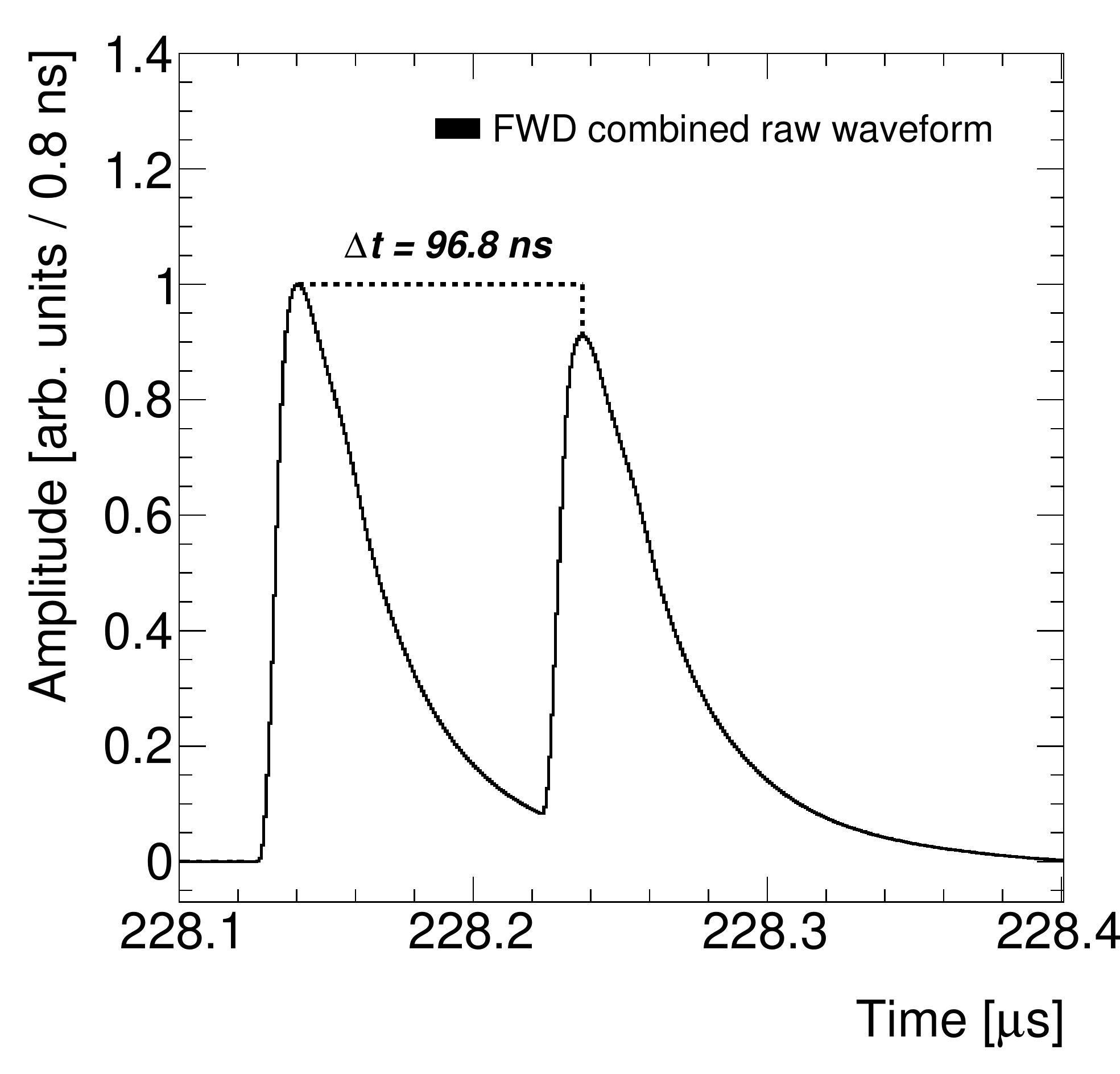}
\caption{Raw combined waveform from the CLAWS FWD sensors for an LER double bunch injection. The time period with the first high background signal is shown. The time difference between the two injection bunches is measured from the peak-to-peak distance in time bins of 0.8 ns. \label{fig:CLAWSInjection:DoubleBunch}} 
\end{figure}
In addition to studies of the overall time pattern of the injection background, CLAWS data allows to investigate individual injection bunches in detail.
Figure \ref{fig:CLAWSInjection:DoubleBunch} shows the raw combined waveform of the inner three FWD sensors for a run recorded with LER double bunch injection for the time region of the first sizable background signal pulse.
The two separated signals from the two injection bunches are clearly visible, smeared by the analog response of the detector.
The peak-to-peak distance of the two signals is 96.8 ns (with a binning precision of 0.8 ns), in excellent agreement with the expected value of 96.285 ns, corresponding to 49 accelerator time buckets.

 % lead author: Sae Yokoyama, Hiro Nakayama
 % file:                 injection_scintillaror.tex
% lead author:  Sae Yokoyama, Hiro Nakayama
%

\subsection{Injection backgrounds in QCSS}
In addition to the Crystals and CLAWS detector systems, the QCSS system discussed in Section~\ref{sec:scintillators} recorded data during the dedicated injection background runs on May 25, 2016. 
The main motivation was to demonstrate that the scintillator-based QCSS system
can provide real-time displays of the injection background time structure.
It is essential for the accelerator operators who tune injection parameters 
to receive such feedback.

The PicoScope oscilloscope of the QCSS system recorded snapshots of waveforms of amplified MPPC signals,
triggered by an injection timing signal, which arrives 106~\si{\micro}s prior to the injected bunch. 
Each waveform is 5~ms long and includes $\sim 10^5$ data points sampled at intervals of 51.2~ns. 

\begin{figure}
        \begin{center}
                \includegraphics[width=\columnwidth]{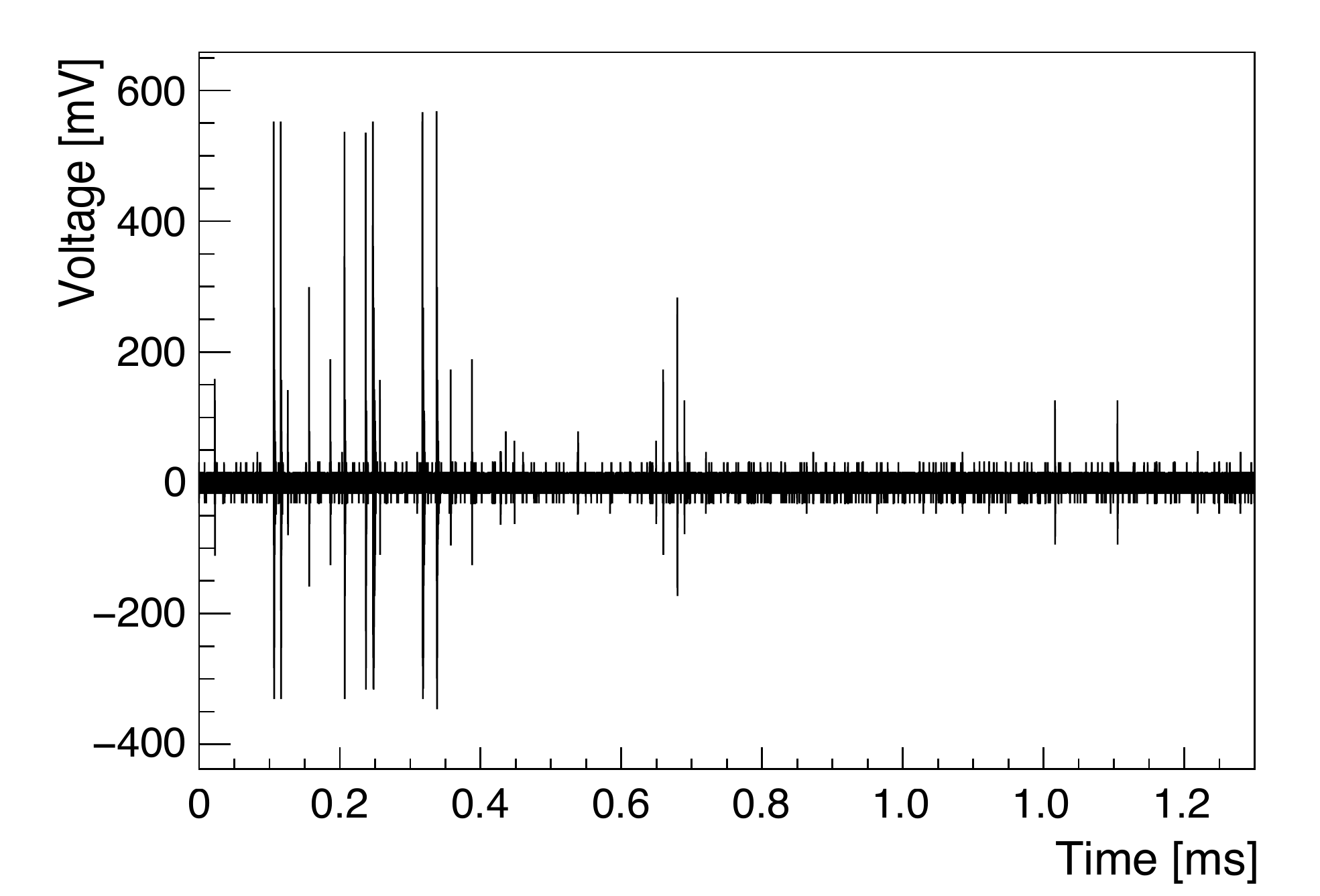}
                \caption{Raw waveform recorded by the QCSS system oscilloscope during an injection background study run,
                         using a sampling interval of 51.2~ns.
                         We recorded 5~ms long waveforms, but only the first 1.5~ms are shown here.
                         The injected bunch arrives at $t \sim 0.1$~ms, and spikes due to injection background are seen after the injection.}
                \label{fig:waveform_8003_4}

                \includegraphics[width=\columnwidth]{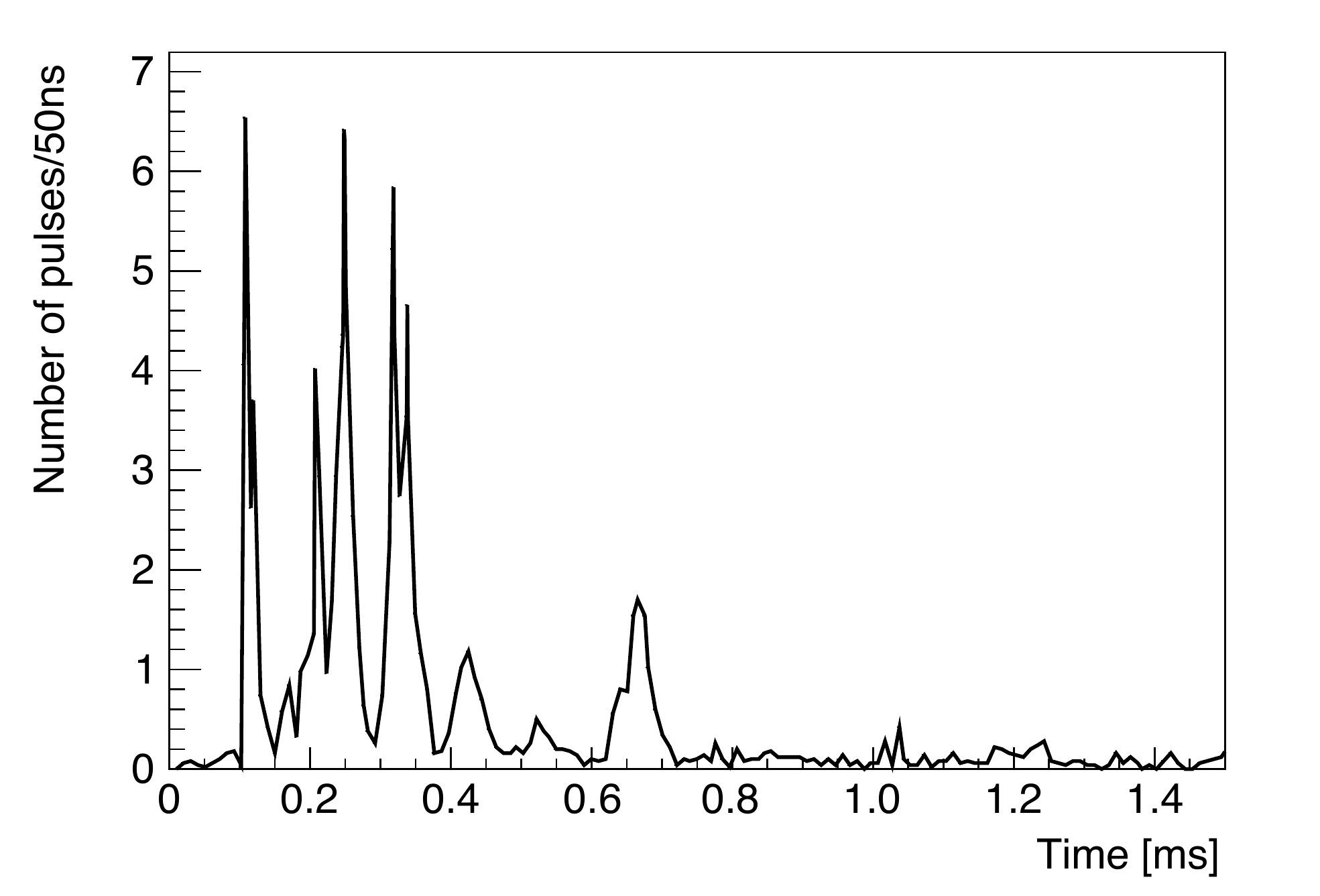}
                \caption{The QCSS system injection background online display, obtained by averaging 50 raw waveforms as described in the text.}
                \label{fig:injection_8003_4}
        \end{center}
\end{figure}

Figure~\ref{fig:waveform_8003_4} shows an example of a raw waveform. %during BEAST II run 8003-4.
Each background hit is observed as a spike, and the injection background timing structure can be obtained from the spike densities.
To make the online injection background display more intuitive, we process the raw waveform data as follows:
1) divide the 5~ms waveform into 10~\si{\micro}s bins,
2) count the number of pulses which exceed the threshold in each bin,
3) calculate the average pulse rate in each bin from 50 waveforms,
and 4) plot the average pulse rate versus bin time.
An example of the resulting, processed waveform is shown in Figure~\ref{fig:injection_8003_4},
which shows the timing structure of injection background hits more clearly the original waveforms.
When the injection rate is 25~Hz, we can update the processed waveform every two seconds in the online display by averaging 50 raw waveforms. This meets the requirement of providing real-time feedback to the SuperKEKB machine operator.

In Figure~\ref{fig:injection_8003_4}, the injected bunch arrives at $t = 0.1$~ms, followed by series of peaks at $\sim$0.1~mc intervals. The peak at $t = 0.65$~ms is larger than neighboring peaks, which implies a larger structure with $\sim$0.5~ms intervals.
The interval of this larger structure corresponds to the synchrotron oscillation of injected bunches,
which was $\sim$2~MHz according to monitor measurements.

 %     file:     injection_summary.tex
%     author:  	Miroslav Gabriel (mgabriel@mpp.mpg.de)
%
%     contents: Summary section for the injection chapter. Since three different 
%		subsystems, all with different measurements and findings, the 
%		summary will try to emphasis commons results and the strongest 
%		messages of all of them.

\clearpage
\subsection{Common patterns in the injection background}
The common measurements of the Crystals and CLAWS allow a combined evaluation of the dominant patterns observed in the time structure of the injection background. 
Both systems observe considerably higher backgrounds during injections in general and larger ones for positrons than for electrons in particular. 
% This can be attributed to the more challenging creation of the positrons and the missing of the damping ring.  
% Furthermore, in general they have less energy and, therefore, larger emittance.
This can be attributed to the higher emittance of the injected positrons due to the absence of the damping ring in this first phase of the commissioning.

The backgrounds from injection decay to the background level of regular circulating bunches after several milliseconds.
Crystals and CLAWS both determine similar decay times, with around 80$\%$ and 90$\%$ of background particles being observed within the first millisecond in the HER and the LER, respectively.

The ns time resolutions of the two detector systems allow us to resolve the time structure of the background due to individual circulating bunches.
%which is clearly confirmed by the presented results of both systems.
Crystals and CLAWS both observe a considerable increase in particle rates during the passage of bunches that received an injection, clearly linking the increase in backgrounds to the injection process. 
This is further demonstrated by the measurements of the background pattern in double bunch injection runs in the LER.
%record considerable rises in background levels during the transition of single bunches which received new particles, unambiguously allocating them to injections.
%The successive injection of particles into two bunches at the same time demonstrates these capabilities.
The measured separation of the injection bunches, 97.1~ns for Crystals and 96.8~ns for CLAWS, agrees within the measurement uncertainty with the value of 96.285~ns predicted by the machine group.

The signals observed from the injection bunches can be used to study recurring timing patterns that are connected to properties of the accelerator. The Crystals data shown in figure \ref{fig:scalerInj-HERLER} and, to a lesser degree also the CLAWS data shown in figure \ref{fig:CLAWSInjection:Injections} both show a modulation of the background rates with a period of approximately 500~\si{\micro}s (50 turns) in the LER and 400~\si{\micro}s (40 turns) in the HER. This is due to synchrotron oscillations in the accelerator ring, matching the expectations from the machine lattice. Beyond this, another superstructure with a period of approximately 100~\si{\micro}s is observed, best visible in figure \ref{fig:CLAWSInjection:Injections} d). This structure originates from betatron oscillations, which have a rate of slightly above 45.5 oscillations per turn. Since the oscillation frequency is approximately 5\% above a half-integer value, this results in a recurring pattern with a period of 10 turns, as seen in the data. Finally, the CLAWS data show an "on-off" pattern in the LER, with background signals occurring every other turn, resulting in a period of 20~\si{\micro}s. This pattern is particularly pronounced for the injections with phase shift. This behavior is attributed to an asymmetry in the energy distribution of the bunches, which is expected to be amplified for injections with a phase shift. 

The presented findings are further strengthened by the measurements of the QCSS detector system, which also observes the longer time patterns discussed above.

 %lead authors: Michael Hedges, Sam de Jong, Sven Vahsen
 \clearpage

 \section{Neutron background measurements}
 % file: neutrons.tex
% lead author: Sven Vahsen

%\subsection{Motivation}

Neutrons have proven to be a pernicious background at B-factories. Neutrons are highly penetrating, difficult to shield, and often not simulated accurately by default Monte Carlo simulation codes. Neutrons degraded the KLM detector performance in Belle \cite{neutrons_in_klm, Hoshi:2006ir} and the DIRC detector performance in the BaBar Experiment \cite{neutrons_in_dirc}. With even higher beam currents and luminosity at SuperKEKB, neutrons from beam backgrounds are expected to be a critical issue at Belle II, as discussed in Section 10.4 of Ref. \cite{Abe:2010gxa}. 

% Production

Neutrons can be copiously produced when high-energy electrons or positrons traverse materials, as is the case when spent beam particles hit the beampipe wall. Both the initial leptons and secondary photons and electrons (produced in subsequent electromagnetic showers) can excite nucleons in the target material. In particular, so-called Giant Resonances \cite{GiantResonances,SLAC-PUB-6628}, which are collective electromagnetic excitations of the nucleons in a particular nucleus, can result in emission of neutrons. 

% Propagation

Being electrically neutral, neutrons are highly penetrating, and can travel from their production points in the beampipe wall into the Belle II detector, leading to background hits. Neutrons lose energy in material primarily via elastic scattering with atomic nuclei, with larger energy-loss per scatter for nuclei with lower atomic numbers. As a result, the typical high-$Z$ materials used to shield against charged particles and gamma rays are less effective at shielding neutrons.

% Belle experience

In Belle, beam background neutrons produced in the central beampipe (inside Belle) led to background hits in the inner barrel layers of KLM detector, which utilized Resistive Plate Chambers (RPCs). Because the RPCs have a long recovery time after a discharge, this neutron background led to significant KLM deadtime and in turn reduced detection efficiency. Neutrons produced in beam background showers outside of Belle also produced hits in the outer layers of the KLM endcaps, reducing their efficiency as well. These neutrons hitting the outer endcaps resulted from radiative Bhabha scattering producing photons at the interaction point. Those photons would travel down the beampipe and produce neutrons via photonuclear interactions in magnets and other material outside of Belle \cite{Hoshi:2006ir}. Polyethylene (i.e.\ low-$Z$) shielding was eventually added on the outside of the Belle KLM endcaps to reduce this background component. Compounding the issue, in the default Belle simulation, the interaction of neutral hadrons (including neutrons) in different detector materials was modeled using GEANT3 \cite{Brun:1987ma}, and did not agree well with measurements. 

% Belle II nominal predictions

The Belle experience coupled with the expectation of even higher neutron rates in Belle II was the motivation for replacing all KLM endcap RPC layers and the two innermost KLM barrel RPC layers with scintillation-based detectors, which have much shorter deadtime, and thus are more robust against beam background neutrons \cite{Aushev:2014spa}. In Belle II, neutron production from beam backgrounds is simulated with Geant4, using the physics list FTFP BERT HP, which contains a high precision neutron package. Using this simulation, the predictions for full luminosity running of SuperKEKB are that neutron rates will be critically high, see Section \ref{implications}.

While most of the BEAST II measurements reported here aim to validate predicted beam background loss distributions from beam-gas and Touschek scattering, for neutrons there are several additional issues that need to be verified experimentally: it is also important to assess if the observed neutron production cross-section via Giant Resonances, the neutron propagation through passive materials, and neutron detection in active detectors, is accurate. This requires accurate and complete production cross-sections in the Geant physics list, an accurate implementation of the beampipe and BEAST II geometries and their respective materials, and a correct neutron propagation model.

% BEAST II measurements

Given the importance of neutrons, we decided to deploy two complementary detector systems in Phase 1: four \heT tubes measure the rate of thermal ($E < 1$~keV) neutrons at four $\phi$ positions, while two TPCs measure the rates of fast neutrons by detecting nuclear recoils with energies of order 10~keV to MeV, at two $\phi$ positions. The TPCs also measure the energy and direction of neutron recoils, which can be compared in detail against Monte Carlo. In principle, these nuclear recoil distributions could be de-convoluted to obtain incident neutron spectra and neutron directional distributions, but that is beyond the scope of the present work, and will be published separately in the future. A description of the \heT rate measurements can be found in Section \ref{sec:beamgas_touschek}. The TPC results are presented below.
 % Sven
 %\input{thermal_neutrons_analysis} 
 \subsection{Fast neutrons: analysis}
\label{neutrons_analysis}
The unique event topology of nuclear recoil signal events in the TPCs, which can be seen in Figure \ref{fig_evt_display}, allows for a simple and effective selection of fast neutron candidates. In short, we required that events have some minimum energy, high ionization density, and no charge at the edge of the fiducial volume. We optimize the event selection with a sample of 13011 Monte Carlo events, corresponding to 5 hours-equivalent of beam background from beam-gas (Coulomb and bremsstrahlung) and Touschek interactions simulated at the conditions described in Section \ref{simulation_generators_sad}.  Events are passed through a dedicated full-detector simulation using a \gF model of the TPCs. The resulting event sample contains both our desired signal (fast-neutron recoils of helium, carbon, and oxygen nuclei), and our expected background (ionization due to any other particles, such as electrons, positrons, and protons).  The TPC simulation will be described in more detail in a separate publication \cite{JaegleTPCSimulator}.

\begin{figure}[h]
	\centering
	\includegraphics[width=\columnwidth]{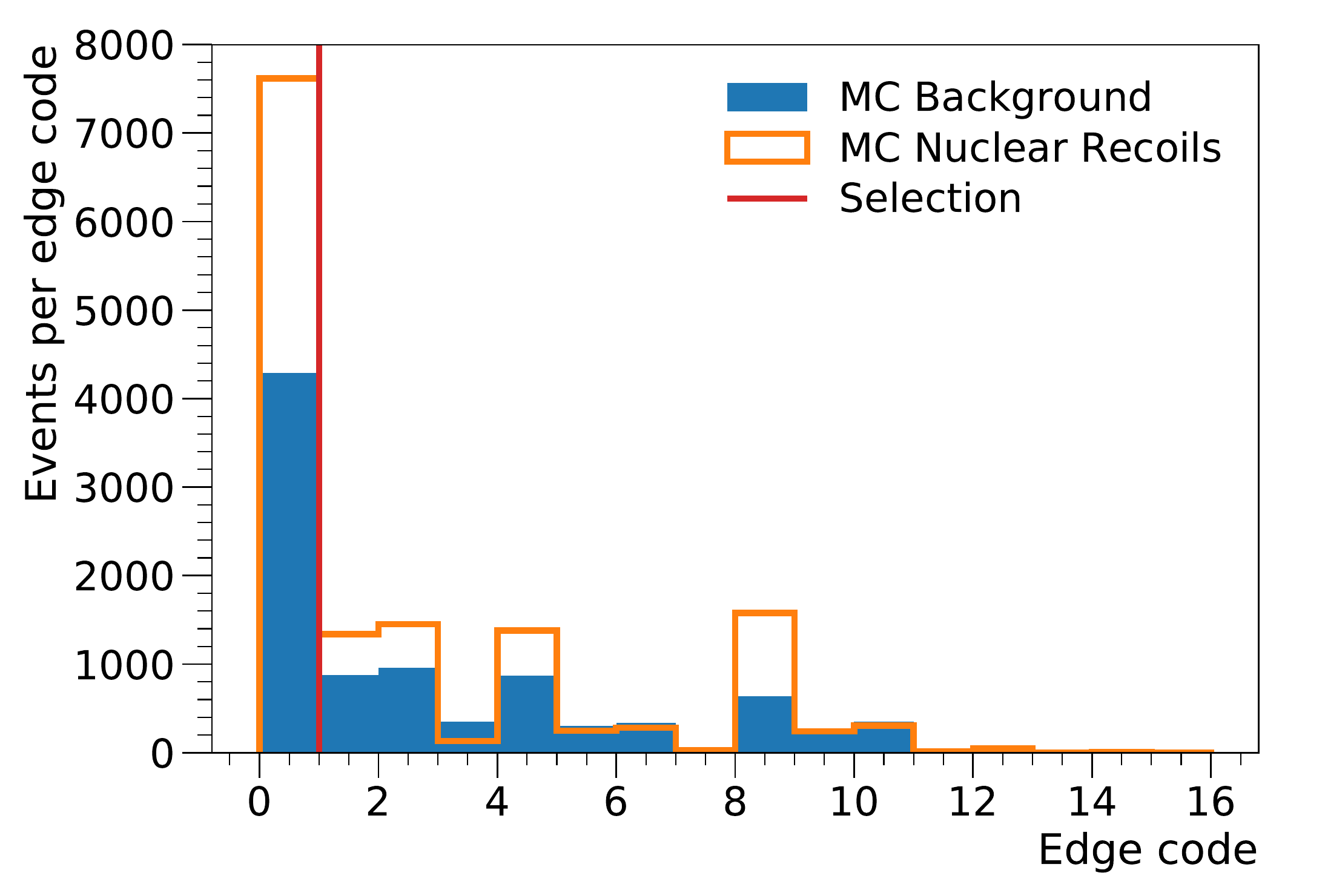}
	\includegraphics[width=\columnwidth]{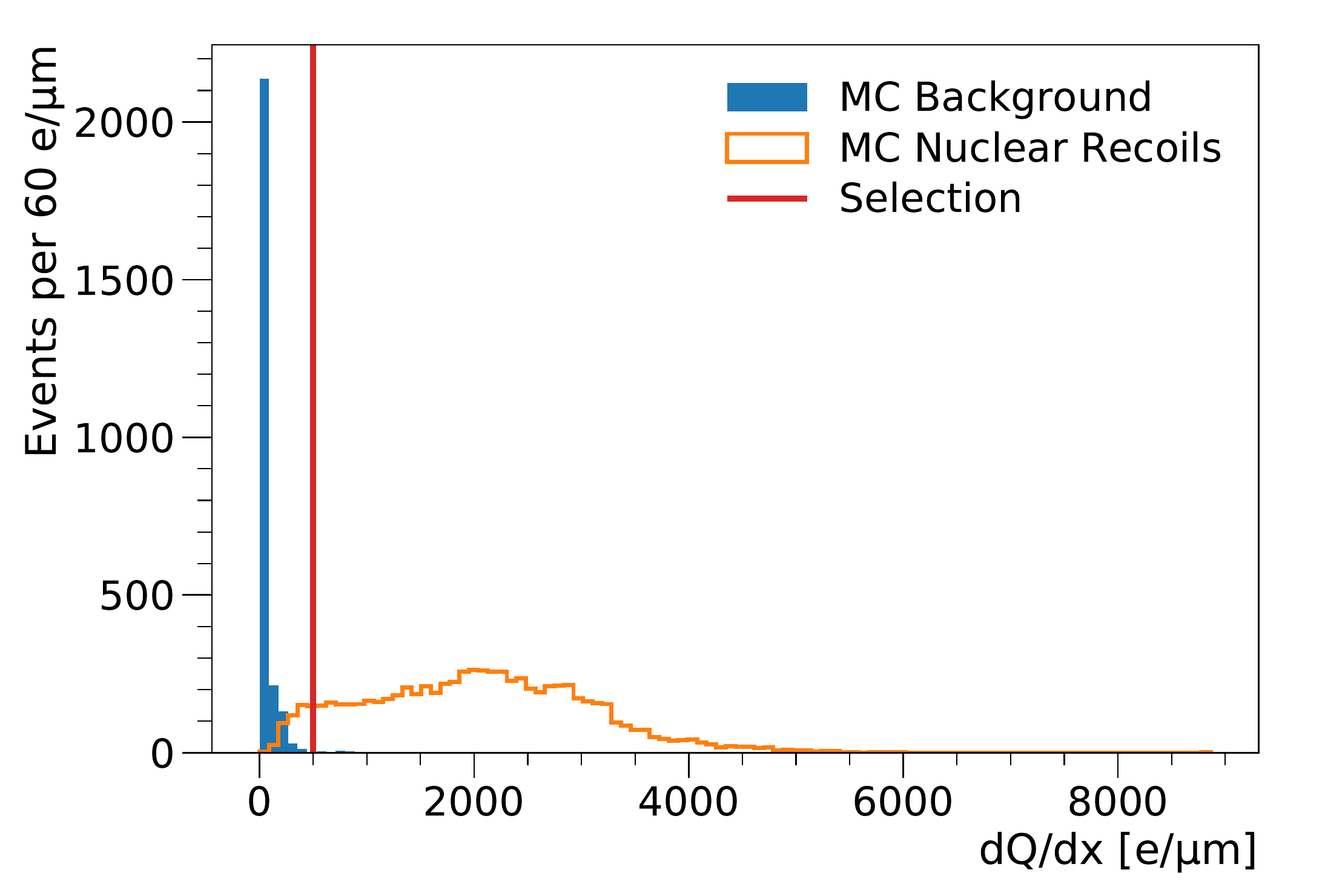}
	\includegraphics[width=\columnwidth]{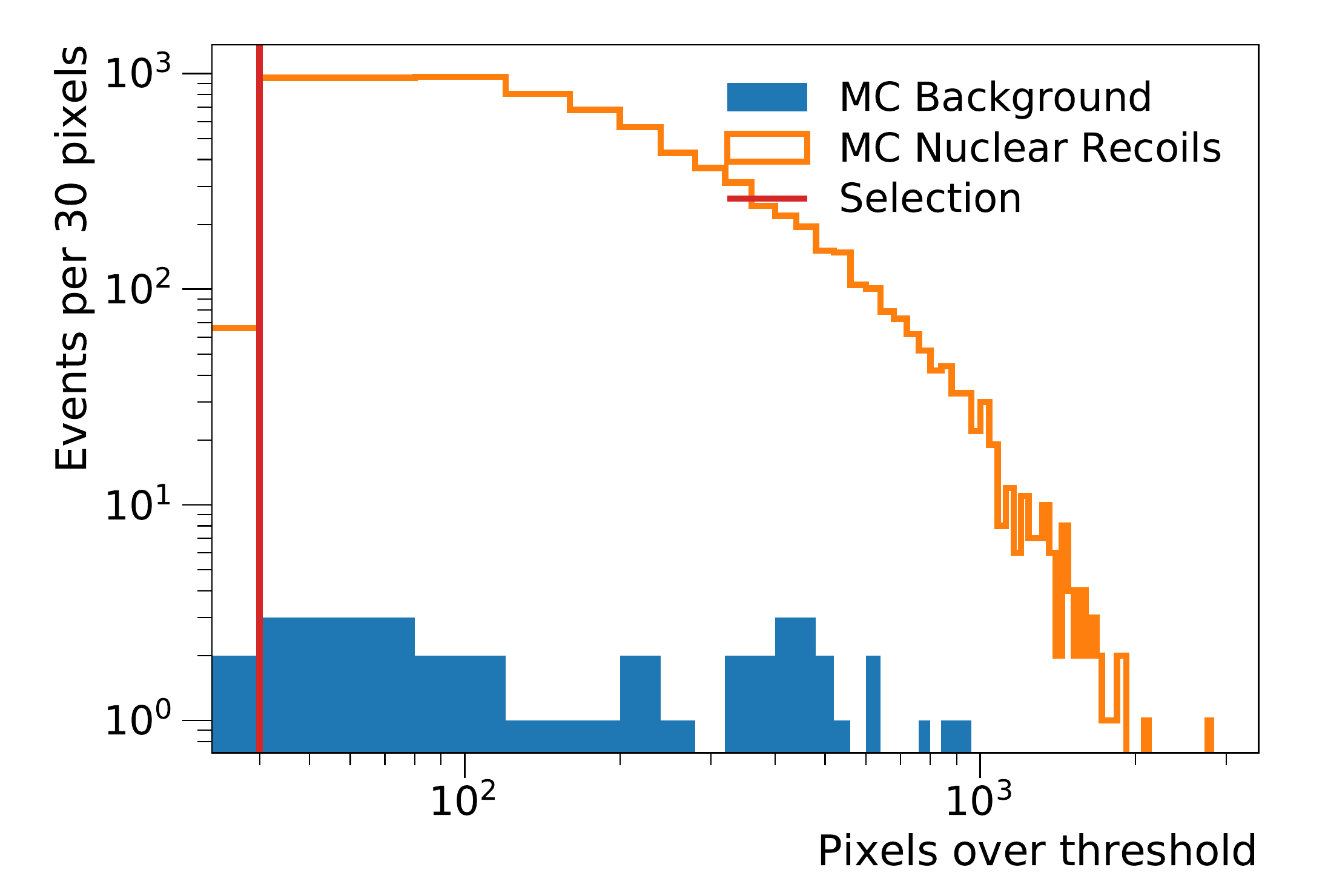}
	\caption[single column]{(color online) Distributions of TPC variables used to select nuclear recoil events in Monte Carlo data, shown in order of application and including previous selections.  The order is the same as is shown in Table \ref{tab:tpc_selections} from top to bottom, starting with the edge veto.  The blue bars correspond to truth level background events, the orange lines correspond to truth level signal events, and the red line represents the threshold value of the applied selection. The edge code (uppermost plot) identifies which edges of the pixel chip the recoil crossed, with a value of zero corresponding to the edge veto.  $dQ/dx$ (middle plot) is the amount of detected charge per \si\micro m in the event. The lowermost plot shows the number of pixels above threshold in the event.}
\label{fig_tpc_cuts}
\end{figure}

\begin{figure}[h]
	\includegraphics[width=\columnwidth]{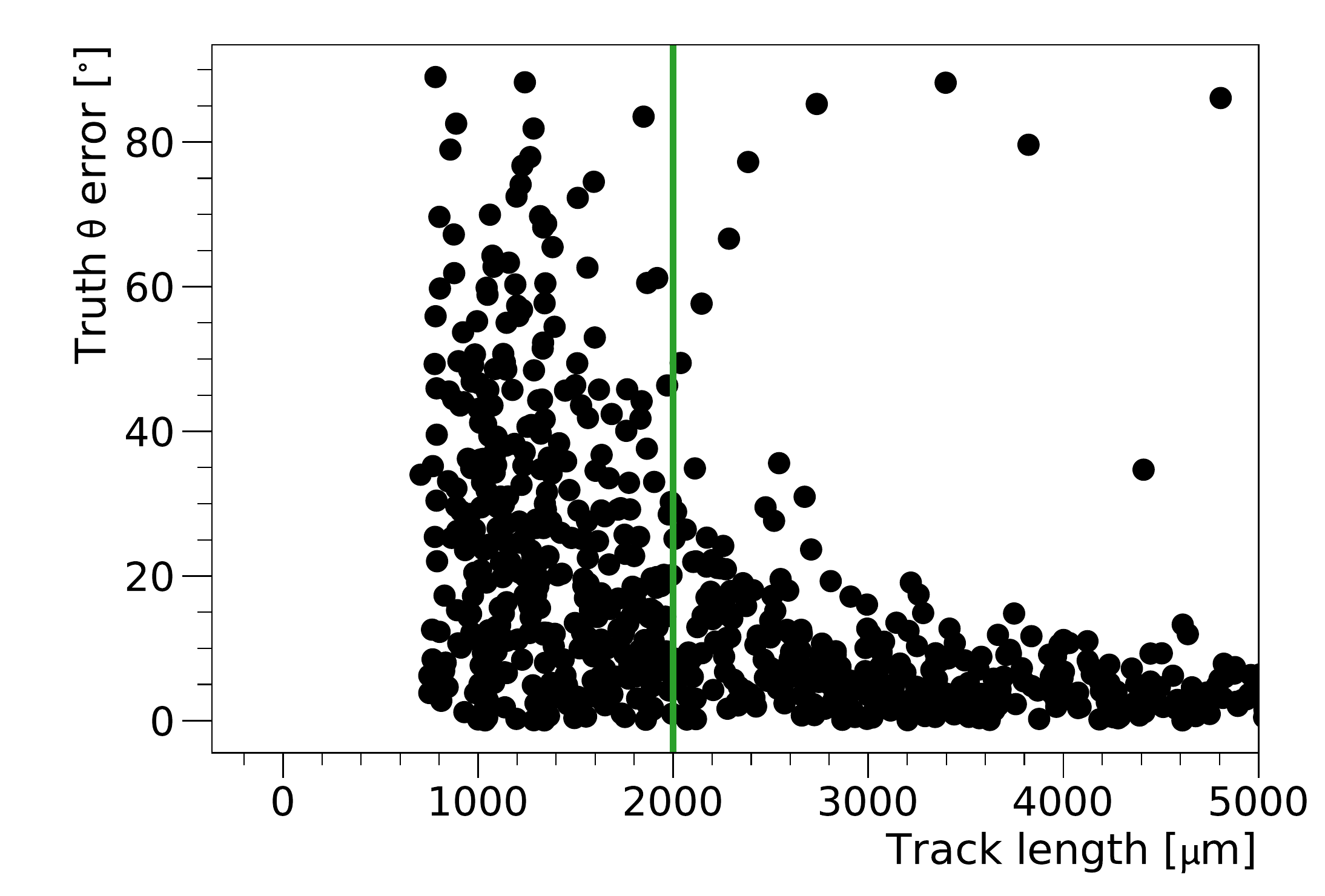}
	\includegraphics[width=\columnwidth]{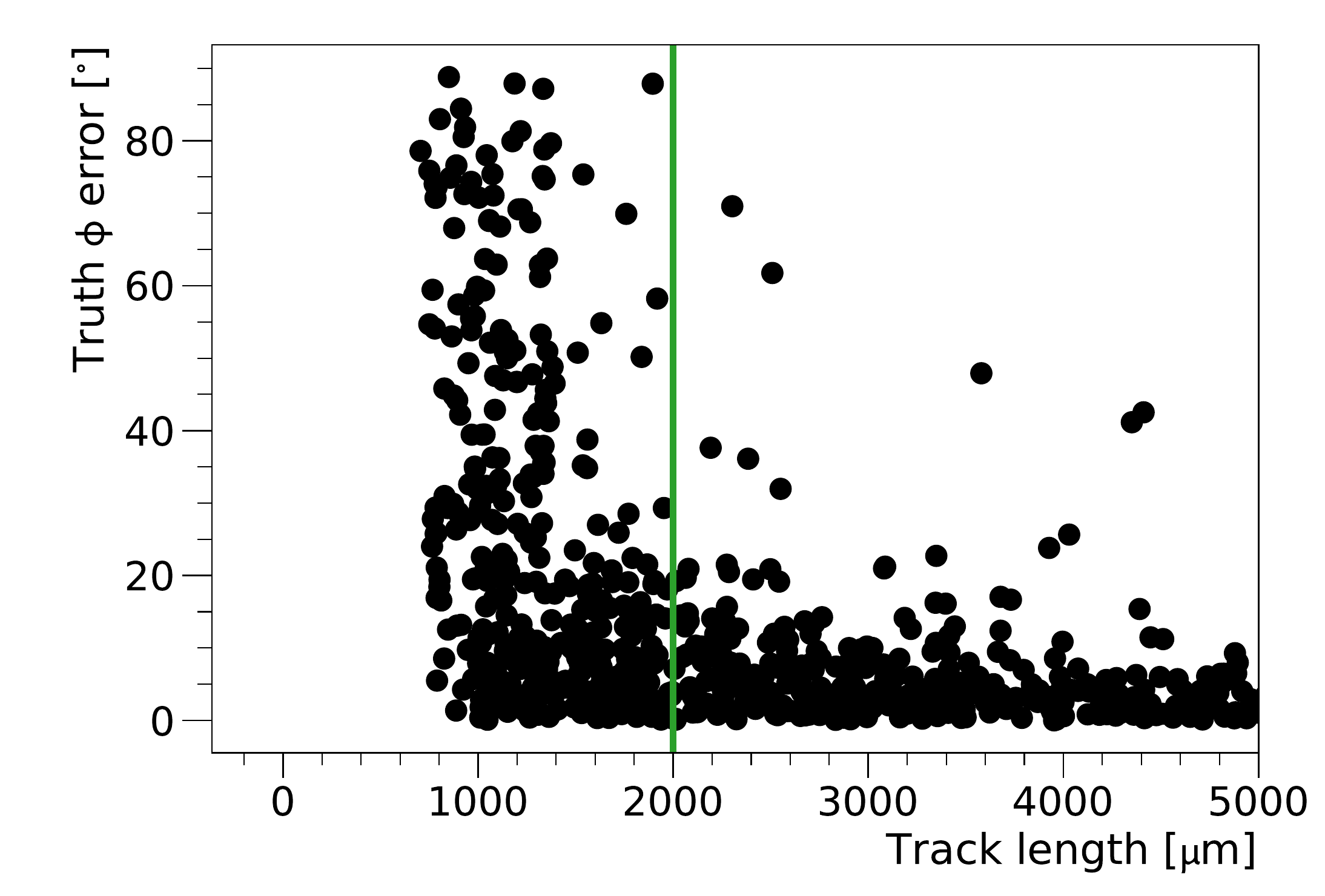}
	\caption[single column]{(color online) Scatter plots of the difference between true and reconstructed $\theta$ and $\phi$ angles in TPC Monte Carlo events versus track length, zoomed in near the selection at $2~\mathrm{mm}$.  This shows that we can achieve, with few outliers, less than $20^{\circ}$ of angular mismeasurement in $\theta$ and $\phi$ by cutting out all events with a track length of less than $2~\mathrm{mm}$.  The cut is represented by the green line. For reference, the bin sizing used in the plots shown in Sec. \ref{fast_neutron_results} is $20^{\circ}$, implying that points lying below $20^{\circ}$ and to the right of the green line in this figure will be reconstructed accurately enough for the resolution required for the results reported in Sec. \ref{fast_neutron_results}.}
\label{fig_tpc_cuts2}
\end{figure}

We find from the Monte Carlo sample that the signal nuclear recoils are readily distinguishable from background with the following event selections: application of an X-ray trigger veto, described in Section \ref{tpc_performance}; a fiducial volume ``edge veto" which requires no pixels triggered within $500~\mathrm{\si\micro m}$ of the four outer edges of the pixel chip in order to veto tracks, including tracks from the calibration alpha sources, originating from outside the fiducial volume; the fitting algorithm used to fit the charge clusters to a straight line converged so that the track length can be properly calculated; the ratio of detected charge to track length ($dQ/dx$) is greater than $500~\mathrm{e/\si\micro m}$, which removes X-ray (electron recoil) event and minimum ionization particles; and a minimum of 40 pixels triggered in the event, which removes the remaining higher energy X-ray backgrounds. We use these selections for all analyses involving the rate of nuclear recoils from fast neutrons. For the studies of recoil angle distributions in Section \ref{fast_neutron_results}, we also require a minimum track length of $2~\mathrm{mm}$ in addition to the previously described selections. This ensures that the true recoil $\theta$ and $\phi$ is measured to an accuracy of approximately $20^\circ$.  The efficiencies of these selections are shown in Table \ref{tab:tpc_selections} and in Figures \ref{fig_tpc_cuts} and \ref{fig_tpc_cuts2}.
			
\begin{table*}[th]
	\centering
	\caption{Selections used to identify nuclear recoils from fast neutrons in the TPCs and the resulting efficiencies of applying each selection in succession for signal Monte Carlo, background Monte Carlo, and experimental data. We note that the experimental data sample (right-most column) includes events from the calibration alpha sources discussed in Sec. \ref{subsubsec:tpc_calibration}, which is by far the largest source of events in our experimental data samples. Simulated events from these calibration sources are not included in the Monte Carlo sample used for selection optimization, as the simulation of the calibration sources is separate from the BEAST II beam background simulation. \newline}
	\begin{tabular}{ lllll }
	\toprule 
	Selection & Signal MC  & Background MC  & Experimental data \\ \midrule
	X-ray trigger veto & 0.9796 &	0.3391 & - \\	
	No pixels triggered on edge of pixel chip & 0.5229 &	0.2214 & 0.1190 \\
	Converged track fit & 0.5200 & 0.2200 & 0.1173 \\
	$dQ/dx > 500~\mathrm{e/\si\micro m}$  & 0.4870 & 0.0007 & 0.0051 \\
	Number of triggered pixels $> 40$ & 0.4803 & 0.0004 & 0.0047 \\
	Recoil track length $> 2~\mathrm{mm}$ & 0.2765 & 0.0004 & 0.0031 \\ \bottomrule
	\end{tabular}
	\label{tab:tpc_selections}
\end{table*}

\begin{figure}[h]
	\centering
	\includegraphics[width=\columnwidth]{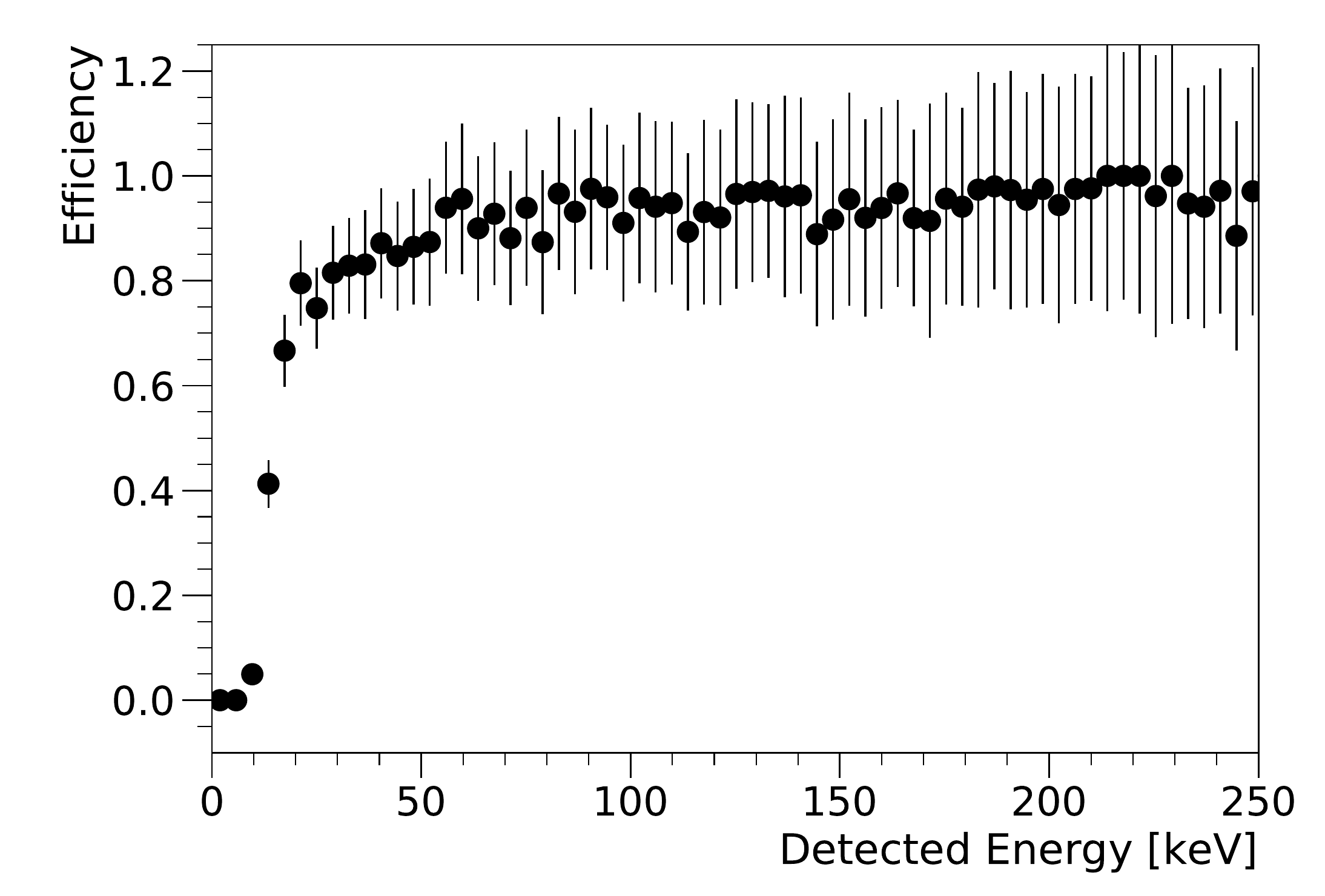}
	\caption[single column]{Efficiency of TPC neutron selections described in Table \ref{tab:tpc_selections} versus detected energy in experimental data. Efficiency of 50\% occurs at approximately $15~\mathrm{keV}$.  The error bars show statistical uncertainties only.}
\label{fig_tpc_n_eff}
\end{figure}

To gauge the effect of our event selection on the recoil energy spectrum, which is one of our final observables, we can calculate the efficiency of event selection as the fraction of events passing the edge veto that pass all neutron selections (except the track length cut), in experimental data. This efficiency versus energy serves to determine which nuclear recoil energies we are sensitive to and whether the selections bias the observed energy spectrum. This efficiency is shown in Figure \ref{fig_tpc_n_eff}. The efficiency becomes 50\% at approximately 15~keV and is near unity and flat for recoil energies larger than $\sim65$ keV.

\begin{figure}[h]
	\centering
	\includegraphics[width=\columnwidth]{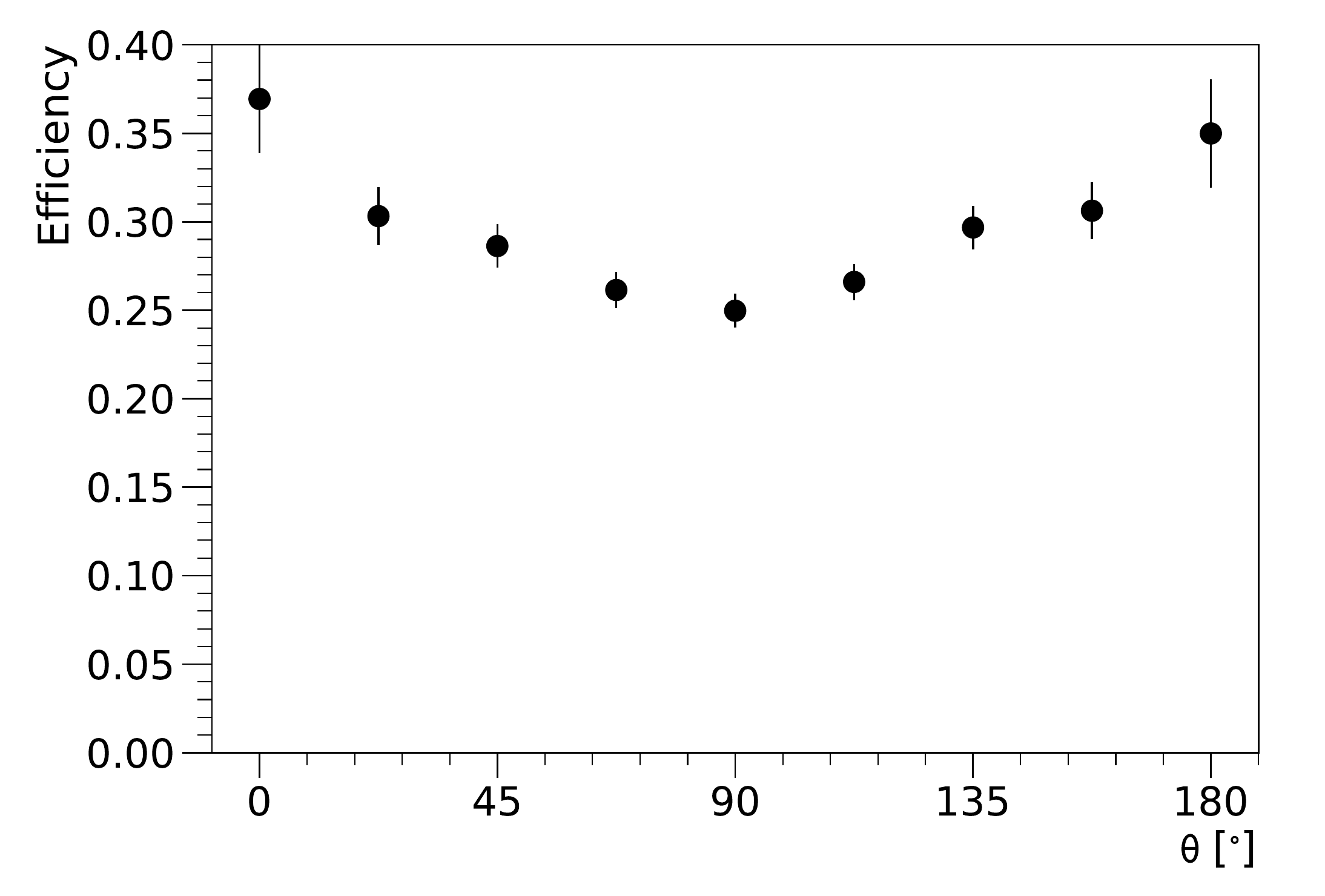}
	\includegraphics[width=\columnwidth]{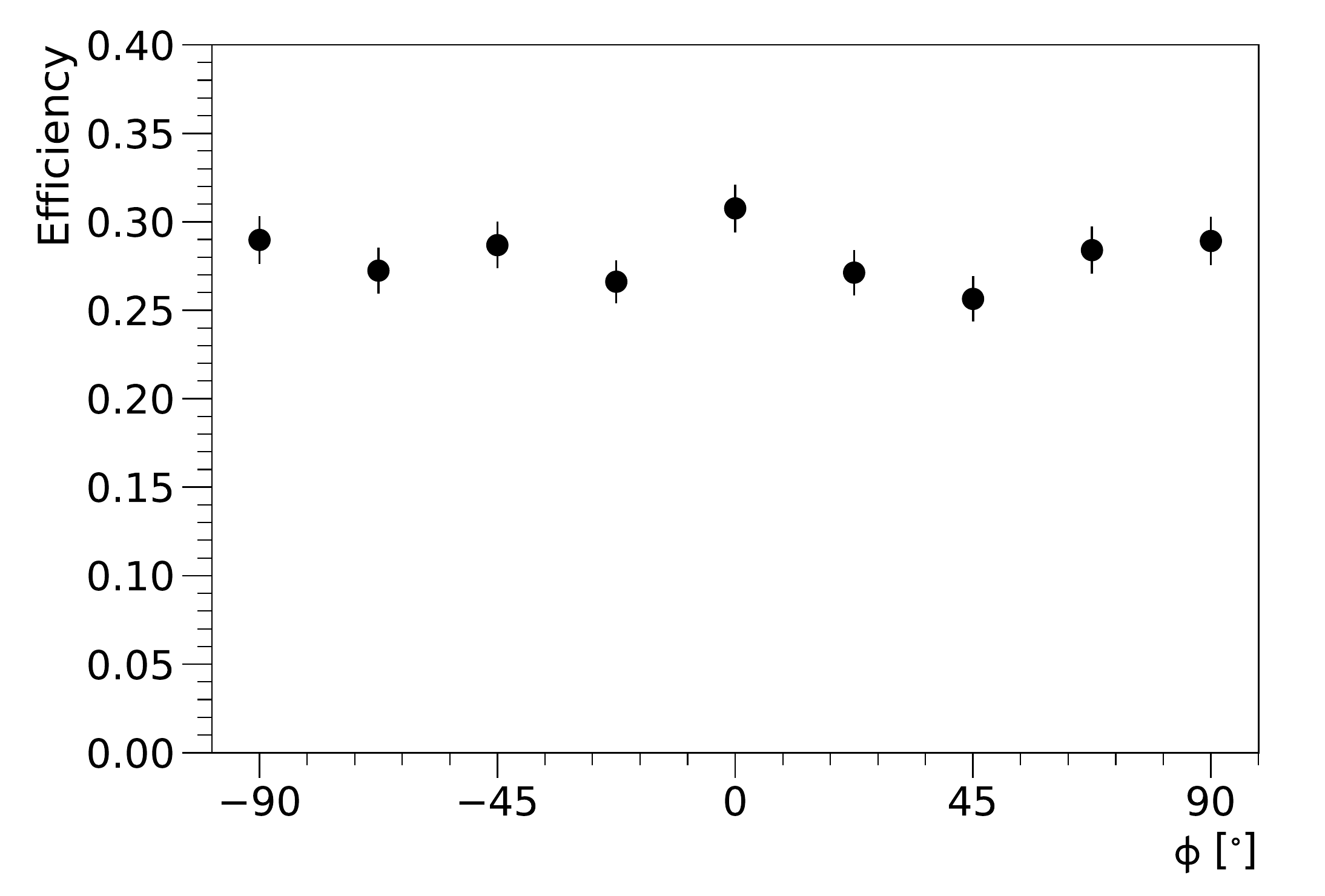}
	\caption[single column]{Efficiency of neutron selections in TPCs described in Table \ref{tab:tpc_selections} versus recoil polar angle $\theta$ (upper plot) and azimuthal angle $\phi$ (lower plot) in both TPCs in simulated nuclear recoils. Each point is a ratio of the number of events at a given reconstructed $\theta$ and $\phi$ passing all neutron selections to the total number of events.  The error bars show statistical uncertainties only.}
\label{fig_tpc_theta_eff}
\end{figure}

Similarly, we check for possible biases in the recoil angle distributions using the signal Monte Carlo data obtained from the TPC simulation. We do so by calculating the ratio of events that pass all selections including the track length cut, to all events, versus $\theta$ and $\phi$, as shown in Figure \ref{fig_tpc_theta_eff}. We note that the efficiency versus $\phi$ is largely flat, while the efficiency versus $\theta$ increases at steep recoil angles, with larger error bars due to smaller statistics in those bins. This means that the measured angular distributions are slightly biased in $\theta$. We also note that for all angular measurements presented here, we reconstruct the axial direction, and not the vector direction of the nuclear recoil candidates.  More specifically, the ``sense", or ``head" and ``tail" of the recoils are not considered, so we present ``folded" angular distributions for the azimuthal angle, i.e. $\phi$ is always reconstructed such that $-90^{\circ} < \phi < +90^{\circ}$.  Vector directionality with head-tail sensitivity is possible, and is currently under development.

\begin{figure}[h]
	\centering
	\includegraphics[width=\columnwidth]{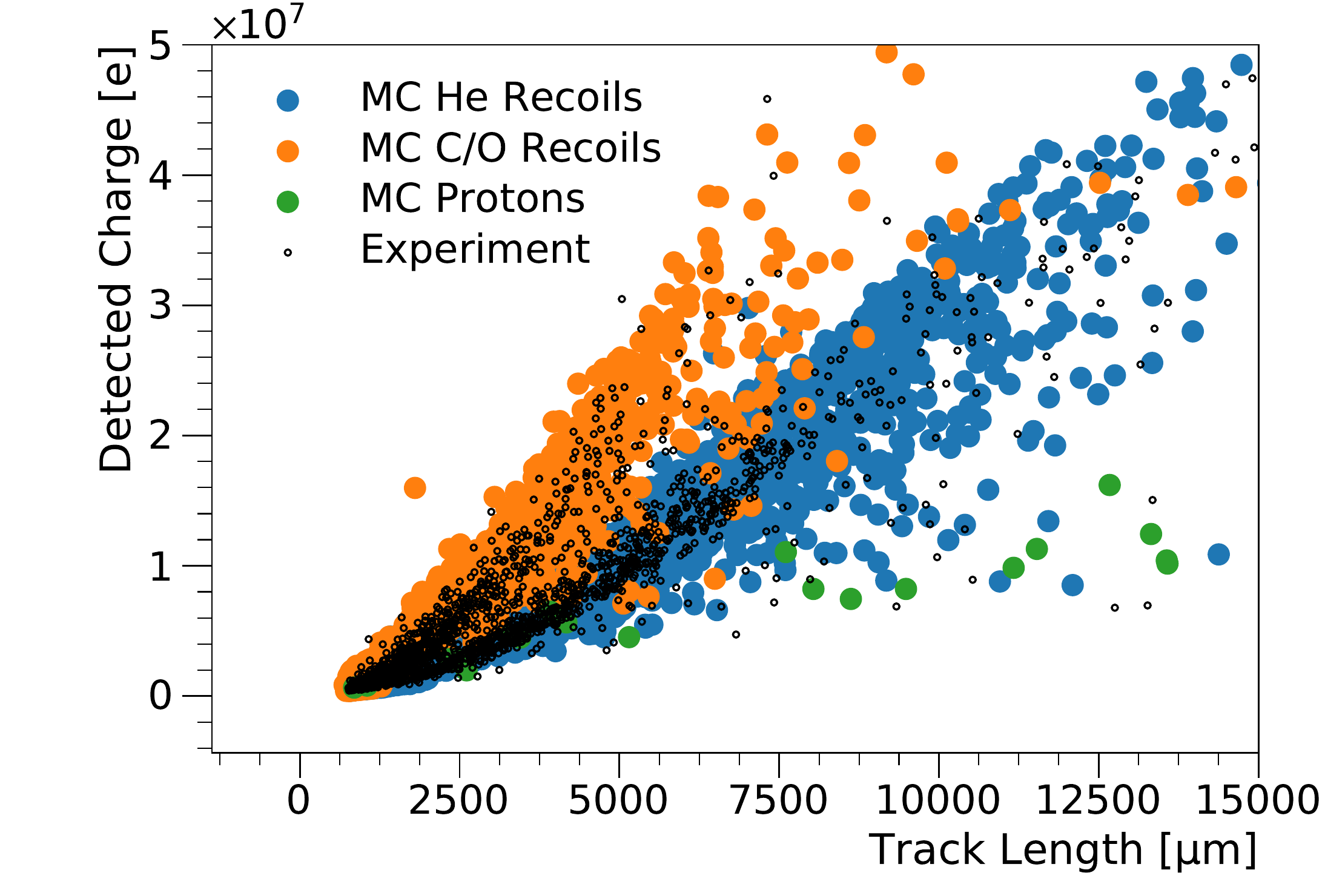}
	\caption[single column]{(color online) TPC recoil charge versus recoil length for neutron event candidates selected in TPCs 3 and 4, for both Monte Carlo and experimental data, combined. This also includes applying the gain correction factors obtained from the methods outlined in Sec. \ref{subsubsec:tpc_calibration} to each TPC. The blue, orange, and green filled circles represent helium recoils, carbon/oxygen recoils, and proton backgrounds in Monte Carlo, respectively.  The open black circles represent the events in experimental data that pass all selections except the track length cut in Table \ref{tab:tpc_selections}. As can be seen, the data points are consistent with the expectations from Monte Carlo, validating our selections and energy calibration procedure.}
\label{fig_tpc_dQdx}
\end{figure}

To further demonstrate the effectiveness of these neutron selections, one can plot the energy versus track length of the neutron candidates, as shown in Figure \ref{fig_tpc_dQdx}.  The Monte Carlo simulation and experimental data both show two distinct bands that are due to fast neutrons scattering with helium, carbon, and oxygen nuclei, and the agreement between simulation and experiment is quite good.

As previously mentioned, we compare the fast neutron distributions in experimental data to those predicted by Monte Carlo using a $5$ hours beam-time equivalent Monte Carlo sample specific to the TPCs.  The beam conditions of this Monte Carlo follow the SAD parameters listed in Table \ref{Table:MachineParamForSim}, as outlined in Section \ref{simulation_generators_sad}.  The $5$ hour-equivalent TPC Monte Carlo sample is then generated such that all neutrons produced by the initial SAD simulation that enter the TPC volume are re-simulated as many times as necessary to extrapolate to an equivalent of 5 hours. For example, for a $1$ second-equivalent beam-time SAD simulation, each TPC neutron would be re-simulated $5 \times 3.6 \times 10^{3}$ times to produce the $5$ hour-equivalent TPC simulation. The rates observed in this $5$ hours-equivalent sample are then rescaled to the experimental data run durations and averages of the beam current $I$, $Z_{e}$, and the local vacuum pressure at the accelerator beam pipe position closest to the largest expected source of beam loss in the each ring.  The rescaling of simulation with the measured beam parameters for beam-gas and Touschek induced backgrounds follows the models outlined in section \ref{sec:combined_heuristic}.

 	\subsection{Fast neutrons: results}
\label{fast_neutron_results}

\subsubsection{Data samples and event yields}
Near the end of the Phase 1 commissioning run, we performed dedicated, longer-duration runs specifically to accumulate a sufficient sample of nuclear recoils in the TPCs to compare the observed number of events against the number of events predicted from reweighting the Monte Carlo via the method described in Section \ref{neutrons_analysis}, applying all selections except for the minimum track length requirement, which is only needed for accurate angular resolution.  The HER run occurred on May 23, 2016 for approximately $1.5~\mathrm{hours}$ at an average beam-size of $\sim40~\mathrm{\si\micro m}$ with initial beam current of $500~\mathrm{mA}$. Table \ref{tab:her_data_mc} shows the number of detected events compared to the reweighted Monte Carlo prediction for this run.  While the total number of detected events in this HER sample is small enough that statistical uncertainties are larger than desired, we find that the Monte Carlo underestimates the observed number of events recorded by the TPCs by approximately a factor of five.  The HER data samples are too small to further separate the experimental data into different background components.

\begin{table}[ht]
	\caption{Number of total events detected compared to the Monte Carlo prediction for the HER run.}
	\label{tab:her_data_mc}
	\centering
	\begin{tabular} {lccc}
	\toprule & MC Beam-gas & MC Touschek & Exp. data \\ \midrule
	TPC 3 & $4 \pm 0$ & $5 \pm 1$ & $ 53 \pm 7$ \\
	TPC 4 & $3 \pm 0$ & $5 \pm 1$ & $39 \pm 6$ \\ \bottomrule
	\end{tabular}
\end{table}

Due to the fact that the Touschek contribution to beam backgrounds in Phase 3 is predicted to be far more problematic in the LER than in the HER and given the very low detection rate of the TPCs, we decided to devote substantially more experiment time to collecting data from the LER than for the HER for fast neutron analysis.  The resulting larger statistics allow us to perform more detailed investigations for the neutron background from the LER, including studies of directional distributions and separating the beam-gas and Touschek contributions to the background in experimental data.  We performed dedicated LER runs on May 29, 2016 for approximately $5.5~\mathrm{hours}$ at a beam current of approximately 600 mA, topping off the beam as required. Using the emittance control knob, the beam size was set at three specific values where each run corresponded to one set beam size.  The beam size was measured using the X-ray monitors, as described in \ref{sec:xrm_corrections}, and was measured to be approximately $40~\mathrm{\si\micro m}$, $60~\mathrm{\si\micro m}$, and $90~\mathrm{\si\micro m}$ for the three runs, respectively. Each run is further divided into sub-runs.  A sub-run is defined as a period of time of stable beam conditions at the desired settings as defined above, specifically excluding injection times.  Table \ref{tab:ler_data_mc} shows the number of detected events compared to the reweighted Monte Carlo prediction for this run. We find that for the LER the agreement between simulation and experimental data is better. On average, the observed number of events is 30\% lower than predicted.

\begin{table}[ht]
	\caption{Number of total events detected compared to the Monte Carlo prediction for the LER run.}
	\label{tab:ler_data_mc}
	\centering
	%\toprule & TPC 3 & TPC 4 \\ \midrule
	%MC beam-gas & $454 \pm 9$ & $ 364 \pm 8$ \\
	%MC Touschek & $724 \pm 19$ & $556 \pm 16$ \\
	%Exp. data & $688 \pm 26$ & $743 \pm 27$ \\ \bottomrule
	\begin{tabular} {lccc}
	\toprule & MC Beam-gas & MC Touschek & Exp. data \\ \midrule
	TPC 3 & $454 \pm 9$ & $724 \pm 19$ & $688 \pm 26$ \\
	TPC 4 & $364\pm 8$ & $556\pm 16$ & $743 \pm 27$ \\ \bottomrule
	\end{tabular}
\end{table}

\subsubsection{Energy spectra of nuclear recoils from fast neutrons}
\label{subsubsec:recoil_energies}
\begin{figure}[hbt]
\begin{center}
\includegraphics[width=\columnwidth]{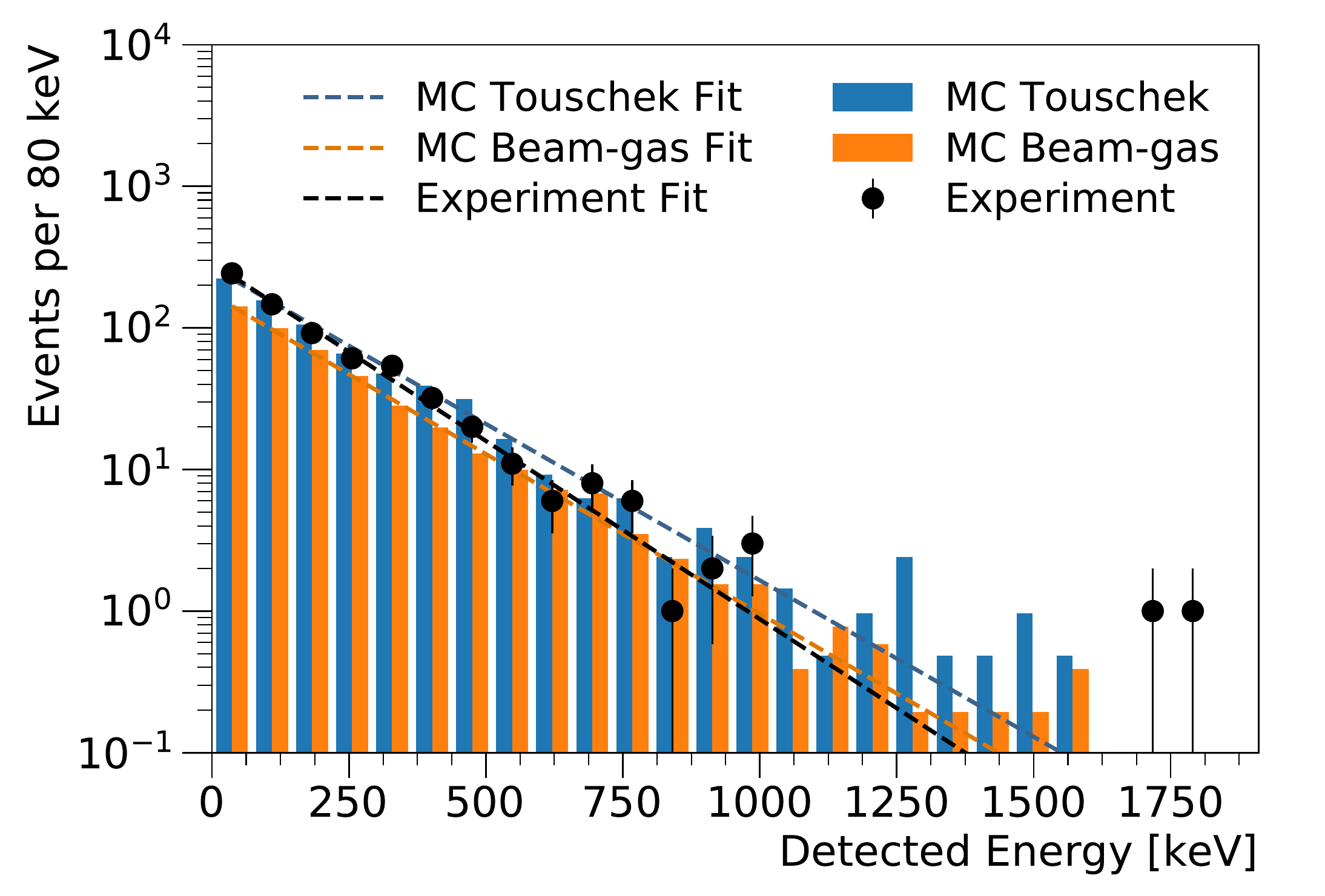}
\includegraphics[width=\columnwidth]{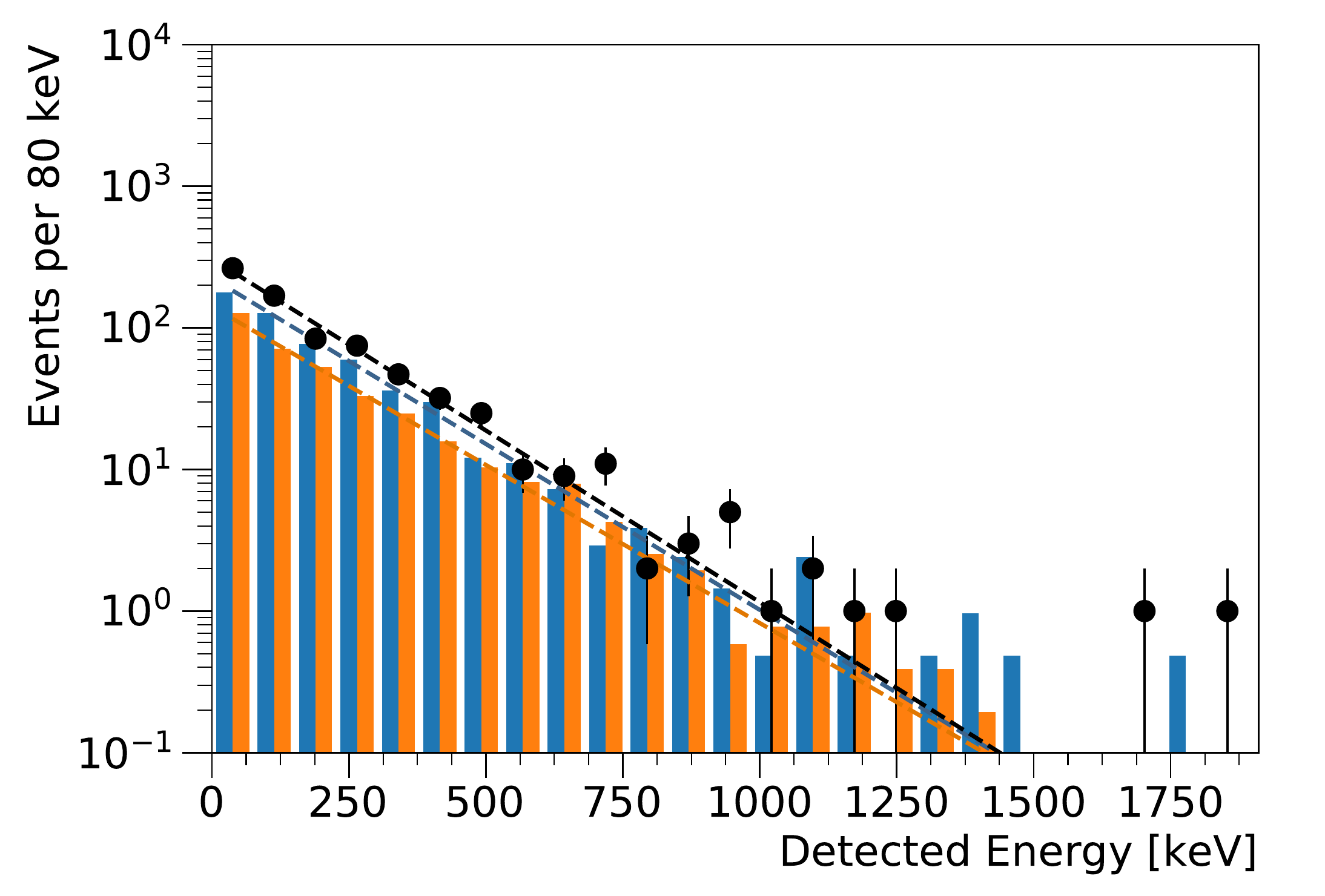}
\caption[single column]{(color online) Detected energy distribution for nuclear recoil candidates in TPCs 3 (upper plot) and 4 (lower plot) for the LER run. The blue and orange bar histograms show the expectations for Touschek and beam-gas (Coulomb and bremsstrahlung) contributions obtained via the reweighted simulation, respectively, and the black points show the measured values in experimental data.  The distributions are fit to a decaying exponential.  The dashed lines show the returned fit functions for the Monte Carlo and experimental data.  The parameters of the fit are shown in Table \ref{tab:recoil_energy_fit_ler}.}
\label{fig_tpc_recoil_energies_ler}
\end{center}
\end{figure}

\begin{figure}[hbt]
\begin{center}
\includegraphics[width=\columnwidth]{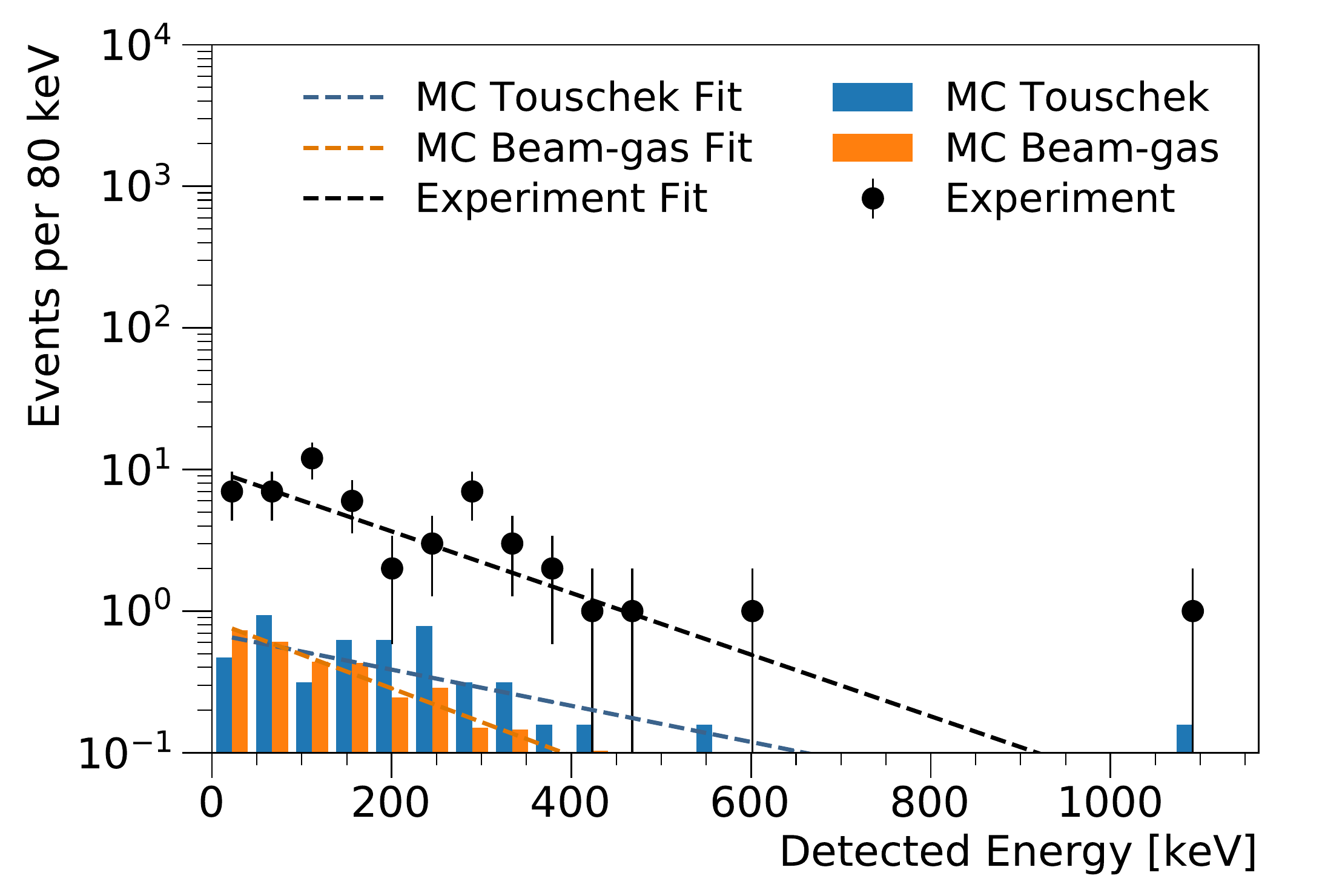}
\includegraphics[width=\columnwidth]{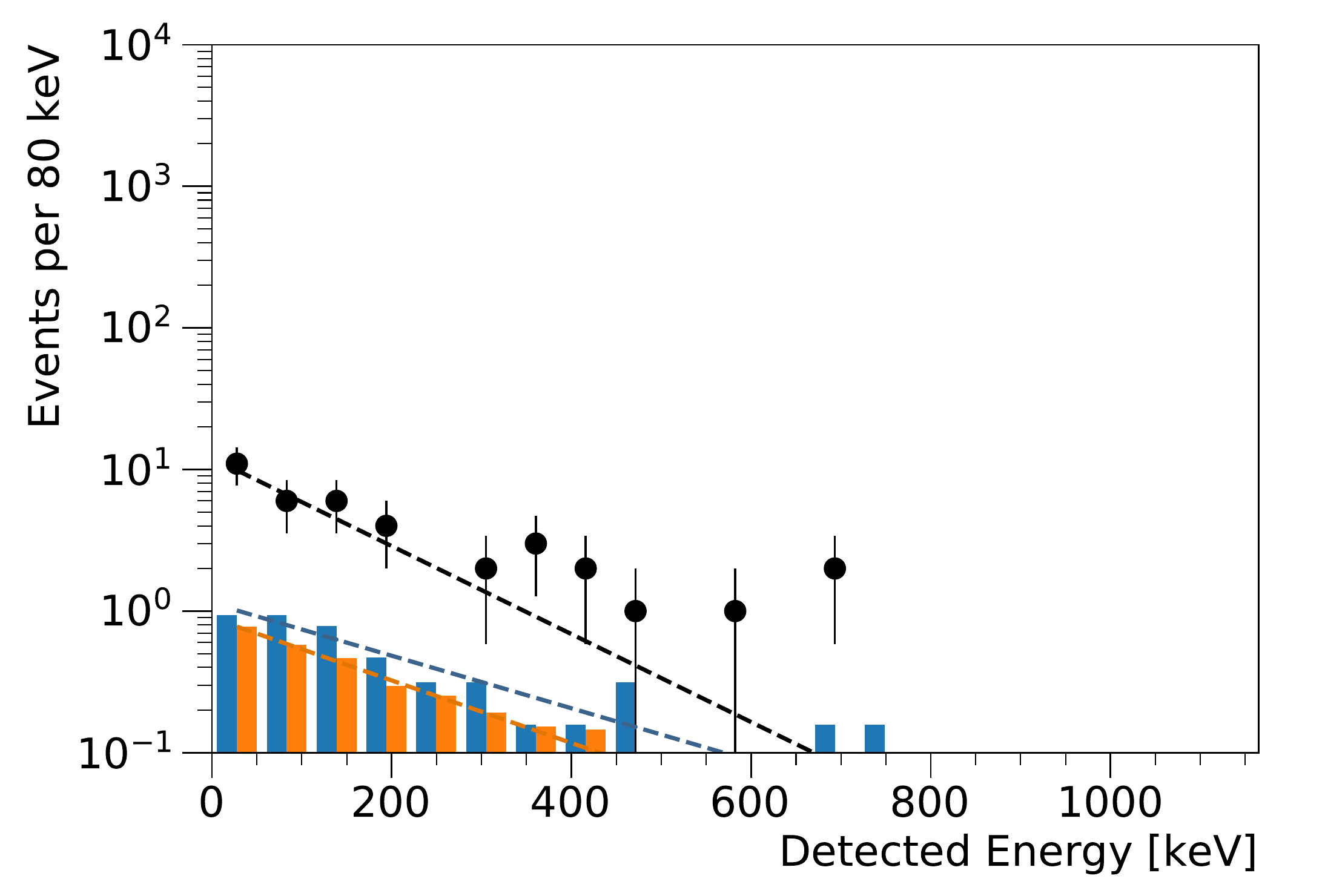}
\caption[single column]{(color online) Detected energy distribution for nuclear recoil candidates in TPCs 3 (upper plot) and 4 (lower plot) for the HER run. The stacked blue and orange bar histograms show the expectations for Touschek and beam-gas (Coulomb and bremsstrahlung) contributions obtained via rescaling the simulation, respectively, and the black points show the measured values in experimental data.  The distributions are fit to a decaying exponential.  The dashed lines show the returned fit functions for the Monte Carlo and experimental data.  The parameters of the fit are shown in Table \ref{tab:recoil_energy_fit_her}.}
\label{fig_tpc_recoil_energies_her}
\end{center}
\end{figure}

Figure \ref{fig_tpc_recoil_energies_ler} shows the recoil energy distributions for all neutron candidates collected in experimental data and the reweighted Monte Carlo simulation for the LER run. The same information is presented for the HER run in Figure \ref{fig_tpc_recoil_energies_her}.  The recoil energy distributions are fit with a decaying exponential of the form $Ae^{-bE}$, where $E$ is the recoil energy in keV.  The fit results are shown in Tables \ref{tab:recoil_energy_fit_ler} and \ref{tab:recoil_energy_fit_her}. We note that the spectral shapes, described by the parameter $b$ of each fit, of the Monte Carlo prediction and the experimental data agree fairly well, considering statistical uncertainties.

\begin{table}[ht]
	\caption{Results of fitting the recoil energy spectra for TPCs 3 and 4 for Monte Carlo and experimental data for the LER runs.}
	\label{tab:recoil_energy_fit_ler}
	%\centering
	\begin{tabular} {lcc}
	\toprule & A & b \\ \midrule
	TPC 3 MC beam-gas & $172.5 \pm 13.7$ & $0.0054 \pm 0.0002$ \\
	TPC 3 MC Touschek & $267.4 \pm 14.5$ & $0.0051 \pm 0.0002$ \\
	TPC 3 Exp. data & $288.8 \pm 16.1$ & $0.0058 \pm 0.0002$ \\
	TPC 4 MC beam-gas & $141.0 \pm 11.2$ & $0.0051 \pm 0.0003$ \\
	TPC 4 MC Touschek & $225.1 \pm 13.7$ & $0.0054 \pm 0.0003$ \\
	TPC 4 Exp. data & $308.0\pm 17.3$ & $0.0056 \pm 0.0002$ \\ \bottomrule
	\end{tabular}
\end{table}

\begin{table}[ht]
	\caption{Results of fitting the recoil energy spectra for TPCs 3 and 4 for Monte Carlo and experimental data for the HER run.}
	\label{tab:recoil_energy_fit_her}
	%\centering
	\begin{tabular} {lcc}
	\toprule & A & b \\ \midrule
	TPC 3 MC beam-gas & $0.85 \pm 0.67$ & $0.0055 \pm 0.0032$ \\
	TPC 3 MC Touschek & $0.69 \pm 0.45$ & $0.0029 \pm 0.0025$ \\
	TPC 3 Exp. data & $10.0 \pm 2.2$ & $0.0050 \pm 0.0010$ \\
	TPC 4 MC beam-gas & $0.89 \pm 0.73$ & $0.0051 \pm 0.0030$ \\
	TPC 4 MC Touschek & $1.1 \pm 0.9$ & $0.0043 \pm 0.0037$ \\
	TPC 4 Exp. data & $12.0 \pm 4.2$ & $0.0071 \pm 0.0023$ \\ \bottomrule
	\end{tabular}
\end{table}

Because the spectral shape of all three background components considered here do not differ significantly, the spectral shape can not be used to separate the different background components. Instead, we attempt to achieve this separation by two other methods: by utilizing the background rate dependence on accelerator beam size, and by utilizing the recoil angle distribution.

\subsubsection{Analysis of fast neutron rates versus beam size}
\label{subsubsec:fastneutrons_heuristic}
\begin{figure}[hbt]
	\centering
	\includegraphics[width=\columnwidth]{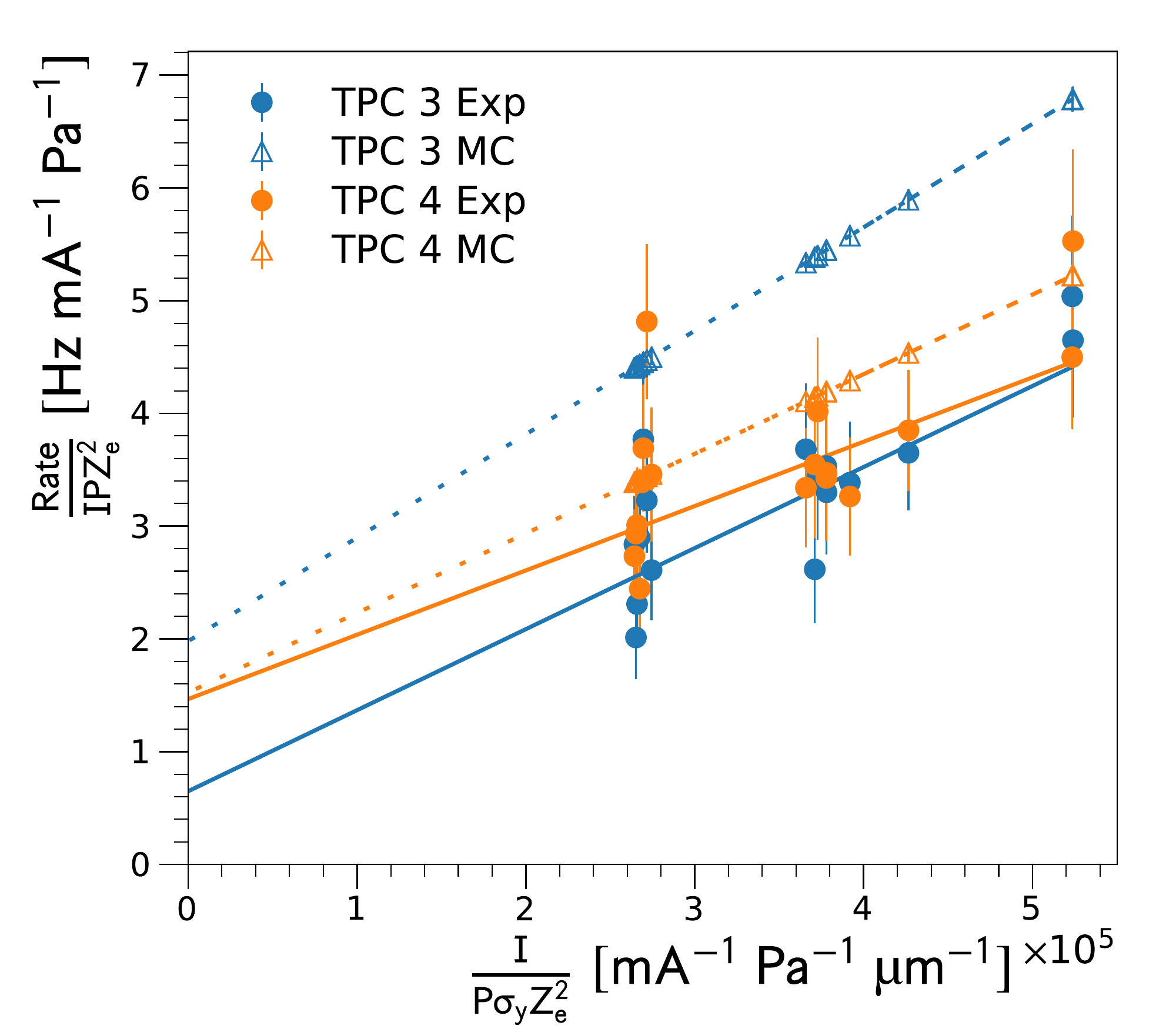}
	\caption[single column]{(color online) Plot of the LER beam-gas and Touschek fast neutron rates in the TPCs, using the same methodology outlined Section \ref{sec:combined_heuristic}. The orange and blue circles  correspond to the results from experimental data, and the orange and blue triangles correspond to the results from Monte Carlo.}

\label{fig:tpc_sensitivities}
\end{figure}

For analyzing the fast neutron background in Phase 1 with the TPCs, we apply the analysis method described in Section \ref{sec:combined_heuristic}, with the observable $O$ being the rate of nuclear recoils detected in the TPCs.  We then compare the rate of nuclear recoils in experimental data to the rate expected from reweighting the Monte Carlo simulation, as described at the end of Section \ref{neutrons_analysis}. The rates versus LER beam size are shown in Figure \ref{fig:tpc_sensitivities}.  For comparison with other methods, we integrate the measured and predicted rates to give a yield, which we denote as $N_{T}$ for the yield of Touschek events or $N_{bg}$ for beam-gas events.  The observed yields are shown in Table \ref{tab:tpc_sensitivities}. We find a significant disagreement between the predictions from the reweighted Monte Carlo and the experimental data in the horizontal plane of the beam-pipe, or in TPC 3, in the beam-gas component. The experimental data and the Monte Carlo prediction for TPC 4, located in the vertical plane of the beam-pipe are in better agreement than for TPC 3, however the Monte Carlo over estimates the rate of nuclear recoils for both beam-gas and Touschek backgrounds by approximately $30\%$.

\begin{table}[ht]
	\centering
	\caption{Calculated yield from the measured rates of nuclear recoils from beam-gas and Touschek backgrounds shown in Figure \ref{fig:tpc_sensitivities} for both experimental data and Monte Carlo in each TPC. \newline}  %For the TPCs, the observable is the number of neutrons detected in the duration of the TPC LER runs. \newline}
	\begin{tabular} {lcc}
	\toprule & $N_{bg}$ & $N_{T}$ \\ \midrule
	TPC 3 Exp.& $113 \pm 18$ & $ 566 \pm 25$ \\
	TPC 3 MC & $431 \pm 20$ & $ 723 \pm 21$ \\
	TPC 4 Exp. & $255 \pm 20$ & $ 450 \pm 27$ \\
	TPC 4 MC & $332 \pm 17$ & $ 556 \pm 18$ \\ \bottomrule
	\end{tabular}
	\label{tab:tpc_sensitivities}
\end{table}

\subsubsection{Analysis of 3D nuclear recoil tracks}
\label{subsubsec:tpc_angular_dists}
\begin{figure}[hbt]
\begin{center}
\includegraphics[width=\columnwidth]{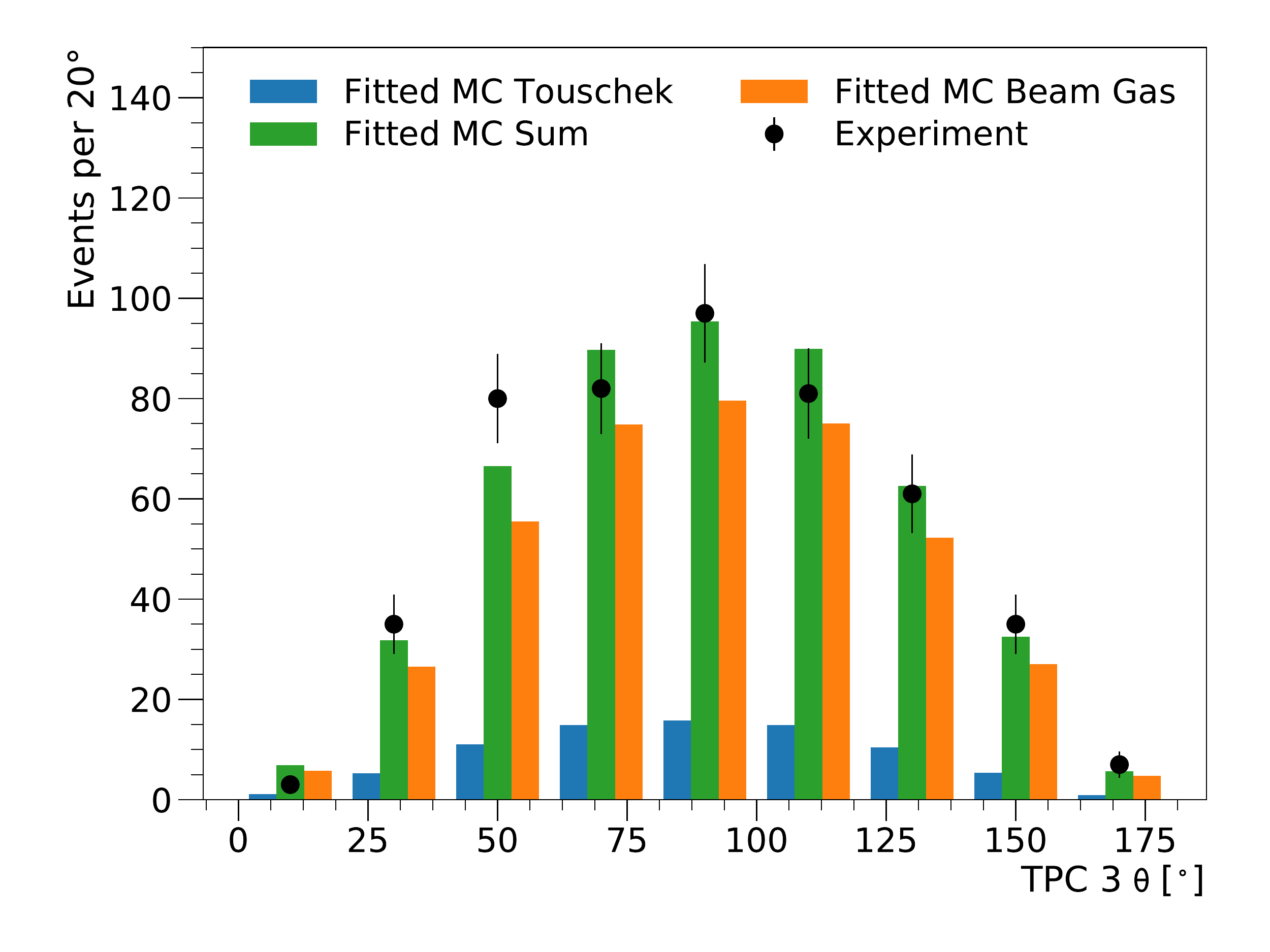}
\includegraphics[width=\columnwidth]{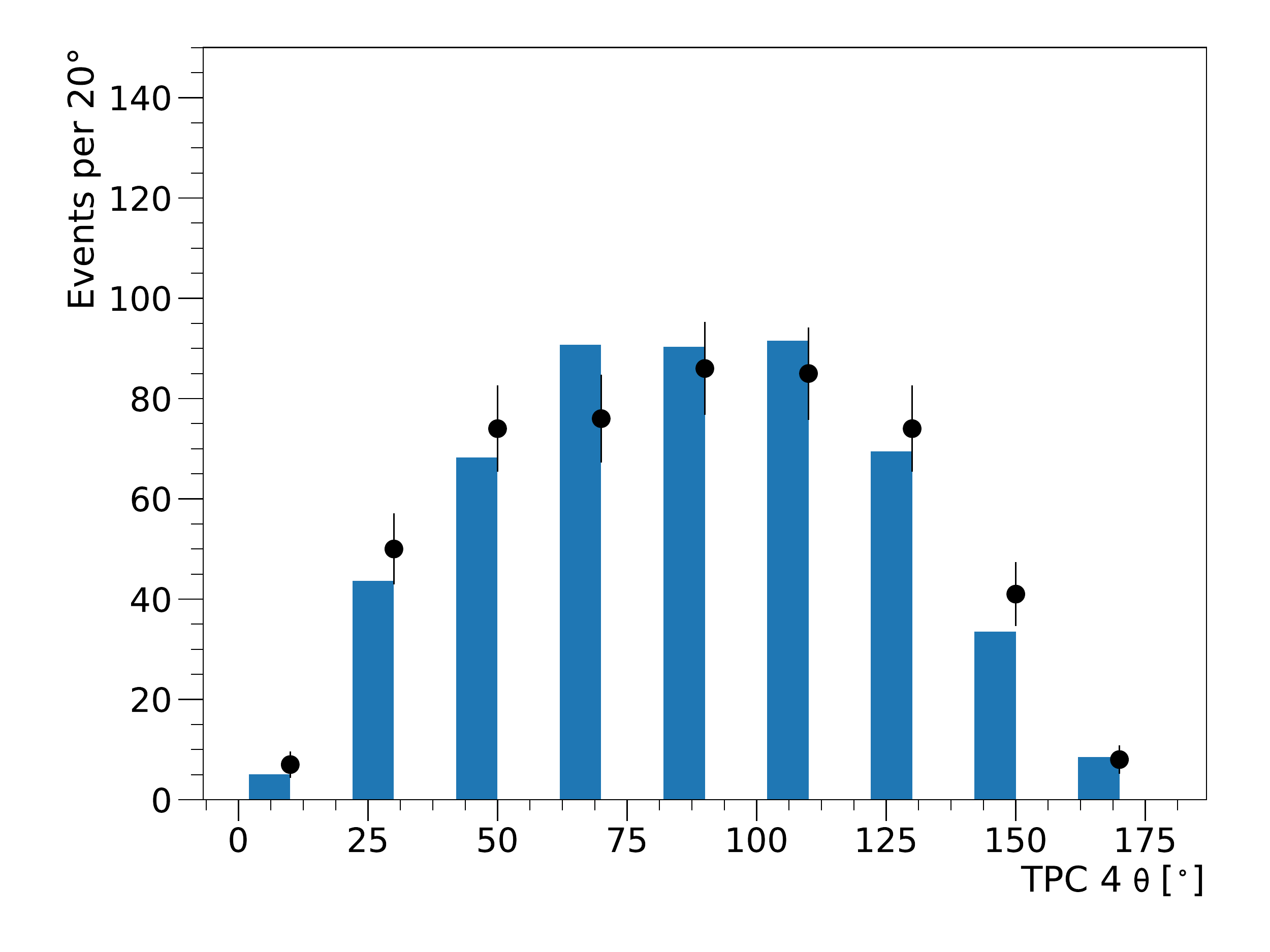}
\caption[single column]{(color online) Results of fitting nuclear recoil $\theta$ distributions in experimental data to a sum of histogram PDFs from LER Monte Carlo Touschek and beam-gas samples.  The individual blue and orange histograms in the upper plot represent the fitted yields of the Monte Carlo fast neutrons generated from Touschek and beam-gas backgrounds, respectively, and the green histogram represents the sum of Touschek and beam-gas backgrounds.  The black points are the values obtained from experimental data. The lower plot shows that for TPC 4, the fit converges on a solution of zero contribution from beam-gas backgrounds.  Therefore, only the Touschek distribution is shown.}
\label{fig:tpc_histogrampdf_theta}
\end{center}
\end{figure}

The TPCs also have a unique method to separate fast neutrons from beam-gas and Touschek backgrounds. We utilize the reconstructed $\theta$ distributions of nuclear recoils in the unweighted $5$ hours-equivalent Monte Carlo sample for beam-gas and Touschek backgrounds in each TPC to construct histogram PDFs for each background type. We then fit the experimental data with the beam-gas and Touschek background histogram PDFs by letting the fractional contribution of each histogram to the experimental data float.  This separately calculates the fraction of events from beam-gas and Touschek backgrounds that would be necessary to best describe the angular distributions measured in experimental data. The fractional fit used in these histograms also considers per-bin statistical fluctuations in the Monte Carlo. The fitted fractional contributions of beam-gas and Touschek backgrounds are shown in Figure \ref{fig:tpc_histogrampdf_theta} and Table \ref{tab:tpc_histogram_pdf}.

We find that statistical uncertainties are quite large. Also, this method has a variety of systematic uncertainties. The values presented in Figure \ref{fig:tpc_histogrampdf_theta} correspond to selecting tracks with a total length of greater than $2~\mathrm{mm}$, as explained in Section \ref{neutrons_analysis}.  The values this fit converges on vary as the track length cut is varied such that the error of the fitted fractional contribution of each background can vary from $0\%$ to $100\%$.  Additionally, independent of variation in selections, this fit will report large errors if the experimental data is in truth represented by additional background components not accounted for in the simulation.  Therefore, we conclude that while this technique does not perform at a robust level for the Phase 1 analyses presented here, further improvements could be made to improve this technique to become a powerful background discrimination tool later, possibly for SuperKEKB commissioning Phase 2, where we will be utilizing eight TPCs and neutron rates will be higher.

\begin{table}[ht]
	\centering
	\caption{Yields obtained from fitting for the fractional yields of beam-gas and Touschek in nuclear recoil $\theta$ distributions in experimental data compared to the yields predicted in the Monte Carlo simulation. $N_{bg}$ represents the yield from the beam-gas contribution, and $N_{T}$ represents the yield from the Touschek contribution.}
	\begin{tabular} {ccc}
	\toprule & $N_{bg}$ & $N_{T}$ \\ \midrule
	TPC 3 Exp. & $404 \pm 113$ & $81 \pm 112$ \\
	TPC 3 MC & $273 \pm 7 $ & $431 \pm 14$ \\
	TPC 4 Exp. & $0.0 \pm 46$ & $503 \pm 490$ \\
	TPC 4 MC & $212 \pm 6$  & $329.0 \pm 13$ \\ \bottomrule
	%\hline & TPC 3 & TPC 4 \\ \hline
	%$BG_{Exp}$ & $188.3  \pm 188.7$  & $0.0 \pm 354.0$ \\
	%$BG_{MC}$ & $140.0 \pm 8.6 $ & $89.0 \pm 6.8$ \\
	%& & \\
	%$T_{Exp}$ & $277.7 \pm 190.1$ & $487.0 \pm 460.7$ \\
	%$T_{MC}$ & $22.0 \pm 2.3$  & $19.0 \pm 2.1$ \\ \hline
	\end{tabular}
	\label{tab:tpc_histogram_pdf}
\end{table}

\begin{figure}[hbt]
	\centering
	\includegraphics[width=\columnwidth]{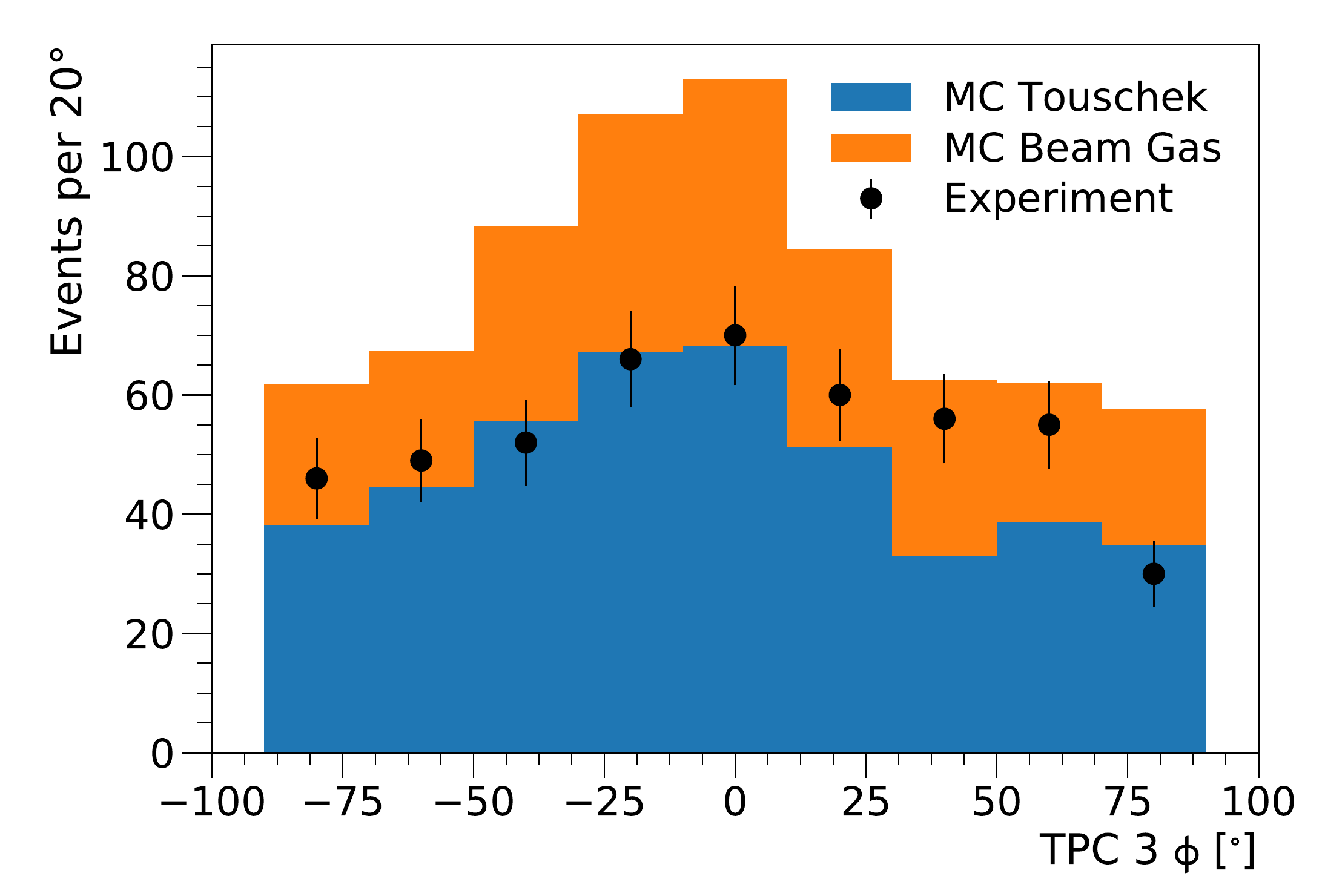}
	\includegraphics[width=\columnwidth]{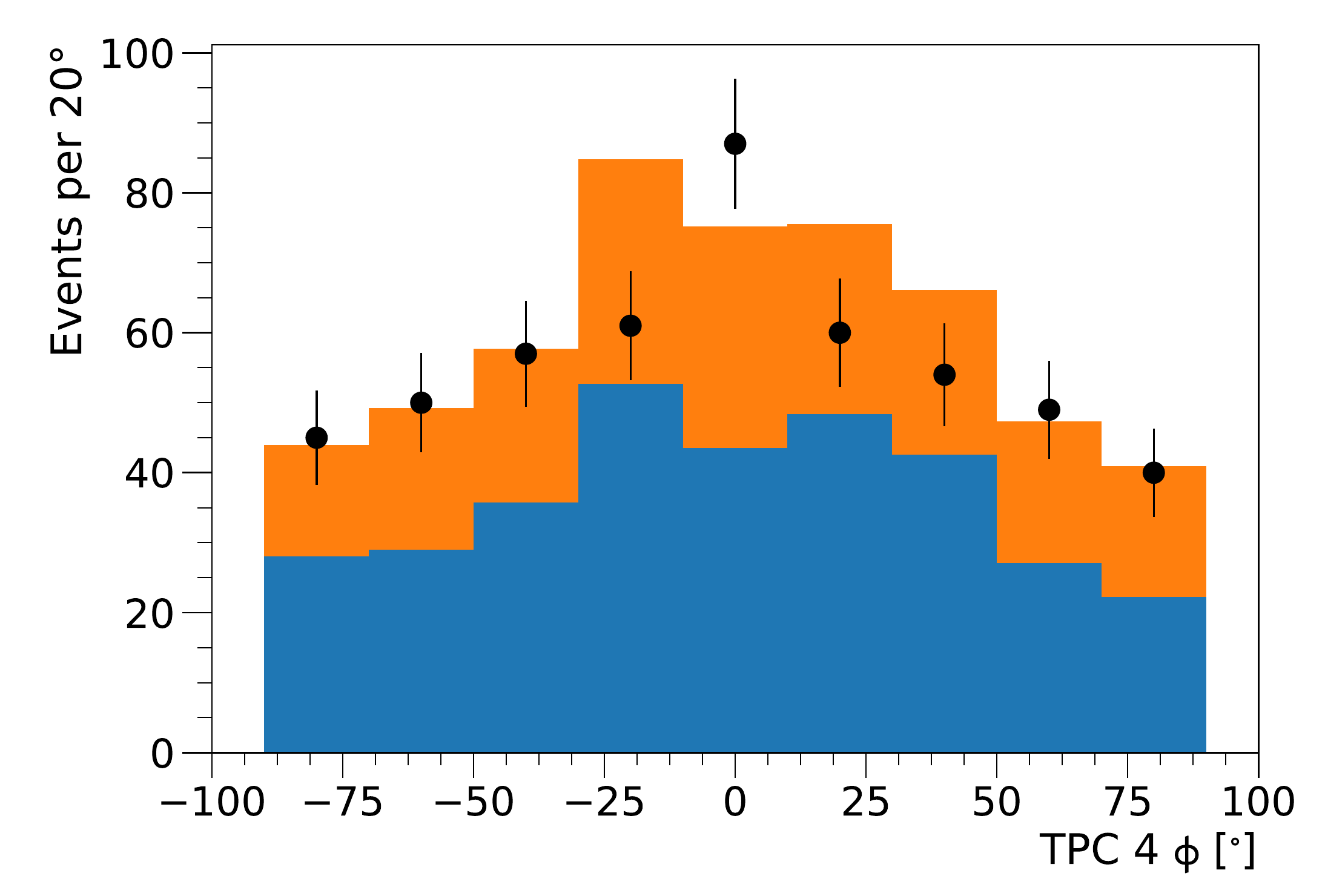}
	\caption[single column]{(color online) Azimuthal angle ($\phi$) distribution for neutron event candidates in TPCs 3 (upper plot) and 4 (lower plot) for Monte Carlo and experimental LER data.  The blue and orange histograms show the the stacked expectations for Touschek and beam-gas contributions obtained via rescaling the simulation, and the black points show the measured values in experimental data.}
\label{fig_tpc_phi_weighted}
\end{figure}

We also present the angular distributions of the recoil tracks for the LER.  Figure \ref{fig_tpc_phi_weighted} shows the raw azimuthal angle, $\phi$, distribution of the nuclear recoils recorded in experimental data compared to  rescaled the Monte Carlo distributions. For both TPCs in experimental data, a peak around $\phi=0^{\circ}$ is observed. As described in Section \ref{neutrons_analysis}, the reconstructed local $\phi$ measurement of a track is ``folded," meaning that it is constrained within the range of $-90^{\circ}$ to $+90^{\circ}$.  We therefore conclude that the peak at $\phi=0^{\circ}$ in both TPCs indicates that most of the fast neutrons interacting in the detectors are originating from the beam-pipe. The relatively broad width of the distributions is due to a combination of the finite recoil angle (i.e.\ individual recoils are not in the same direction as the incoming neutrons) and some scattering of neutrons.

\begin{figure*}
	\centering
	\begin{tabular}{cc}
	\includegraphics[width=\columnwidth]{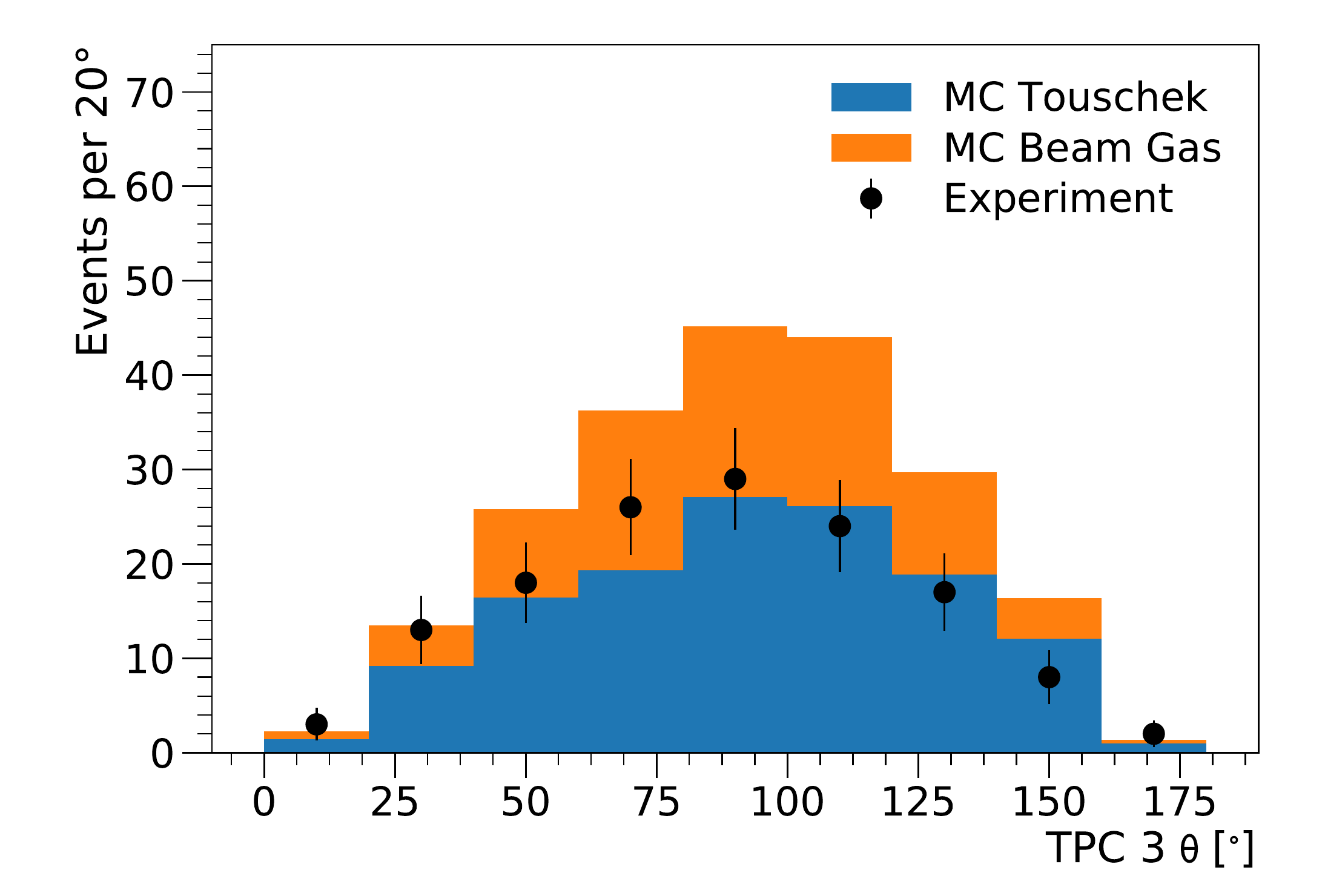} & \includegraphics[width=\columnwidth]{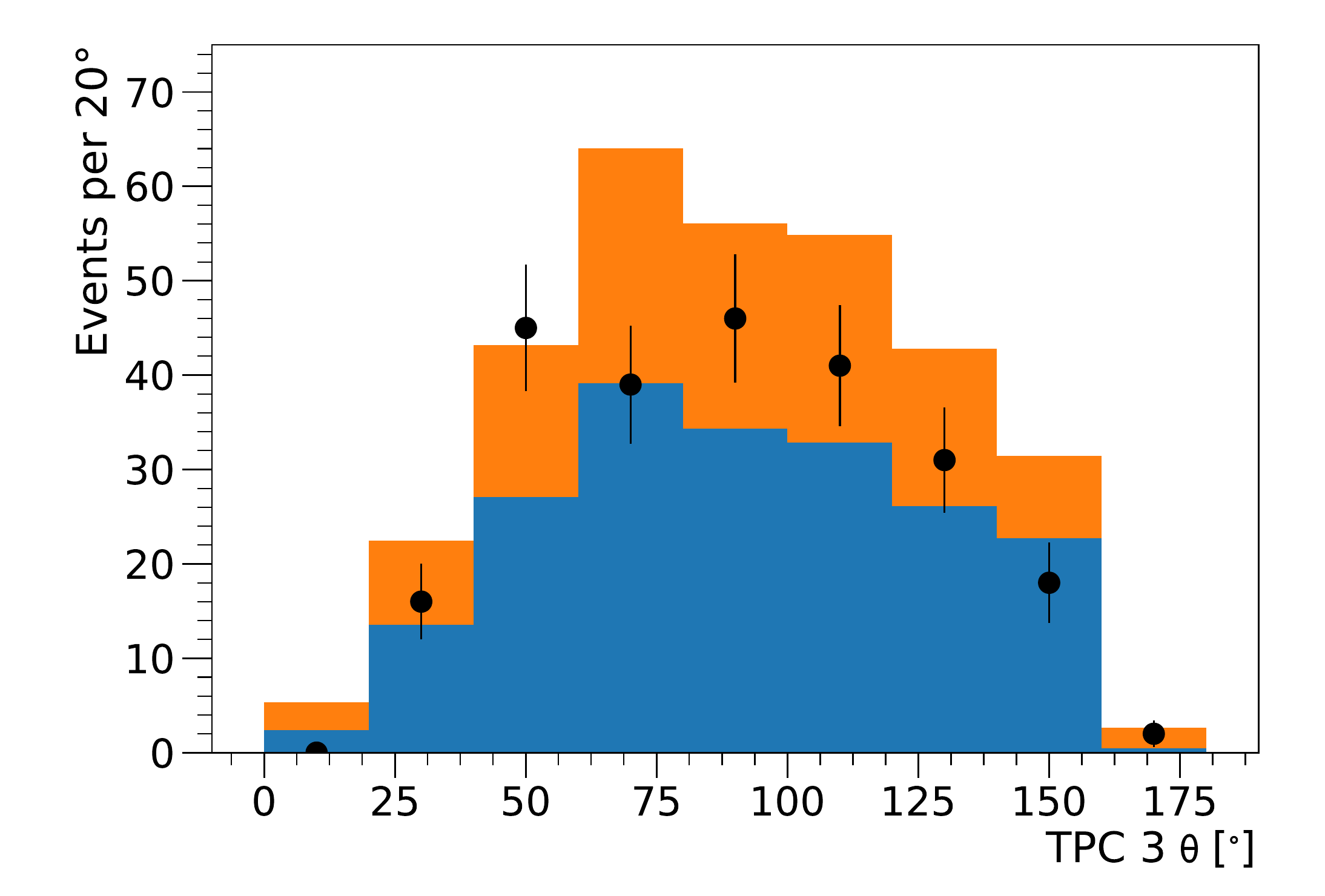} \\
	\includegraphics[width=\columnwidth]{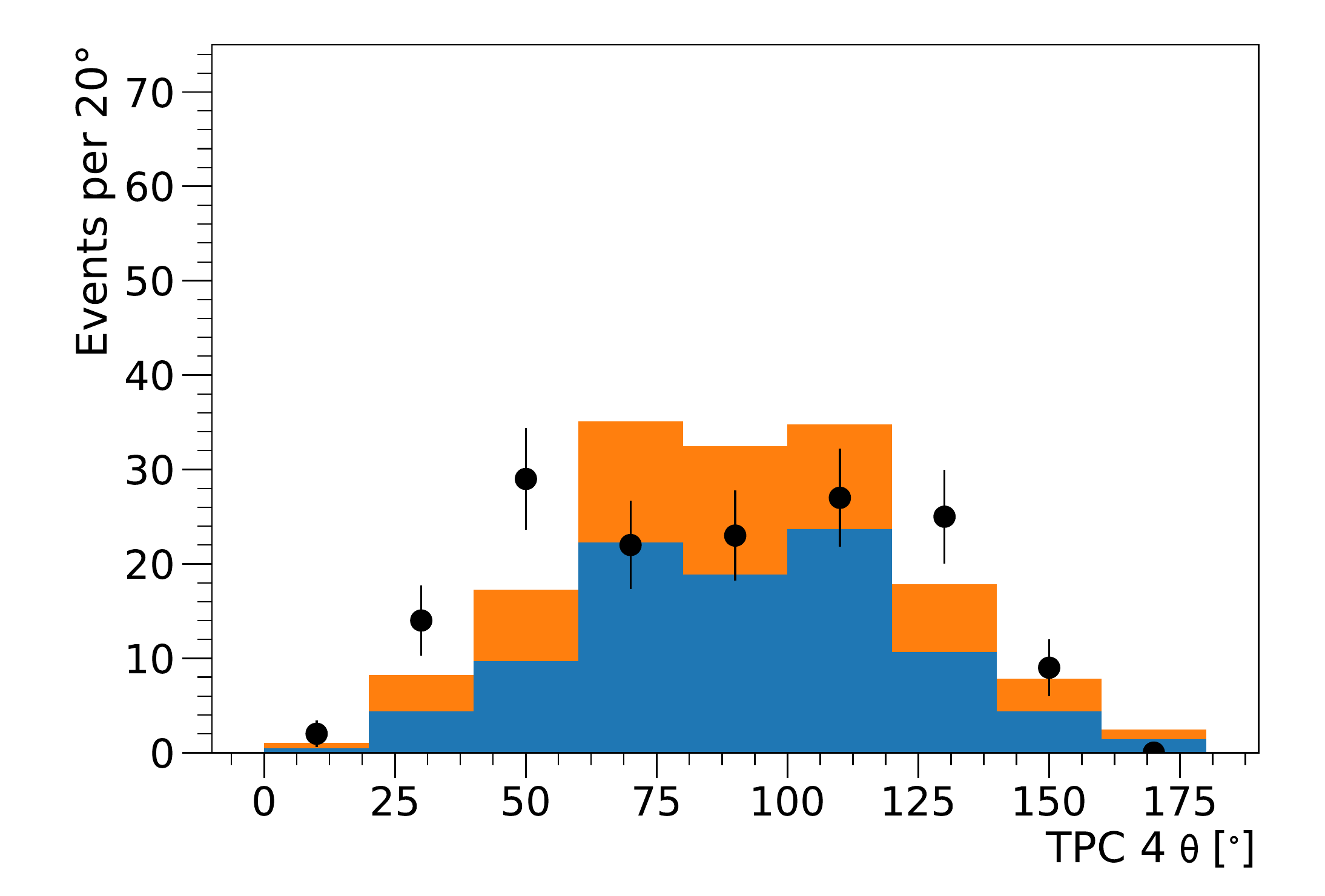} & \includegraphics[width=\columnwidth]{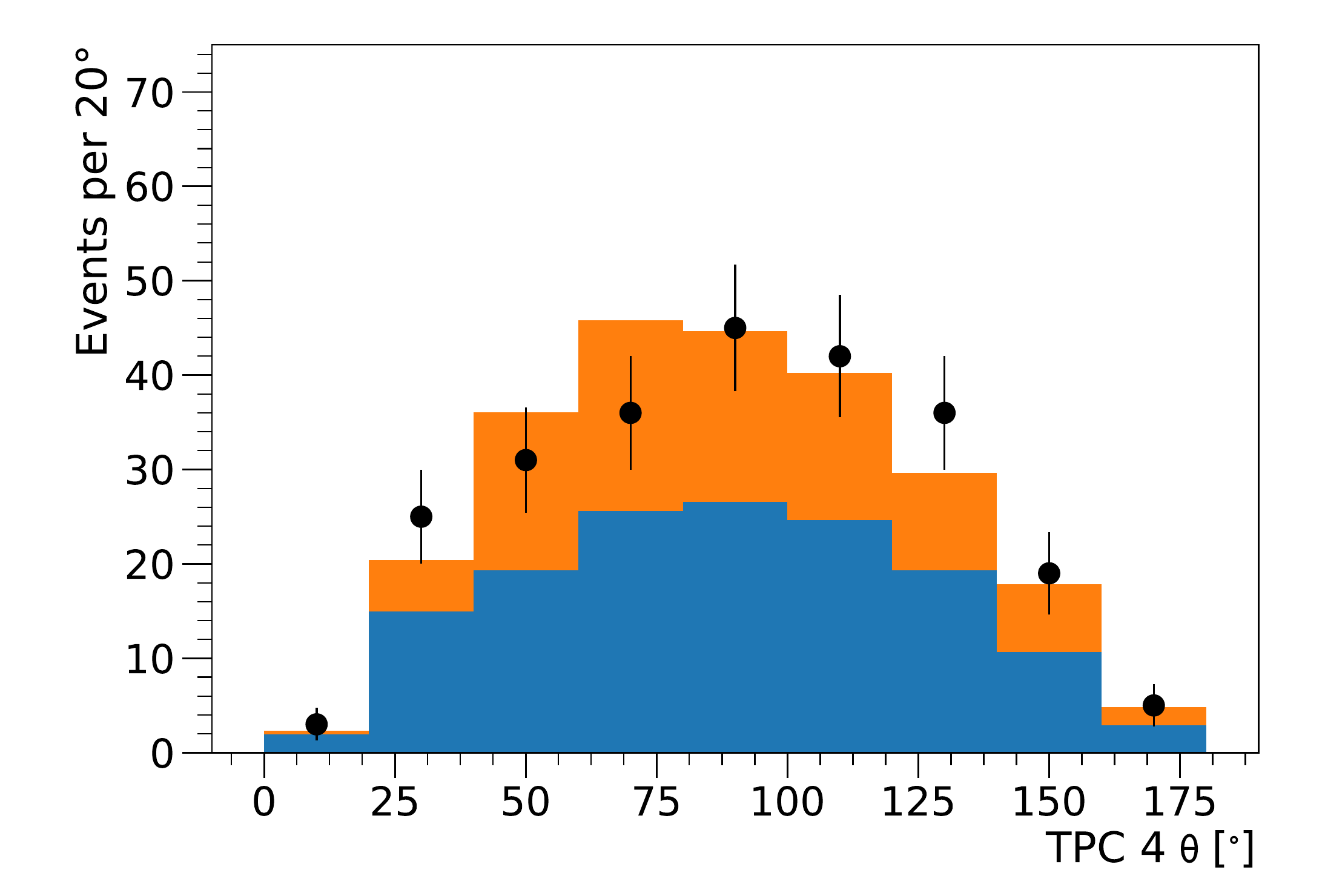} \\
	Prompt neutron sample & Re-scattered neutron sample \\
	
\end{tabular}
	\caption[double column]{(color online) Polar angle ($\theta$) distribution for prompt (left column) and rescattered (right column) selected neutron candidates in TPCs 3 (upper row) and 4 (lower row) for experimental and Monte Carlo data. The blue and orange histograms show the stacked expectations for Touschek and beam-gas contributions obtained via reweighting the simulation, and the black points show the measured values in experimental data.}
\label{fig_tpc_theta_unweighted_bp}
\end{figure*}

By selecting events with $|\phi| < 20^{\circ}$, we obtain a ``prompt'' sample, where neutrons coming directly from the beam-pipe to the TPC are enhanced. Conversely, by selecting events with $|\phi| > 40^{\circ}$, we obtain a ``re-scattered'' sample, where events with neutron scattering are enhanced. This allows us to compare the scattered and non-scattered distributions separately against Monte Carlo. Both neutron production in showers and neutron scattering in material needs to be simulated accurately to get good experiment/simulation agreement in both samples. Figure \ref{fig_tpc_theta_unweighted_bp} shows the polar angle, $\theta$, distributions for these two samples in experimental data and the reweighted TPC Monte Carlo.  Statistics are really too low to draw firm conclusions, but the agreement between experiment and simulation appear slightly better for the re-scattered sample.

\subsubsection{Fast neutrons: conclusions}
In concluding the discussion of fast neutron backgrounds, we note the following conclusions:

\begin{itemize}
\item{While limited in statistics, the HER Monte Carlo prediction underestimates the experimentally measured fast-neutron yield by a factor of five.}
\item{For the LER, the observed number of events is 30\% lower than predicted, on average. The Monte Carlo predicts a larger rate in the TPC in the horizontal plane than the TPC in the vertical plane. The experimental data shows that both TPCs are more accurately described by the rates predicted by the vertical plane.}
\item{The analysis of neutron rates versus beam size suggests the rate discrepancy between the two TPCs in the LER simulation is due to the measured beam gas component in the horizontal plane of the beam-pipe being smaller than the prediction.}

\item{The observed energy spectrum of nuclear recoils agrees fairly well with simulation. This suggest the neutron production, neutron scattering, and energy loss mechanisms, such as material of the experimental setup, are accurately simulated. These factors are unlikely to be the cause of the observed rate discrepancies.}

\item{The angular recoil distributions indicate that the observed backgrounds do indeed originate from the beam pipe.}

\item{There is more information contained in the angular recoil distributions. We have demonstrated how this can provide a complimentary method to separate backgrounds, and to separate direct from rescattered neutrons. The results are still hampered by low statistics, but these new methods may be quite useful in the coming SuperKEKB commission stages, where background rates will be much higher.}
\end{itemize}

%\begin{figure}[hbt]
%	\centering
%	\includegraphics[width=\columnwidth]{TPCimages/TPC3_theta_datavsmc_sim_weighted.pdf}
%	\includegraphics[width=\columnwidth]{TPCimages/TPC4_theta_datavsmc_sim_weighted.pdf}
%	\caption[single column]{Polar angle ($\theta$) distribution for neutron event candidates in TPCs 3 (upper plot) and 4 (lower plot) for Monte Carlo and experimental data.  The blue and orange histograms (color available online) show the the stacked expectations for Touschek and beam-gas contributions obtained via rescaling the simulation, and the black points show the measured values in experimental data.}
%\label{fig_tpc_theta_weighted}
%\end{figure}

 \clearpage

 %lead author: Igal Jaegle
 \section{Dosimetry}\label{sec:dosimetry}
 % File:		dosimetry.tex
% Lead author:	Igal Jangle

It is critical to have a good understanding of the integrated dose near the interaction point. Otherwise, we risk damaging or reducing the longevity of the Belle II detectors in SuperKEKB Phases 2 and 3. The vertex detectors, which reside closest to the beam pipe, are particularly vulnerable. To keep Belle II safe, we must ensure that the simulation accurately predicts the integrated dose, that our dose measurements are accurate, and that the beam abort system is sufficiently sensitive. Despite the background distributions being quite different in each of the SuperKEKB commissioning phases, these three issues can be studied already in Phase 1, as described in this section.

\subsection{Measured versus predicted integrated dose}

\begin{figure}[ht!]
\centering
\includegraphics[width=1.0\columnwidth]{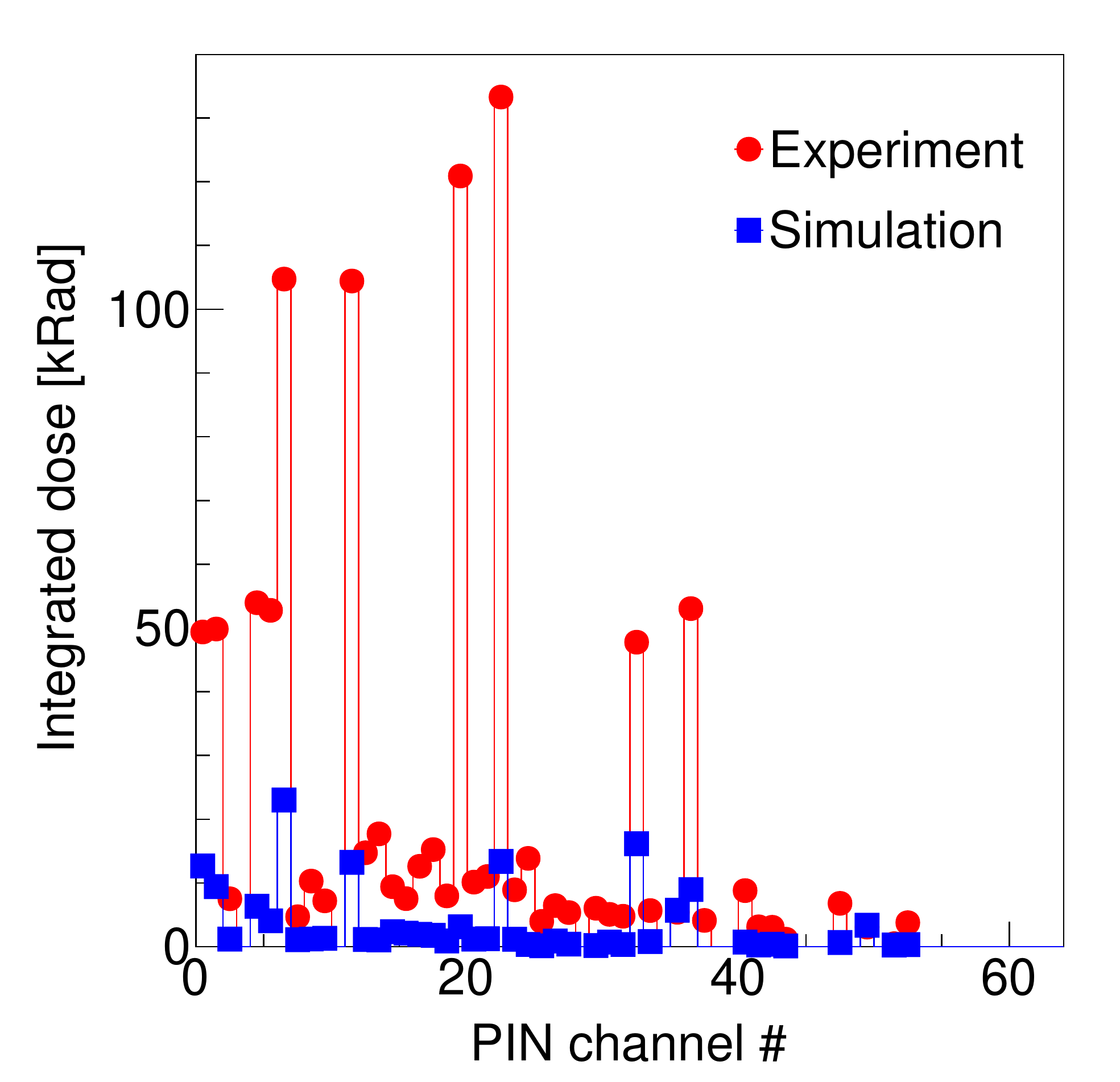}
\caption{Measured and simulated integrated dose for each channel of the PIN detector system, with periods of injection excluded.}
\label{fig:PINDosimetrySummary}
\end{figure}

\begin{figure}[ht!]
\centering
\includegraphics[width=1.0\columnwidth]{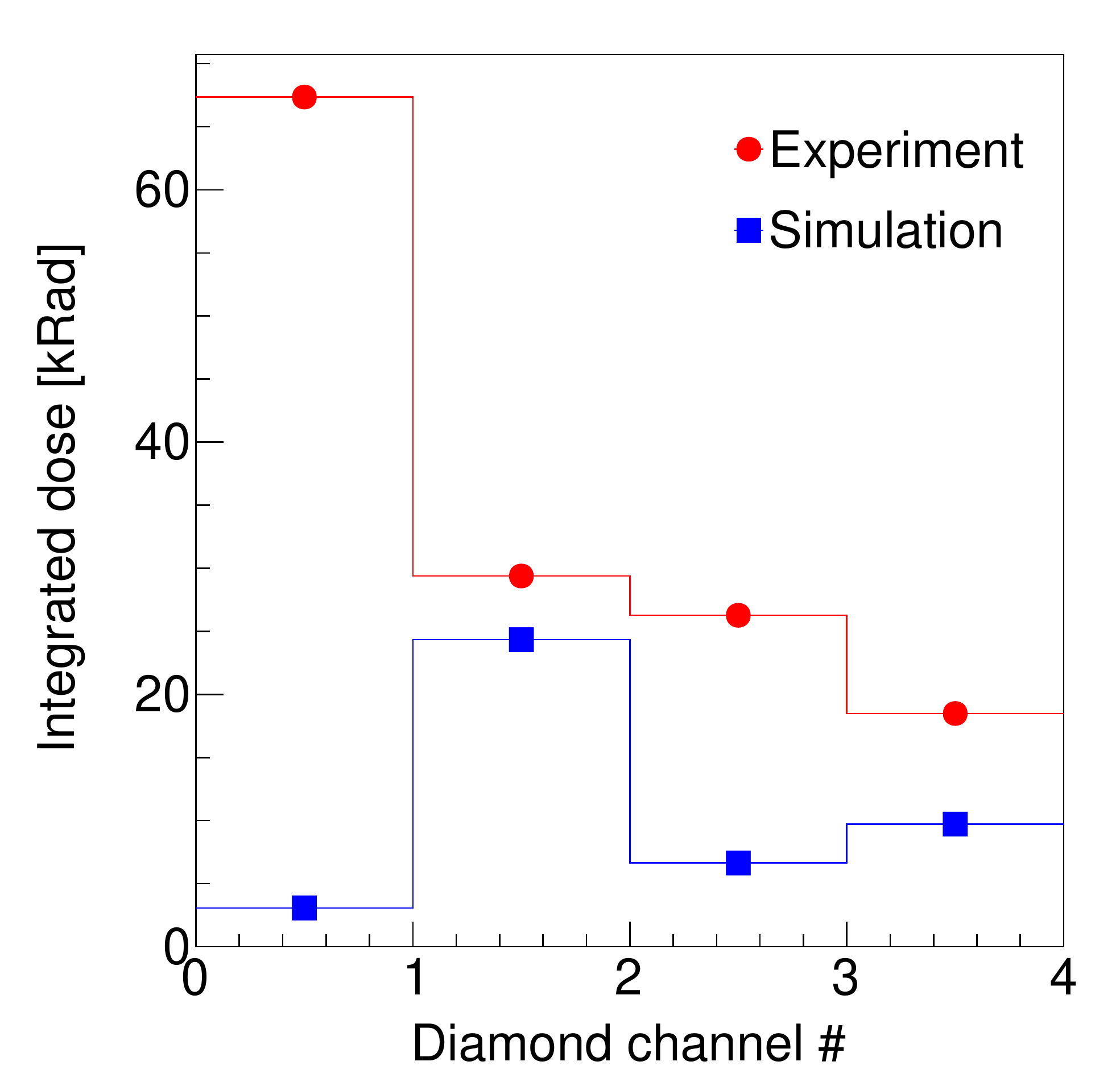}
\caption{Measured and simulated integrated dose for each channel of the Diamond detector system, with periods of injection excluded.}
\label{fig:DIADosimetrySummary}
\end{figure}

\begin{figure}[ht!]
\centering
\includegraphics[width=1.0\columnwidth]{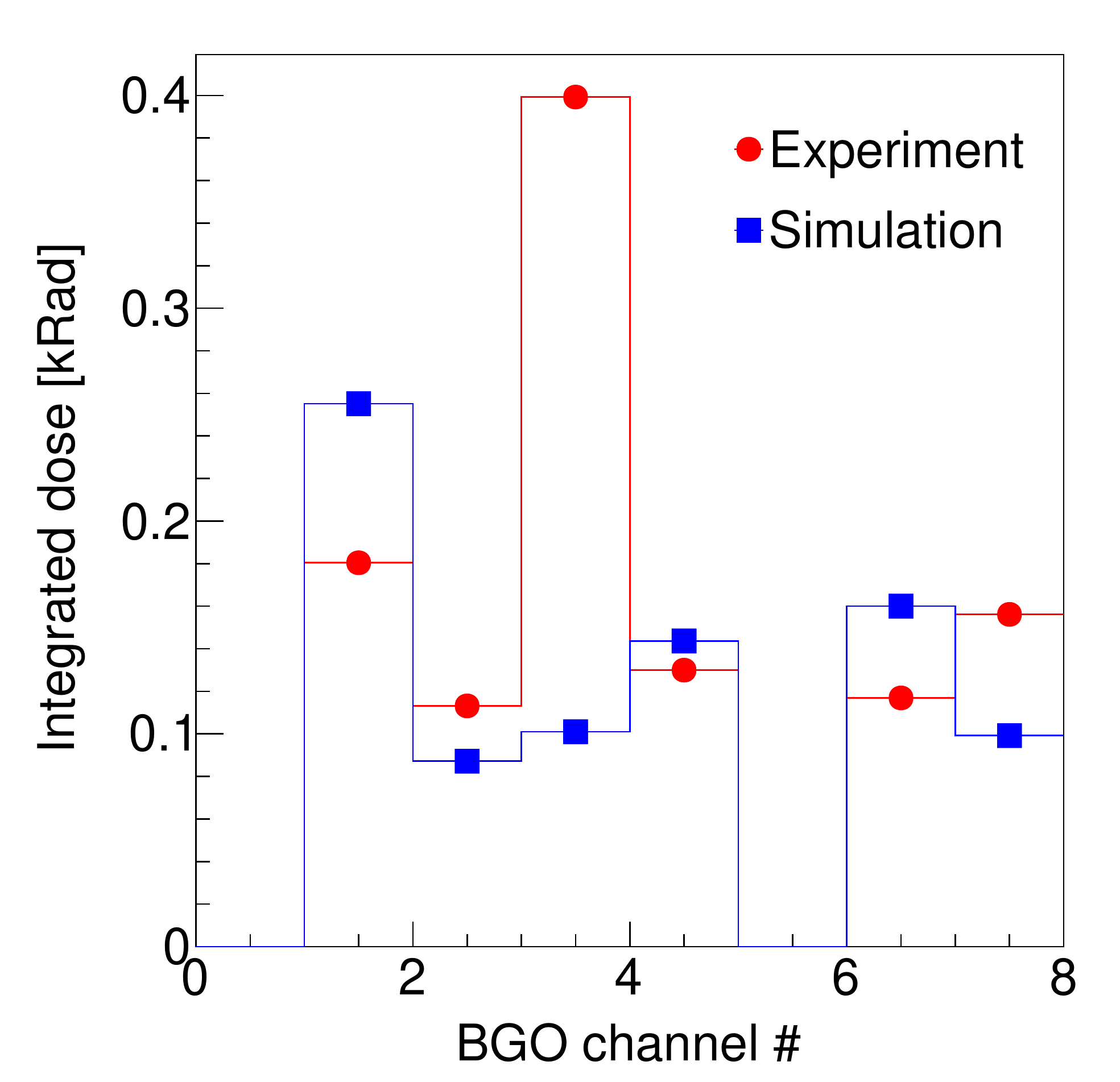}
\caption{Measured and simulated integrated dose for each channel of the BGO detector system, with periods of injection excluded.}
\label{fig:BGODosimetrySummary}
\end{figure}

\begin{figure}[ht!]
\centering
\includegraphics[width=1.0\columnwidth]{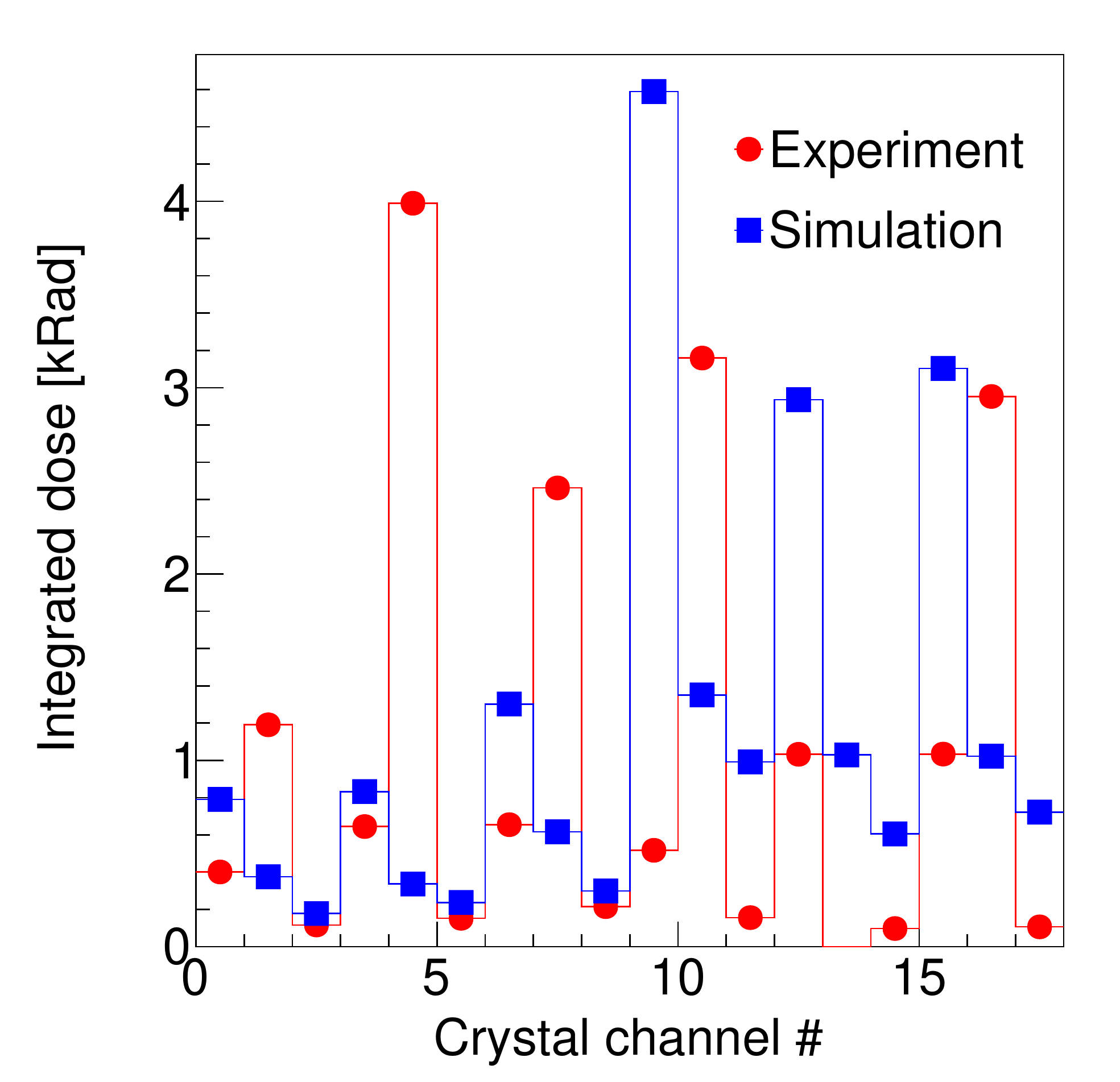}
\caption{Measured and simulated integrated dose for each channel of the Crystals detector system, with periods of injection excluded.}
\label{fig:CSIDosimetrySummary}
\end{figure}

\begin{figure}[ht!]
\centering
\includegraphics[width=\columnwidth]{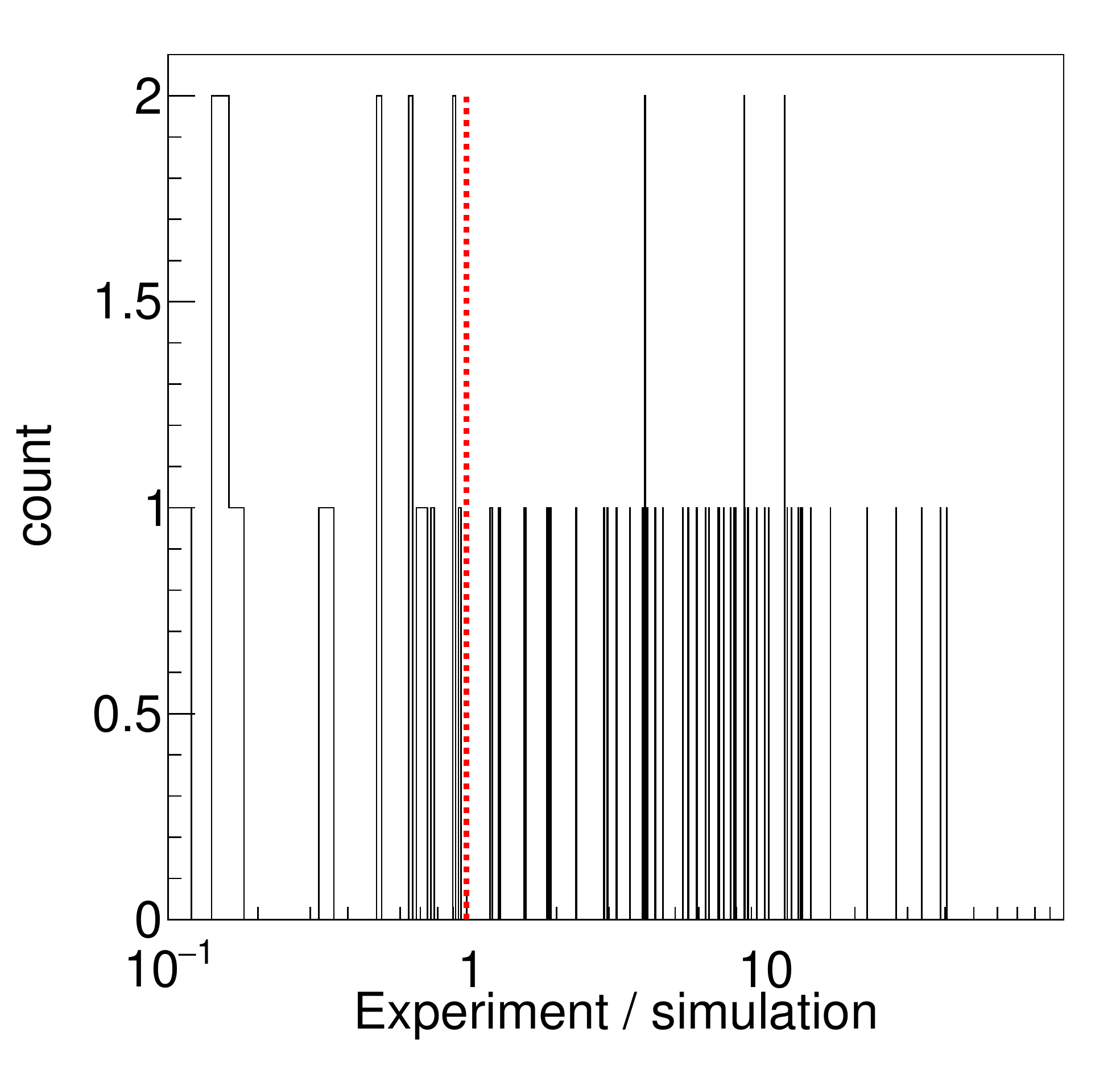}
\caption{Ratio of the measured and simulated integrated dose in four of the BEAST II detector systems, with periods of injection excluded.}
\label{fig:DoseRatio}
\end{figure}

The integrated dose in BEAST II is measured by the PIN, Diamond, BGO, and Crystals detector systems. The integrated dose is updated each second. Since injection backgrounds are not simulated, any data recorded during periods of either LER or HER injection are excluded from the analysis when the experimental dose is compared against simulation. Figures~\ref{fig:PINDosimetrySummary}, ~\ref{fig:DIADosimetrySummary}, ~\ref{fig:BGODosimetrySummary}, and ~\ref{fig:CSIDosimetrySummary} show the measured and predicted integrated dose for each channel of the Diamond, PIN, BGO, and Crystals systems, with periods of injection excluded. The resulting ratios of the measured to the simulated integrated dose are summarized in Figure~\ref{fig:DoseRatio}. The simulation underestimates the measured dose by a factor of 2 to 40 for sensors insensitive to HER and overestimates the the measured dose by a factor of 2 to 10 for sensors with non-negligible HER sensitivity, which are mostly the Crystals located at $\sim$ -118 cm (see Figure~\ref{fig:CSIDosimetrySummary}, channel 8 to 17). In simulation, most of the dose is due to LER Touschek and LER beam-gas events, in roughly equal proportion, while the dose contributed by HER Touschek and HER beam-gas events is negligible.

\begin{figure}[ht!]
\centering
\includegraphics[width=1.0\columnwidth]{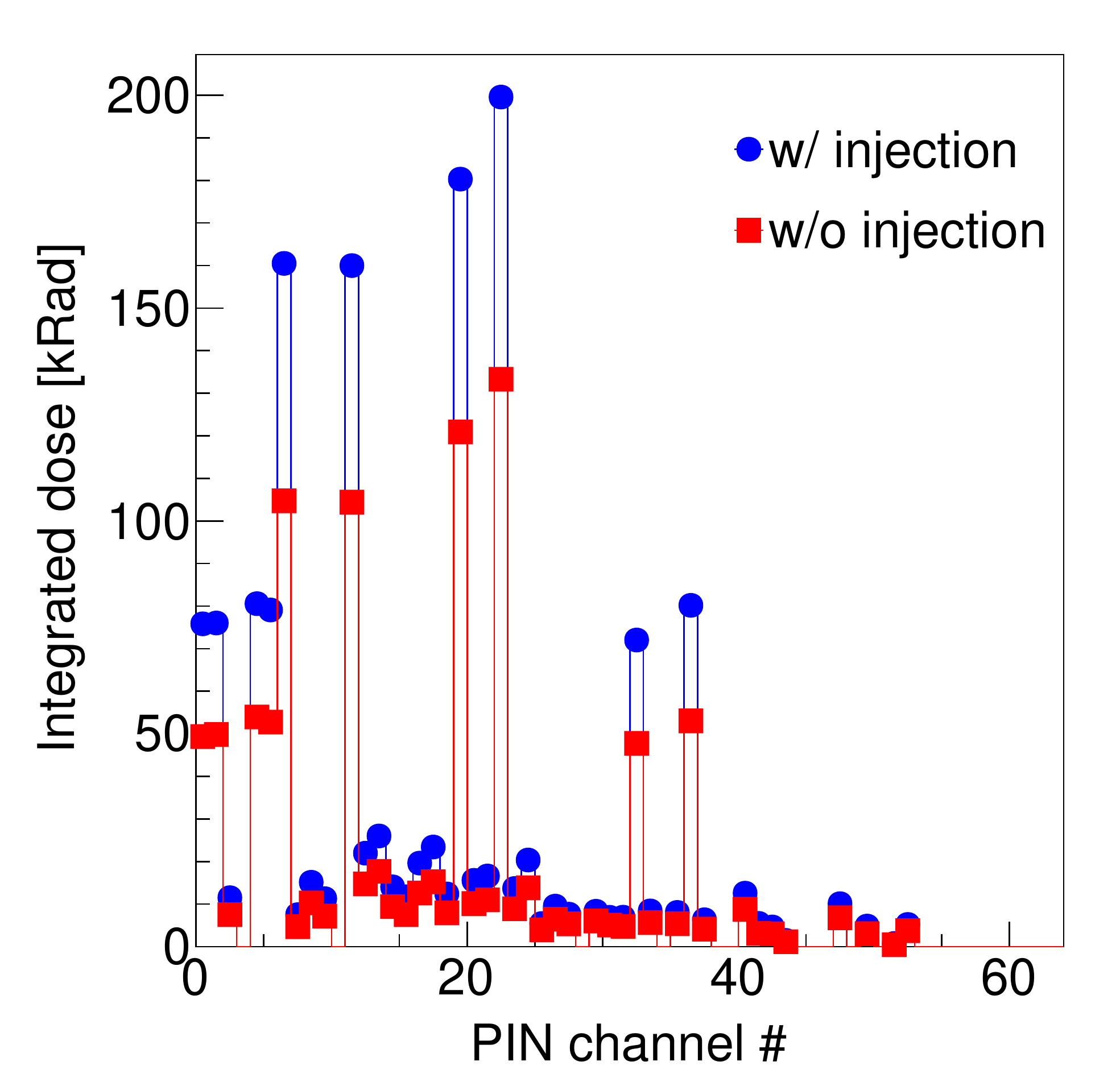}
\caption{Integrated experimental dose per channel for the PIN system, with periods of injection included (with injection) and excluded (without injection).}
\label{fig:PINDosimetryInjection}
\end{figure}

\begin{figure}[ht!]
\centering
\includegraphics[width=1.0\columnwidth]{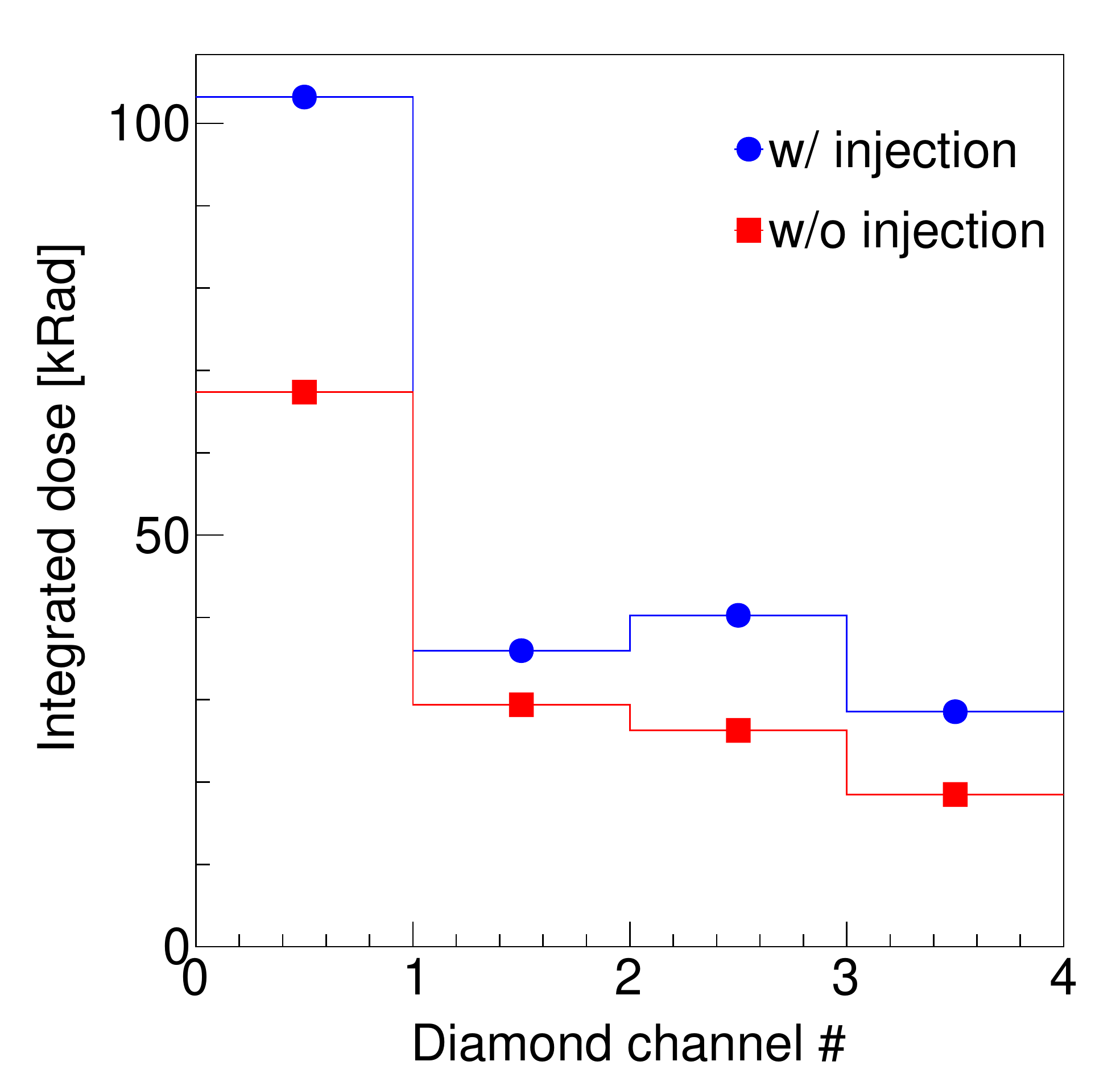}
\caption{Integrated experimental dose per channel for the Diamond system, with periods of injection included (with injection) and excluded (without injection).}
\label{fig:DIADosimetryInjection}
\end{figure}

\begin{figure}[ht!]
\centering
\includegraphics[width=1.0\columnwidth]{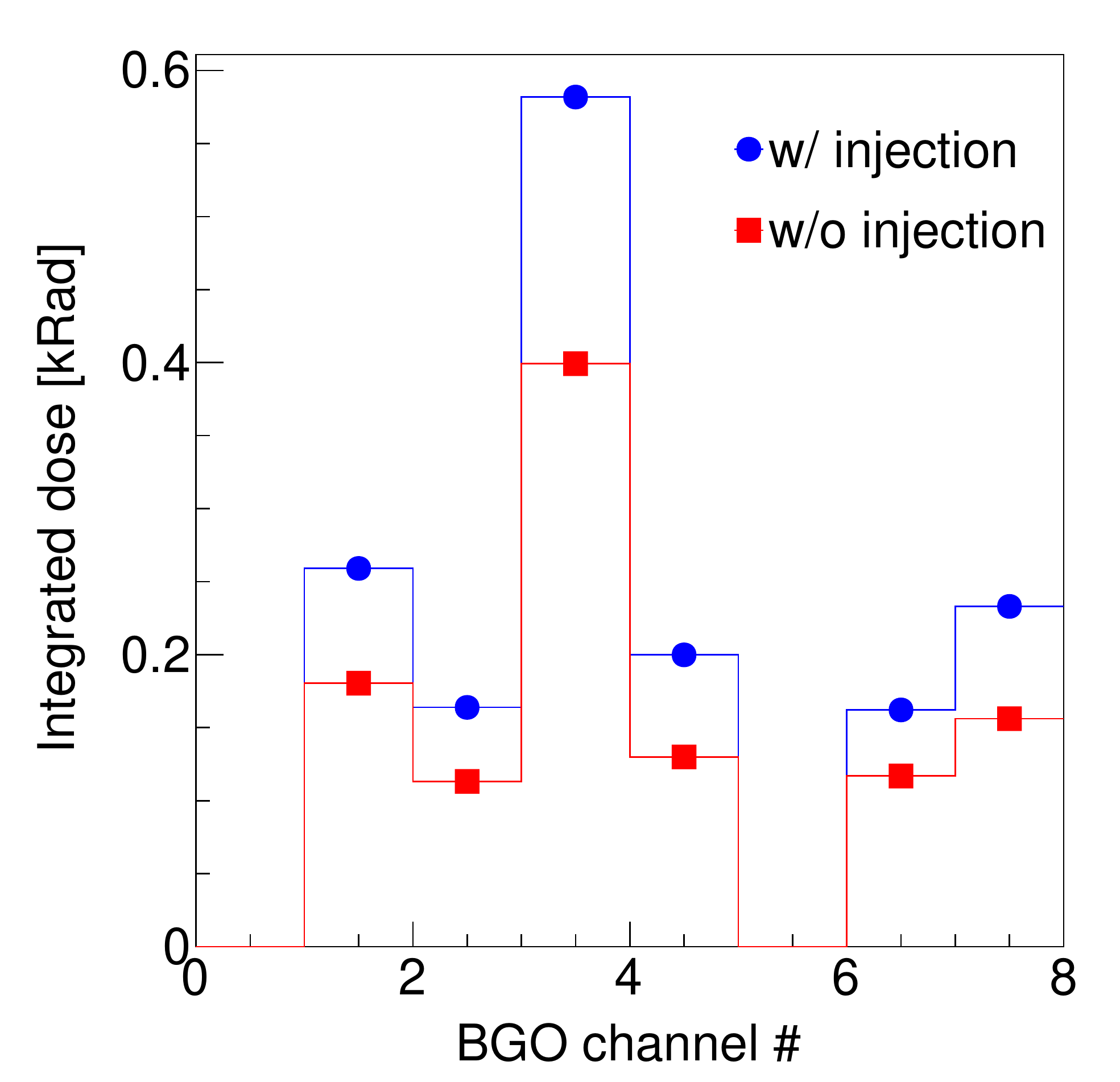}
\caption{Integrated experimental dose per channel for the BGO system, with periods of injection included (with injection) and excluded (without injection).}
\label{fig:BGODosimetryInjection}
\end{figure}

\begin{figure}[ht!]
\centering
\includegraphics[width=1.0\columnwidth]{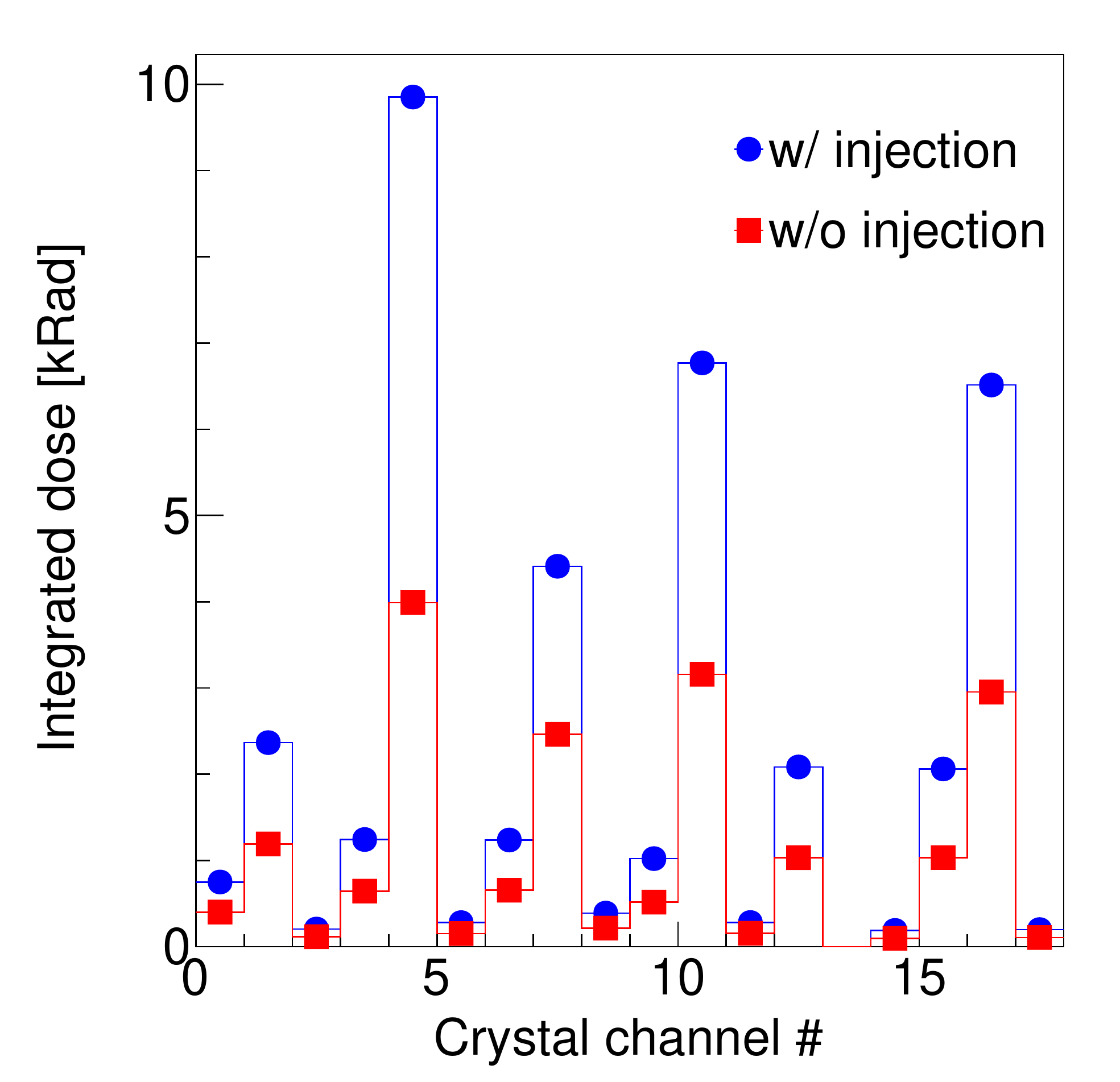}
\caption{Integrated experimental dose per channel for the Crystals system, with periods of injection included (with injection) and excluded (without injection).}
\label{fig:CSIDosimetryInjection}
\end{figure}

Figures~\ref{fig:PINDosimetryInjection}, ~\ref{fig:DIADosimetryInjection}, ~\ref{fig:BGODosimetryInjection}, and ~\ref{fig:CSIDosimetryInjection} show how the integrated dose increases when periods of injection are included in the integral. For most channels and detectors 40\% or less of the dose is accumulated during times of injection. Since the dose during times of injection is a combination of non-injection and injection backgrounds, 40\% is an upper limit, but not yet an estimate of the total dose fraction due to injection background. We can learn more, and distinguish injection background from other backgrounds occurring during injection by using a time-dependent analysis, which is presented in section \ref{sec:instant_injection_dose}.

\subsection{Integrated dose with commercial dosimeters}

To validate the experimental dose measurements in BEAST II, we performed a cross-check using commercial dosimeters. Five Optically Stimulated Luminescence (OSL) dosimeters were placed on the beam pipe, near five of the PIN diode modules. These dosimeters are sensitive to X-rays and $\gamma$-rays 
between 5\,keV and 10\,MeV and to $\beta$-rays between 100\,keV and 10\,MeV; the dosimeters become saturated at 10\,Sv. The conversion from Sv to Rad used is 1 Sv = 100 Rad; this conversion factor carries a $50\%$ uncertainty. Two of the dosimeters, located at ($z = -126.5$~cm, $\phi = 0^{\rm o}$) and 
($z = -71$~cm, $\phi = 0^{\rm o}$) saturated. The other three dosimeters recorded an integrated dose within a factor of two of the PIN diode
measurements and simulation. This seems reasonable, as the discrepancy is of a magnitude that could been explained by the difference in the PIN diode and dosimeter positions, and/or the large uncertainty on the conversion factor.     

\begin{figure}[ht!]
\centering
\includegraphics[width=\columnwidth]{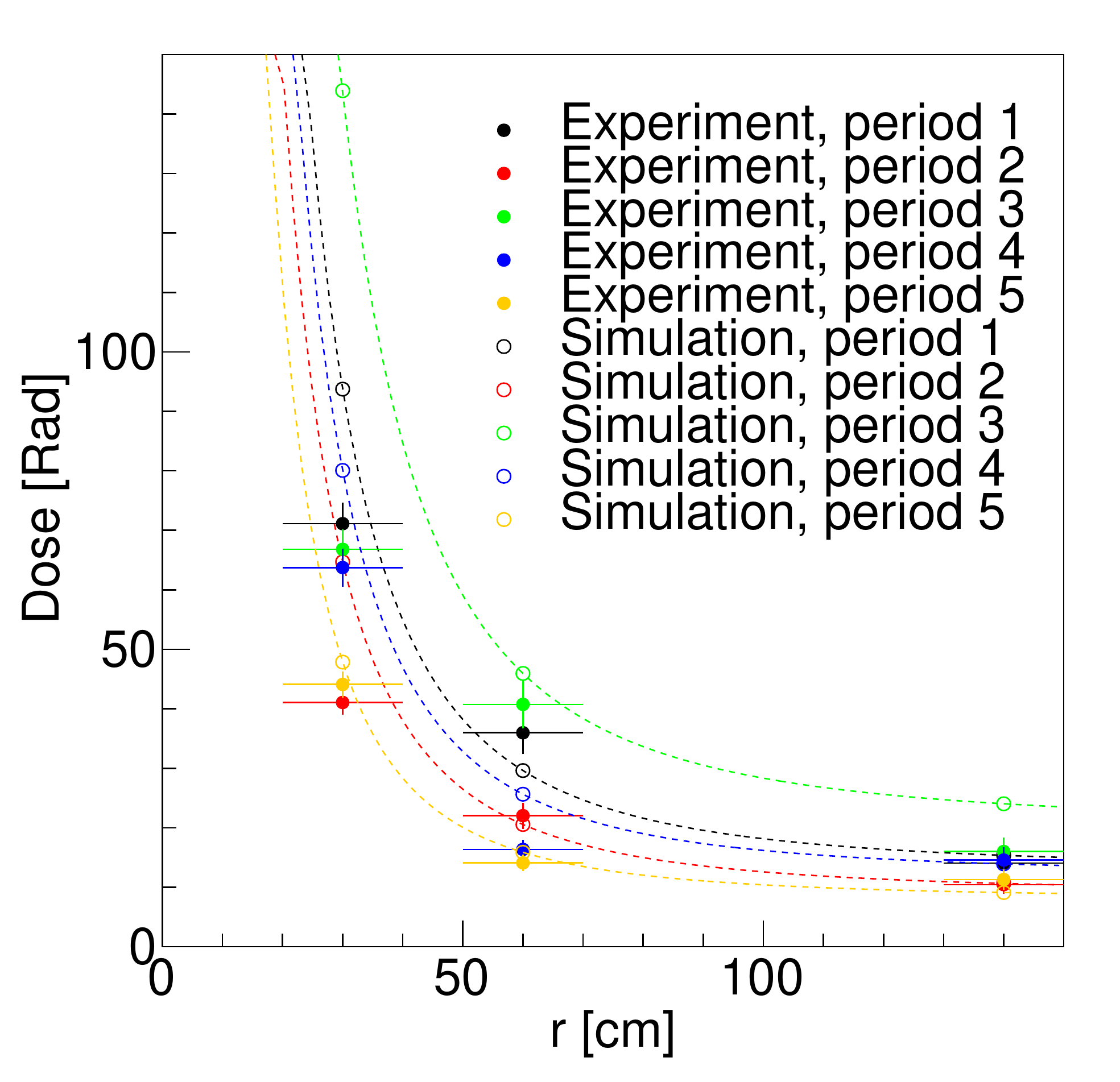}
\caption{Integrated dose versus radius, for dosimeters located at $z = 0$~cm, $\phi = 45^o$ and for different time period. Each time period corresponds approximately to two weeks. Filled data points: experimental data. Open data points: simulation.}
\label{fig:DosimetrySummarUFy}
\end{figure}

In addition, three dosimeters were placed at ($z = 0$~cm, $\phi = 45^o$), and at three different radii ($r = 40, 60$ and 140~cm), by attaching them to a rope anchored to the concrete wall and the central beam pipe. These three positions correspond to the locations of the Belle II Central Drift Chamber, Electromagnetic Calorimeter, and TOP. Figure~\ref{fig:DosimetrySummarUFy} shows the measured and simulated integrated dose versus radial position. The simulated dose falls off as $1/r^2$, which indicates a source-like origin of the radiation. Given the large uncertainty on the dosimeter radial position, of the order of 10~cm, the measured and predicted radial dependence seems to agree reasonably well.

\subsection{Integrated dose on the beampipe}
 
Most of the BEAST II PIN diodes and all of the diamond sensors are located on the beam pipe. These two types of sensors represent very different compromises of price and performance: the PIN diodes are much less expensive, but also less sensitive and less stable than the diamond sensors.  Together, the two systems allowed us to instrument a large number of locations, and provide a comprehensive picture of dose levels near the beam pipe.
 
 % Igal Jaegle
 \subsubsection{PIN diode dose}
 
\begin{figure}[hbt]
 \centering
  \includegraphics[width=\columnwidth]{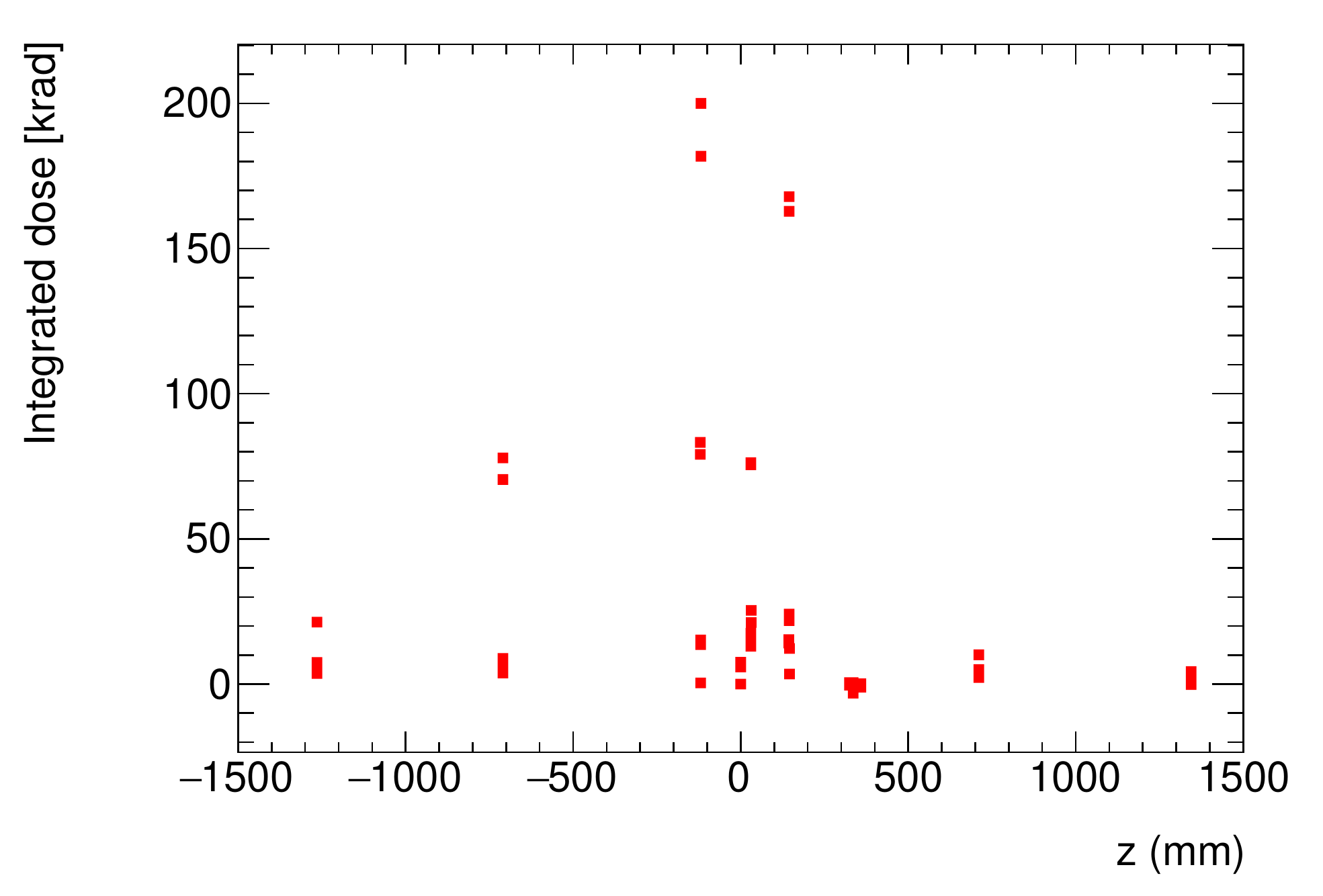}
    \caption{PIN system Phase 1 total integrated dose vs $z$-distance from IP for all 64 PIN diodes.}
      \label{fig:PIN_integrateddose}
\end{figure}

Figure~\ref{fig:PIN_integrateddose} shows the total Phase 1 integrated dose on all 64 PIN diodes versus the $z$-distance from the interaction point (IP). The full coordinates of each diode are provided in Table~\ref{table:pinlocation_beampipe} in Section \ref{sec:pins}. The maximum measured dose is of order 200~krad, near the IP, and falls off quickly with $z$. Eight of the 64 diodes are not located on the beam pipe, but instead mounted near the TPCs. These diodes do not measure any significant dose above noise, and we conclude that there is less than 0.1~krad of total ionizing radiation in the region of the TPCs. This also provides a good estimate of the minimum dose rate visible in the PIN system; approximately $10^{-4}$~rad/sec.  This estimate agrees well with another estimate based on the noise level of the diodes observed in times of no beam.

\subsubsection{Synchrotron radiation dose}
\begin{figure}[hbt]
 \centering
  \includegraphics[width=\columnwidth]{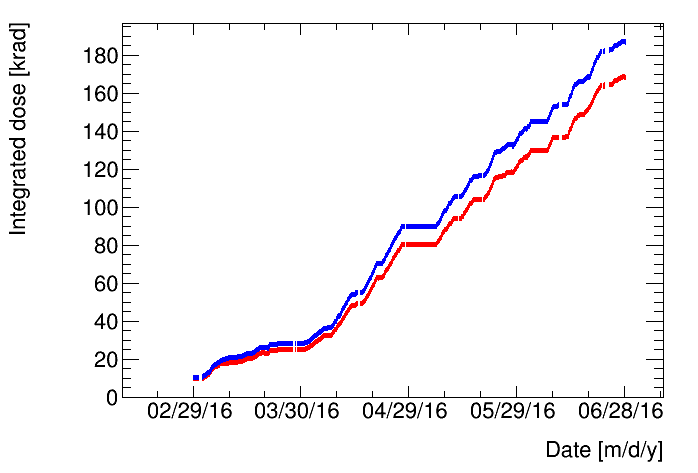}
    \caption{(color online) BEAST II Phase~1 integrated dose versus time for the two PIN diodes in a single
module in the horizontal plane of the ring that had the highest integrated dose. 
The gold foil shielded diode is shown in red, and the aluminum shielded is shown in blue.}
      \label{fig:integrateddosevstime}
\end{figure}

 A unique capability of the PIN system is the ability to measure synchrotron
radiation background.  Our expectation in Phase~1 running is that
no synchrotron radiation signal would be visible in the BEAST II Phase~1 interaction region.
Figure~\ref{fig:integrateddosevstime} shows the Phase~1 integrated dose versus
time for the diode pair in the horizontal plane of the ring 
with the largest total dose.
The difference between the two is smaller than the 20\% uncertainty
we estimate on the integrated dose, which is based on the scatter of 
individual channel response to the same source. The aluminum shielded diode has a higher total dose, as expected due to the small shielding effect
of the high~$Z$ gold layer if X-rays are a negligible component of the radiation. 

\begin{figure}[hbt]
 \centering
  \includegraphics[width=\columnwidth]{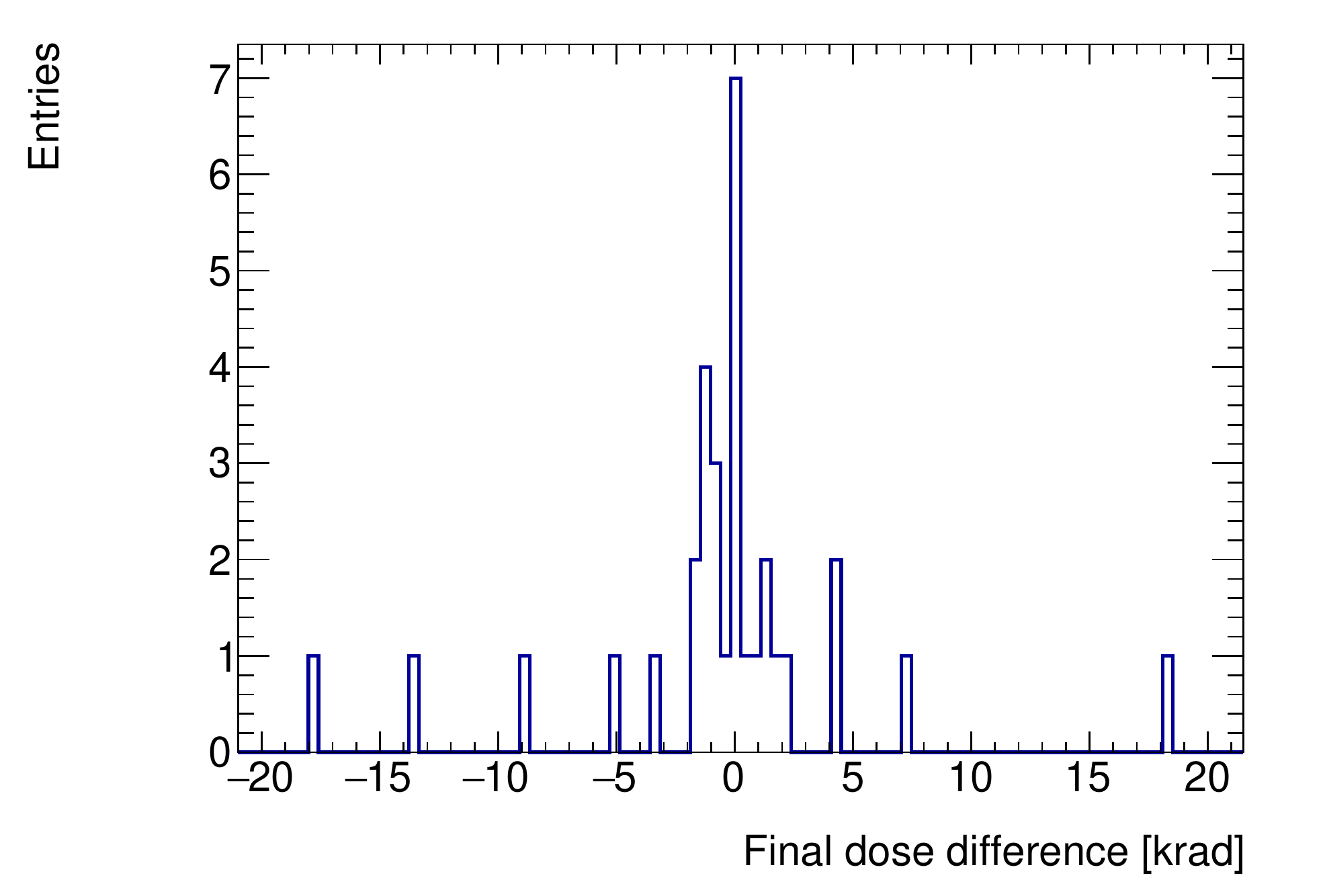}
    \caption{Phase 1 integrated dose difference between aluminum-shielded and gold-shielded diode pairs located in the same module.} 
      \label{fig:PIN_integrateddoseratio}
\end{figure}

The Phase 1 integrated dose difference between all unshielded and shielded diode pairs is shown in Figure~\ref{fig:PIN_integrateddoseratio}.  The mean of the distribution is consistent with zero, and there are no more positive than negative outliers in the tails of the distribution. This means we observe no significant difference between the dose in adjacent aluminum and gold covered diodes. 

We therefore have no clear evidence for synchrotron radiation X-ray beam background
during the BEAST II Phase~1 running.  
We also see no clear synchrotron radiation
signal during any daily dose on any of the diode pairs.  

The sensitivity to X-rays in the Phase 1 integrated dose via the method described here is clearly limited by systematic effects that broaden the dose difference distribution. These effects include cumulative errors in the daily pedestal subtraction, and the fact the two diodes in one PIN diode module are not in exactly the same location. 

Given these systematic uncertainties, and the observed daily dose differences for gold shielded and aluminum
diode pairs, we estimate that the integrated synchrotron radiation dose is less than 17~krad out of the 180~krad that
we observed on the diode pair with the largest dose. This upper limit is three times the standard deviation of the difference between the total dose on shielded and unshielded pairs over all diode pairs.

\subsubsection{Diamond sensor dose}
\begin{figure}[ht!]
	\centering
		\includegraphics[width=\columnwidth]{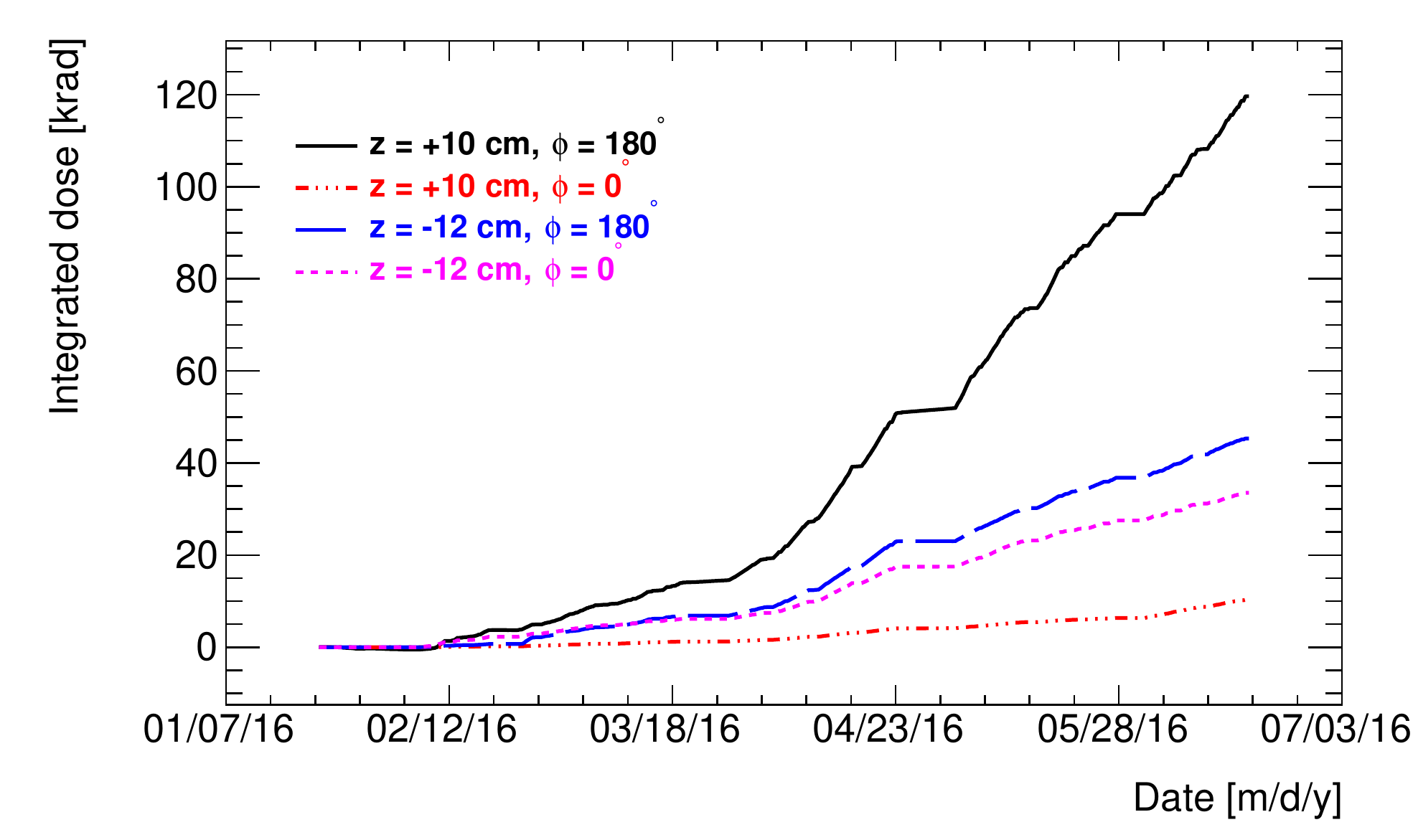}
	\caption{(color online) The integrated dose measured in this first commissioning phase. The black, red, blue and magenta lines show the results obtained in the four different diamond positions on the beampipe. The relative uncertainty of about 17\% is dominated by systematic errors in the calibration procedure, described in Section~\ref{sec:Diamond:Current-Dose-Calibration}.}
	\label{fig:integratedDose}
\end{figure}

 The main goal of the radiation monitoring system in the next phases will be the measurements of both instantaneous and integrated doses in the inner part of the Vertex Detector of Belle II in order to protect it from large beam losses, with the capability to abort the electron and positron beams.
In this first commissioning phase the total integrated dose has been estimated (fig.~\ref{fig:integratedDose}).
The four lines refer to the results obtained in the four different diamond positions on the beampipe, the FW-180 ($z$=+9.5, $\phi$ = 180), FW-0 ($z$=+9.5, $\phi$ = 0), BW-180 ($z$=-13.2, $\phi$ = 180) and BW-0 ($z$=-13.2, $\phi$ = 0) positions.
These results have been obtained relating the current measurements to the deposited radiation doses using the conversion factors estimated in the calibration procedure described in Section~\ref{sec:Diamond:Current-Dose-Calibration}. The corresponding systematic error, estimated to be about 17\%, dominates the uncertainty in integrated dose measurements. 

  % Chiara, Livio 
\subsection{Dose rate sensitivity of the beam abort system}
 %     file:		beam_abort.tex
%     author:  	Livio Lanceri
%
%     contents:  Description of the averaging and buffering of the measurements of the diamond-sensor currents
%                      in their read-out system, and description of preliminary results on their fluctuations
%                      during stable beam conditions, that are compatible with their use to generate a beam abort signal.

The Belle II radiation monitoring system, based on diamond sensors, has the primary goal of detecting beam conditions that might be damaging for the PXD+SVD sensors and front-end electronics. These could either correspond to sudden large increase in backgrounds and the corresponding received instantaneous radiation dose, or a lesser increase, that however brings to an unacceptable integrated dose over some longer time period. The corresponding actions should be an immediate trigger signal to the SuperKEKB beam-abort system in the first case (\emph{``fast" abort trigger}), and a warning signal followed after some time by a beam-abort trigger signal in the second case (\emph{``slow" abort trigger}).

Appropriate radiation thresholds for these actions will be set based on future operational experience, however minimum requirements can be specified based on previous experience from Belle and BaBar~\cite{ref:babar_radmon}, on the present status of the SuperKEKB project, and on the available simulations.

The \emph{``fast" abort trigger} system, protecting against radiation bursts, should be able to measure instantaneous dose-rates up to about $50$~krad/s with a precision of $50$~mrad/s, on the time scale set by the beam revolution period of about $10$~\si{\micro}s. Typically, the trigger signal should be generated whenever a total dose of about $2 \div 3$~rad is integrated above a dose rate threshold of about $1$~rad/s. The above precision requirement ensures that the uncertainty on the measurement of the radiation levels is small with respect to the radiation levels at which beams should be aborted. Two separate trigger signals should be provided to allow separate aborts for the two circulating beams (Low Energy and High Energy), in cases where the beam losses are clearly correlated with only one beam.

The \emph{``slow" abort trigger}, protecting against long-term radiation damage, should be able to measure instantaneous rates with an accuracy of about $5$~mrad/s, allowing a 10\% accuracy at a dose rate threshold of $50$~mrad/s.

During Phase 1 commissioning of SuperKEKB, noise levels in the prototype readout system of the diamond sensors were measured as a preliminary test of the beam abort function, that will be activated in Phase 2.

\begin{figure}
        \centering
                \includegraphics[width=\columnwidth]{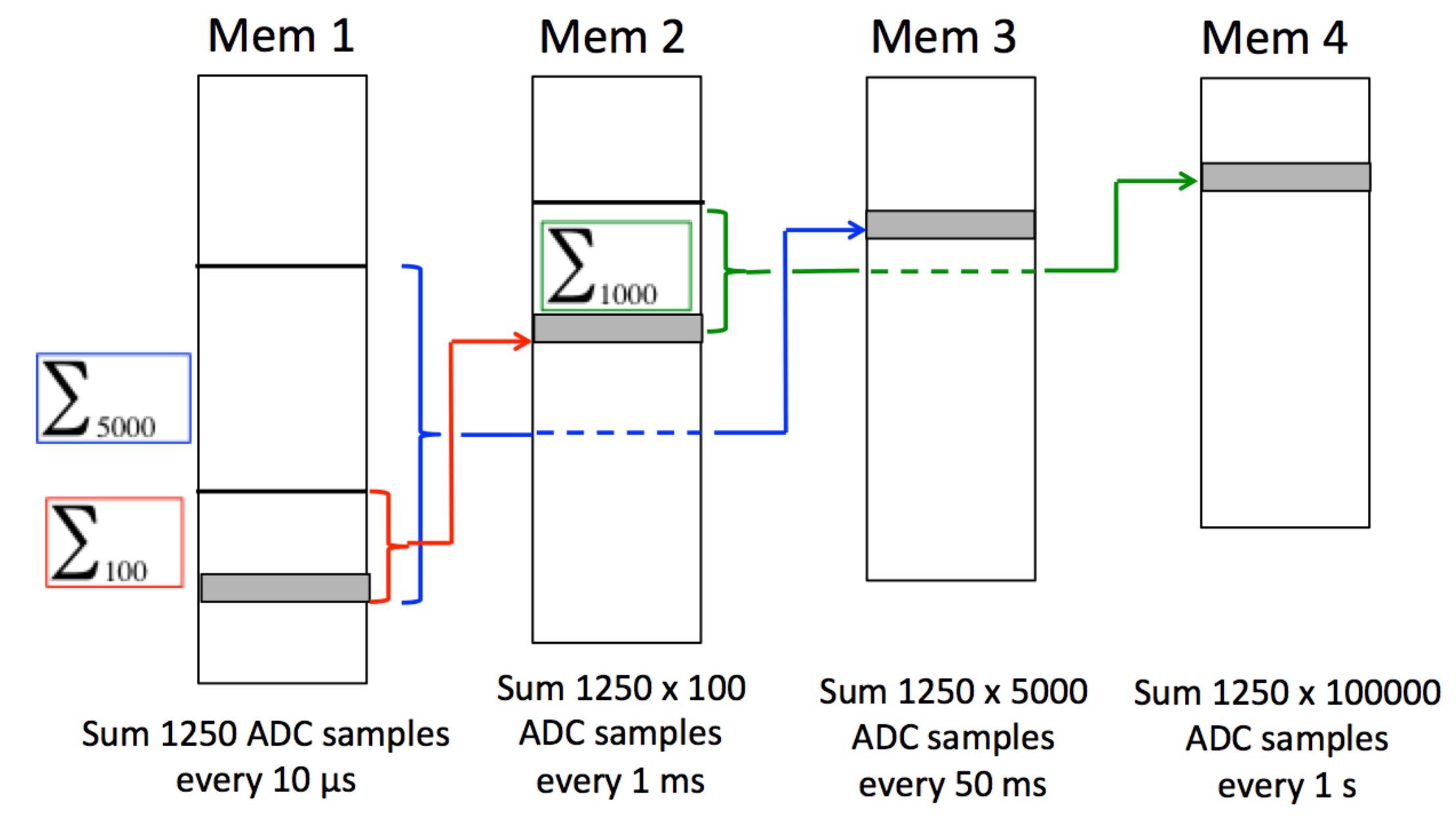}
        \caption{Block diagram of the averaging and buffering process after the digitisation and readout of the diamond-sensor currents. The leftmost block corresponds to the buffer storing data obtained from the sum of $1250$ ADC samples every $10$~\si{\micro}s; the rightmost block represents the memory containing data summed over $1250 \times 10^{5}$ ADC samples.}
        \label{fig:diamonds_memories}
\end{figure}

\begin{figure}[h]
        \centering
                \includegraphics[width=\columnwidth]{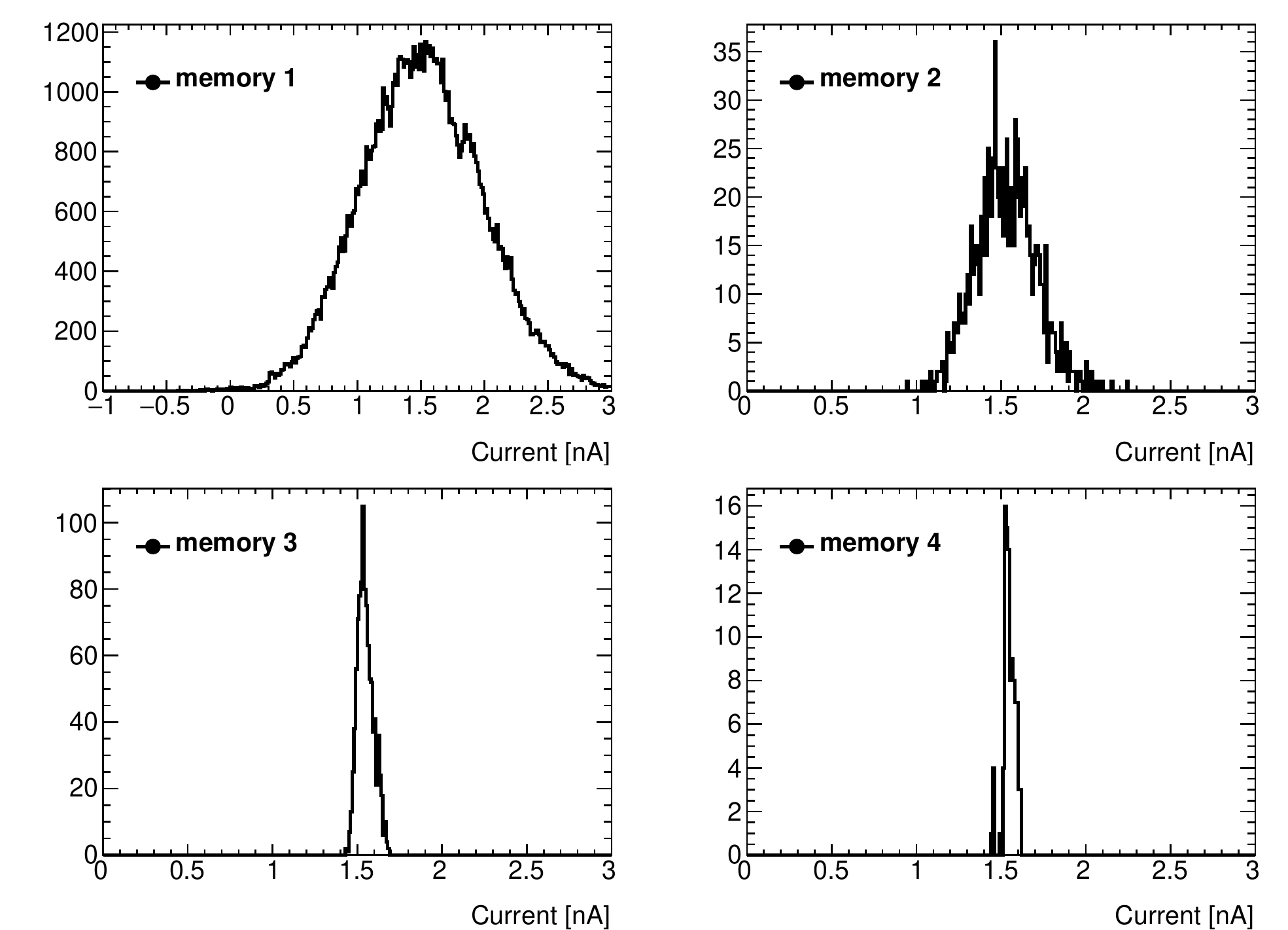}
        \caption{Histograms of data extracted from snapshots of the contents of the four buffer memories: (top left) first memory, $100000$ entries, corresponding to a time interval of $1$~s; (top right) second memory, $1000$ entries, corresponding to the same $1$~s ; (bottom left) third memory, $1000$ entries, corresponding to $50$~s; (bottom right) fourth memory, $100$ entries, corresponding to a time interval of $100$~s.}
        \label{fig:diamonds_memories_histograms}
\end{figure}

In Figure~ \ref{fig:diamonds_memories_histograms} the progressively reduced fluctuations in the averages are demonstrated by histograms of data from the BW-0 sensor, taken in short time intervals, with stable beam conditions during vacuum scrubbing, that induced a current of about $1.5$~nA in this sensor. RMS fluctuations ranged from about $0.47$~nA ($10$~\si{\micro}s, first buffer) to $0.04$~nA (fourth buffer). Taking into account the calibration, the RMS fluctuation in the fastest range corresponds to a dose rate of about $5$~mrad/s, an order of magnitude better than the required $50$~mrad/s. This result gives confidence in the safe operation of the beam abort feature in Phase 2, with negligible probability of generating fake aborts.

As described in Section~\ref{sec:diamond:signal_processing}, four successive levels of averages of the digitized currents, computed in moving time windows of $10$~\si{\micro}s, $1$~ms, $50$~ms and $1$~s, are stored in four revolving buffers for each diamond sensor; Figure~\ref{fig:diamonds_memories} shows a simplified block diagram of the averaging and buffering process. Beam abort signals are generated comparing the digitized and averaged data with predefined thresholds. The fastest abort, on the $10$~\si{\micro}s time scale, uses the data in the first buffer memory, while slower aborts may be generated using data averaged on longer time scales.
 % Chiara, Livio 
 \subsection{Injection dose\label{sec:instant_injection_dose}}
The Touschek plus beam-gas heuristic of Eq.~\ref{eqn:combined_heuristic} accounts for effectively all of the beam background during coasting of the beam --- when no current is being injected. However, a substantial fraction of the integrated dose may come from injection background, which is not simulated and therefore must be measured. Here we extend the heuristic to include injection backgrounds and measure injection doses during Phase 1.

\subsubsection{Identifying injection background}
To isolate HER and LER injection in a stable configuration, we use two test periods in Phase 1 during which one beam completed ten consecutive cycles of fill-coast while the other beam was off. The first period consists of 3700 seconds of continuous LER scrubbing at 723~mA on May 13th, and the second of 18000 seconds of continuous HER scrubbing at 828~mA on June 12th. 

\begin{figure*}[p]
	\centering
        \subfigure{
          \label{fig:instant_inj_dose_LER}
	  \includegraphics[width=\columnwidth]{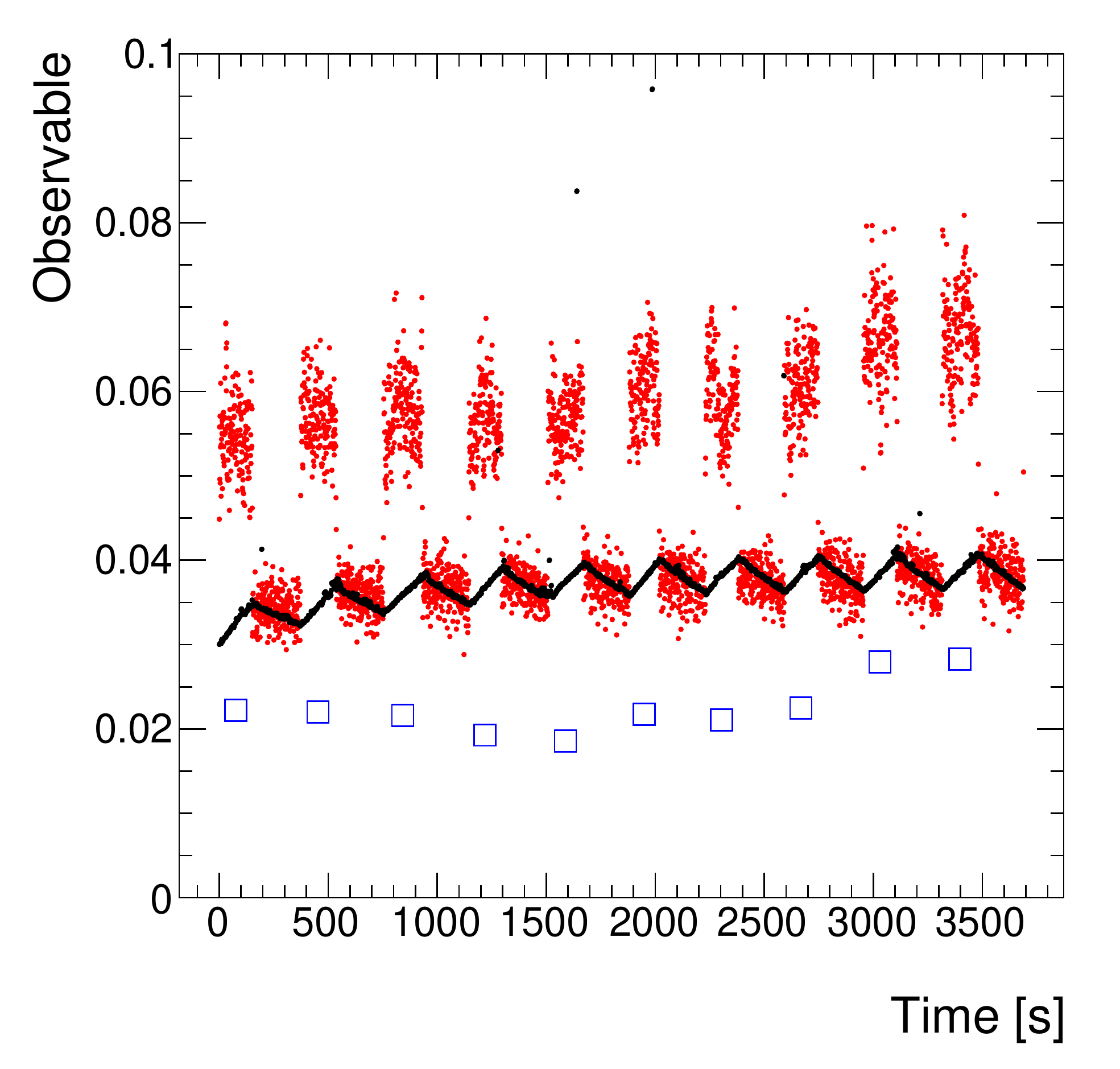}
        }
        \subfigure{
          \label{fig:instant_inj_dose_HER}
	  \includegraphics[width=\columnwidth]{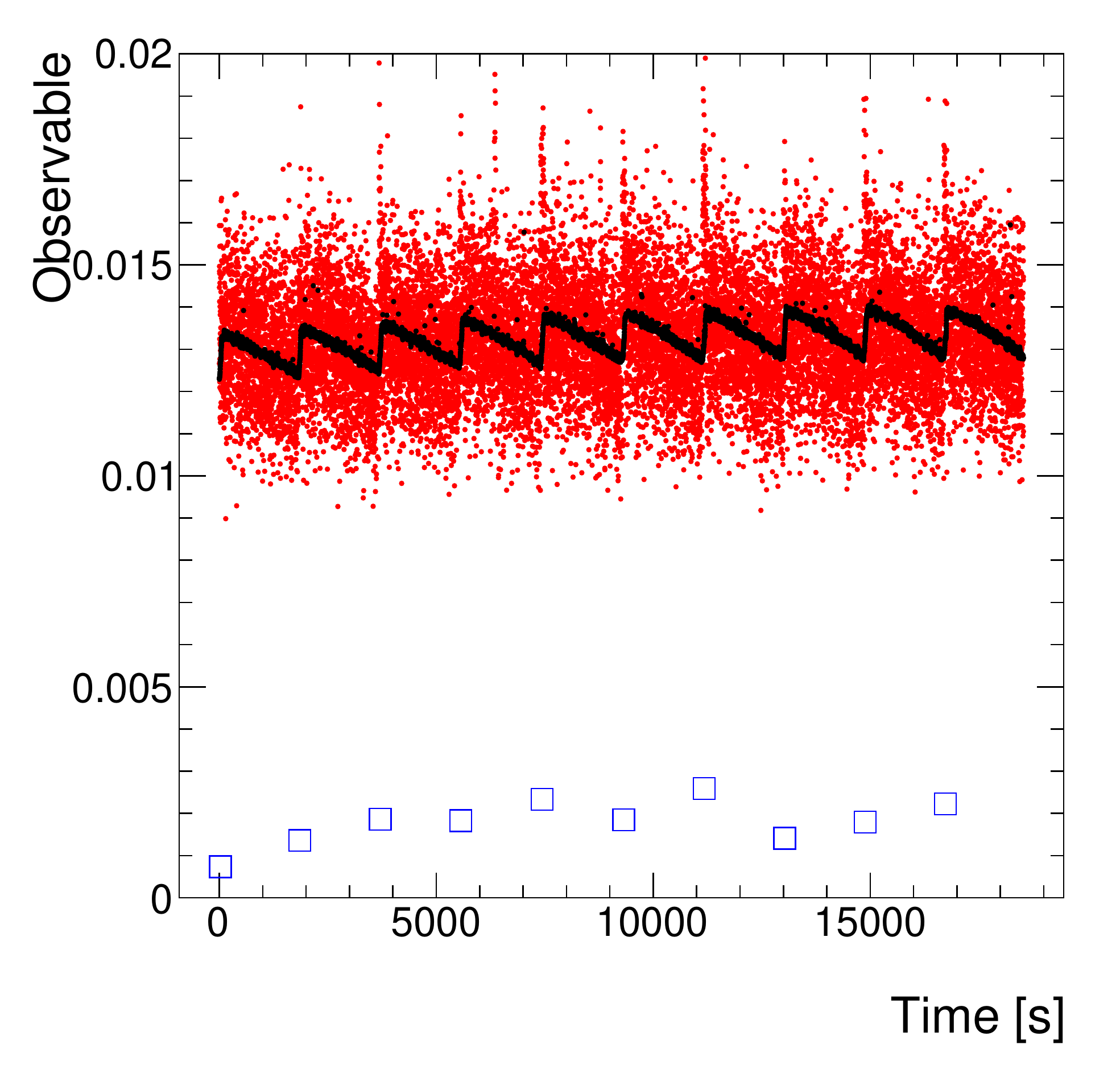}
        }
	\caption{(color online) A demonstration of the procedure for measuring the instantaneous injection dose using a single BGO channel. The time periods correspond to ten consecutive fill-coast cycles in the LER, left (injecting to 723~mA) with the HER off, and similarly for the HER, right (injecting to 828~mA) with the LER off. Black points are the observable expectation based on the heuristic equation Eq.~\ref{eqn:combined_heuristic}, red points are measurements and blue squares are the difference between the two averaged over each injection period. During injection, the difference between the observable and the predicted observable based on the Touschek and beam-gas heuristic is large. This is a measure of the instantaneous dose rate from injection background only.}	
\end{figure*}

We determine Touschek and beam-gas sensitivities using a fit similar to Fig.~\ref{fig:touschek_demo_fit} on all non-injection data in each period. We then predict the Touschek and beam-gas contributions during injection based on the heuristic. The difference between the observable and this predicted value is what we call injection background. Figs.~\ref{fig:instant_inj_dose_LER} and ~\ref{fig:instant_inj_dose_HER} show the contribution of injection background during the two test periods.

\subsubsection{Injection dose parameterization}
We expect the instantaneous injection dose to be proportional to the rate $R$, charge $Q$ and efficiency $\epsilon$ of the injection. This suggests a third term in the combined heuristic:
\begin{equation}\label{eq:heuristic_inj}
  \mathcal{O}_{inj} = S_{inj} \cdot R Q (1-\epsilon).
\end{equation}
By measuring the injection sensitivity $S_{inj}$ in units $[\mathcal{O}]\text{Hz}^{-1}\text{nC}^{-1}$ instead of the raw observable, our results can easily be reinterpreted for different injection conditions.

\subsubsection{Injection dose results}
Table~\ref{tab:inj_params} summarizes injection sensitivities for BEAST II channels extracted from the two test periods. We found that the injection efficiency measurements are unreliable and therefore we assume a completely inefficient injection, with $\epsilon=0$. Quoted uncertainties come from variation of the extracted sensitivity only and do not reflect uncertainties of the injection parameters, which are unknown. For completeness, we also show the average injection observable in these test periods, $\mathcal{O}_{inj}$ .

These sensitivities can be used to generate estimates of the absolute injection dose during Phase 1. For example, the BGO system recorded a sensitivity of roughly $2$~\si{\micro}rad~Hz$^{-1}$~nC$^{-1}$ to LER injection, which corresponds to a dose rate of 7~rad per month given a time-averaged injection rate of 5~Hz, charge of 0.5~nC and efficiency of 50\%, all reasonable estimates for continuous top-off running. The expected dose from HER injection backgrounds is an order of magnitude lower. 

\begin{table*}[p]
	\caption{Results from the instantaneous injection dose measurements of Sec.~\ref{sec:instant_injection_dose}. We show the injection sensitivities $S_{inj}$ in units $[\mathcal{O}]\text{Hz}^{-1}\text{nC}^{-1}$ in order to provide general results that can be easily extrapolated to different injection parameters using Eq.~\ref{eq:heuristic_inj}. Additionally, we show the average injection observable during injection, $\mathcal{O}_{inj}$, as a measurement of the typical instantaneous dose due to injection seen during Phase 1. Blank entries are due to missing data for the analyzed period. The PIN diodes, He$^3$ tubes, CsI(Tl) crystals and TPCs do not see any injection backgrounds in the test periods and are excluded from the table. An entry reading ``Not seen'' means that the detector channel was operational during the test period but we find no significant injection dose.}
	\centering	
	\begin{tabular}{ llllll }
        \toprule
	Observable                 & Ch	 & $S_{inj}^{LER}$     & $S_{inj}^{HER}$     & $\mathcal{O}_{inj}^{LER}$  & $\mathcal{O}_{inj}^{HER}$ \\\relax
        [units]                    &     &                     &                     &                            & \\
        \midrule
	BGO dose rate              & 1	 & $2.35 \pm 0.22$     & $0.103 \pm 0.017$   & $14.06 \pm 0.63$           & $0.826 \pm 0.069$ 	   \\\relax
        [$\si{\micro}$rad/s]       & 2   & $0.725 \pm 0.068$   & $0.126 \pm 0.020$   & $4.34 \pm 0.18$            & $1.009 \pm 0.065$      \\
                                   & 3   & $2.07 \pm 0.21$     & $0.234 \pm 0.038$   & $12.38 \pm 0.74$           & $1.87 \pm 0.14$        \\
                                   & 4   & $3.79 \pm 0.36$     & $0.226 \pm 0.039$   & $22.7 \pm 1.0$             & $1.81 \pm 0.17$         \\
                                   & 5   & $2.06 \pm 0.19$     & $0.0289 \pm 0.0056$ & $12.35 \pm 0.41$           & $0.231 \pm 0.030$       \\
                                   & 6   & $2.24 \pm 0.21$     & $0.0275 \pm 0.0064$ & $13.43 \pm 0.53$           & $0.220 \pm 0.041$       \\
                                   & 7   & $3.02 \pm 0.28$     & $0.132 \pm 0.023$   & $18.05 \pm 0.68$           & $1.05 \pm 0.10$         \\
        Diamond dose rate          & 1   & $14.5 \pm 1.4$      & Not seen            & $86.5 \pm 3.9$             & Not seen                \\\relax
        [$\si{\micro}$rad/s]       & 3   & $131 \pm 16$        & Not seen            & $784 \pm 68$               & Not seen                \\
        Pure CsI hit rate          & 1   & $14.7 \pm 1.4$      &                     & $88.3 \pm 4.4$             & \\\relax
        [kHz]                      & 4   & $104 \pm 10$        &                     & $621 \pm 28$               & \\
                                   & 7   & $79.7 \pm 7.1$      &                     & $477 \pm 13$               & \\
                                   & 10  & $54.2 \pm 6.2$      &                     & $324 \pm 25$               & \\
                                   & 13  & $74.8 \pm 6.6$      &                     & $448 \pm 11$               & \\
                                   & 16  & $42.9 \pm 4.0$      &                     & $257 \pm 11$               & \\
        LYSO hit rate              & 2   & $0.180 \pm 0.019$   &                     & $1.08 \pm 0.07$            & \\\relax
        [kHz]                      & 5   & Not seen            &                     & Not seen                   & \\
                                   & 8   & $0.175 \pm 0.056$   &                     & $1.05 \pm 0.32$            & \\
                                   & 11  & Not seen            &                     & Not seen                   & \\
                                   & 14  & Not seen            &                     & Not seen                   & \\
                                   & 17  & $0.086 \pm 0.067$   &                     & $0.52 \pm 0.40$            & \\
        \bottomrule
	\end{tabular}
	\label{tab:inj_params}
\end{table*}

\subsubsection{Injection dose discussion}
To understand the relative importance of injection backgrounds, we evaluate the ratio of injection to non-injection backgrounds during injection, $\mathcal{O}_{inj}/(\mathcal{O}_{bg} + \mathcal{O}_{T})$. In the LER, we find that this ratio is roughly $0.3$, $0.1$, $6$, and $0.02$ for BGO, Diamond, pure CsI, and LYSO, respectively. The pure CsI crystal hit rate has by far the largest increase during injection, but this may be exaggerated compared to the increase in dose rate if the injection background is dominated by low-energy photons. The LYSO crystals, despite having a faster response, may saturate during injection due to a lower energy threshold, suggesting a high rate of very low-energy photons. This may explain the very small increases in the LYSO hit rates during injection. Consequently, we consider the BGOs and Diamonds to be the most reliable sources of injection dose rate measurements.

We make three broad conclusions about injection backgrounds in Phase 1: (1) injection doses in the HER are very small and may be negligible, (2) injection background may be dominated by low-energy photons, and (3) the dose rate due to injection in the LER is roughly 20\% of the total dose rate during injection and 10\% overall.

 % Peter
 \clearpage

 %lead author: Sven Vahsen
 \section{Summary and discussion of findings}\label{sec:findings}
 In this section we summarize and discuss the main results and lessons learned from the work described above. The resulting implications for the Belle II experiment will be discussed in section \ref{implications}.

\paragraph{Measurement program} BEAST II recorded data during SuperKEKB commissioning Phase 1 in spring 2016, virtually continuously. The diverse suite of deployed background detectors allowed bread-and-butter measurements of dose rates, as well as novel measurements of beam backgrounds, including bunch-by-bunch measurements of charged particles and direction- and energy-sensitive neutron measurements, near the interaction point. On February 10th and 26th, the CLAWS detector system observed the very first beam bunches that were successfully circulated in the SuperKEKB positron and electron rings, respectively \cite{firstbeams}. Following this, we monitored the integrated dose and time-varying background levels during beam commissioning and subsequent periods of beam background scrubbing. Towards the end of Phase 1, we also carried out dedicated machine study runs, where we varied the beam parameters systematically, to separate different background contributions. In summary, in Phase 1 we observed beam-gas, Touschek, beam-dust, and injection backgrounds, and set limits on the dose rate from synchrotron radiation near the IP. The remaining beam background components, due to colliding beams, will be studied in commissioning Phase 2.

\paragraph{Dosimetry} The integrated radiation dose on the outer surface of the Phase 1 aluminum beampipe was measured with PIN diodes and diamond sensors, and found to be of order 200~krad or less, with a strong dependence on position. BEAST II integrated dose measurements were cross checked in several locations with commercial personnel dosimeters, and are consistent within systematic uncertainties. We measured no significant difference in the integrated dose from adjacent PIN diodes shielded with aluminum and gold, consistent with the expectation from simulation that no synchrotron radiation would be observed outside the beampipe in Phase 1. The observed integrated dose in BEAST II detectors was often one order of magnitude (and sometime more) higher than the dose predicted by simulation. For most channels and detectors 40\% or less of the dose is accumulated during times of injection, the majority of which was due to LER injections, which were found to be less clean. The dose during times of injection is a combination of injection-related and non-injection backgrounds. While the former is not included in the simulation, it must obviously be less than 40\% of the total dose, and hence is insufficient in magnitude to explain the observed discrepancy between the observed and predicted integrated dose. The measured dose falls off quickly with radius. For instance, for $r>50$~cm the total integrated Phase 1 dose is always less than 50~rad. The diamond sensor beam abort system was commissioned, and has a dose-rate resolution of 5 mrad/s per 10~\si{\micro}s of measurement, which is an order of magnitude more sensitive that the design requirement.

\paragraph{Vacuum scrubbing} During vacuum scrubbing, the gas composition and vacuum levels in the beampipe evolved. Several vacuum bump studies were performed, which showed that background levels at the IP were generally sensitive to the beam-gas scattering positions predicted by simulation. Improved agreement between measured and predicted background rates during such studies was obtained by scaling the simulation to account for the time-varying gas composition in the beampipe, which was measured at two LER positions using residual gas analyzers (RGAs).  Past Belle II background predictions have assumed a beampipe pressure of $10^{-9}$~torr, which would require a dynamic pressure, $dP/dI$, below $5.1\times 10^{-8}$~Pa/A and $3.7\times 10^{-8}$~Pa/A at HER and LER design beam currents, respectively. Past background predictions have also assumed a beampipe residual gas mixture with atomic number squared $Z^2$=49. At the end of Phase 1, the observed dynamic pressure in the HER already satisfied the requirement, while that in the LER was of order $2\times 10^{-6}$~Pa/A. An estimated additional 384~ampere hours of LER vacuum scrubbing will be required to satisfy the dynamic pressure requirement. The RGA data showed an LER beampipe residual gas mixture with approximate effective atomic number squared $Z^2_{eff}$=5 at the end of Phase 1, and $Z^2$ was monotonically decreasing with time. (The HER is not equipped with RGAs.) During vacuum scrubbing, the observed backgrounds at the IP initially decreased as expected from the evolution of the dynamic pressure alone. Towards the end of Phase 1, however, the background reduction slowed, relative to the improving dynamic pressure. This has not been investigated further. It could simply be a reflection of the Touschek background relatively larger compared to the reduced beam-gas scattering after vacuum scrubbing. It could also be related to machine work that occurred around that time, which included NEG conditioning and addition of magnets to reduce electron multipacting.

\paragraph{Beam-dust events} During Phase 1, we observed of order ten short-duration increases in BEAST II background levels per 48-hour period. These events are hypothesized to be caused by collisions between dust and beam particles, and hence dubbed beam-dust events. While the contribution of these events to the total dose is negligible, they can abort the beam and are thus an operational nuisance. Some fraction of these events are correlated with sudden, localized pressure bursts in the ring. Our tentative conclusion is that both the BEAST II background spikes and the SuperKEKB pressure burst are due to beam-dust events, but only weakly correlated due to their localized effects on pressure and backgrounds. More data are required to establish whether the rate of beam-dust events is decreasing, or constant, in time.

\paragraph{Integrated beam-gas and Touschek losses} {We carried out a number of background studies, where we scanned the LER and HER beam sizes, beam currents, fill patterns, and other machine parameters, in order to separate contributions from beam-gas and Touschek scattering. The normalization of the beam-gas and Touschek background components was measured both by analyzing the beam lifetime, and by analyzing BEAST II rates, both versus machine parameters. The lifetime analysis shows large variations in the level of agreement between experiment and simulation throughout Phase 1, with a consistent overall excess of beam-gas background in experiment, in agreement with the results of the BEAST II measurements at the IP, discussed below. The total ring loss rates versus beam parameters are not always well described by the heuristic used to fit the data. This suggests that using the average ring pressure and single-RGA $Z_e$ may be insufficient for total ring losses. We expect this to be the case if gas conditions are highly localized so that the ring beam-gas loss rate is dominated by isolated pockets of high-$P$ or high-$Z$ gas.

\paragraph{Beam-gas and Touschek backgrounds near the IP}
The Touschek background observed in BEAST II is on average $1.4^{+1.8}_{-1.1}$ times higher than the prediction for the LER, and $4.8^{+8.2}_{-2.8}$ times higher than predicted for HER. Observed LER beam-gas background is high by a factor of $2.8^{+3.4}_{-2.3}$. For most detectors, count rates or dose rates attributed to these three background components are within one order of magnitude of expectation, for all channels. Larger discrepancies between measurement and simulation are mostly observed for the PIN diodes and diamond sensors, which are closest to the beam pipe. This may be a result of the finite number of scattering positions that are used in the Phase 1 accelerator tracking simulation to keep the computational requirements manageable. The HER beam-gas background observed in BEAST II is on average $108^{+180}_{-64}$ larger than expected. The reported discrepancies with respect to the predictions from simulation are observed after incorporating all presently known relevant corrections, i.e.\ after correcting for the measured pressure distributions in the rings, and after rescaling the gas composition in simulation based on the RGA-measurements of the LER. The BEAST II beam-gas results are rather sensitive to which pressure gauge is used for the analysis of a particular BEAST II detector channel, so it is conceivable that the apparent excesses are systematic effects of the pressure measurement. To accurately simulate beam-gas losses may require more fine grained measurements of the beam pipe gas composition and pressure.

\paragraph{Neutron backgrounds} In both the direct and constrained analyses, the observed rate of thermal neutrons was on average approximately three to four times higher than expected. An excess is seen for both beam-gas and Touschek backgrounds from both the LER and the HER.

One caveat is that the constrained analysis was analyzed by setting the ratio of beam-gas to Touschek losses equal to the expected ratio from Monte Carlo simulation. As a result, any discrepancy in this ratio between data and Monte Carlo was instead absorbed into correction factors for the beampipe gas pressure and composition. If one instead were to use the nominal pressure correction factor of 3.0 for the LER (consistent with other beam-gas and Touschek analyses in this article), where the gas composition was measured, then the observed thermal neutron rate from LER background becomes eight times higher than expected. 

For {\it fast} neutrons we observe a factor of five times higher counting rate than predicted from the HER. For LER backgrounds, the count rate agrees with the prediction in the vertical plane, but is about forty percent lower than expected in the horizontal plane. The discrepancy appear to be primarily in the LER beam gas component. This is the same background component in which the largest discrepancies are seen in the direct and constrained analyses of other detectors, although an excess is observed there, while a deficit is observed here. 

The measured fast neutron recoil energy spectrum appears consistent with simulation. This suggests that the neutron production, material budget, and neutron propagation in Geant4 are sufficiently accurate, compared to the level of precision probed here.

There are multiple possibilities for the observed thermal neutron rate discrepancies, and more work is required to understand them fully. One possibility is a contribution from ``cavern neutrons'', which are produced further away from the IR, but then scattered to our detectors in the IR. We are planning to extend the nuclear recoil analysis from axial tracking to vector tracking, which should improve the separation of neutrons coming directly from the beampipe, and neutrons that have re-scattered. This may allow us to determine if the observed excess is due to cavern backgrounds or not.

\paragraph{Injection background}The time-dependent injection background was measured with the Crystals, CLAWS, and the QCSS detector systems. We clearly observed higher background levels at the IP after injection, coincident with the injected bunch passing the IP. The injection background typically decayed by two to three orders of magnitude within $1 - 1.5$~ms after injection for nominal, optimized injection parameters, and within 3 ms with modified (non-optimal) injection parameters. The injection background level was oscillatory, and the period was measured to be consistent with the predicted period of synchrotron oscillations. We found that LER injection produces significantly higher backgrounds than HER injection, which is consistent with our integrated dose measurements. The HER injection background was also found to ramp up earlier, about 6 turns (60 \si{\micro}s), after injection, versus 14 turns (140 \si{\micro}s) after LER injection.

 % lead author: Sven Vahsen
 \section{Phase 3 predictions and implications for Belle II}
 %     file:		implications.tex
%     authors:  	Sven Vahsen
%
%     contents:  

In this section, we discuss our expectations for beam backgrounds in the Belle II detector during SuperKEKB Phases 2 and 3, as predicted by simulation, and comment on our confidence in these predictions in light of our measurements in Phase~1. We also summarize other lessons learned regarding beam backgrounds that will be important going forward.

\begin{table*}
\caption{Belle II detectors most vulnerable to beam backgrounds in SuperKEKB Phase~3. Upper limits and safety factors assume ten years of SuperKEKB operation at full luminosity. Only detectors with safety factors less than five are included. Although all limits have been converted into rates, in several cases the detector degradation is a cumulative, rather than rate-dependent effect. Neutrons flux numbers are in units of $10^{11}/{\rm cm}^2/{\rm yr}$ and NIEL-damage weighted. See text for further explanation and discussion. }
\label{Table:BelleIILowSF}
\begin{center}
\begin{tabular}{llcccl}
  \toprule
  Belle II detector           &  quantity      			& expected value 		& upper limit value 	& safety factor  & dominant process(es)  \\ 
  \midrule
  PXD			&  occupancy			& 1.1\%					& 3 \%				& 3				& two-photon, synchrotron radiation \\
  CDC			& wire hit rate			& 400~kHz				& 200~Hz			& 0.5				& radiative Bhabha, two-photon\\
 CDC			& electr. neutron flux 	&2.5 					&1 					&	0.3					& radiative Bhabha, Touschek			\\
  CDC			& electr. dose rate			& 250 Gy/yr				& 100				& 0.3				& radiative Bhabha, two-photon \\
  TOP 			& PMT hit rate			& 5-8 MHz 				& 1 MHz			& 0.2			     & radiative Bhabha, two-photon\\
  TOP 			& PCB neutron flux 		& 0.35					& 0.5				& 3		     & radiative Bhabha, Touschek\\
  ARICH 		& HAPD neutron flux		& 0.3					& 1.0					& 3				& radiative Bhabha \\
  ECL			& crystal dose rate		& 6 Gy/yr in BWD		& 10 Gy/yr		& 2				& radiative Bhabha, two-photon\\
  %ECL			& diode neutron flux 		&					&				&					&\\
  %ECL			& pile-up noise			&					&				&					&		\\
  %KLM 			& ?					&					&				&					& \\
  \bottomrule 
\end{tabular}
\end{center}
\end{table*}

At the end of SuperKEKB Commissioning Phases and 2 and 3, the accelerator is anticipated to reach the operational parameters shown in Table \ref{Table:MachineParamAllPhases}. The resulting expected beam background doses and hit rates in Belle II have been estimated using the same software tools discussed in this report, and already incorporate a number of improvements resulting from the Phase~1 work.

Table \ref{Table:BelleIILowSF} lists the Belle II sub-detectors predicted to be vulnerable to beam backgrounds in Phase~3 at full luminosity. Note that all sub-detectors except the SVD and the KLM appear in this list, illustrating the severity of the challenge posed by beam backgrounds. These estimates are being regularly updated by the Belle II collaboration. The values shown are for ``version 15'' of the beam background simulation. The Time Of Propagation detector (TOP) has the lowest safety factor, resulting from a cumulative decrease in the MCP-PMT efficiency as the photocathode detects light and emits photo-electrons. In simulation, the integrated PMT charge in the TOP is dominated by radiative Bhabha and two-photon events. For instance, in the case of radiative Bhabha events, photons generated at the interaction point travel nearly parallel to the beampipe and create showers inside the final focusing quadrupole magnets. Gamma rays from such showers can travel outward radially and penetrate the TOP counter, where they Compton scatter. Cherenkov light from the resulting electrons travels down the quartz bars, where it is detected by the PMTs. Because the PMT degradation is cumulative, it is expected that the TOP PMTs will need to be replaced after a few years of high-luminosity running with atomic layer deposition (ALD) MCP-PMTs, which have much larger background tolerance than the currently used conventional MCP-PMTs. The predicted Central Drift Chamber (CDC) hit rate from beam backgrounds is also high, up to 400~kHz in the 8th layer, corresponding to a 20\% occupancy, dominated by radiative Bhabha and two-photon events. Being a true rate-dependent, rather cumulative, effect this will lead to excess fake hits and consequently reduced tracking performance at higher luminosities. In the worst case, this CDC layer would be turned off, at some cost in tracking efficiency that has not yet been evaluated. 

We note that in all detectors except the pixel detector (PXD), the dominant beam background process is radiative Bhabha scattering, followed by either two-photon or Touschek scattering. The first two of these are well-known QED processes at colliders that presumably should be accurately simulated. Since they are luminosity dependent, they should be manageable in Phase~2, and it will be important to validate their prediction experimentally at the end of Phase~2, before increasing the luminosity in Phase~3. 

The Touschek scattering rate has already been confirmed to be within one order of magnitude of simulation in Phase~1. It is unlikely to become problematic in Phase~3, as long as the beam collimators are as effective in practice as in the Phase~3 simulation. More accurate simulation of collimators, including simulation of collimator-tip-scattering, and validation of this simulation, will be important in Phase~2. 
 
None of the Belle II detectors are predicted to be especially vulnerable to beam-gas events in Phase~3, after collimation. Even though we measure a slightly elevated LER beam-gas rate, and on average two orders of magnitude higher HER beam-gas backgrounds than predicted, our other Phase 1 results somewhat mitigate these factors: the current Phase~3 Belle II backgrounds projections assume a vacuum pressure of 1~nTorr, and an effective nuclear charge squared of the gas of $Z^2=49$. The beam-gas backgrounds are proportional to both these quantities. Our Phase~1 measurements of vacuum scrubbing show that the dynamic pressure in the LER is not yet low enough to reach the 1~nTorr target at full beam current. The required dynamic pressure should be reached at $10^4$~Ah of accumulated beam dose, which may correspond to about a year of SuperKEKB running. However, this temporary increase in the beam-gas rate will be compensated for by the measured gas composition, $Z^2=7$ in the LER at the end of Phase~1 for 1~A of beam current, which is more favorable than the assumed value of $Z^2=49$ in the existing Phase 3 simulation. In the long term, after the completion of vacuum scrubbing, the LER beam-gas rate is therefore expected to be lower than originally predicted. For the HER, the dynamic pressure already meets the simulated Phase~3 target, but unfortunately there are no RGAs installed in the HER that can verify the gas composition in that ring. One might naively guess that the HER gas composition is dominated by similar gases as the LER. It would be satisfying and prudent to install RGAs in the HER in Phase~2, to verify that this bears out. If the HER gas composition is indeed found to be similar to that of the LER, it would imply that the observed HER beam-gas background excess (a factor 108 in Phase 1) will lead to only a factor fifteen increase in the Phase 3 prediction, which is still manageable for this sub-dominant background.

For the pixel detector (PXD), synchrotron radiation and injection backgrounds are also critical. The expectation for these backgrounds in Phase~3 is not constrained by our Phase~1 work, due to the difference in the machine-detector interface (IP chamber and QCS magnets) and the forthcoming damping ring. Measuring these backgrounds is therefore a priority for Phase~2, and we have developed dedicated detectors (e.g. \cite{ahlburg, heuchel}) that will be installed in the vertex detector volume for that purpose. 

Finally, we note that the predicted rate of neutrons in Phase~3 is above or near the acceptable level for a number of Belle II sub-detectors. These neutrons mainly originate from radiative Bhabha events. While the rate of Bhabha events is unlikely to be incorrect in simulation, the simulation of neutron production and propagation is thought to have greater systematic uncertainties. This, coupled with the excess neutron production already observed in Phase~1, makes it critical to measure neutrons from radiative Bhabha events in Phase~2. We will install He-3 tubes and TPC fast neutron detectors outside of the QCS magnets in Phase~2, which will allow us to directly measure this neutron background component. This way we can assess if further mitigation of neutrons will be required. It is possible that the Phase~1 excess of neutrons is due to ``cavern" backgrounds, i.e.\ neutrons originating from the beam line away from the Belle II detector. These neutrons are currently not part of our simulation, but will be added. This component would presumably be observed in the Belle II KLM endcaps in Phase~2.

\label{implications}
 \section{Conclusions}
 We have performed extensive measurements of beam backgrounds during SuperKEKB Phase 1, the first of three commissioning stages. Beam gas, Touschek, beam-dust, and injection backgrounds were individually measured with dedicated detectors and have been discussed extensively in this paper. Synchrotron radiation background was also searched for, but not found, as expected in the Phase 1 configuration. Most beam backgrounds appear safe for Belle II when extrapolating to Phase 3, but the safety factors are small. The TOP MCP-PMT integrated charge is a known problem, and will require replacement of about half of the PMTs at some point. Neutrons from radiative Bhabha events may also turn out to be problematic, and will require more study in Phase 2. Given the small safety factors and strong dependence of background rates on collimators and alignment of magnets, it is prudent to proceed with caution. To protect against unforeseen circumstances, accidents, and mistakes, the diamond sensor beam abort system has already been tested in Phase 1. All remaining background components will be individually measured in Phase 2, and the vertex detectors will not be installed until Phase 3. We hope that this plan will allow us to proceed without accidents and allow us to detect and mitigate any unexpectedly large beam background components before commencing Phase 3.

\label{conclusions}

 \section{Acknowledgements}

We thank the following current and former Belle II collaboration members who provided the Phase 3 beam background estimates for Belle II: Andreas Moll, Martin Ritter, Sebastian Skambraks, Pit Vanhoefer, Peter Kvasnicka, Makoto Uchida, Xionhong He, Dong van Thanh, Tara Nanut, Luka Santelj, Poyuan Chen, Samuel de Jong, Andrea Fodor, Christopher Hearty, Leo Piilonen, Timofey Uglov, Anselm Vossen, Yuri Soloviev, Hiroshi Nakano, and Ushiki Itaru.

%% CLAWS Acknowledgements
We thank Yong Liu from the University of Mainz for radiation-hard scintillator tiles. The MPP group acknowledges support from the DFG Excellence Cluster "Origin and Structure of the Universe" of Germany and by the H2020 project AIDA-2020, GA no.\ 654168.

%% TPC & Mechanics Acknowledgements
We thank Marc Rosen and Kamaluoawaiku Beamer of the University of Hawaii for their assistance in designing and installing the BEAST II mechanical support structure. We thank Tommy Lam and Annam L\^{e} for their help in constructing and testing the TPC fast-neutron detectors. We thank David-Leon Pohl and Jens Janssen from the University of Bonn for contributing the data acquisition software for the ATLAS FE-I4 pixel chips utilized in the TPC fast-neutron detectors. The Hawaii group acknowledges support from the U.S. Department of Energy under Award Numbers DE-SC0007852 and DE-SC0010504.

%% UVictoria ack.
We would like to acknowledge CMC Microsystems for the provision of CAD tools  that facilitated this research.

%% INFN Crystal Acknowledgements
We thank A. Russo of the Laboratori Nazionali di Frascati and G.~Scolieri of INFN Perugia for their technical skills and dedication during the assembly and installation of the Crystal boxes. 

\clearpage
%%%%%%%%%
%  APPENDIX
%%%%%%%%%
%\appendix
%\appendix
%\addcontentsline{toc}{section}{Appendices}
\renewcommand{\thesection}{\Alph{section}}

%%%%%%%%%
%  REFERENCES
%%%%%%%%%

% NOTE: PLEASE ADD REFERENCES BY SECTION / BEAST SYSTEM
%             AS IS CURRENTLY DONE BELOW

\section*{References}
\bibliographystyle{elsarticle-num}
\bibliography{%
references_detectors,%
references_findings,%
references_injection,%
references_intro,%
references_xrm,%
references_neutrons,%
references_simulation,%
references_vacuum_scrubbing}
\end{document}